# A room temperature optomechanical squeezer

by

Nancy Aggarwal

Submitted to the Department of Physics
in partial fulfillment of the requirements for the degree of

Doctor of Philosophy

at the

MASSACHUSETTS INSTITUTE OF TECHNOLOGY

February 2019

Author . . . . . . . . . . . . . . . . . . . . . . . . . . . . . . . . . . . . . . . . . . . . . . . . . . . . . . . . . . . . . . . . . .
Department of Physics
January 18, 2019

Certified by . . . . . . . . . . . . . . . . . . . . . . . . . . . . . . . . . . . . . . . . . . . . . . . . . . . . . . . . . . . . . .
Nergis Mavalvala
Curtis and Kathleen Marble Professor of Astrophysics
Thesis Supervisor

Accepted by . . . . . . . . . . . . . . . . . . . . . . . . . . . . . . . . . . . . . . . . . . . . . . . . . . . . . . . . . . . . .
Nergis Mavalvala
Associate Department Head of Physics

**A room temperature optomechanical squeezer**
by
Nancy Aggarwal

Submitted to the Department of Physics
on January 18, 2019, in partial fulfillment of the
requirements for the degree of
Doctor of Philosophy

## Abstract

Decades of advancement in technologies pertaining to interferometric measurements have made it possible for us to make the first ever direct observation of gravitational waves (GWs). These GW emitted from violent events in the distant universe bring us crucial information about the nature of matter and gravity. In order for us to be able to detect GWs from even farther or weaker sources, we must further reduce the noise in our detectors. One of the noise sources that currently limits GW detectors comes from the fundamental nature of measurement itself. When a certain measurement reaches very high precision, the Heisenberg uncertainty principle comes into play. In GW detectors, this uncertainty manifests itself in the quantum nature of the light. Due to its quantum nature, light (or electromagnetic field) has an uncertain amplitude and phase. Since the interferometric measurement is directly measuring the phase of light, this uncertainty poses a limit on the precision of GW measurements. Additionally, this measurement is also subject to quantum back-action, which arises due to the radiation pressure force fluctuations caused by the amplitude uncertainty (QRPN). In order to lower this quantum noise, GW detectors plan to use squeezed light injection. Squeezed light is a special quantum state of light which has lower uncertainty in a certain quadrature, at the expense of higher uncertainty in the orthogonal quadrature.

In this thesis, I focus on using radiation-pressure-mediated optomechanical (OM) interaction to generate squeezed light. Creating squeezed states by using optomechanical interaction opens up possibilities for engineering truly wavelength-independent squeezed light sources that may also be more compact and robust than traditionally used non-linear crystals. Additionally, this project inherently involves studying the OM interaction, which is the mechanism for back-action noise in GW detectors. Our basic setup is a Fabry-Perot cavity with a movable mirror. We start by understanding the physics of this system in the presence of realistic imperfections like losses and classical noise. This study furthers the previous work done on OM squeezing in an ideal Fabry-Perot cavity. We use this understanding of the system to optimize the experimental parameters to obtain the most possible squeezing in a broad audio-frequency band at room temperature. This optimization involves choosing the optical properties of the cavity, and the mechanical properties of the oscillator. We then present the experimental implementation of this design, and subsequent observation of QRPN as well as OM squeezing from the optimized design.

These observations are the first ever direct observation of a room temperature oscillator's motion being overwhelmed by vacuum fluctuations. More so, this is also the first time it has been shown in the low frequency band, which is relevant to GW detectors, but poses its own technical challenges, and hence has not been done before. Being in the back-action dominated regime along with optimized optical properties has also enabled us to observe OM squeezing in this system. That is the first direct observation of quantum noise suppression in a room temperature OM system. It is also the first direct evidence of quantum correlations in a audio frequency band, in a broadband at non-resonant frequencies.

Thesis Supervisor: Nergis Mavalvala
Title: Curtis and Kathleen Marble Professor of Astrophysics

# Acknowledgements

I have had the fortuitous opportunity to be able to work on many interesting problems during my PhD, making my research experience quite enjoyable. But graduate school has also been a long and arduous journey, which would have been even more challenging without the support I had from all the people in my life.

First of all, none of this would be possible without the mentoring provide by my advisor, Nergis. I am grateful to you for taking me under your wing, for believing in me and providing me with the academic freedom that led me to work on a myriad of things. Thanks for letting me continue working on them even when they all looked like numerous loose ends. You have been a constant source of inspiration and aspiration throughout graduate school. You have taught me many life lessons, within and outside of science, which would be a chapter of their own if I were to list them here. You are the only advisor I know, who would encourage their own students to skip a conference session to go on a hike! Thanks for seeing the light at the end of the tunnel even when things got murky and motivating me to keep going. I hope to be able to put all the knowledge and experiences I gained under your guidance to good use.

Thomas, soon after starting graduate school I had become a fan of yours from all the Corbitt et al papers. It was only later that I got to know you as a person. You are a phenomenal physicist and a great person. Thanks for letting me pester you with various questions throughout graduate school and always taking the time to answer them, even if sometimes it involved excruciatingly long emails. Brainstorming physics with you has been a gratifying experience for me, even though I'm sure you're probably fed up of my questions by now. I have truly enjoyed working with you through telecons, conferences, and even more so in your lab at LSU.

As every soul at LIGO MIT knows, Marie, you are exceptional. You are immensely organized and your efficiency baffles me, but most importantly you are just an amazing individual! The lengths to which you are willing to go to help us are unheard of, even in the book of kindness! Thanks for all your support and recurring kind words of encouragement.

I would like to recognize all the wise opto mechanics whom I have had the pleasure of working with, Adam, Bobby, Haixing, and Jon. I have learned from each one of you. Adam and Bobby, thanks a lot for your help in setting up the experiment and numerous engaging conversations some of which would start at quantum physics but end in metaphysics. Haixing, thanks for the stimulating discussions on vegetarianism, the "dichotomy" between being an experimentalist and a theorist, and of course quantum mechanics. Jon, thanks for welcoming me on your experiment, I know it can be daunting to let new people touch your setup. Also many thanks for hosting me in Baton Rouge, your friendship made the work and the city much more exciting.

I also would like to give credits to the brilliant MIT LIGO team. I thank Fred and Myron for all their technical support, without which this PhD would have been at least another year longer. I thank all the lab members for the warm atmosphere and educational group meetings; Slawek, Evan, Aaron, Maggie, Ryan, Alvaro, Georgia, Lisa, Matt, Eric, Tim, Sheila, Paul, Patrick, Tomoki, John, Mike, Lee, Peter, Fabrice, Erik, Rich, Seb, Arnaud, Antonios, Vivishek, Reed, Ben, Nick, and Chris. Thanks especially to Rai, you are a living inspiration for thinking big as well as having fun in the everyday execution of the project. One moment you are thinking about big questions like the beginning of the universe, next moment you are immersed in soldering an electric field meter. I thoroughly enjoyed all the science-history tales and trivia. Thanks for constantly checking on me, for your continuous curiosity about the experiment's progress, and for giving me advice on various matters.

The brunt of my madness was borne by my friends; Lina, Gabriel, Alex L., Rohan, Evan, John H, Jenny, Alex J, Cody, Tom M, Tom C, Elise, Sabrina, Liang, John M, Brenda. You guys provided me the whole package – studying for quals, board-games, proofreading my painful writing, grad school rants, and physics, politics and philosophy discussions. Thanks for not letting me graduate without a certain amount of familiarity with American pop culture. Thanks guys for being my pillars and letting me lean on you. Especially my beloved addiction-enablers, Lina for Flour, Slawek for coffee, and Evan for procrastination, I would have gone insane without you guys! Thanks also to Sabrina and Liang for the 019 powwows and instilling the painting fever in me. I would be remiss to not mention GWIP, Anna Frebel, Wolfgang Ketterle, Roop Goyal, Edgerton House, and the MIT thesis group for being an important part of my wholesome support system here at MIT.

Finally, I write this as I suffer from terrible lack of words that fully encompass my gratitude towards my family. But I try nonetheless. Thanks Fancy for being the caring person you are, especially your habit of making sure we get everything that I could ever need on my India trips. Your energy and enthusiasm surpass everyone I know. Penzy and Sonu, loving you guys teaches me that love is irrational. Jokes aside, thanks for filling that Venn-diagram area at the intersection of pals and family for me. My heart is heavy as I think of all the sacrifices my parents have made in order to give us a comfortable life. Mummy, thanks for standing by me throughout this whole journey, accepting me even with my peculiar ideals, and believing in me. Your hard work, generosity, and sacrifice is with me every day, and its reminder makes me constantly strive to be a better person. This thesis is as much yours as it is mine.

# Contents

# List of Figures

# List of Tables

# Chapter 1

# Introduction

## Contents

Recent advances in technology have enabled unprecedented precision in measurements of physical quantities. Precision measurements are the backbone of fundamental science, e.g. gravitational wave detection [1, 2, 3, 4], force sensing [5], and optical clocks [6, 7, 8]. When the uncertainty on a given measurement is eliminated from all classical effects, fundamental quantum mechanical uncertainty must be encountered. The minimum amount of uncertainty on a measured variable is bounded by the Heisenberg uncertainty relation, which requires that the product of uncertainties on measurements of two conjugate variables has a non-zero minimum. For standard measurements, the uncertainty is equally distributed in the two normalized orthogonal observables [9, 10, 11]. In order to improve precision measurement beyond this standard limit, we must employ exotic quantum techniques.

In this thesis, we focus on designing and building one such quantum device; an optomechanical (OM) system to generate squeezed light. The precision measurement application we will be focusing on is gravitational wave (GW) detection. Ground-based GW detectors [1, 2, 3] are based on a Michelson interferometer (Fig. 1-1a) and look for GWs that range from 10 Hz to a few kHz in frequency (Fig. 1-1b). The astrophysical sources that are most likely to produce GWs in this frequency range and sensitivity are compact binary mergers, at distances up to hundreds of megaparsecs. In recent years ground-based GW detectors have directly recorded gravitational waves from several such compact binary mergers [12, 13, 14, 15, 16], revealing numerous facts about our universe. The first direct observation is shown in Figs. 1-1c and 1-1d. These waves were generated by the merger of a pair of black-holes, each about as massive as thirty suns, hundreds of megaparsecs away. The observation of gravitational-waves not only confirms Einstein's general theory of relativity [13], but also provides direct evidence of existence of black-hole pairs and of intermediate mass black-holes. The observation of the first gravitational-waves from a binary neutron star merger [16] has provided similar verification of general relativity, as well as insights to interiors of stars and kilo-novae.

The current generation ground-based GW detectors are limited by the aforementioned quantum imprecision, usually referred to as quantum noise. In a laser interferometer,this imprecision can be understood as a result of uncertainty in the electromagnetic field (i.e. light). The most classical state of laser light is called a coherent state [9], which is an eigenstate of the annihilation operator. This state is a minimum uncertainty state, i.e. it equalizes the Heisenberg uncertainty relation. Consequently, it also has same uncertainty in all quadratures. In a ball-and-stick representation [9], the probability distribution is circularly symmetric, as depicted in Fig. 1-2a. Here we have represented the EM field as a phasor. The size of the arrow can be interpreted as related to the field's magnitude. Its angle from the x-axis can be interpreted as related to the phase of the sinusoidal wave. For a coherent state, the location of

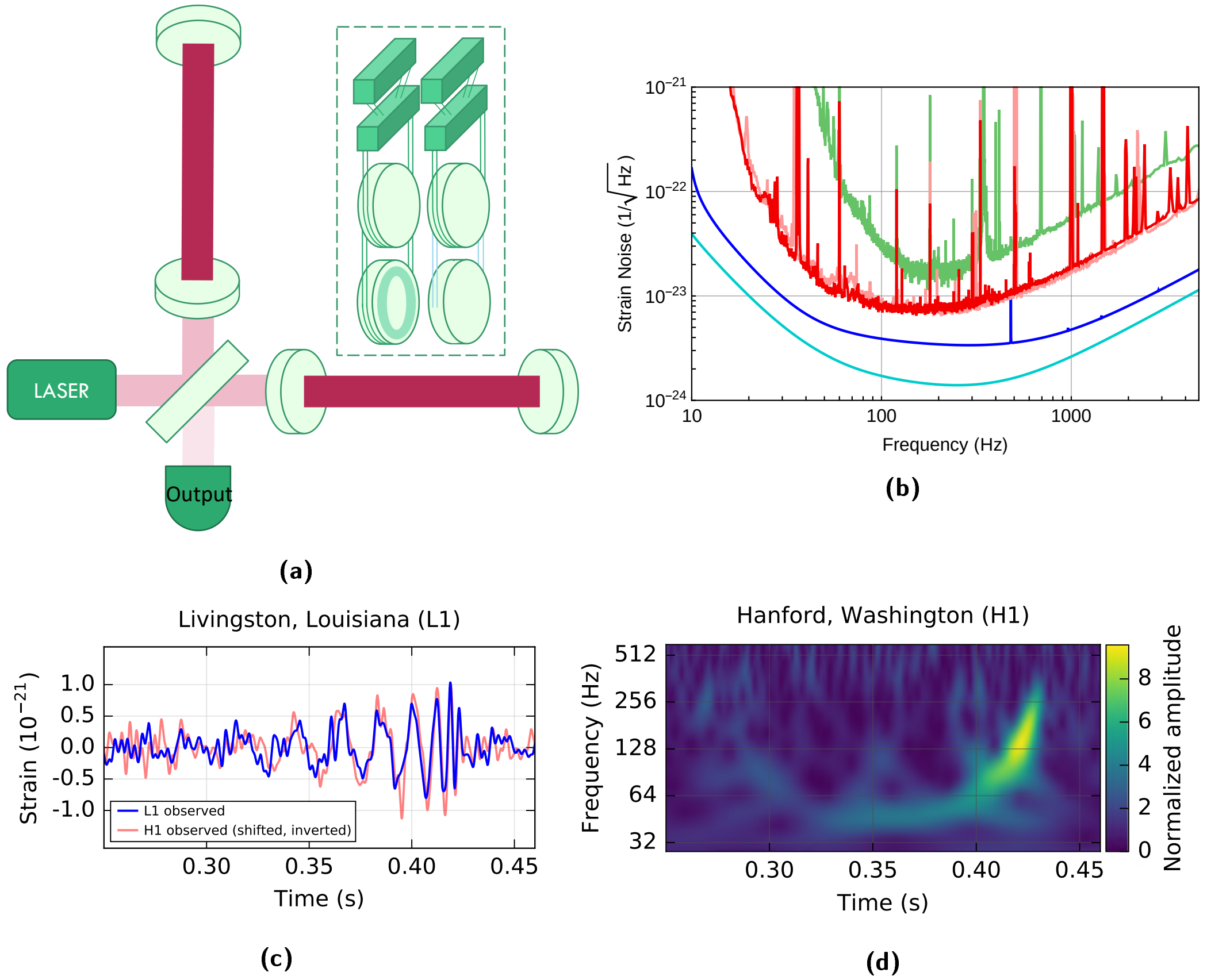

**Figure 1-1:** Ground-based GW detectors: **(a)** A basic schematic of a Michelson interferometer based GW detector. The light from a laser is split into two components on a beam splitter and sent to two several kilometer long arms. The light that comes back from these arms interferes on the beam splitter and is measured on the output port. If the arms are of exactly the same length, the output port has complete destructive interference. When a GW passes by, it changes the lengths of the two arms relative to each other, changing the interference condition, which leads to a measurable change in intensity on the output. Each of the four test masses are mechanically suspended as shown in the inset. **(b)** Baseline noise of ground based detectors (credit `https://doi.org/10.1103/PhysRevLett.116.131103`). The red curves are the sensitivity of the two Advanced LIGO detectors during their first observing run in 2015. The dark blue is the design sensitivity of Advanced LIGO, and the cyan curve corresponds to a planned upgrade to Advanced LIGO. **(c)** Strain as a function of time as observed by the Advanced LIGO detectors on September 14, 2015 from gravitational waves from a binary black-hole merger. **(d)** Spectrogram of the same signal at the Hanford detector. Credits `https://doi.org/10.1103/PhysRevLett.116.061102`.

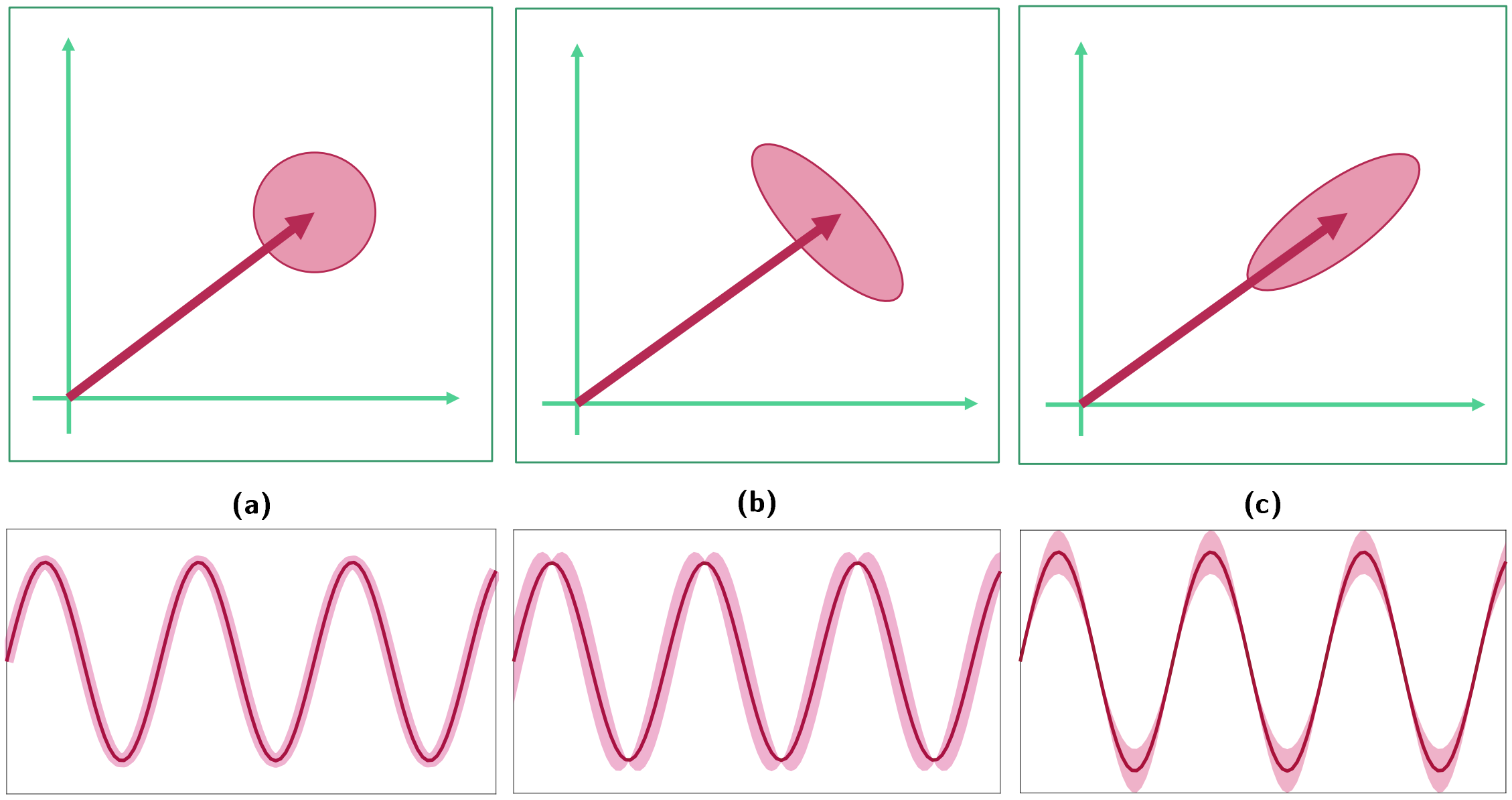

**(a)** **(b)** **(c)** **(d)** **(e)** **(f)**

**Figure 1-2:** A pictorial representation of the quantum uncertainty on light. We represent the electromagnetic field as a phasor, such that its amplitude is the size of the phasor, and phase is the angle of the phasor. The standard deviation from quantum mechanical uncertainty is represented as the shaded region. **(a)** and **(d)** show the uncertainty on a coherent state, which is the most classical quantum state. The area of the uncertainty region for coherent states is minimum, given by the Heisenberg uncertainty relation, and it has equal uncertainty in all quadratures. **(b)** and **(e)** show an amplitude squeezed state. The uncertainty in amplitude is lowered, while the phase becomes more uncertain. **(c)** and (f) show a phase squeezed state instead.

this phasor follows a Poisson distribution, with equal uncertainty in all directions. This uncertainty in the observables of light is what leads to quantum noise in GW detectors, as shown in the illustration in Fig. 1-3a. In nominal operation, the state measured by the output photodetector is a coherent state. The uncertainty on this coherent state can be fully attributed to the vacuum entering the dark port of the interferometer [17, 18]. The GW signal is measured by measuring the intensity of the output light, which will have Poissonian noise on it. The predicted noise for the advanced LIGO detectors from this effect is shown in Fig. 1-1b in dark blue.

One way to lower the quantum noise is to redistribute the uncertainty, by using squeezed light instead of coherent states [18, 19]. Squeezed states still satisfy the Heisenberg uncertainty relation, but have lower uncertainty in the quadrature of interest, at the expense of higher uncertainty in the orthogonal quadrature. For example, an amplitude squeezed state has minimum uncertainty in amplitude, but higher in phase, as shown in Fig. 1-2b. Similarly, a phase squeezed state is shown in Fig. 1-2c. GW detectors are limited by quantum noise in the phase quadrature at frequencies higher than the arm cavity linewidth; we usually refer to this noise as just shot noise. This can be lowered by injecting a phase squeezed vacuum at the dark port, leading to the final output being amplitude squeezed, as shown in Fig. 1-3b.

At frequencies lower than the arm-cavity linewidth, though, the amplitude quadrature uncertainty limits their sensitivity. This is because the interferometer mirrors are mechanically suspended and and can move due to radiation pressure from the light. The amplitude uncertainty leads to an uncertainty in this radiation pressure force, which is transduced into a distance and phase uncertainty at frequencies below the arm cavity linewidth, leading to quantum noise in the GW measurement. We refer to this noise as quantum radiation pressure noise (QRPN) or radiation pressure shot noise (RPSN). In order to lower QRPN, amplitude squeezed light must be injected at low frequencies. The cyan trace in Fig. 1-1b shows the noise of the advanced LIGO detector's design with injection of such frequency-dependent squeezing. The right frequency dependence of the squeezing angle can be obtained by first generating frequency independent squeezed light, and then filtering it through a filter cavity. The goal of this work is to utilize

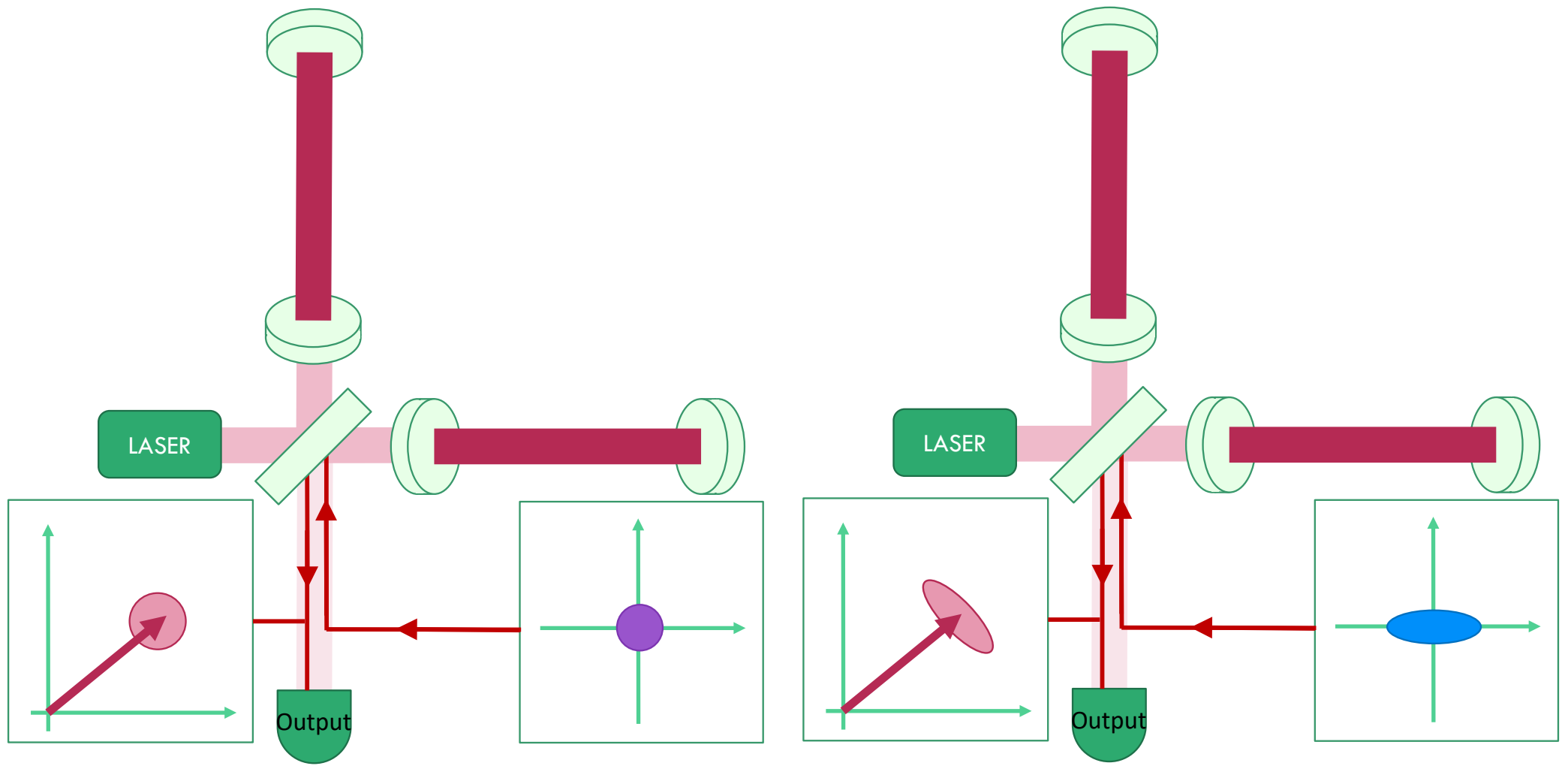

**(a)** **(b)**

**Figure 1-3:** Quantum noise in a Michelson interferometer: **(a)** shows the nominal operation, where no special quantum tricks have been used. The output state is then a coherent state whose uncertainty is seeded by the vacuum entering the dark port. The expected noise performance of advanced LIGO under this nominal operation is shown in dark blue curve in Fig. 1-1b Instead, if we replace the dark port vacuum by a squeezed state as shown in **(b)**, the output state will be squeezed, improving the precision with with which we can measure gravitational waves. The expected noise performance for advanced LIGO design with frequency-dependent squeezed light injection is shown in light blue in Fig. 1-1b.

the aforementioned OM interaction that leads to QRPN as a resource to generate frequency-independent squeezed light.

## 1.1 What is optomechanical squeezing?

OM interaction refers to interaction between the state of light and that of a mechanical oscillator, mediated via radiation pressure. The simplest OM squeezer is a Fabry-Perot cavity with one movable mirror as shown in Fig. 1-4. The radiation pressure due to the photon number uncertainty (amplitude quadrature, $\Delta N$) creates an uncertain force $\Delta F$, which in turn leads to a change in cavity length $\Delta L$, which in is related to the phase of the light $\Delta\phi$. This interaction, mediated by the radiation pressure force, and seeded by the quantum fluctuations on the light, leads to correlations between the amplitude and phase quadrature of light, which in turn results in squeezed light. For a full mathematical description, see Refs. [19, 20, 21] and Appendix C.

## 1.2 Motivation for OM squeezing

The current generation GW detectors plan on using non-linear crystals to generate squeezed light [22, 23, 24, 25, 26]. For reasons of thermal noise and detection efficiency, future generation GW detectors may use a different wavelength laser [27, 28, 29] . While crystal squeezer technology has been perfected at 1064 nm, OM squeezers are inherently wavelength-independent. Since photons of all wavelengths exert radiation pressure, there is no need to search for an optical material that has a transparency window as well as adequate non-linearity at the chosen wavelength. Moreover, there is no need to know the wavelength a-priori. As long as there are good lasers, mirror coatings, and detectors (all of which are already required to build an interferometer), one can make a good OM squeezer. Secondly, there is no fundamental limitation on how much squeezing can be generated in an OM squeezer. There is no concept of threshold or maximum non-linear gain. Finally, OM squeezers can be made to be fairly compact, as can be seen in Chapter 5, thus could be made be more robust against physical environment noises.

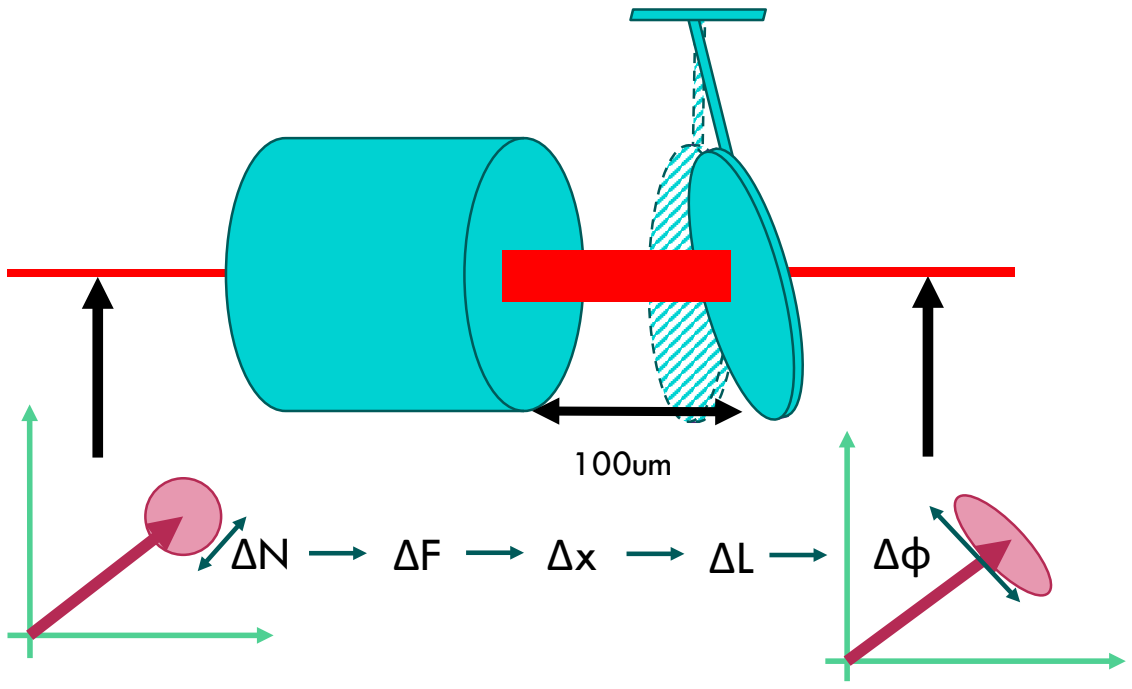

**Figure 1-4:** A basic OM squeezer: A Fabry-Perot cavity whose one mirror is mechanically fixed (left), and the other mirror is a mechanical oscillator (right). The photon number fluctuations in the electromagnetic field lead to a fluctuating radiation-pressure force on the oscillator mirror. This force is then transduced into displacement of the mirror, or cavity length by the mirror's susceptibility. A change in length constitutes a change in the phase acquired by the light. This interaction between light and the mechanical oscillator (OM interaction) adds correlations between number and phase of the light, which in turn leads to squeezing of the uncertainty on the light.

## 1.3 Main hurdles

Since OM squeezing is a result of the correlations between the two quadratures of light, anything that lowers these correlations destroys squeezing. A major hurdle in all experiments is classical noise. Classical noise can enter at various ports of the sequence shown in Fig. 1-4. This classical noise creates parallel paths of classical correlations, which lower the relative contribution from the quantum correlations, hence destroying squeezing. For example, any classical displacement noise will create phase fluctuations that are due to the classical noise and not due to the quantum fluctuations in amplitude. Examples of such noise are thermal noise, seismic noise, acoustic noise, and actuator/feedback-injected noise. Similarly, there can be classical contributions to $\Delta N$ itself, either coming directly from the laser as relative intensity noise, or due to other technical noise sources throughout the experiment. This part of $\Delta N$ also goes through the same transduction chain and creates $\Delta\phi$ that is classical, hence lowering squeezing. Historically, the two most challenging hurdles to quantum OM systems are thermal noise and laser noise. The challenges of laser noise and physical environment noise have driven previous efforts [30, 31, 32, 33] towards measurements at high frequencies, where lasers are usually shot noise limited and seismic and acoustic motion is negligible (MHz to GHz). Thermal occupation has motivated operation at cryogenic temperatures in addition to higher mechanical resonance frequencies, which also help lower thermal occupation at a given temperature.

## 1.4 Previous work

Due to the above stated reasons, previously, OM squeezing has been observed [30, 31, 32, 33] at higher frequencies, mostly in cryogenic systems. Additionally, for the same noise reasons, these systems operate close to the mechanical resonance and show resonant OM squeezing. OM squeezing near the resonance has a highly frequency-dependent squeezing amount and quadrature. As we will see throughout this thesis, here we focus on creating a frequency-independent squeezed light source, so we must operate far from the OM resonance, which is key to obtaining frequency-independent squeezing. Most of the above systems have a mechanical spring frequency ($\Omega_m$) higher than the optical spring (OS) frequency ($\Omega_{\rm OS}$), leading to the OM spring ($\Omega_{\rm OM} = \sqrt{\Omega_m^2 + \Omega_{\rm OS}^2}$) being close to the mechanical spring. In contrast, we use a system with a mechanical resonance much smaller than the OS, which leads to the OM spring being dominated by the OS (see Ref. [20], Chapter 2).

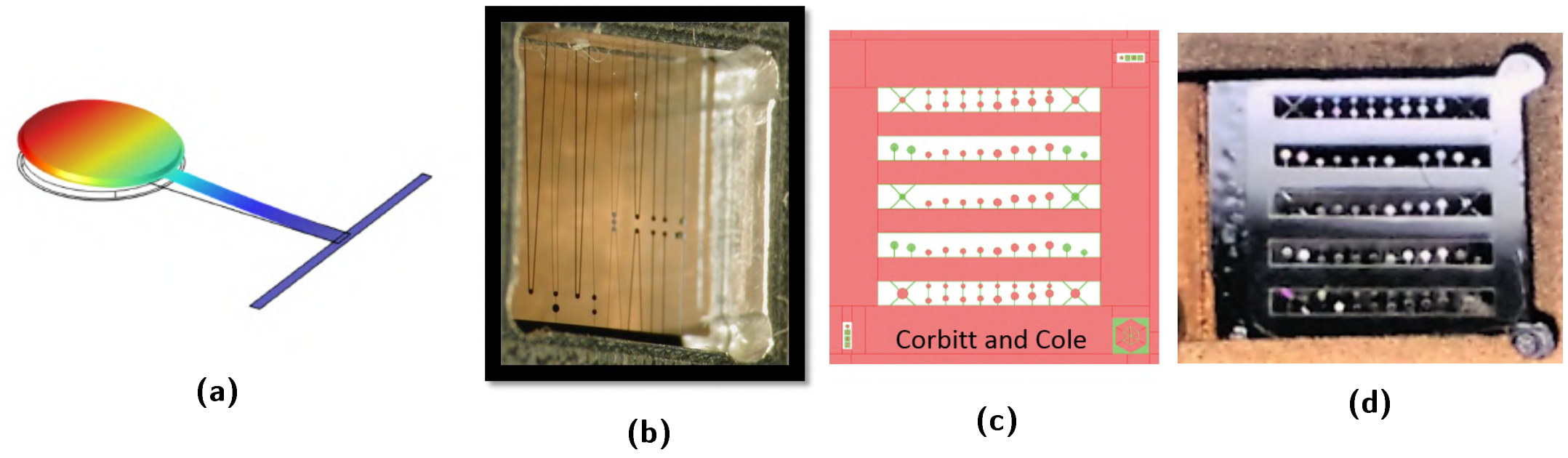

**Figure 1-5:** **(a)** Illustration of the mechanical oscillator in this experiment. The coloring represents the displacement due to the fundamental mode of the cantilever. The fundamental and other mechanical modes relevant in this experiment are due to the elasticity of the material, not to be confused with a pendulum mode due to gravity. The red is region of maximum displacement, and blue is minimum displacement. **(b)** devices in the first generation chip. The tethers were a few millimeters long, while the mirrors were hundreds of microns. Both the cantilevers and the mirrors had the same number of layers (few microns thick). **(c)**, **(d)** Second generation chip design and implementation: This time the cantilevers were made much shorter (hundreds of microns), and were also made much thinner (hundreds of nanometers). The mirror stack was a few microns to obtain the high reflectivity. This change in the cantilever geometry lowered the thermal noise, as shown in Fig. 1-6b.

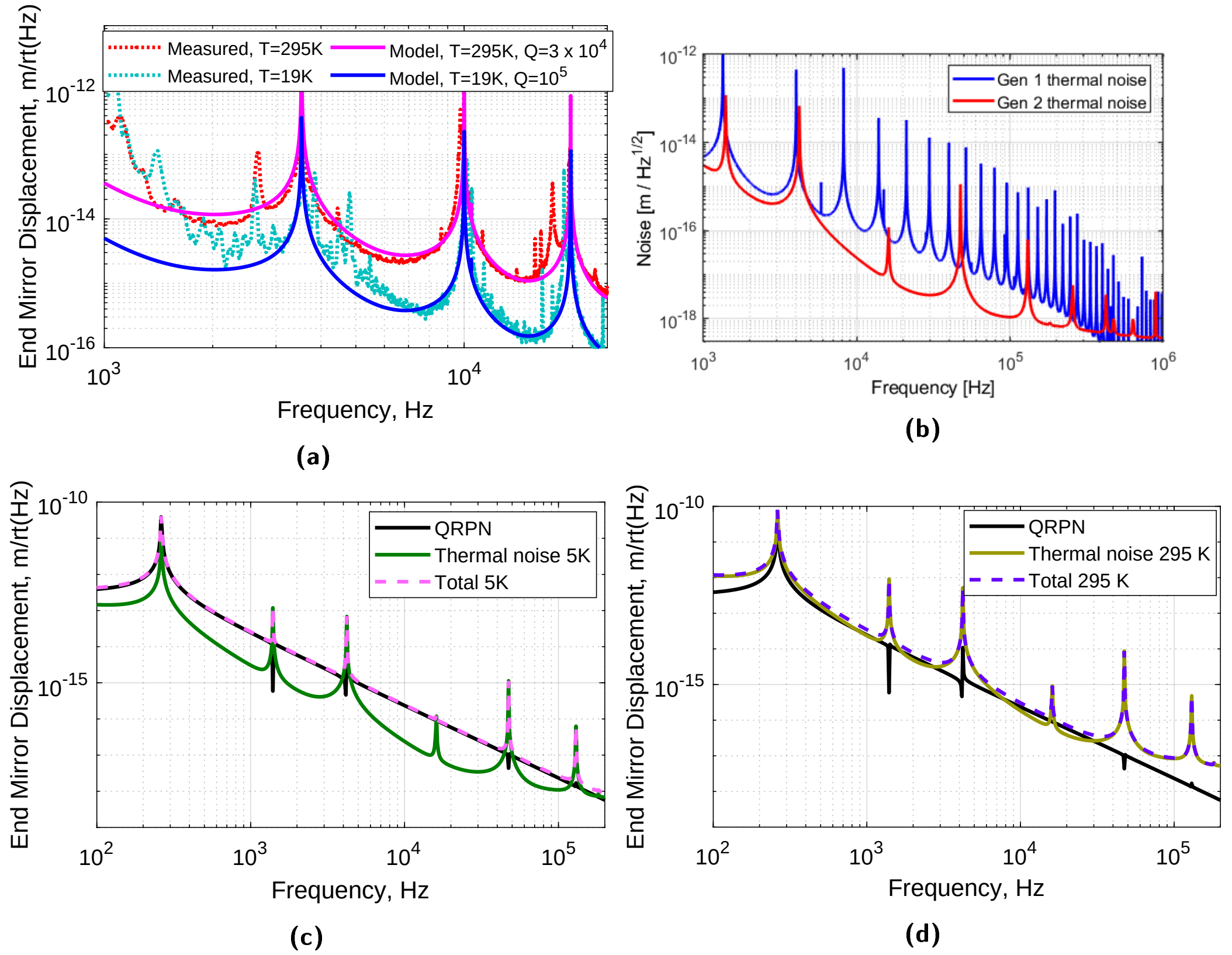

**Figure 1-6:** Evolution of thermal noise through the generations: **(a)** Measurement of thermal noise for first generation devices (credit: Thomas Corbitt, Shannon Sankar, Garrett Cole, Nergis Mavalvala [34]) **(b)** With the second generation design, the lowered cantilever length and thickness reduced the number of higher order modes in the band and hence showed promise for much lower thermal noise (credit Thomas Corbitt). **(c)** Comparison of QRPN and thermal noise at 5 K for the second generation design. **(d)** Comparison of QRPN and thermal noise at room temperature for the second generation design.

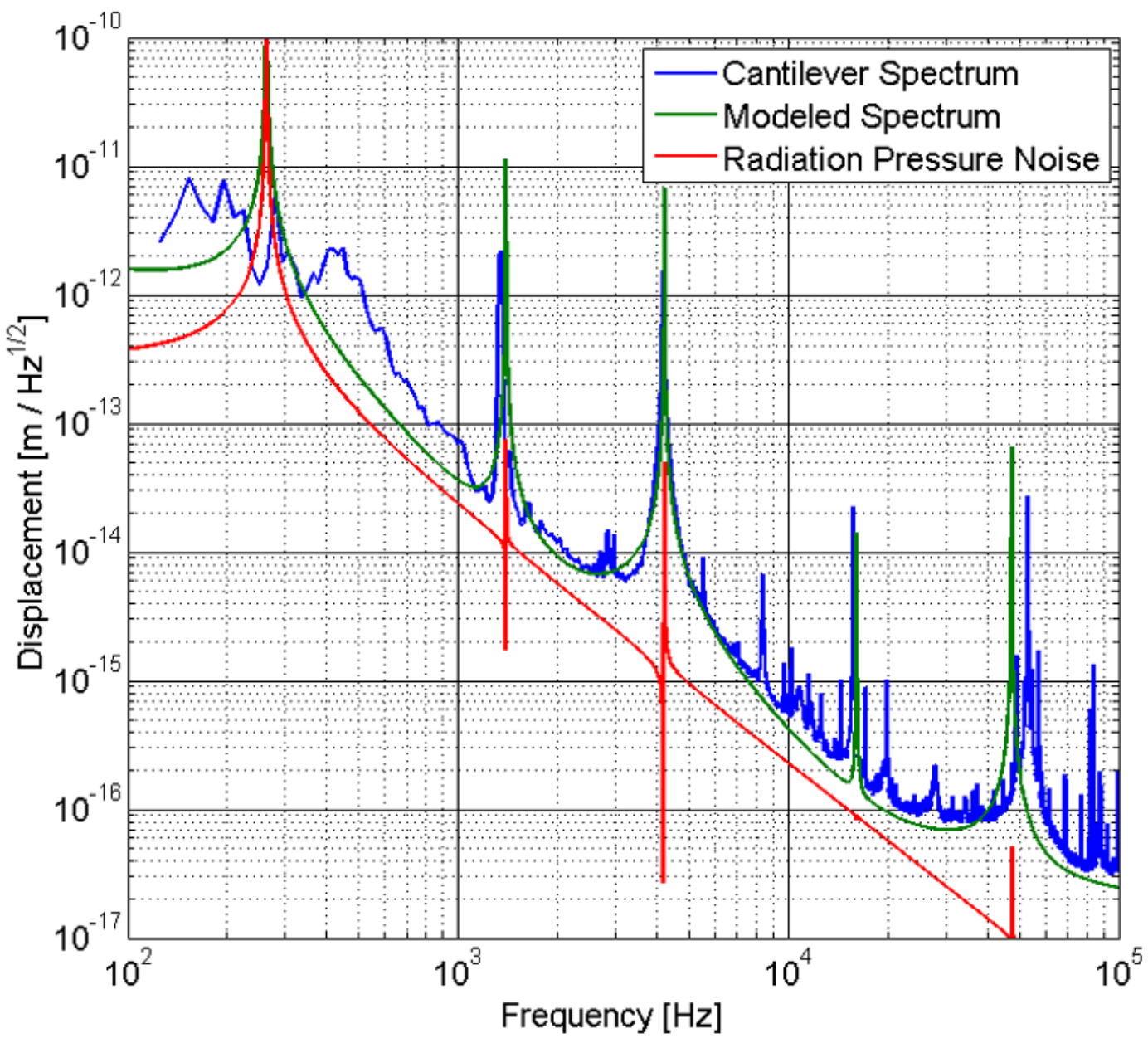

**Figure 1-7:** Room temperature thermal noise measurement and updated model for second generation design.

In this limit, the off-resonant QRPN must be significant compared to the off-resonant thermal noise. This has historically been the toughest roadblock in this experiment. The mechanical resonator in this experiment has gone through multiple iterations of improvements to attempt to achieve this. Through all the generations, the basic geometry has stayed somewhat similar, and is shown in Fig. 1-5a. In each case, there was a high-reflectivity "mirrorpad" supported by a cantilever "tether". The first generation devices are shown in Fig. 1-5b, and the thermal noise measurement from one of them is shown in Fig. 1-6a. The tethers on these devices were millimeters in length, and as thick as the mirror pads, which is several microns. We see that the room temperature thermal noise matches the model pretty well. At cryogenic temperatures, the models predicted QRPN to be higher than thermal noise at 4 to 6 kHz, but the measurement had extraneous technical noise which masked thermal noise as well as QRPN at those frequencies.

The next generation devices were then designed to aim for a broader band of QRPN, which could be achieved by lowering the number of higher order mechanical modes in the measurement band. This design is shown in Fig. 1-5c, and the expected thermal noise compared to first generation is shown in Fig. 1-6b. As we see, the second generation had many fewer mechanical modes, which was achieved by making the cantilevers much thinner and shorter (shown in green in Fig. 1-5c). This iteration showed promise of having higher QRPN than thermal noise at cryogenic temperatures, as seen in Fig. 1-6c. In fact, this design also predicted thermal noise lower than QRPN even at room temperature, as shown in Fig. 1-6d. The fabricated chip from this design is shown in Fig. 1-5d, and a measurement of its thermal noise at room temperature is shown in Fig. 1-7. This thermal noise measurement showed higher thermal noise than in the design-stage model. In order to understand the measurement, we then developed a more sophisticated model that included thermoelastic damping from the mirrorpad modes in addition to the structural damping from all the modes. This model explained the measured thermal noise, as can be seen in Fig. 1-7, and also revealed that at room temperature the thermal noise exceeds expected QRPN. If we had cooled this device to 5 K, QRPN would be higher than thermal noise. Nevertheless, the amount of squeezing expected from this generation was minimal even upon cooling, as shown in Fig. 1-8. We later learn that this is due to the mirror reflectivities of the Fabry-Perot. Hence, there was also a need to optimize the optical design of the Fabry-Perot cavity. Since that meant manufacturing new OM devices, we also take the lessons learnt from the thermal noise measurement in Fig. 1-7 to further improve the mechanical design.

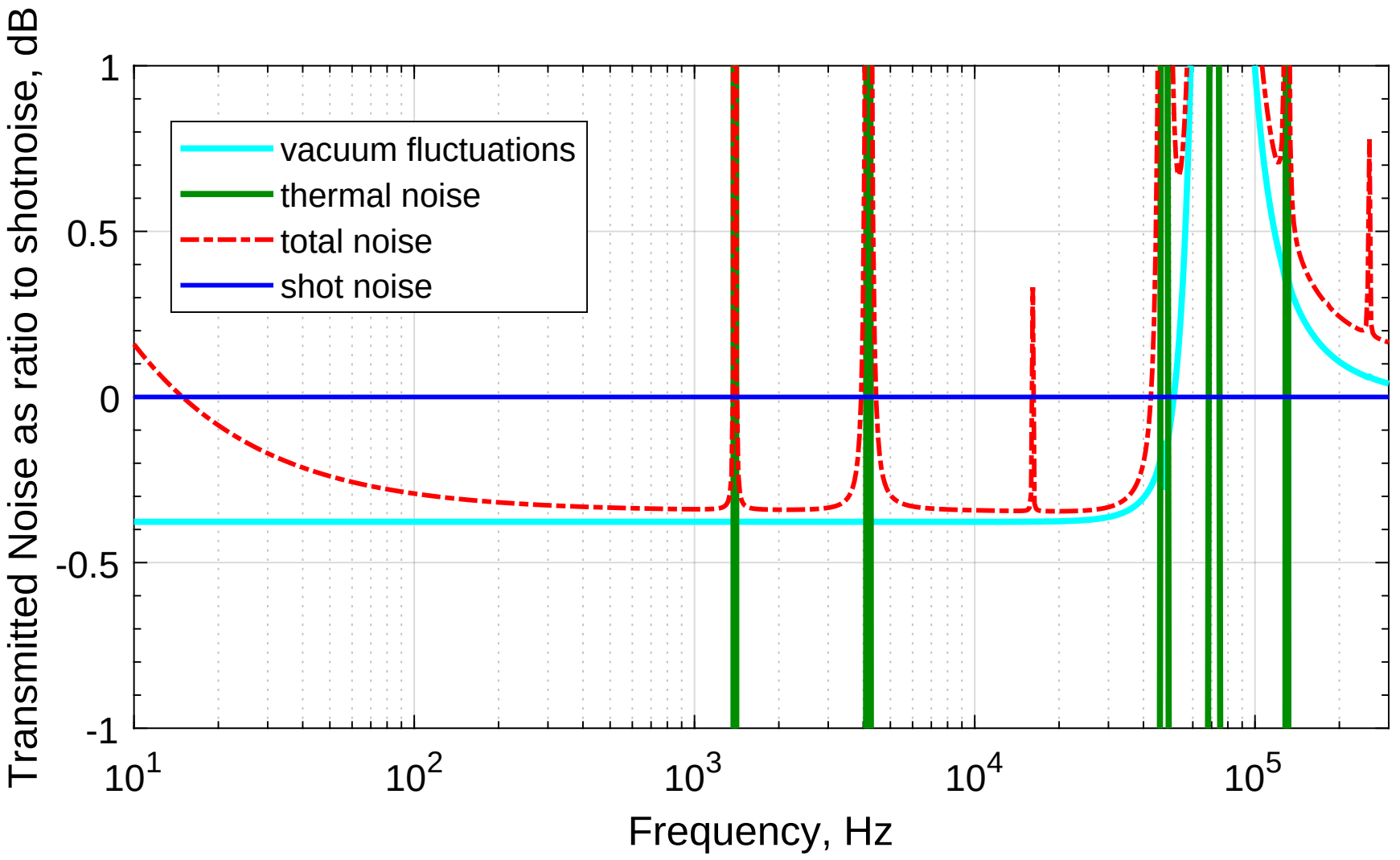

**Figure 1-8:** Expected squeezing from second generation device at 5 K. We see that in this high finesse cavity, even after cooling the chip, this device was only predicted to show less than 0.5 dB squeezing. This called for a need to rethink the experiment design.

## 1.5 Summary of this work

This thesis is a journey from Figs. 1-7 and 1-8 (Figs. 1-5c and 1-5d) to Fig. 1-9 (Figs. 1-11a and 1-11b). Fig. 1-9 shows the expected squeezing from one of the designs from this work, and the full chip is shown in Fig. 1-11. We provide the physics behind designing the system, combined with first observations from one of the many designed devices. To achieve this goal of building an OM squeezer that generates broadband, audio-frequency squeezing at room temperature, we have developed low thermal noise devices, whose displacement at room temperature can have a significant contribution from QRPN at the frequencies of interest. In our experiment, this is aided by the thermal noise being predominantly from structural damping. The force density of Brownian fluctuations for a structurally damped system is inversely proportional to frequency, while that of QRPN is white. We leverage this feature in our design to achieve a frequency band where motion due to quantum fluctuations is larger than that due to Brownian fluctuations.

We start in Chapter 2 with laying down the formalism that defines the fields as we will be using them, and provides a mathematical description of the fields interacting with various optical elements. Next is the optimization of all the experimental parameters for the OM system. This includes optimizing the optical parameters of the Fabry-Perot cavity, as well as the mechanical design of the oscillator. This optimization is performed by looking at the properties of the squeezed field as a function of the mechanical and optical properties of the system (Fig. 1-10). We address these two optimizations independently; first we go through the optical optimization in Chapter 3, and next we look at the mechanical optimization in Chapter 4. We approach the optimization of optical parameters both analytically as well as numerically. The analytical calculation is done in Wolfram Mathematica, and is attached in Appendix C with results summarized in Section 3.5, while the numerical calculation is done in MATLAB and the results are summarized in Section 3.6[1]. From both these approaches, we learn about how each optical parameter contributes to the quantum behavior of the system. For example, contrary to intuition, it is not good enough to maximize the cavity finesse. Instead, one must also think about the escape efficiency. We also learn that while increasing the escape efficiency might increase the maximum squeezing, it lowers the bandwidth of squeezing due to a lower finesse. So, a balance must be sought.

We understand how each classical noise couples to the squeezed field, and its relation to the coupling to the displacement spectrum. For example, we find that the coupling of thermal noise to the amount

[1]The packages can be obtained by contacting me.

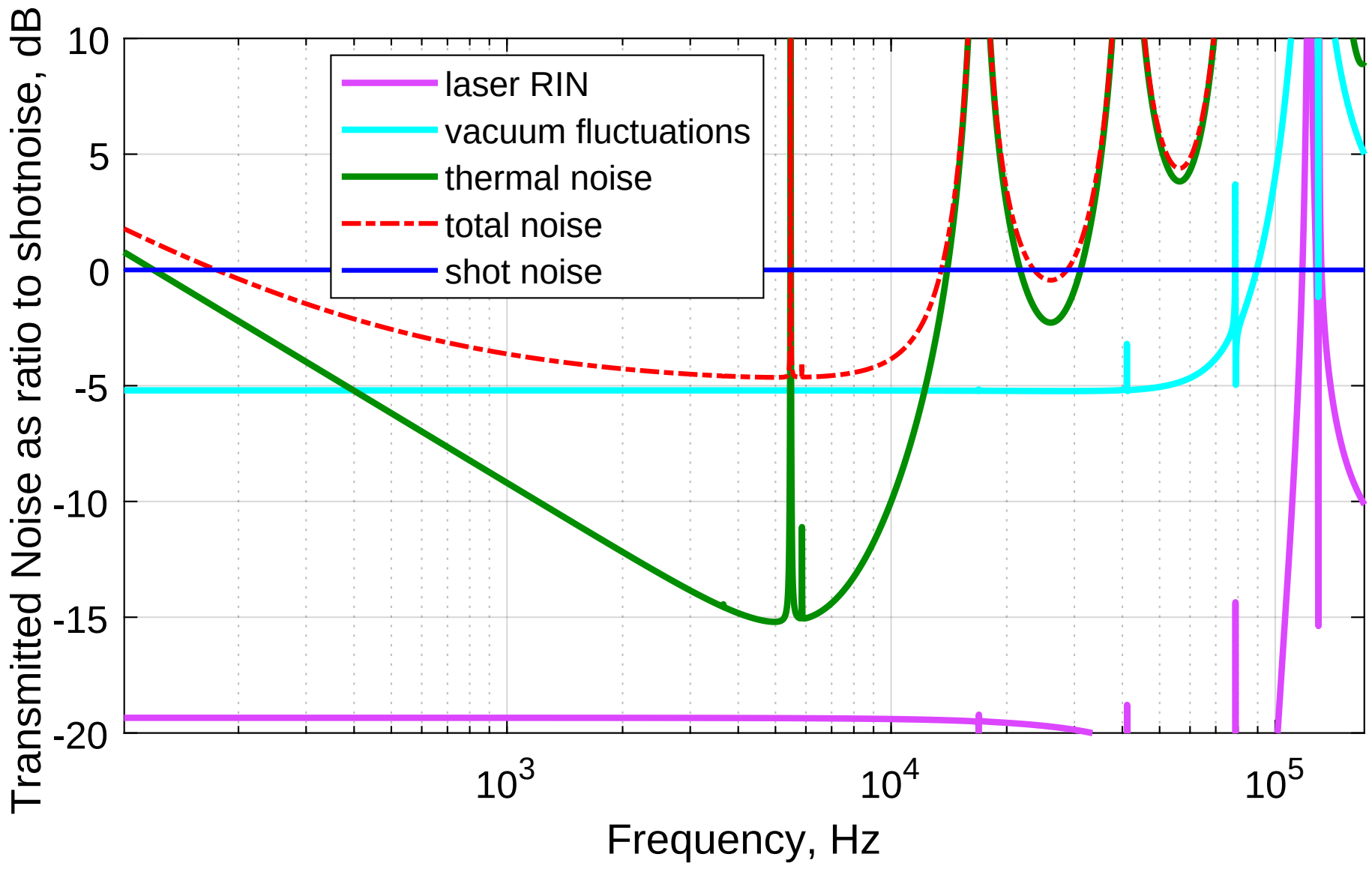

**Figure 1-9:** Expected squeezing after optimizing the design: As a result of the optimization study, we are able to design devices that could produce up to 5 dB of squeezing at room temperature with their full thermal noise including the extra loss from thermoelastic damping.

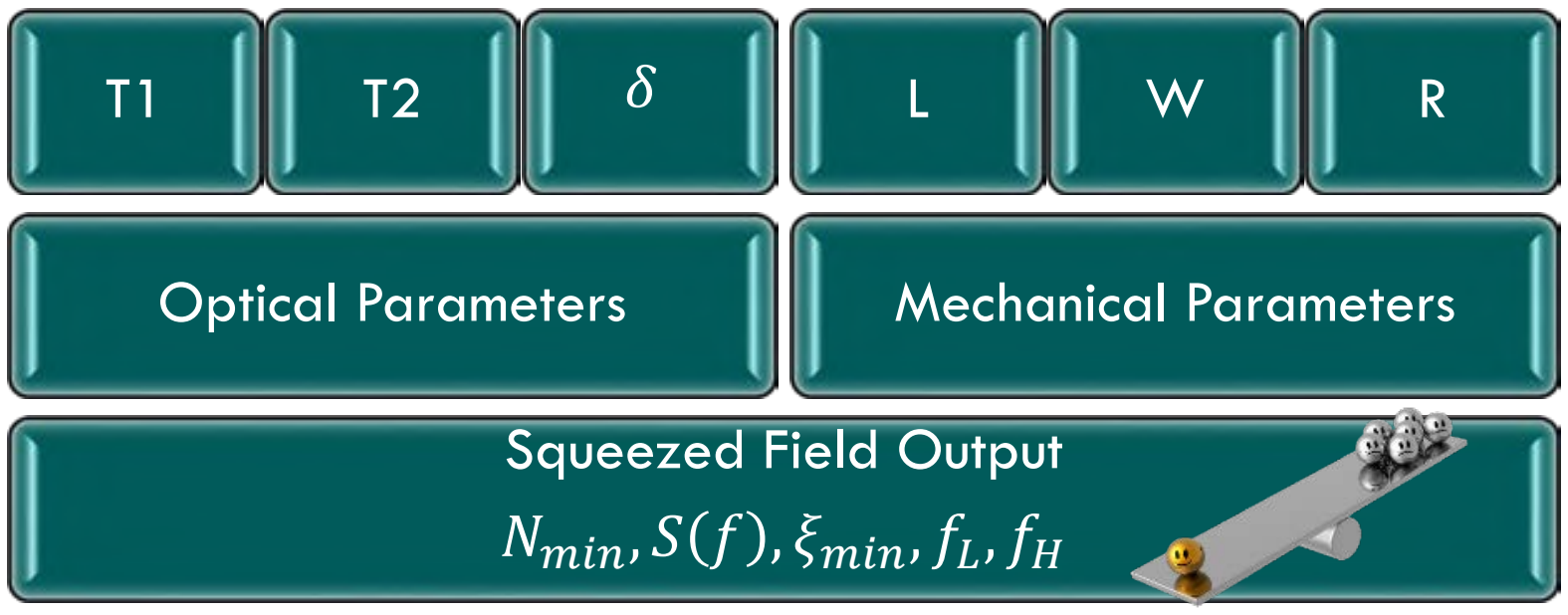

**Figure 1-10:** Input and output parameters for the optimization study: To optimize the squeezer, we look at the squeezed field as a function of the optical and mechanical properties of the system. For this first iteration of optimization, we do this in two parts, first optimizing the optical properties for a fixed mechanical design, and then optimizing the mechanical properties for a fixed optical design.

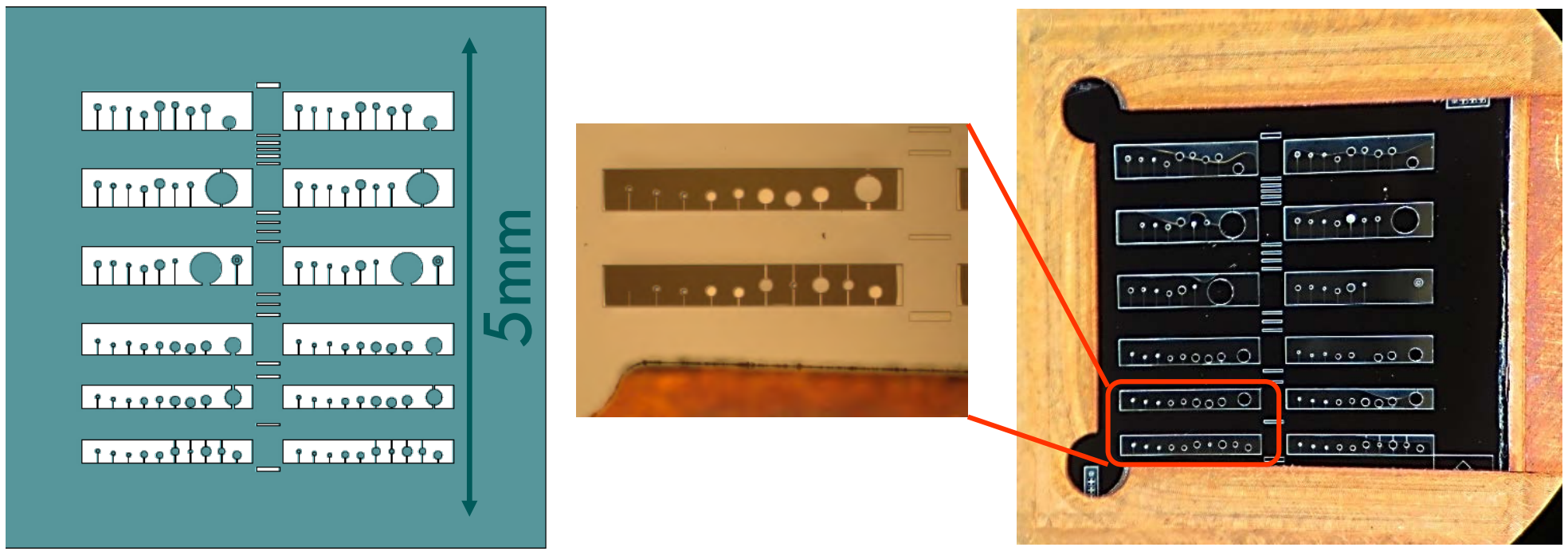

**(a)** **(b)**

**Figure 1-11:** Design and implementation of third generation devices: devices based on basic principles from the second generation, but this time each dimension has been fine-tuned and optimized for best performance.

of measurable squeezing can be parametrized in a dimensionless parameter $\lambda_{\mathrm{th}}$, which is proportional to the power spectral density of the force due to thermal fluctuations, and inversely proportional to the intra-cavity power and finesse:

$$\lambda_{\mathrm{th}} = \frac{S_{\mathrm{F,th}} c \mathcal{T} \lambda_0}{16\pi\hbar P_{\mathrm{cav}}},$$

where $\mathcal{T}$ is the total round-trip loss of the cavity, c is the speed of light, $\lambda_0$ is the optical wavelength, $\hbar$ is reduced Planck's constant, and $P_{\mathrm{cav}}$ is the intracavity optical power. In Chapter 3, we derive this parameter and show how $\lambda_{\mathrm{th}}$, cavity detuning, and escape efficiency each play a role in the amount of squeezing. Here, $S_{\mathrm{F,th}}$ is the power spectral density of force due to thermal noise, which for a single mode oscillator of resonance frequency $\Omega_m$, mass $m$ at temperature $T$, structurally damped due to internal loss factor $\phi(\Omega)$ can be written as [35]:

$$S_{\mathrm{F,th,str,SM}}(\Omega) = 4k_{\mathrm{B}}T\frac{m\ \Omega_m^2\ \phi(\Omega)}{\Omega}.$$

This force spectral density for thermal noise is more complex to write down for a multi-mode oscillator, so we resort to numerical simulations for that. Nonetheless, this allows us to write an effective quantity that governs the relative contribution of thermal motion of a single mode oscillator to QRPN inside the cavity:

$$\lambda_{\mathrm{th,str,SM}} = 4k_{\mathrm{B}}T\frac{m\ \Omega_m^2\ \phi(\Omega)}{\Omega}\frac{c\mathcal{T}\lambda_0}{16\pi\hbar P_{\mathrm{cav}}}. \tag{1.1}$$

This parameter is different from the standard cavity optomechanics treatment because it pertains to the comparison of thermal motion to QRPN at off-resonant frequencies, unlike the standard treatment in cavity optomechanics, which are done for resonant systems [21]. While this parameter is useful to understand whether an OM system is dominated by QRPN, the amount of squeezing generated depends also on the cavity detuning and the escape efficiency, as shown in Section 3.5. This is because an OM system will still generate squeezing even if the classical noise is higher than QRPN.

The full off-resonant and multi-mode response as well as thermal noise of the system are studied in detail in Chapter 4, where we show how each of the dimensions of the mechanical oscillator impacts squeezing. There are again trade-offs to be made. For example, a higher mass gives a low fundamental frequency and hence low thermal noise, but a lower mass gives more QRPN. Since our mechanical oscillator is a multi-mode device, we have to take into account the higher-order mechanical modes. These higher-order mechanical modes contribute to a complex susceptibility to external forces like radiation pressure force, and also contribute extra Brownian thermal noise. For ease of mathematical handling, we assign each mode an effective mass that signifies its contribution to the dynamics. Lighter modes are easily excitable, so contribute more displacement per unit force, and have higher thermal noise. This effective mass is derived for common forces like radiation pressure and thermal fluctuations in Appendix A.

Next, we show the experimental implementation of the above design in Chapter 5. Finally, in Chapters 6 and 7 we present experimental observation of QRPN [36] and OM squeezing [37], respectively. We report these observations at room temperature, in the audio-frequency band, many mechanical linewidths away from the mechanical resonance. In order to be in the regime of frequency-independent squeezing, we detune the cavity, creating a strong OS. This OS is many orders of magnitude stronger than the mechanical spring of the oscillator, and we make use of it in controlling our system [38]. The OS is inherently unstable because it has a negative damping constant, so we stabilize it with an external electronic feedback system. The experimental details of the locking are given in Appendix B, and the various plant transfer functions useful for understanding the locking are calculated analytically in Chapter 2. The device that we fabricate has imperfections and is usually not completely identical to the design. In order to then understand its behavior viz. thermal noise and susceptibility, we need to model the designed device again with expected fabrication errors. This iterative process allows us to converge on the final mechanical properties of the device in use. The details of the finite-element modeling of the mechanical devices are given in Chapter 4.

# Chapter 2

# Optical spring system

## Contents

## Abstract

We calculate the effect of an optomechanical (OM) cavity on the laser field that interacts with it. In particular, we relate the optical power transmitted by the cavity to amplitude or phase modulation of the incident light. This analysis is done using classical fields and can be easily extended to a semi-classical approach to add the effect of quantum fluctuations. This description gives an intuitive understanding of the OM cavity as a subsystem in a bigger experimental system. To that end, the expressions connect quantities that can be directly measured in a laboratory, and are written out from a control systems perspective. Having this analytical model for the cavity and its transfer functions allows us to handily design feedback loops. Finally, in addition to the exact forms, we provide simplified expressions under the approximation that the cavity is high finesse and the frequency of operation is much smaller than the cavity linewidth, as is the case for most current systems.

## 2.1 Introduction

Optomechanical (OM) systems are increasingly playing a central role in precision measurements, including the measurement of gravitational waves [39, 4, 40], temperature [41], force [5], etc. In addition, they provide promising prospects as a toolkit for implementing quantum computation by cooling the mechanical oscillators to the ground state, for example Refs. [42, 43]. One important characteristic of OM systems when they are operated in a detuned regime is their internal feedback arising due to the optical spring (OS) effect [44, 45, 39]. In this work, we present a simple derivation for the full optical spring constant $K_{\mathrm{OS}}$ (including both the restoring and the damping part), and the open loop gain $G_{\mathrm{OL}}$ of the internal feedback loop pertaining to the basic OM system : a Fabry-Pérot cavity with one movable mirror.

Expressions for $K_{\mathrm{OS}}$ and $G_{\mathrm{OL}}$ are essential for calculations for locking OM systems [38] (Appendix B), or extracting squeezed light from them [20, 37] (Chapters 3 and 7). Since radiation pressure is a classical phenomenon, its effects can be understood without quantizing the electromagnetic field or the mechanical oscillator. Hence, in this work we will use classical electromagnetism from first principles. The quantum fluctuations of the field or mechanical displacement can later be incorporated in this formalism by a semi-classical approach, or by following Refs. [46, 45, 47] for a fully quantum mechanical treatment.

## 2.2 Formalism and notation

Let's consider an electromagnetic plane wave of frequency $\omega$ traveling in the z-direction, with an electric field $\mathcal{E}(x, y, t)$. We assume that the spatial dependence $f(x, y)$ of the electric field can be separated from its time dependence. We then define a new electric field $E(t)$ normalized by $A_0 = \int f^2(x, y)\,\mathrm{d}x\mathrm{d}y$ which allows us to simplify the optical power $P(t)$.

$$\begin{aligned}
\mathcal{E}(x, y, t) &= f(x, y) \times g(t) \\
E(t) &= \sqrt{\frac{\epsilon_0 c}{A_0}}\,\frac{\mathcal{E}(x, y, t)}{f(x, y)} \\
P(t) &= \epsilon_0 c \int \mathcal{E}^2(x, y, t)\ \mathrm{d}x\mathrm{d}y \\
P(t) &= E^2(t).
\end{aligned} \tag{2.1}$$

Now we split the electric field into cosine and sine quadratures (subscripts 1 and 2, respectively). We then further split each quadrature into its DC (constant) and AC (fluctuating) components, denoting them by uppercase and lowercase letters, respectively:

$$\begin{aligned}
E(t) &= E_1(t)\cos(\omega t) + E_2(t)\sin(\omega t) \\
E_i(t) &= E_i + e_i(t)
\end{aligned} \tag{2.2}$$

The Laplace transform of the time-dependent (fluctuating) part, with $s = j\omega$, is

$$e_i(s) = \int_{-\infty}^{\infty} e_i(t)\exp(-st)\mathrm{d}t \tag{2.3}$$

We define an AC column vector $\vec{e}(s)$, the DC component $\vec{E}$ and another DC vector $\vec{E}_\perp$ as

$$\vec{e}(s) := \begin{pmatrix} e_1(s) \\ e_2(s) \end{pmatrix},\ \vec{E} := \begin{pmatrix} E_1 \\ E_2 \end{pmatrix},\ \vec{E}_\perp := \begin{pmatrix} E_2 \\ -E_1 \end{pmatrix} \tag{2.4}$$

for later use, and note that

$$\vec{E}_\perp \cdot \vec{E} = 0. \tag{2.5}$$

The time-dependent power in a given field can be written by inserting Eq. (2.2) in Eq. (2.1). We only keep the slowly varying terms that constitute the change in the field amplitude envelope because the terms oscillating at $2\omega$ get averaged out in the integration by the photo-diodes We also assume that the AC terms $e_i(t)$ are much smaller than the DC terms $E_i$, hence allowing us to ignore the terms that are second order in $e_i(t)$. The constant (DC) power is then separated from the fluctuating component, i.e.

$$\begin{aligned} P(t) &\approx \frac{1}{2}\left(E_1^2+E_2^2\right)+E_1 e_1(t)+E_2 e_2(t) \\ &=: P+\delta P(t) \\ P &= \frac{1}{2}(E_1^2+E_2^2) \qquad (2.6) \\ \delta P(t) &\approx E_1 e_1(t)+E_2 e_2(t)=\vec{E}\cdot\vec{e}(t) \\ \delta P(s) &\approx \vec{E}\cdot\vec{e}(s) \qquad (2.7) \end{aligned}$$

The $s$ dependence of the fluctuating terms is implied, but will be dropped from now on to reduce clutter.

Finally, the quantities in our formalism that are proportional to the drive of an amplitude or phase modulator are

$$V_{\mathrm{AM}} := \frac{\vec{E}\cdot\vec{e}}{\beta_{\mathrm{AM}}},\ V_{\mathrm{PM}} := \frac{\vec{E}_{\perp}\cdot\vec{e}}{\beta_{\mathrm{PM}}} \qquad (2.8)$$

where $\beta_{\mathrm{AM}}$ and $\beta_{\mathrm{PM}}$ represent the strength of modulation provided by the modulators.

## 2.3 Interaction of EM field with optical components

The formalism in Section 2.2 can be used to derive the effect of various optical components the field encounters in our system. The results we derive here agree with those stated without derivation in Ref. [46]. Our results are obtained by using a purely classical description of the field.

### 2.3.1 Free space

*Result*: When laser light (denoted by $\vec{E}^{\,\mathrm{in}}$ and $\vec{e}^{\,\mathrm{in}}$) passes through a free space of length $L$, its DC and AC components experience a rotation and phase shift (giving a field denoted by $\vec{E}^{\,\mathrm{out}}$ and $\vec{e}^{\,\mathrm{out}}$):

$$\vec{E}^{\,\mathrm{out}} = \mathbb{R}(kL)\,\vec{E}^{\,\mathrm{in}} \qquad (2.9a)$$

$$\vec{e}^{\,\mathrm{out}}(s) = \mathbb{R}(kL)\ \exp\left(\frac{-Ls}{c}\right)\ \vec{e}^{\,\mathrm{in}}(s) \qquad (2.9b)$$

Here $k$ is the wave number, $c$ is the speed of light, and $\mathbb{R}(\alpha)$ is a rotation matrix for an angle $\alpha$, as defined in Eq. (2.10).

*Derivation*: When light travels through free space of fixed length $L$, the phase front of the outgoing light at time $t$ is the same as the phase front of the incoming light at time $L/c$ earlier.

$$\begin{aligned} E^{\mathrm{out}}(t) &= E^{\mathrm{in}}\left(t-t_0\right) \\ t_0 \equiv L/c &\Rightarrow\ \omega t_0 = kL \end{aligned}$$

We substitute this in the field form in Eq. (2.2), and then expand the time argument inside the sinusoidal terms. Thus we have,

$$\begin{aligned} E^{\mathrm{out}}(t) &= \left[E_1^{\mathrm{in}}+e_1^{\mathrm{in}}\left(t-t_0\right)\right]\cos\left(\omega\left(t-t_0\right)\right) \\ &\quad+\left[E_2^{\mathrm{in}}+e_2^{\mathrm{in}}\left(t-t_0\right)\right]\sin\left(\omega\left(t-t_0\right)\right) \\ &= \left[E_1^{\mathrm{in}}+e_1^{\mathrm{in}}\left(t-t_0\right)\right]\left[\cos(\omega t)\cos(kL)+\sin(\omega t)\sin(kL)\right] \\ &\quad+\left[E_2^{\mathrm{in}}+e_2^{\mathrm{in}}\left(t-t_0\right)\right]\left[\sin(\omega t)\cos(kL)-\cos(\omega t)\sin(kL)\right] \end{aligned}$$

After rearranging the terms, recollecting cosine and sine quadratures and defining $\mathbb{R}(\alpha)$ as a matrix that rotates a vector by the angle $\alpha$:

$$\mathbb{R}(\alpha) = \begin{pmatrix} \cos\alpha & -\sin\alpha \\ \sin\alpha & \cos\alpha \end{pmatrix},$$

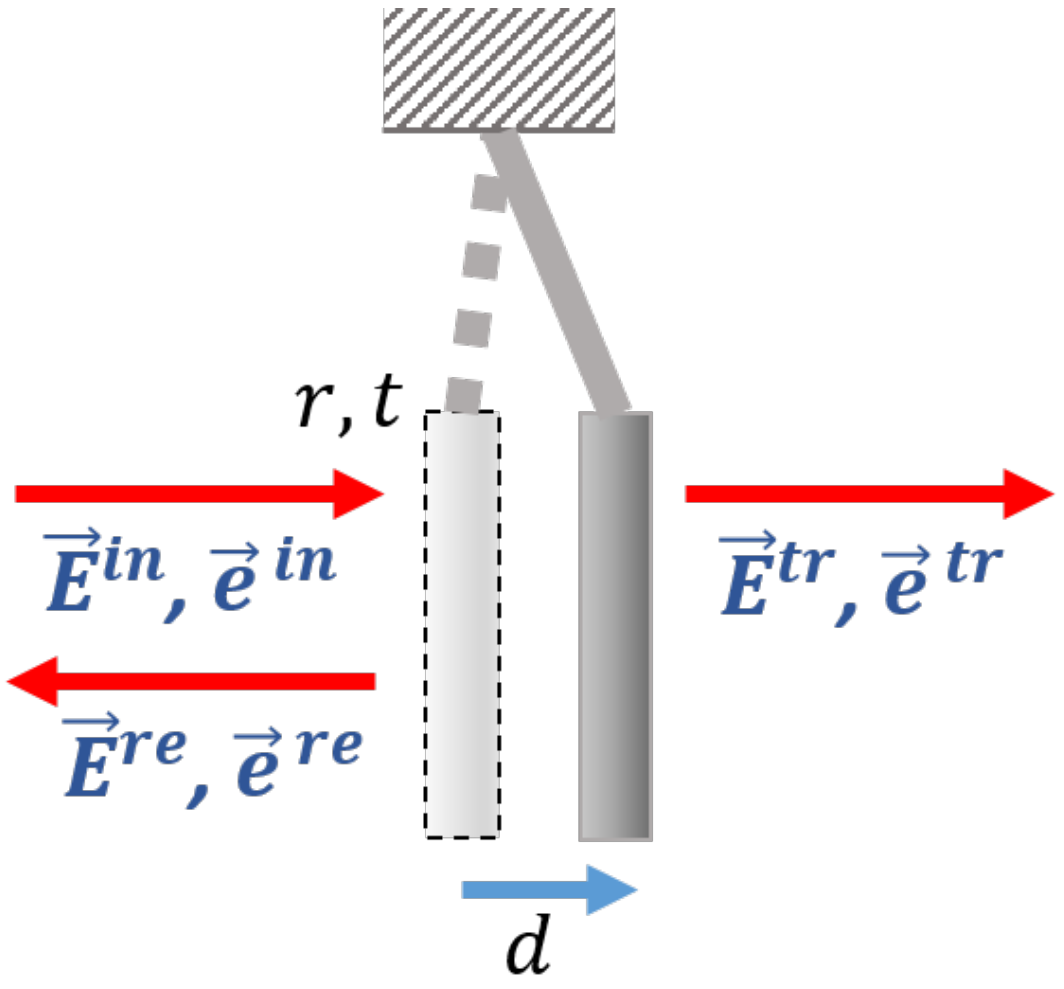

**Figure 2-1:** A schematic labeling the reflected and transmitted fields from a moving mirror.

we get the DC part as in Eq. (2.9a) and the AC component as:

$$\vec{e}^{\,\mathrm{out}}(t) = \mathbb{R}(kL)\, \vec{e}^{\,\mathrm{in}}\left(t - \frac{L}{c}\right) \tag{2.10}$$

After Laplace transformation of the time-dependent part according to Eq. (2.3) we get Eq. (2.9b).

### 2.3.2 Reflection and transmission by a moving mirror

*Result*: Let's consider a laser field denoted by $\vec{E}^{\,\mathrm{in}}$ and $\vec{e}^{\,\mathrm{in}}$ incident on a non-relativistically moving mirror of amplitude reflectivity $\rho$ and transmittivity $\tau$. The displacement spectrum of the mirror is $d(s)$, where $d(t)$ is assumed to be positive in the direction of the incident beam (Fig. 2-1). Then the reflected light from this mirror is given by:

$$\vec{E}^{\,\mathrm{re}} = \rho\, \vec{E}^{\,\mathrm{in}} \tag{2.11a}$$

$$\vec{e}^{\,\mathrm{re}}(s) = \rho\left(\vec{e}^{\,\mathrm{in}}(s) - 2kd(s)\vec{E}^{\,\mathrm{in}}_{s}\right) \tag{2.11b}$$

The light that is transmitted through this mirror does not get affected by the mirror motion, and is simply given by

$$\vec{E}^{\,\mathrm{tr}} = \tau\, \vec{E}^{\,\mathrm{in}} \tag{2.12a}$$

$$\vec{e}^{\,\mathrm{tr}}(s) = \tau\, \vec{e}^{\,\mathrm{in}}(s) \tag{2.12b}$$

*Derivation*: The amplitude of the phase front of the reflected light at any time $t$ is the reflection coefficient times the amplitude of the phase front of the incoming light at time $\frac{2d(t)}{c}$ earlier.

$$E^{\mathrm{re}}(t) = \rho\, E^{\mathrm{in}}(t - t_0)$$

$$t_0 \equiv \frac{2d(t)}{c} \Rightarrow \omega t_0 = 2kd(t)$$

We expand the cosine and sine as we did in the case of free-space propagation, but this time we also Taylor expand the time-dependent field terms because the mirror is moving slowly with respect to speed of light. Thus we get,

$$
\begin{aligned}
E^{\rm re}(t) &= \rho \left[ \left[E_1^{\rm in} + e_1^{\rm in}(t-t_0)\right] \cos(\omega(t-t_0)) \; + \left[E_2^{\rm in} + e_2^{\rm in}(t-t_0)\right] \sin(\omega(t-t_0)) \right] \\
&= \rho \Bigg[ \left( E_1^{\rm in} + e_1^{\rm in}(t) - \frac{2d(t)}{c} \frac{\partial e_1^{\rm in}(t)}{\partial t} + ... \right) \Big( \cos(\omega t)\cos(2kd(t)) + \sin(\omega t)\sin(2kd(t)) \Big) + \\
&\quad \left( E_2^{\rm in} + e_2^{\rm in}(t) - \frac{2d(t)}{c} \frac{\partial e_2^{\rm in}(t)}{\partial t} + ... \right) \Big( \sin(\omega t)\cos(2kd(t)) - \cos(\omega t)\sin(2kd(t)) \Big) \Bigg]
\end{aligned}
$$

Under the approximation that the fluctuations are smaller than the DC field ($e_i \ll E_i$), the field is changing slowly w.r.t. the optical frequency ($\partial e_i/\partial t \ll e_i\,\omega$), and the mirror is moving non relativistically, we can simplify the above relation by keeping only first order terms in $d(t)$ and $e_i(t)$:

$$
\begin{aligned}
E^{\rm re}(t) = \rho \Big\{ & \Big( E_1^{\rm in} + e_1^{\rm in}(t) - 2kd(t)E_2^{\rm in} \Big) \cos(\omega t) \\
& + \Big( 2kd(t)E_1^{\rm in} + E_2^{\rm in} + e_2^{\rm in}(t) \Big) \sin(\omega t) \Big\}
\end{aligned}
\tag{2.13}
$$

We can separate the DC component of the above equation to give Eq. (2.11a), leaving us with the AC component:

$$
\vec{e}^{\,\rm re}(t) = \rho \left( \vec{e}^{\,\rm in}(t) - 2kd(t)\vec{E}_s^{\,\rm in} \right)
$$

Here, $d(t)$ is assumed to be positive in the direction of incident light. Converting to the $s$ domain via Laplace transformation, we get Eq. (2.11b).

### 2.3.3 Radiation pressure force

Momentum carried by a field is given by energy per unit $c$, where c is the speed of light. Consequently, the force that the field imparts to an object that absorbs this field will be power per unit $c$. By momentum conservation, the AC component of the radiation pressure force on a mirror from the incident, reflected and transmitted fields will then be given by the AC power in each field divided by speed of light, i.e.,

$$
\delta F_{\rm RP}(s) = \frac{1}{c}(\delta P^{\rm in}(s) + \delta P^{\rm re}(s) - \delta P^{\rm tr}(s)) \tag{2.14}
$$

Here again, the mirror displacement is assumed to be positive in the direction of incident light.

## 2.4 System Description

We start by writing the equations of motion in the Laplace domain for a Fabry-Pérot cavity with one moving mirror as shown in Fig. 2-2. The Laplace domain can be easily converted into the frequency domain by setting $s = i\Omega$. We use the notation where vectors with uppercase letters denote the DC component, and vectors with lowercase letters denote the AC component of a time-varying electric field (defined in detail in Section 2.2, see Eq. (2.4)). The incoming laser field with DC component $\vec{L}$ and AC component $\vec{l}(s)$ enters the cavity via the input mirror (IM) with amplitude transmission $\tau_1$ and reflection $\rho_1$. From here on, the $s$ dependence will be implied and is dropped from the notation to reduce clutter.

DC and AC components of the intra-cavity fields are similarly represented by $\vec{W}$-$\vec{Z}$ and $\vec{w}$-$\vec{z}$. $\vec{R}$, $\vec{r}$ represent the field reflected from this cavity, and $\vec{T}$, $\vec{t}$ represent the field transmitted by this cavity via the end mirror (EM) which has amplitude transmission $\tau_2$ and reflection $\rho_2$. The IM is assumed to be mechanically fixed. The scalar $d$ is the displacement spectrum of the EM.

In order to describe our system by a set of linear equations, we consider the effect of each optical element – a fixed mirror (IM), free space (cavity length) and a movable mirror (EM) on the laser field. These effects are derived in Section 2.3. The field $\vec{w}$ is a combination of $\vec{l}$ transmitted via the IM and $\vec{x}$ reflected from it. Hence, we use Eqs. (2.11b) and (2.12b) with no displacement since the IM is fixed, to give,

$$\vec{w} = \rho_1 \vec{x} + \tau_1 \vec{l} \tag{2.15a}$$

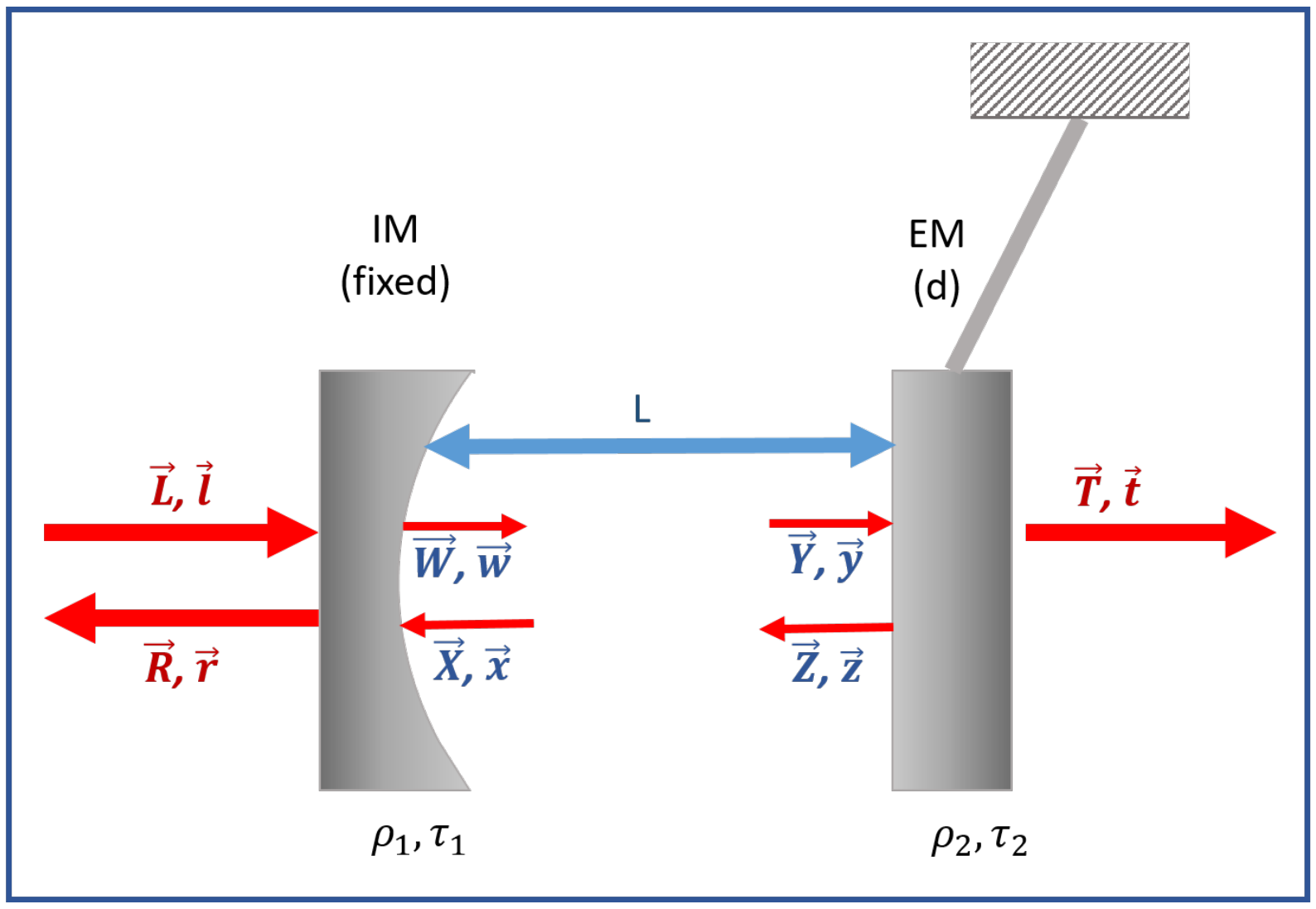

**Figure 2-2:** Model: A Fabry-Pérot cavity of length $L$ with a fixed input mirror (IM) and a moveable end mirror (EM). Uppercase and lowercase vectors are the DC and AC components of the respective fields, as described in the text. $\rho_1$, $\tau_1$ and $\rho_2$, $\tau_2$ are the amplitude reflectivities and transmittivities of the IM and EM respectively. $d$ is the displacement flcutuations of the EM.

Similarly, we use Eqs. (2.11b) and (2.12b) to obtain the field reflected and transmitted by the EM, this time keeping the displacement. We also note that the term incorporating the mirror movement is proportional to $\vec{Y}_\perp$ which is a transformed form of the DC field $\vec{Y}$, defined in Eq (2.4).

$$\vec{z} = \rho_2 \left( \vec{y} - 2kd\,\vec{Y}_\perp \right) \tag{2.15b}$$

$$\vec{t} = \tau_2 \vec{y} \tag{2.15c}$$

Finally, the propagation through free space going from the IM to the EM and back incurs some phase acquisition by the field, using Eq (2.9b) we have,

$$\vec{y} = \exp\{(i\phi)\}\ \mathbb{R}(\theta)\ \vec{w} \tag{2.15d}$$

$$\vec{x} = \exp\{(i\phi)\}\ \mathbb{R}(\theta)\ \vec{z} \tag{2.15e}$$

Here $\mathbb{R}$ is a rotation matrix as defined in Eq. (2.10); $\theta$ and $\phi$ are the phase acquired by the carrier and the sidebands in a single trip:

$$\theta = \frac{L\omega}{c} \equiv \frac{L(\omega - \omega_C)}{c} = \frac{L\Delta}{c}, \quad \phi = \frac{-Ls}{ic}.$$

We have defined $\Delta := \omega - \omega_C$, the amount by which the laser frequency is detuned from the cavity's resonance frequency. Since $\omega_C$ is set by the resonance condition of the cavity, $\omega_C L/c = 2n\pi$, $\theta$ inside a sinusoidal function can be written as just a function of the detuning from the cavity length. Using Eqs. (2.9a), (2.11a) and (2.12a), we can also write similar equations for the DC components. It is straightforward to check that they follow the same form as Eq. (2.15) with $\phi = 0$ and $d = 0$[1,2].

The solution of the system described by Eq. (2.15) and the respective DC fields is given by

$$\vec{y} = 2kd\,(1 - \mathbb{N}^{-1})\vec{Y}_{\perp} + \tau_1 \exp(i\phi)\,\mathbb{N}^{-1}\mathbb{R}(\theta)\vec{l} \tag{2.16a}$$

$$\vec{Y} = \tau_1\,\mathbb{M}^{-1}\mathbb{R}(\theta)\vec{L}, \tag{2.16b}$$

where, the matrices $\mathbb{N}$ and $\mathbb{M}$ are scaled rotations:

$$\mathbb{N} = \mathbb{I} - \exp(2i\phi)\,\rho_1\rho_2\mathbb{R}^2(\theta) =: a\,\mathbb{R}(-b) \tag{2.16c}$$

$$\mathbb{M} = \mathbb{I} - \rho_1\rho_2\mathbb{R}(\theta)^2 =: A\,\mathbb{R}(-B). \tag{2.16d}$$

Above we have defined four new scalars that capture the dynamics controlled by the cavity parameters: $a, b, A, B$ for use in the next section, the exact expressions for them are

$$a = \sqrt{(1 - \rho_1\rho_2 e^{2i\phi}\cos(2\theta))^2 + \rho_1^2\rho_2^2 e^{4i\phi}\sin^2(2\theta)} \tag{2.17a}$$

$$\tan b = \frac{\rho_1\rho_2 e^{2i\phi}\sin(2\theta)}{1 - \rho_1\rho_2 e^{2i\phi}\cos(2\theta)}, \tag{2.17b}$$

and

$$A = \sqrt{(1 - \rho_1\rho_2\cos(2\theta))^2 + \rho_1^2\rho_2^2\sin^2(2\theta)} \tag{2.18a}$$

$$\tan B = \frac{\rho_1\rho_2\sin(2\theta)}{1 - \rho_1\rho_2\cos(2\theta)}. \tag{2.18b}$$

In order to capture the radiation pressure dynamics we also need the radiation pressure force on the EM. We obtain this by summing the RP force exerted by all the fields in contact with the EM, and simplify to write the total force as a function of $\vec{Y}$ and $\vec{y}$ only. Using Eqs. (2.5), (2.7) and (2.14) and the DC version of Eqs. (2.15b) and (2.15c), we get

$$\begin{aligned}\delta F_{RP} &= \frac{1}{c}\left(\vec{Y}\cdot\vec{y} + \vec{Z}\cdot\vec{z} - \vec{T}\cdot\vec{t}\right)\\ &= \frac{1 + \rho_2^2 - \tau_2^2}{c}\vec{Y}\cdot\vec{y}.\end{aligned} \tag{2.19}$$

The above radiation pressure force moves the EM, which changes the fields (see Eq. (2.16a)), in turn changing the RP force (Eq. (2.19)); thereby creating an internal cavity feedback loop. We discuss the effect of this feedback in the next section.

## 2.5 System Transfer Function

One of the important characteristics that describe the physical dynamics of the OM system is the optical spring (OS) effect. This OS arises because the radiation pressure force has a component that is proportional to the displacement of the moving mirror. In order to show this, we use the solution in Eq. (2.16a) to substitute $\vec{y}$ in the expression for radiation pressure force in Eq. (2.19). This yields the radiation pressure force as a function of $\vec{l}$ and $d$, from which we obtain the OS constant $K_{OS}$, which is

[1] This system of equations has been derived from classical Maxwell description. To include the effects of quantum fields, include vacuum fields in Eqs. (2.15b) and (2.15c). To describe the transfer functions relevant to locking this cavity, which is the main concern of this work, the vacuum fields can and have been ignored.

[2] To include the full back-action effect, $d$ should be written as a sum of displacement due to radiation pressure force and any external displacement noise in Eq. (2.15). But that treatment is not required to for the purposes of this work, and complicates the analytical solutions to the system.

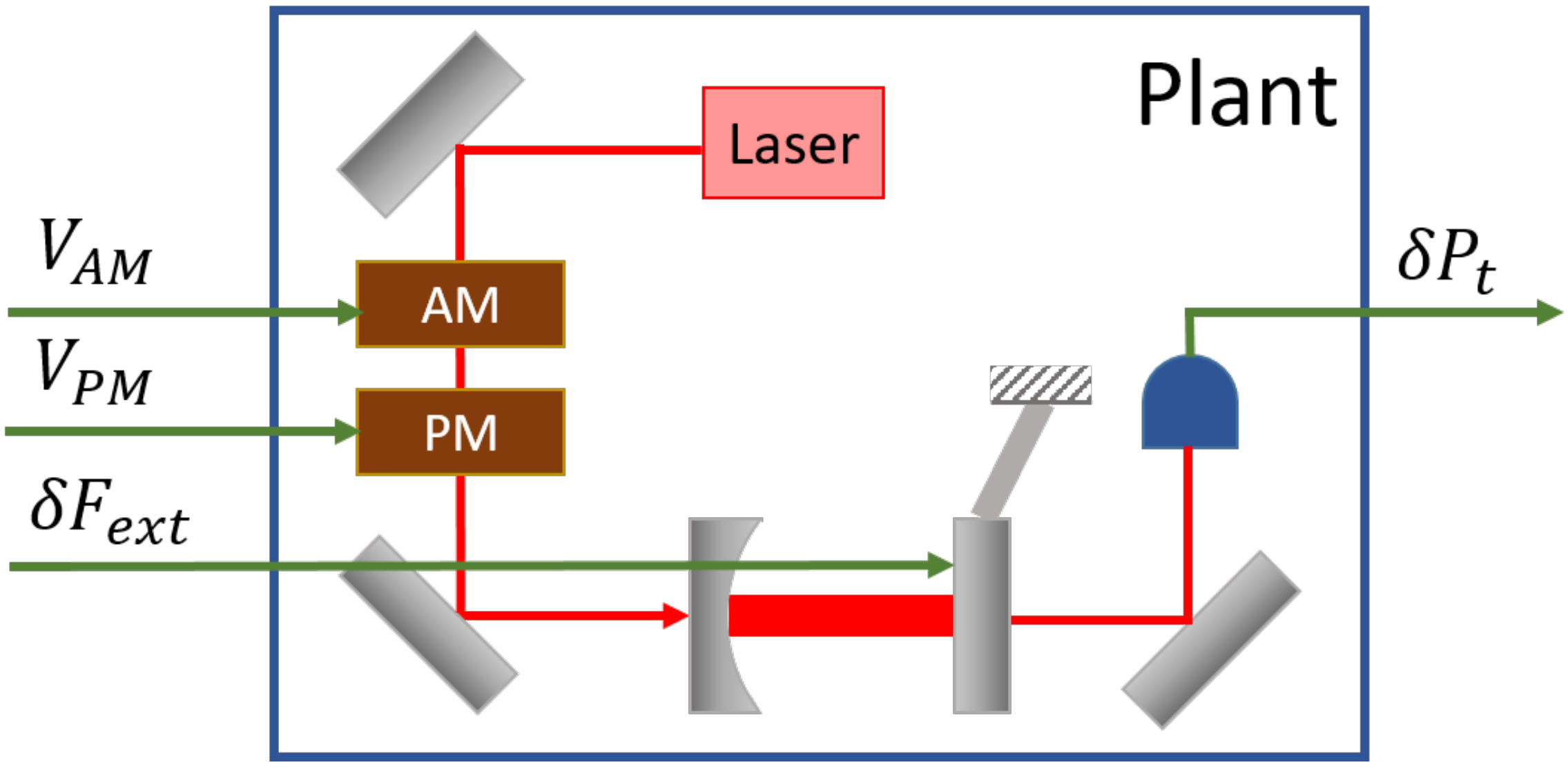

**Figure 2-3:** Plant under study: Change in the transmitted power from the cavity as a function of external force on the oscillating mirror and drive to amplitude or phase modulator on the incoming laser.

the negative of the coefficient of $d$ in $\delta F_{\mathrm{RP}}$:

$$K_{\mathrm{OS}} = \frac{2k}{c}\left(1+\rho_2^2-\tau_2^2\right)\vec{Y}\cdot\mathbb{N}^{-1}\vec{Y}_\perp \tag{2.20}$$

We would now like to see how the internal cavity feedback via radiation pressure plays a role on a quantity measurable in the laboratory, e.g. the transmitted optical power from the cavity ($\delta P_{\mathrm{t}} = \vec{T}\cdot\vec{t}$). To do so, we derive a relationship from $\delta P_{\mathrm{t}}$ to the drive on an amplitude or phase modulator on the input field, i.e. $V_{\mathrm{AM},I}$ and $V_{\mathrm{PM},I}$, respectively. The goal is to derive a relation that looks like

$$\delta P_{\mathrm{t}} = \frac{G_{\mathrm{out}}\left(G_{\mathrm{in}}^{\mathrm{A}}V_{\mathrm{AM},I} + G_{\mathrm{in}}^{\mathrm{P}}V_{\mathrm{PM},I} + G^{\mathrm{F}}\,\delta F_{\mathrm{ext}}\right)}{1+G_{\mathrm{OL}}}, \tag{2.21}$$

where we have denoted various gains that we would like to derive as $G$: $G_{\mathrm{OL}}$ is the open-loop gain of the internal cavity feedback, $G_{\mathrm{out}}$ is the gain from the intracavity power to the transmitted power, $G_{\mathrm{in}}^{A}$ and $G_{\mathrm{in}}^{P}$ are the gains corresponding to amplitude and phase modulation on the input light respectively, and $G^{F}$ is the gain factor for an external force $\delta F_{\mathrm{ext}}$ applied to the movable mirror.

In order to derive this, we first use Eq. (2.15c) and its DC counterpart to obtain $\vec{T}\cdot\vec{t} = \tau_2^2\vec{Y}\cdot\vec{y}$, giving Eq. (2.23a) and implying we can just calculate $\vec{Y}\cdot\vec{y}$. Next, we pre-multiply Eq. (2.16a) by $\vec{Y}$, simplify using $\vec{Y}\cdot\vec{Y}_\perp = 0$ (Eq. (2.5)), and substitute $K_{\mathrm{OS}}$ from Eq (2.20) in the term proportional to $d$. We also split the mirror displacement $d$ into displacement from radiation pressure force and that from any other external forces, assuming the mirror has a mechanical susceptibility $\chi_{\mathrm{mech}}$ to all forces.

$$d = \chi_{\mathrm{mech}}(\delta F_{\mathrm{ext}} + \delta F_{\mathrm{RP}}) \tag{2.22}$$

The radiation pressure force in the above displacement can be obtained from Eq. (2.19), which gives another term proportional to $\vec{Y}\cdot\vec{y}$, thus giving a closed loop. Finally, we substitute Eq. (2.16b) in the term proportional to $\vec{l}$ and simplify using the fact that $\mathbb{N}$ and $\mathbb{M}$ commute with $\mathbb{R}$ and $\mathbb{R}^{-1}$. This gives two terms, each proportional to $\vec{L}\cdot\vec{l}$ and $\vec{L}_\perp\cdot\vec{l}$, which are related to $V_{\mathrm{AM},I}$ and $V_{\mathrm{PM},I}$ (Eq. (2.8)) respectively. This allows us to write a closed form solution for Eq. (2.15). This solution follows the form of our target

Eq. (2.21) if the gains are as follows:

$$G_{\text{out}} = \tau_2^2 \tag{2.23a}$$

$$G^{\text{F}} = -\frac{c\, K_{\text{OS}}\, \chi_{\text{mech}}}{1+\rho_2^2-\tau_2^2} \tag{2.23b}$$

$$G_{\text{OL}} = \chi_{\text{mech}}\, K_{\text{OS}} \tag{2.23c}$$

$$G_{\text{in}}^{\text{A}} = \frac{\tau_1^2 e^{i\phi}\beta_{\text{AM}}}{Aa}\cos(b-B) \tag{2.23d}$$

$$G_{\text{in}}^{\text{P}} = \frac{\tau_1^2 e^{i\phi}\beta_{\text{PM}}}{Aa}\sin(b-B) \tag{2.23e}$$

Using the DC power Eq. (2.6), we can get the intra-cavity power as the power in the field $\vec{Y}$ i.e.

$$P_{\text{cav}} = \frac{1}{2}(Y_1^2+Y_2^2), \tag{2.24}$$

which then allows us to simplify $K_{\text{OS}}$ in Eq. (2.20) to

$$K_{\text{OS}} = \frac{4k}{c}(1+\rho_2^2-\tau_2^2)P_{\text{cav}}\frac{\sin b}{a} \tag{2.25}$$

The exact expressions for the cavity gains without any approximation, are then as follows:

$$G_{\text{in}}^{A} = \tau_1^2\frac{(1+\rho_1^2\rho_2^2\exp(2i\phi))\exp(i\phi)-2\rho_1\rho_2\exp(2i\phi)\cos(2\theta)\cos(\phi)}{(1+\rho_1^2\rho_2^2-2\rho_1\rho_2\cos(2\theta))(1+\rho_1^2\rho_2^2\exp(4i\phi)-2\rho_1\rho_2\exp(2i\phi)\cos(2\theta))} \tag{2.26a}$$

$$G_{\text{in}}^{P} = \tau_1^2\frac{2i\exp(2i\phi)\rho_1\rho_2\sin(2\theta)\sin(\phi)}{(1+\rho_1^2\rho_2^2-2\rho_1\rho_2\cos(2\theta))(1+\rho_1^2\rho_2^2\exp(4i\phi)-2\rho_1\rho_2\exp(2i\phi)\cos(2\theta))} \tag{2.26b}$$

$$K_{\text{OS}} = \frac{4k(1+\rho_2^2-\tau_2^2)P_{\text{cav}}\rho_1\rho_2\exp(2i\phi)\sin(2\theta)}{c(1+\rho_1^2\rho_2^2\exp(4i\phi)-2\rho_1\rho_2\exp(2i\phi)\cos(2\theta))} \tag{2.26c}$$

Now we make the high-finesse approximation i.e. both mirrors are high reflectors $(1-\rho_1)\ll 1$ and $(1-\rho_2)\ll 1$. In order to do so, we first write out the angles $\theta$ and $\phi$ as a function of the cavity linewidth $\gamma$ and the total power loss from the cavity $T$, including transmission and losses due to scattering, diffraction etc.

$$\gamma = \frac{c(1-\rho_1\rho_2)}{2L\sqrt{\rho_1\rho_2}} \approx \frac{cT}{4L} \tag{2.27a}$$

$$\theta \approx \frac{\Delta T}{4\gamma},\ \phi \approx -\frac{sT}{4i\gamma} \tag{2.27b}$$

The lowest order transfer functions of the cavity after we make the above high finesse approximation are:

$$G^{\text{F}} = -\frac{1}{2}\,c\,G_{\text{OL}} \tag{2.28a}$$

$$K_{\text{OS}} = \frac{16kP_{\text{cav}}}{cT}\frac{\Delta_\gamma}{1+\Delta_\gamma^2}\frac{1}{(1+s/\gamma_+)(1+s/\gamma_-)} \tag{2.28b}$$

$$G_{\text{in}}^{\text{A}} = \frac{4\,t_1^2\,\beta_{\text{AM}}}{T^2}\frac{1}{1+\Delta_\gamma^2}\frac{(1+s/\gamma_0)}{(1+s/\gamma_-)(1+s/\gamma_+)} \tag{2.28c}$$

$$G_{\text{in}}^{\text{P}} = -\frac{4\,t_1^2\,\beta_{\text{PM}}}{T^2}\frac{\Delta_\gamma}{1+\Delta_\gamma^2}\frac{s/\gamma_0}{(1+s/\gamma_-)(1+s/\gamma_+)} \tag{2.28d}$$

Here we have defined a normalized detuning $\Delta_\gamma$ scaled by the cavity linewidth $\gamma$ (half-width half maximum), and modified cavity poles $\gamma_\pm$ for a detuned cavity:

$$\Delta_\gamma := \Delta/\gamma \tag{2.29a}$$

$$\gamma_\pm := \gamma(1 \pm i\Delta_\gamma) \tag{2.29b}$$

$$\gamma_0 := |\gamma_+|^2/\gamma = |\gamma_-|^2/\gamma = \gamma_+\gamma_-/\gamma = \gamma(1+\Delta_\gamma^2) \tag{2.29c}$$

Both amplitude and phase modulations are filtered by the cavity, and the transfer functions are as expected for a Fabry-Pérot cavity without any optomechanics. The amplitude modulation has a frequency independent gain to the cavity power at frequencies lower than the cavity linewidth, and the gain decreases at frequencies greater than $\gamma$ as $1/f$. On the other hand, the transfer function of phase modulation to cavity power has a zero at DC, and two poles near the cavity linewidth, which give the maximum gain at frequencies near the cavity linewidth.

The OM effects of the cavity are seen by looking at $G_{\mathrm{OL}}$ and $G^{\mathrm{F}}$, which in turn depend on the OS constant $K_{\mathrm{OS}}$. The OS constant is a frequency dependent quantity with two poles at $\gamma_\pm$. This can be understood as the delay in conversion from displacement to cavity power. The OS constant's DC value gives the characteristic OS frequency of the system.

$$m\,\Omega_{\mathrm{OS}}^2 := K_{\mathrm{OS}}\big|_{s=0} = \frac{16kP_{\mathrm{cav}}}{cT}\frac{\Delta_\gamma}{1+\Delta_\gamma^2} \tag{2.30}$$

Here $m$ is the mass of the oscillator mirror. The dependence of magnitude of $\Omega_{\mathrm{OS}}$ on cavity power, finesse and detuning is clear from the above equation. When the laser frequency is greater than the cavity's resonance frequency, $K_{\mathrm{OS}}$ is positive, giving rise to a restoring force. When the laser is red detuned w.r.t. the cavity, the radiation pressure force is positively proportional to the cavity displacement, or is anti-restoring. We have chosen to parametrize the cavity gains in terms of the DC cavity power because experimentally it is easier to constantly measure cavity power by measuring the transmitted power and scale by the end mirror transmission. On the other hand, scaling with input power will require knowledge of quantities like mode-matching, losses etc. in addition to the individual mirror transmissions.

Finally, we note one trivial limit of these results ie a rigid cavity. This can be obtained by making the EM infinte mass. When $m \to \infty$, we have $\chi_{\mathrm{mech}} \to 0$, giving $G_{\mathrm{OL}}, G^{\mathrm{F}} \to 0$ and $\Omega_{\mathrm{OS}} \to 0$, implying there is no internal OM feedback, the transmitted power is insensitive to forces applied to the end mirror, and no OS (even though the optical rigidity $K_{\mathrm{OS}}$ is finite).

### 2.5.1 Example : Control of a single mode oscillator

As an example, we will apply this treatment to a high-finesse Fabry-Pérot cavity with a single mode mechanical oscillator as end mirror whose optical linewidth is larger than the mechanical resonance as well as the OS frequency. The oscillator's resonance frequency is $\Omega_{\mathrm{m}}$, it is viscously damped, providing it a linewidth of $\Gamma_{\mathrm{m}}$. Then, we can obtain the closed loop response by using the mechanical susceptibility, to give us the full plant transfer function:

$$\chi_{\mathrm{mech}} = \frac{1}{m\,(s^2+\Omega_{\mathrm{m}}^2+2\,s\,\Gamma_{\mathrm{m}})} \tag{2.31a}$$

$$G_{\mathrm{CL}} = \frac{1}{1+G_{\mathrm{OL}}} = \frac{1}{1+\chi_{\mathrm{mech}}K_{\mathrm{OS}}} \tag{2.31b}$$

The closed loop gain $G_{\mathrm{CL}}$ is obtained by substituting Eqs. (2.28b) and (2.31a) in Eq. (2.31b). We simplify it by using a Padé approximant with the assumption that the cavity linewidth $\gamma$ is the largest frequency in the transfer function, expanding it to zero-th order in the numerator and first order in the denominator.

$$G_{\mathrm{CL}} \approx \frac{s^2+2\,s\,\Gamma_{\mathrm{m}}+\Omega_{\mathrm{m}}^2}{s^2+2\,s\,(\Gamma_{\mathrm{m}}-\Gamma_{\mathrm{OS}})+\Omega_{\mathrm{m}}^2+\Omega_{\mathrm{OS}}^2} \tag{2.32}$$

Here we have defined $\Gamma_{OS} = \Omega_{OS}^2/\gamma_0$, which can be seen as an effective linewidth of the OS, essentially arising from the cavity delay. We can similarly Padé approximate $G_{in}^A$, and obtain the full plant transfer functions:

$$G_{\rm in}^{\rm A} \approx \frac{4\,\tau_1^2\,\beta_{\rm AM}}{T^2(1+\Delta_\gamma^2)}\,\frac{1}{(1+s/\gamma_0)} \tag{2.33a}$$

$$G_{\rm OL} \approx \frac{\Omega_{\rm OS}^2}{\Omega_{\rm m}^2}\,\frac{(1-2s/\gamma_0)}{1+s^2/\Omega_{\rm m}^2+2\,s\,\Gamma_{\rm m}/\Omega_{\rm m}^2} \tag{2.33b}$$

$$G^{\rm F}G_{\rm CL} \approx \frac{1-2\,s/\gamma_0}{1-2s/\gamma_0+(\Omega_{\rm m}^2+s^2+2\,s\,\Gamma_{\rm m})/\Omega_{\rm OS}^2} \tag{2.33c}$$

In the above example system, the plant transfer function will be proportional to $G_{CL}$, which will have at least one pole in the right-half plane for a red detuned system, since $\Omega_{OS}^2$ will be negative. In the blue detuned case, the system could be intrinsically stable if $\Gamma_m > \Gamma_{OS}$, failing which there will be again a pole in the right half plane. For most unresolved systems, if the mechanical oscillator is a high Q oscillator, this stability will be achievable if the OS is weaker than the mechanical spring, rendering the whole OM effect weak. If strong OM effects are desirable, this instability must be overcome via external control.

Finally, $G_{CL}$ also provides us with a new effective oscillator for the entire system. From Eq. (2.32), we can see that this effective oscillator has a spring frequency $\Omega_{OM}$, given by

$$\Omega_{\rm OM}^2 = \Omega_{\rm m}^2 + \Omega_{\rm OS}^2, \tag{2.34}$$

and a damping constant $\Gamma_{OM}$, given by

$$\Gamma_{\rm OM} = \Gamma_{\rm m} - \Gamma_{\rm OS}. \tag{2.35}$$

Since $\Omega_{OM}$ defines the effective oscillator for the closed loop system, it provides a characteristic frequency that is important in the OM squeezing behavior. Frequency independent squeezing is observed at frequencies well below $\Omega_{OM}$. At frequencies close to $\Omega_{OM}$, OM squeezing has a highly frequency dependent quadrature and suppression.

## 2.6 Comparison with previous work

While the result is widely used in the field, *ab initio* no derivation exists to our knowledge. Hence we provide a coherent, self-consistent and simple derivation of OM feedback for the case of a Fabry-Pérot cavity, which is the most ubiquitous OM system. The references (e.g. [39, 20, 46, 45, 47]) except Ref. [48] approach it using quantized fields.

Ref. [39] uses the quantum formalism to derive $R_{FF}$ (Eq. 3.26) which is effectively a similar OS constant for a signal recycled Michelson interferometer. They also mention $1 - \chi K$ form explicitly and the impact on system stability.

Corbitt et al derive the OS constant in Ref. [20] using the quantum two-photon formalism, and derive only the real part of the OS. They consider the end mirror to be perfectly reflective, which is not the case for all experiments, e.g. Ref. [38]. In addition, Ref. [46] uses the same formalism to calculate the optical springs of differential and common mode motion of a Michelson interferometer, but again does not provide the damping of that OS.

Perreca et al in Ref. [48] derive $K_{OS}$ in the Appendix using classical fields. They consider the system in the time domain by iterating every reflection from the cavity – they use this approach to calculate the intracavity power, which they convert into radiation pressure force (caution: the force is erroneously $r_2^2$ times what it should be, $r_2$ being the amplitude reflectivity of the moving mirror). They also make the approximation that $\frac{L\Delta}{c} < 1$ before the high-finesse approximation, which is technically incorrect because this approximation assumes the cavity detuning is much smaller than its free spectral range, which implicitly implies the cavity has a high finesse.

We take a different approach that addresses some of the shortcomings of Ref. [48]. We solve the problem in the frequency domain directly, providing a simple connection to the time domain; the radiation pressure force takes into account force from transmitted beam; and we derive the exact amount of damping associated with the OS (rather than adding it ad-hoc, as in Ref. [48]). Our analysis also differs in that we approach everything from a feedback system perspective – which beautifully yields the spring constant as well other required physical quantities.

Finally, we derive the basic interaction between the field and optical elements in Section 2.3. Ref. [46] states the two-photon quantum mechanical counterparts of the same effects. Our results agree with theirs, and we provide simple proof for it, which is missing in Ref. [46].

## 2.7 Conclusion

In conclusion, we show that the OM response of a Fabry-Pérot cavity can be easily understood as a typical control loop, using classical electromagnetism. The OM open-loop gain is the product of the mechanical susceptibility of the oscillator and the optical rigidity of the cavity. Finally, the effects of quantum fluctuations of the field or the oscillator displacement can be easily incorporated for such a system by using the quantum mechanical description of the field amplitude and/or the oscillator displacement and inserting them in the cavity equations.

# Chapter 3

# Squeezer design: optical

# Contents

## Contents

## Abstract

In this work we present a method for optimizing the experimental parameters of an optomechanical squeezed light source that can improve the sensitivity of interferometric gravitational wave detectors. Our analysis goes beyond the standard case of an ideal cavity to include practical realizations that account for experimental imperfections. Our goal is to design an optomechanical squeezed light source from which we can extract the most possible squeezing within a given set of experimentally realizable parameters. We focus on generation of broadband squeezed light suitable for broadband measurements like gravitational wave detection. We find that for a fixed amount of laser power allowed inside the optical cavity, we not only need high cavity finesse but also a high escape efficiency. The high finesse reduces the effect of thermal fluctuations and allows for generation of higher squeezing inside the cavity, whereas the escape efficiency determines the fraction of the squeezed light that can be made available for use outside the cavity. We also report on the impact of such realistic imperfections on the squeezing quadrature and bandwidth.

## 3.1 Overview: optical optimization of squeezer design

In this work, we present an optimization of the optical properties of an optomechanical (OM) squeezer. The goal is to design an OM squeezed light source from which we can extract the most possible squeezing in the required frequency band. When talking about the frequency band, we will usually refer to the frequencies at which the system is a frequency-independent squeezing source, i.e. the amount and quadrature of squeezing are frequency-independent. We also aim to do this within experimentally realizable parameters.

The OM system under consideration is a Fabry-Perot cavity with one movable mirror. The squeezing performance of this system depends both on the optical properties of the Fabry-Perot cavity, as well as the mechanical properties of the oscillator mirror. In this chapter, we focus on the optical properties, while keeping the mechanical behavior of the oscillator fixed. The mechanical properties decide the susceptibility of the oscillator to all forces, as well as its motion due to Brownian thermal fluctuations. One can also optimize the mechanical properties to improve the squeezer design, which is addressed in Chapter 4.

Our analysis goes beyond the standard case of an ideal cavity to include practical realizations that account for experimental imperfections. Realistically, an OM system has losses and classical noises like thermal noise and classical laser noises. We include these effects in performing our optimization. Previously, there have been studies of OM squeezing from ideal Fabry-Perot cavities [20], without considering classical noise, and restricted to squeezing measurement in reflection. Instead, we do this optimization for a realistic system that has classical noises, and that could be operated in transmission. Such optimization has been pursued numerically to maximize the squeezing output from a Michelson interferometer based OM squeezer in the presence of classical noises and losses [20], but has not been reported for a Fabry-Perot based system. While some of our conclusions are broadly useful, wherever we need to make assumptions, we choose to focus on the unresolved sideband regime in order to generate broadband squeezed states suitable for the audio frequency band in which GW detectors operate. These assumptions will be clearly stated wherever they are made.

We have developed a both an approximate analytical formulation, as well as an exact numerical simulation to perform this optimization. The analytical approach allows us to intuitively understand the physics, albeit for a simplified case. The numerical analysis assures us that the full complicated behaviour of the system is taken into account before making the design decisions. This is especially required since our real system can get close to not obeying the assumptions that are needed to simplify the analytical results. For example, in the analytical results we assume that the impact of thermal noise is much smaller than that of quantum noise. While they are comparable in our current system, this approximation is likely to be true for future generations. Hence we need both analytical as well as numerical analysis to tackle this problem.

While most of our attention is focused on maximizing the amount of squeezing, we also report on the role of realistic imperfections and their impact on the squeezing quadrature and bandwidth. [1]

1. In the ideal case of a Fabry-Perot cavity with a movable end mirror of 100 % reflectivity and no cavity losses, the squeezed magnitude and angle both depend only on the dimensionless detuning parameter. This result has been analytically shown in [20].

2. For a Fabry-Perot cavity with end transmission as well as cavity losses, but without any contribution from classical noises, the squeezing behaviour is very similar to that of an ideal squeezer but with added effect of escape efficiency of the measurement port. The bandwidth of squeezing is limited by the optomechanical spring frequency $\Omega_{\rm OM}$ (Eq. (2.32)), same as in [20].

3. With the inclusion of classical noises, finesse becomes important along with the escape efficiency. And hence for a fixed output port, there is an optimal reflectivity that produces maximum amount of squeezing. The role of finesse becomes more important as we diverge from the ideal system in the presence of classical noises, i.e. increase in classical noise, increase in loss etc. The bandwidth of squeezing in the presence of classical noises is decided by the region of frequencies where the total displacement of the oscillator has a significant contribution from quantum noise. If the higher end of this region is above $\Omega_{OM}$, $\Omega_{OM}$ forms the cutoff. The squeezed quadrature depends not only on detuning and transmissions but also on the classical noises since they have a quadrature dependent contribution.

[1] We have concentrated on squeezing due to the traditionally studied dispersive interaction, one can also study squeezing due to dissipative interaction.

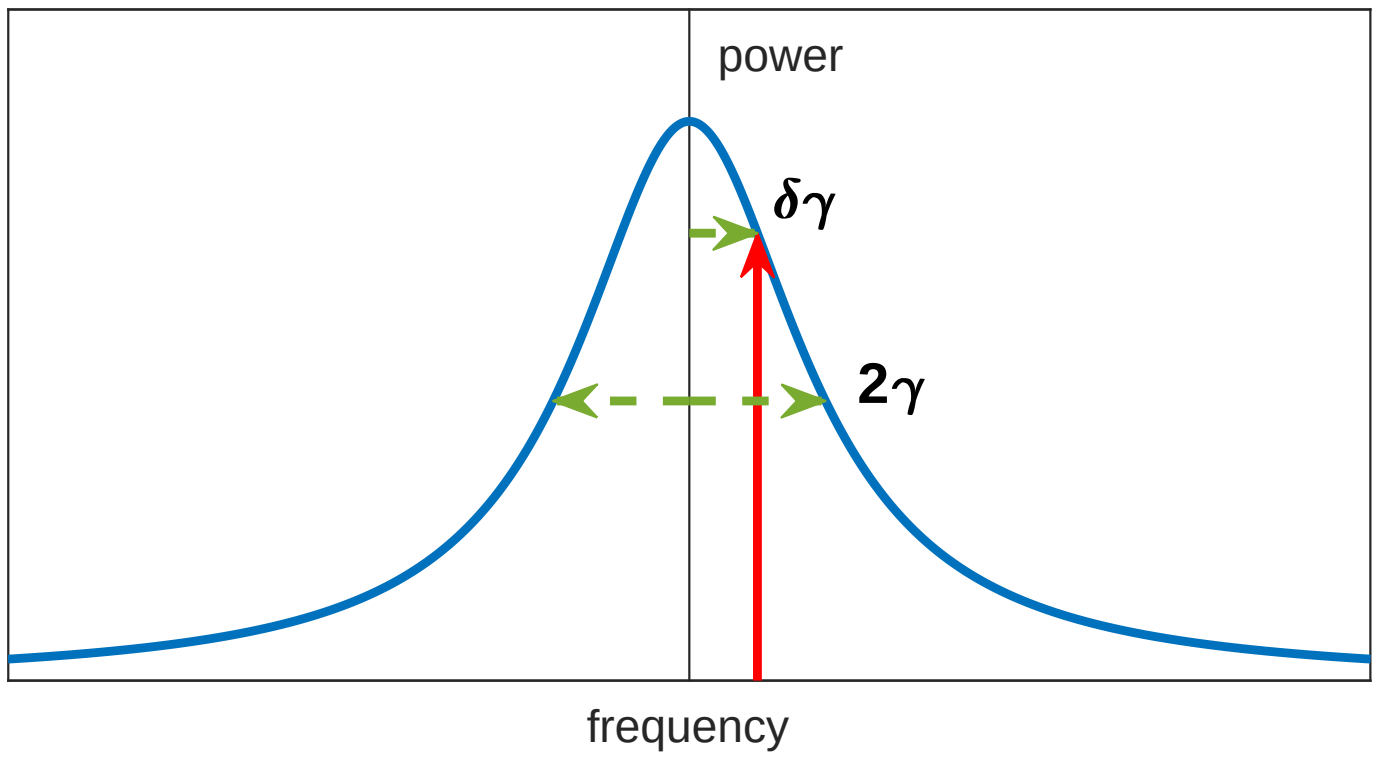

**Figure 3-1:** Definition of linewidth and detuning as used in the entire manuscript. The cavity's HWHM is $\gamma$, and $\delta$ is a dimensionless number that specifies the frequency difference of the laser from the cavity's resonance frequency, in units of $\gamma$. It is important to note that when we optimize the optical parameters like reflectivities, $\gamma$ is different for different parameters. That is inherently taken into account in how the calculation is setup, so whenever $\gamma$ is mentioned, it is w.r.t. the cavity parameters under consideration. Similarly, when we look at how $\delta$ impacts squeezing from two different cavities, we assume that $\delta$ is the dimensionless detuning scaled to each configuration's individual linewidth, and is not a measure of absolute detuning.

4. In the presence of thermal noise, higher intracavity power also helps get more squeezing. Additionally, at high intracavity power, the contribution of thermal noise is smaller than QRPN, and can be handled in a perturbative manner. At low powers though, thermal noise can become comparable or even larger than QRPN, and the optimization regime changes, requiring full numerical analysis to decide the best optical parameters.

5. An OM squeezer is a frequency independent source of squeezing at frequencies below its OM spring (Eq. (2.32)). If the the thermal noise at frequencies near or higher than the OM spring is low enough, it will have squeezing with a frequency dependent quadrature and noise.

## 3.2 Setup

The setup considered in this optimization is depicted in Fig. 3-2 . Here $\vec{l}$ is the input laser field, $\vec{r}$ and $\vec{t}$ are the fields reflected and transmitted from the cavity respectively, and $\vec{v}$ denotes the various open ports where vacuum enters the system. All the different vacuum fields are treated as uncorrelated. The measurement is depicted on $\vec{t}$, but can be similarly done on $\vec{r}$. We analyse the properties of $\vec{t}$ and $\vec{r}$ as a function of all the inputs of the system as previously described.

## 3.3 Basic optical quantities

### 3.3.1 Cavity parameters

We first start with establishing some basic cavity parameters that will be recurring throughout this thesis. Using basic field propagation with classical Maxwellian physics, we can obtain the cavity linewidth, finesse, circulating, reflected and transmitted powers, and the optical spring (OS) constant. All these quantities are listed in their exact[2] as well as approximate forms in Table 3.2. The approximate form is obtained under the assumption that the cavity finesse is high, or in other words the total cavity loss $\mathcal{T}$ is much smaller than 1. All the cavity variables are defined in Fig. 3-2 and Table 3.1.

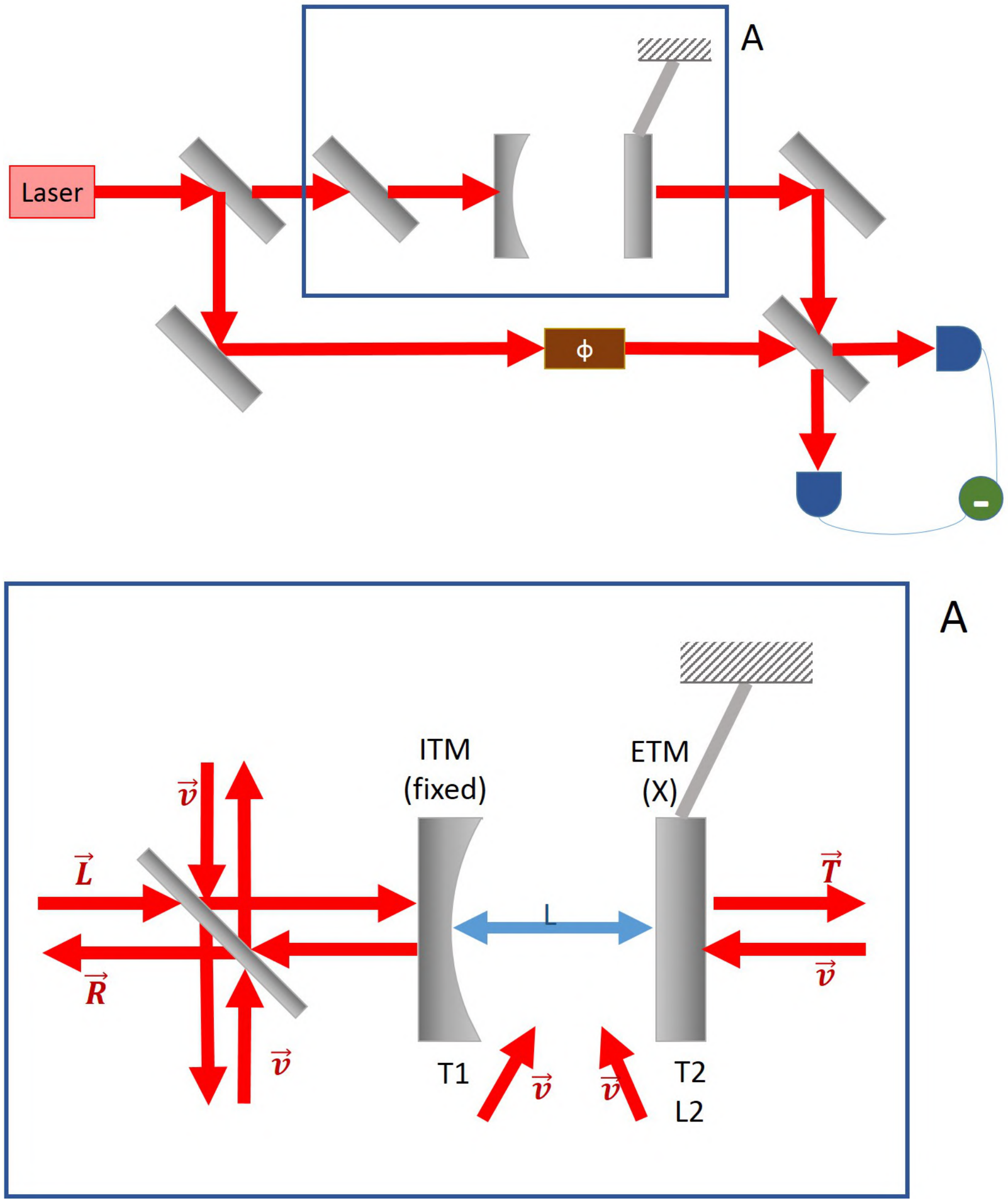

**Figure 3-2:** The first beam splitter picks off an LO beam from the laser, which is then mixed with the field exiting the cavity in a balanced homodyne detection. The inset box A shows the OM system in detail. The beam splitter here models the loss from mode matching inefficiency or any other loss that is common to the input and reflected fields. The input mirror, labeled ITM is fixed and has a variable transmission $T_1$. The end mirror, labeled ETM is the mechanical oscillator whose position is denoted by $X$. This mirror has a variable transmission $T_2$. All cavity losses are grouped together in the variable $L_2$, which is modeled as the loss at the end mirror without the loss of generality. All the open ports have been assigned uncorrelated vacuum fields ($\vec{v}$) entering the system.The two arrows below the cavity labeled $\vec{v}$ depict the vacuum that enters the system due to the loss $L_2$. The arrow pointing to the end mirror from the right represents the vacuum that enters the cavity due to the ed mirror's finite transmission. Lastly, there are two more vacua that enter on the beam splitter on left due to mode matching losses. It is to be noted that these vacuum state inputs are treated as uncorrelated noise sources, but in this figure we have chosen to denote them all as just $\vec{v}$ for ease of readability.

| Symbol | Quantity | Dependencies |
|---|---|---|
| $\hbar$ | reduced Planck's constant | $h/2\pi$ |
| $\omega, c, \lambda$ | optical frequency, speed, wavelength | |
| $L$ | cavity length | |
| $T_1$ | Input mirror transmission | |
| $T_2$ | Output mirror transmission | |
| $L_1$ | Input mirror loss | |
| $L_2$ | Output mirror loss | |
| $\rho_1$ | Input mirror amplitude reflectivity | $\sqrt{1-T_1-L_1}$ |
| $\rho_2$ | Output mirror amplitude reflectivity | $\sqrt{1-T_2-L_2}$ |
| $\mathcal{T}$ | Total loss | $T_1+T_2+L_2$ |
| $\delta$ | detuning to linewidth ratio | $\Delta/\gamma$ |
| $E^{\mathrm{R}}$ | escape efficiency in reflection | $T_1/\mathcal{T}$ |
| $E^{\mathrm{T}}$ | escape efficiency in transmission | $T_2/\mathcal{T}$ |
| $\xi_0$ | nominal squeezing quadrature | $1/2\arctan(-\delta)$ |

**Table 3.1:** Definitions for recurring variables

### 3.3.2 Covariance matrix definition

Since we will be dealing with states that have Gaussian probability distributions, we can work directly with their covariance matrices. This matrix gives the variance of the field operators in the linearized regime. The 1 represents the cosine quadrature, and the 2 represents the sine quadrature (also referred to as the amplitude and phase quadrature respectively). Each of these variances are also a function of frequency. Now, given a state that is described by the covariance matrix

$$\sigma(f)=\begin{pmatrix}\sigma_{11}(f) & \sigma_{12}(f)\\ \sigma_{21}(f) & \sigma_{22}(f)\end{pmatrix} \tag{3.1}$$

by performing a balanced homodyne detection on this state we measure the effective noise in a generalized quadrature $\xi$:

$$\begin{aligned}\mathcal{N}(\xi,f)&=\begin{pmatrix}\cos\xi & \sin\xi\end{pmatrix}\begin{pmatrix}\sigma_{11}(f) & \sigma_{12}(f)\\ \sigma_{21}(f) & \sigma_{22}(f)\end{pmatrix}\begin{pmatrix}\cos\xi\\ \sin\xi\end{pmatrix}\\ &=\cos^2\xi\,\sigma_{11}(f)+\sin^2\xi\,\sigma_{22}(f)+\cos\xi\sin\xi\,(\sigma_{12}(f)+\sigma_{21}(f))\end{aligned} \tag{3.2}$$

The covariances are normalized such that the covariance matrix for coherent states and vacuum states is identity. So a particular state is squeezed if $\mathcal{N}$ is less than 1 for any quadrature.

### 3.3.3 Normalization of classical laser noises

In this formalism, we have normalized all quantities such that the shot noise power spectral density is 1. For easy conversion, here we will describe how to convert classical laser noise as normally described in the lab into these normalized units.

[2] The 'exact' form also has an underlying approximation, which is usually satisfied in optical cavities. It is that the cavity detuning is much smaller than the free spectral range.

| Quantity | Symbol | Units | Exact | Approximate |
|---|---|---|---|---|
| Linewidth (HWHM) | $\gamma$ | Hz | $\frac{c}{2L}\frac{1-\rho_1\rho_2}{2\pi\sqrt{\rho_1\rho_2}}$ | $\frac{c\mathcal{T}}{8\pi L}$ |
| Finesse | $\mathcal{F}$ | | $\frac{1-\rho_1\rho_2}{\pi\sqrt{\rho_1\rho_2}}$ | $\frac{2\pi}{\mathcal{T}}$ |
| Intracavity Power | $P_{\mathrm{cav}}, P_0$ | $P_{\mathrm{in}}$ | – | $\frac{4E^{\mathrm{R}}}{\mathcal{T}}\frac{P_{\mathrm{in}}}{(1+\delta^2)}$ |
| Reflected Power | $P_{\mathrm{R}}$ | $P_{\mathrm{in}}$ | – | $P_{\mathrm{in}}\left(1-\frac{4E^{\mathrm{R}}(1-E^{\mathrm{R}})}{(1+\delta^2)}\right)$ |
| Transmitted Power | $P_{\mathrm{T}}$ | $P_{\mathrm{in}}$ | $\frac{T_1T_2}{(1-\rho_1\rho_2)^2}\frac{P_{\mathrm{in}}}{(1+\delta^2)}$ | $4E^{\mathrm{T}}E^{\mathrm{R}}\frac{P_{\mathrm{in}}}{(1+\delta^2)}$ |
| Carrier rotation | $\tan\theta_\delta$ | | $-\frac{2\pi L\Delta}{c}\frac{1+\rho_1\rho_2}{1-\rho_1\rho_2}$ | $-\delta$ |
| Optical spring constant | $K_{\mathrm{OS}}$ | N/m if $P_{\mathrm{cav}}$ is in W | $\frac{8\sqrt{\rho_1\rho_2}\omega P_{\mathrm{cav}}\delta}{c^2(1-\rho_1\rho_2)(1+\delta^2)}$ | $\frac{16\omega P_{\mathrm{cav}}\delta}{c^2\mathcal{T}(1+\delta^2)}$ |
| Optical spring damping | $\Gamma_{\mathrm{OS}}$ | Hz | – | $\frac{K_{\mathrm{OS}}}{2\pi^2 M\gamma(1+\delta^2)}$ |

**Table 3.2:** Fabry-Perot cavity parameters: the column 'exact' refers to making the approximation that cavity detuning is smaller than the free spectral range. The column 'approximate' refers to the cavity being a high finesse cavity. The exact expression for the intracavity power is omitted here because depending on which field is being considered, there will be a factor of $\rho_2^2$ difference. The exact expression for reflected power is not easily usable, so has also been omitted. For detailed derivation of the OS parameters, see Chapter 2.

$$S_{\mathrm{RIN}} = \mathrm{RIN}^2\frac{P_{\mathrm{in}}}{2\hbar\omega} \tag{3.3a}$$

$$= \mathrm{RIN}^2\frac{P_{\mathrm{cav}}}{2\hbar\omega}\frac{\mathcal{T}^2(1+\delta^2)}{4T_1} \tag{3.3b}$$

$$S_{\mathrm{PN}} = \frac{S_{\mathrm{ff}}}{f^2}\frac{P_{\mathrm{in}}}{2\hbar\omega} \tag{3.4a}$$

$$= \frac{S_{\mathrm{ff}}}{f^2}\frac{P_{\mathrm{cav}}}{2\hbar\omega}\frac{\mathcal{T}^2(1+\delta^2)}{4T_1} \tag{3.4b}$$

Here RIN is the relative intensity noise, $\delta P/P$ in amplitude spectral density $(1/\sqrt{Hz})$, $S_{ff}$ is the power spectral density of classical frequency noise of the laser, $f$ is the sideband frequency in Hertz. For example, for a free running NPRO laser at 1064 nm, the frequency noise is usually 100 $\mathrm{Hz}/\sqrt{\mathrm{Hz}}$ at 100 Hz, and has the shape 1/f. So if we extrapolate, we can write the noise as $10^4/f\ \mathrm{Hz}^2/\sqrt{\mathrm{Hz}}$. Which we can then use to write

$$S_{ff}^{\mathrm{NPRO}} = \frac{10^8}{f^2}\,\mathrm{Hz}^3 \tag{3.5}$$

## 3.4 Optimization layout

In order to optimize the design, we look at the amount of squeezing, and the quadrature and frequency dependence of squeezing as a function of the optical experimental parameters. We first describe an approximate analytical method, and later move on to exact optimization using numerical simulations.

We divide the various experimental parameters into constraints and variables. Constraints are most likely fixed for a given setup, and variables are the ones that need optimizing. While this particular classification is motivated by our own experimental setup, we anticipate that it is applicable in most other situations.

### 3.4.1 Constraints

- **Intracavity power**: Since the micromechanical devices are fragile, small amount of optical absorption can be enough to heat up and damage the devices at high powers. In our case, this limit was expected to be roughly near 400 mW. For our setup and most of the configurations tested in this simulation, the input power is much lower than the power limits of the lab. Hence we could do our analysis at fixed intracavity power, increasing or decreasing the input power as required upon changing the cavity parameters.

- **Loss and modematching inefficiency**: All optical systems have a certain amount of optical power lost to scattering, diffraction or absorption. We combine all of these losses inside our cavity into one number. Without the loss of generality, this loss has been implemented as a loss at the end mirror, and is called $L_2$ in the analysis. Similarly, all optical systems have a finite mode mismatch from the input laser mode to the cavity mode as well as from the cavity mode to the output mode. In our case, the input mirror is formed by a fiber mirror, which means that not only the light coming into the cavity suffers a mode mismatch loss, but the light being reflected from the cavity also is subject to the same loss as it propagates back into the fiber. This leads to the outgoing light interfering with a vacuum state and hence degrading its squeezing. Since the incoming light is not squeezed, that mode mismatch only requires an increase in the input power, without altering the quantum state of the incoming light. This is a unique case where mode mismatch becomes an important parameter in the consideration of squeezer performance, especially since we have measured our mode matching efficiency to be only 40%. For other configurations, the most important mode mismatch is most likely the mode mismatch between the local oscillator and the signal beam, which is usually lower, and is present at both the reflected and transmitted ports. For those cases, this mode matching consideration can be ignored. But since fibers allow us to build compact and modular optical systems, we foresee this being a non-negligible effect. In the future generations we hope to be able to increase the efficiency to around 80 to 90%.

  While loss and mode matching efficiency change with each fiber mirror or chip, they are not tunable by the experimenter. One will try to make them as small as possible, and then optimize all other parameters around them. We have included these parameters in our simulation as variables to understand their effects, but did most of the optimization at fixed values within the expected range.

- **Mechanical Susceptibility**: We used our knowledge from previous measurements and finite element modeling (FEM) to determine the eigenmodes, modal masses and quality factors for the oscillator. We have about 15 different mechanical designs which were chosen by running a combined optimization code including FEM and OM analysis. For our purposes here, we have fixed the mechanical oscillator, and its eigenmodes and modal masses are shown in Fig A-1 in Appendix A. Once we have those, we can write down a mechanical susceptibility for the oscillator as follows:

$$\chi(f) = \sum_{i=1}^{N} \frac{1}{(2\pi)^2 m_i (f_i^2 - f^2 - \frac{i f_i^2}{Q})} \tag{3.6}$$

  Here $m_i$ and $f_i$ are the effective mass and eigenfrequency of the i-th eigenmode, and $Q$ is the mechanical quality factor, assumed to be the same for all modes. We assume structural damping with constant loss angle, $1/Q$. A detailed prescription viscous damping and calculation of modal masses can be seen in Appendix A.

- **Classical Noises** : In real world applications, one has to deal with classical disturbances alongside quantum noises. The biggest contributors are thermal noise of the oscillator and classical laser noise. The thermal noise estimate was obtained by using the fluctuation dissipation theorem with the susceptibility in Eq. (3.6) and room temperature. This is also shown in the Appendix A. We fixed the laser RIN to a measured performance of the intensity stabilization from our lab, $10^{-8}/\sqrt{\text{Hz}}$. We also include the laser frequency noise for a free-running Mephisto NPRO laser, but it is usually a small effect. The reason is that since the local oscillator is also drawn from the same source laser, the LO and signal have a common phase noise that does not show up in the homodyne current.

- **Cavity Length** In most cases, cavity length is fixed by the frequency band of interest and the fundamental mode of the oscillator. It changes the cavity linewidth and the optical spring damping constant ($\Gamma_{\text{OS}}$). Since we're looking to design a broadband squeezer here, we chose a small length such that we were in a highly unresolved sideband limit, which gives us a wide band of low frequencies to work with. This also allows our OS to have a high Q, which again, increases the high frequency cutoff of squeezing. Another major consideration in deciding cavity length is the optimization of spatial modes and their separation in the cavity.

### 3.4.2 Variables

Once the above constraints are fixed, we're left with a few knobs that can be tuned to maximize the squeezing performance.

- **Measurement Port** : One can decide to use either the reflection or transmission port of the cavity as the main measurement port (or both!). We will mostly be focusing on squeezing in transmission. Most results can be intuitively generalised for reflection. Key differences are the squeezing angle in reflection, and the contribution of classical laser noises. If the mode-matching from input light to the cavity mode and back is low, it degrades the squeezed light exiting at the reflection port. Since in our case we have measured mode-matching efficiency of 40%, 60% of the reflected squeezed light is lost because of mode mismatch on its way back to the detector through the fiber.

- **Input Mirror Reflectivity** : The fiber mirrors can be coated for a desired reflectivity. However, for the purpose of this simulation we already had fibers that were coated. So some portion of the optimization will be done at fixed input transmission of 50 ppm. Nonetheless, we have included the dependence on input transmission for generality.

- **Output Mirror Reflectivity** : The chip reflectivity is decided by the number of Bragg layers in the mirror. This can be finely controlled during the process of growing the material via molecular beam epitaxy. A higher transmission implies fewer layers and hence lesser mass. However, we have assumed the mechanical properties do not change with reflectivity for this optimization. This is because the range in which the transmission is varied (50 to 1000 ppm) did not change the mass significantly to alter the mechanical response of the oscillator substantially.

- **Cavity Detuning** : We assumed that the cavity detuning is an easily tunable parameter in the lab, so the simulation spanned a lot of detunings. Practically speaking though, our locking scheme uses the radiation pressure back-action effect Appendix B which means we always lock on the blue side of the resonance, and higher OS frequencies increase our lock bandwidth. So detuning near $1/\sqrt{3} \approx 0.5\gamma$ is ideal. While the detuning was freely varied during the analysis; in making final decisions, we favored detunings that would facilitate ease of experiment[3].

Finally we look at how squeezing performance changes as a function of the above variables and constraints. For each cavity configuration, we look at the noise properties of reflected and transmitted light (Fig 3-16d). We also look at the contribution of individual noise sources in the transmitted and reflected light in each quadrature (Fig 3-9), as well as the individual noise contributions to the displacement spectrum (Fig 3-10) of the end mirror.

[3]We did the simulation at both fixed intracavity power irrespective of the detuning ($P(\delta) = fixed$) and fixed intracavity power at resonance ($P(\delta = 0) = fixed$). Here we are focusing on the latter. This means if you change detuning but keep all the optical losses and reflectivities same, the input power stays the same (and the intracavity power changes). The reason behind this approach is an experimental one – since the detuning in lab is easily changeable, we want to cap the input power such that if the cavity detuning wanders off to go near resonance by chance, the chip does not get damaged. This is why the best OS occurs at $\delta = 1/\sqrt{3}$, if we discuss the case of fixed detuned-cavity-power instead, the best OS occurs at $\delta = 1$.

## 3.5 Analytical approach

In this section we approach the task of optimization of an OM squeezer's optical properties analytically. In order to access this problem analytically and get the most intuition out of the results, we have to make the approximation that the OS frequency is much larger than the fundamental resonance frequency of the oscillator, as well as the measurement sideband frequency. We also have to assume that the measurement frequency is much smaller than the cavity linewidth. In addition to these approximations, in the second half of this section, we will assume that the effect of classical noises is subdominant as compared to shot noise. This allows us to treat the classical noises as small perturbations to the system. Here we will summarize the methodology and the final results, the full proof is provided in Appendix C, it is based on extending the formalism developed by Korth et al in [49].

### 3.5.1 Ideal case

We already understand quite a lot about the nature of OM squeezing in the broadband limit from Corbitt et al [20]. They analysed the amount and quadrature of squeezing for an ideal cavity – a cavity with no loss, measured in reflection, and no classical noises. For such an ideal cavity, they obtained the squeezing and squeezing quadrature to be:

$$S_{\text{ideal}}(f) = \frac{\delta^2}{2 + \delta^2 + 2\sqrt{1+\delta^2}} \tag{3.7a}$$

$$\xi_{\text{min}}(f) = \frac{1}{2}\arctan(\delta) \tag{3.7b}$$

In this work, we go beyond the ideal cavity, both from the perspective of adding losses, measuring in transmission, and having losses; as well as adding most common classical noises. Ref. [20] also consider all these effects in detail numerically, focusing on optimizing the squeezing from a Michelson interferometer configuration instead of a Fabry-Perot.

### 3.5.2 Open system

First we start with the effect of opening up the cavity, that is allowing vacuum fluctuations to enter from loss and end mirror, as well as allowing light to exit to these two ports. We denote this system with the subscript Q, for quantum, since this system has no classical noise yet. In this case, it is intuitive to think of the effect of an open cavity as a simple beam splitter. A beam splitter of loss factor $\mathcal{L}$ will change the variance of a Gaussian state from $e^{-2r}$ to $(1-\mathcal{L})e^{-2r} + \mathcal{L}$. Similarly, the effect of an open cavity will be equivalent to a loss factor of $1 - E$, where $E$ is the escape efficiency of the measurement port:

$$S_Q = S_{\text{ideal}}\, E + 1 - E \tag{3.8a}$$

$$\mathcal{T} = T_1 + T_2 + L_2 \tag{3.8b}$$

$$E^{\text{T}} = \frac{T_2}{\mathcal{T}},\ E^{\text{R}} = \frac{T_1}{\mathcal{T}} \tag{3.8c}$$

Here we have defined $S_{\text{ideal}}$ as the amount of squeezing in a cavity with only one open port and no classical noises, $\mathcal{T}$ as the total loss of the cavity, including mirror transmissions and losses. $E^{\text{T}}$ is then the escape efficiency in transmission, and $E^{\text{R}}$ is the escape efficiency in reflection. We see that when $S_{\text{ideal}}$ is less than 1, $S_{\text{Q}}$ is lower for higher $E$.

The detailed proof for this result is given in Appendix C. We start with writing all the transfer functions from all input ports to the cavity, from the cavity to output ports, from intracavity fields to radiation pressure force, from force to displacement, and from displacement to fields. We can then solve this linear system to get transfer functions from each input port to each output port. Now, since the covariance matrices are known for the input states, we obtain the covariance matrices for the output states. The covariance matrices in the absence of classical noise are:

$$\sigma_{\text{ideal}} = \begin{pmatrix} 1 & -\frac{2}{\delta} \\ -\frac{2}{\delta} & 1+\frac{4}{\delta^2} \end{pmatrix} \tag{3.9a}$$

$$\sigma_{\text{Q}} = \begin{pmatrix} 1 & -\frac{2E}{\delta} \\ -\frac{2E}{\delta} & 1+\frac{4E}{\delta^2} \end{pmatrix} \tag{3.9b}$$

Here we see that the 11 term is 1. This has a very important physical meaning, that the amplitude quadrature of an OM system will always be at the shot noise level, and won't reveal any radiation pressure effects.

### 3.5.3 Open system with classical noises

We now add in the classical noise sources to our system. We first look at how the classical noises get added to the covariance matrices. At this point no approximation about the amount of classical noises has been made. Later on, we assume the classical noise is smaller than shot noise, and apply perturbation theory to assess its impact on the variance and squeezing quadrature.

#### 3.5.3.1 Full covariance matrices

In this section, we will obtain the full covariance matrices including all contributions from classical noises, both in transmission and reflection. We will denote the additional terms to the covariance by $\delta\sigma$, with a subscript indicating the particular noise source:

$$\sigma^{\text{T}} = \sigma^{\text{T}}_{\text{Q}} + \delta\sigma^{\text{T}}_{\text{th}} + \delta\sigma^{\text{T}}_{\text{RIN}} + \delta\sigma^{\text{T}}_{\text{PN}}, \tag{3.10}$$

and superscript indicating the measurement port. The absence of superscript will mean the result is applicable for both ports.

1. Thermal Noise

   The effect of thermal noise is identical in reflection and transmission, so we'll just write it here for a general measurement port. We define a new dimensionless parameter $\lambda_{\text{th}}$ which is proportional to the force spectral density of thermal fluctuations, and allows us to simplify the covariance matrices in terms of this parameter.

$$\lambda_{\text{th}} = \frac{S_{\text{F,th}} 2\pi\gamma}{\hbar^2 g^2}, \tag{3.11}$$

   Here $\gamma$ is the cavity linewidth, (HWHM) in Hertz, and $g$ is the OM coupling constant:

$$g^2 = \frac{4\pi P_{\text{cav}}}{\hbar L\lambda}. \tag{3.12}$$

   Here, $L$ is the length of the cavity, and $\lambda$ is the optical wavelength. Using Eq. (3.12), we can rewrite $\lambda_{\text{th}}$:

$$\lambda_{\text{th}} = \frac{S_{\text{F,th}} c\mathcal{T}\lambda}{16\pi\hbar P_{\text{cav}}} \tag{3.13}$$

   Physically, we can interpret $\lambda_{\text{th}}$ as effect of thermal force in comparison to radiation pressure force. To see this, let's first write the power spectral density of radiation pressure force due to shot noise of light of power $P$:

$$\delta F_{\text{RP}} = {}^{2}\!/_{c}\,\delta P \tag{3.14a}$$

$$\delta P = \sqrt{2\hbar\omega P} \tag{3.14b}$$

$$S_{\text{F,RP}} = {}^{4}\!/_{c^2}\,(2\hbar\omega P) \tag{3.14c}$$

Barring the effects of cavity finesse, we can write an equivalent force spectral density for $P_{\rm cav}$, and using that we can rewrite $\lambda_{\rm th}$ as:

$$\lambda_{\rm th} = \mathcal{T}\frac{S_{\rm F,th}}{S_{\rm F,RP(P_{cav})}} \tag{3.15}$$

Here $S_{\rm F,RP(P_{cav})}$ is the power spectral density of single-pass radiation pressure force due to shot noise of a beam of power $P$cav. Inside the cavity though, it gets amplified by the finesse, as seen by the factor $1/\mathcal{T}$ in Eq. (3.15). On the other hand, the thermal force is not dependent on the cavity at all, so gets no effect from finesse.

Now, using the $\lambda_{\rm th}$ in Eq. (3.11), we can write the covariance matrix for the light at either measurement port, in the presence of thermal noise.

$$\sigma_{\rm th} = \begin{pmatrix} 1 + E\,\lambda_{\rm th} & -\frac{E}{\delta}\left(2 + \lambda_{\rm th}\right) \\ -\frac{E}{\delta}\left(2 + \lambda_{\rm th}\right) & 1 + \frac{E}{\delta^2}\left(4 + \lambda_{\rm th}\right) \end{pmatrix} \tag{3.16a}$$

$$= \sigma_{\rm Q} + \delta\sigma_{\rm th}, \tag{3.16b}$$

where E is the escape efficiency of the measurement port, and we have defined a matrix that carries thermal noise as

$$\delta\sigma_{\rm th} = E\,\lambda_{\rm th}\begin{pmatrix} 1 & -1/\delta \\ -1/\delta & 1/\delta^2 \end{pmatrix}. \tag{3.17}$$

Finally, notice here that for a fixed cavity power, the thermal noise term is proportional to $E\lambda_{\rm th}$, which is proportional to $T_i$, and doesn't depend on $\mathcal{T}$. So in order to improve squeezing via reducing the effect of thermal noise, one should lower the transmission of the measurement port. This will be more clearly seen in Eq. (3.27).

2. Classical Laser Noises

Next, we look at the effect of classical laser noise on the covariance matrices in transmission and reflection. Classical laser noise includes the classical relative intensity noise on the laser, as well as classical frequency noise. Since the reflection port involves interference with the promptly reflected laser field, the effect of classical laser noise in transmission is not the same as in reflection. So unlike the thermal noise, we need to look at the effect of laser noise in transmission and reflection separately.

2 (a) **Transmission**

In transmission, all of the classical noise from the laser can be parametrized into a single parameter $\lambda_{\rm CLN}$. This parameter encompasses both relative intensity noise, as well as frequency noise on the incoming laser. The covariance matrix for the field in transmission with this effect is:

$$\sigma^{\rm T}_{\rm CLN} = \begin{pmatrix} 1 & -\frac{2E^{\rm T}}{\delta} \\ -\frac{2E^{\rm T}}{\delta} & 1 + \frac{4E^{\rm T}}{\delta^2}\left(1 + E^{\rm R}\lambda_{\rm CLN}\right) \end{pmatrix} \tag{3.18}$$

where we have defined the $\lambda_{\rm CLN}$ parameter:

$$\lambda_{\rm CLN} = \lambda_{\rm RIN} + \lambda_{\rm PN} \tag{3.19a}$$

$$\lambda_{\rm RIN} = S_{\rm RIN}\frac{1}{1+\delta^2} \tag{3.19b}$$

$$= {\rm RIN}^2\frac{P_{\rm cav}}{2\hbar\omega}\frac{\mathcal{T}^2}{4T_1} \tag{3.19c}$$

$$\lambda_{\rm PN} = S_{\rm PN}\frac{\delta^2}{1+\delta^2} \tag{3.19d}$$

$$= \frac{S_{\rm ff}}{f^2}\frac{P_{\rm cav}}{2\hbar\omega}\frac{\mathcal{T}^2\delta^2}{4T_1} \tag{3.19e}$$

$$\lambda_{\mathrm{CLN}} = \frac{P_{\mathrm{cav}}}{2\hbar\omega}\frac{\mathcal{T}^2}{4T_1}\left(\mathrm{RIN}^2 + \delta^2\frac{S_{\mathrm{ff}}}{f^2}\right) \tag{3.20}$$

The perturbation term depends on $\lambda_{\mathrm{CLN}}\ E^{\mathrm{T}}E^{\mathrm{R}}$ , which is proportional to just $T_2$, and not $\mathcal{T}$ and $T_1$. So to reduce the impact of classical laser noise, one should also decrease $T_2$. Interestingly though, the effect of classical laser noise in transmission will be worse for higher circulating power in the cavity. This is simply because the contribution of classical laser noise w.r.t. shot noise ($S_{\mathrm{RIN}}$ and $S_{\mathrm{PN}}$, Eqs. (3.3b) and (3.4b)) is proportional to power. For a fixed intracavity power and $T_2$, a lower detuning will reduce the impact of frequency noise.

Now let's also understand the reason behind the covariance matrix in transmission being simply depending on $\lambda_{\mathrm{CLN}}$, and only in the 22 term. The transfer function matrix from the input port to the transmission port is:

$$\begin{bmatrix} a^{\mathrm{T}}_{\mathrm{out},1} \\ a^{\mathrm{T}}_{\mathrm{out},2} \end{bmatrix} = \begin{pmatrix} 0 & 0 \\ \frac{2\sqrt{E^{T}E^{R}}}{\delta} & 0 \end{pmatrix} \begin{bmatrix} a_{\mathrm{in},1} \\ a_{\mathrm{in},2} \end{bmatrix}, \tag{3.21}$$

The only non-zero term is the 21 term, which means that only the 11 term of the input field's covariance matrix couples to the transmitted field, and it only couples to the 22 term of the transmitted field. Now we can find out what this 11 term for the input field should be. We know that when the cavity is detuned, the carrier in the cavity is rotated with respect to the carrier of the incoming laser. Since this calculation is done in the basis of the cavity, we need to project the known laser noise in the laser basis on to the cavity basis.

$$S_{\mathrm{in}}\big|_{\text{laser basis}} = \begin{pmatrix} 1+S_{\mathrm{RIN}} & 0 \\ 0 & 1+S_{\mathrm{PN}} \end{pmatrix} \tag{3.22a}$$

$$S_{\mathrm{in}}\big|_{\text{cavity basis}} = \mathbb{R}(\theta_\delta)\cdot S_{\mathrm{in}}\big|_{\text{laser basis}}\cdot \mathbb{R}(\theta_\delta)^{\dagger}, \tag{3.22b}$$

where $\theta_\delta$ is the rotation caused by the cavity detuning, $\tan\theta_\delta = -\delta$, giving

$$S_{\mathrm{in}}\big|_{\text{cavity basis}} = \mathbb{I} + \frac{S_{\mathrm{RIN}}}{1+\delta^2}\begin{pmatrix} 1 & -\delta \\ -\delta & \delta^2 \end{pmatrix} + \frac{S_{\mathrm{PN}}}{1+\delta^2}\begin{pmatrix} \delta^2 & \delta \\ \delta & 1 \end{pmatrix} \tag{3.23a}$$

$$= \frac{1}{1+\delta^2}\begin{pmatrix} 1+S_{\mathrm{RIN}}+(1+S_{\mathrm{PN}})\,\delta^2 & \delta\,(S_{\mathrm{PN}}-S_{\mathrm{RIN}}) \\ \delta\,(S_{\mathrm{PN}}-S_{\mathrm{RIN}}) & 1+S_{\mathrm{PN}}+\delta^2\,(1+S_{\mathrm{RIN}}) \end{pmatrix}. \tag{3.23b}$$

This shows that the 11 term in the cavity basis for the classical laser noise is $1+\lambda_{\mathrm{CLN}}$.

It is crucial to note that this simplification is because of all the 0 terms in Eq. (3.21), which are in turn a result of the approximation that the OS frequency is larger than the measurement frequency and the fundamental resonance frequency of the oscillator. In the absence of this approximation, the 11, 12, and 22 terms of that transfer function are proportional to $\Omega^2$ [4], while the 21 term has a constant plus a term proportional to $\Omega^2$. So all these transfer functions show basically a suppression of all fields coming from the laser into the cavity, upto the OS, which is easily understandable as the expected response of the OS, as is also seen in detail in Chapter 2.

2 (b) **Reflection**

In reflection, due to the interference of the cavity field with the promptly reflected laser field, classical laser noise becomes a little more complicated. We don't have a fixed universal combination of $S_{\mathrm{RIN}}$ and $S_{\mathrm{PN}}$, so we need to keep them separate and look at each of their effects individually.

Effect of intensity noise in reflection:

[4] Or $\Omega_0^2$ for measurement frequencies below the fundamental frequency, or $\Omega_{\mathrm{eff}}^2 = \Omega_0^2 - \Omega^2 + i\Omega_0^2/Q$ for a generic oscillator. The important approximation is $\Omega_{\mathrm{eff}} \ll \Omega_{\mathrm{OS}}$.

$$\sigma^{\mathrm{R}}_{\mathrm{RIN}} = \sigma^{\mathrm{R}}_{\mathrm{Q}} + \delta\sigma^{\mathrm{R}}_{\mathrm{RIN}} \tag{3.24a}$$

$$\delta\sigma^{\mathrm{R}}_{\mathrm{RIN}} = \lambda_{\mathrm{RIN}} \begin{pmatrix} 1 & -\frac{2E^{\mathrm{R}}+\delta^2}{\delta} \\ -\frac{2E^{\mathrm{R}}+\delta^2}{\delta} & \left(\frac{2E^{\mathrm{R}}+\delta^2}{\delta}\right)^2 \end{pmatrix} \tag{3.24b}$$

Effect of phase noise in reflection:

$$\sigma^{\mathrm{R}}_{\mathrm{PN}} = \sigma^{\mathrm{R}}_{\mathrm{Q}} + \delta\sigma^{\mathrm{R}}_{\mathrm{PN}} \tag{3.25a}$$

$$\delta\sigma^{\mathrm{R}}_{\mathrm{PN}} = \lambda_{\mathrm{PN}} \begin{pmatrix} 1 & \frac{1-2E^R}{\delta} \\ \frac{1-2E^R}{\delta} & \left(\frac{1-2E^R}{\delta}\right)^2 \end{pmatrix} \tag{3.25b}$$

#### 3.5.3.2 Squeezing in the presence of classical noises

Now we use the covariance matrices from the previous section to inform us about the nature of squeezing in the presence of classical noises. The eigenvalues of the covariance matrix give the level of squeezing and anti-squeezing, and the corresponding eigenvectors give the quadrature of squeezing and anti-squeezing. In order to find the effect of classical noises on the amount and quadrature of squeezing, we apply first order perturbation theory to get the new eigenvalues and eigenvectors.

For example, we can write the full covariance matrix in transmission, which includes the effects of non-unity escape efficiency and classical noise:

$$\sigma^{\mathrm{T}}_{\mathrm{th,CLN}} = \sigma^{\mathrm{T}}_{\mathrm{Q}} + \delta\sigma^{\mathrm{T}}_{\mathrm{CLN}} + \delta\sigma^{\mathrm{T}}_{\mathrm{th}} \tag{3.26a}$$

$$\sigma^{\mathrm{T}}_{\mathrm{Q}} = \begin{pmatrix} 1 & -\frac{2E^{\mathrm{T}}}{\delta} \\ -\frac{2E^{\mathrm{T}}}{\delta} & 1+\frac{4E^{\mathrm{T}}}{\delta^2} \end{pmatrix} \tag{3.26b}$$

$$\delta\sigma^{\mathrm{T}}_{\mathrm{CLN}} = 4E^{\mathrm{T}}E^{\mathrm{R}}\lambda_{\mathrm{CLN}} \begin{pmatrix} 0 & 0 \\ 0 & 1/\delta^2 \end{pmatrix} \tag{3.26c}$$

$$\delta\sigma^{\mathrm{T}}_{\mathrm{th}} = E^{\mathrm{T}}\lambda_{\mathrm{th}} \begin{pmatrix} 1 & -1/\delta \\ -1/\delta & 1/\delta^2 \end{pmatrix} \tag{3.26d}$$

Applying perturbation theory to each of the classical noise contributions, we can then find the new eigenvalues. The smaller eigenvalue gives the PSD in the squeezed quadrature:

$$S^{\mathrm{T}} = S_{\mathrm{Q}}^{\mathrm{T}} + \delta S_{\mathrm{th}}^{\mathrm{T}} + \delta S_{\mathrm{RIN}}^{\mathrm{T}} + \delta S_{\mathrm{PN}}^{\mathrm{T}} \tag{3.27a}$$

$$S_{\mathrm{Q}}^{\mathrm{T}} = 1 - E^{\mathrm{T}}\left(1 - S_{\mathrm{ideal}}\right) \tag{3.27b}$$

$$= 1 - \frac{T_2}{\mathcal{T}} \frac{2(-1 + \sqrt{1+\delta^2})}{\delta^2} \tag{3.27c}$$

$$\delta S_{\mathrm{th}}^{\mathrm{T}} = E^{\mathrm{T}} \lambda_{\mathrm{th}} \frac{1+\delta^2}{2\left(1+\delta^2+\sqrt{1+\delta^2}\right)} \tag{3.27d}$$

$$= T_2 \frac{S_{\mathrm{F,th}} c \lambda \left(1+\delta^2\right)}{32\pi \hbar P_{\mathrm{cav}} \left(1+\delta^2+\sqrt{1+\delta^2}\right)} \tag{3.27e}$$

$$\delta S_{\mathrm{RIN}}^{\mathrm{T}} = 4E^{\mathrm{T}} E^{\mathrm{R}} \lambda_{\mathrm{RIN}} \frac{1}{2\left(1+\delta^2+\sqrt{1+\delta^2}\right)} \tag{3.27f}$$

$$= T_2 \mathrm{RIN}^2 \frac{P_{\mathrm{cav}}}{2\hbar\omega} \frac{1}{2\left(1+\delta^2+\sqrt{1+\delta^2}\right)} \tag{3.27g}$$

$$\delta S_{\mathrm{PN}}^{\mathrm{T}} = 4E^{\mathrm{T}} E^{\mathrm{R}} \lambda_{\mathrm{PN}} \frac{1}{2\left(1+\delta^2+\sqrt{1+\delta^2}\right)} \tag{3.27h}$$

$$= T_2 \frac{S_{\mathrm{ff}}}{f^2} \frac{P_{\mathrm{cav}}}{2\hbar\omega} \frac{\delta^2}{2\left(1+\delta^2+\sqrt{1+\delta^2}\right)} \tag{3.27i}$$

These results show us a few important things. In order to improve squeezing in the absence of classical noises, one should maximize the escape efficiency and minimize the detuning. Once the contribution from classical noises is added, things become more complicated. First of all, both thermal noise and classical laser noise have a higher contribution for a higher $T_2$. This means if we increase $T_2$ to improve the squeezing from the quantum fluctuations, we will also increase the coupling of classical noises. So the optimum scenario would be to minimize $T_2$, but maximize the escape efficiency. The limits on this would be placed by the technical abilities to build up an optical power of $P_{\mathrm{cav}}$ inside the cavity for the available amount of input power. A high intracavity power lowers the coupling of thermal noise to squeezing due to higher ratio of quantum radiation pressure (QRP) displacement to thermal noise. It is interesting to note that the absolute amount of QRP induced displacement does not matter, as can be seen in the absence of thermal noise (or other classical noises) in the system. This is because everything is scaled to the shot noise. So for higher powers we have a higher QRPN, but also higher shot noise. This means when the final noise is scaled to its shot noise, we won't see more squeezing. On the other hand, the coupling of classical laser noises will increase with intracavity power. This effect can be understood by comparing the classical noise to quantum noise for the laser. Classical noise PSD varies as the square of power, whereas the shot noise PSD varies linearly with power, so in the end increasing the power increases the fractional contribution of classical noise w.r.t. shot noise. Since the current state of materials is such that the thermal noise is still a limiting factor, a better strategy is to lower the RIN or the frequency noise by prestabilization, and have the maximum power possible to combat thermal noise. Finally, increasing the detuning of the cavity while keeping all other parameters fixed, will decrease the relative coupling of RIN to squeezing, but increase the coupling of the other three sources. So if the system is prestabilized with low enough RIN, again working at smaller detunings will lead to more squeezing. In all of this, one must also keep in mind the frequency of the OS, which will be the high frequency limit of the squeezing bandwidth. While increasing the cavity finesse and power will make the squeezing and the OS better, decreasing the detuning will increase squeezing, but decreasing it below 0.5 will lead to a decrease in the squeezing bandwidth due to the OS.

Further treatment of this subject is easier done in a new basis. This basis is formed by the eigenvectors of the unperturbed covariance matrix $\sigma_{\mathrm{Q}}$, and diagonalizes the unperturbed covariance matrix. We then must also transform the perturbation matrices to this new basis. Since the squeezing quadrature is given by Eq. (3.7b), the unperturbed covariance matrix can be diagonalized by setting $\delta$ to $\tan(2\xi_0)$:

$$\delta \to \tan(2\xi_0) \tag{3.28a}$$

$$\sigma_{\mathrm{Q}} \to \begin{pmatrix} 1 & -2E\cot 2\xi_0 \\ -2E\cot 2\xi_0 & 1+4E\cot^2 2\xi_0 \end{pmatrix} \tag{3.28b}$$

Now the basis transformation is just a rotation $\mathbb{R}(\xi_0)$ (i.e. going in the $(\cos\xi_0, \sin\xi_0)$ basis):

$$\sigma'_{\mathrm{Q}} = \mathbb{R}(\xi_0)^\dagger\ \sigma_{\mathrm{Q}}\ \mathbb{R}(\xi_0) \tag{3.29a}$$

$$= \begin{pmatrix} 1-E+E\tan^2\xi_0 & 0 \\ 0 & 1-E+E\cot^2\xi_0 \end{pmatrix}. \tag{3.29b}$$

The eigenvectors for this new basis are now just unit vectors, while the eigenvalues stay the same as before. We rewrite the eigenvalues also as a function of $\xi_0$:

$$S_{\mathrm{Q,min}} = 1-E+E\tan^2\xi_0;\ |0\rangle = \begin{pmatrix} 1 \\ 0 \end{pmatrix} \tag{3.30a}$$

$$S_{\mathrm{Q,max}} = 1-E+E\cot^2\xi_0;\ |1\rangle = \begin{pmatrix} 0 \\ 1 \end{pmatrix} \tag{3.30b}$$

Now, let's apply perturbation theory to get the perturbation added by classical noise. We are interested in the change in eigenvalues (i.e. change in minor and major axis of the squeezing ellipse), as well as the change in eigenvectors (i.e. rotation of the squeezing ellipse). In the limit that the classical noises only cause a small rotation $\delta\theta^{(1)}$ of the eigenvectors, the perturbation to the eigenvectors can be expressed as :

$$\delta\mathbf{v^{(1)}} = \mathbf{v^{(1)}} - \mathbf{v_0} \tag{3.31a}$$

$$= \mathbb{R}(\delta\theta^{(1)})\mathbf{v_0} - \mathbf{v_0} \tag{3.31b}$$

$$\approx \begin{cases} \begin{pmatrix} 0 \\ \delta\theta^{(1)} \end{pmatrix}, & \text{if } \mathbf{v_0} = |0\rangle \\ \begin{pmatrix} -\delta\theta^{(1)} \\ 0 \end{pmatrix}, & \text{if } \mathbf{v_0} = |1\rangle \end{cases} \tag{3.31c}$$

With these simplifications, we then obtain the change in eigenvalue and eigenvectors and express that with three quantities: change in squeezing, change in anti-squeezing, and rotation angle. We do this for each noise source at both reflection and transmission ports. We first calculate the effect of thermal noise, find the perturbation matrix from thermal noise, and then use the unperturbed eigenvectors and eigenvalues to calculate the perturbed ones.

$$\delta\sigma'_{\mathrm{th}} = \mathbb{R}(\xi_0)^\dagger\ (\sigma_{\mathrm{th}} - \sigma_{\mathrm{Q}})\ \mathbb{R}(\xi_0) \tag{3.32a}$$

$$= \frac{1}{4}E\lambda_{\mathrm{th}} \begin{pmatrix} \sec^2\xi_0 & -\csc\xi_0\sec\xi_0 \\ -\csc\xi_0\sec\xi_0 & \csc^2\xi_0 \end{pmatrix} \tag{3.32b}$$

$$\delta S_{\mathrm{th,i}} = \mathbf{v_{0,i}}^\dagger \delta\sigma'_{\mathrm{th}} \mathbf{v_{0,i}} \tag{3.32c}$$

$$= \frac{1}{4}E\lambda_{\mathrm{th}}\sec^2\xi_0;\ \frac{1}{4}E\lambda_{\mathrm{th}}\csc^2\xi_0 \tag{3.32d}$$

$$\delta\mathbf{v}_{\mathrm{th,i}}{}' = \frac{\mathbf{v_{0,j}}^\dagger \delta\sigma'_{\mathrm{th}}\mathbf{v_{0,i}}}{S_{0,i} - S_{0,i}}\mathbf{v_{0,j}} \tag{3.32e}$$

$$= \begin{pmatrix} 0 \\ \delta\xi_{\mathrm{th}} \end{pmatrix};\ \begin{pmatrix} -\delta\xi_{\mathrm{th}} \\ 0 \end{pmatrix} \tag{3.32f}$$

| Port | Noise | Squeezing | Antisqueezing | Rotation |
|---|---|---|---|---|
| T | $\lambda_{\text{th}}$ | $\frac{1}{4}E^{\text{T}}\sec^2\xi_0$ | $\frac{1}{4}E^{\text{T}}\csc^2\xi_0$ | $\frac{1}{8}\tan 2\xi_0$ |
| T | $E^{\text{R}}\lambda_{\text{CLN}}$ | $E^{\text{T}}\cos^2 2\xi_0\sec^2\xi_0$ | $E^{\text{T}}\cos^2 2\xi_0\csc^2\xi_0$ | $-\frac{1}{4}\sin 4\xi_0$ |
| R | $\lambda_{\text{th}}$ | $\frac{1}{4}E^{\text{R}}\sec^2\xi_0$ | $\frac{1}{4}E^{\text{R}}\csc^2\xi_0$ | $\frac{1}{8}\tan 2\xi_0$ |
| R | $\lambda_{\text{RIN}}\left(E^{\text{R}}\to 1\right)$ | $16\csc^2 4\xi_0\sin^6\xi_0$ | $\cos^2\xi_0\cot^2\xi_0\sec^2 2\xi_0$ | $-\frac{1}{8}\tan^3 2\xi_0$ |
| R | $\lambda_{\text{PN}}\left(E^{\text{R}}\to 1\right)$ | $\frac{1}{4}\sec^2\xi_0$ | $\frac{1}{4}\csc^2\xi_0$ | $\frac{1}{8}\tan 2\xi_0$ |

**Table 3.3:** Effect of classical noises on squeezing and squeezing quadrature: This table summarizes the effects of classical noise on squeezing, anti-squeezing, and the rotation of the squeezing ellipse due to each noise source, both in reflection (R) and transmission (T). The column 'Noise' shows the classical noise factor for the respective noise source, each quantity in the next three columns must be multiplied by this factor. $\xi_0$ is the squeezing quadrature in the absence of all classical noises, given by $\tan 2\xi_0 = \delta$; $E^{\text{T}}$ and $E^{\text{R}}$ are the escape efficiencies in transmission and reflection. For the last two rows, the expressions have been simplified by assuming full escape efficiency in reflection. The full expressions for a non-unity escape efficiency in reflection can be found in Appendix C.

where we have defined $\delta\xi_{\text{th}} = \dfrac{\lambda_{\text{th}}\tan 2\xi_0}{8}$. This angle represents the small rotation that occurs to the ellipse due to the addition of a small thermal noise. Another interesting question we can ask is, which are the quadratures of minimum and maximum thermal noise. In order to do this, we simply find the eigenvalues and eigenvectors of $\delta\sigma_{\text{th}}$. We find that in fact all the thermal noise is located one quadrature, $2\xi_0 + \pi/2$, and that there is zero thermal noise in the quadrature $2\xi_0$. The total power spectral density of thermal noise is $E\lambda_{\text{th}}\csc^2(2\xi_0)$. So all the thermal noise is in a quadrature that is $\xi_0$ counterclockwise from the anti-squeezing quadrature, and there should be no thermal noise at twice the squeezing quadrature. Similarly, from Eq. (3.26c) it's evident that all the laser classical noise would be in the phase quadrature for transmitted light.

Following the same procedure, we find the change in squeezing and squeezing quadrature for all other contributions, and the final results from this exercise are listed in Table 3.3. We see that the rotation of the squeezing ellipse in transmission due to thermal noise is positive, while due to classical laser noises is negative. This is understandable, because thermal noise is zero at $2\xi_0$, so it pulls the squeezing quadrature towards $2\xi_0$, whereas classical laser noise is zero in the amplitude quadrature, and hence pulls the squeezing quadrature towards 0. It is also interesting to note that while the rotation on squeezing in reflection from RIN is negative for a full escape efficiency in reflection, it is positive for lower escape efficiencies. Similarly, the rotation from phase noise is positive for a full escape efficiency, but will be negative for lower escape efficiencies. Both these effects are shown in Fig. 3-3, these are the only quantities that will switch sign of rotation as one goes towards lower escape efficiencies. The exact analytical expressions for coupling of classical laser noises for non-unity escape efficiency in reflection can be found in Appendix C, but we have included here a plot (Fig. 3-4) showing how each noise couples as a function of escape efficiency.

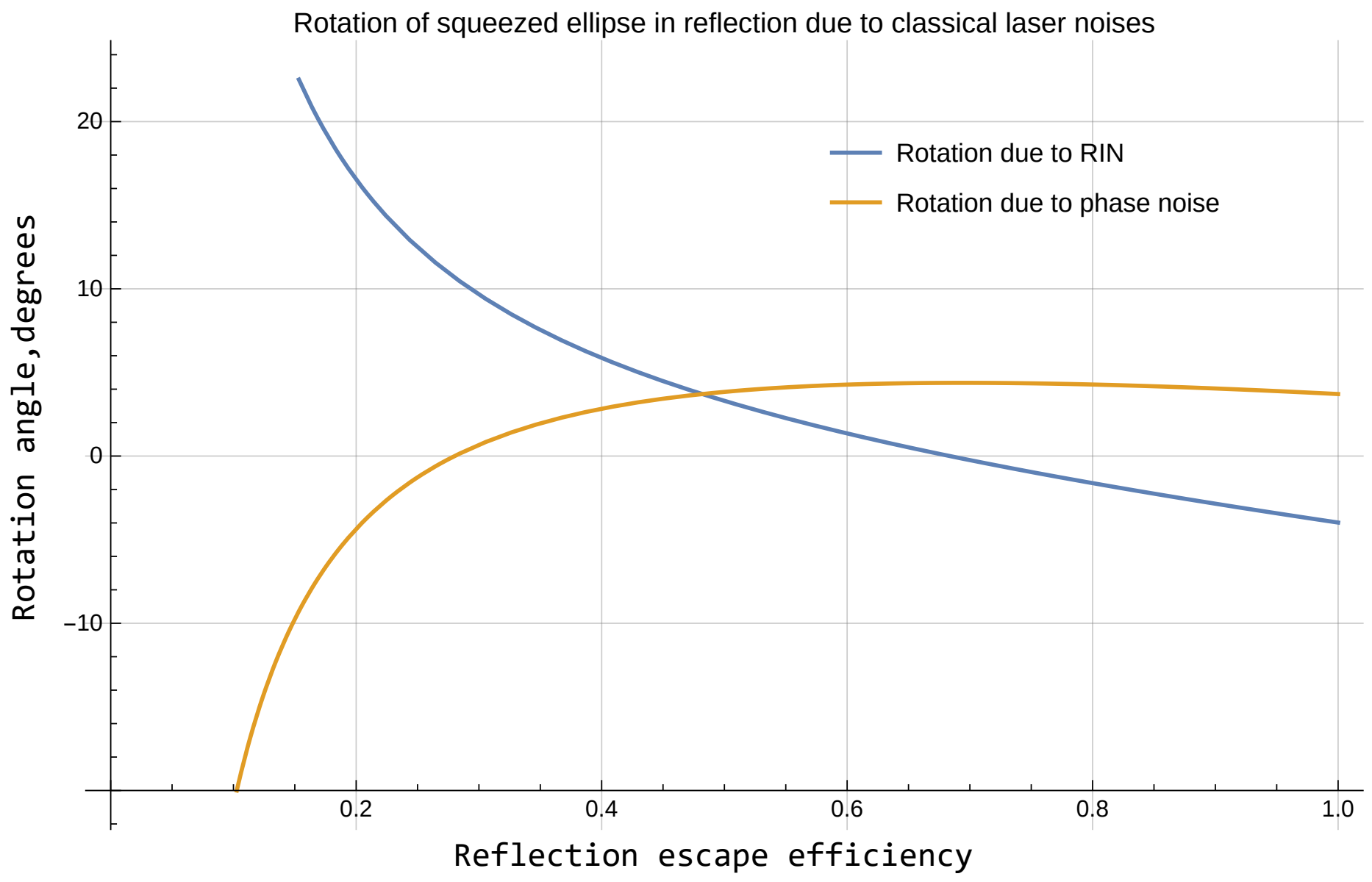

**Figure 3-3:** Rotation of squeezing ellipse in reflection due to classical laser noises: This plot is made for a base squeezing angle of 23 degrees (thus a detuning of 1), and both $\lambda_{RIN}$, $\lambda_{PN}$ of 0.5. The rotation due to RIN switches from positive to negative as the escape efficiency is lowered, and vice versa for phase noise.

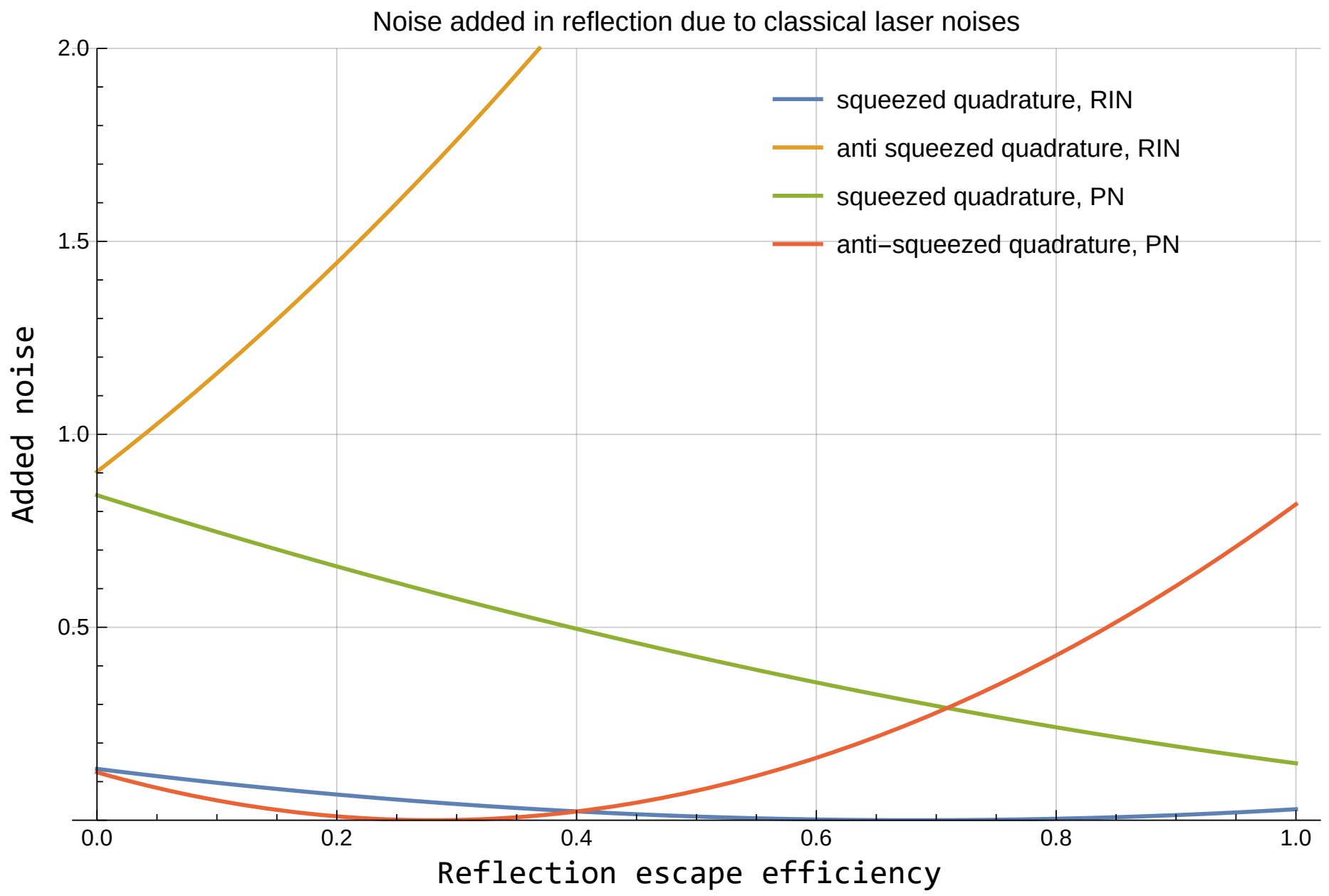

**Figure 3-4:** Enlargement of the squeezing ellipse in reflection due to classical laser noises: This plot shows the noise added due to classical noise relative to shot noise for a base squeezing angle of 23 degrees (thus a detuning of 1), and both $\lambda_{RIN}$, $\lambda_{PN}$ of 0.5.

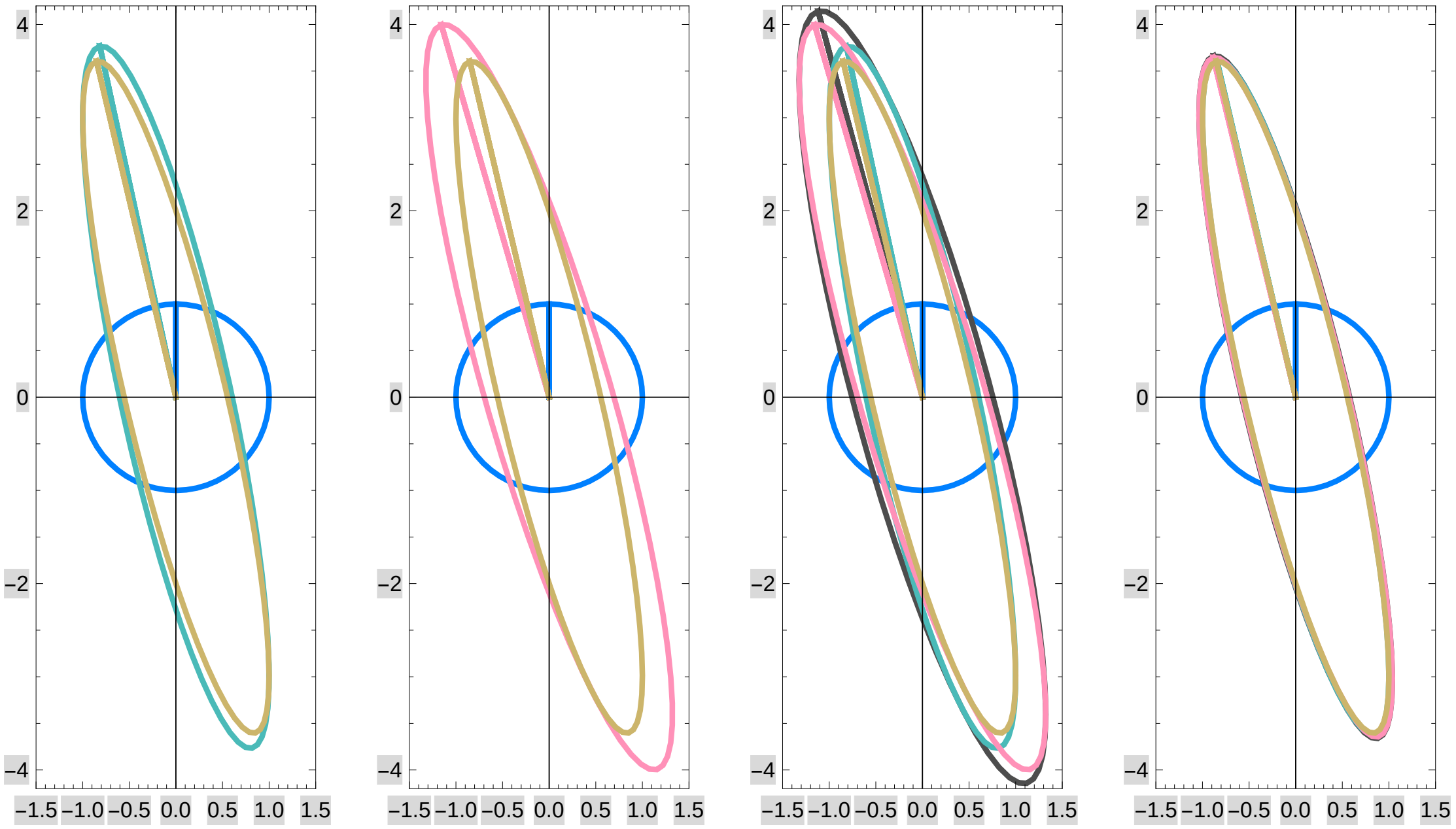

**(a)** $\lambda_{\mathrm{CLN}} = 1$
$\lambda_{\mathrm{th}} = 0$

**(b)** $\lambda_{\mathrm{CLN}} = 0$
$\lambda_{\mathrm{th}} = 1$

**(c)** $\lambda_{\mathrm{CLN}} = 1$
$\lambda_{\mathrm{th}} = 1$

**(d)** $\lambda_{\mathrm{CLN}} = .1$
$\lambda_{\mathrm{th}} = .1$

**Figure 3-5:** Effect of classical noises on squeezing in transmission: here we show the distribution of the variance of squeezed states in transmission as a function of added classical noises. The blue circle is the vacuum state, the inner dark-yellow ellipse is the squeezed state with no classical noise, for a detuning $\delta = 0.5$ and escape efficiency $E^{\mathrm{T}} = 0.75$. We then add classical noises. These noise ellipses are made without perturbation theory, for the full covariance matrix, so $\lambda$ parameters don't have to be less than 1. **(a)** the teal ellipse has classical laser noise with $\lambda_{\mathrm{CLN}} = 1$ and $E^{\mathrm{R}} = 0.1$. We can see that the noise increases in squeezing and anti-squeezing quadrature, as well as the ellipse rotates clockwise towards the amplitude quadrature. **(b)** the pink ellipse has thermal noise with $\lambda_{\mathrm{th}} = 1$. There is increased noise, and the ellipse rotates towards the phase quadrature. **(c)** Combined effect of thermal noise and laser noise, the dark-gray ellipse includes both thermal noise and laser noise. **(d)** full noise ellipse with more realistic estimates for $\lambda's$, as opposed to (a - c), which are made for large $\lambda's$ to show the effect.

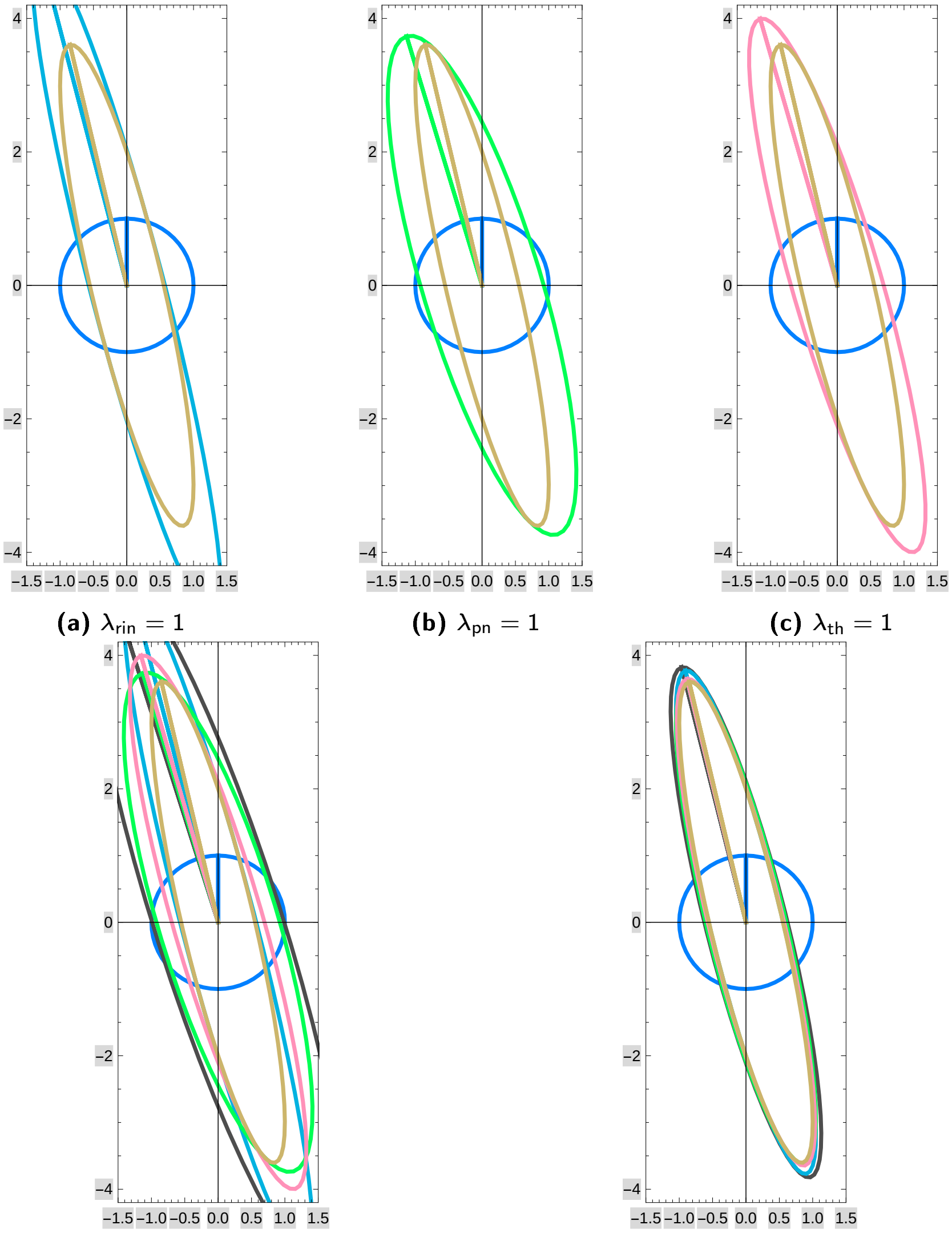

**(a)** $\lambda_{\text{rin}} = 1$ **(b)** $\lambda_{\text{pn}} = 1$ **(c)** $\lambda_{\text{th}} = 1$

**(d)** $\lambda_{\text{rin}} = 1,\ \lambda_{\text{pn}} = 1,\ \lambda_{\text{th}} = 1$ **(e)** $\lambda_{\text{rin}} = 0.1,\ \lambda_{\text{pn}} = 0.1,\ \lambda_{\text{th}} = 0.1$

**Figure 3-6**: Effect of classical noises on squeezing in reflection: Similar to Fig. 3-5, this figure shows the effect of classical laser noise and thermal noise on squeezing. These ellipses are made for a reflection escape efficiency $E^{\text{R}} = .75$, and a detuning $\delta = 0.5$. As explained in text, the contribution of phase noise and RIN is analysed separately in reflection squeezing. Here the color scheme is same as Fig. 3-5. Additionally, light-blue ellipse is with the contribution of RIN, and the bright-green ellipse is with the contribution of phase noise.

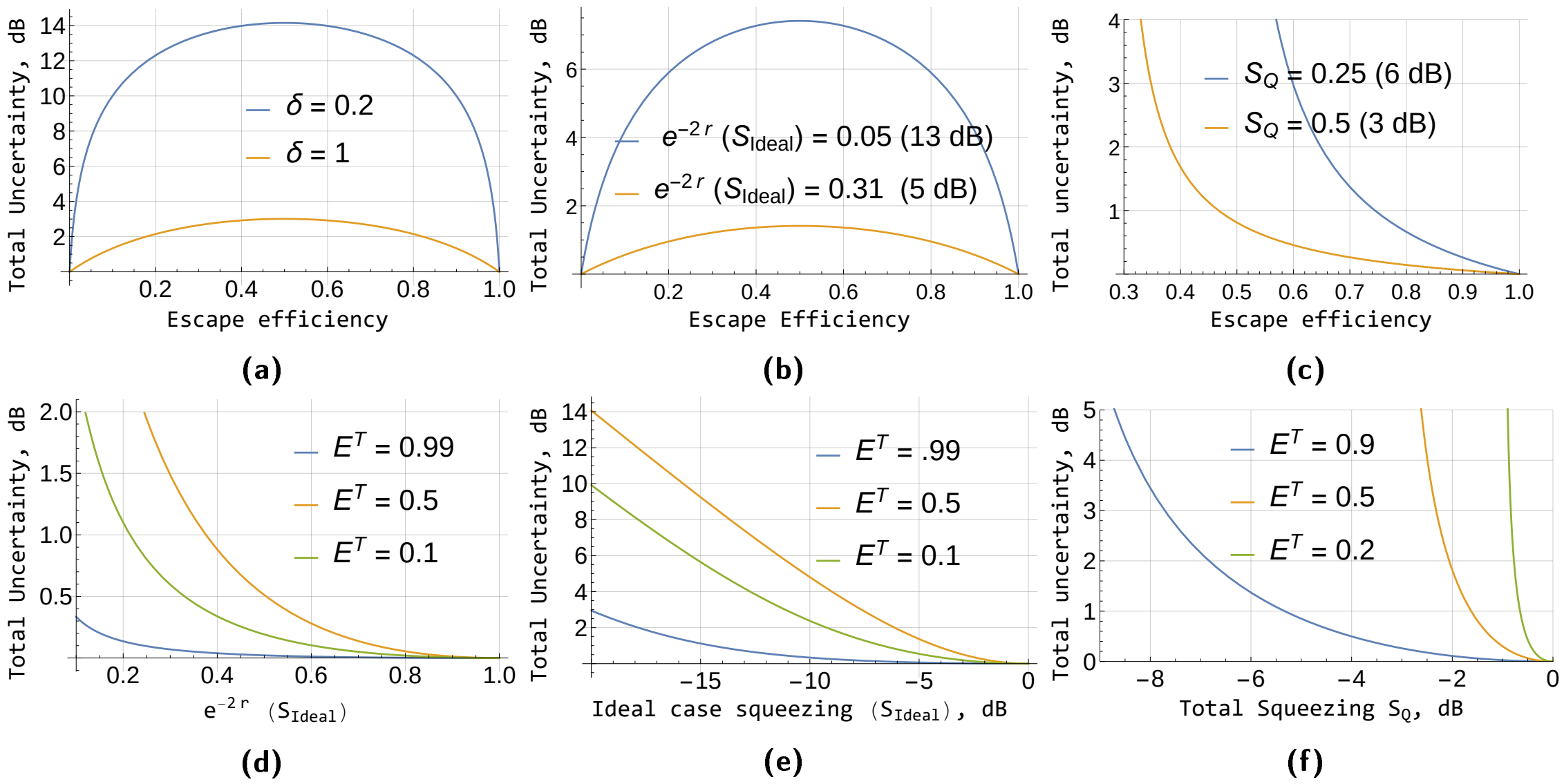

**Figure 3-7:** Total uncertainty of squeezed state in transmission as a function of escape efficiency and amount of squeezing: **(a)** shows that as one goes from maximum escape efficiency to lower, the total uncertainty increases, it increases more for the state with more squeezing. But, after the escape efficiency is lower than 50%, the total uncertainty starts decreasing again. This is because at lower escape efficiencies, the state is neither highly squeezed, nor highly anti-squeezed, and eventually converges to a vacuum state for zero escape efficiency. **(b)** shows similar effects, this time made at fixed maximum squeezing instead of detuning. **(c)** shows total uncertainty as a function of escape efficiency at fixed actual squeezing(i.e. squeezing after taking escape efficiency into account). We again see that the uncertainty increases faster for a higher squeezed state. **(d)** and **(e)** show the total uncertainty at fixed escape efficiency as a function of maximum squeezing. More squeezing leads to higher total uncertainty, and the increase in uncertainty is higher as escape efficiency gets closer to 0.5. **(f)** we see uncertainty as a function of actual squeezing, we see that more squeezing leads to more uncertainty for a fixed escape efficiency. We also see that for lower escape efficiency, the achievable squeezing asymptotes to a maximum available squeezing (e.g. 3 dB for 50% escape efficiency), but to achieve that one must have infinite anti-squeezing.

### 3.5.4 Total uncertainty

Given that we have the full covariance matrix of the squeezed states, we can also ask how much is the total uncertainty in a given state, and how does the total uncertainty change with the optical parameters. This is useful if one desires to perform quantum operations with states. The determinant of a $2 \times 2$ covariance matrix gives the total uncertainty in that state. The total uncertainty in the transmitted or reflected light in the absence of any classical noises is

$$\frac{4E(1-E)+\delta^2}{\delta^2}, \tag{3.33}$$

showing that a non-unity escape efficiency increases the total uncertainty of the states. Additionally, the more squeezed a state is (i.e. lower detuning), the worse is the total uncertainty. This is a direct consequence of entanglement between the various ports of the cavity (i.e. transmission, reflection, and loss). Finally, for a comparable amount of decrease in escape efficiency, a higher squeezed state will see more increase in the total uncertainty. This can be seen easily in Fig. 3-7. This is the total uncertainty when one looks at one individual port at a time.

On the other hand, if the full Hilbert space is kept, keeping the cross-covariances of all the ports, one can find the total uncertainty of this multi-mode state by performing the determinant of this full multimode covariance matrix. We show in Appendix C that in the absence of classical noises this total

uncertainty is

$$\frac{4E^{\mathrm{L}}\left(1-E^{\mathrm{L}}\right)+\delta^2}{\delta^2}, \tag{3.34}$$

where $E^{\mathrm{L}}$ is the fraction of signal lost to irrecoverable losses. As expected, this uncertainty tends to 1 for $E^{\mathrm{L}}=0$ or 1. This implies that if all the ports of the system are kept track of (i.e. $E^{\mathrm{L}}=0$ ), the system stays in a minimum uncertainty state. Or if all the light is lost, there is no squeezing, and both transmission and reflection stay at vacuum level ($E^{\mathrm{L}}=1$, $E^{\mathrm{R}}$=$E^{\mathrm{T}}$=0). At all intermediate losses, the total uncertainty is made higher by the losses, which is equivalent to tracing a density matrix over certain degrees of freedom.

We also calculate this total uncertainty in the presence of classical noises. While the full expression can be found in Appendix C, the simplest case is treating each classical noise as a standalone noise source, assuming $E^{\mathrm{L}}=0$. In this case we find that the added uncertainty by laser noises is just $S_{\mathrm{RIN}}$ and $S_{\mathrm{PN}}$, while the uncertainty added by thermal noise is $\lambda_{\mathrm{th}}\left(1+\delta^2\right)/\delta^2$.

## 3.6 Numerical Approach

### 3.6.1 Limitations of analytical approach

While the analytical approach gives a lot of intuition about the working of an OM squeezed light source, the assumptions made in the analytical calculation lead to some important effects being ignored. These assumptions were needed to simplify the problem to get understandable analytical solutions. To verify the accuracy of the analytical results for real life scenarios, a full numerical model was made in addition. The numerical model addresses the following limitations of the analytical model:

1. First, we assumed that the measurement frequency is much higher than the fundamental frequency of the oscillator (ie $\Omega \gg \Omega_0$). This leads to a simplification of the force to displacement relation in the analytical model ($F=-m\Omega^2 x$ instead of $F=m(\Omega_0^2(1+i\phi)-\Omega^2)x$), which in turn leads to simplified analytical results. We make no such approximation in the numerical model, giving reliable predictions even when the measurement frequency is comparable to the fundamental frequency of the oscillator.

2. Second, we also make the assumption that the measurement frequency is much smaller than all other frequencies of interest (achieved by setting $\Omega \to 0$ after the OM transfer functions are calculated). This assumption leads to an interesting effect in the analytical case – the coupling of classical laser noise only shows up in the 22 element of the covariance matrix. This happens because all the other matrix elements that couple classical laser noise to the measured squeezing go as $\Omega^2$ or $\Omega^4$. They then get set to zero by the $\Omega \to 0$ approximation. These matrix elements are also a direct consequence of the $\Omega \gg \Omega_0$ assumption above, and would be non-zero for a rigid cavity.

3. We also ignored the contribution of the higher-order modes (HOMs) in the mechanical susceptibility. The HOMs make analytically estimating $S_{\mathrm{F,th}}$ in closed form impossible. It can be written as a ratio of two summations, but cannot be simplified further.

### 3.6.2 Overview: numerical optimization of optical parameters

In the numerical model, we simulate the full mechanical response of the multimode oscillator without any of the assumptions mentioned above. We show an example of such a realistic system in Figs. 3-8 to 3-10, showing the squeezing behavior and the contributions of each noise source to the total noise in transmission and in the displacement of the mechanical oscillator.

In order to analyze this type of a system, it serves well to use the input-output formalism of continuously quantized two-photon states [50, 51, 52, 53]. In particular we have used the framework developed by Corbitt et al in Ref. [46]. . The input-output relations provide us with transfer function matrices that connect each output field to each input port. These matrices are then combined with the noise spectral densities for the input states to obtain the covariance matrices for the output states. . This approach is qualitatively different from the commonly used Hamiltonian type formulations [21]. Instead of focusing on the (states of the) system of the oscillator plus field inside the cavity, we propagate the field density matrices at all mirrors, beam splitters and free space. The mechanical oscillator's motion is dominated by classical displacements because we are looking at squeezing of light at high temperatures, far from the quantum ground state of the oscillator. Nevertheless, this model can easily be used for a quantum oscillator – in addition to the thermal noise of the oscillator (which will probably be very small in such cases), one can add a quantum noise spectrum from the quantization of the position of the oscillator.

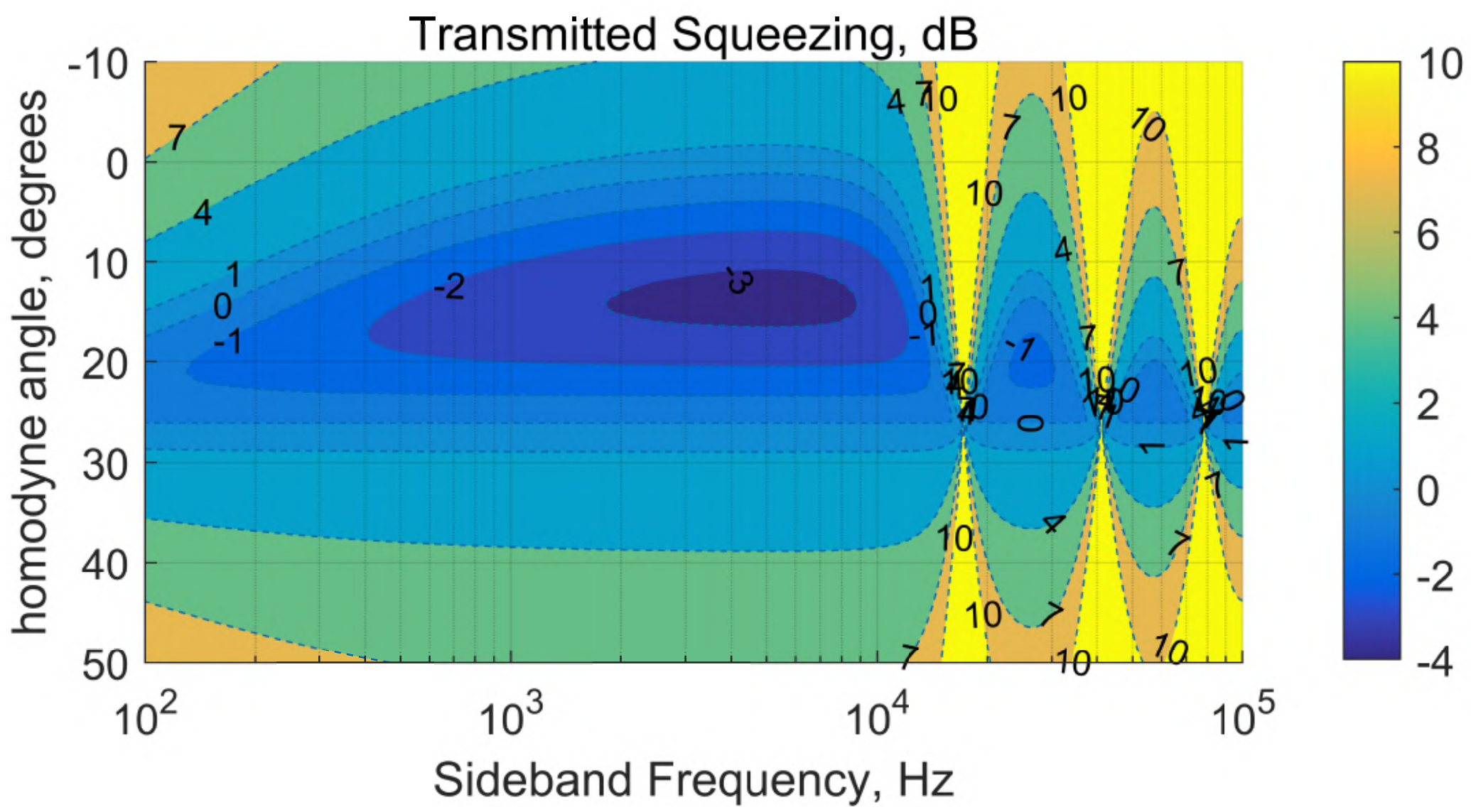

**Figure 3-8:** Total noise in transmission as a function of frequency (log scale) and quadrature, each contour is a contour of constant squeezing level. This plot is made for the baseline parameters given in Table 3.4. It includes thermal noise at room temperature with a mechanical Q of 20000 and a laser RIN of $10^{-8}/\sqrt{\text{Hz}}$. The dark blue region is where the noise is below the shotnoise level. The narrow vertical bands are mechanical modes as can also be seen in Fig. 3-9. The OS is to the right of maximum x-axis value.

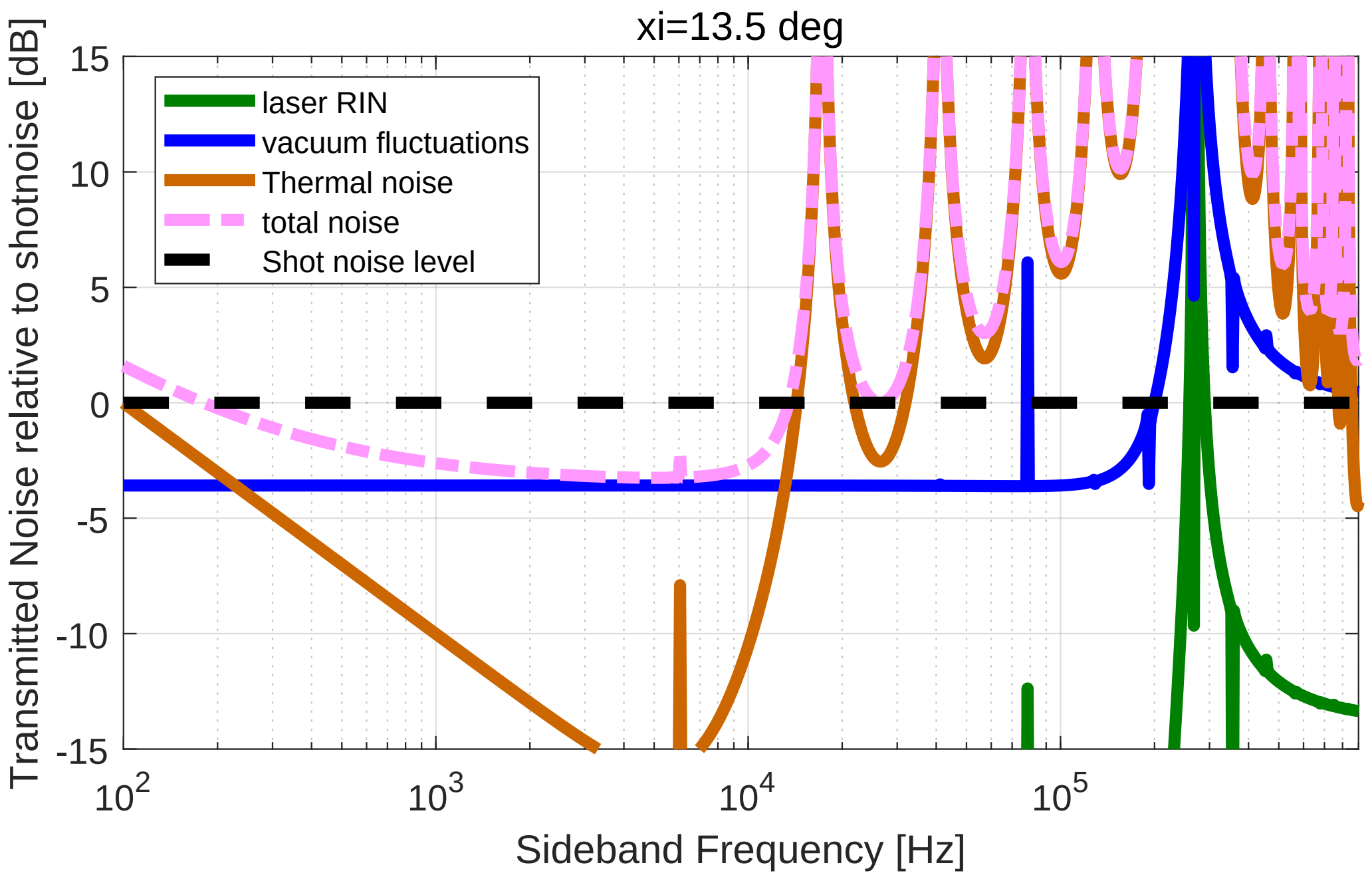

**Figure 3-9:** Noise budget in dB in transmission for the same cavity parameters as in Fig. 3-8 at the quadrature of maximum squeezing, 13.5°. The dashed pink trace represents the total noise as sum of the contribution from quantum noise (blue), thermal noise (brown) and laser amplitude noise(green).

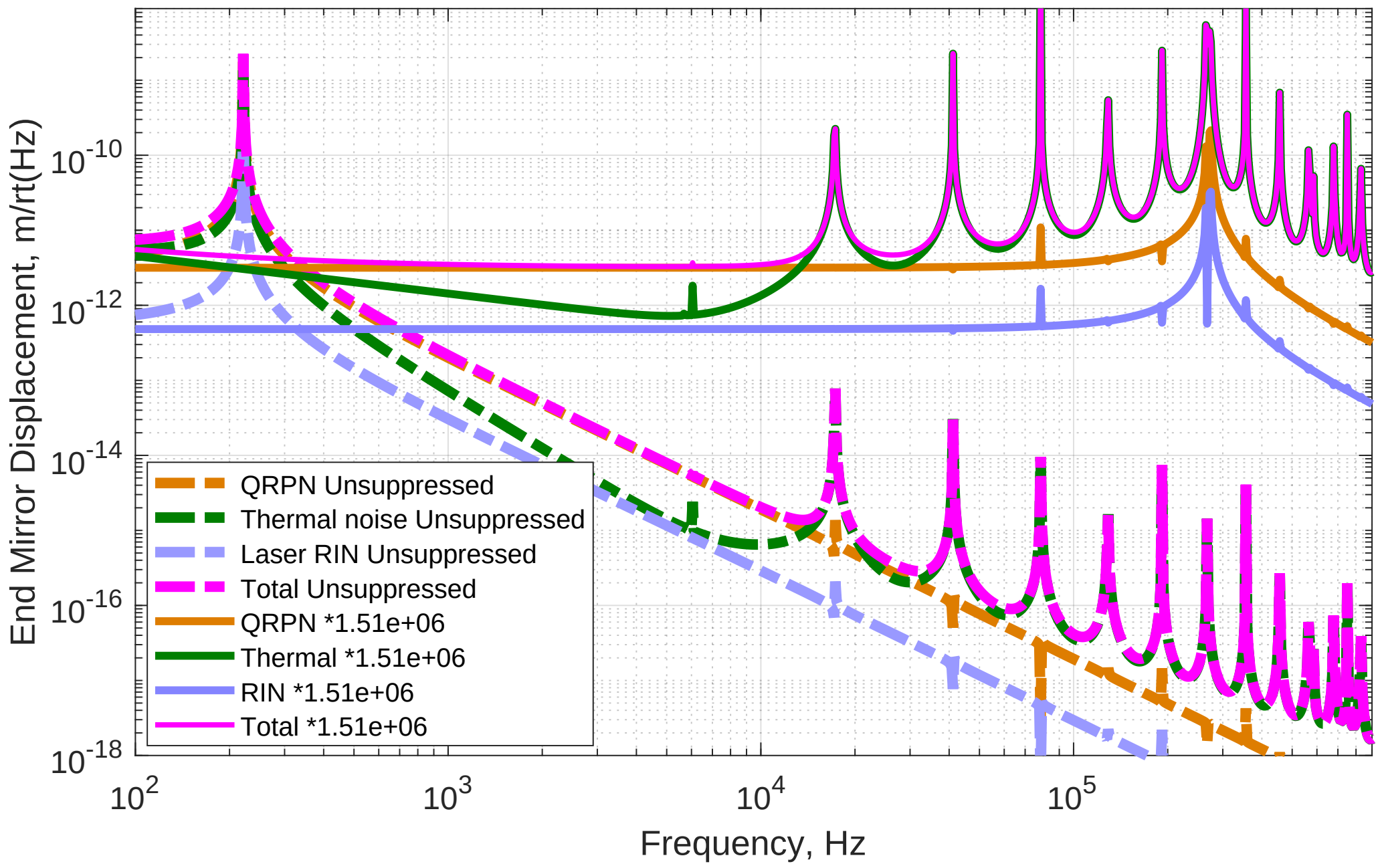

**Figure 3-10:** Amplitude spectral density of displacement of the end mirror, with and without the OS, for the same cavity parameters as in Fig. 3-8 The dashed traces represent the displacement spectrum for an undetuned cavity. The fundamental mechanical mode is at 221 Hz, above which frequency the mechanical response of the oscillator suppresses all noises as $1/f^2$. The thermal noise force spectral density goes as $1/f$ for a structurally damped oscillator, so it has a steeper slope than QRPN or RIN, both of which have a white force spectrum. The solid traces show the displacement spectrum for the same system but in the presence of a strong OS ($\delta = 0.5$). This OS ($f_{OS} = 275\,\text{kHz}$) suppresses the oscillator motion below 275 kHz by $f_{OS}^2/f^2$. In order to show this highly suppressed motion on the same scale, it has been scaled up by $1.5 \times 10^6$, which is the ratio of the two displacements at the low frequencies.

| Symbol | Quantity | Baseline value |
|---|---|---|
| $L$ | cavity length | 100 µm |
| $T_1$ | Input mirror transmission | 50 ppm |
| $T_2$ | Output mirror transmission | 250 ppm |
| $L_1$ | Input mirror loss | 0 |
| $L_2$ | Output mirror loss | 120 ppm |
| $\delta$ | detuning to linewidth ratio | 0.5 |
| $MM$ | Input mode matching efficiency | 40 % |
| $P_0^{\text{res}}$ | Intracavity power on resonance | 400 mW |

**Table 3.4:** Baseline optimization parameter values

We code the input-output relations for each optical element in MATLAB, and combine them to get the transfer functions and spectral densities. This code is then parallelised over numerous (700,000 or so) cavity configurations of varying optical parameters. From each configuration we pick the following numbers, both in reflection and transmission to design the optimum experiment. These are shown in Fig. 3-11.

- **Maximum Squeezing ($N_{\text{min}}$):** We extract one data point from the total noise at the output as a function of frequency and quadrature angle at which the noise is the minimum (Fig. 3-11). We call this quantity maximum squeezing and is most of the times plotted on a logarithmic scale in decibels (dB).

- **Best Frequency:** The frequency value of the above point. In order to make sure that the maximum is chosen from a frequency-independent region, we artificially impose a maximum on this frequency for the search function. This is especially needed in cases where there is no classical noise, because there the squeezing near the OS is usually stronger than the frequency-independent squeezing. In the presence of classical noises as per our estimates, this is usually not the case. [5]

- **Best Angle ($\xi_{\text{min}}$):** The quadrature angle at which the above maximum squeezing is found. This is referred to as the squeezing quadrature in the rest of this work.

- **Start Frequency ($f_L$):** At $\xi_{\text{min}}$ , the lowest frequency at which the total noise goes from greater than one to less than one.

- **Stop Frequency ($f_H$):** At $\xi_{\text{min}}$, the highest frequency at which the total noise goes from less than one to more than one.

The start and stop frequency are used to define the bandwidth in which there is squeezing. It must be kept in mind that sometimes there are peaks of classical noise from mechanical modes inside this band, and that is not conveyed by this number.

[5] This is also for extremely low finesse cavities even with classical noise – but they are usually pretty bad for squeezing anyways.

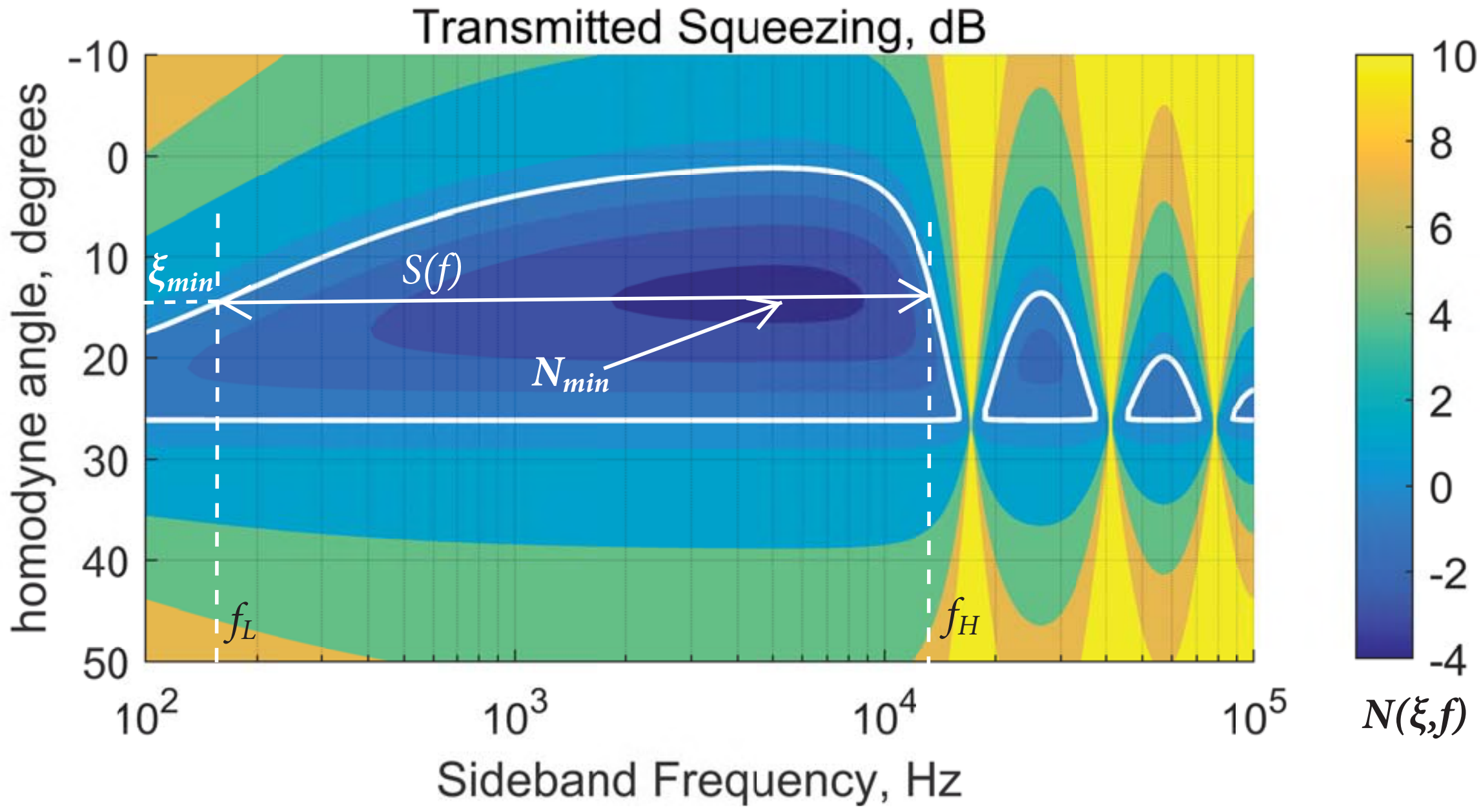

**Figure 3-11:** This figure shows the definitions of the main output parameters from the OM system that are then used to optimize the squeezer. From each such noise plot, we pick the point that has minimum noise and its frequency and quadrature and use those to optimize the cavity parameters. The squeezing is not quite frequency independent in magnitude because of relatively high classical noise. At the squeezing quadrature, we also look at the frequency band of the squeezing by looking at frequencies where this spectrum stays below shot noise.

### 3.6.3 Suppression (squeezing) magnitude

The amount of squeezing depends on the cavity detuning, the reflectivity of each mirror, and the contributions of classical noises. We address each separately here.

#### 3.6.3.1 Role of mirror transmissions

Intuitively, one would think that the amount of squeezing generated inside the cavity depends on the OM gain. This OM gain increases with the increase in OS frequency. For a fixed cavity power, this reduces to finesse and detuning. So a higher finesse should generate more squeezing. As we find in our simulations, this is not exactly the case.

A higher finesse will generate more suppression of the classical noises inside the cavity as compared to quantum noise, even at a fixed intracavity power, as per intuition. This would generate more squeezing inside the cavity, but on the other hand, the amount of squeezing that is observable outside the cavity is higher for higher escape efficiency. Hence one needs high escape efficiency as well as high finesse. For example, in the case where $T_1$ and $L_2$ are fixed, increasing $T_2$ increases the escape efficiency but decreases the finesse, which means there is an optimum $T_2$ that produces the most observable squeezing. If the classical noises are increased, this optimum $T_2$ will occur at a lower value (higher finesse). Similarly, if the available intracavity power is lowered, the ratio of quantum noise to classical noises decreases, lowering the squeezing. This can be compensated with a higher finesse. An increase in optical losses or the transmission of the un-used port implies decrease in both finesse and escape efficiency. The effect of losses on the optimum $T_2$ will depend on the relative contribution of classical and quantum noise.

As an example we show the optimization of EM transmission for our particular estimates of classical noise in Figs. 3-12 and 3-13a. Fig. 3-12 shows $N_{\rm min}$ as a function of the EM transmission $T_2$ with everything else fixed, i.e. each point in the figure represents the maximum squeezing over all frequencies and quadratures for a given optical configuration, power and classical noises. The baseline assumption case is the expected experimental parameters, ie IM transmission $T_1$ is 50 ppm, cavity losses are 120 ppm, mode matching efficiency of 40%, RIN of $5 \times 10^{-8}/\sqrt{\rm Hz}$ and thermal noise corresponding to room temperature (298K), and a mechanical Q of 20000 (Table 3.4). From this baseline, we start modifying these fixed conditions one by one to make them worse. A similar optimization for performing the squeezing measurement on reflection is shown in Fig. 3-13b.

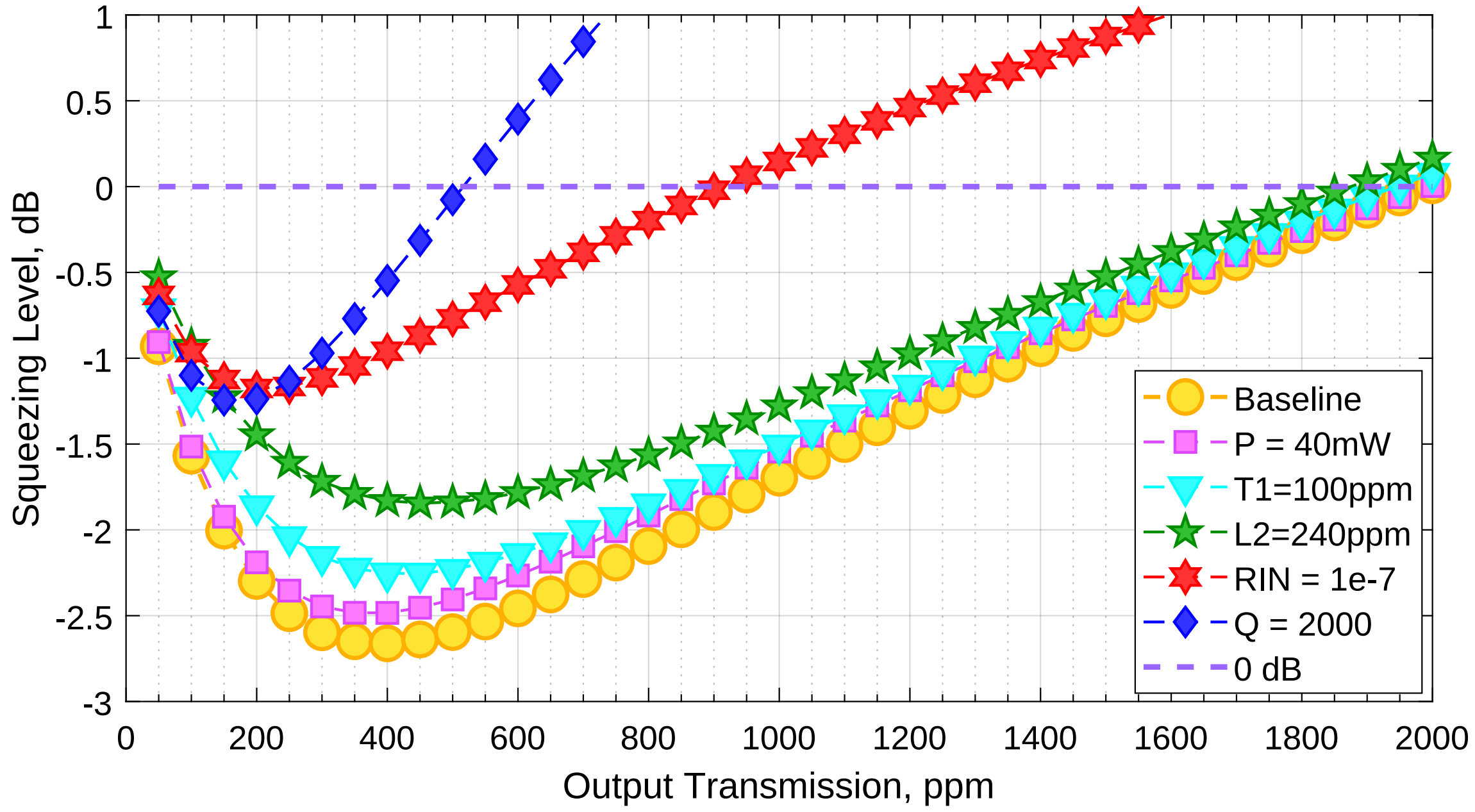

**Figure 3-12:** This figure shows the best squeezing ($N_{\mathrm{min}}$) as a function of varying end mirror transmission ($T_2$), with all other parameters fixed. The baseline case is our best estimate for the expected parameters in the experiment, listed in Table 3.4, RIN $= 5 \times 10^{-8}/\sqrt{\mathrm{Hz}}$ and thermal noise corresponding to room temperature and Q of 20000. For comparison, we also show similar optimizations for more classical noise. We can see that the optimum $T_2$ shifts towards lower values. Another noteworthy observation is that the decrease in squeezing due to a factor of 10 decrease in the mechanical Q is far worse than a factor of 10 decrease in the cavity power, and is comparable to just a factor of 2 increase in RIN. Also worth noting is how the slopes change by changing RIN vs. changing losses or thermal noise.

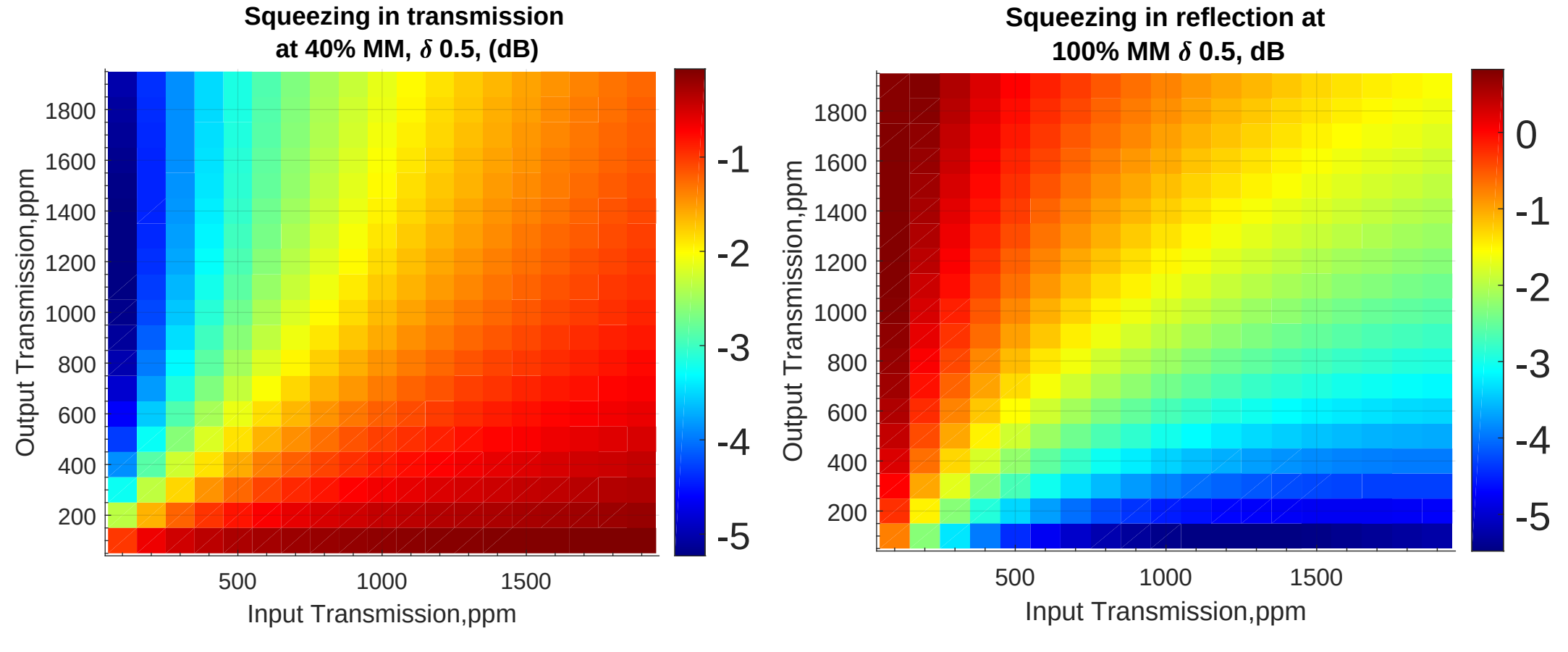

(a)

(b)

**Figure 3-13:** Effect of IM and EM transmissions on squeezing in reflection and transmission: **(a)** shows $N_{\mathrm{min}}$ in transmission as a function of $T_1$ on x-axis and $T_2$ on the y axis. We see that the best squeezing is obtained for minimum $T_1$, but an intermediate value of $T_2$. **(b)** Similarly, for reflection squeezing, best case is to have as low $T_2$ as possible, and find an optimum $T_1$.

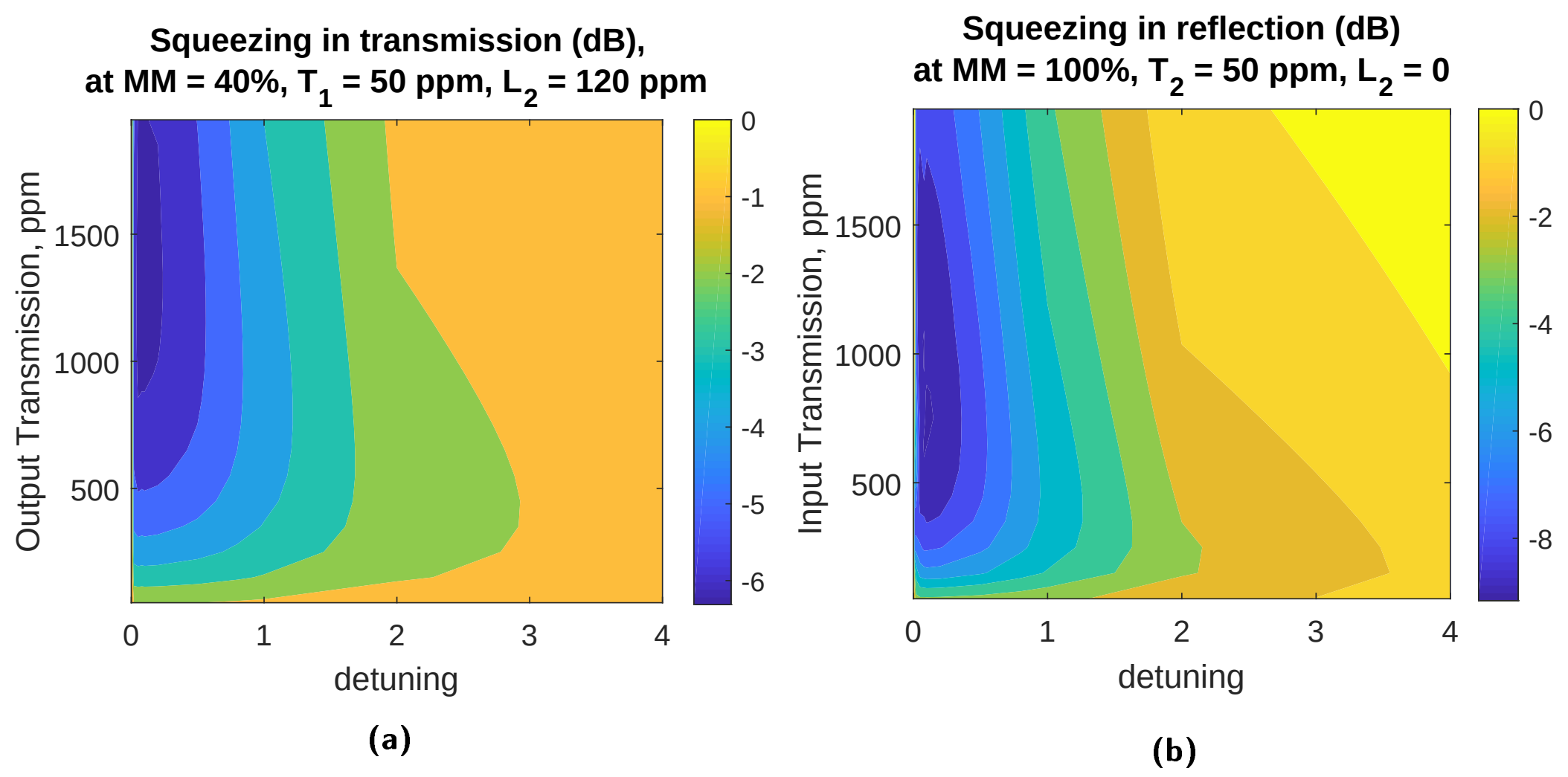

**Figure 3-14:** Effect of detuning $\delta$ and measurement port transmission on squeezing in reflection and transmission. The detuning is normalized to the cavity HWHM linewidth. **(a)** shows the amount of squeezing in transmission as a function of $\delta$ and $T_2$, with all other parameters fixed. Only positive detunings are shown because the amount of squeezing is symmetric in detuning. We see that the best squeezing is obtained for small detunings, but higher transmission. **(b)** shows the same plot, but for measurement in reflection. In this plot we have chosen to set $L_2$ to zero, leading to higher total squeezing.

#### 3.6.3.2 Role of detuning

Detuning plays a slightly more convoluted role. Changing the detuning changes the OM gain, and hence the amount of squeezing but also changes the squeezing quadrature, which in turn changes the amount of squeezing in the presence of classical noises. As an example, the contribution of quantum noise, thermal noise and classical laser intensity noise in transmission is shown in Fig. 3-16 as a function of frequency and quadrature. In the absence of classical noises, a change in squeezing angle does not contribute to extra change in squeezing besides the change in gain. In most cases the system will be designed such that classical noises are smaller than quantum noise, so lower detunings will still be better even with classical noises because of the basic suppression $S_{\text{Ideal}}(\Omega)$ that is applied to noise in all quadratures. The net effect of detuning can be seen in Fig. 3-14, where we show the maximum squeezing as a function of detuning and the measurement port transmission, for a fixed level of classical noises [6].

#### 3.6.3.3 Role of classical noises

As one would imagine, a higher classical noise to quantum noise ratio would degrade the amount of squeezing. This is not to be confused with the fact that classical noise inside the cavity is also experiencing the same OM mixing of quadratures. The classical noise is also being squeezed, but for our purposes, when we talk of squeezing we talk about suppression as compared to the shot noise of a perfect coherent state. Hence, the addition of classical noises would increase the noise per unit shot noise. As long as the classical noises enter in a quadrature that has a non-zero projection on $\xi_0$, they will lower squeezing and shift the effective squeezing quadrature. Given that the quadrature of classical noise is inherently a part of the OM interaction and cannot be engineered, the best approach to improve squeezing is to just decrease the ratio of classical noise to quantum noise as much as possible.

[6] For numerical precision reasons, the code cannot be trusted for detunings below 0.1.

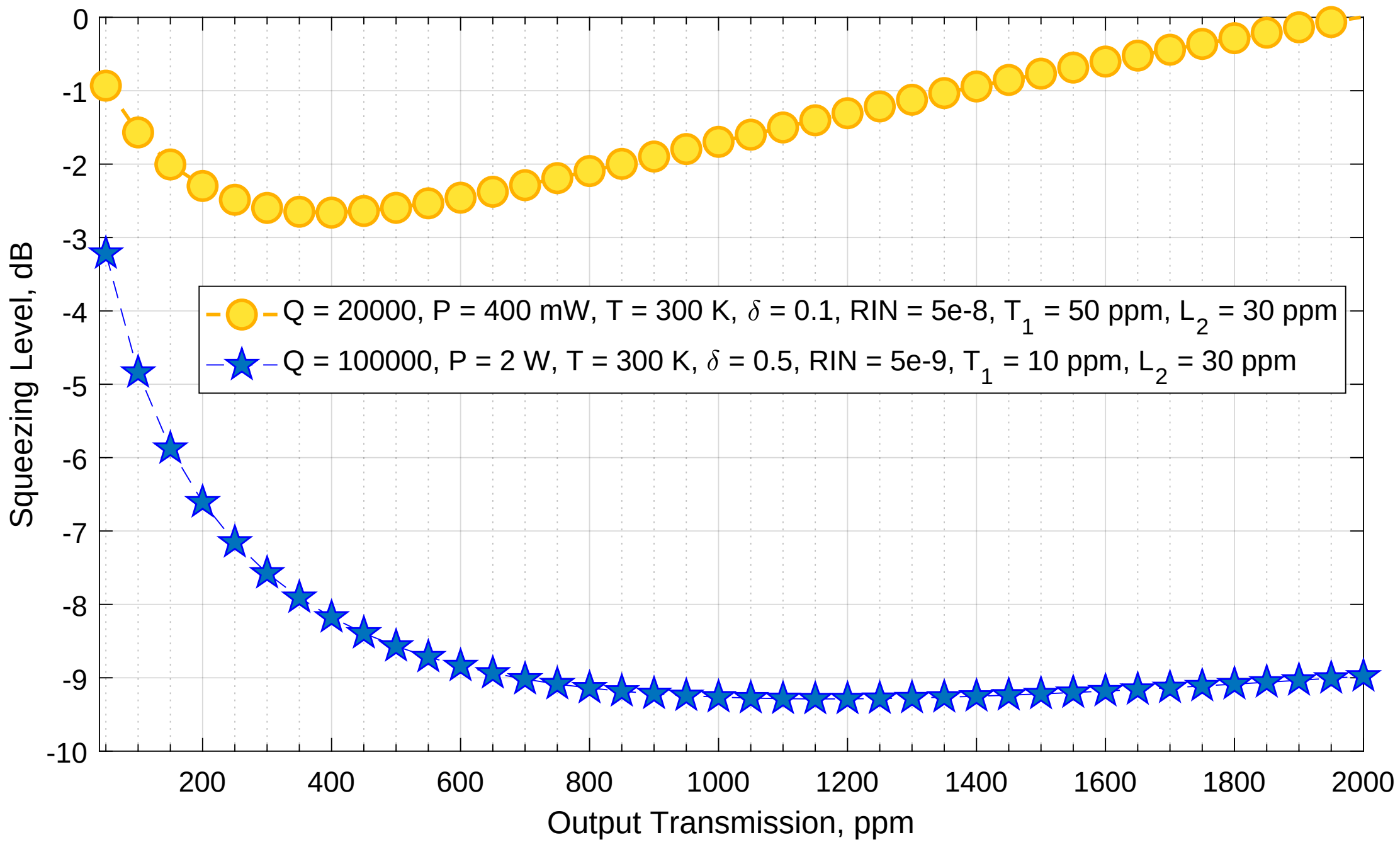

**Figure 3-15:** In this figure we show the baseline case from Fig. 3-12 along with an improved version of the experiment. Here we assume the technological advances in materials will get us better quality factors, better power handling of the devices, lower classical intensity noise in the laboratory, and reduced roundtrip losses due to scattering and diffraction. In that case, to measure squeezing in transmission, we would pick a very highly reflective input mirror, and around 99% reflective end mirror.

### 3.6.4 Squeezing quadrature

Change in detuning mixes the amplitude and phase quadratures differently. One can think of change in detuning modifying which quadrature noise gets correlated to its orthogonal quadrature.

Moreover, since the classical noises contribute differently in different quadratures, $\xi_{\rm min}$ gets shifted by the addition of this quadrature-dependent noise. We show the contribution of classical noise separately from the contribution of quantum noise in Fig. 3-16. We see that the minimum quantum noise quadrature is different from the minimum RIN and minimum thermal noise quadrature. Just as predicted by the analytical approximations, the laser noise minima is at the amplitude quadrature, and the thermal noise minima is at roughly twice the quadrature at which the quantum noise has its minima.

We can also understand the complicated contribution of classical noises by looking at Figs. 3-17 and 3-18. These figures show the contribution of classical and quantum noises in the amplitude and phase quadrature of the same cavity configuration as in Figs. 3-9, 3-10 and 3-16d. We see in Fig. 3-17 that in the amplitude quadrature, at frequencies where the quantum noise is higher than the classical noise (which can be inferred from Fig. 3-10 to be from about 300 Hz to 10 kHz), the amplitude quadrature only measures 0 dB. This is because in this regime, the fluctuations that carry the information of mirror motion are canceled by back-action, and only the promptly reflected vacuum survives [54]. At higher and lower frequencies where classical noise is dominant, we see it show up in the amplitude quadrature. On the other hand, Fig. 3-18 shows the contribution of various noises in the phase quadrature of transmission. Here we see that the frequencies at which quantum radiation pressure noise was higher show a flat noise which is 10 dB. This is the contribution of all the quantum fluctuations that enter the cavity through various ports ($\vec{I}$, $\vec{v_i}$). We also see the contribution of thermal noise and RIN are different from that in the amplitude quadrature.

Another way to understand this is to look at all quadratures but restrict to one noise source at a time. We show the contribution from only quantum noise in Fig. 3-16c. Here we see a nice clean squeezed field at $\xi_0 = 13^{\circ}$. Now if we see Fig. 3-16a, we see the contribution to the transmitted field from the

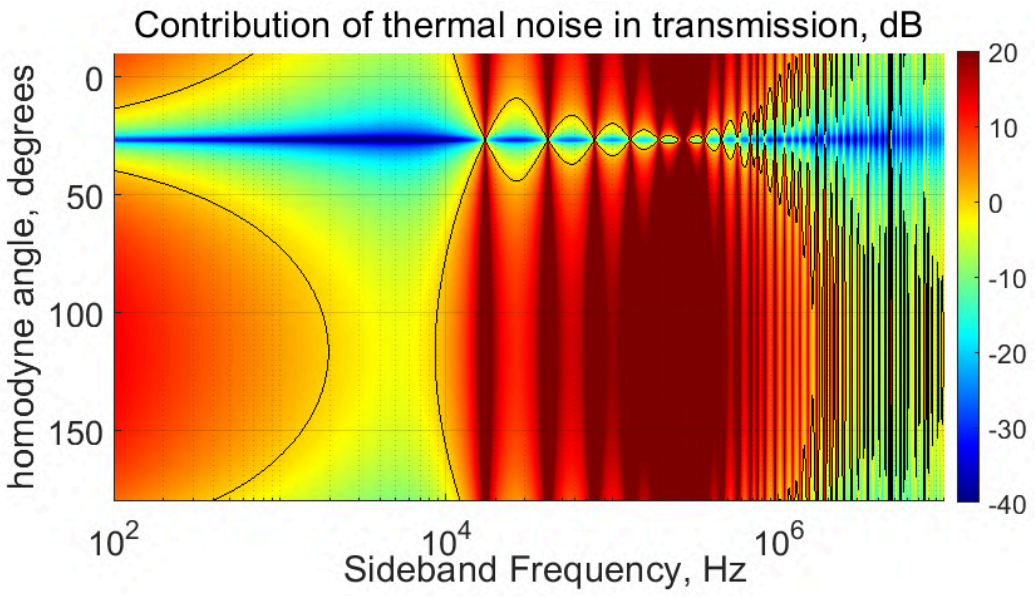

**(a)**

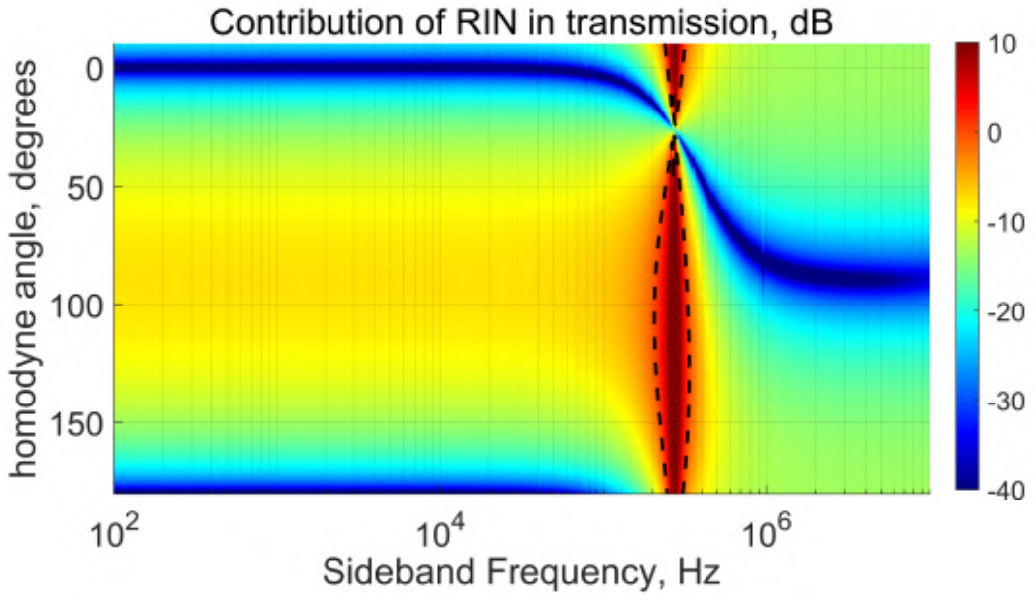

**(b)**

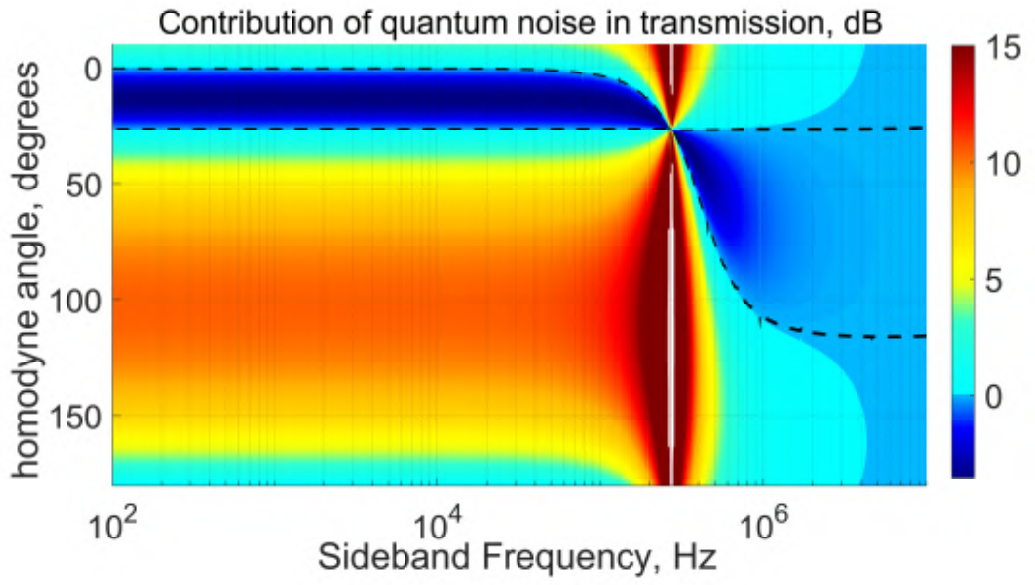

**(c)**

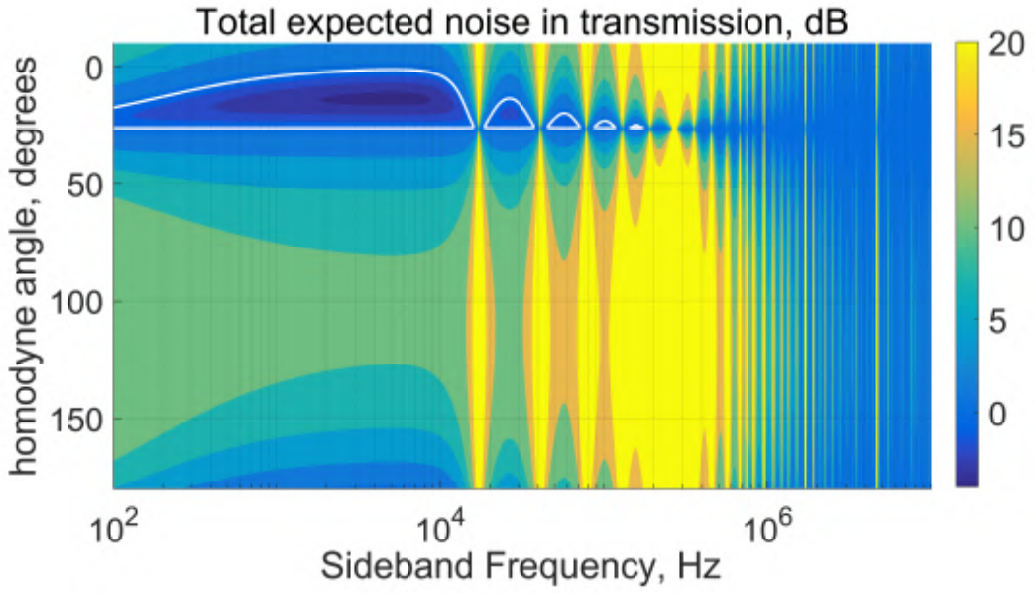

**(d)**

**Figure 3-16:** Individual contributions to total noise in transmission as a function of frequency and quadrature: The colorbars for each plot are shown on the right of each plot, the black contour line is the 0 dB level. **(a)** Contribution of thermal noise: Here we can see that the thermal noise is minimum at 27°. Also, the thermal noise is high in the red regions, due to the HOMs as well as the OS. **(b)** Contribution of RIN: Since the RIN is a white noise, the frequency dependence is dominated by the cavity's response. Here we can see that RIN's contribution to transmission is minimum in the amplitude quadrature, which agrees with the approximate analysis done in the analytical approach. **(c)** Contribution from quantum noise: Here we can see the frequency-independent squeezing goes upto the OS frequency at about 200 kHz. The squeezing quadrature is near 13°, which agrees with the analytical calculation. In the absence of classical noises, we see that squeezing after the OS frequency continues till the end of the x axis, with a decreasing amount. The squeezing in absence of classical noises stops at the cavity linewidth, which is higher than 10 MHz. **(d)** Total noise, sum of the above three. The white contour is 0 dB level. We see that the addition of thermal noise has reduced the amount of squeezing and the squeezing bandwidth. There is also no squeezing above the OS frequency, which is between 200 and 300 kHz. It is the combination of the quadrature dependence of classical noises and the purely quantum mechanical quadrature of 13° that give us the minimum at 13.5° for total noise below the OS frequency. The quadrature of minimum quantum noise and minimum RIN shifts at the OS frequency, unlike the minimum of thermal noise which stays at the same quadrature even at frequencies higher than the OS. At these frequencies, the quadratures at which quantum noise is lower than shot noise, the thermal noise is large, so in our system we would not see squeezing above OS frequency.

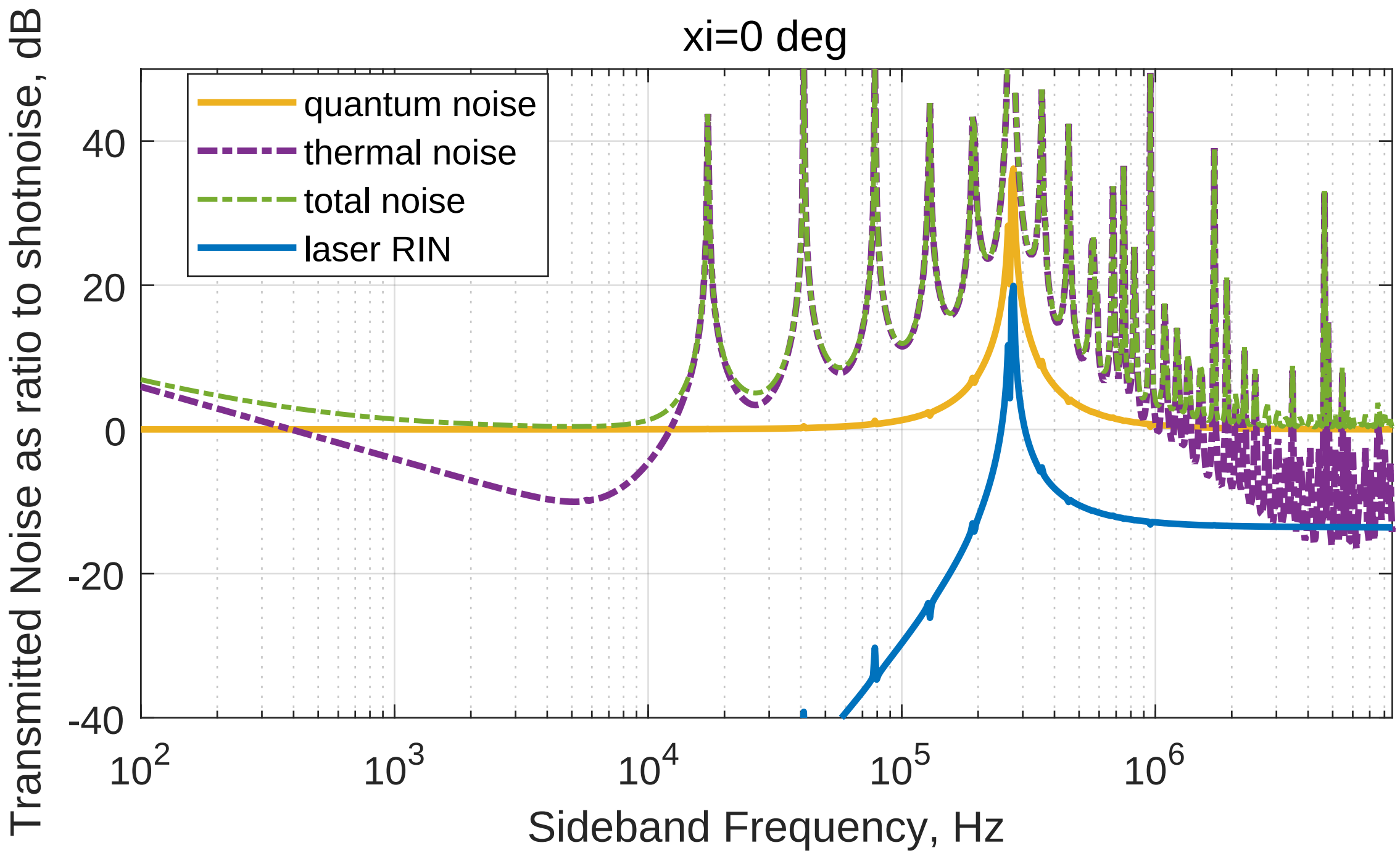

**Figure 3-17:** Contribution of each source in the amplitude quadrature of transmitted light for the token cavity configuration. If the classical noises in the cavity are lower than quantum noise, we only observe the shotnoise in transmission. Otherwise the contribution from classical noises can be seen in the amplitude quadrature. This implies that in the quantum dominated regime the amplitude quadrature does not carry any information of the displacement spectrum of the mechanical oscillator.

thermal noise due to the Brownian motion of the oscillator. We see the narrow blue horizontal patch; it's where the thermal noise contribution is minimum, at around 27° quadrature. We add this quadrature (and frequency) dependent noise to the quantum noise from Fig. 3-16c, and we should expect this noise to shift the squeezing quadrature from 13° towards the minimum in thermal noise(27°). Similarly, the RIN would shift the squeezing quadrature to a lower angle (towards 0°). Their combination makes the squeezing quadrature $\xi_{min}$ around 13.5° as seen in Figs. 3-8 and 3-16d. The shift is very small, and would be more intuitive if we also paid attention to the magnitude scale on Fig. 3-16c and Fig. 3-16a. We see that at frequencies near 4-5 kHz, at any given angle, thermal noise is much lower than quantum noise (by design, as discussed in Chapter 4). So it pulls the angle by a small amount at those frequencies. At lower or higher frequencies where its magnitude is higher, it starts to pull the quadrature of minimum noise more as is seen more clearly in Fig. 3-8 – the top part of the 0 dB contour is curving down. We can now also look at Fig. 3-16b, here we show the contribution of RIN for this cavity configuration as a function of quadrature and frequency. The RIN when showing up in the transmitted field contributes the lowest noise in amplitude quadrature. But the entire scale of magnitudes is much smaller than the quantum noise or thermal noise (again, by design.) so the shift is towards higher quadratures. If the RIN is higher (say 1e-6 $/\sqrt{\text{Hz}}$), we see a lot more character from it.

The quadrature at which thermal noise is minimized is special. Since at frequencies below OS, thermal noise is the only frequency dependent noise, at that quadrature, the total noise in that quadrature is frequency independent. As shown in both approximate (analytical) and exact (numerical) work, this quadrature is twice the squeezing quadrature (ie $2\xi_0$). Since the contribution of quantum noise has to be symmetric around the squeezing quadrature ($\xi_0$), and since it has to be shot noise limited in the amplitude quadrature, it is also shot noise limited in the quadrature $2\xi_0$. Since the contribution of thermal noise is minimal in this quadrature, the only classical contribution is from RIN, which is also frequency independent. Hence the noise in the quadrature $2\xi_0$ should be frequency independent[7]. This is seen clearly in Fig. 3-8, where we see that the lower side of the 0 dB contour is perfectly horizontal.

[7] One exception to this will be when the RIN or laser frequency noise is frequency dependent, and not negligible comparable

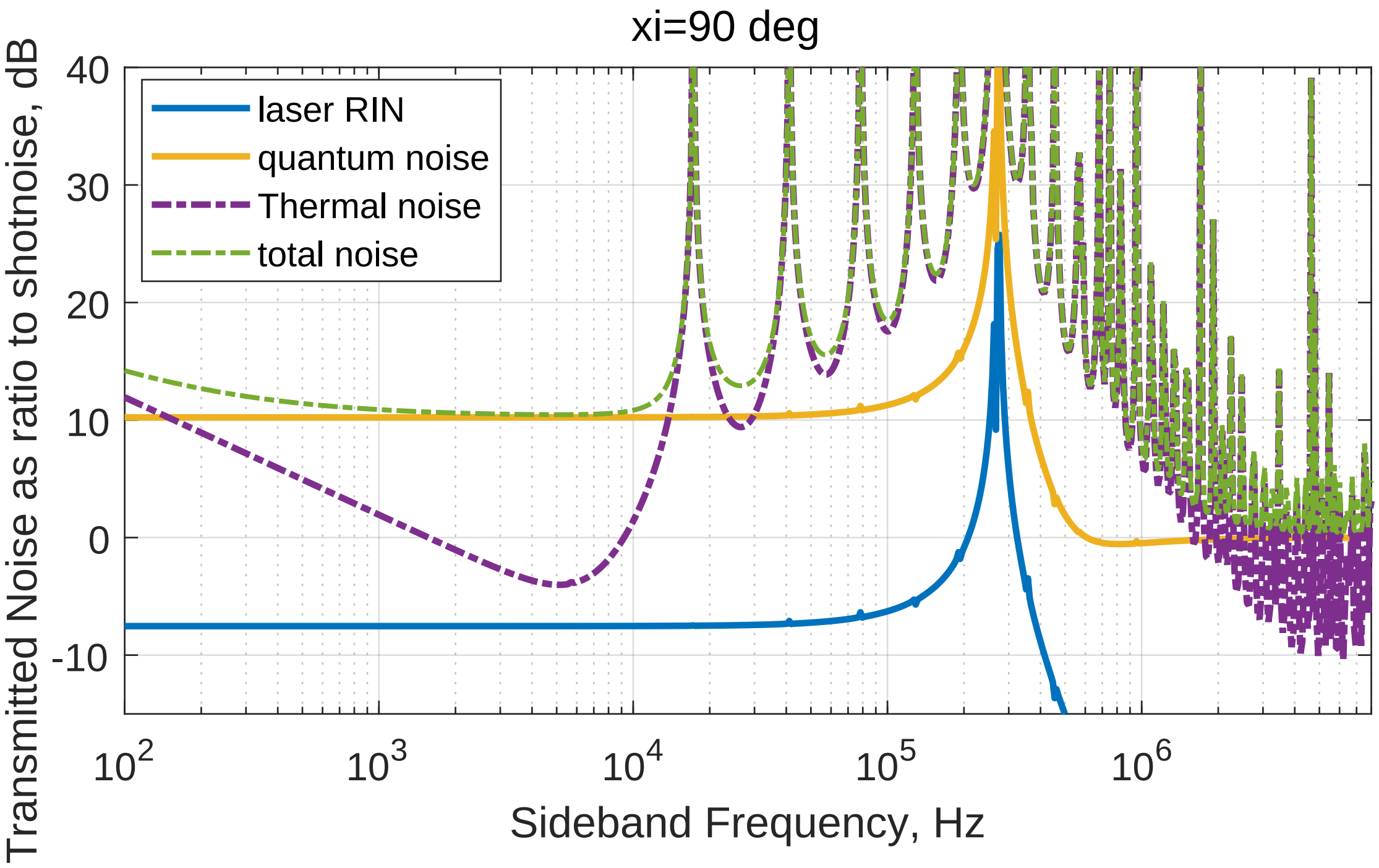

**Figure 3-18:** Contribution of each noise source in the phase quadrature of transmitted light for the token cavity configuration. Here we see that both thermal noise and RIN are higher than their counterparts in Fig 3-17, owing to their quadrature dependence. Also the quantum noise is no more zero dB, and it is in fact correlated to the mirror motion in the phase quadrature. RIN looks flat at 90° quadrature, but is $\Omega^2$ in Fig. 3-17. This behaviour matches the prediction from the analytical calculation before making the $\Omega \to 0$ approximation.

Since an increase in the cavity finesse will lower the contribution of classical noises (quadrature dependent), change in finesse also causes change in quadrature.

To summarize, everything that contributes to quadrature :

(a) Detuning

(b) Finesse

(c) Classical noises

This is the story in transmission. In reflection, the squeezing angle depends not just on detuning and finesse but also on the individual reflectivities of the two mirrors because of the interference with the input beam. In transmission, one can concentrate on what's going on with the field inside the cavity, and the quadrature angle remains the same when the squeezed intracavity field beats against the vacuum outside the cavity. We show the net effect of mirror reflectivities and cavity detuning on the squeezing quadrature in transmission and reflection in Fig. 3-19. We see that the squeezing quadrature in transmission increases with $T_1$ as well as $T_2$, which can be understood as an increased contribution from classical noises due to lowering of the finesse. The squeezing quadrature in reflection is more complicated due to the effect of the DC rotation of the carrier being dependent on the individual mirror reflectivities. Overall, the situations with $T_1 > T_2$ have negative $\xi_{\text{min}}$ (which is equivalent to greater than 90°). The squeezing quadrature increases weakly with increasing finesse, but changes strongly with escape efficiency. Finally, the change of quadrature with detuning is the strongest. We also see that the squeezing quadrature is the same sign as detuning in transmission, but is the opposite sign of detuning in reflection.

Finally, at zero detuning there is no squeezing because there is no transduction of phase fluctuations to amplitude fluctuations. All the QRPN is in the phase quadrature, and the amplitude quadrature is shot noise limited (which is also the case for finite detunings.) But since there is no squeezing, the quadrature of minimum noise becomes 0°, because all other quadratures have extra combination from QRPN.

### 3.6.5 Squeezing frequencies

In an ideal case (no classical noises) the broadband squeezing will occur at all frequencies from DC to the OS frequency. For the realistic case, we show the start and end frequencies of squeezing in reflection and transmission as a function of the optical parameters in Fig. 3-20. We see that the higher finesse configurations have better bandwidth in transmission, both at start as well as stop frequencies. The classical noises dominate the low-frequency end of the noise spectrum, so higher finesse makes the low-frequency cutoff lower. Towards the high frequencies – it is a competition between OS and thermal noise from some offensive HOMs – whichever is lower determines the high frequency cutoff. The higher finesse helps increase the high-frequency cutoff both because of OS as well as classical noise contribution. Using our FEM models, we optimized our cantilever designs to have fewest possible offensive modes, so we expect to be limited by the OS, even at room temperature. We also notice in reflection there is a wall of high start frequency to the left, and a wall of low stop frequencies also on the left. This is because at those optical configurations, $T_1$ is so much lower than $T_2$ that there is no squeezing. [9]

To summarize, the squeezing frequency band depends on

(a) Classical noises for low frequencies

(b) OS for high frequency cutoff

(c) Thermal noise for high frequency cutoff

to shot noise.

[9] This is an artifact of the numerical algorithm that searches for the start and stop cutoff frequencies. The algorithm assumes there is some squeezing, and in cases where there is no squeezing, it returns the maximum and minimum frequency in the frequency array.

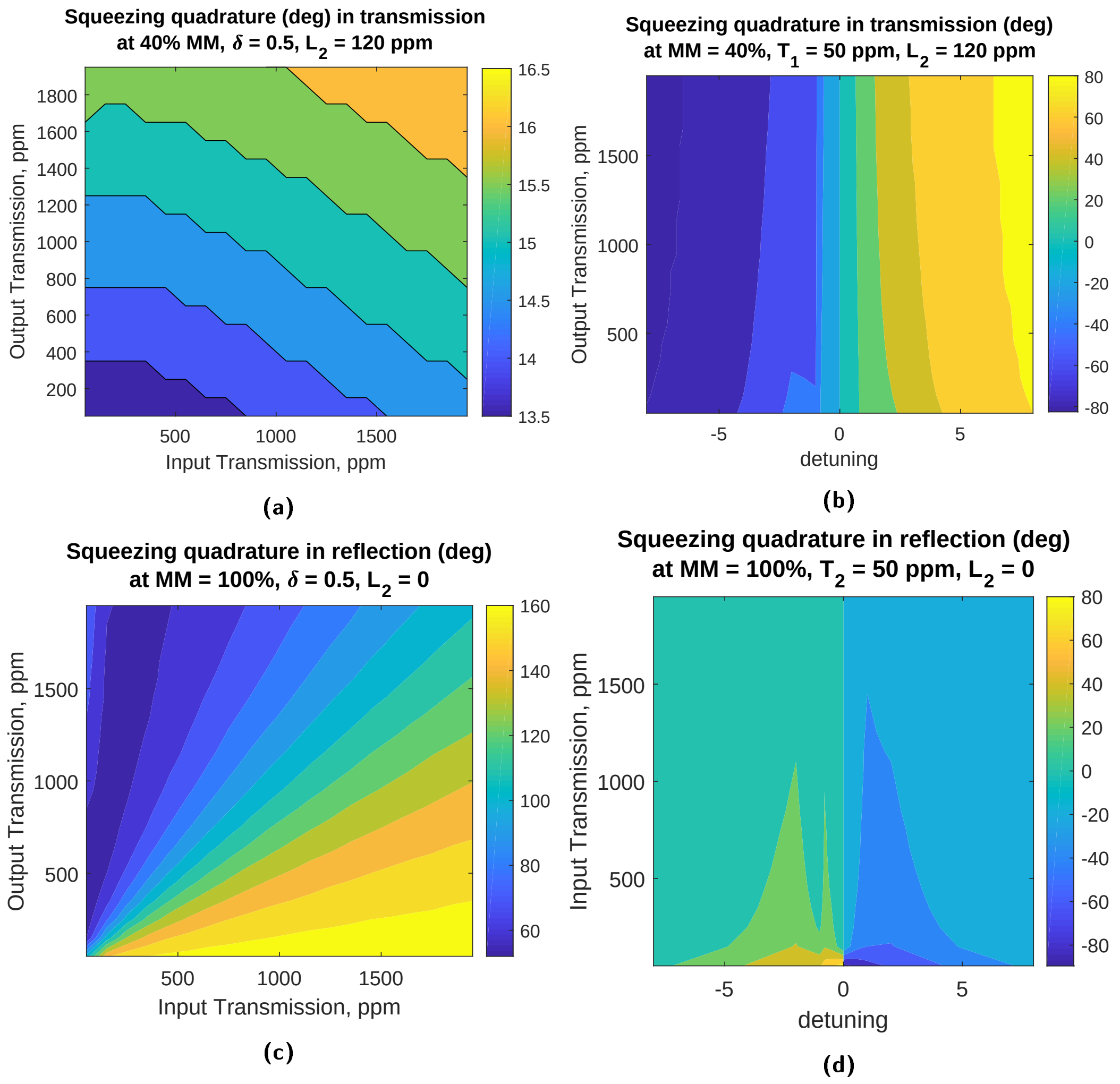

**Figure 3-19:** Squeezing quadrature ($\xi_{\rm min}$) as a function of optical properties: **(a)** In transmission, increasing $T_1$ or $T_2$ increases $\xi_{\rm min}$. **(c)** In reflection, in addition to the squeezing quadrature of the intra-cavity field, $\xi_{\rm min}$ also depends on whether the cavity is over-coupled or under-coupled. **(b)** $\xi_{\rm min}$ in transmission as a function of $\delta$ and $T_2$, shows that the dependence of $\xi_{\rm min}$ on detuning is far stronger than the dependence on $T_2$. Also, $\xi_{\rm min}$ has the same sign as $\delta$. **(d)** $\xi_{\rm min}$ in reflection as a function of $\delta$ and $T_2$: The sign of $\xi_{\rm min}$ is opposite to the sign of $\delta$. The overall dependence is more complicated, $\xi_{\rm min}$ gets closer to phase quadrature at lower detunings, and for lower $T_1$. [8].

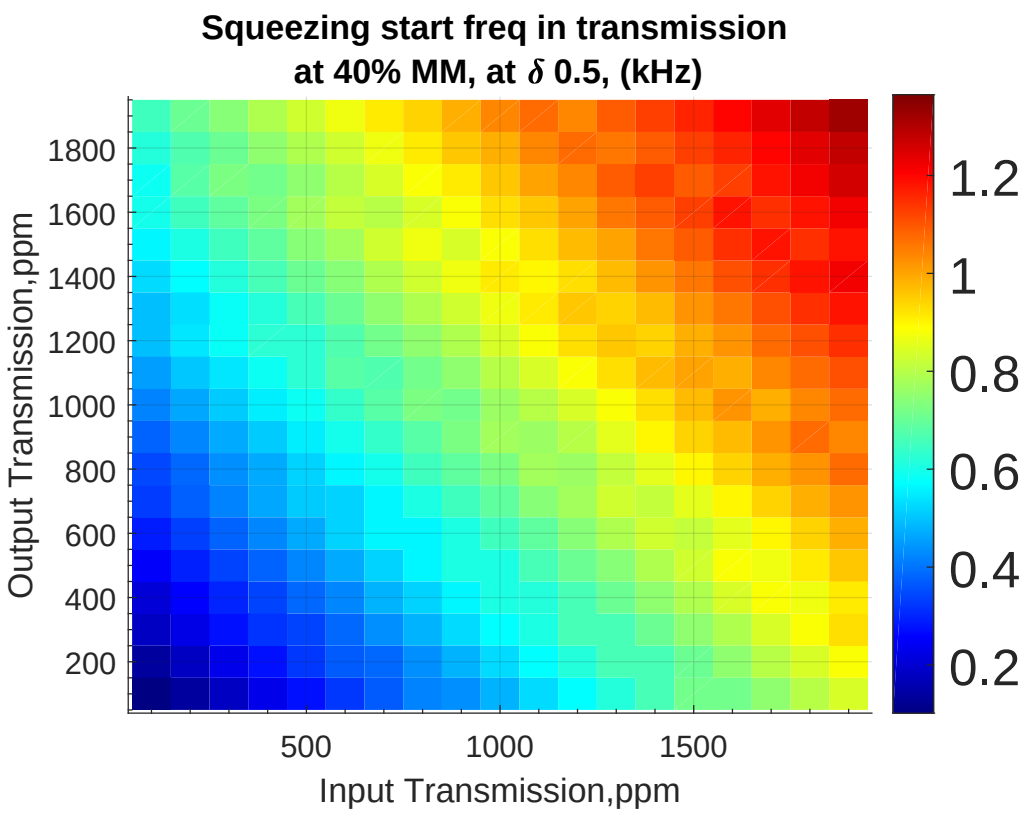

(a)

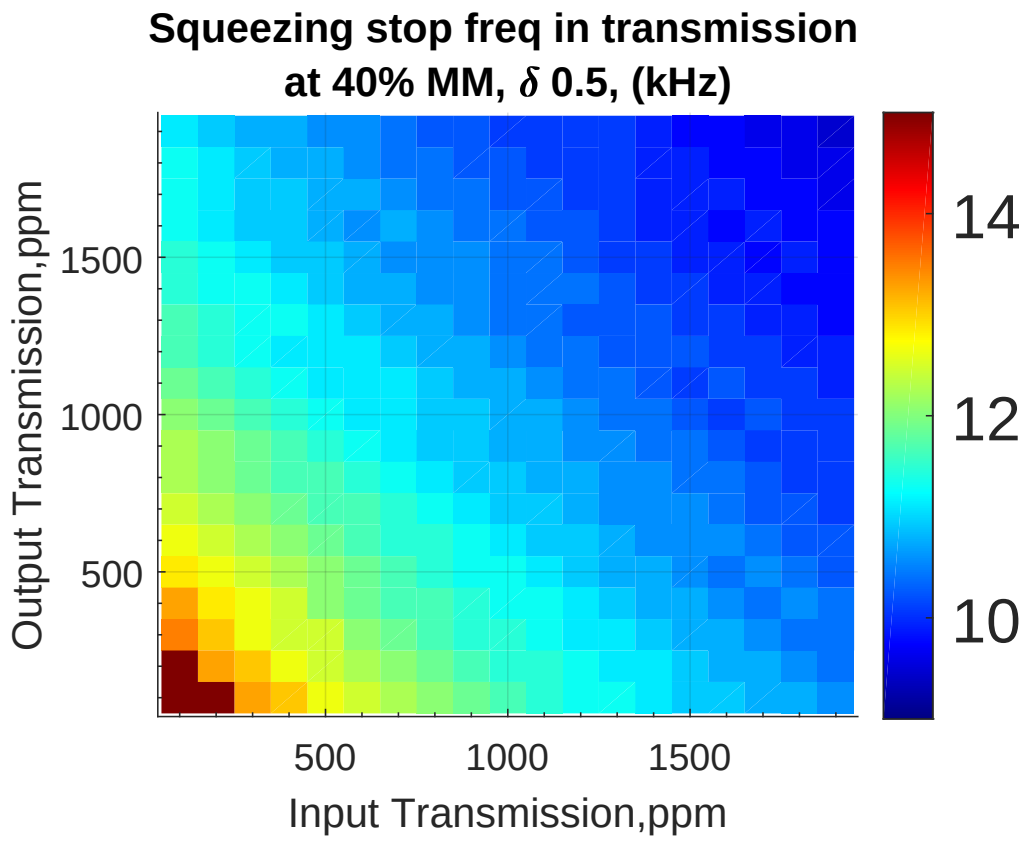

(b)

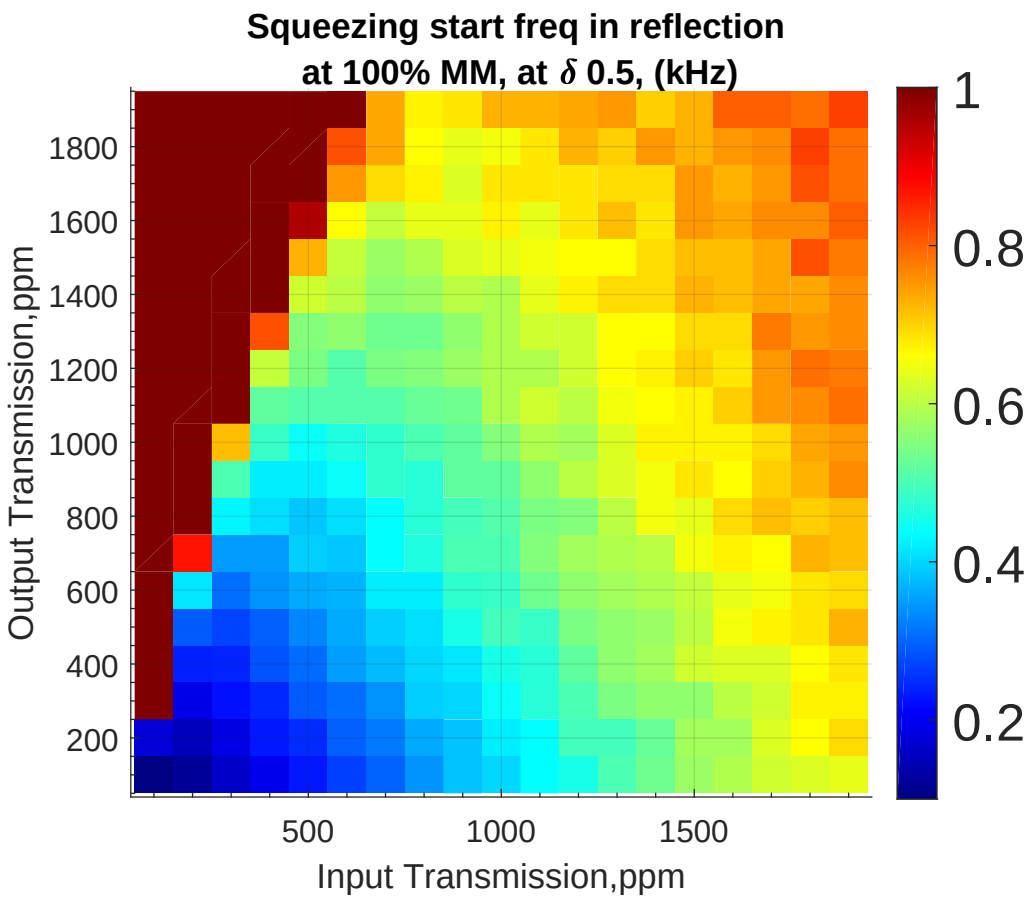

(c)

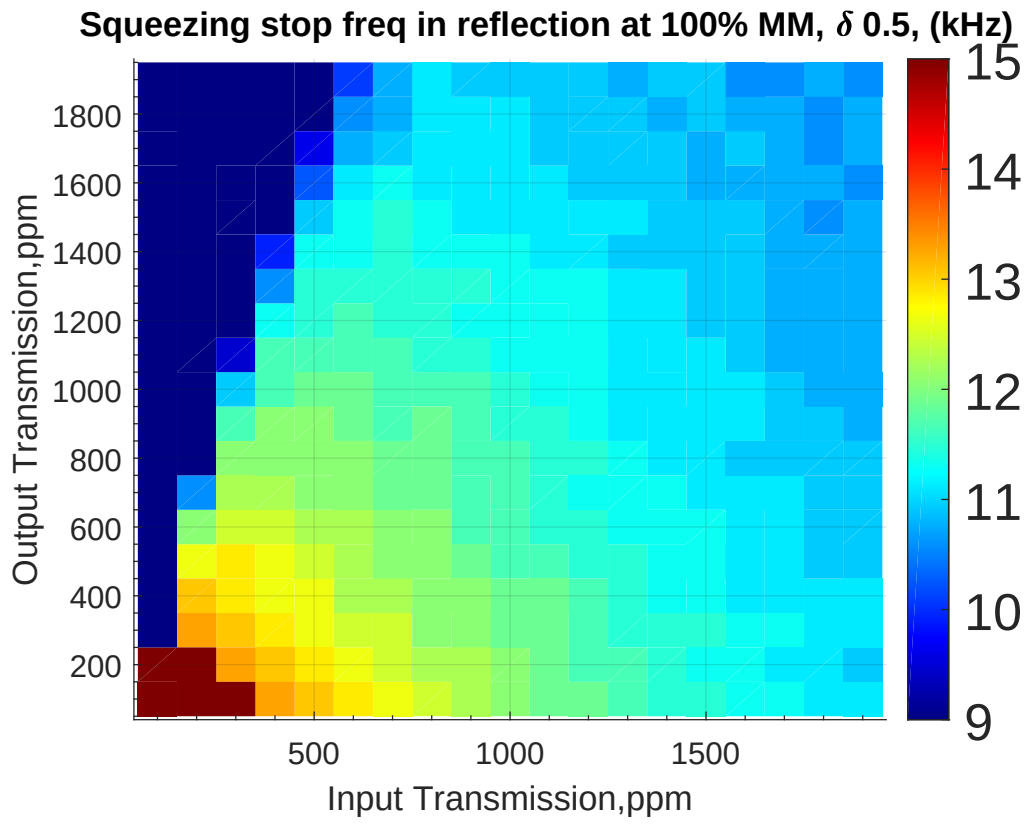

(d)

**Figure 3-20:** Squeezing bandwidth as a function of mirror reflectivities: Here we show the start and stop frequency of squeezing in reflection and transmission as a function of $T_1$ and $T_2$, at a fixed $L_2 = 120$ ppm, and $\delta = 0.5$. Overall, the bandwidth is better in each direction for a lower $T_1$ as well as lower $T_2$.

## 3.7 Conclusion

Using our numerical model, bolstered with intuition from our analytical model, we have designed an OM squeezer with the maximum squeezing output in the presence of imperfections like losses and classical noises. We use an approximate analytical approach, as well as an exact numerical approach to tackle the optimization. We focus on maximizing the frequency-independent squeezing at frequencies lower than the optomechanical spring and its bandwidth. When dealing with classical noises, a higher finesse is desired but there is a trade-off with the escape efficiency. The squeezing bandwidth is usually limited by classical displacement noise towards lower frequencies, and the OS at the high frequency. With advancement of better materials and better devices, a lower thermal noise will allow us to obtain OM squeezing in the GW band.

# Chapter 4

# Squeezer design: mechanical

## Contents

## Abstract

We present a design process for micro-mechanical mirrors to be used in an optomechanical (OM) cavity to perform quantum measurements. The goal of this design is to measure effects of quantization of EM field in a broadband audio frequency band at room temperature. These requirements put constrains on the geometrical design of the micro-devices. We mainly look at the geometry of a rectangular cantilever beam supporting a circular disk. As an example, we study the effect of changes of this geometry on the squeezed light output from this OM system. We find that the geometrical changes affect thermal noise, in turn affecting measurable squeezing. They also affect the mechanical susceptibility by changing the frequency and effective mass of each higher-order mode (HOM). And finally they change the amount of mirror motion caused by quantum fluctuations of light. Combining all the geometrical parameters, we present the design considerations and the best design outcomes for the particular use in our experiments.

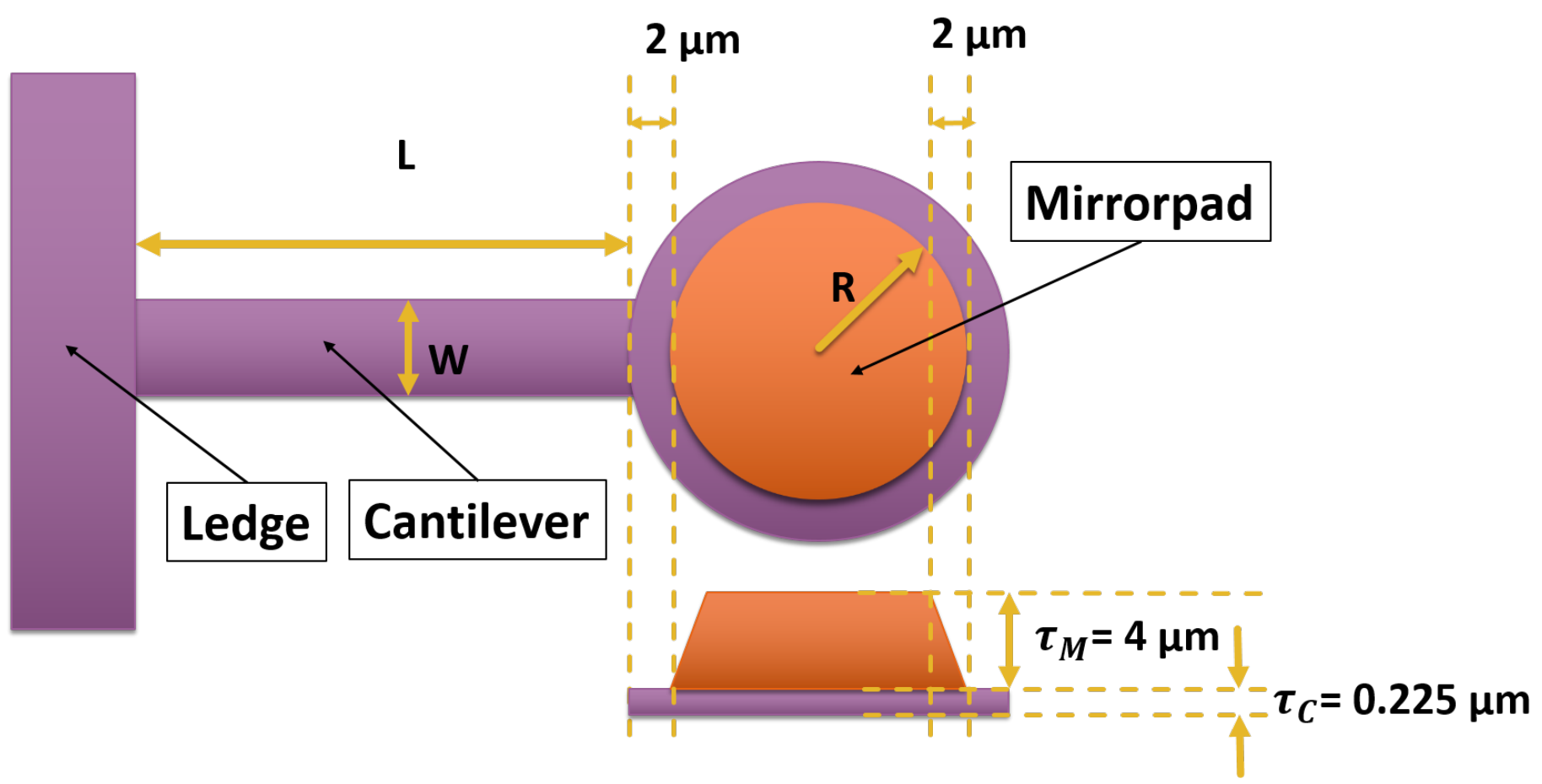

**Figure 4-1:** Basic device geometry definitions: a mirror of radius $R$ is attached to a cantilever of length $L$ and width $W$. The orange represents a 4 µm thick DBR mirrorpad made of alternating layers of GaAs and AlGaAs, and the purple represents a 225 nm thick GaAs layer. The mirrorpad has a taper of 2 µm in its radius from top to bottom, $R$ being the radius of the bottom of the mirrorpad, and $R$-2 µm being the radius of the top surface of the mirrorpad. Underneath the mirrorpad, there is a layer of the 225 nm GaAs, which is $R$+2µm in radius.

## 4.1 Basic structure

### 4.1.1 Geometry

The main device geometry under consideration is a high reflector mirror which is mechanically suspended and has a resonance frequency of a few hundreds of Hertz. A schematic for our device is shown in Fig. 4-1. A disc of radius $R$ and thickness $\tau_M = 4$ µm forms the highly reflective mirror (shown in orange in Fig. 4-1), which is used as one of the mirrors of a Fabry-Pérot cavity. The mirror is made of alternating layers of crystalline AlGaAs and GaAs, forming a Bragg reflector with the goal of obtaining high finesse, low thermal noise, crystalline coatings. In our case, the mirror in the final device does not have a substrate, just the coating layers. The thickness of the mirror is determined by the desired reflectivity of the mirror, which is optimized in Chapter 3. This mirror is attached to the bulk of the material via a thin cantilever of length $L$, width $W$, and thickness $\tau_C = 225$ nm (shown in Fig. 4-1 in purple).

The bulk of the material is around a 150 µm thick layer of GaAs. The layer structure of the full device is shown in Table 4.1. Another side view is shown in Fig. 4-3. This figure shows that there is a ledge of the 225 nm GaAs layer. This ledge is also shown in Fig. 4-1 to the left of the cantilever. The width of this ledge varies between each fabrication cycle. The length of it extends throughout the extent of the chip, and the other devices in the same row are also attached to the same ledge (see Fig. 4-2).

### 4.1.2 Fabrication

Our devices are fabricated by Crystalline Mirror Solutions in Santa Barbara. The detailed procedure on how the devices are fabricated can be found in the supplementary material in Ref. [55]. Here, we will provide a short summary of the process. First a wafer is grown with the desired layers using molecular beam epitaxy. This comprises the substrate, the cantilever layer, the mirror layer, and various stop layers that are used to control the etching process. This layer structure is shown in Table 4.1. The substrate thickness that was grown in the wafer was 675 µm originally. First, photolithography masks are made for each step. One that defines mirrorpads, and another that has both mirrorpads and cantilevers. The first (dry) etching starts from the mirror side, where the mirror material is removed from everywhere but the mirrorpads and chip edges, using mask 1 and a combination of inductively-coupled plasma (ICP) and wet etching. This etching is stopped by the InGaP layer. Next the InGaP layer is removed.

Once the mirrorpads and windows are defined, the cantilever material is removed in another ICP etching process, using mask 2. This removes the cantilever material, and the AlGaAs stop layer that would be in the gap between two cantilevers etc., and leaves the cantilever layer and the AlGaAs stop layer under the mirror layers and where the cantilevers are supposed to be. The final step is to remove the substrate from under the cantilevers and mirrors, but leave it on the chip edges. In order to do this, the backside is lapped down to 150 µm. The back surface of the wafer is repolished for backside lithography and etching. Since the next step is etching the back side, the front side of the chip is protected by mounting it in a glass handle using a high temperature wax. In order to define windows on the back side, a new SiN layer is deposited on the back, and converted into a mask defining the windows using lithography. A wet etch is then used to etch out the substrate. This etches out the substrate completely from the space between cantilevers in windows, but stops at the AlGaAs stop layer under the devices. This etch stop is removed using hydroflouric acid, and then the protective wax is cleaned by dipping the chip in acetone. In order to prevent breakage of the devices by surface tension when the acetone dries from their surface, they are dipped in ethanol and dried inside a critical point dryer.

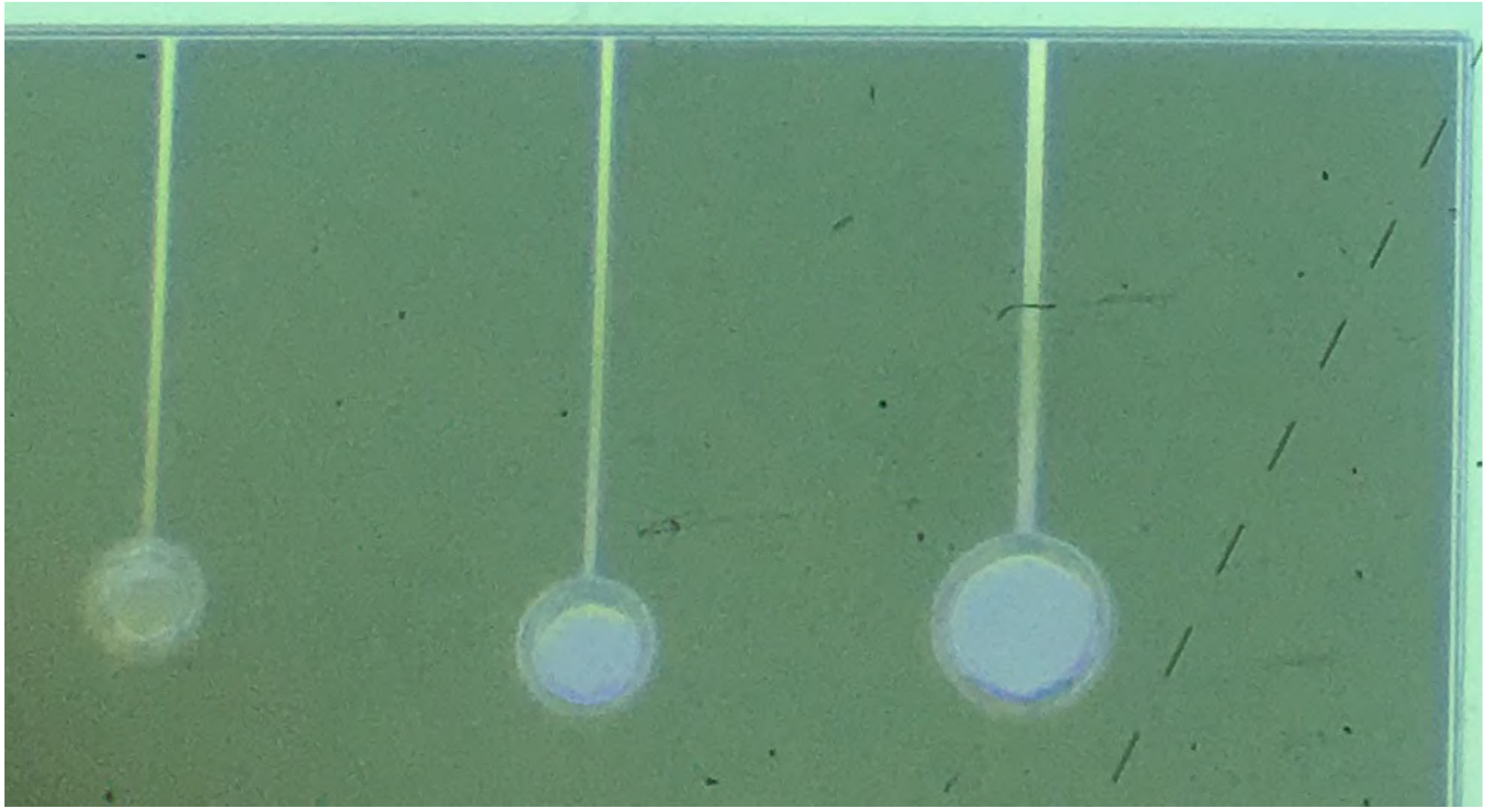

**Figure 4-2:** GaAs Shelf/Ledge: A zoomed in picture of the chip showing three devices and the shared GaAs shelf between them. The cantilevers are all attached to the shelf, which is connected to the bulk of the material. This shelf extends through the entire window.

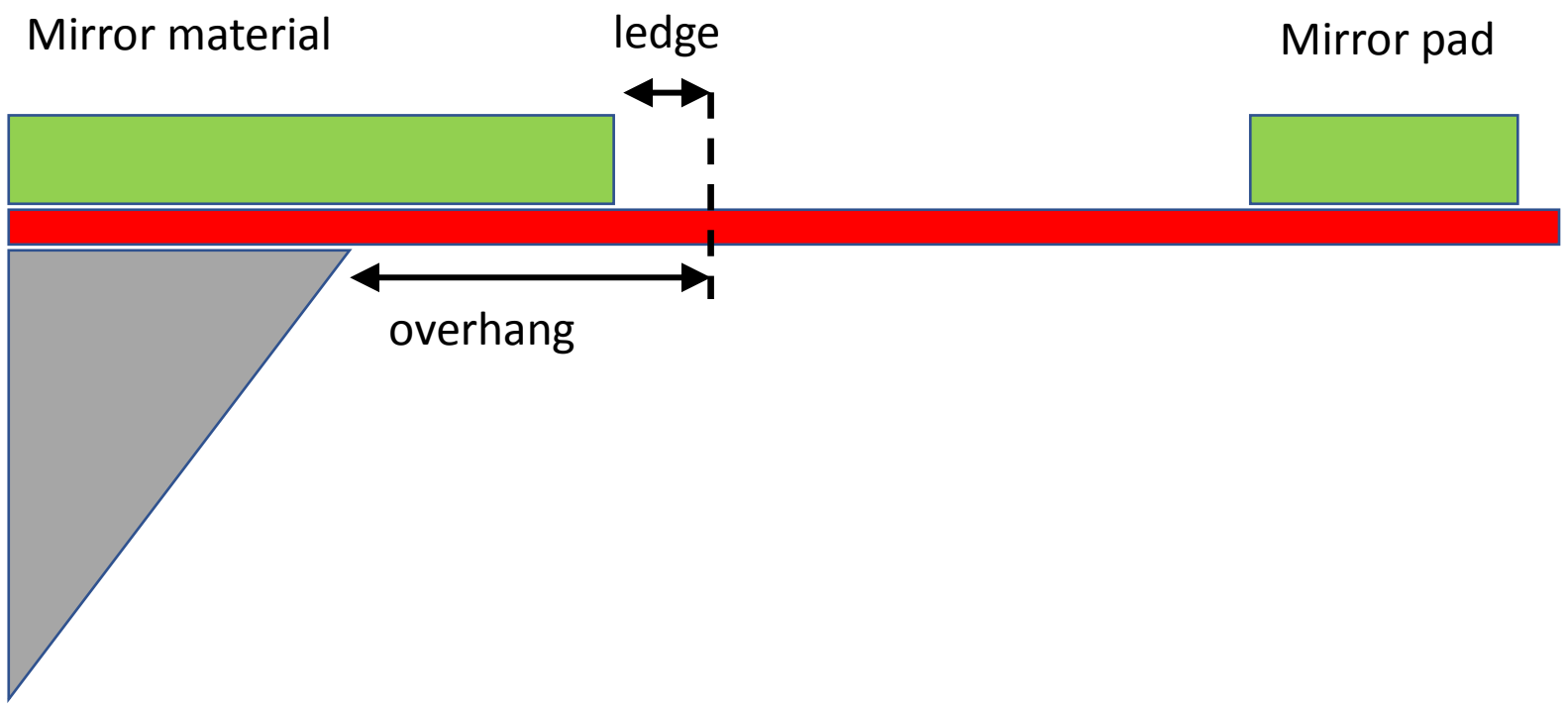

**Figure 4-3:** A side view schematic of a device on the chip. The green layer is the mirror material (AlGaAs and GaAs DBR), the red is the 225 nm cantilever material, and gray is the bulk substrate. This image shows the vocabulary we use, i.e. "overhang" vs "ledge". The COMSOL model includes the mirror, the cantilever and the ledge (which is also called "shelf"). Credit:Garrett Cole

| Layer # | no. of repeats | Al content | In content | description | thickness (nm) | index of refraction (300 K) | comments |
|---|---|---|---|---|---|---|---|
| 0 | 1 | – | – | GaAs substrate | 150e3 | 3.48041 | |
| 1 | 1 | – | – | GaAs Buffer layer | – | 3.48041 | |
| 2 | 1 | 0.92 | – | AlGaAs | 268.0 | 2.97717 | subs. removal etch stop |
| 3 | 1 | – | – | GaAs | 225.0 | 3.48041 | |
| 4 | 1 | – | 0.51 | InGaP (lattice matched) | 30.0 | 3.22330 | mirror pad etch stop |
| 5 | 1 | – | – | GaAs | 135.8 | 3.48041 | |
| 6 | 23 | 0.92 | – | AlGaAs | 90.8 | 2.97717 | |
| | | – | – | GaAs | 77.7 | 3.48041 | |
| | | | | Total thickness w/o buffer | 4534.3 | | |
| | | | | ignoring etch stop | 4266.2 | | |
| | | | | mirror pad only | 4041.2 | | |
| | | | | cantilever only | 225.0 | | |

**Table 4.1:** Layer structure of the wafer: This table shows various layers that are a part of the wafer. The GaAs layer highlighted in purple is the layer that forms the mechanical oscillator, and the AlGaAs/GaAs layers highlighted in orange form the mirrorpad. The rest are various stop layers for fabrication, and the bulk substrate on which everything is grown.

## 4.2 Modeling

We model the device using finite element analysis (FEA) in COMSOL®, interfacing to it via Matlab. We introduce a basic device geometry, reduced material properties, and appropriate boundary conditions in the COMSOL model, and then use its "Solid Mechanics" physics package to obtain the eigenmodes of the device. This package also provides the displacement profile (eigenvector) associated with each eigenmode, and we use this displacement profile to obtain an effective mass for each mode (called modal mass). This modal mass is the ratio of a density weighted norm (Norm) and a laser weighted displacement (LWD). The expression used to obtain the modal mass with a certain optical beam position and for each mechanical mode is given in Eq. (4.1), and a detailed explanation to how and why this is the effective mass is provided in Appendix A.

$$\mathrm{LWD_{ni}} = \frac{1}{\pi r^2}\int_S \mathrm{Re}(w_n)\mathrm{e}^{-\frac{(x-x_i)^2+(y-y_i)^2}{r^2}} \tag{4.1a}$$

$$\mathrm{Norm_n} = \int_V \rho\, |d_n|^2 \tag{4.1b}$$

$$m_{\mathrm{ni}} = \frac{\mathrm{Norm_n}}{\mathrm{LWD_{ni}^2}} \tag{4.1c}$$

Here, the first integral is calculated over a patch $S$ on the mirror surface, which is of 15 µm radius. $w_n$ is the z-component of the displacement field of the $n$-th mechanical mode, $r$ is the 1/e radius of the expected spot size of the optical beam, $(x_i, y_i)$ are the coordinates of the $i$-th beam position, $n$ is the index over various mechanical modes, and $i$ is an index over various positions of the laser beam. The second integral is calculated over the entire volume $V$ of the device, $\rho$ is the material density and $d_n$ is the magnitude of the total displacement of the $n$-th mode. While we have one LWD calculation at a fixed beam position set up in the GUI model, we usually set the LWD using Matlab to enter new values of beam positions $(x_i, y_i)$. The Norm is calculated in the COMSOL® model for each position and imported into Matlab.

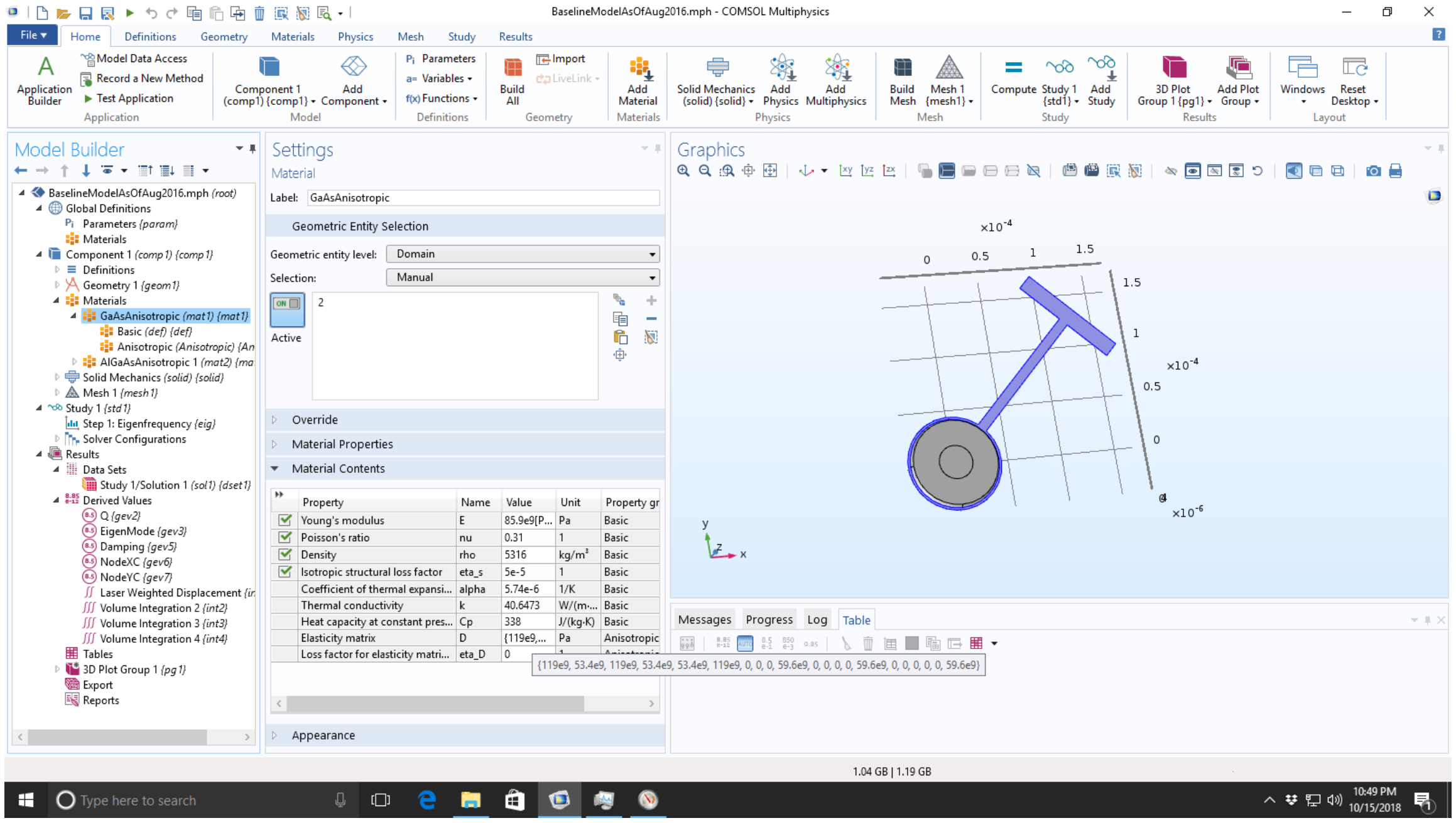

**Figure 4-4:** GaAs material properties: A screenshot showing the material properties of the GaAs layer as entered in the COMSOL model. The mouse cursor, highlighting the elasticity matrix, shows the full composition of the elasticity matrix.

### 4.2.1 Material properties

The COMSOL model contains the mirrorpad, the cantilever, and the free hanging ledge. We have two different materials in COMSOL, GaAs and AlGaAs. Even though the mirrorpad is made of alternating layers of GaAs and AlGaAs, we use just a single material in COMSOL to model it. This allows us to reduce the computation time by not including twenty-three 90 µm and 77 µm layers. The mirrorpad material is shown in orange in Fig. 4-1. We include the density, and the elasticity matrix for each material, as shown in Figs. 4-4 and 4-5. The elasticity matrix governs properties like the effective Young's modulus, sheer modulus of each of the materials. We also include thermal properties: coefficient of thermal expansion, heat capacity, and heat conductivity to model thermoelastic effects, but as explained in Section 4.2.4, that has resulted in problems, so the physics is currently set up to ignore those properties.

Finally, it is important to note that while the compliance matrix includes the material's anisotropy, COMSOL's default settings in the physics ignore this anisotropy. But to get correct results, we must change those settings manually to have COMSOL not ignore the material's anisotropy. A screenshot of the relevant settings is shown in Fig. 4-6.

### 4.2.2 Geometry

As with the material properties, simplifications must also be made in the device geometry to capture the essential physics in the least amount of computation time and minimal complications. Fig. 4-6 also

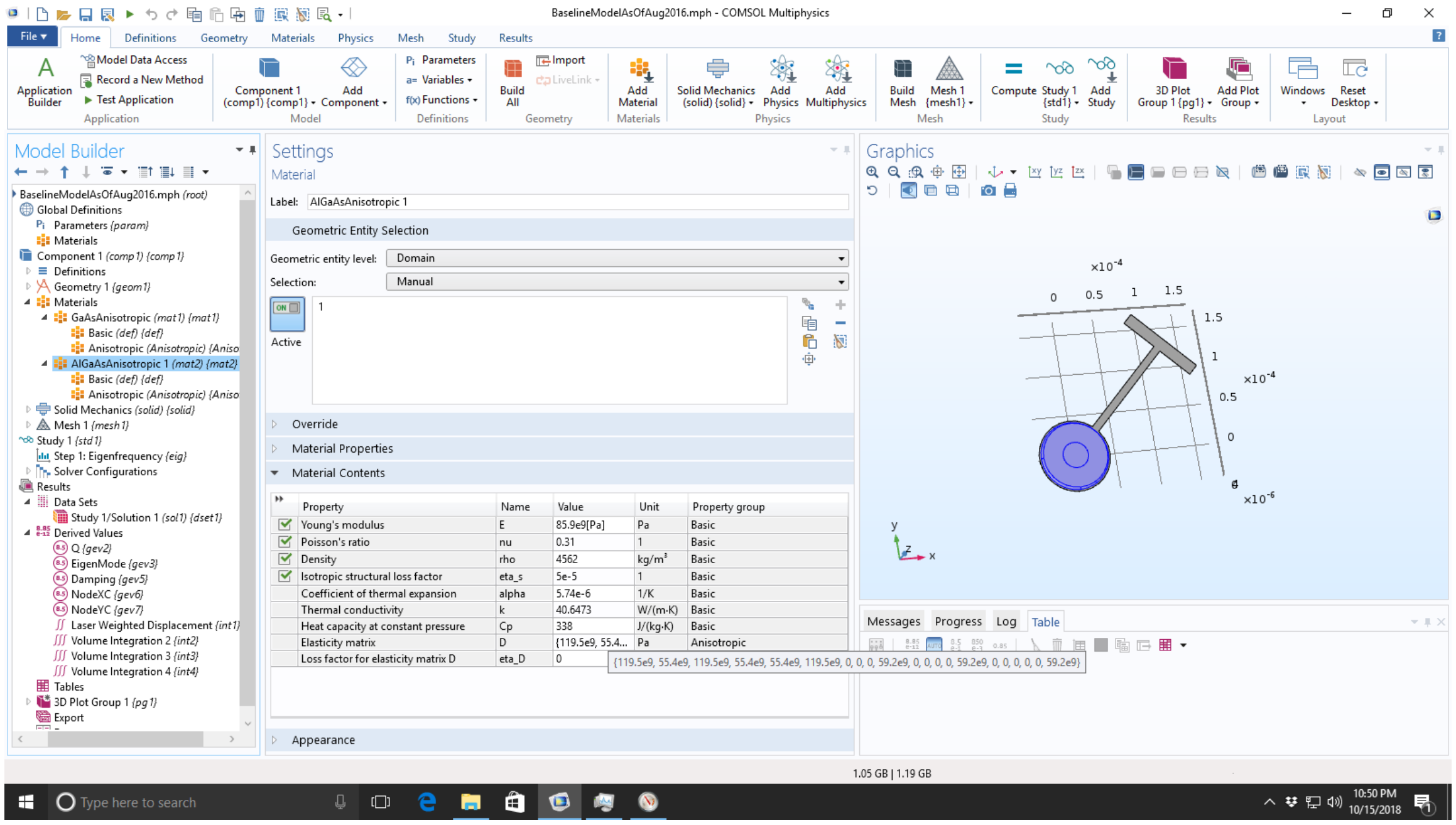

**Figure 4-5:** AlGaAs material properties: A screenshot showing the material properties of the mirror layer as entered in the COMSOL model. The mouse cursor, highlighting the elasticity matrix, shows the full composition of the elasticity matrix. It can be slightly confusing that it is called AlGaAs in the COMSOL model, but in reality represents the DBR stack made of both AlGaAs and GaAs.

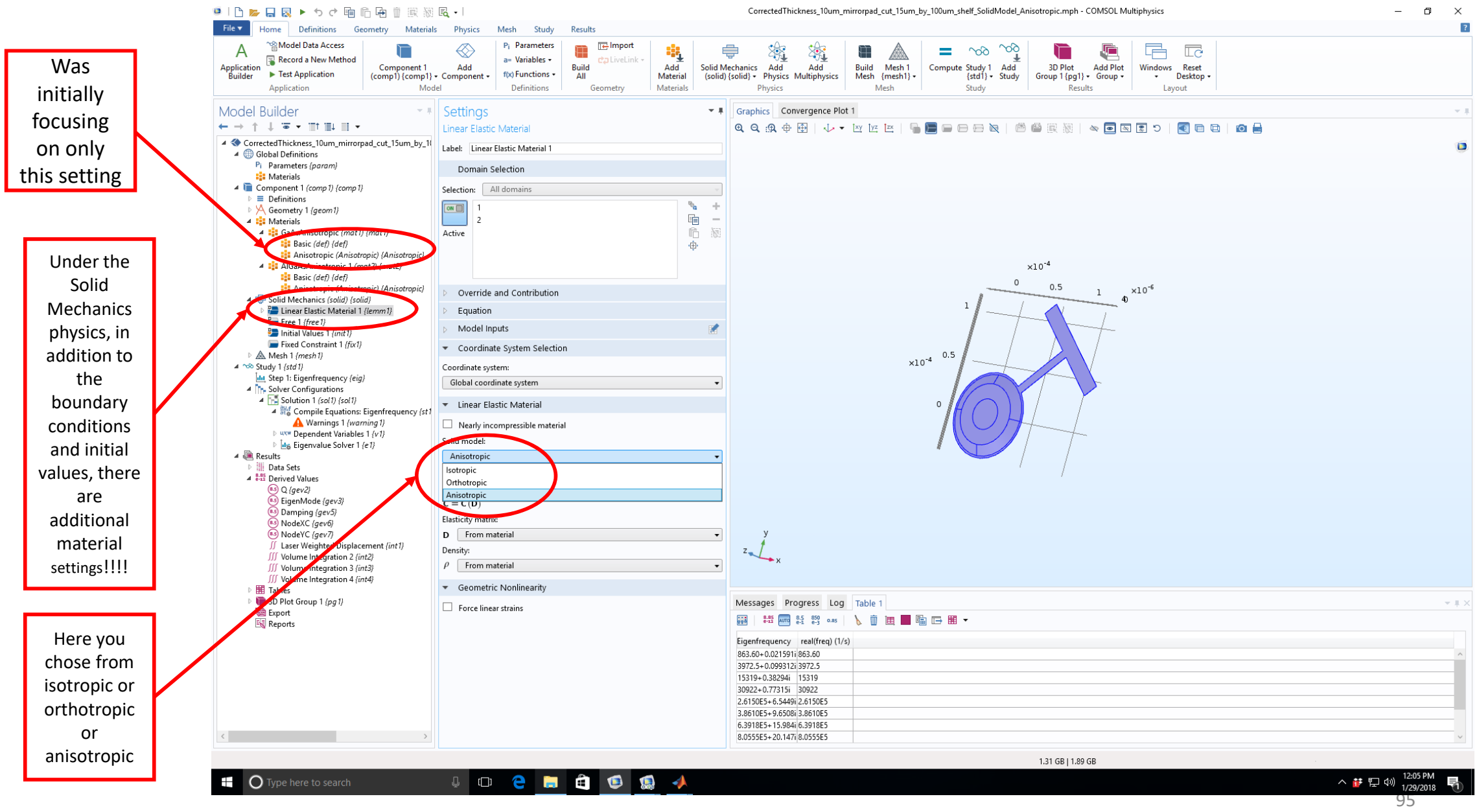

**Figure 4-6:** Anisotropy settings in COMSOL: A screenshot showing additional anisotropy settings in COMSOL. It is not enough to just set a given material to be anisotropic, one also needs to select anisotropy in the physics.

shows the geometry as seen in the COMSOL window. The main features of the geometry can be seen in Fig. 4-1:

- **GaAs Cantilever** The main cantilever is a 225 nm thick layer of GaAs, of length $L$ and width $W$. The length and width vary from device to device, but the layer thickness is fixed.

- **GaAs Pad Under Mirror** When etching the backside of the chip, the GaAs layer under the mirrorpad is intentionally left as is. Also, to allow for more wiggle room in the fabrication process, this GaAs layer is made to be 2 µm bigger in radius, by design. This is to ensure that the back etch does not cut the mirror.

- **AlGaAs Mirror Pad** The mirrorpad is modeled as a truncated cone. This is because during the etch process, there is a slope to the cut (modeled to be 2 µm).

- **GaAs ledge** As described in the fabrication section, there is usually a little leftover GaAs material from the chip edge during the etching process. The width of this ledge varies from fabrication to fabrication. We used 15 µm as a ballpark number in our models, but fine tuned this later to fit the model to thermal noise measurements. The length of this ledge usually extends to the entire window. But there are also other cantilevers attached to the same ledge in the same window. This means that the entire chip length is not a good length to model the ledge. We chose 100 µm to model the ledge, since that seems a good order-of-magnitude effective length that includes the effect of free hanging ledge but also the fact that it is clamped to other cantilevers. The ledge length is also usually fine-tuned when trying to fit the model to the data.

- **Mirror Surface** The surface of the mirror pad that sees the laser beam is also drawn separately, in order to allow for more nodes there, as explained in the Meshing section.

Finally, the crystalline wafer is cleaved at 45° from the symmetry axis of the crystal to get a clean cut. This means that the chip is rotated 45° w.r.t. the crystal axis. In order to take this into account, the COMSOL basis is chosen to be the same as the crystal basis in order for the compliance matrix to be correct. This means the cantilever is at 45° from COMSOL axis. But for ease of computation and notation from the laboratory point of view, we define a rotated coordinate system in the cantilever frame: $X_C$ and $Y_C$ that lines up with the cantilever (Fig. 4-7). This allows us to interpret various physical effects in our system, for example, all the eigenmodes, and interpreting the effect of beam misalignment as "pitch" or "yaw". The convention for beam misalignment is shown in Fig. 4-8.

### 4.2.3 Meshing

Meshing in COMSOL refers to dividing up the model into smaller pieces for the finite element computation. The relevant partial differential equation is then solved for each unit cell, and later the solution is patched together. While COMSOL offers automatic meshing, it was not suitable for our device due to the extreme geometrical proportions. For example, the cantilever thickness can be as much as 1000 times smaller than the length. This extreme nature of our device required us to make our own custom mesh, which was derived from multiple automatic meshes offered by COMSOL. So here we will describe what are the considerations to be kept in mind when meshing a system, and what mesh finally did the job for us.

When meshing a device, a finer[1] mesh gives more numerical resolution in the computation. But that also results in larger computation time and resources. So the basic idea is to test a mesh on a model by going finer and finer till the results converge. If the results of the $n$-th, $(n+1)$-th, and the $(n+2)$-th meshes are same within desired error (but $(n-1)$-th mesh's results were outside the desired error bars), the $n$-th mesh is good enough. This is the ideal case scenario. When we did this study for our system, we found that the accuracy first increased, and then decreased again as the mesh was

[1] The names of COMSOL's various meshes are also "fine", "finer" etc. Each of these meshes changes the size of the unit cell, which can be found in the COMSOL manual (or if you change from "fine" to manual, you can see the element size that was chosen by the "fine" setting.) Here though, when we say finer, we mean a mesh with smaller and more numerous elements in general, and not the particular "finer" mesh setting in COMSOL.

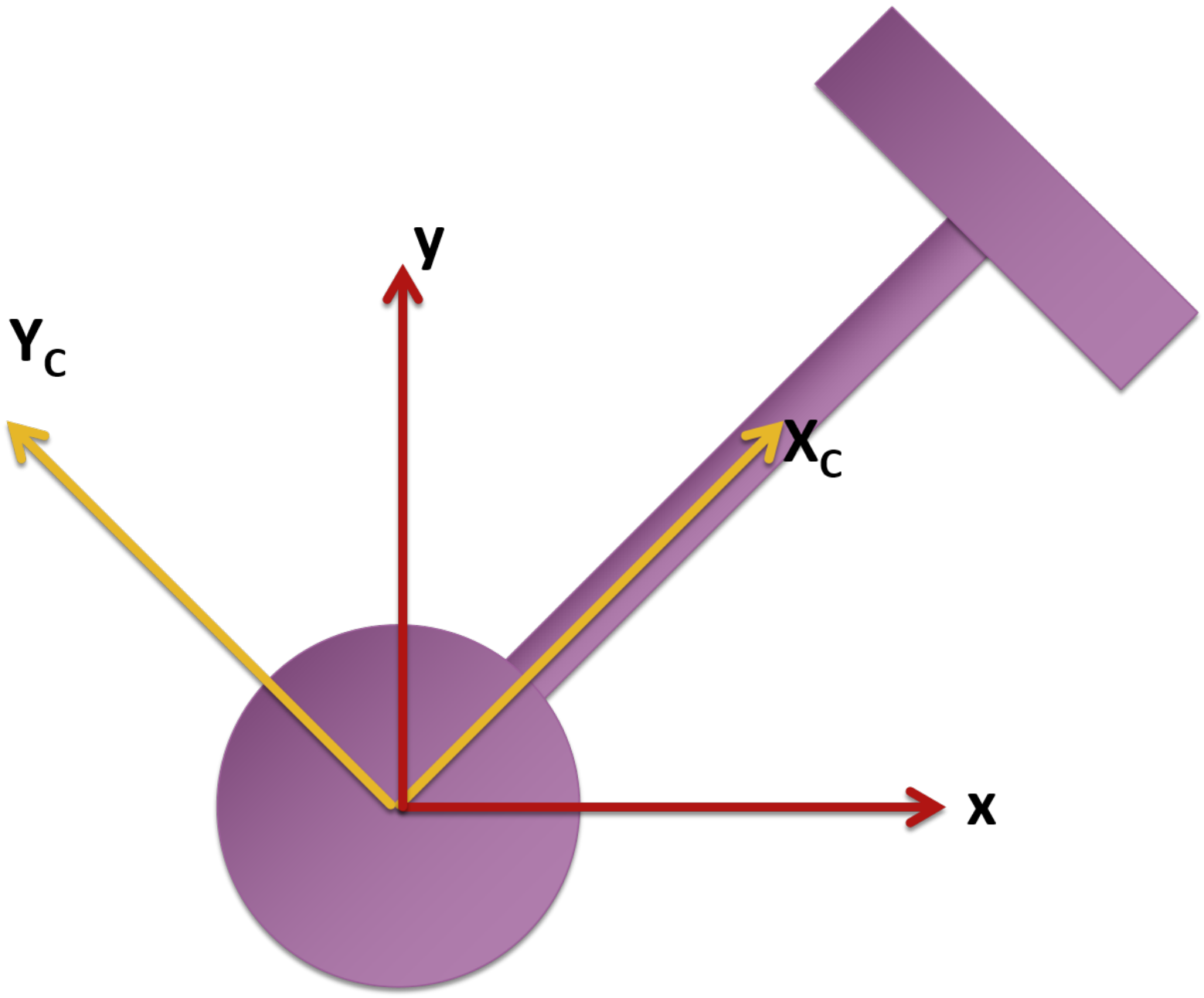

**Figure 4-7:** Axes definition in COMSOL Model w.r.t. crystal axis. The crystal axis is shown in red, which is the main axis in COMSOL, i.e. the axis for which the elasticity matrix is provided. We then define a new "cantilever" axis in COMSOL, shown in yellow, which makes drawing and result interpretations easier.

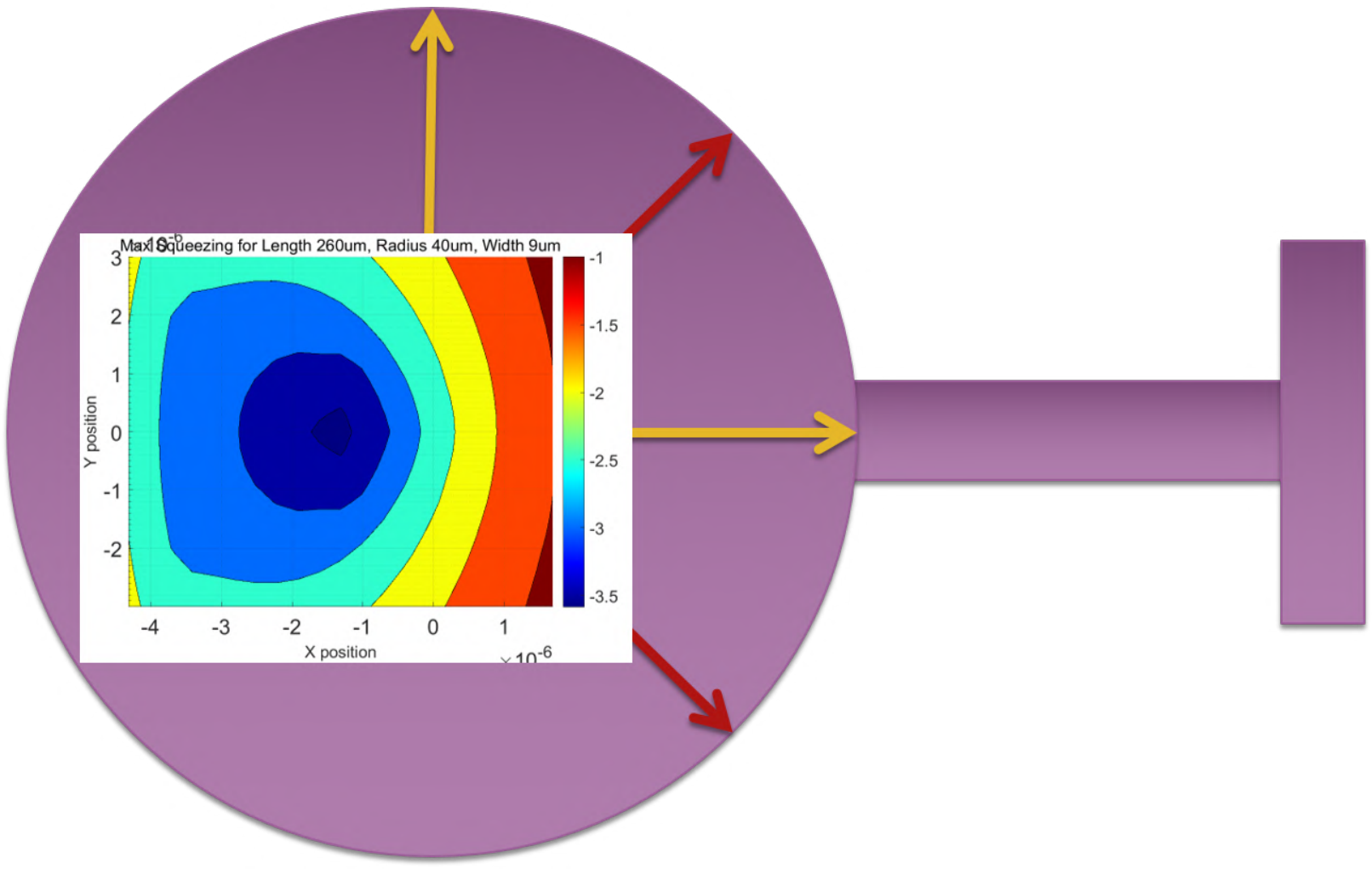

**Figure 4-8:** Beam alignment axis definition: This figure shows the definition of $X$ and $Y$ as used in the alignment sensitivity plots, with respect to the device.

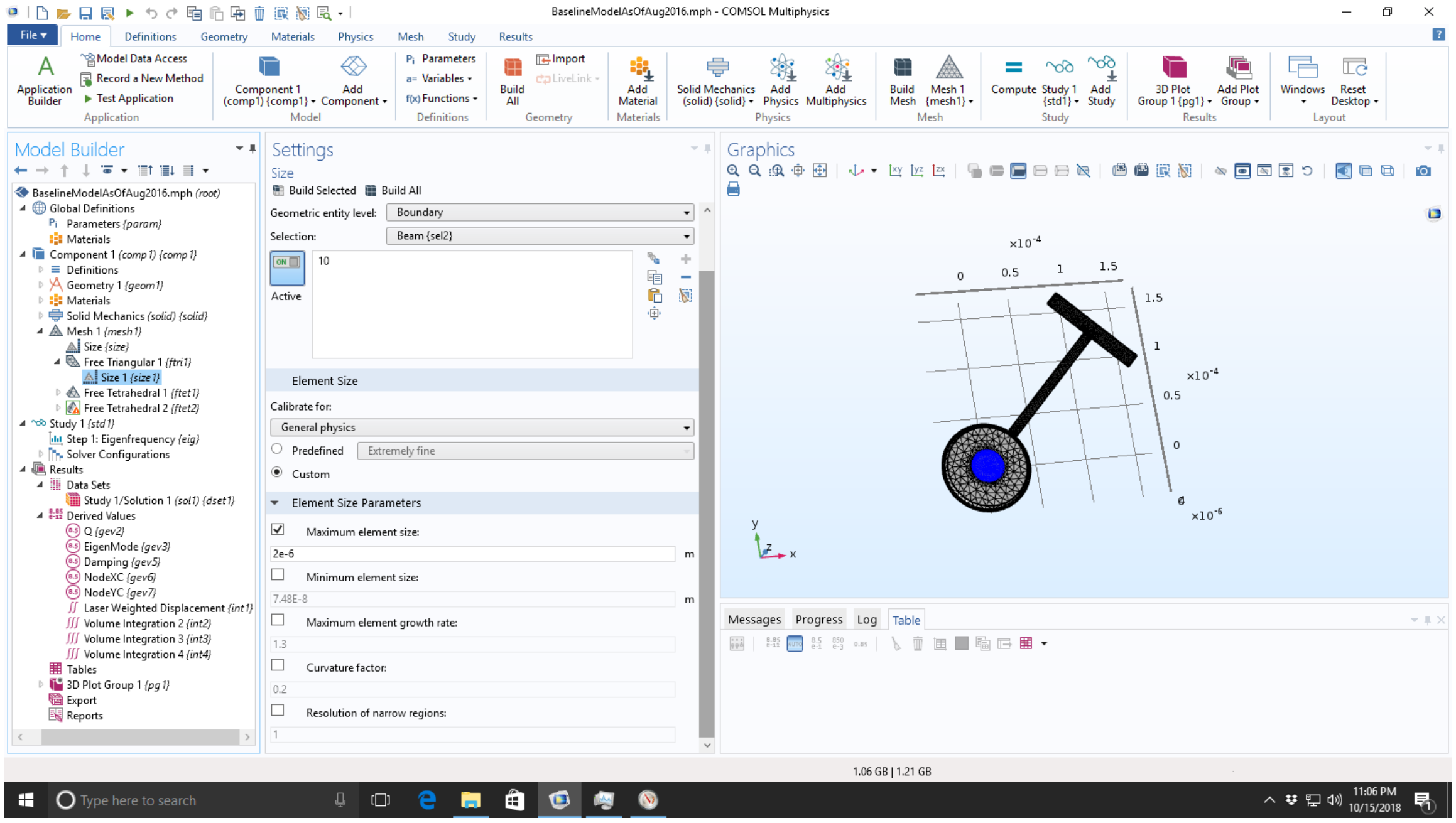

**Figure 4-9:** Meshing the mirror surface in COMSOL: This screenshot shows the central area on the mirrorpad that gets meshed before everything. This patch is 15 µm in radius, and is meshed extremely finely with triangular elements to get a good numerical precision on the overlap integral between the laser's intensity profile and the displacement field of the device. In the window second to the left one can see that we use the "maximum element size" setting to force COMSOL to mesh this patch is small elements.

made finer. Our current understanding is that this has to do with the mesh being too fine, and the GPU not being able to handle a numerical computation of that size, giving us such a result. We also noted the computation time along with the mesh size, and saw that while the computation time increased at first (as expected) with a finer mesh, it then decreased. This is consistent with our hypothesis that some meshes were just too fine and the computer could not handle those, albeit it would have been nice if that came with an error message. The details of this study can be found in the repository at `Cantilever\2016GenChips\Mesh_Studies\ComparisonOfCantiMeshes.xlsx`

In addition to controlling the element size, we also need to pay attention to the Krylov dimension and the solver tolerance [56, 57]. These settings can be found in the solver, although we usually set the Krylov dimension programmatically through MATLAB. In short, the Krylov dimension is related to how sparse the numerical matrix that COMSOL is trying to invert is. It is hence coupled to the number of eigenmodes one wants to solve for. In general if we want to solve for more modes, increasing the Krylov dimension along with the number of modes gets better results. Now, finally the mesh that was used in our model :

- **GaAs Layer** We used the "fine" mesh setting in COMSOL to mesh the GaAs layer (Fig. 4-11), giving element sizes between $2.33 \times 10^{-6}$ m and $1.86 \times 10^{-5}$ m. This is shown in Fig. 4-11.

- **AlGaAs Mirror Pad** We used the "normal" mesh setting in COMSOL to mesh the mirrorpad (Fig. 4-10), which has elements of size range $4.19 \times 10^{-6}$ m to $2.33 \times 10^{-5}$ m, as shown in Fig. 4-10.

- **Optical Beam patch on mirror pad** Initially we used to just have the mirror pad be the "fine" mesh and that's it. While this is capable of calculating the eigenfrequencies correctly, it is not the right way of estimating the modal mass. That is because the calculation of modal mass consists of integration of an overlap integral between optical intensity and the mirror displacement (Eq. (4.1), see details in Appendix A). This integration is performed in COMSOL via interpolation. So if the element size is comparable to the laser beam size (which it was), we get inaccurate results. In order to fix this, we added a 2D free-triangular mesh to the patch of the mirror that is relevant for the optical beam. Since the optical beam is expected to be roughly 2.5 µm in radius, we chose this area to be 15 µm in radius. We mesh this disc as "extremely fine" with "custom" with a maximum element size of 2 µm. Since this is now the finest mesh of all, it has to be implemented before any other mesh is implemented, and that way we can mesh the remaining mirror afterwards (shown in Fig. 4-9).

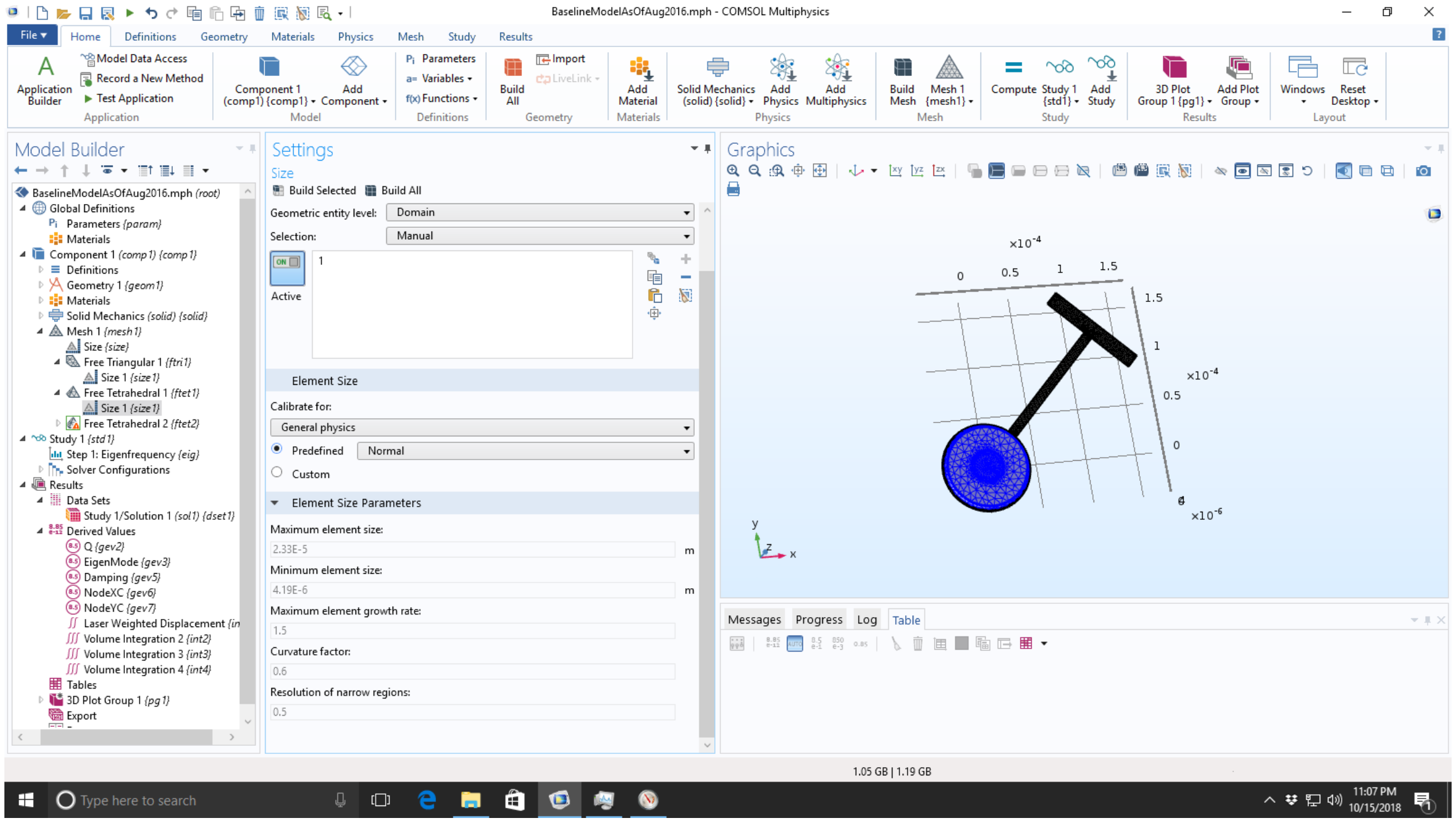

**Figure 4-10:** Meshing of the mirrorpad in COMSOL: This screenshot shows the next step in meshing – the mirrorpad. Since the mirrorpad is big, we do not need this mesh to have a lot of elements. We use tetrahedral elements to mesh this.

## 4.2.4 Thermoelasticity

Thermoelastic effects are basically an interplay between thermal and the elastic nature of a material: let's say the material has a mechanical stress, in addition to the mechanical strain, this stress also results into heating of the material governed by its heat capacity ($C_P$). This heating can cause additional strain, which is in turn governed by the coefficient of thermal expansion ($\alpha$). But since the speed of sound in the material is different from the heat conductivity ($k$), there is a larger delay in the heat-induced strain from the mechanical strain. This effect can cause extra damping which then leads to a lower effective quality factor.

In our device, thermoelastic damping (TED) effects are more relevant for the modes of the mirrorpad than the modes of the cantilever. We think this is because the mirrorpad has a bigger thickness and area which requires a larger time for the heat flow and temperature to reach equilibrium.

Initially, we used the "Thermoelasticity" physics in COMSOL to model the device. This allowed us to use the full differential equations including the heat flow and its effects on the mechanical stress together. The system of equations solved in COMSOL is [58]:

$$\rho \frac{d^2\mathbf{u}}{dt} - \nabla \cdot \sigma = \mathbf{F}_V \tag{4.2a}$$

$$\rho c_P \frac{dT_0}{dt} = \nabla \cdot (\kappa \nabla T_0) + Q - T_0 \alpha : \frac{d\sigma}{dt} \tag{4.2b}$$

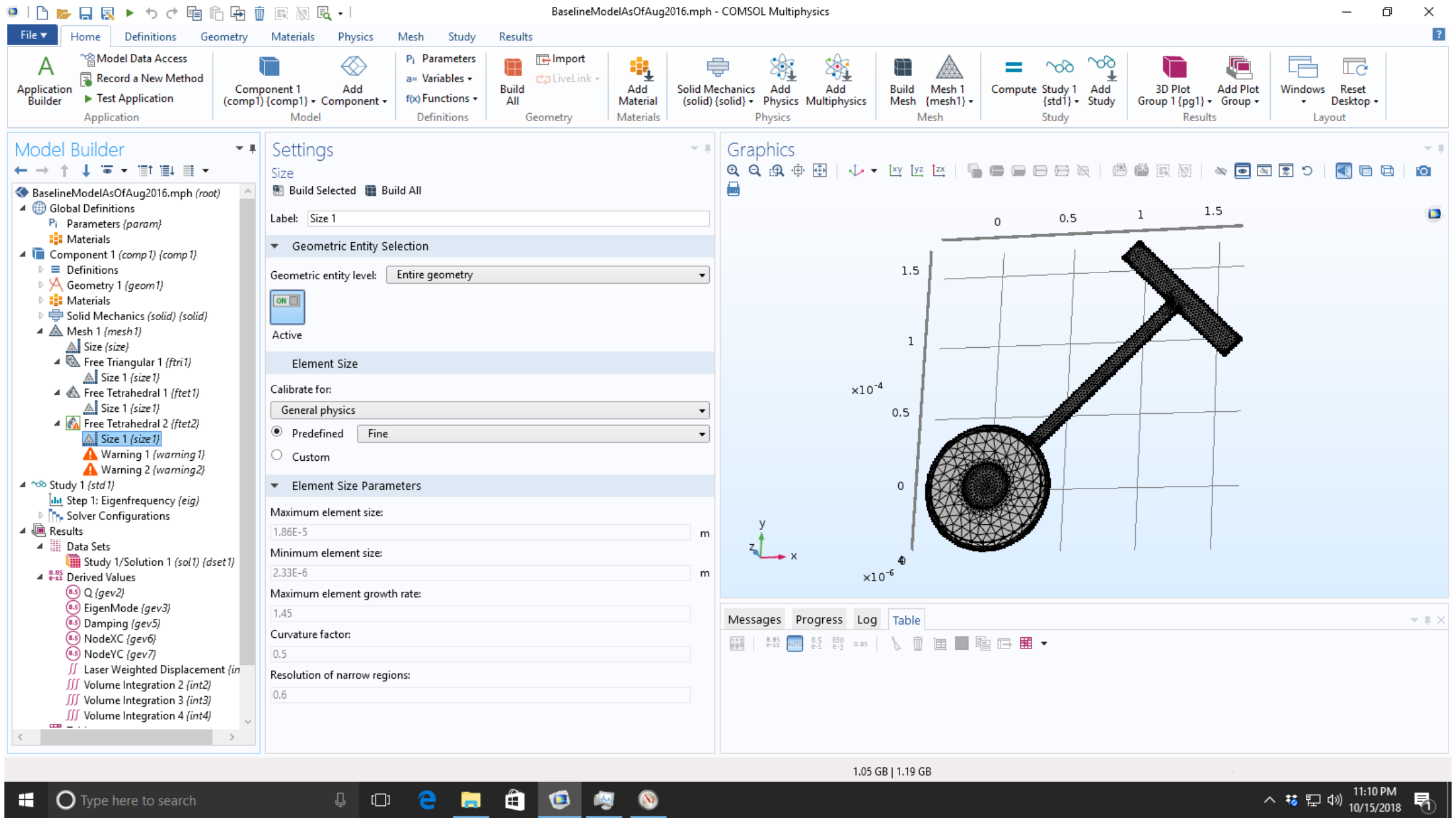

**Figure 4-11:** Meshing the GaAs layer in COMSOL: In order to mesh the GaAs layer, it is important to select 'remaining domains' setting in the 'domains' setting in 'Free Tetrahedral 2'. Next, since the GaAs layer is thin, we select a "fine" mesh.

Here, in the mechanics equation 4.2a $\rho$ is the material density, $\mathbf{u}$ is the displacement field, $\sigma$ is the stress tensor, and $\mathbf{F}_V$ is the force density of an external applied force. Similarly, in the heat transfer equation 4.2b, $c_P$ is the heat capacity at constant stress per unit volume, $T_0$ is the absolute temperature, $\kappa$ is the thermal conductivity, $\alpha$ is the coefficient of thermal expansion, or rate of change of strain per unit temperature at a fixed stress. $Q$ is an external applied heat. These equations are solved by COMSOL for given material properties, and $\mathbf{F}_V$ and $Q$ get decided by user-provided boundary conditions. These boundary conditions are also provided in the physics settings.

The above equations are solved in the frequency domain, which also allows one to include a structural damping phase factor in the mechanical equation. We then use the imaginary part of the eigenvalues found by COMSOL to get the effective quality factor for each mode. We discovered that for a few modes in some device geometries these Q's were negative. When we investigated this closely, we also found that for some modes these Q's were higher than 20000 (going up to 100000). Both these results are unphysical, since the isotropic structural loss of the material allows a maximum Q of 20000. After a lot of investigating, we could not find the source of this error, but this error meant we were not convinced that the real parts of the eigenvalues were correct.

We then moved to using just solid mechanics physics. This physics only solves the first equation. While it can use the loss factor to calculate the imaginary part of the eigenvalues, that calculation is now of little use because this loss factor is isotropic. So we just used COMSOL to calculate the eigenvalues and assigned them a fixed quality factor of 20000. This treatment was not the best way to deal with the modes of the mirror-pad though. In order to do that, we found the modes of the mirrorpad and assigned them a lower Q of 1000. This value was informed by a previous measurement of thermal noise of similar devices, which was then fit to a model. While this Q=1000 would have large uncertainties with respect to what the actual TED Q would be, it was still our best guess scenario and allowed us to perform a first order optimization of the devices. In order to find the drumhead modes programmatically, without manually looking at each of their displacement fields, we look at the modes with the lowest modal mass. To ensure that these modes are drumhead modes, this cut needs to be made with the beam position at the nodes of the first two HOMs. The topic of modal masses and nodes is explained in more details in the next section.

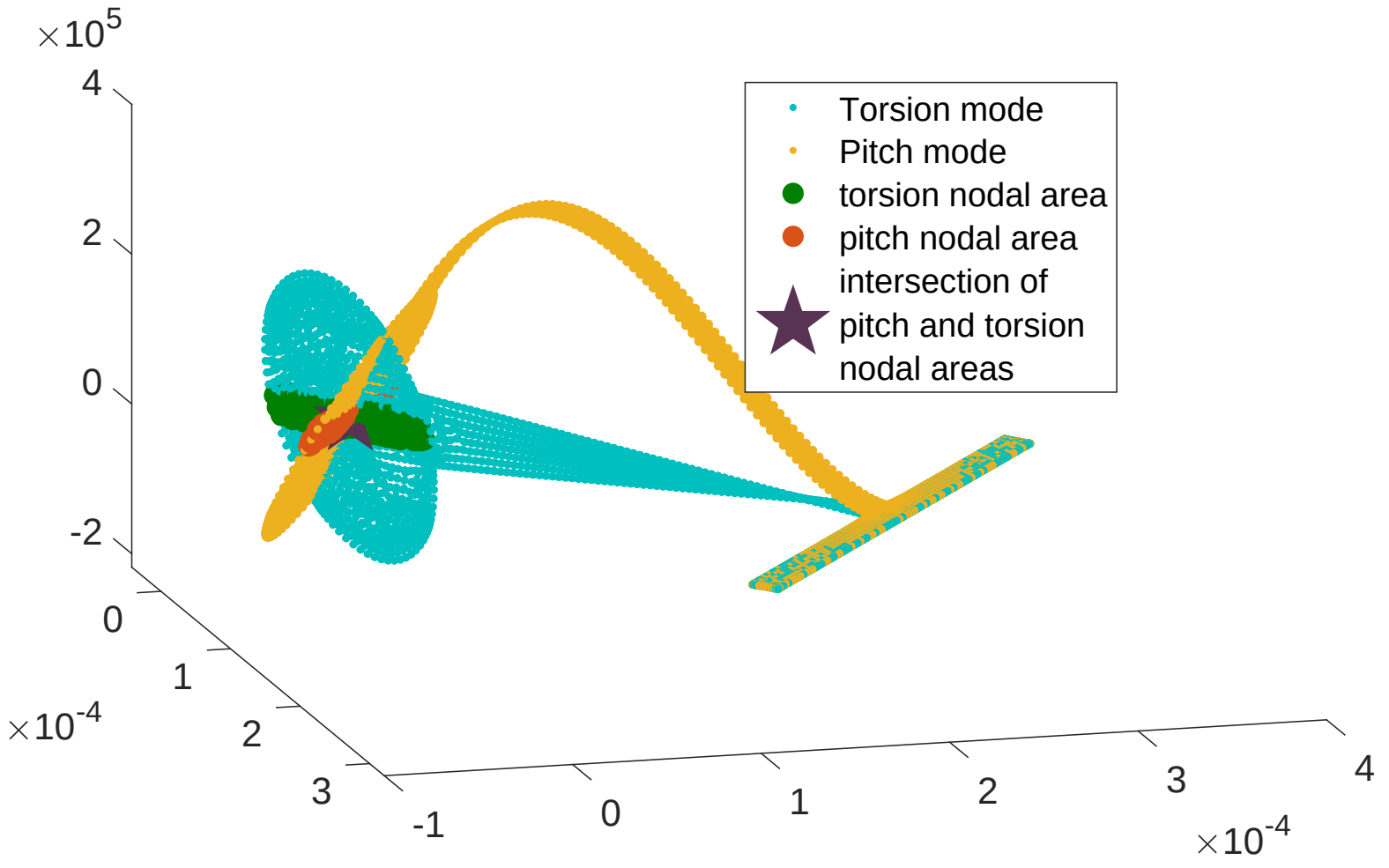

**Figure 4-12:** Beam centering for thermal noise cancellation: Each mechanical mode adds thermal noise into the system. Here we show the displacement profile of the first two HOMs of our device, the torsion mode in teal and the first bending mode in yellow. The green circles are where the torsion mode has close to zero displacement, and the orange circles are the same for the pitch mode. There is one overlapping point between the two regions of zero displacement, which is shown by the big star. If we perfectly align our cavity so as to have the center of the cavity mode coincide with this star, we minimize the thermal noise injected by these two modes.

### 4.2.5 Modal mass and nodes

Each mechanical mode has a particular displacement field ($\vec{\psi}_n(\mathbf{r})$) associated with it. Some of these modes are shown in Figs. 4-13 and 4-14. As derived in Appendix A, we can assign an effective mass to each mode based on the overlap between the displacement of the mirror and the intensity profile of the optical beam. This allows us to model in the system in one dimension.

$$m_n = \frac{M_n}{N_n^2} \tag{4.3a}$$

$$M_n = \int \rho(\mathbf{r})\vec{\psi}_n(\mathbf{r}) \cdot \vec{\psi}_n^*(\mathbf{r})\mathrm{d}^3\mathbf{r} \tag{4.3b}$$

$$N_n = \frac{1}{\mathcal{I}_0}\int \psi_n^z(\mathbf{r})\mathcal{I}(x,y)\delta(z-z_0)d^3\mathbf{r} \tag{4.3c}$$

Here, $m_n$ is the effective mass of the $n$-th mechanical mode, $\rho(\mathbf{r})$ is the material density at a position $\mathbf{r} = (x, y, z)$, $\mathcal{I}(x, y)$ is the intensity profile of the optical beam, with average intensity $\mathcal{I}_0$.

The nodes for a given eigenmode are defined as the positions where the displacement is zero at all times. In general we can minimize the impact of a given mode on the device's thermal noise by aligning the cavity such that the center of the optical beam falls on the node of the mechanical mode. This minimizes the overlap integral ($N_n$), in turn maximizing the modal mass of that mode. A higher mass of any given mode means that that mode contributes less to the thermal noise. In our system, we gain the most amount of benefit by performing such centering for the first pitch mode (Fig. 4-13c) and the torsion mode (Fig. 4-13b) of the cantilever. We can perform this simultaneous centering because their nodal lines intersect. Thus we can place the optical beam at the intersection of their nodal lines, as shown in Fig. 4-12.

## 4.3 Optimization

Now, we wish to design devices that will give us the best squeezing. While most parameters in the geometry are not available to be optimized (they are either fixed or not in control during fabrication), the parameters that we can control are $L$, $W$, and $R$. Thus, in order to optimize the design of the device, we vary $L$,$R$, and $W$, and look at their impact on the performance of the OM system. After performing a finite element analysis of the device's mechanical modes to get the eigenfrequencies and modal masses for each set of $L$,$R$, and $W$, we use the eigenfrequencies and modal masses to compute the mechanical susceptibility and thermally driven displacement fluctuations for the device. The mechanical susceptibility and thermal noise are then used in our optical model to calculate the squeezing performance of the given geometry.

While optimizing the design, we encounter some constraints on the device geometry and material. These constraints come from fabrication and optical optimization. They are already described in the text before this, so here we will just list them.

- Thickness of mirror pad
- Thickness of cantilever
- size of shelf
- size of mirror pad bevel/taper
- larger radius of GaAs layer under mirror

We look at the squeezing performance of a particular design by mainly looking at the amount of squeezing in the squeezed quadrature, and its frequency dependence. The best goal is most possible squeezing, at lowest possible frequencies, in the widest possible band. The best way to achieve this is to have a low fundamental resonance, low mass and push the HOMs out of the spectrum to as high frequencies and/or high modal masses as possible. In addition to looking at the squeezing performance, there are a few other things that need to be kept in mind. For locking stability, it is crucial that the HOMs of the cantilever be separate from the optical spring (OS). The frequency of the first two HOMs is less important than the others because we can lower their thermal noise by beam centering, so it is okay if they are inside the spectrum. Finally, it is also important to pay attention to the sensitivity of squeezing to cavity alignment, i.e., if the cavity were slightly misaligned, how much impact would that have on the squeezing performance. This is to allow for some room in the optimization for unforeseen experimental limitations.

### 4.3.1 Pertinent eigenmodes

Given that the normal modes of this structure play a central role in optimizing its design for quantum measurements, let's look at the mechanical modes that play the most important role. Their displacement functions are shown in Figs. 4-13 and 4-14, blue being the minimum displacement, and red being the maximum displacement.

1. Fundamental mode ($f_0^Z$), shown in Fig. 4-13a. The fundamental mode frequency is the most important frequency. A lower fundamental mode implies a lower thermal noise at any given frequency. This decides the amount as well as the lower cutoff frequency of squeezing.

2. Pitch modes ($f_n^Z$), shown in Figs. 4-13c and 4-13d. These and any other HOMs add additional thermal noise in the measurement band. In some cases the added noise can be high enough that the pitch mode will be the higher frequency cut-off for squeezing.

3. Drumhead modes, shown in Fig. 4-14. The eigenfrequencies of these modes are $f_{DH} \propto 1/R^2$. The drumhead modes, as shown in the figures, have a large displacement of the mirror-pad. Since the mirror-pad is thick, these modes can have additional loss due to thermoelastic effects. While the eigenfrequencies of these modes are usually higher than the measurement band of interest, the additional loss in these modes can widen their peaks enough to add thermal noise within the measurement band.

4. The mechanical modes in the X direction usually have high effective mass due to their displacement being perpendicular to the laser, so it is less important to push these to higher frequencies.

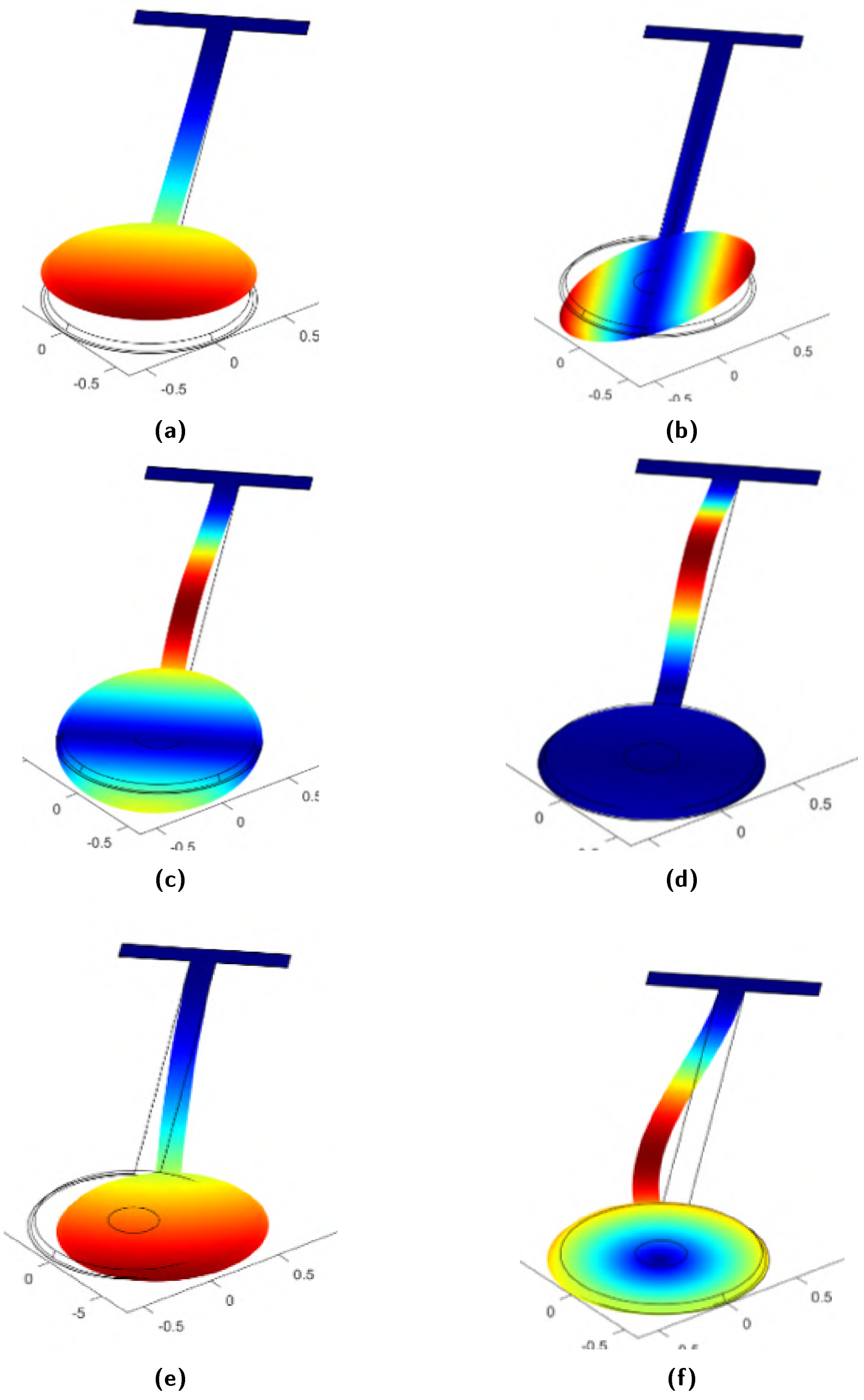

**Figure 4-13: Mechanical modes of the cantilever:** Displacement profile of various mechanical eigenmodes of the cantilever. The color profile is such that red is maximum displacement amplitude and blue is minimum displacement amplitude. The axes are the coordinates $X, y, Z$ in the COMSOl simulation frame. **(a)** The fundamental mode of the cantilever in the Z direction. This mode usually is the lowest frequency, ranging from 300 Hz - 1 kHz, so is also referred to just as the fundamental mode. **(b)** Usually the second eigenfrequency of the cantilever. This mode is a torsion mode of the cantilever, and is usually referred to as the yaw mode for historical reasons. **(c)** The first bending mode of the cantilever (we usually refer to this mode as the pitch mode). One of the nodes of this mode goes through the mirror pad (blue region). **(d)** The second bending mode of the device. **(e)** The fundamental mode of the cantilever in the Y direction. This mode has little displacement in the axis of the cavity, and so typically has a high modal mass. We refer to the Y-direction modes as side-side modes. **(f)** Second mode in the Y- direction. All the modes shown here including the fundamental mode are due to elasticity of the material, not to be confused with the pendulum mode due to gravity.

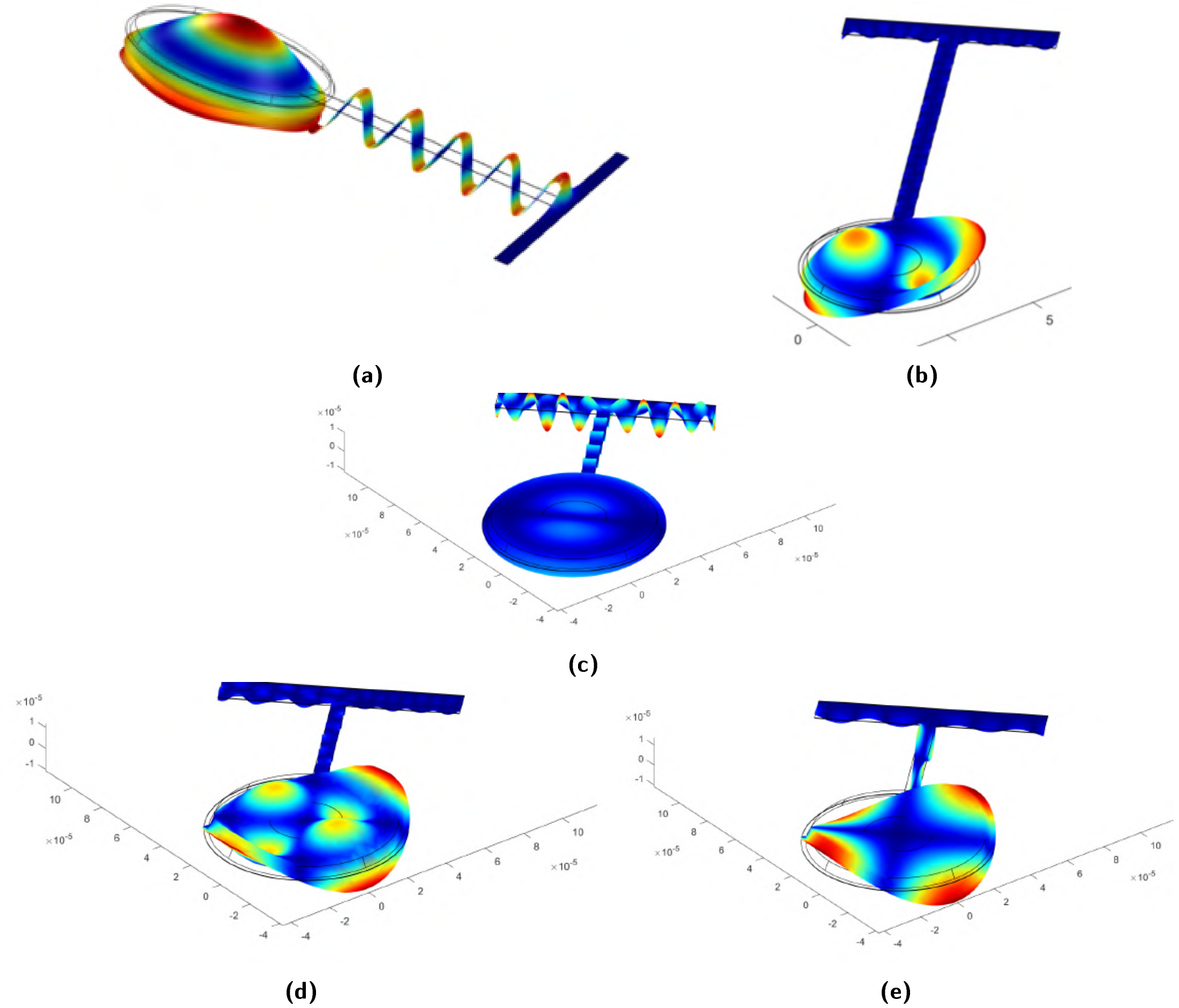

**Figure 4-14: Modes of the mirrorpad: (a)** First drumhead mode of the mirror, we see the entire mirrorpad goes through a drum-like displacement. **(b), (c),(d)** 10, 01, and 11 modes of the mirrorpad. **(e) butterfly mode of the mirrorpad.** Because of the large displacement of the mirrorpad, the drumhead modes couple very well to the optical beam profile of TEM00 cavity mode. Additionally, due to the large displacement of the thick mirrorpad, all the mirrorpad modes involve extra delay associated with thermoelastic damping. The increased coupling leads to a mode having a low modal mass. The low modal mass combined with increased damping, results in higher thermal noise displacement.

5. The rest of the mechanical modes are usually at frequencies higher than frequencies of interest.

So, in order to optimize the device design, we want to increase the frequency and modal mass of the HOMs, which increases the bandwidth and lowers the thermal noise. Of all the HOMs, we focus on ones that are prone to having the lowest modal mass and are in or near the squeezing frequency band. Next, when it comes to the fundamental mode frequency, a lower fundamental frequency increases the squeezing at a given frequency by lowering the thermal noise. This also results in a higher bandwidth of squeezing (mostly on the low frequency side because the high frequency limit probably comes from the OS or a HOM). Wider cantilevers worse for thermal noise due to higher fundamental frequency, but better for durability in chip transport and handling. Especially when it comes to the high mass designs, it becomes important for their structural stability to have wider support. A larger mirror radius provides a lower diffraction, a lower thermal noise from the fundamental mode, but higher thermal noise from the lowering of drumhead frequency. Higher fundamental mode devices are easier to handle. When looking at displacement thermal noise, a high mass results in low displacement thermal noise as well as low fundamental, which also leads to a lower thermal noise. But a high mass also lowers quantum radiation pressure displacement noise for a fixed optical design.

#### 4.3.1.1 Analytical expressions for normal modes

Conducting this full simulation takes a significant amount of time for each geometry and the scale up for three free parameters is pretty steep if we just vary them all freely. In order to address this problem, it helps us to get some good initial guesses from intuition on where we'd like the eigenfrequencies to be. Once we know that, then in principle we can start to design a device from that point. To get a good set of initial guesses for the geometries once we know what is desirable for eigenfrequencies, the most intuitive way we found was to derive analytical expressions for eigenfrequencies depending on the geometry.

While there are no exact analytical expressions for the complicated geometry under consideration, a few simplifications allow one to get approximate expressions. To find the modes of the cantilever, we assume that the mirror-pad is a point mass, located at its center or mass, at the end of a cantilever of length $L + R$. The cantilever thickness is much smaller than the mirror thickness, which allows further simplification that the mass of the entire device only depends on $R^2$. Next, the effect of the shelf is ignored because it is much shorter than the cantilever length, which provides most of the lever arm. These simplified expressions are shown in Table 4.2. The frequencies of the drumhead modes are usually proportional to $R^2$, and don't depend on $L$ and $W$.

We use the approximate analytical expressions to have an informed pool of initial guesses going by our above intuition. We then simulate these geometries in COMSOL and look at their squeezing performance. Once we have the numerically obtained data for the thermal noise and squeezing performance, we can update the input geometries accordingly. (For example, there might be a geometry which is overall pretty good but a HOM gets close to the OS. In that case, we can try changing the width to see if it will change the frequency enough without changing any other behavior.) After iterating this process a few times, we pick geometries that not only provide us good squeezing, but also provide us with diverse options. For example, we want to have devices of big and small mirror pads in case our estimates for diffraction are inaccurate. In addition, we also want to have devices that produce more squeezing at higher frequencies in addition to the (preferred) low frequency squeezers. This is in order to have back-ups in case there is a problematic noise source in a given frequency band. Taking all of this into account, we picked eight base geometries for room temperature squeezing. Their squeezing performance is shown in Fig. 4-28. Corresponding thermal noise is shown in Fig. 4-29, and the ratio of QRPN to thermal noise for each is shown in Fig. 4-30. In Sections 4.3.2 to 4.3.4, we demonstrate the major realizations we learned during this optimization process.

### 4.3.2 Role of length

In Fig. 4-15 we show two of the chosen designs, where they have the same radius and width, but only differ in the cantilever length. (In all of these plots, the legend convention is the fundamental mode frequency followed by dimensions in the order $L/R/W$ in micrometers.) The device with the longer cantilever length (shown in red curve) has a lower fundamental frequency, and hence also a lower displacement thermal noise than the one with the shorter cantilever (blue curve). But the longer cantilever also results in lowering of the HOM frequencies. This has lesser effect on the first two HOM due to the beam being placed on their nodes, but the next mechanical mode (which is the next pitch mode in this case) does not have the same cancellation. So lowering its frequency causes an increased thermal noise from this mode, as can be seen starting at around 10 kHz. We can similarly see the next few HOMs are also at lower frequencies as compared to their counterparts in the shorter cantilever device. So while a longer cantilever can decrease the low frequency thermal noise, a cantilever too long will have its entire band full of HOMs. This was indeed the case with the first generation designs, where the cantilevers were as long as a few millimeters.

While a low thermal noise gives us intuition for a good squeezer, we need to ultimately look at the squeezing performance to decide if the lower thermal noise was good enough. For that let's look at the yellow and green traces in Fig. 4-28. As we see, the longer cantilever has more squeezing as well as squeezing that goes all the way to 300 Hz on the low frequency side, as compared to 1 kHz for the shorter one. This can be attributed to the low thermal noise from the fundamental mode. But we also find that the yellow trace stops squeezing around 10 kHz, as compared to above 30 kHz for the shorter cantilever. This is because the HOMs inject thermal noise that is high enough to overwhelm squeezing at the high frequencies for the shorter cantilever.

| Mode | Symbol | Full Expression | Relevant parameters |
|---|---|---|---|
| Fundamental, Z | $f_0^Z$ | $\frac{1}{4}\sqrt{\frac{YW\tau_C^3}{\rho_M\pi^3R^2(L+R)^3\tau_M}}$ | $\propto\sqrt{\frac{W}{R^2(L+R)^3}}$ |
| Fundamental, Y | $f_0^Y$ | $\frac{1}{4}\sqrt{\frac{YW^3\tau_C}{\rho_M\pi^3R^2(L+R)^3\tau_M}}$ | $\propto\sqrt{\frac{W^3}{R^2(L+R)^3}}$ |
| Fundamental, torsion | $f_0^T$ | $0.1\sqrt{\frac{GW\tau_C^3}{LR^4\rho_M\tau_M}}$ | $\propto\sqrt{\frac{W}{LR^4}}$ |
| n-th order, Z | $f_n^Z$ | $\frac{\lambda_n^2}{4\pi}\sqrt{\frac{\tau_C^2Y}{3L^4\rho_C}}$ | $\propto\sqrt{\frac{1}{L^4}}$ |
| n-th order, Y | $f_n^Y$ | $\frac{\lambda_n^2}{4\pi}\sqrt{\frac{W^2Y}{3\rho_CL^4}}$ | $\propto W\sqrt{\frac{1}{L^4}}$ |

**Table 4.2:** Dependence of relevant mechanical modes on geometrical and material properties. Here the material properties are obtained from Ref. [59, 60]. $Y$ is the material's Young's modulus, around $85\times10^9$ Pa (exact value depends on the crystal axis orientation). $G$ is the sheer modulus, around $60\times10^9$ Pa. $\rho_M$ is the density of the mirror, 4562kg/m$^3$ (obtained by a weighted average of GaAs and AlGaAs), $\rho_C$ is the density of the cantilever material, 5316 kg/m$^3$. $L, W, R, \tau_C, \tau_M$ are the geometrical dimensions as defined in Fig. 4-1. The constants $\lambda_n$ specify the n-th bending mode of the cantilever, and their values are $\lambda_3 = 4.7, \lambda_4 = 7.9, \lambda_5 = 11.0$. Note that the second bending modes (Figs. 4-13c and 4-13f) have high modal-mass and have significant motion of the mirrorpad, thus their frequencies have not been approximated here.

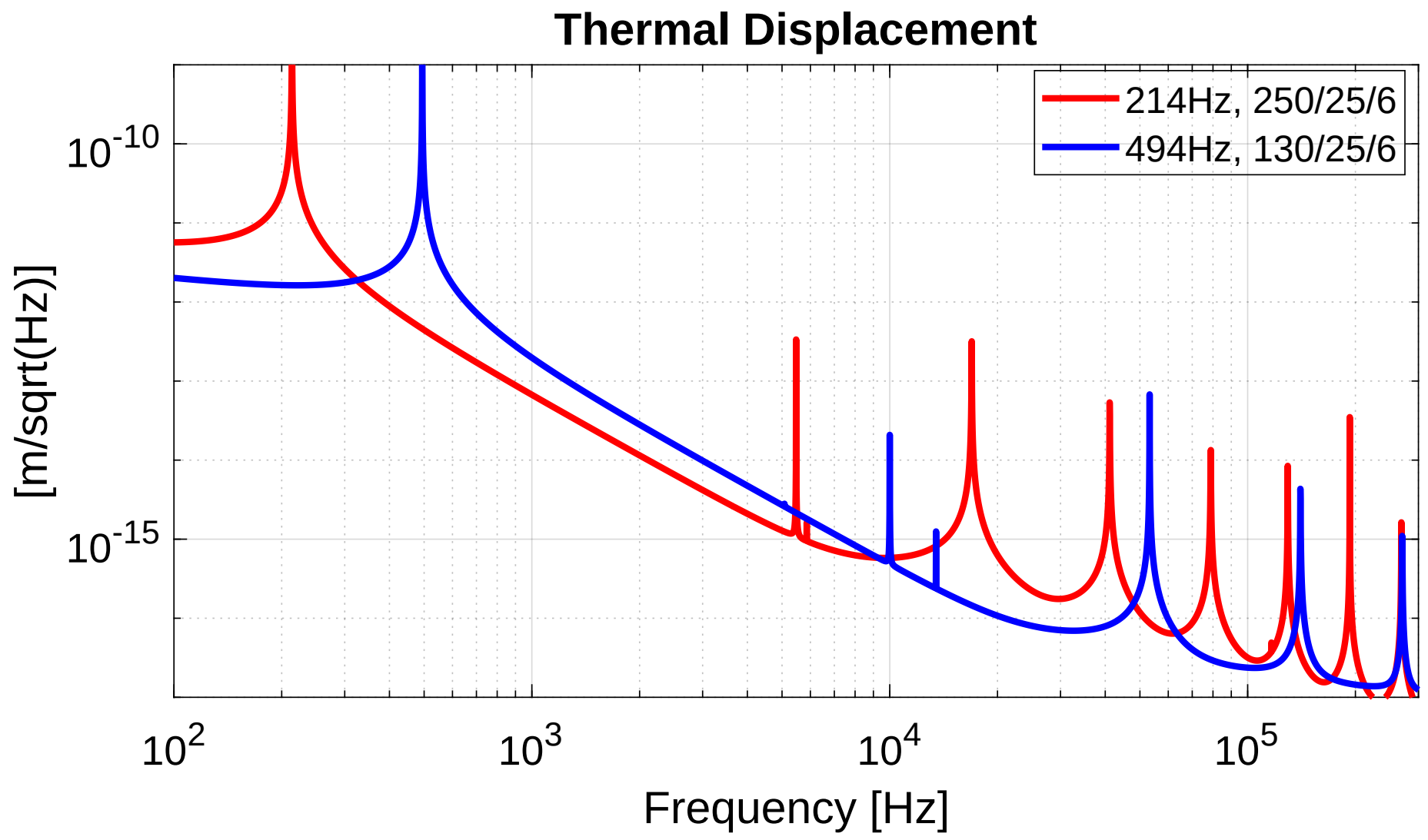

**Figure 4-15:** Effect of length on thermal noise: This plot shows thermal noise of two oscillators that are different only in their length. In blue is the shorter oscillator (130 µm long), resulting in a higher fundamental frequency. The higher fundamental frequency also leads to this oscillator having a higher thermal noise. On the other hand, in addition to having a low fundamental and low thermal noise, the longer oscillator (shown in red, 250 µm long) also has lower frequency HOMs which inject more thermal noise. So the choice of length is a trade-off between lower broadband thermal noise vs thermal noise from HOMs.

### 4.3.3 Role of radius

Next, we look at the effect of the mirror-pad radius by comparing two devices with same cantilever length but different mirrorpad radius, as can be seen in Fig. 4-16. Here, we have a bigger (60 µm) radius mirror pad in blue, leading to a lower fundamental frequency and higher mass; and a smaller mirrorpad (30 µm) in black. As we can see, the higher mass and lower fundamental frequency results in a lower thermal noise for the 114 Hz oscillator (when compared above their respective fundamental frequencies). Due to their same lengths, the HOMs are at similar frequencies. But we also see that the overall slope of thermal noise for the blue curve flattens out a little after around 20 kHz, whereas the black curve experiences a less drastic change in slope. This is due to the effect of the drumhead mode – the larger radius mirrorpad has a much lower drumhead mode frequency (by a factor of about $R^2$). Since the drumhead modes are known to have a lower Q and low modal masses, they can inject thermal noise at frequencies as low as a decade before the mode itself, and that is what we see in Fig. 4-16.

The comparison in radius is also a great example of how thermal noise displacement is not enough to get the full story. Let's look at Fig. 4-30 to see what this difference in radius means for the ratio of QRPN to thermal noise. The same devices are shown in bright green and pink in this plot. As we can see, the pink (222 Hz), which is the smaller mirrorpad device has a slightly higher QRPN to thermal noise ratio than bright green (114 Hz). This is despite of the lower thermal noise of the 114 Hz device. We can attribute this effect to the difference in their masses. The device with a lighter mass is able to produce more radiation pressure induced displacement, which cancels out with the effect of low thermal noise. The slight difference observed at low frequencies between pink and green is due to their difference in width. If we look around 5 to 6 kHz, green starts going back down earlier than pink. This can be attributed to its lower drumhead mode frequency as well as the lower frequency of the bending mode. Finally, we look at Fig. 4-28 to understand the combined effect of mass and fundamental frequency. Here we can see that the bright green device (114 Hz) has less squeezing than the yellow device (214 Hz) at a few kHz frequency. This can be understood as an effect of the ratio of QRPN to thermal noise. But we also see that they behave similarly at low frequencies. This is because the laser phase noise is the dominant noise source at those frequencies, and both these devices' squeezing performance is limited by the same noise.

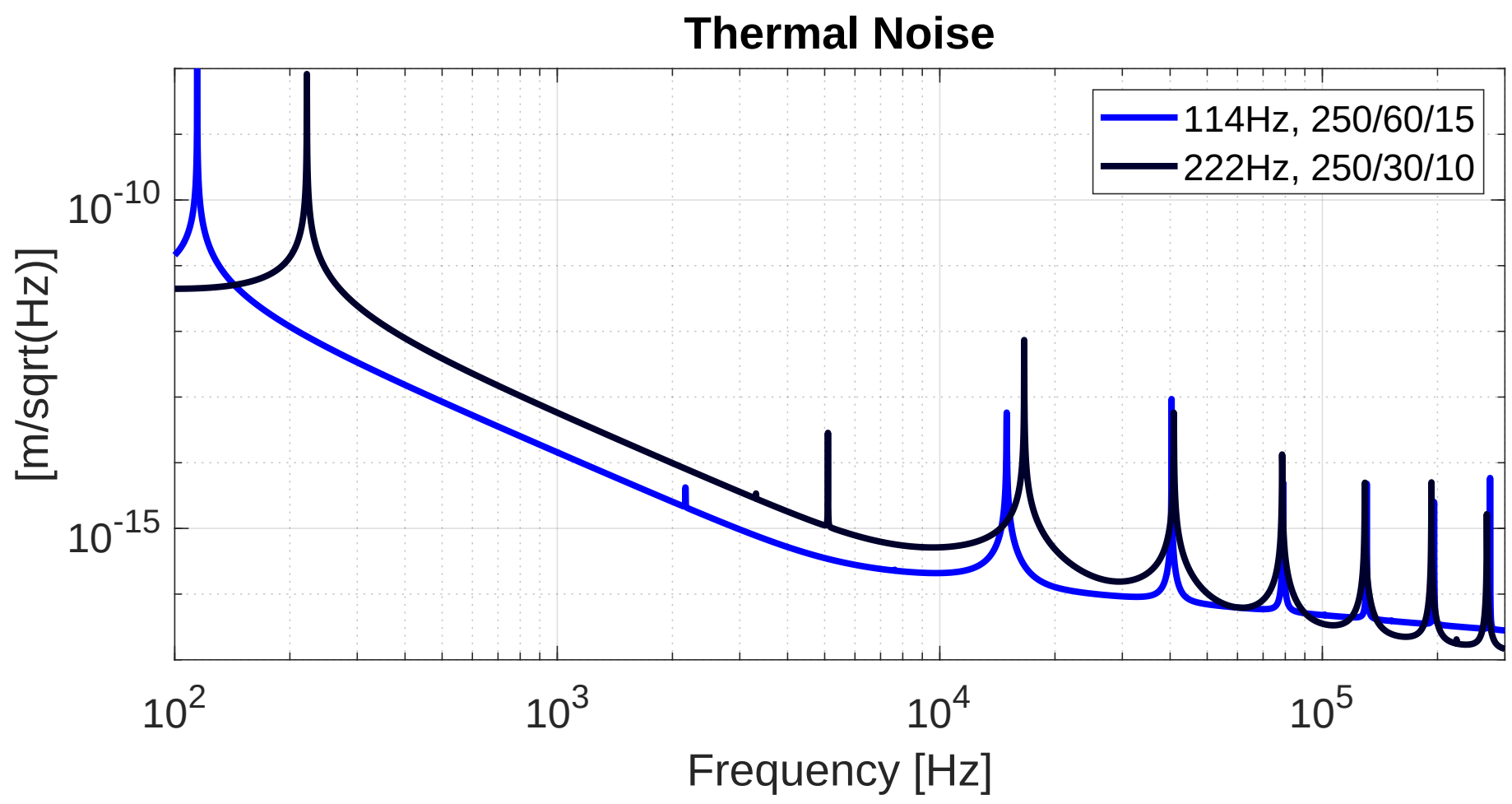

**Figure 4-16:** Effect of radius on thermal noise: This figure compares two oscillators with the same length but different radii. The larger radius oscillator (blue, 60 µm radius) has a smaller mass and a lower fundamental frequency which result in a lower displacement thermal noise than the smaller radius (black, 30 µm radius). Since they have the same length, their HOMs are close to each other. In addition, we also see more broadband noise showing up in the larger mirrorpad device towards the higher frequencies. This can be understood as extra noise from the drumhead mode, which is lower in frequency for a larger mirrorpad. This phenomenon is explained in more detailed in Section 4.2.4, and Figs. 4-17 and 4-19.

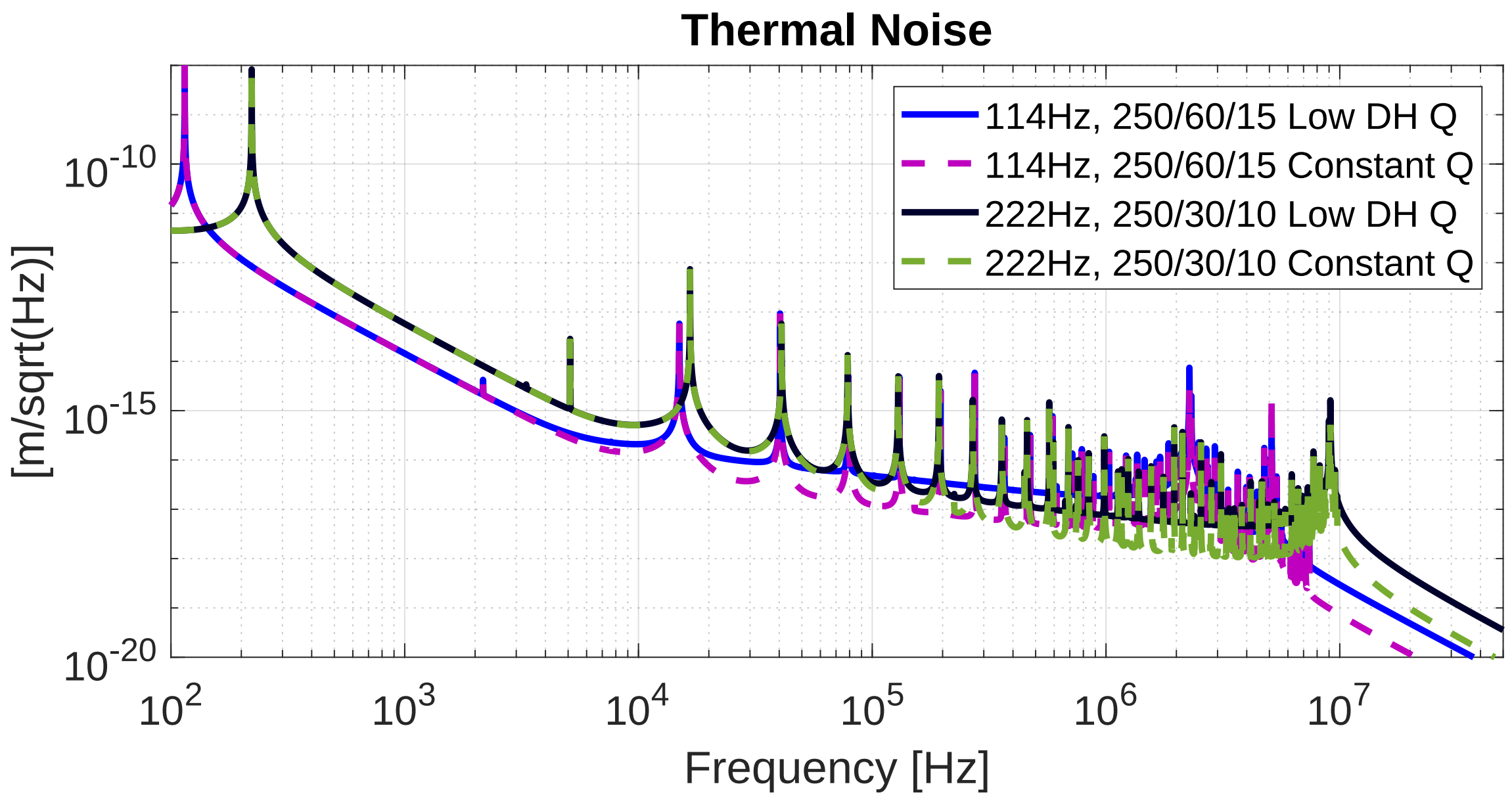

**Figure 4-17:** Effect of drumhead mode on thermal noise: Here we compare two devices of the same length, different radii, with and without lower Q for their respective drumhead mode. The larger radius device (shown in solid blue and dashed pink) has a lower drumhead mode frequency, near 2 MHz. The smaller mirrorpad device (shown in solid black and dashed green) has a higher drumhead frequency, near 9 MHz. The solids use a constant Q of 20000, whereas the dashed ones have a lower Q of 1000 for the drumhead modes (shown in Fig. 4-18). We see that while the smaller radius device's thermal noise starts deviating noticeably after 100 kHz (compare solid black and dashed green), the larger radius device's thermal noise starts deviating around 20 kHz (compare solid blue and dashed pink). This shows that the TED from drumhead mode increases the noise more for larger mirrorpad devices at a given frequency.

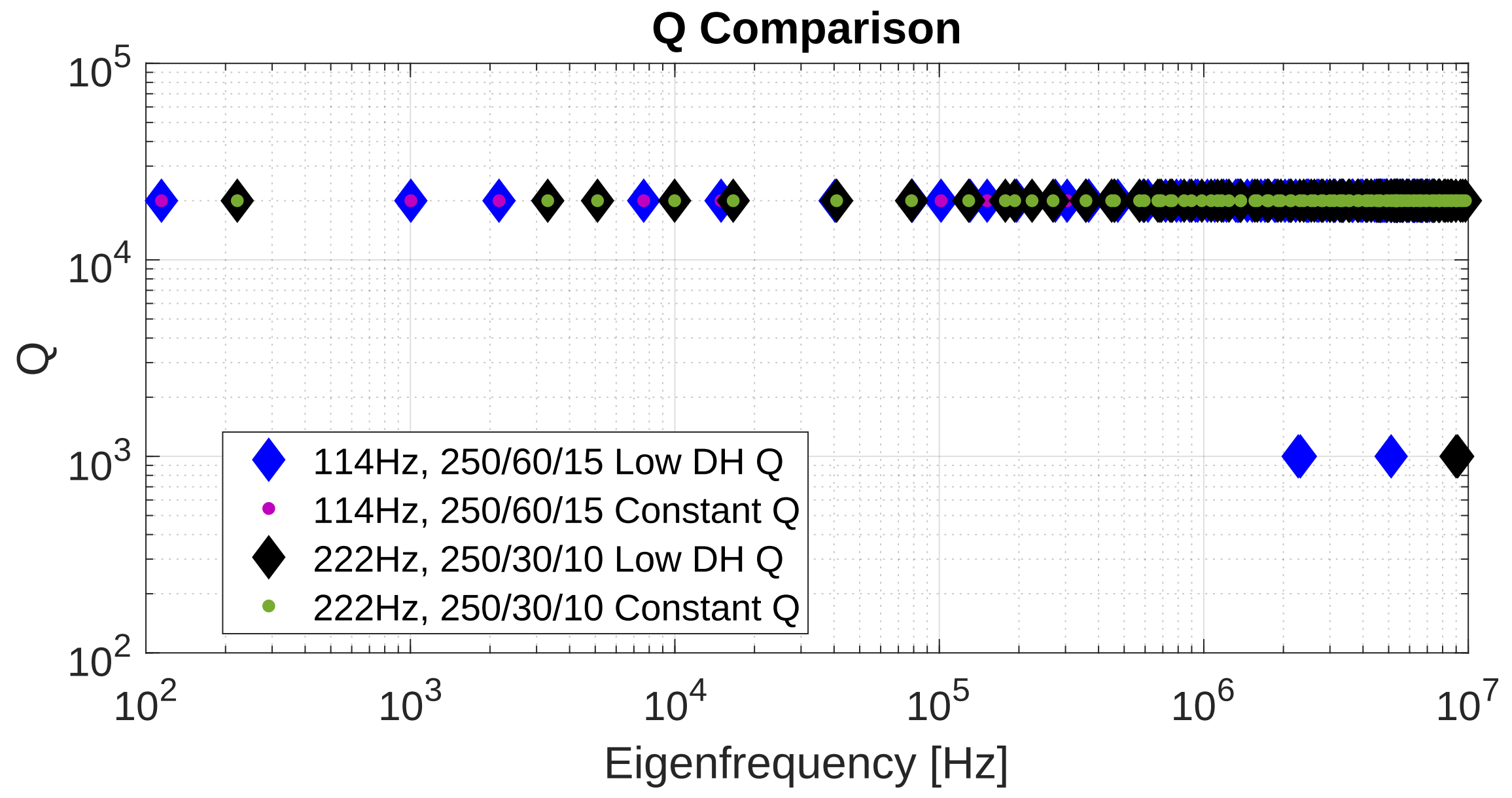

**Figure 4-18:** Plot showing the quality factors used for various devices in Figs. 4-17 and 4-19 to 4-23. The diamonds are Qs used after lowering the drumhead Q, and the dots are the Qs used for a constant Q for all eigenmodes. This plot also clearly shows that the larger radius device has its first drumhead mode near 2 MHz, while the smaller radius device has its first drumhead mode near 9 MHz.

By far the most important role of radius can be summarized via its contribution to the fundamental mode and the drumhead mode. To understand this with more depth, we expand on the case study of these two oscillators in Figs. 4-17 to 4-23.The complications added by the HOMs make intuitive understanding for some of the physics harder, so we also show the case of a single mode oscillator for comparison, in Fig. 4-31.

### 4.3.4 Role of width

The strongest effect of cantilever width is seen on the frequencies of the side-to-side modes. Increase in width also increases the fundamental mode by $\sqrt{W}$. In addition, change in width has a minimal effect on the mass of the device. Thus, theoretically speaking, lower width always gives better performance. We also verified this using simulations. That said, all of the effects of cantilever width are pretty small because:

1. the side-side modes do not have strong coupling with the cavity length because they have little displacement in the direction of the cavity axis.
2. The change in fundamental frequency is small because of its square-root dependence on width.
3. Overall, within our range of optimization, the changes made to the cantilever width are fractionally small. If we go too high, we degrade the squeezing quality, and if we make them too narrow we risk higher breakage.

So, the best optimization for width is to keep it small, except increase when attaching to a very floppy and/or heavy device.

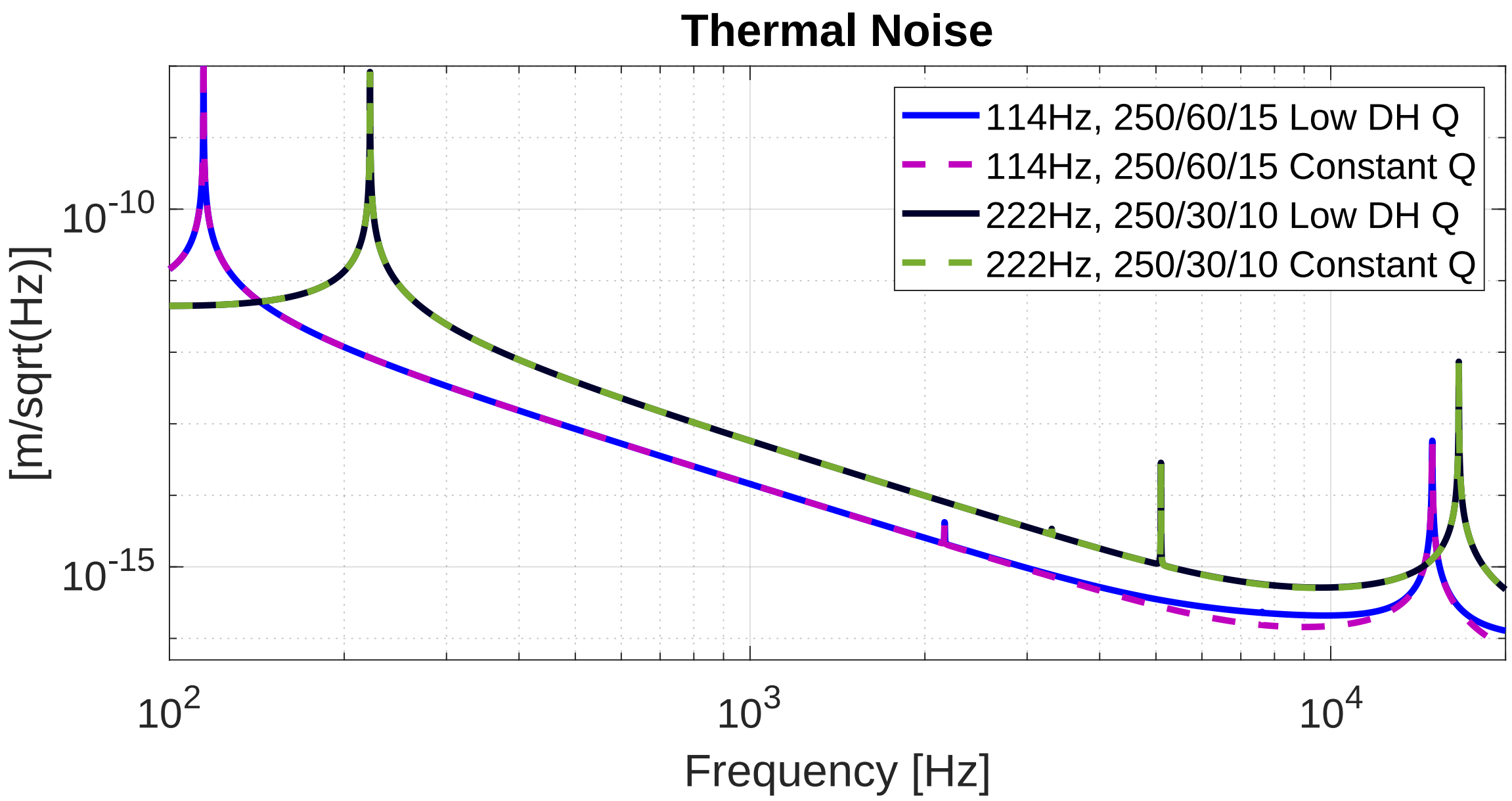

**Figure 4-19:** Effect of drumhead mode on thermal noise in measurement band: This plot is a zoom in of Fig. 4-17 at the squeezing frequencies. As we can see, the effect of TED is visibly increasing the thermal noise for the larger mirrorpad device (solid blue), as compared to the same device with no TED (dashed pink). When we look at the smaller radius device, the thermal noise in the measurement band remains similar both with and without TED.

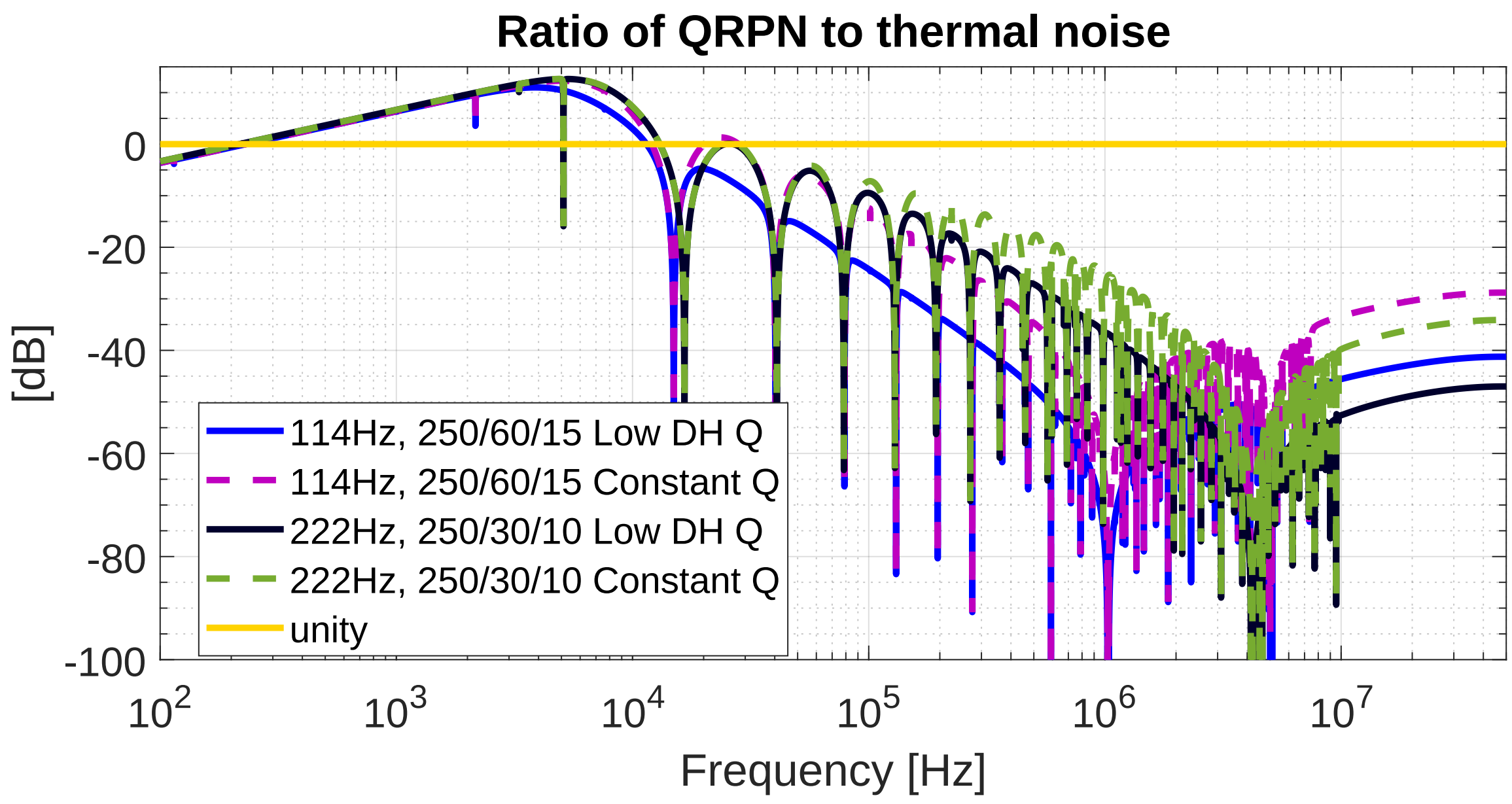

**Figure 4-20:** Effect of drumhead mode on ratio of QRPN to thermal noise: Here we show the effect of TED on the ratio of QRPN to thermal noise for the same two oscillators. We can see that the envelop of the drumhead mode is decreasing this ratio a lot more for the bigger mirrorpad than the smaller mirrorpad, starting at around 20 to 30 kHz. The effect at low frequencies can be seen more easily in Fig. 4-21, which is zoomed in.

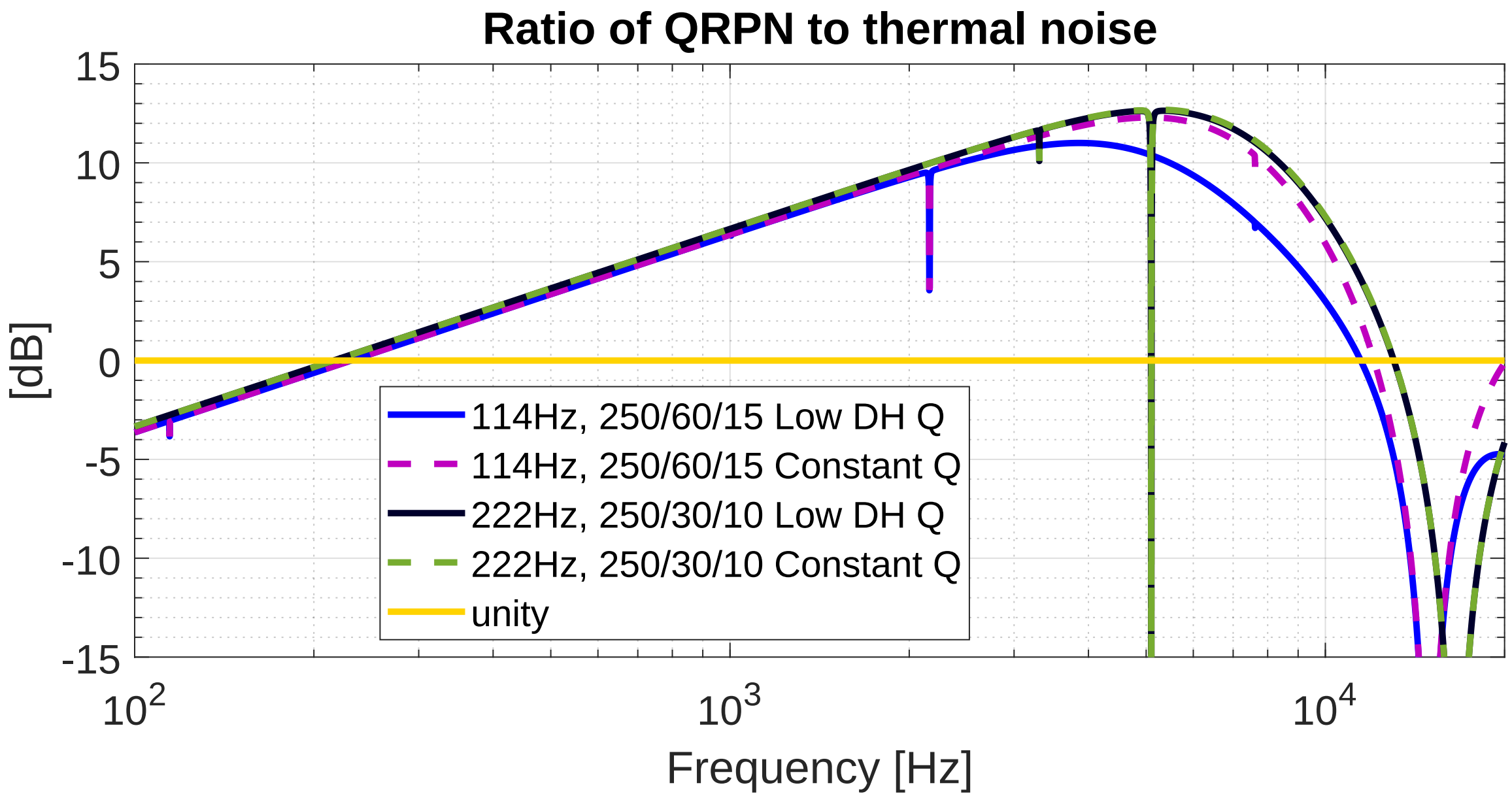

**Figure 4-21:** Effect of drumhead mode on ratio of QRPN to thermal noise in measurement band: first of all, let's look at the dashed pink and dashed green (i.e. no TED) at frequencies below 1 kHz. We see that the ratio of QRPN to thermal noise for these two devices is similar. This is because they have the fractional change in $L+R$ for their difference in $R$ is small, and they have different widths, leading to comparable $W/(L+R)^3$. We can simplify the physics in these frequencies by considering each of the devices as a single mode oscillator at their respective fundamental frequencies, allowing us to compare force spectral density. The force spectral density of QRPN is white, and that of thermal noise is proportional to $W/(L+R)^3$ if this were a single mode oscillator. This means that if two devices have comparable $W/(L+R)^3$, they will have comparable ratio of QRPN to thermal noise (in the simple case of a single mode oscillator). Now, we see that adding TED lowers this ratio for the larger mirrorpad, starting visibly at about 2 to 3 kHz, whereas the smaller mirrorpad gets that lowering at frequencies higher than the squeezing band.

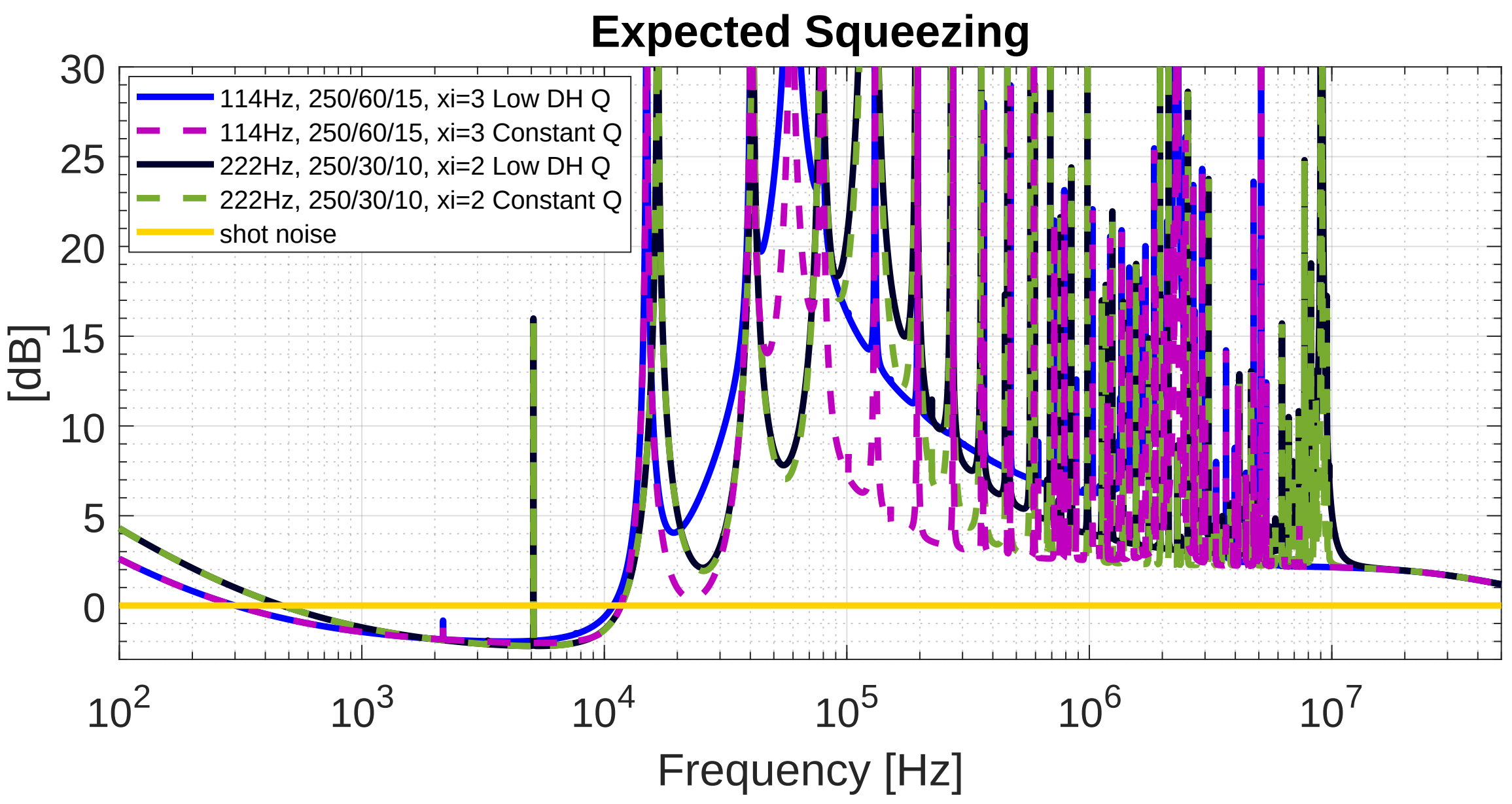

**Figure 4-22:** Effect of drumhead mode on squeezing: In this plot we show the expected squeezing obtainable from the two oscillators in the presence and absence of TED. Looking at a plot like this is not the best way to ascertain where the drumhead mode is. For example, the resonance around 60 kHz has a big difference in the dashed pink and blue curves, which are the same device, without and with TED. This might lead one to interpret that as the drumhead mode. But if we look at Figs. 4-17 and 4-18, we see that there is no drumhead mode at that frequency, and the drumhead mode for this device is at 2 MHz instead. The resonance at 60 kHz is actually the OS, which has a broad linewidth. So what we are seeing here is the effect of the added damping from the drumhead mode on the broadband thermal noise. The OS for the smaller radius is near 120 kHz. As we can see in both these devices, the bending mode just below 20 kHz causes the higher frequency cut off for squeezing, in conjunction with the broader effect from the OS. In the absence of the thermal noise from the bending modes, the smaller mirrorpad would have squeezing till higher frequencies (in this quadrature) because of its higher OS. For the larger mirrorpad, if the ratio of QRPN to thermal noise still was greater than 1 at frequencies higher than the OS, the squeezing would shift to other quadratures at those frequencies. The effect of TED in the squeezing band is more clearly noticeable in the zoomed in version – Fig. 4-23.

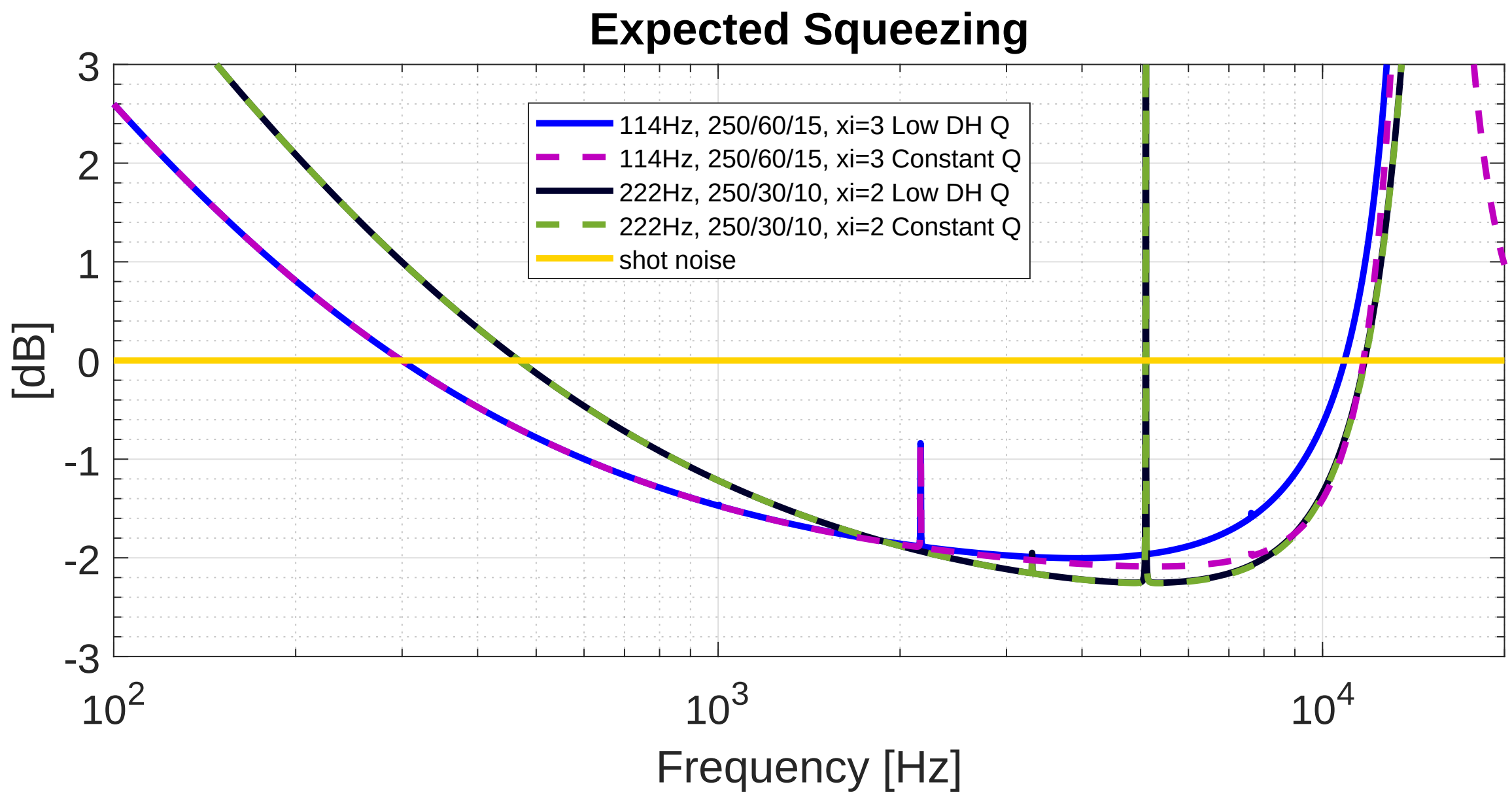

**Figure 4-23:** Effect of drumhead mode on squeezing in measurement band: the addition of TED makes the larger radius device create less squeezing at few kHz to 10 kHz (compare solid blue to dashed pink), but has negligible impact on the smaller mirrorpad device (compare solid black to dashed green).

## 4.4 Novel geometries

Apart from the basic geometries, we also considered some different designs to see if we could modify some of the most problematic patterns in our basic cantilever geometry. In short, most of these designs either ended up having a worse performance than the basic geometry, or ended up looking close to what we already have. One of the ideas we looked into was rounding all the sharp edges in the device, as shown in Fig. 4-26. This would moderate the concentration of energy at the high stress points and make the device overall better. When consulting with Garrett Cole on the fabrication aspect of this, he indicated that all our devices will already have some rounding effect because of the way the device is fabricated. Let's look at some of the other designs one-by-one.

### 4.4.1 Trampolever

The trampolever is a device where the mirror-pad was doubly suspended from the bulk: first stage by the cantilever, and the second stage by a trampoline (shown in Fig. 4-24). The basic idea behind this geometry was to help the trade-off when optimizing radius. A larger mirrorpad leads to a lower thermal noise due to its higher mass, but also leads to a lower drumhead mode and hence higher thermal noise at high frequencies. So, the idea is to make an inner mirror-pad that has a small radius (and hence a higher drumhead mode frequency), but suspend it by an outer ring, which is also made of the thick mirror material. This outer ring would enable keeping the mass high, resulting in a lower thermal noise.

We simulated this device in COMSOL with fixed cantilever length and width, varying the trampoline geometry. We found that most designs were worse than the base cantilever geometry, with a tiny improvement seen in one of the designs. That design has been included on the chip in the hopes to test out some of these ideas.

### 4.4.2 Tapered cantilevers

We also looked at the distribution of stress in the baseline cantilever devices. In order to try having a more uniformly distributed stress in the cantilever, we tried making a taper to the cantilever, as shown in Fig. 4-25. We found that this had little to no impact on the squeezing performance. This was good news, because it meant we can make the devices tapered and perhaps lower the chances of breakage. We have done this for some devices on the chip.

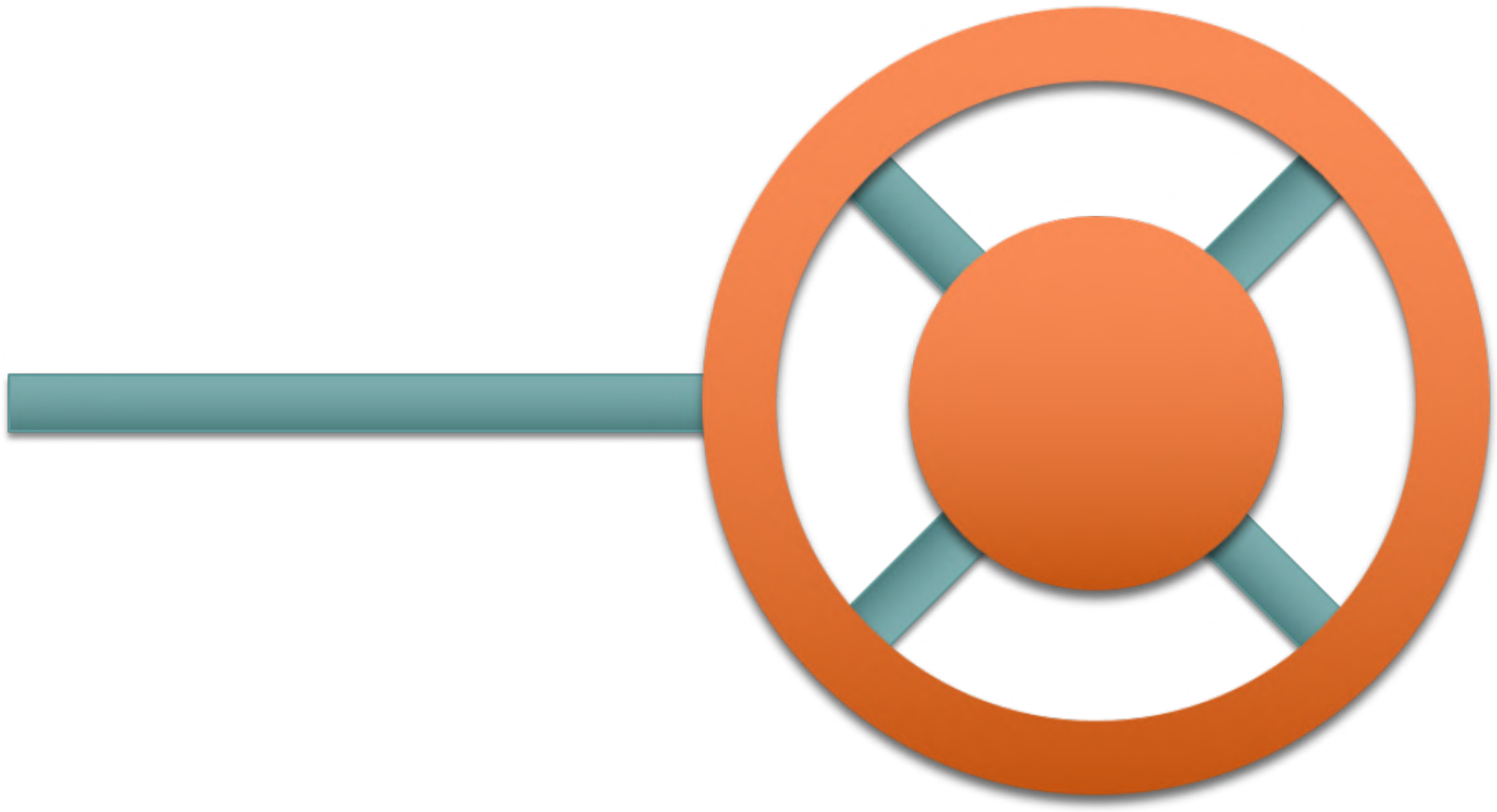

**Figure 4-24:** Trampolever: Here we have a GaAs cantilever connected to a trampoline The trampoline consists of an outer annular ring made of the mirror material, connected to an inner mirror disc by four supports made of just the thin GaAs layer. This geometry was derived from the basic geometry to address two concepts. First, it would be interesting to test if a double suspension system is able to give us less thermal noise by altering the slope. Secondly, even if the slope of the thermal noise stays the same, this allows us to have a more massive device with a small radius mirrorpad. The higher mass helps lower the fundamental mode frequency, in turn lowering the thermal noise, and the smaller mirrorpad helps increase the drumhead mode frequency.

### 4.4.3 TED testers

As explained in Section 4.2.4, we were unable to produce models that resulted in physical outcomes when we simulated thermoelastic effects in these devices in COMSOL. While we can use measured thermal noise of some of the squeezing-optimized cantilevers to give us a rough idea of the thermoelastic damping (TED) in these devices, it would be more optimum to measure it directly. To that end we intentionally made some devices with very large mirrorpads (100 to 250 µm in radius). These devices will have too high thermal noise to measure squeezing, but their thermal noise will have a significant contribution from TED in the frequency band of our measurements. Some of them will even have the drumhead mode close to the measurement band, allowing us to fully map out its effective quality factor. To keep their spectrum clean of the cantilever modes, the cantilever length of these devices is chosen to be very short. These devices are on the chip to allow us to experimentally understand TED effects in this material in this regime, and hopefully inform future designs and applications to other fields like high reflectivity and low thermal noise coatings [61, 62].

### 4.4.4 Doubly clamped mirrors

We have also included some doubly clamped mirrors on the chip. These mirrors are attached to the bulk via two cantilevers, one on the top and other on the bottom. The idea behind this is to try stress dilution to increase the effective $Q$ for the device. Since we did not trust COMSOL for the quality factors, we just picked some of our best designs and added copies of them doubly clamped on the chip.

### 4.4.5 Others

In addition to all the above non-standard geometries, we also tried a few others shown in Fig. 4-27. They also had motivations focused on HOMs, but simulations and careful revisions showed that they were worse than the basic design.

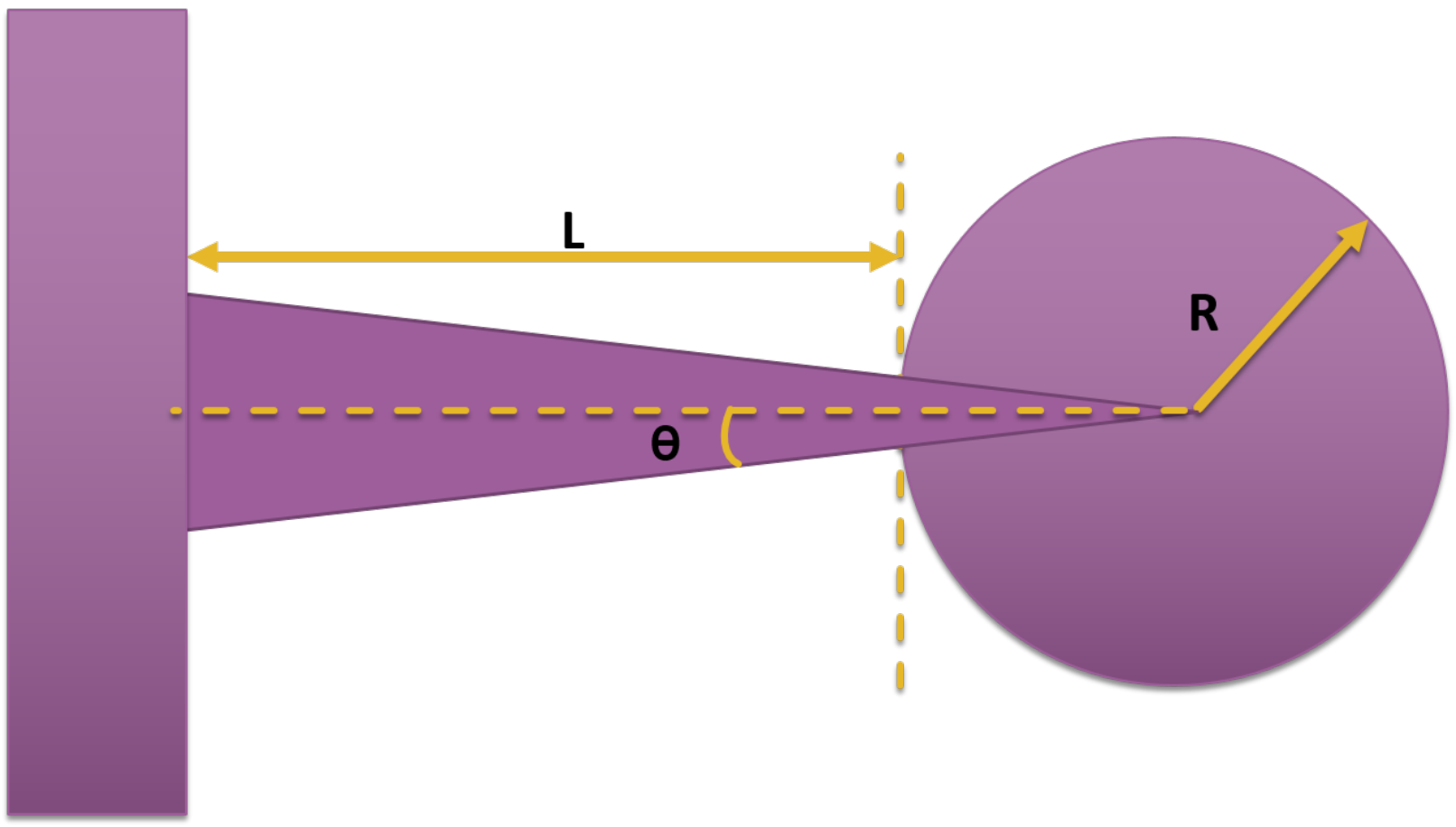

**Figure 4-25:** Tapered Cantilever: Here we try to address the problem of potential breakage due to unequal stress distribution. This design is inspired by the LIGO blade springs design. In order to get a perfectly uniform stress distribution in the GaAs layer, $\theta$ should be 14°.

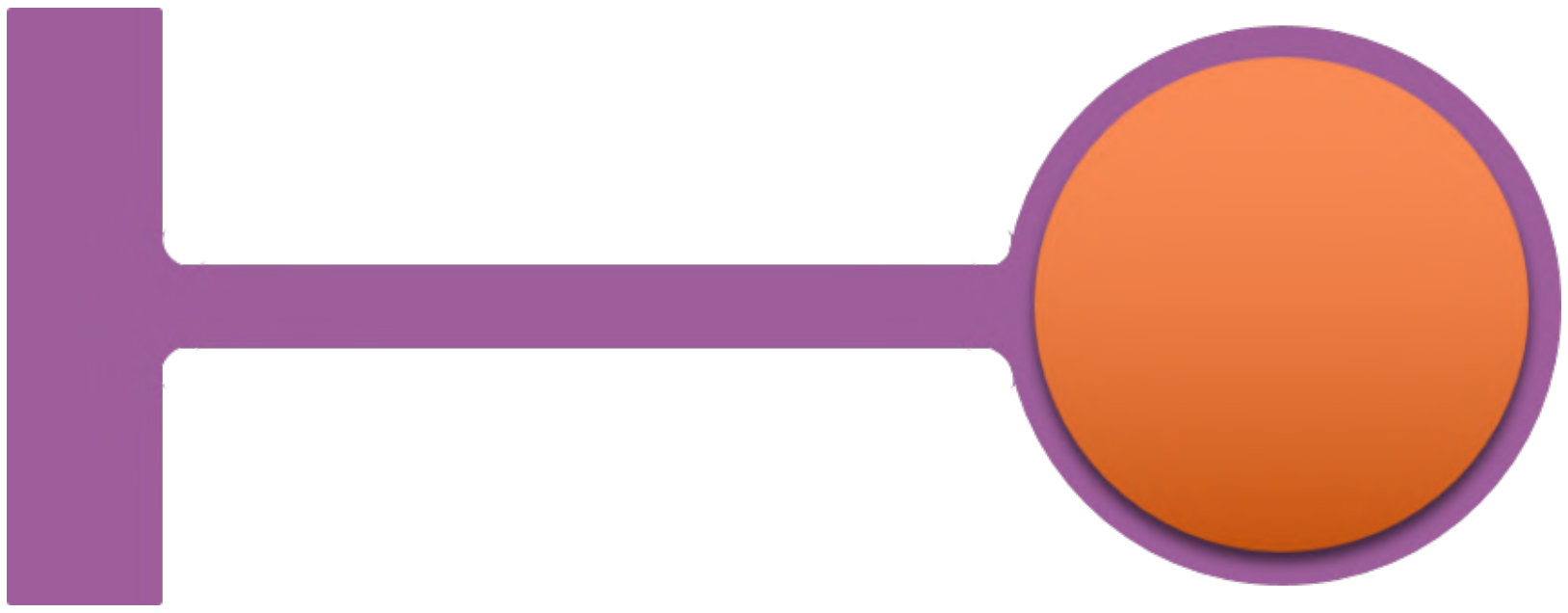

**Figure 4-26:** Rounded Edges: Having sharp edges leads to a high stress concentration at those edges. On the other hand smoothing those edges, for example by rounding them helps getting a more smoother stress distribution.

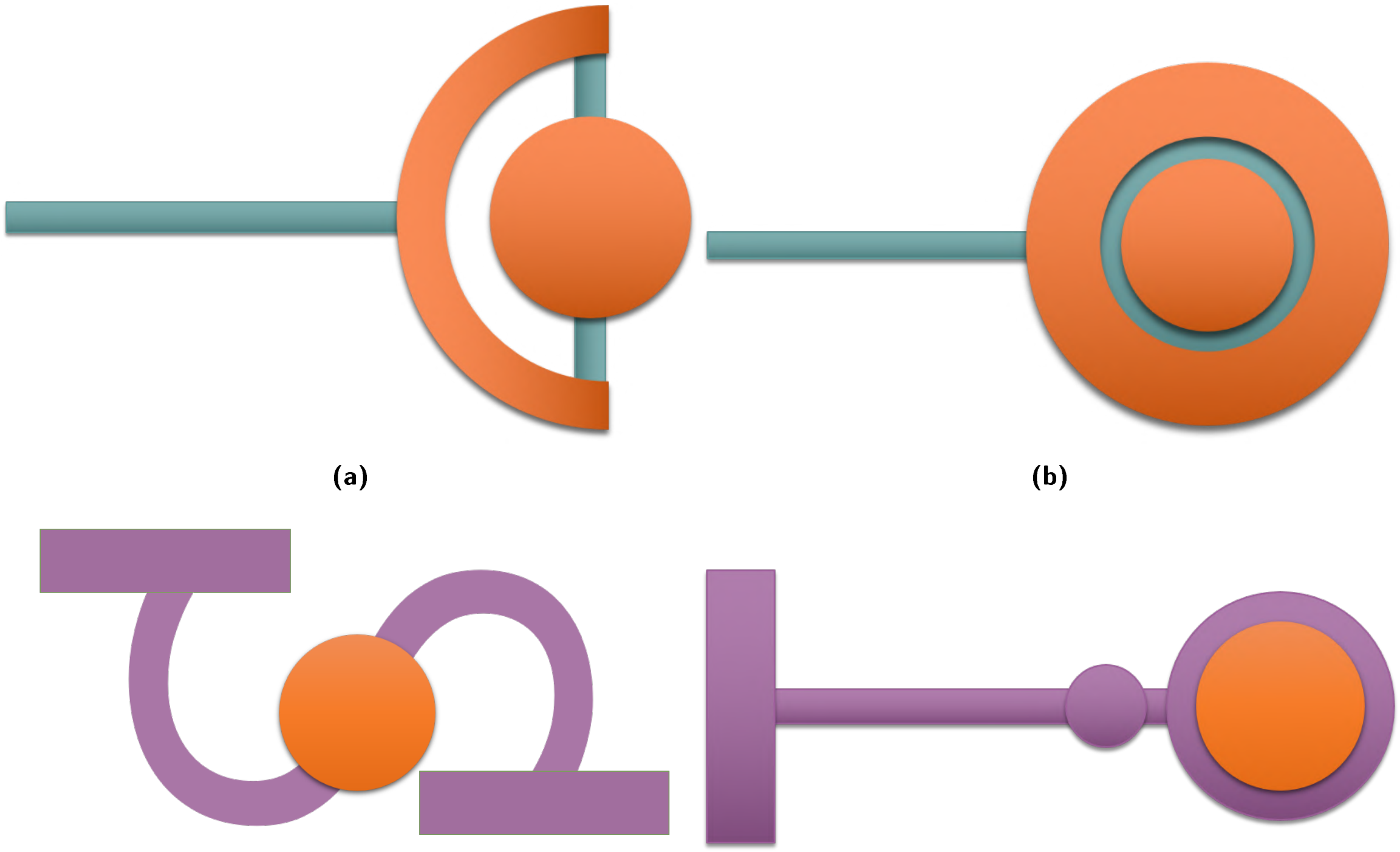

(a) (b) (c) (d)

**Figure 4-27:** Some other geometries that were considered, but simulations showed they were worse than the vanilla designs.

## 4.5 Final chip design

Finally, we combine all the above knowledge and come up with a final set of cantilevers that were expected to be good for squeezing. This includes eight room temperature designs, and six cryogenic temperature designs. Since thermal noise is reduced at cryogenic temperatures (both due to a lesser temperature, as well as a decreased isotropic structural loss factor, and lower thermoelastic effects), the focus for those designs is to push the HOMs as far out as possible and maximize the clean region of the spectrum. These fourteen base designs, some trampolevers, doubly-clamped mirrors, and TED testers are then combined on a single chip.

### 4.5.1 Room temperature designs

After performing the optimization, we came up with eight designs that were best candidates for room temperature squeezing. They are listed in Table 4.3. As we can see, they have a diverse set of radii, lengths, widths and fundamental frequencies to allow for various unforeseen experimental circumstances.

#### 4.5.1.1 Squeezing performance

In Fig. 4-28 we show the squeezing performance of the above eight designs. We can see that in addition to having diverse geometries, these devices also cover a wide variety of squeezing performances. Some have better squeezing performance at low frequencies, while others have a better squeezing at high frequencies, and some are optimized for mid-band frequencies. While the low frequency squeezers run into their respective HOMs for their high frequency squeezing cut-off, the high frequency squeezers are usually limited by their OS frequency. All these squeezers can potentially achieve somewhere around 4 to 5 dB of squeezing at room temperature.

| Fundamental Frequency (Hz) | Length (µm) | Radius (µm) | Width(µm) |
|---|---|---|---|
| 114 | 250 | 60 | 15 |
| 142 | 260 | 40 | 9 |
| 214 | 250 | 25 | 6 |
| 222 | 250 | 30 | 10 |
| 238 | 170 | 40 | 9 |
| 494 | 130 | 25 | 6 |
| 616 | 100 | 25 | 5 |
| 790 | 80 | 25 | 5 |

**Table 4.3:** Room temperature squeezers

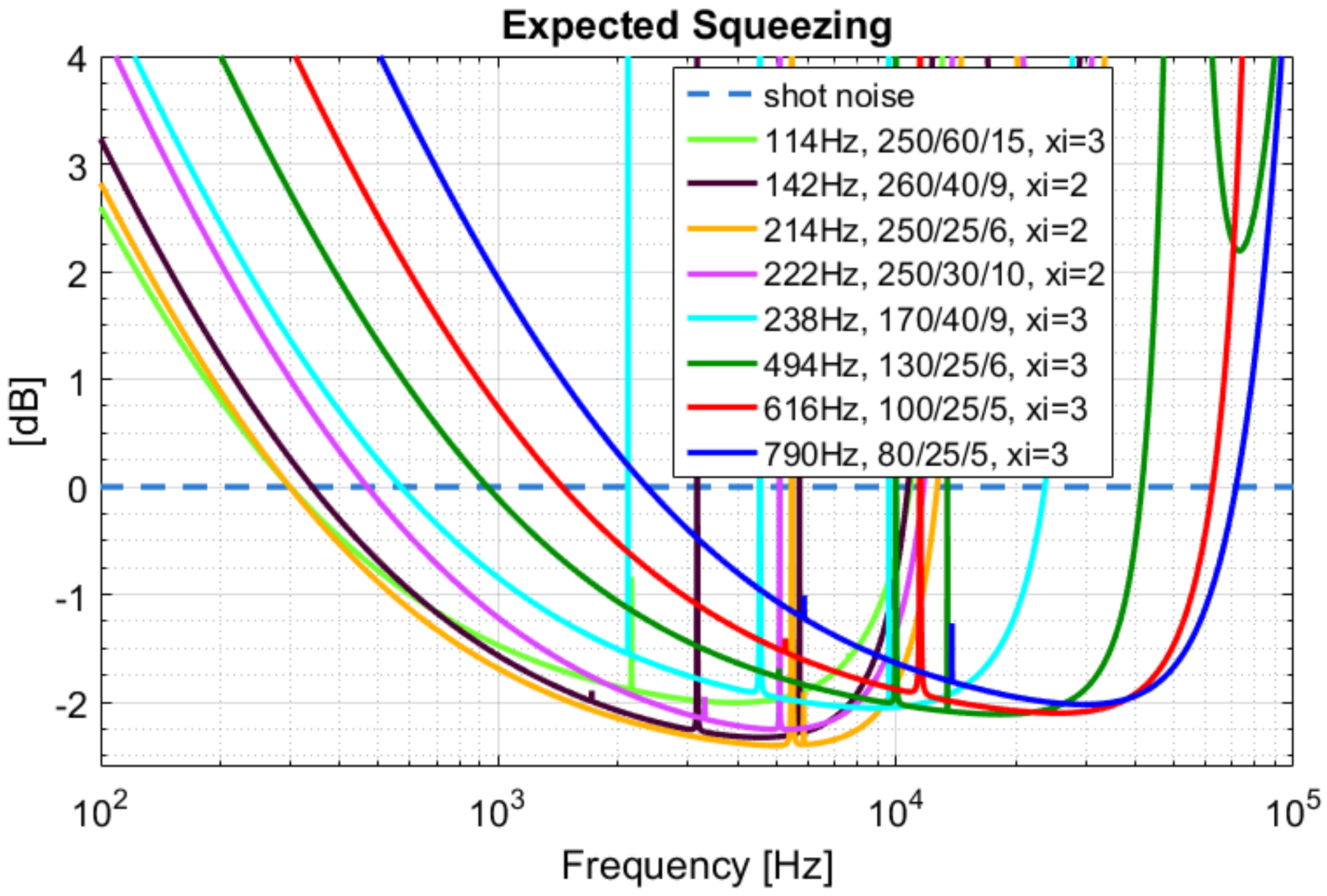

**Figure 4-28:** Expected squeezing at 295 K: Here we show the expected squeezing from all the room temperature squeezer devices, assuming a lower drumhead mode $Q$ of 1000. The first number on the legend is a device's fundamental frequency, the next three numbers are its geometry in the order $L/W/R$, and $xi$ is the quadrature in which this device produces the most squeezing. We see that most of the devices can reach more than 2 dB of squeezing. While some of the devices are good at low frequencies, others do better at higher end of the spectrum, as explained in the text.

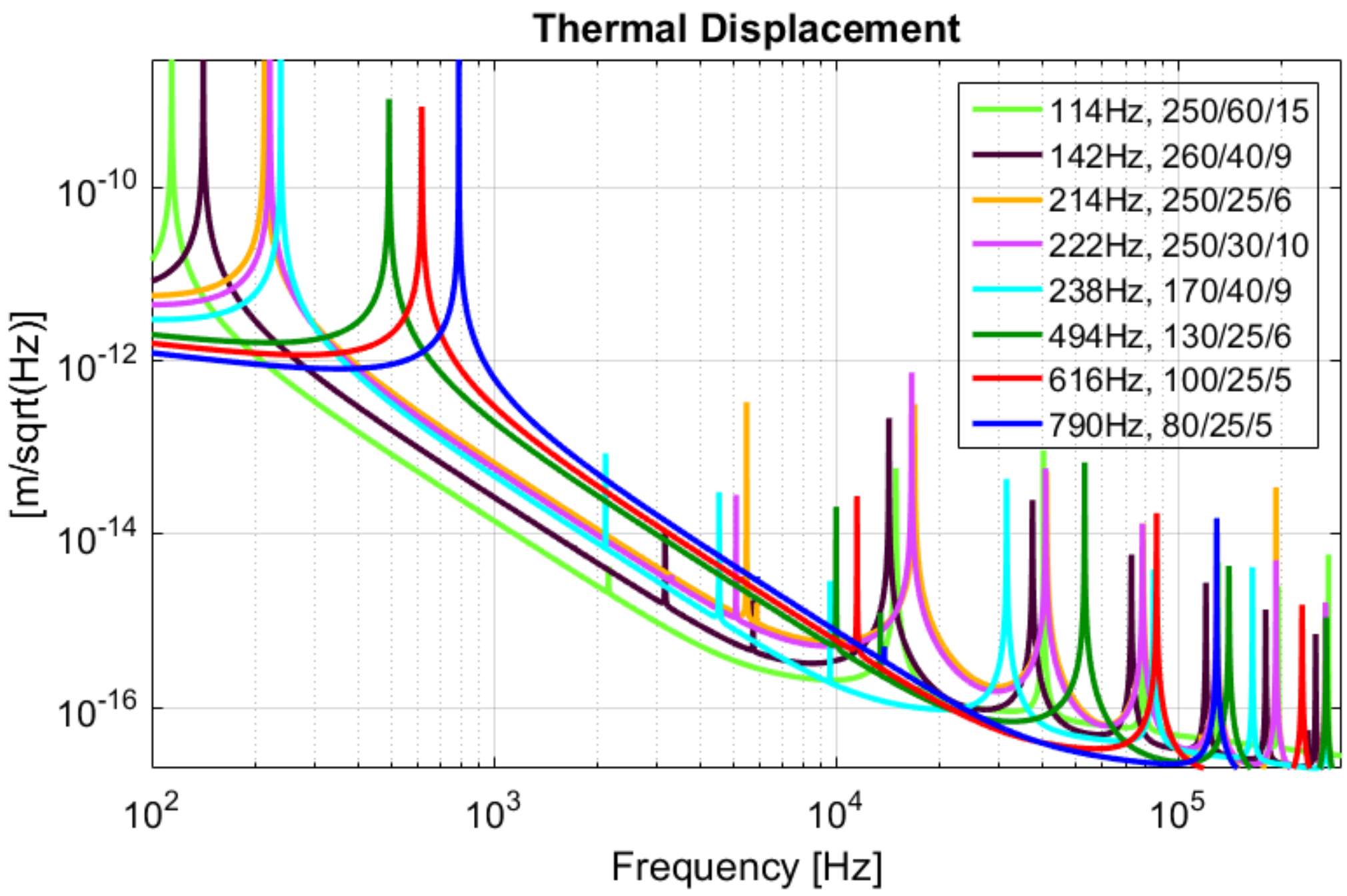

**Figure 4-29:** Thermal Noise at 295 K: This plot shows the thermal noise from the room temperature squeezers shown in Fig. 4-28. We see that the lowest fundamental mode devices have the lowest thermal noise, as expected. We see that the lower fundamental can be caused either by a higher length, or more strongly by a larger radius. The higher order bending modes are at lower frequencies for the longer devices, allowing the shorter devices to have lower thermal noise after 10 kHz, making them good candidates for high frequency squeezing.

#### 4.5.1.2 Thermal noise

In Fig. 4-29 we show the expected displacement thermal noise for the room temperature squeezer designs. They have fundamental mode frequencies as low as around 100 Hz, and as high as 800 Hz. They all have thermal noise that goes as $f^{-5/2}$ above the fundamental frequency due to structural damping [35]. Occasionally we can see the thermal noise being increased by a HOM. The HOMs add noise near the resonance, but do not have a big effect on the spectrum away from their resonance. We also notice systematically, how the shorter cantilever devices have cleaner bands (due to higher frequency HOMs) than the longer cantilevers, but still have higher thermal noise due to higher fundamental frequency.

#### 4.5.1.3 QRPN to thermal noise ratio

We also show the ratio of QRPN to thermal noise for each of the designs in Fig. 4-30. As we can see, the devices that have a higher ratio of QRPN to thermal noise produce more squeezing, as is intuitively expected. The effects of HOMs are also easily visible in this plot. First this ratio increases because the slope of thermal noise after the fundamental mode is steeper than QRPN. Then the slope starts decreasing locally due to HOMs. The envelop slope decreases due to the cavity linewidth, this effect is shown more clearly in Fig. 4-31. Here we have simulated another OM system, but this time without any HOMs of the oscillator. We show the ratio of QRPN to thermal noise for such a system detuned as well as undetuned. As we can see, this ratio first increases but then starts to decrease. The frequency where it starts to decrease is the optical linewidth of this cavity, adjusted for the relevant detuning (see Eq. (2.29)).

#### 4.5.1.4 Other metrics

In addition to the amount of squeezing, we also looked at how alignment sensitive a particular device would be. For instance, if we were only able to align the beam to within 2 µm in the lab, would that

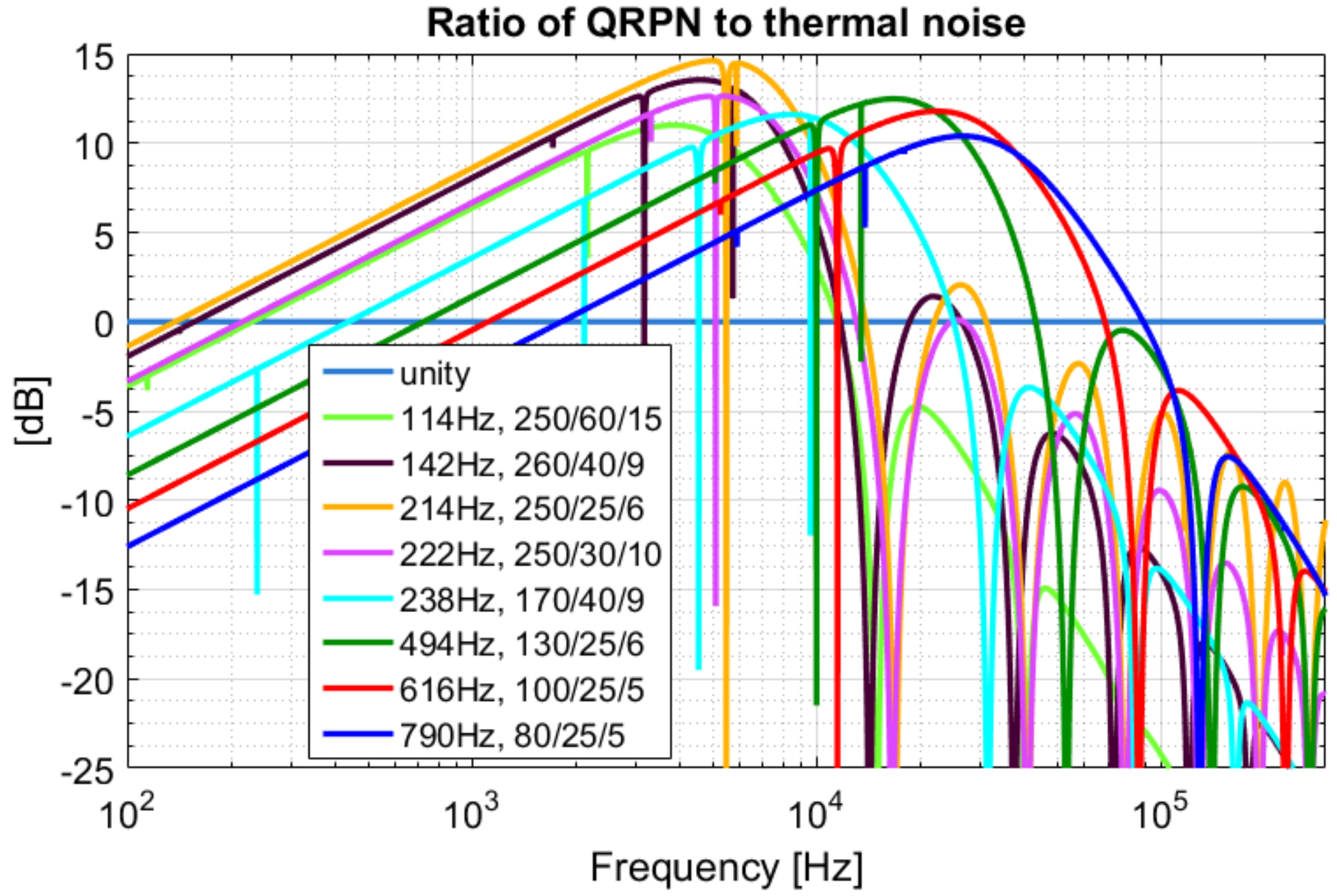

**Figure 4-30:** Ratio of QRPN to thermal noise at 295 K: Here we show how much of the displacement of the cantilever device is dominated by QRPN vs thermal noise, in dBs. This ratio is related to the quantity $\lambda_{th}$ in Chapter 3, so in other words influences how much squeezing a particular device can produce. A simple mapping is made hard by the fact that these are all multi-mode devices. Nevertheless, at low frequencies, a single mode approximation does a good job of explaining the physics. The low frequency behaviour in this plot can be understood as a consequence of the quantity $W/(L+R)^3$, while the high frequency ratios get influenced by the HOMs. In the absence of HOMs, the inflexion point would be at the cavity linewidth, as shown in Fig. 4-31.

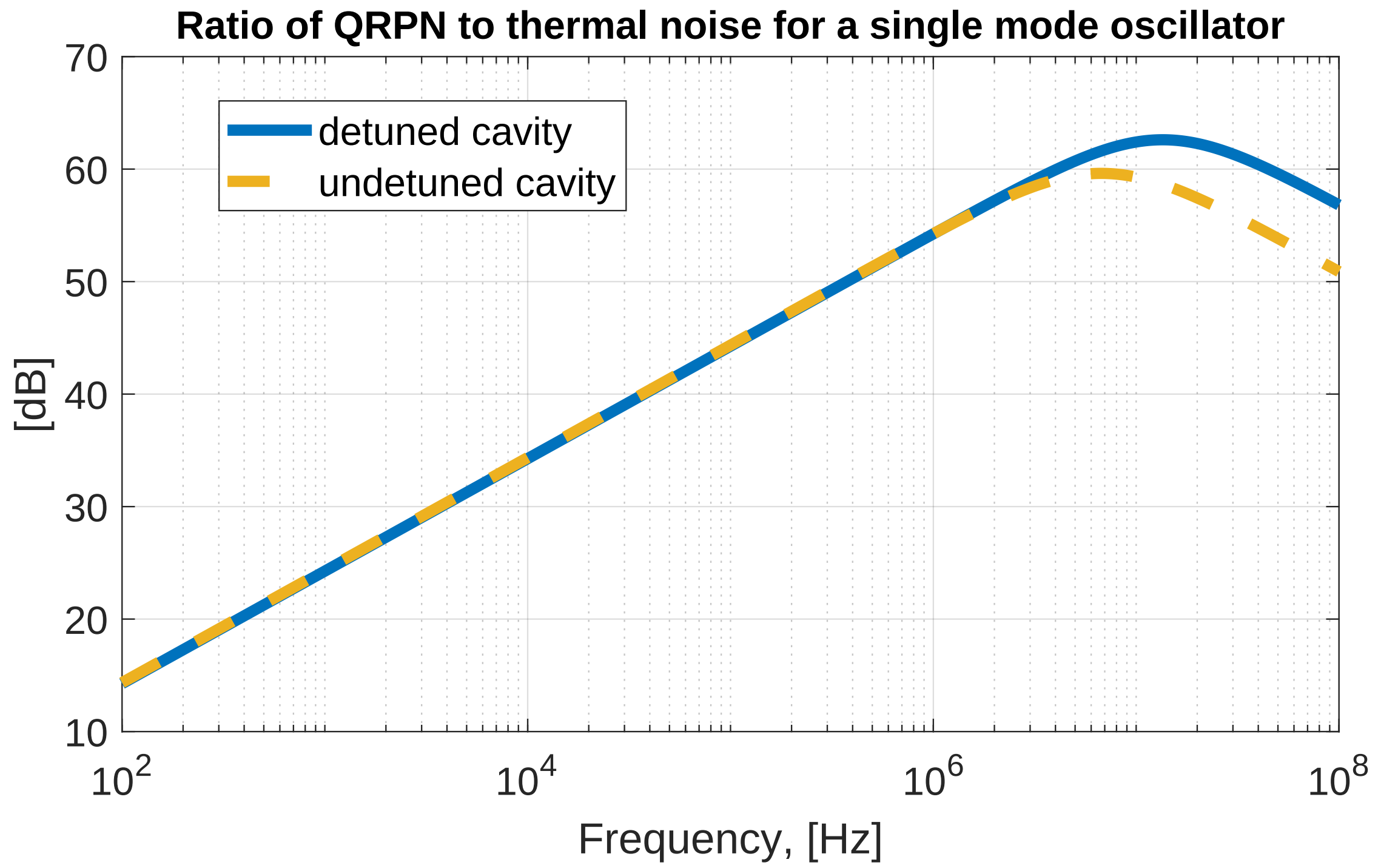

**Figure 4-31:** Ratio of QRPN to thermal noise for a single mode oscillator: We want to understand what is the ultimate limit to the QRPN to thermal noise ratio. In the absence of HOMs, would it keep increasing as a function of frequency? It should not depend on the OS frequency because both QRPN and thermal noise get affected by the same suppression TF from the OS response. Here we have simulated a high $Q$ single mode oscillator to understand this. In dashed yellow we see the ratio of QRPN to thermal noise when the cavity is operated on resonance, and in solid blue we see this ratio when it is operated detuned. The inflexion point happens close to the cavity pole frequency, adjusted for the detuning (see Eq. (2.29)). This is because the radiation pressure force depends on the cavity's linewidth – it is white only within the cavity linewidth.

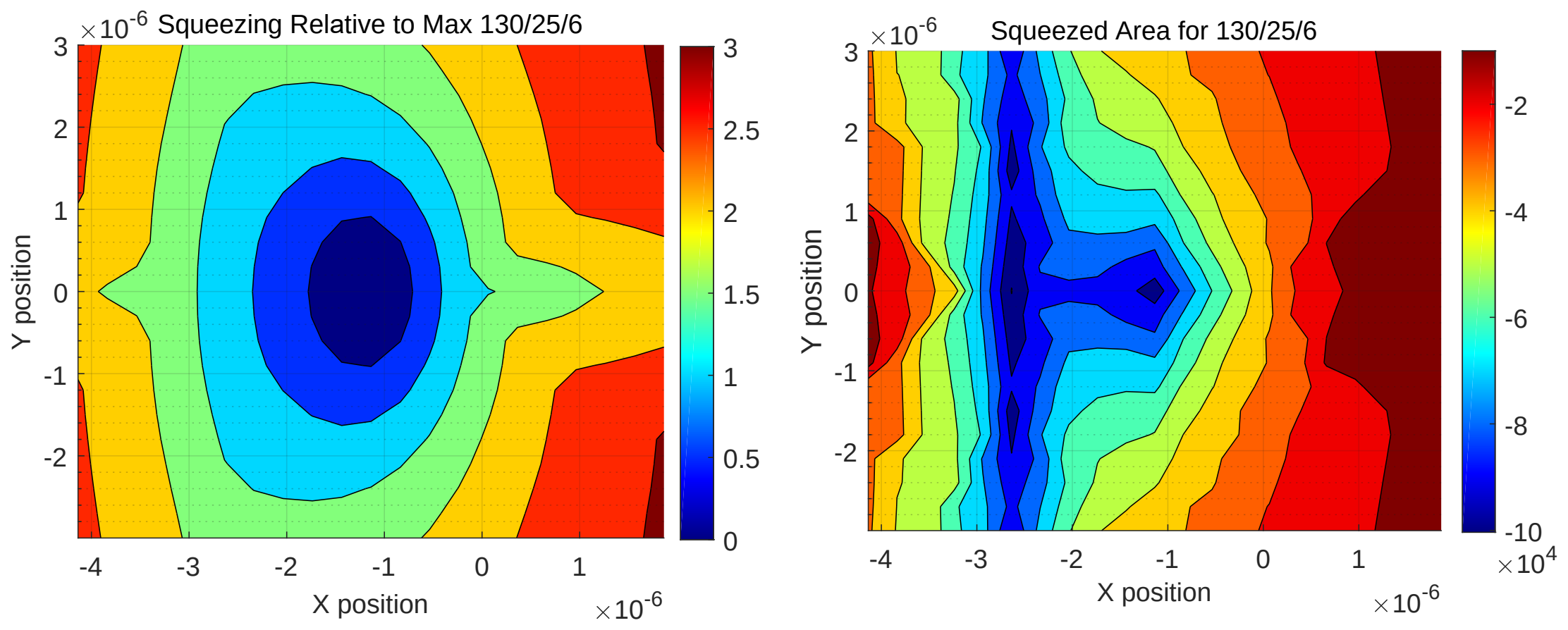

**(a)** **(b)**

**Figure 4-32:** Alignment sensitivity of squeezing: Here we show how the squeezing behaviour would change if we were misaligned from the node position (Fig. 4-12) that minimizes thermal noise, for the device 130/25/6 (see Table 4.3). The $X$ and $Y$ conventions in these plots are shown in Fig. 4-8. The origin $(0, 0)$ is the center of the mirrorpad, and these plots span a range of $\pm 3$ µm from the node position (which is usually a negative number in $X$, in this case close to $(-1.2, 0)$ µm). **(a)** shows the maximum squeezing obtainable (over all frequencies and quadratures) as a function of beam position on the mirror, normalized to the maximum squeezing for perfect cancellation. We see that if we are misaligned by 1 µm in Y (yaw) in either direction, we lose 0.5 dB of squeezing, whereas we lose more squeezing for a similar misalignment in X. Additionally, since the displacement field for the pitch mode is not symmetric around its node, these plots are not perfectly symmetric in $X$. **(b)** Since the first plot only shows the maximum squeezing and no information about the impact of misalignment on the frequency dependence of squeezing, here we show the area under the squeezing curve as a function of misalignment (in dB Hz). We notice that a negative misalignment in X gives similar results as the aligned case. We see that even though the maximum squeezing is reduced there, the area under the squeezed curve can still be maximum.

severely impact the squeezing for some designs but not others? This could be the case especially for the small radii mirrors because the same absolute mis-centering would be a larger fractional mis-centering on a smaller mirrorpad. In order to investigate this, we calculated the modal masses for each mode when the beam was not on the node. We did this in a grid, and looked at the maximum squeezing for each misaligned point. One such plot it shown in Fig. 4-8. Following the same axis convention, the beam alignment sensitivity for one of the final designs is shown in Fig. 4-32. This figure shows the area under the squeezing curve (in dB Hz) as a function of beam position on the mirrorpad. The origin is the node of first yaw and pitch mode. As we can see, this design will still produce squeezing even after misalignment. Due to a reduced thermal noise, a given amount of misalignment has an even smaller effect at cryogenic temperatures. Similarly, we checked alignment sensitivity of all the devices and selected the ones that showed the most robustness against misalignment.

Just like the beam alignment example, a lot of times there is a need to reduce each design to one number. While one can look at the maximum squeezing, it does not provide information about the frequency dependence of the squeezing. For this reason, we developed some metrics like area under the squeezing curve and lowest frequency of squeezing. Plotting these metrics against numerous design iterations was useful in screening out bulk of the designs quickly without looking at full frequency dependence for each design.

### 4.5.2 Chip layout

The last step after deciding various device geometries is to design the full layout of the chip. The layout is shown in Fig. 4-34. To be space efficient and fit as many devices on a single chip as possible, the windows are chosen to be of different sizes. In order to avoid cross-talk between various devices, we intentionally make their dimensions slightly different from each other. A back of the envelop calculation to see how much a 1 µm change in length changes the frequency of a given mode shows that it is enough to not have any overlap given a $Q$ of 20000. So when making copies of each design, we vary their dimensions slightly to keep things non-degenerate. In addition to making them non-degenerate, we also try to put similar devices well separated from each other to further minimize cross-talk. To make life easier in the laboratory when trying to locate a particular device, the left and right are copies of each other (except for the small variations needed to make each device different.) Lastly, since the longer devices can only fit in the longer windows, the window size and number of devices has to alternate a few times to get the optimization right. For easy row recognition, we added slits between the left and right window, the number of slit corresponds to the row number. The idea is that one can translate the chip vertically, and observe clipping and appearance of the laser through the slits, counting of which would allow one to decipher which row we are on. Additionally, we had decided to have alphabets inscribed on the right side of the chip for easy recognition of right vs left window and row number. This was supposed to be added to the photo-lithography mask, but got omitted as a mistake. The slits are not etched out all the way like the windows. They just have the mirror-material etched out, but still contain the GaAs substrate. A list of devices is shown in Table 4.4.

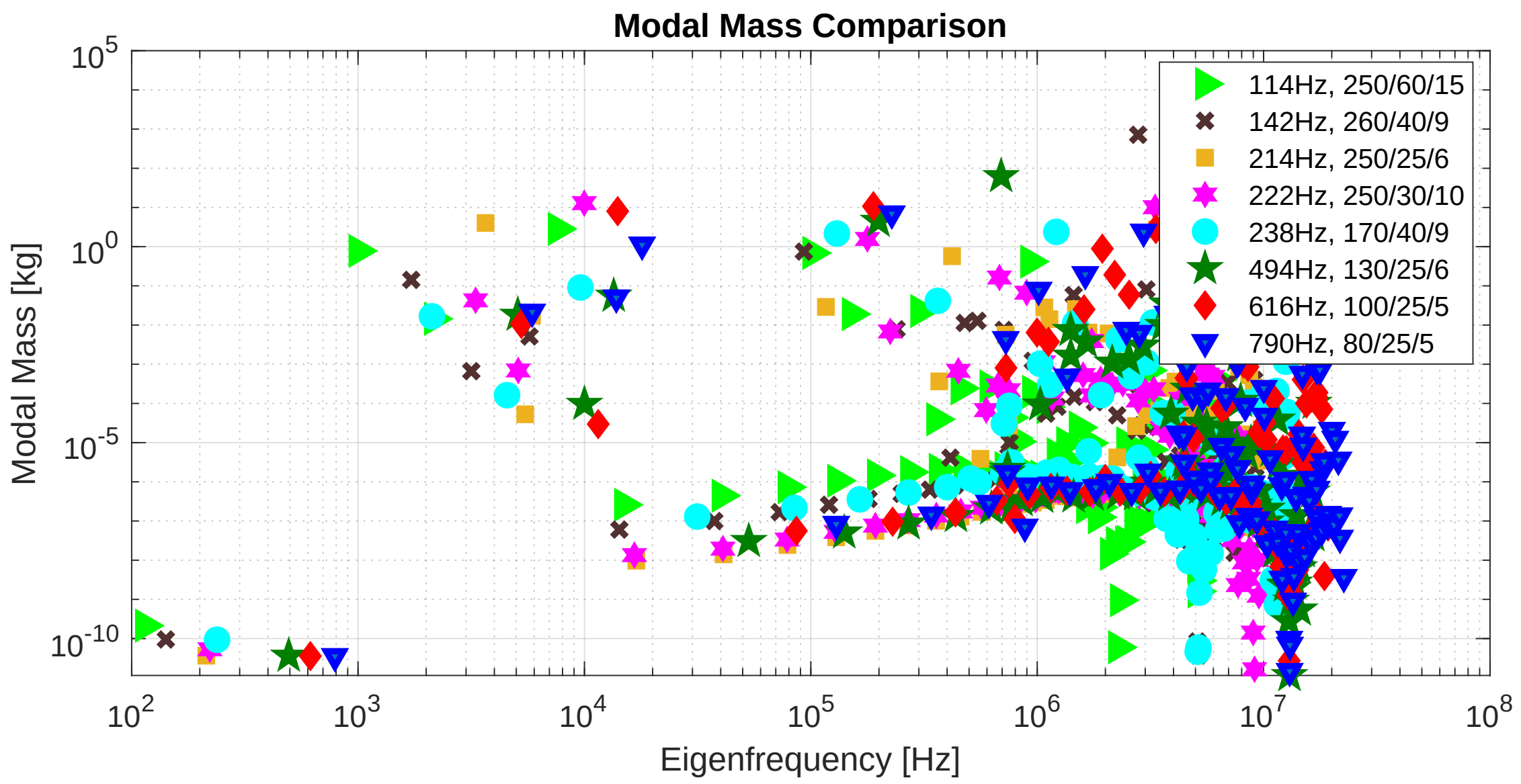

**Figure 4-33:** Modal masses for room temperature squeezers: this figure shows the modal mass of all the eigenmodes of the room temperature squeezers (for the beam position aligned to the node of pitch and yaw). The fundamental modes usually have the physical mass of the device. Here we also see that the devices have low modal mass for their drumhead modes, usually at frequencies above 1 MHz, while most other modes are orders of magnitude heavier. This figure also makes it easy to see that the larger mirrorpads have lower fundamental frequency, higher physical mass, and lower drumhead mode frequency. Next, if one looks at the HOMs of the cantilever in the 1 kHz to 1 MHz band, the longer cantilevers have lower frequency cantilever modes.

## 4.6 Conclusion

In conclusion, we have described the things that need to be kept in mind when building one of these quantum OM experiments. We present eight designs that have been optimized to generate 4 dB squeezing in the audio band at room temperature. These designs are chosen to have some redundancy in performance but not in geometry to allow for oversights. Three of the designs are good squeezers below 2 kHz and five are above. Some possible improvements that can be made in the future :

- The algorithm of using modal mass to identify mirrorpad modes is not accurate since it will miss modes that have symmetric displacement around the beam center (example Figs. 4-14b and 4-14e).

| | | | Left | | | Right | | | Comment |
|---|---|---|---|---|---|---|---|---|---|
| # | Unique Identifier | f0 (Hz) | L | R | W | L | R | W | |
| 1 | A | 142 | 260 | 40 | 9 | 247 | 40 | 9 | |
| 2 | B | 222 | 250 | 30 | 10 | 238 | 30 | 10 | |
| 3 | C | 214 | 250 | 25 | 6 | 238 | 25 | 6 | |
| 4 | D | 238 | 160 | 40 | 9 | 152 | 40 | 9 | |
| 5 | E | 114 | 250 | 60 | 15 | 238 | 60 | 15 | |
| 6 | A | | 285 | 40 | 9 | 271 | 40 | 9 | |
| 7 | E | | 200 | 50 | 10 | 200 | 50 | 8 | |
| 8 | E | | 230 | 50 | 8 | 230 | 50 | 6 | |
| 9 | F | | 20 | 80 | 30 | 19 | 80 | 30 | TED tester |
| 1 | A | | 260 | 40 | 7 | 247 | 40 | 7 | |
| 2 | B | | 250 | 30 | 8 | 238 | 30 | 8 | |
| 3 | C | | 250 | 25 | 4 | 238 | 25 | 4 | |
| 4 | D | | 200 | 40 | 9 | 190 | 40 | 9 | |
| 5 | E | | 250 | 60 | 10 | 238 | 60 | 10 | |
| 6 | B | | 250 | 30 | 6 | 238 | 30 | 6 | |
| 7 | A | | 260 | 35 | 9 | 247 | 35 | 9 | |
| 8 | F | | 50 | 200 | 30 | 50 | 200 | 35 | Doubly clamped TED tester |
| 1 | A | | 230 | 40 | 9 | 219 | 40 | 9 | |
| 2 | B | | 250 | 30 | 6 | 238 | 30 | 6 | |
| 3 | C | | 230 | 25 | 6 | 219 | 25 | 6 | |
| 4 | D | | 170 | 40 | 6 | 162 | 40 | 6 | |
| 5 | E | | 200 | 60 | 12 | 190 | 60 | 12 | |
| 6 | C | | 285 | 25 | 6 | 271 | 25 | 6 | |
| 7 | F | | 20 | 200 | taper | 19 | 200 | taper | Tapered TED tester |
| 8 | G | | 260 | Trampolever | 15 | 247 | Trampolever | 15 | |
| 1 | H | | 150 | 25 | 5 | 143 | 25 | 5 | |
| 2 | I | | 100 | 25 | 6 | 95 | 25 | 6 | |
| 3 | J | | 120 | 25 | 5 | 114 | 25 | 5 | |
| 4 | K | 573 | 60 | 40 | 6 | 57 | 40 | 6 | |
| 5 | L | 558 | 80 | 40 | 10 | 76 | 40 | 10 | |
| 6 | M | 401 | 40 | 60 | 6 | 38 | 60 | 6 | |
| 7 | N | 755 | 20 | 60 | 10 | 19 | 60 | 10 | |
| 8 | O | 413 | 50 | 60 | 9 | 48 | 60 | 9 | |
| 9 | F | | 20 | 100 | taper | 19 | 100 | taper | Tapered TED tester |
| 1 | H | 494 | 130 | 25 | 6 | 124 | 25 | 6 | |
| 2 | I | 616 | 100 | 25 | 5 | 95 | 25 | 5 | |
| 3 | J | 790 | 80 | 25 | 5 | 76 | 25 | 5 | |
| 4 | K | | 60 | 40 | 8 | 57 | 40 | 8 | |
| 5 | L | | 80 | 40 | 8 | 76 | 40 | 8 | |
| 6 | M | | 40 | 60 | 8 | 38 | 60 | 8 | |
| 7 | N | | 20 | 60 | 8 | 19 | 60 | 8 | |
| 8 | O | | 50 | 60 | 10 | 48 | 60 | 10 | |
| 9 | F | | 50 | 100 | 30 | 50 | 100 | 25 | Doubly clamped TED tester |
| 1 | H | | 130 | 25 | 4 | 124 | 25 | 4 | |
| 2 | I | | 100 | 25 | 4 | 95 | 25 | 4 | |
| 3 | J | | 80 | 25 | 4 | 76 | 25 | 4 | |
| 4 | K | | 70 | 40 | 6 | 67 | 40 | 6 | |
| 5 | L | | 60 | 40 | 10 | 57 | 40 | 10 | |
| 6 | P | | 100 | 50 | 10 | 100 | 50 | 8 | doubly clamped squeezer |
| 7 | P | | 120 | 30 | 10 | 120 | 30 | 8 | doubly clamped squeezer |
| 8 | P | | 90 | 60 | 10 | 90 | 60 | 8 | doubly clamped squeezer |
| 9 | P | | 110 | 40 | 10 | 110 | 40 | 8 | doubly clamped squeezer |
| 10 | O | | 50 | 50 | 10 | 50 | 50 | 8 | |

**Table 4.4: Final Chip Layout:** Table showing the final layout of the chip, detailed device dimensions in each window. Here we have assigned a unique identifier color and alphabet to each device. Every copy of a given device is intentionally made slightly different to avoid overlapping eigenmodes. $f0$ is fundamental frequency, which is only shown for the base devices, and not for their copies. The trampolever's radius is not given, its dimensions are: an inner mirrorpad disc of 25 µm radius, trampoline arms 15 µm wide and 5 µm long, and an annular ring outside of 32 µm width. The tapers are 14°.

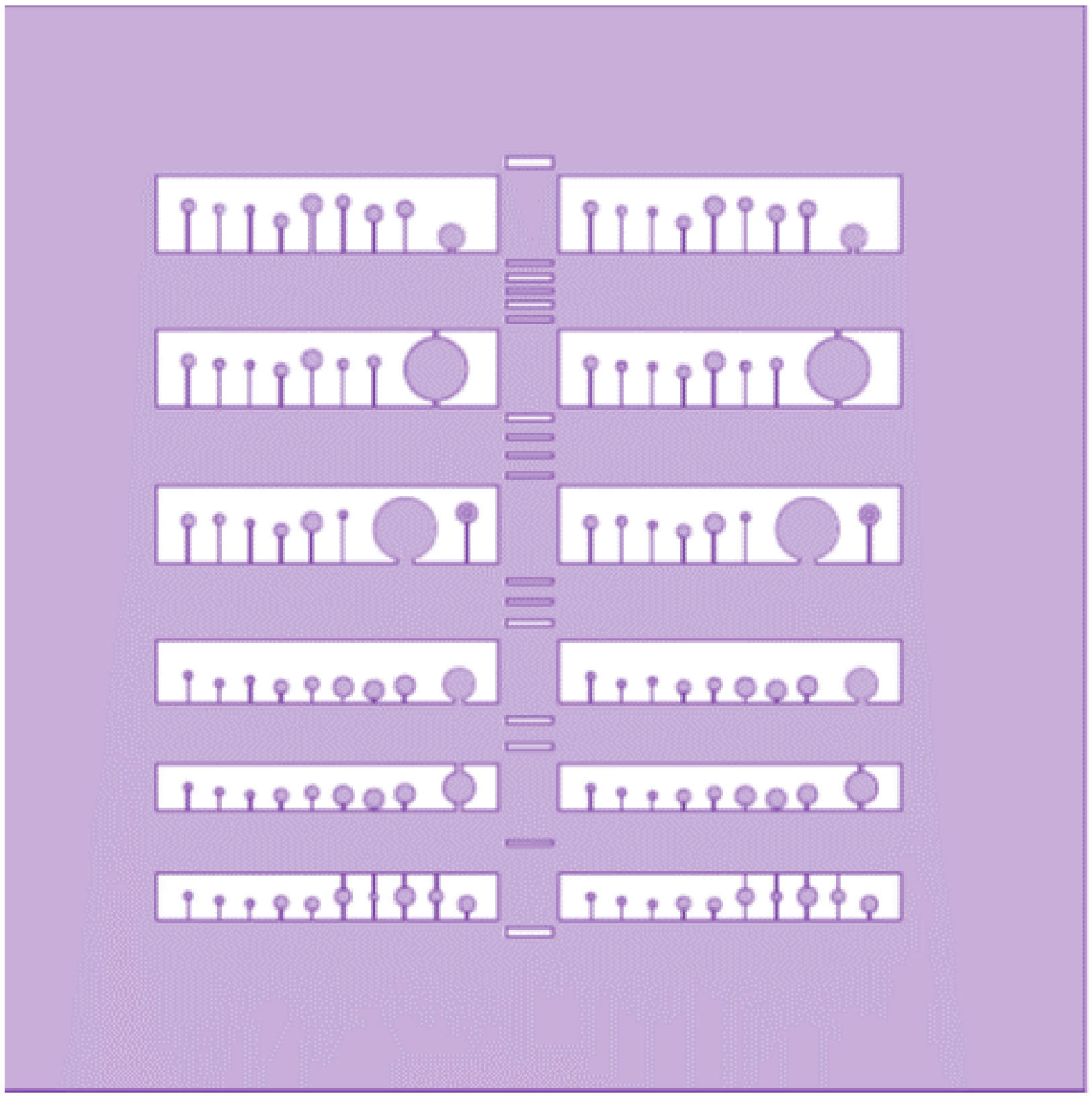

**Figure 4-34:** Final Layout of 3rd generation chips

So instead, another clever way should be used to accurately identify these modes and assign them the lower quality factors.

- Throughout this modeling we have assumed a sum-of-modes approach to the problem. One can also instead use Levin's approach [63], which does not make some of the assumptions that go into the current approach. One could measure the frequency-dependence of the quality factors and that could lead us to better than current models, even with the sum-of-modes approach.
- We tried modeling in ANSYS to try to solve the unphysical damping issues, but the first results given by ANSYS (about just the eigenfrequencies) didn't match our COMSOL results (which did match the analytical results for simple geometries.) This means that the ANSYS model needed more tweaking, which we abandoned in the interest of time. It is indeed possible that once modeled correctly in ANSYS, one might have better luck in solving the TED equations with physical results.
- Another piece in the optimization could be to try to optimize not just for the eigenfrequencies, but also the modal masses. For instance, one could try designing a device that had a higher modal mass for the drumhead mode, or the second bending mode, both of which are the dominant source of thermal noise above tens of kHz.
- It is also possible to optimize the size of the beam for the size of the device. That would involve using different cavity length/input coupler for each different device, but might result in better performance.

# Chapter 5

# Experimental implementation

## Contents

In this chapter, we describe the experiment we built at MIT. Key features of the experiment were a Fabry-Perot cavity comprising a fiber mirror [64] as an input mirror, and a crystalline mirror attached to a micro-mechanical cantilever [65] as the end mirror. The fiber mirror is mechanically fixed, while the end mirror is movable, allowing radiation pressure interaction. The laser light is coupled into the cavity via the fiber mirror. A fiber mirror can be machined to have a small radius of curvature (roughly 200 µm), allowing for stable cavities of small length (100 µm), which is not possible with macro sized mirrors. The light reflected from the cavity propagates back in the fiber, while the transmitted light is detected in free space. In this work, we describe the preparation of low-noise infrastructure, laser system, detection, and locking of this setup. The measurements described in Chapters 6 and 7 are taken on a similar setup at Louisiana State University, with a macro mirror as the input mirror instead of a fiber mirror.

## 5.1 Infrastructure setup

A major technical noise considerations in a precision measurement experiment is vibrations. Since we are looking for effects of motion at the scale of femtometers, we must isolate the experiment from all environmental sources of displacement noise, like seismic motion and acoustic vibrations. In this section we lay out our approach to building a low-noise foundation for the experiment.

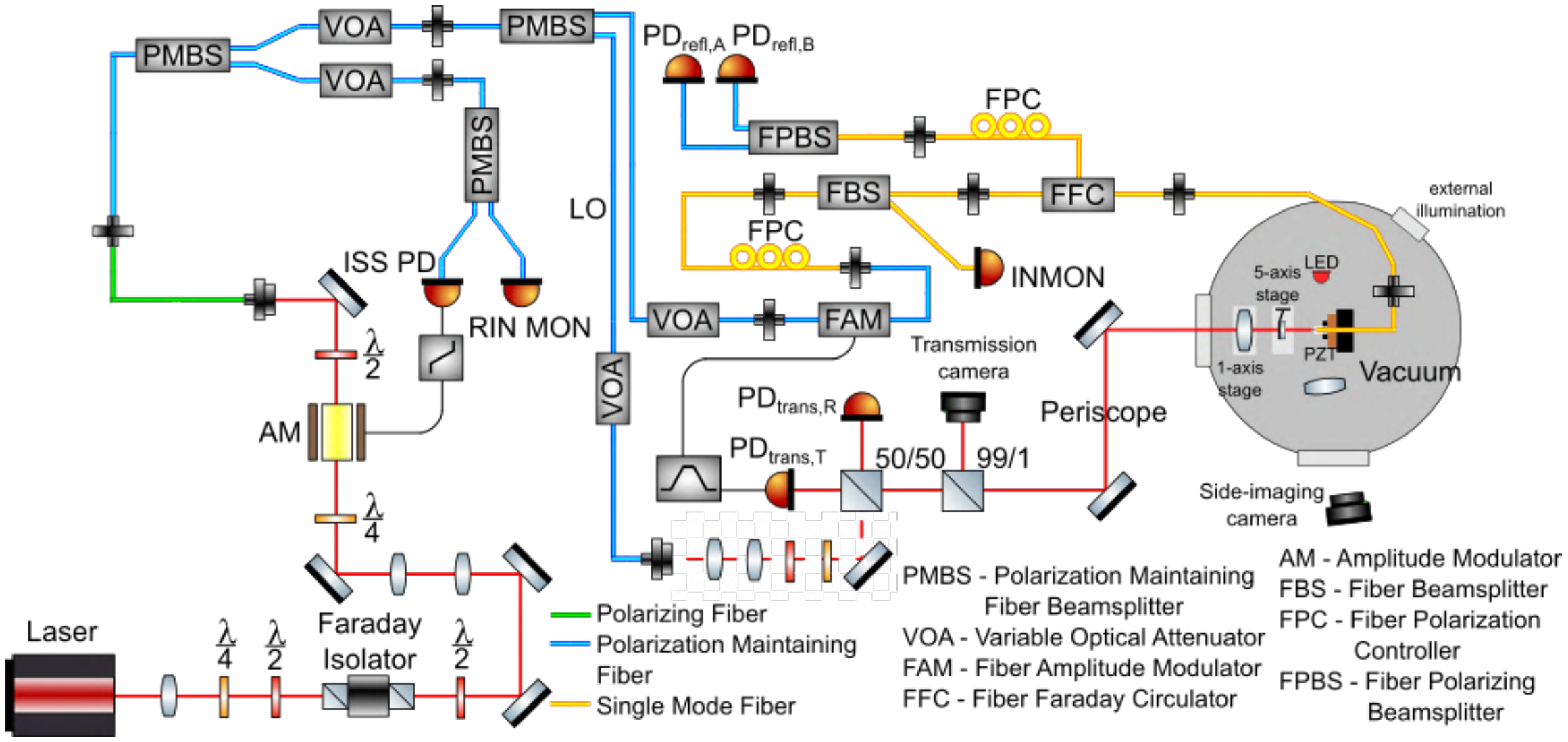

**Figure 5-1:** Experimental setup at MIT: a Fabry-Perot cavity with a fiber-mirror and a cantilever mirror is housed in the vacuum chamber. The laser light is coupled into the cavity via the fiber mirror. The laser beam from an Nd:YAG Mephisto laser is mode-matched and coupled to a polarizing fiber (shown in green). This polarizing fiber is used to clean up any polarization misalignment with the fiber's slow axis at the input. Next, the light is split for an intensity stabilization servo (ISS) readout, and a local oscillator (LO). The rest of the light is sent to the cavity via an in-fiber amplitude modulator. Each path has its own power control via an in-fiber variable optical attenuator, and the total optical power is controlled in free space by a wave-plate and PBS. All fibers till the amplitude modulator are polarization maintaining fibers, which ensures the polarization going into the FAM is vertical. The fibers after the AM are single mode, with fiber polarization control paddles. This allows us to rotate the polarization of the light going into the cavity. The fiber circulator ensures easy separation of input light from the light reflected by the cavity. The reflected light is fed through another set of polarization control paddles, and an in-fiber polarizing beam splitter to rotate the cavity eigenmode and split it into two linear polarizations. The cavity is locked via radiation pressure by using the transmitted light to obtain the error signal and an amplitude modulator as the actuator ([38], Appendix B). The ISS loop had not been closed for the data presented in this chapter. Credits: Ben Lane.

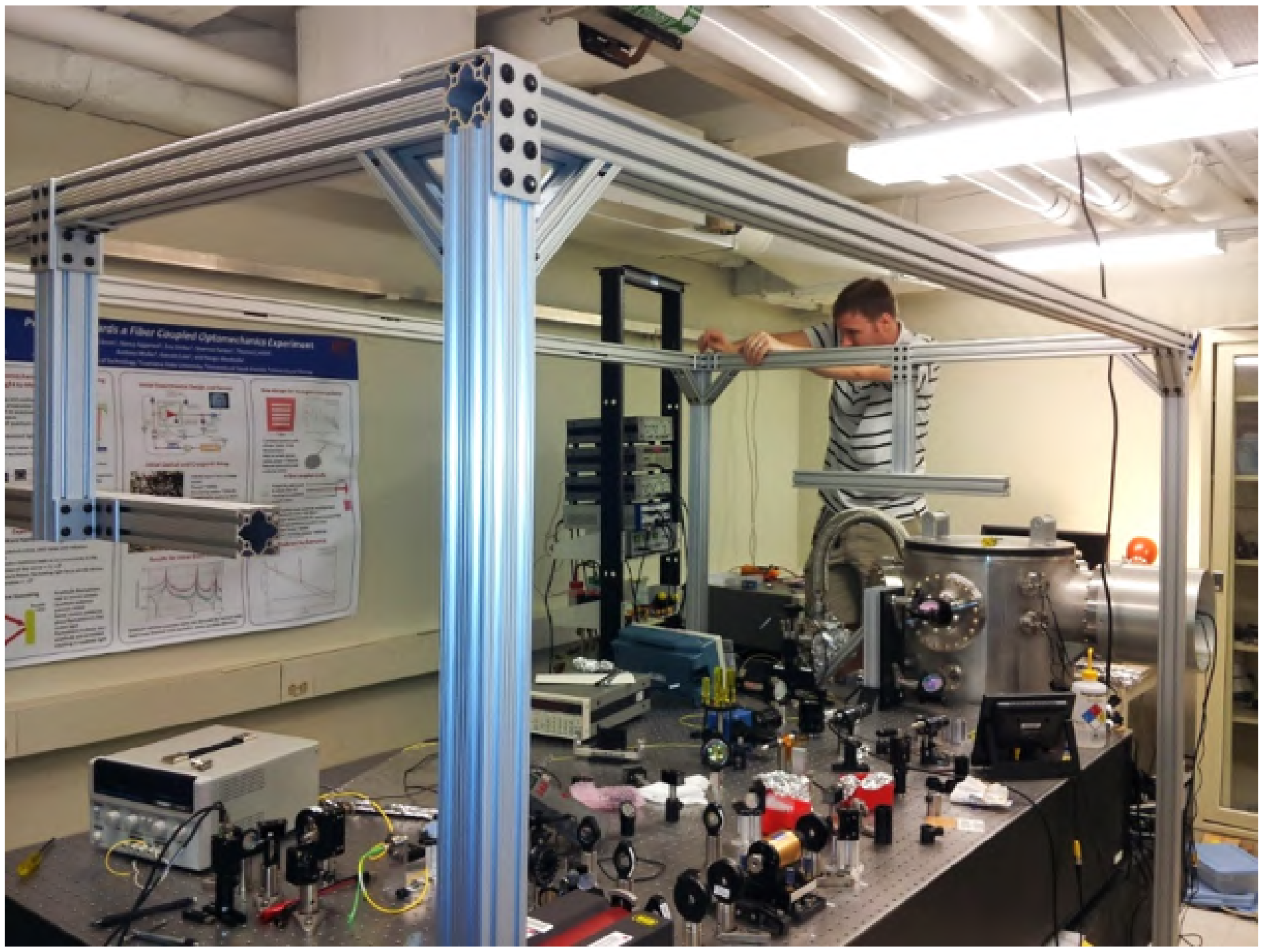

**Figure 5-2:** The full optics table and the vacuum chamber commonly sit on a slab of granite. The entire system is supported by Newport's air-pressurized legs, and can be floated to go in low-noise mode.

### 5.1.1 Floating Table

Our experiment has multiple stages of isolation from ground motion as we will lay out in this section. The first stage of those is air-floating the entire setup. The optics table containing the laser preparation and the readout system, and the vacuum chamber containing the OM cavity both rest on a massive granite slab, as shown in Fig. 5-2. The combined mass of these three systems, around 7000 lbs is floated using air-pressurized legs, making a first-order passive isolation stage. This stage is at around a 1 Hz resonance frequency. Next, we show the placement of the vacuum chamber relative to the granite slab and the optics table in Fig. 5-3.

### 5.1.2 Vacuum chamber

The main experiment is housed inside a vacuum chamber. Fig. 5-3 shows this chamber from the outside, while Fig. 5-4 shows the experiment inside the chamber from a top view. The chamber is situated on top of multiple high-mass, high-damping stacks. The damper is a Sorbothane type material, which can be seen as blue disks in Fig. 5-3. The high-mass is formed by steel discs. This stack contains three mass-damper stages below the chamber. This stack achieves a dual purpose; it provides passive isolation from ground vibrations, as well as raises the chamber up to expose the bottom flange, which could be useful in the future for electrical feedthroughs or plumbing.

In addition to the external passive isolation, we add three more passive isolation stages inside the vacuum system. They are much higher mass steel blocks as shown in Fig. 5-5, and are separated with Viton corks for damping. The optics breadboard is then clamped to the top steel mass, as shown in Fig. 5-4.

### 5.1.3 Fiber optics

The optics table has its own internal vibration isolation stages, but is supported by one extra stage of the same steel mass and Sorbothane damper that is used under the vacuum chamber. The setup in relation to the vacuum chamber is shown in Fig. 5-6. The laser preparation and control is done all via in-fiber components. The first stage is an intensity stabilization servo (ISS), followed by polarization control and pick-off for the local oscillator. The light is then sent to the cavity via a fiber circulator which allows for separation of reflected light. The circulator is connected to a fiber feedthrough to the vacuum chamber, with the other end of the feedthrough connected to the fiber mirror.

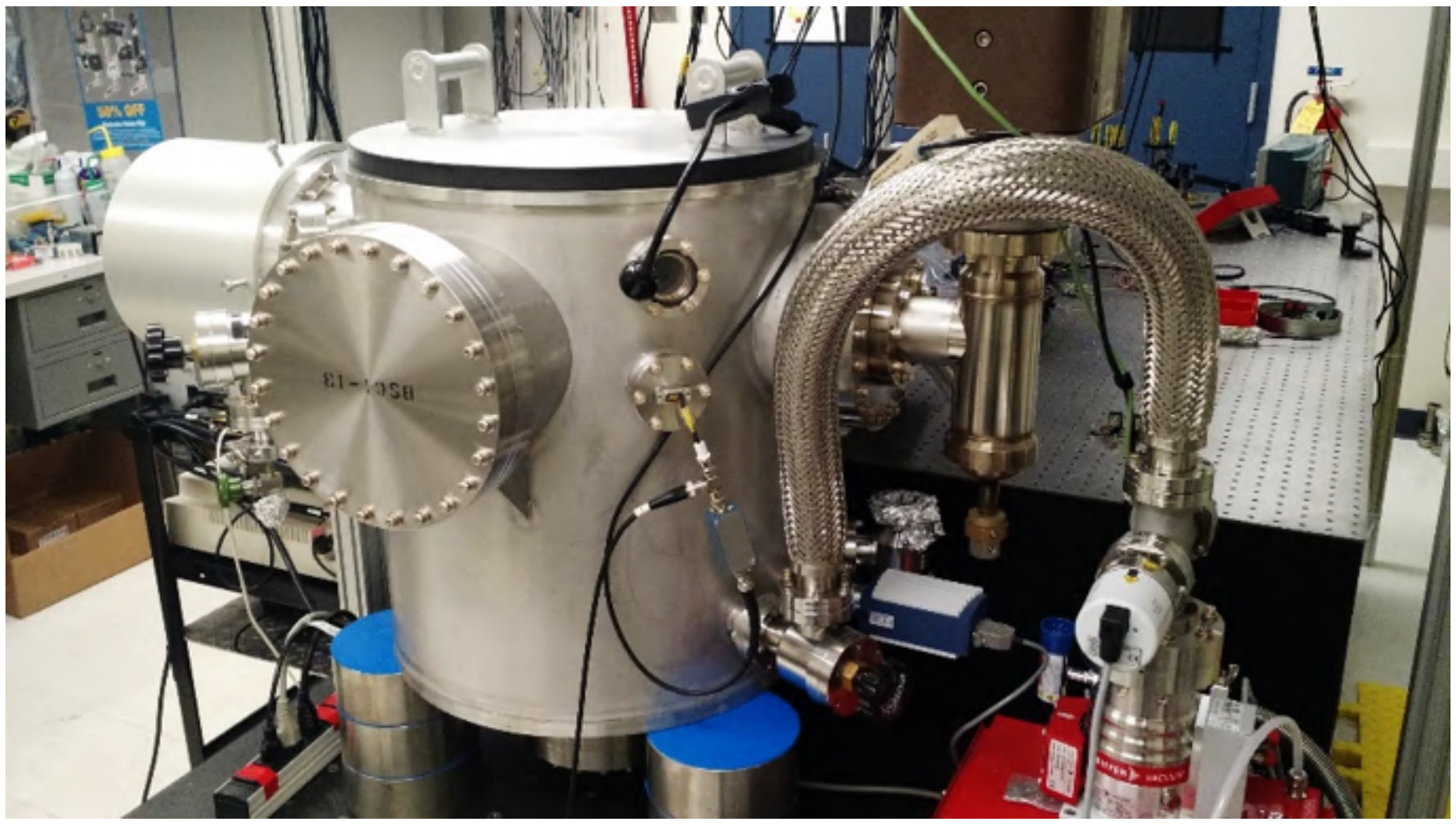

**Figure 5-3:** Vacuum chamber that contains the main Fabry-Perot cavity of the experiment. The vacuum chamber is placed on top of three stacks of steel masses sandwiched with high damping material (blue disks). This view shows the side-imaging port on the left of the chamber. The 2.75 inch flange supporting the window for external illumination for the side imaging can be seen above the flange connected to the turbo pump.

## 5.2 Fabry-Perot cavity

At the heart of the experiment is the Fabry-Perot cavity which resides inside the vacuum chamber. The layout of the in-vacuum breadboard is shown in Fig. 5-4. The green path is sketched to draw attention to the in-vacuum fibers (transparent) which couple the laser to the cavity via a fiber-mirror. The fiber mirror faces the micro-mechanical device on the chip. The transmitted light from the cavity is sketched in red. The transmitted light is routed back to the air table via a periscope.

There is a 50 mm focal length lens right behind the chip to collimate the highly diverging mode that exits the cavity. The fiber-mirror has a 200 micron radius of curvature. The micro-mirror is flat, with 100 microns cavity length. This gives roughly a 6 $\mu$m waist size at the chip, hence requiring a lens close by to collimate it.

The lens to the left of the cavity is used to image the cavity from the side, and the PEEK holder on the right side of the breadboard holds an in-vacuum infrared LED to illuminate the cavity for side imaging.

### 5.2.1 Chip setup

The chip is clamped in a copper holder. Copper is chosen for this holder because it is a thermal conductor. If there is a need to cool the chips, cooling braids can be attached directly to this copper holder. The copper holder is made to be one inch in diameter, and is secured in an adapter that allows to mount the copper holder in a gimball mount meant for two-inch mirrors. Securing the chip in a two inch mirror mount allows for space for the fiber mount as well as side imaging. The gimball mount is connected to a picomotor driven translation stage for x,y, and z actuation, and two picomotors are attached to the mount directly for angular actuation.

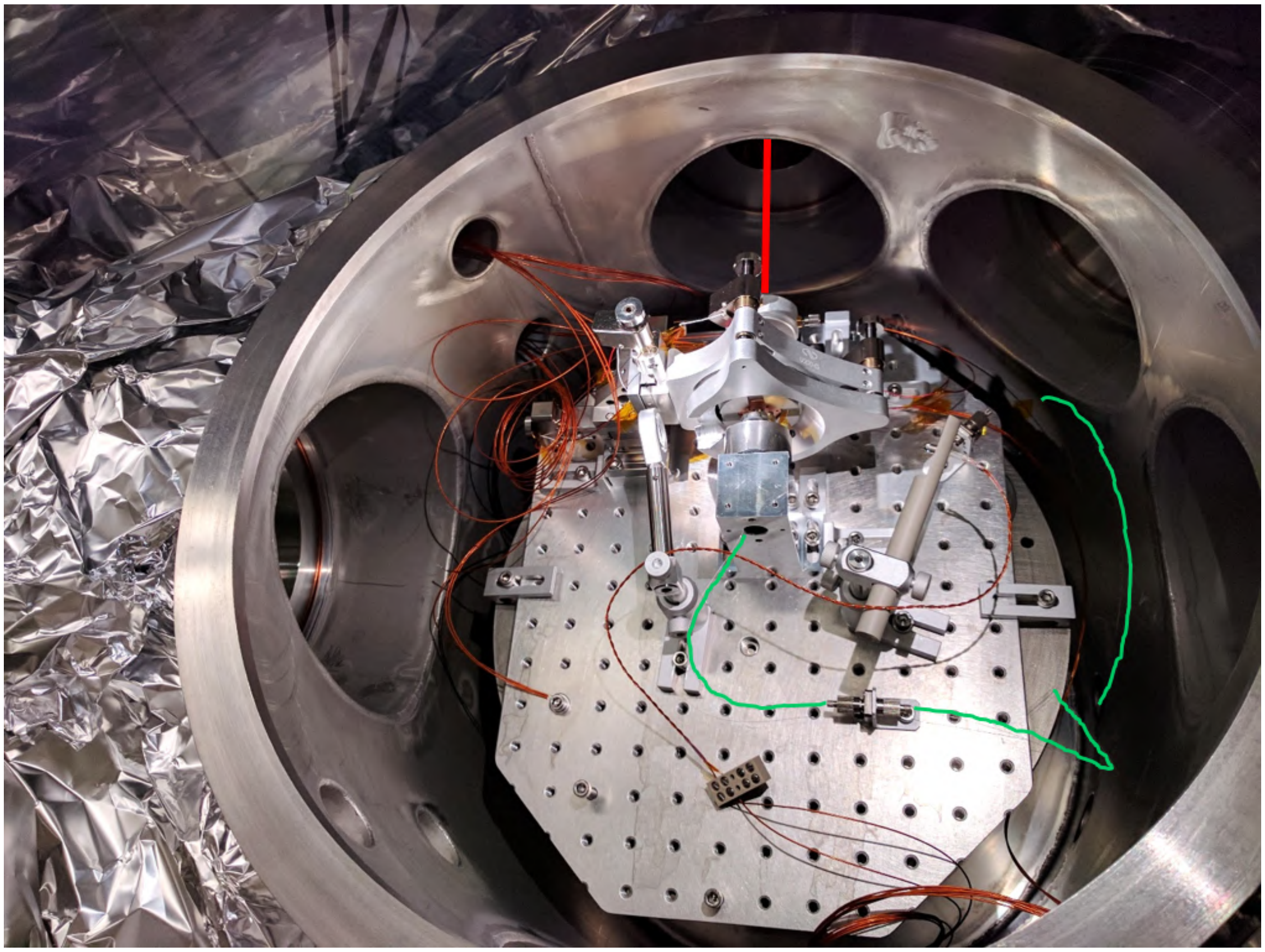

**Figure 5-4:** Optical layout inside the vacuum chamber. The green path is sketched to highlight the single-mode glass fiber (transparent) that brings laser light into the vacuum chamber. The fiber mirror is held in the cuboidal aluminum mount, which faces a gimball mount that secures the chip containing mirco-resonators. The gimball mount is on a 3-axis translation stage actuated by picomotors, which along with the pitch and yaw actuation via picomotors allows for cavity alignment. The transmitted light from the cavity exits at the north side, highlighted in red, via a mode-matching lens that is placed behind the gimball mount. The lens to the left of the cavity is used to magnify and focus the cavity to an external camera to image the cavity length. The PEEK holder to the right of the cavity is used to position a 1 µm LED to illuminate the cavity from inside the vacuum for better side-imaging.

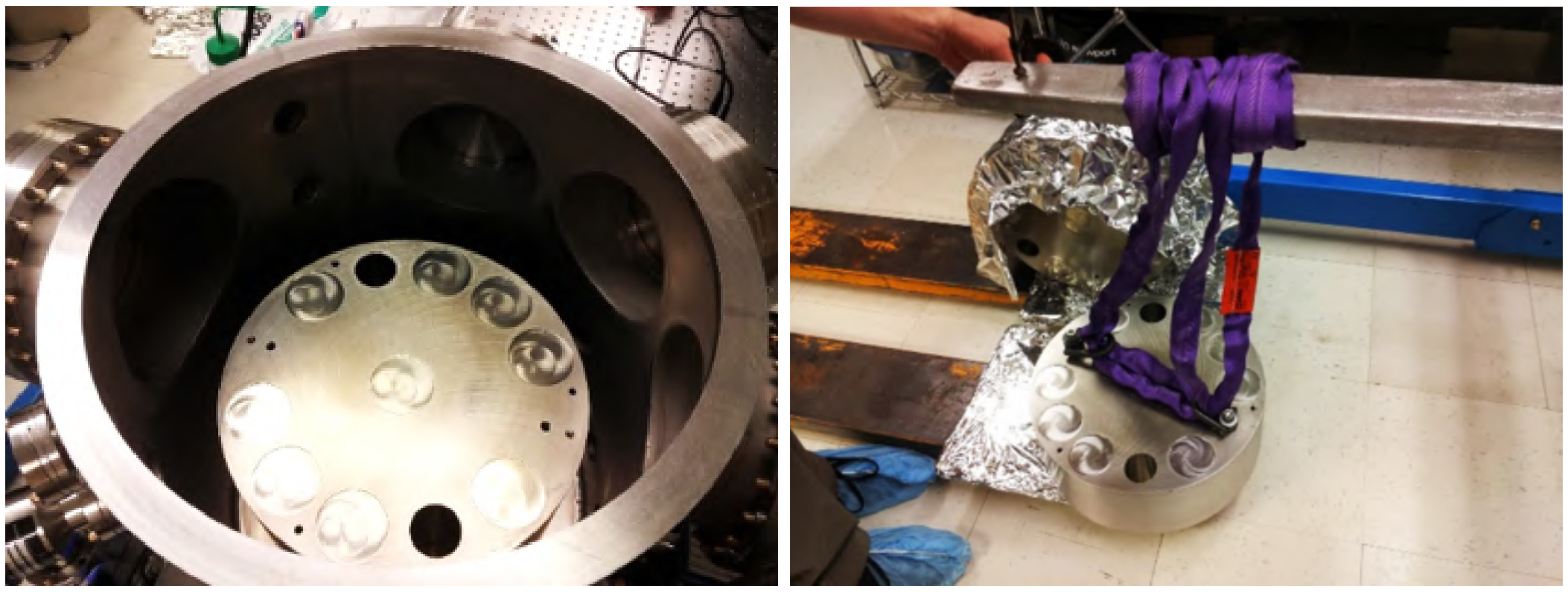

**(a)** **(b)**

**Figure 5-5:** **(a)** Foundation inside the chamber for the optics breadboard. This foundation provides three extra stages of passive isolation for the optics breadboard. **(b)** Single stage steel mass being crane-lifted for installation.

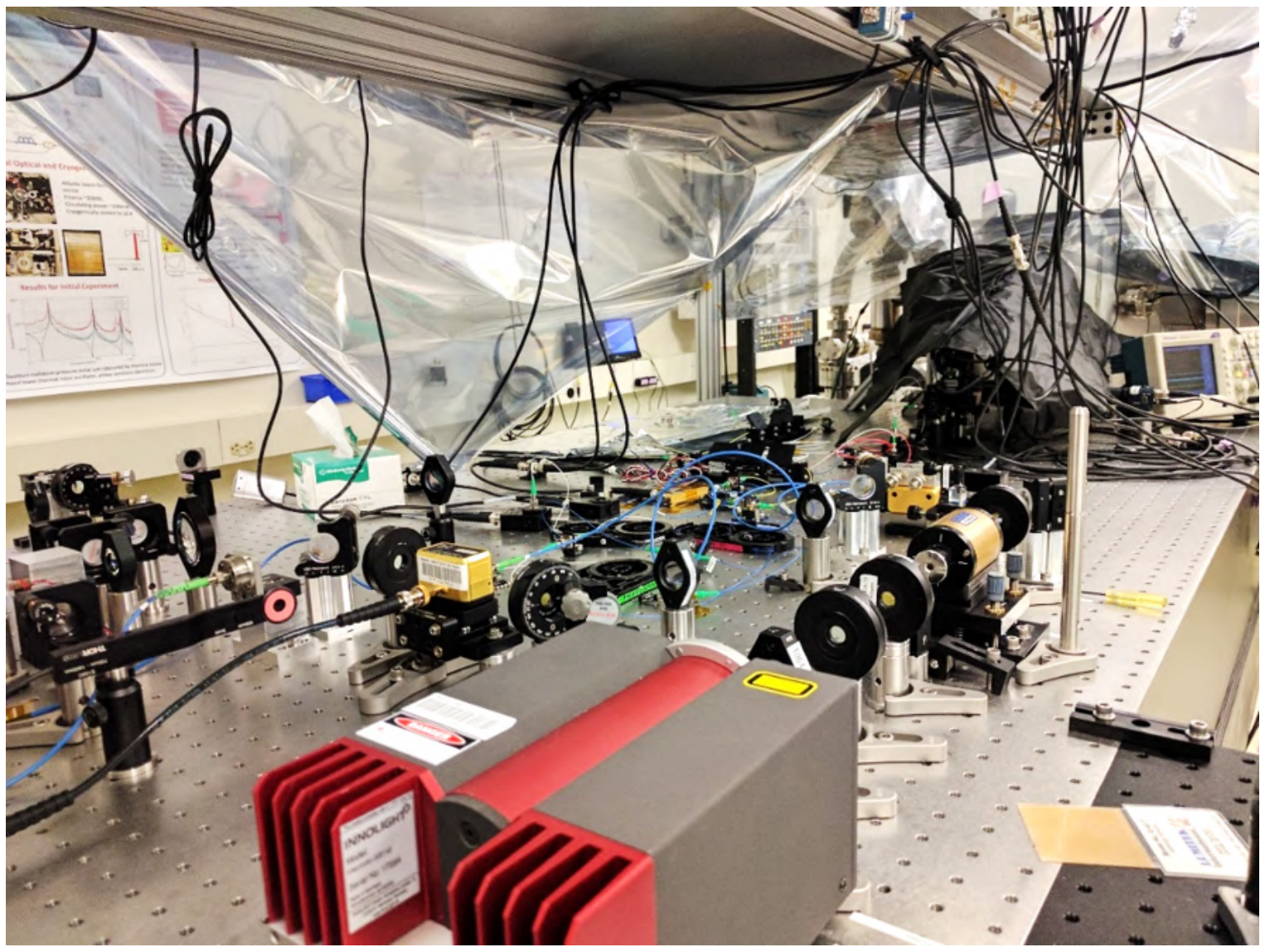

**Figure 5-6:** Optical table, see Fig. 5-1 for a detailed layout.

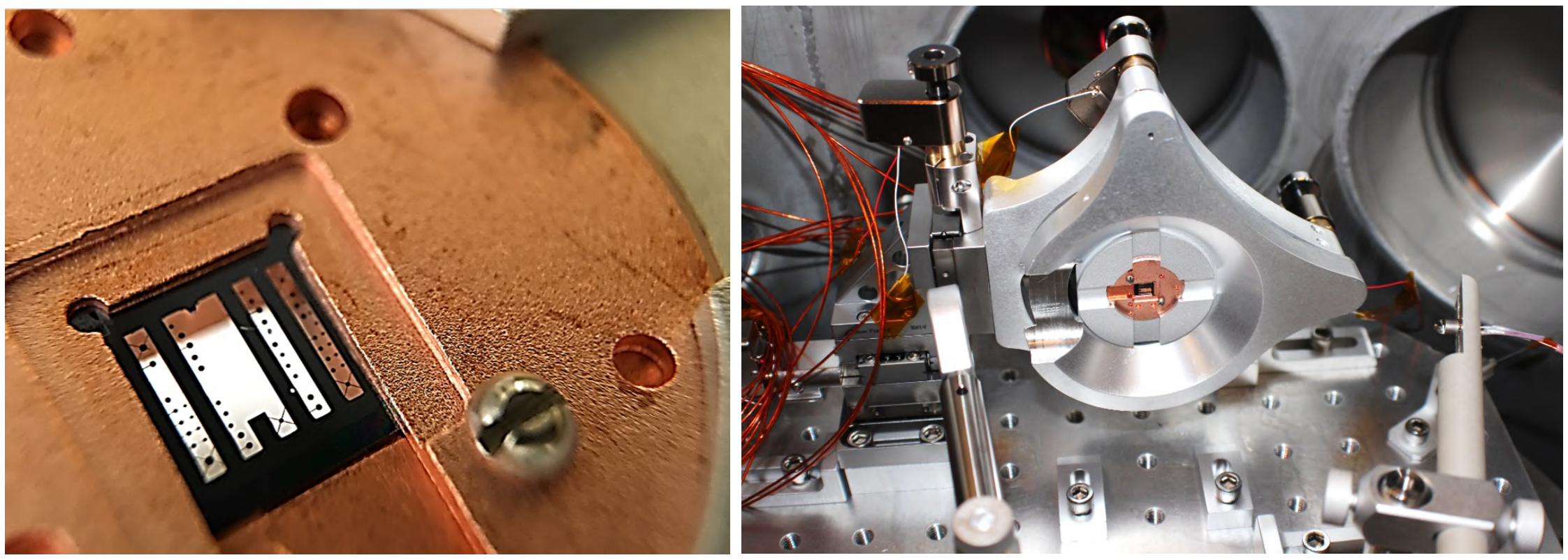

(a) (b)

**Figure 5-7:** Details of chip mounting: **(a)** The second generation chip in the copper holder **(b)** The copper holder containing the chip is mounted in a gimball mount for a 2 inch mirror, with five axes for alignment actuation. Also can be seen on the right the 1 µLED, and on the left the lens used for imaging the cavity-length from the side.

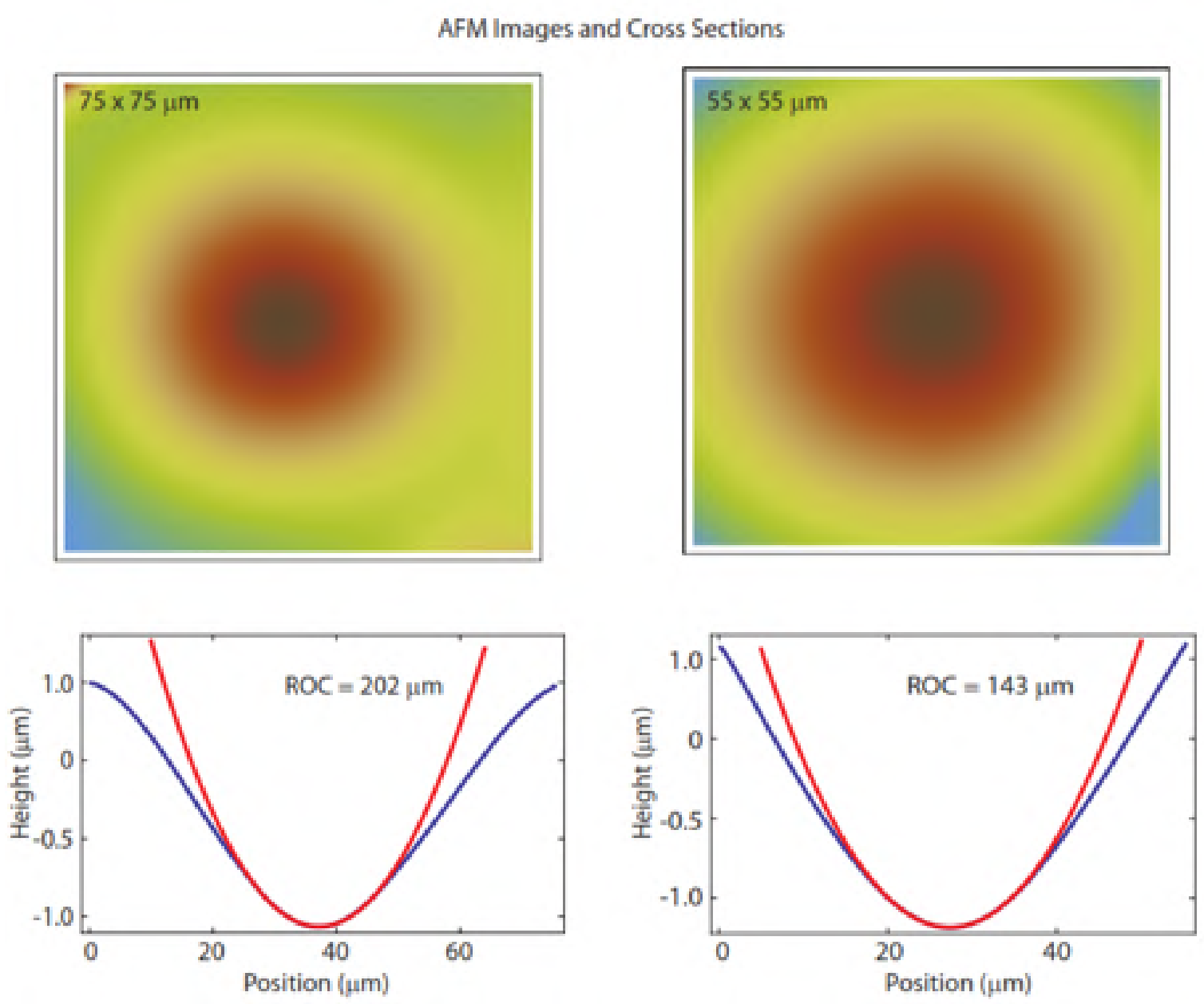

**Figure 5-8:** Surface profile of a sample ablated fiber that makes the fiber-mirrors measured in an AFM. Credits: Andreas Muller.

### 5.2.2 Fiber mirror setup

The input mirror to our Fabry-Perot cavity is chosen to be a fiber mirror [64]. Fiber mirrors allow for a small radius of curvature, hence enabling small cavities. The fiber mirror is a small dimple mirror created on the tip of a fiber. It is made by first ablating the tip of a fiber using a $CO_2$ laser. This ablation is done to get the right aperture size of the dimple and the desired radius of curvature. This process is highly variable because it depends on the relative alignment of the fiber with the ablating laser, as well as the mode of the laser at the tip of the fiber. The fibers being ablated are usually secured in a ceramic ferrule such that they can be easily mounted for ablation as well as later use. A few millimeters of the bare, uncoated glass fiber (i.e. just the core and cladding) sticks out in front of the ferrule, making them quite prone to breaking. Due to the variability of the ablation process and the fragile nature of the fibers, we make a batch of many of them together. Next the fibers are sent to a coating house to deposit high reflective coating on them via ion-beam sputtering.

Our application of fiber mirrors is similar to [64], with a few alterations. Firstly, we use them with a flat mirror on the other side (the cantilever mirror) to form a semi-confocal cavity, rather than a confocal cavity with two such fiber mirrors. Secondly, the smaller the spot size on the cantilever mirror, the better, which inspired us to go for even smaller RoCs than in Ref. [64]. We aim for a 200 µm RoC, allowing for a 100 µm length, giving a spot size of around 6 µm on the cantilever. This gives a spot size of roughly 10 µm on the fiber mirror, which has a decent theoretical mode coupling efficiency with the mode confined in the fiber core. The fiber mirrors are custom made by our collaborators in the group of Prof. Andreas Muller at University of South Florida. The surface profile data taken by an atomic force microscope of one of our fiber mirrors is shown in Fig. 5-8. The fiber core is used to couple light into the cavity, as well as to carry the reflected light from the cavity for readout.

In our setup, the fiber mirror is attached to a piezoelectric crystal to allow for length control of the cavity. The final assembly looks as shown in Fig. 5-9. The piezo is glued on the back to a heavy steel reaction mass, and on the front to an annular adapter. This adapter has tapped screw holes on it which are used to connect it to another adapter that is glued to a fiber mirror. Multiple such adapters holding fiber mirrors are shown in Fig. 5-10 from the back.

During our first few iterations of the assembly, we noticed that the piezo and glue joints were getting compromised in the process of assembly. We learned that this was caused by the shear stress that was

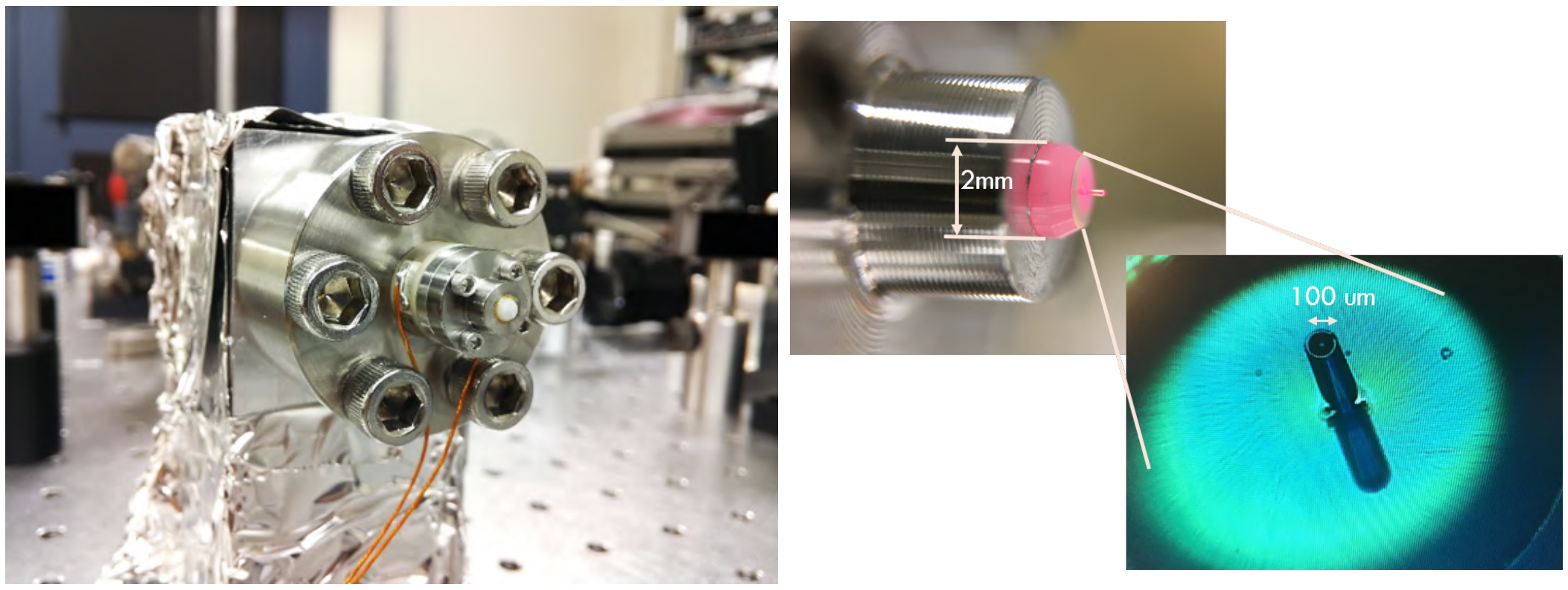

**(a)** **(b)**

**Figure 5-9:** Fiber Mirror: **(a)** Front view of the fiber mirror held on a reaction mass via a piezo. **(b)** Side and isometric closeups of the fiber mirror tip.

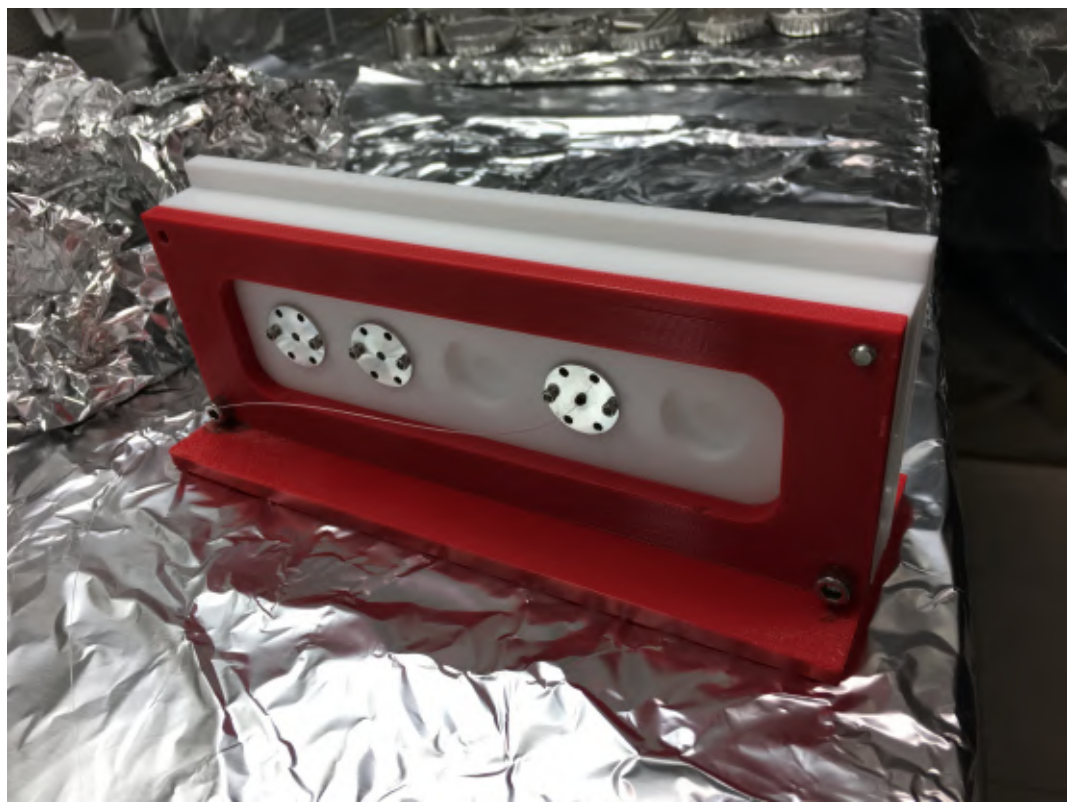

**Figure 5-10:** Fiber mirror assembly and storage stage. Each fiber mirror is glued to an aluminum adapter. The aluminum adapter has 0-80 through holes, which can be used to secure spare fibers, as in this picture, or to attach the fiber to an adapter that is glued to a PZT (as shown in Fig. 5-9a).

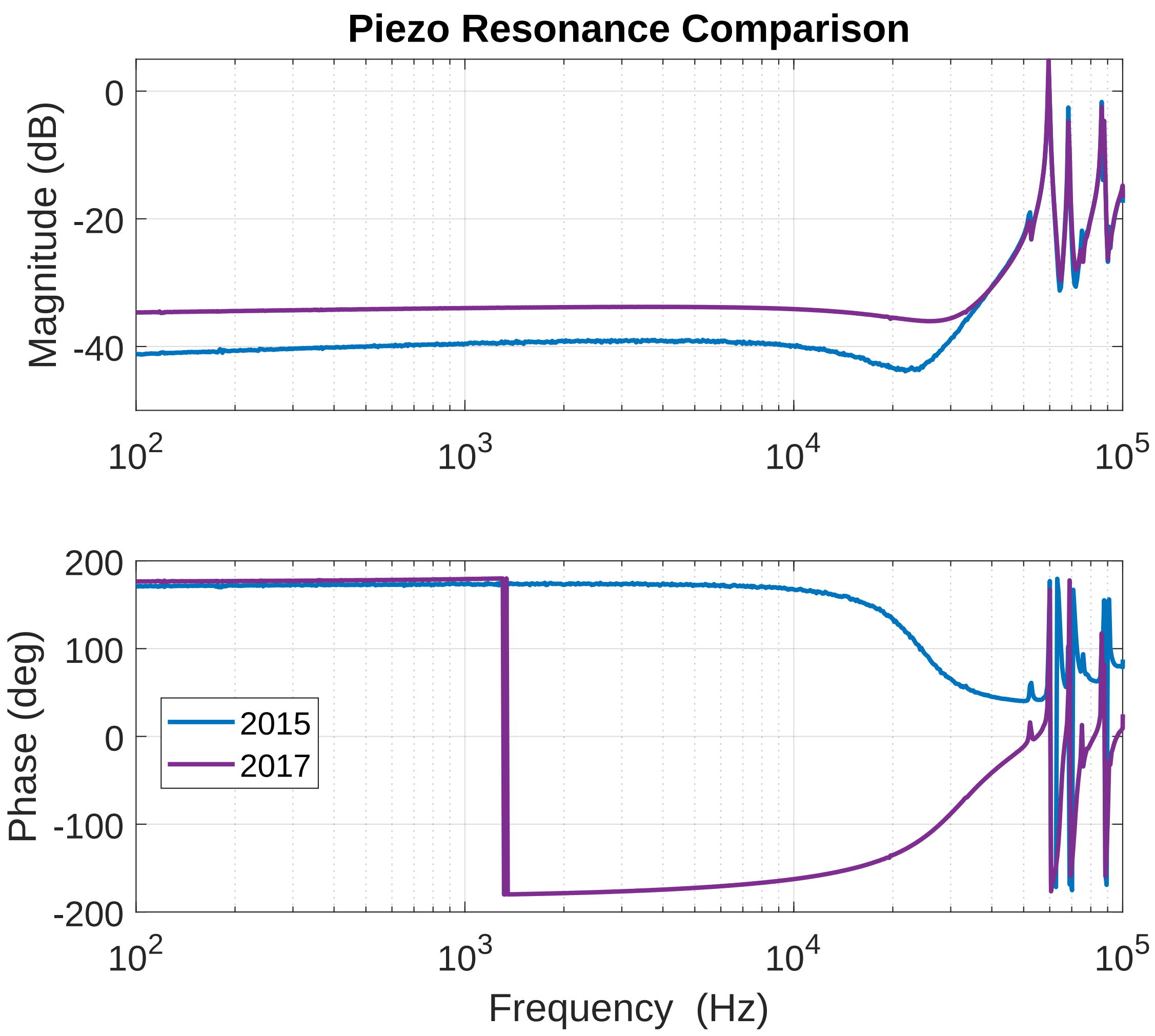

**Figure 5-11:** Measurements of piezo resonance for fiber mirror actuation taken using a capacitive bridge circuit to excite the piezo's mechanical resonance leading to an electrical resonance in the circuit. The two datasets are for the same pzt, first in 2015 when we first installed it, and next in 2017 after we changed the fiber mirror on it.

induced on the piezo and the epoxy during tightening of the screws. In order to eliminate this stress, we now tighten the screws by countering the torque required for tightening. This can be done by wedging a levered arm between two screws opposite to the screw being tightened, and applying slight torque on the lever as the screw is tightened. We measured the piezo resonance at various intermittent steps in between. When the old fiber mirror was taken off (i.e. the piezo was unloaded), we measured about 100 kHz resonance. When we started putting the new fiber on – the resonance frequency was low at about 45 kHz. But then by tightening the screws more we were able to get it up to 60 kHz. At that point the screws felt pretty tight so we didn't go further. The piezo resonance measurements are taken by making a capacitive bridge circuit balanced for the piezo's measured capacitance. The measurement are shown in Fig. 5-11.

## 5.3 Cavity length imaging

Since the cavity is designed to be around hundred micrometers in length, there is a need of real-time imaging system to safely manipulate the cavity. This is to both gauge if the cavity length is in the stability region to resonate a mode, as well as to make sure the chip and the fiber mirror do not collide.

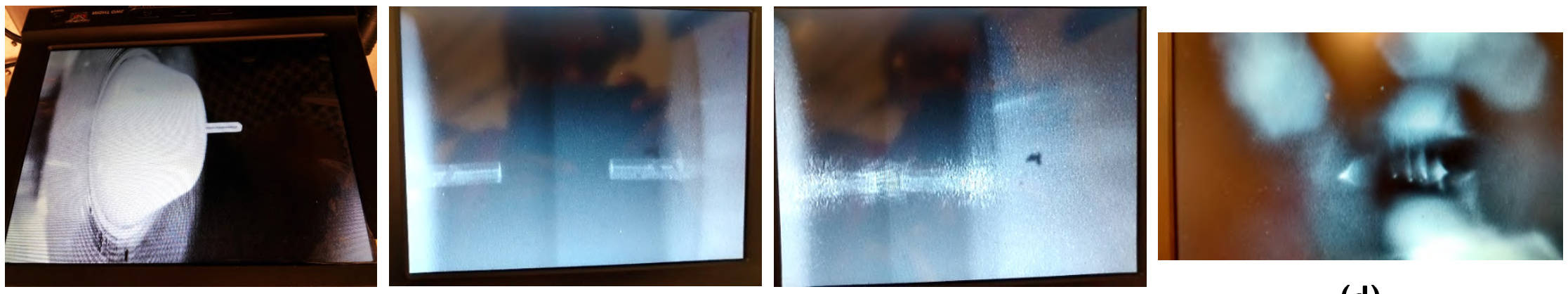

(a) (b) (c) (d)

**Figure 5-12:** Cavity length imaging: **(a)** shows the intermediate, lower magnification stage to obtaining the final side-imaging of the fiber. The fiber is held in a ceramic ferrule, which is glued in an aluminum mount. We slowly increase the magnification in order to keep the camera aligned to the fiber tip. The magnification is increased by adding distance between the camera and the composite lenses, reducing the distance between the fiber and the lens, and if required, changing the lens to a larger focal length. **(b)** Side-imaging of the fiber-mirror cavity obtained after the above method. In this image the flat mirror is served by a glass witness sample. The camera is placed at a slightly less than perpendicular angle to the cavity-axis, which allows us to see the fiber-mirror and its reflection in the witness sample. The distance between the image of the fiber and its reflection is used to get a measure of approximate cavity length. **(c)** Using the translation stage, the flat mirror is brough closer to the fiber, and the laser is turned on. We see the light scattered from this fiber into all directions. **(d)** Same side image, except the witness sample has now been replaced by the chip. Since the chip is held in a copper holder, that drastically lowers the lighting near the cavity, making the side-image hard to decipher. The tip of the fiber mirror is seen on the left, while a few cantilever mirrors are seen on the right in this picture.

In order to image the cavity, we place a camera at an angle close to, but not exactly at 90° from the cavity axis. There is also a 50 mm focusing lens between the camera and the cavity to get the image in the right distance and magnification. The camera sees the fiber mirror, as shown in Fig. 5-12a. This image is taken at a much lower magnification, and hence also shows the ferrule and the mount.

We then increase the magnification and bring in the flat mirror closer to the fiber mirror. The image then can be seen in Fig. 5-12b. Since the camera is situated at an angle, it is able to see the fiber as well as the fiber's image in the flat mirror. We can calibrate the cavity length as half the distance between the images of the fiber and its image, which in turn can be known because the fiber's diameter is known to be 125 µm. Sometimes, the fiber mirrors would have excess scattered light, which can be seen in Fig. 5-12c. The side imaging is best if there is diffused illumination at the cavity. For our test setups with a glass flat mirror, this imaging technique was successful. But when the glass witness sample was replaced by the chip, the opaque copper holder blocks most illumination, and the images become much harder to decipher. One such image is shown in Fig. 5-12d, where this time the chip is being used instead of a flat witness sample. We see the lack of overall illumination, which is responsible for masking most of the fiber length. We do manage to see a small fraction of the fiber top on the left, as well as a row of cantilever mirrors on the right.

## 5.4 Readout

As explained in Chapter 3, we designed this generation of the experiment so that the squeezed light could be measured in transmission from the cavity. So, our most sensitive readout is in transmission (by design). The transmitted light exits the cavity, and since it is a highly diverging mode, it is collimated by an in-vacuum super-polished lens. The light is then directed to the in-air optical table for detection. All optics in this chain are superpolished to minimize loss. A small fraction of the light is sent to a camera for imaging, and the rest of it is sent to photodetectors. While transmitted light is the squeezed light, we still use the reflected light for diagnostics, troubleshooting, as well as control. The reflected light that is in the fiber goes through the circulator and is sent to a polarization controller, followed by a polarization beam splitter, which is connected to two photodetectors.

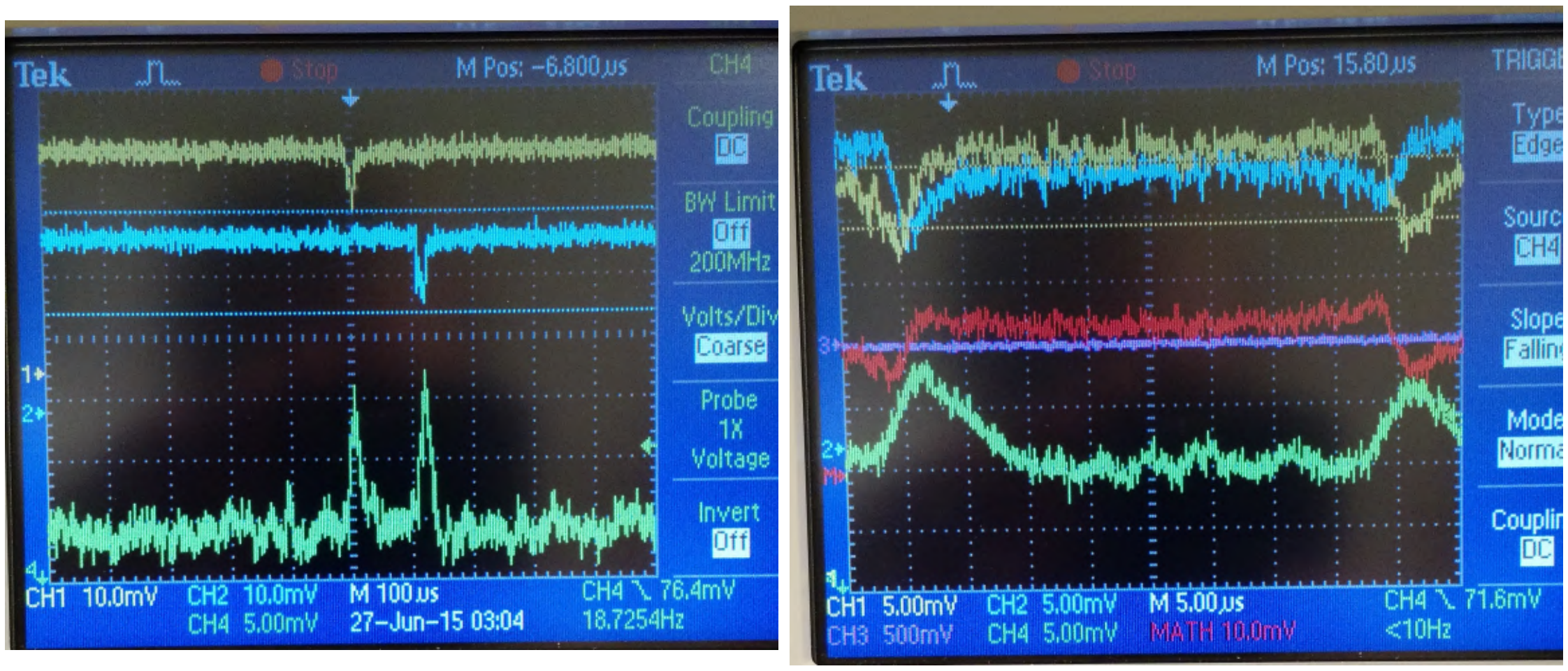

**(a)** **(b)**

**Figure 5-13:** Cavity birefringence: In **(a)** we show the transmitted signal from the cavity in green, and the reflected signal, split onto two PDs (Refl A and Refl B) using a PBS (Fig. 5-1) in yellow and blue. In **(b)** we show the error signal obtained by subtracting the two polarization-separated reflection signals in red, along with the corresponding transmission and reflection signals with the same color scheme as a.

## 5.5 Birefringence

The fiber mirror, as well as the cantilevers are known to have birefringence, which means that the Fabry-Perot cavity is also birefringent. In order to investigate this, we intentionally send a mixture of the two eigen-polarizations of the cavity and look at the flashes in transmission and reflection from them. This is shown in Fig. 5-13. In Fig. 5-13a we show the transmitted signal from the cavity along with the reflected signal. The reflection has been split into its two orthogonal polarization components by using a PBS. We see both eigenpolarizations of the cavity in transmission, and we see them separately in reflection (once the FPC is properly set). Next, by changing the polarization of the in-going light, we are able to transfer the intensity from one eigen-polarization to the other. In order to do this, we leave the reflection FPC in its previous (diagonalized) configuration, and minimize the dip in one of the reflection signals, while increasing the coupling into the other.

The error signal for a typical (i.e. non optomechanical) birefringent locking scheme is shown in Fig. 5-13b. Here we have not arranged the polarization controller at the reflection to separate the two eigen-polarizations. Instead we let both polarizations go to both the photodetectors. Now, in red we show the difference between the two reflected photodiodes, and as we can see, the two modes have a different sign in their error signal. This is a consequence of the delay between the two polarization modes.

Next, we attempt to use this birefringence to auto-lock the cavity. The idea would be to tune the power and detuning of the two eigen-polarizations such that one provides an optical spring (OS), and the other provides the damping [44]. The short length of our cavity makes the cavity linewidth large. Given the frequency separation of the two eigen-polarizations, our calculations showed that an auto-locking scheme using birefringence will not be possible for a short cavity like ours. This scheme has been demonstrated to work for a 1 cm long cavity [55].

## 5.6 Radiation pressure locking

We lock our cavity using the OS effect. We park on the blue-detuned side of the cavity fringe, which creates an OS with a positive spring constant but a negative damping constant [20]. This implies the inherent dynamics of this cavity are unstable (Chapter 2, Ref. [66]). We stabilize the system by deploying an external feedback, which adds a phase delay to stabilize the unstable dynamics. This feedback is actuated by using radiation pressure, i.e. modulating the power going into the cavity to change the radiation pressure force on the cantilever mirror. The details of the locking scheme are presented in Appendix B and Ref. [38]. The measured closed loop transfer functions of the stabilized OS lock for the 100 µm cavity at MIT are shown in Fig. 5-14 for two different OSs.

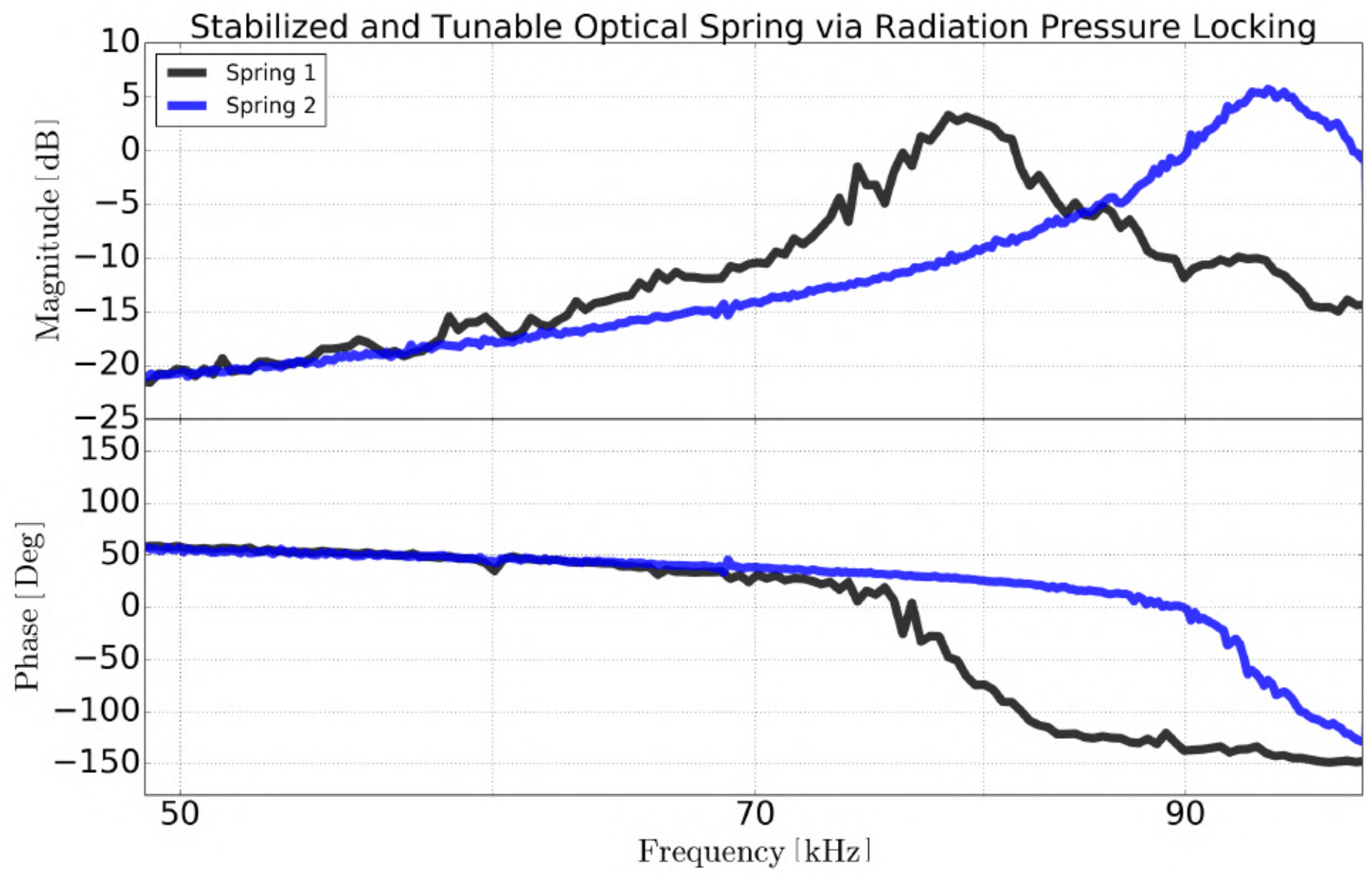

**Figure 5-14:** Locking transfer function: Closed loop transfer function of the radiation pressure lock for the fiber-mirror cavity taken at two different OSs.

## 5.7 Length Spectrum

We used the piezo we have on our fiber to calibrate the cavity's noise spectrum to displacement. We know from our measurements of the free spectral range, and from the manufacturer's specifications that the piezo has a meters to volts conversion of 532 nm/40 V. This could be off by perhaps 10%. We started by taking the transfer function from the voltage on the piezo (which is not actually in our loop) to the transmission PD. We only took the transfer function out to 20 kHz, because we did not want to get close to the piezo resonance at 58 kHz which would mess with our calibration. This transfer function is show in Fig. 5-15a.

Having taken the above transfer function, we then took the spectrum of the voltage on the transmission PD. This is shown in Fig. 5-16a. Taking this spectrum and dividing by the transfer function (and applying the conversion from volts to meters) gives a calibrated spectrum from 1 kHz to 20 kHz. From here, we took a closed loop transfer function from 15 kHz to 100 kHz, shown in Fig. 5-15b.

We then took the transmission PD voltage spectrum divided by the transfer function, and used the overlap from 15 to 20 kHz to calibrate this spectrum to meters. The complete calibrated displacement spectrum of our cantilever is shown in Fig. 5-16b. We put the direct PZT calibrated spectrum in green, while the indirectly calibrated spectrum computed from the closed loop TF is in blue.

Comparing it to the thermally limited spectrum observed in a previous iteration, we are about a factor of 5 above this measurement. We can certainly see what we think is sensor noise above about 50 kHz. We also have not optimized the position of the beam on the cantilever, so we may be able to reduce the apparent motion by positioning the beam on the nodes of some of the modes seen in the spectrum.

### 5.7.1 Calibrated spectrum free of detector noise and shot noise

The spectrum shown in Fig. 5-16b is limited by detector noise on the transmission PD. In order to surpass the PD-noise-limited sensitivity, we installed a second transmission PD and 50:50 beamsplitter, so that the transmitted power is equally split on the two PDs. Averaging the cross-spectrum between the two

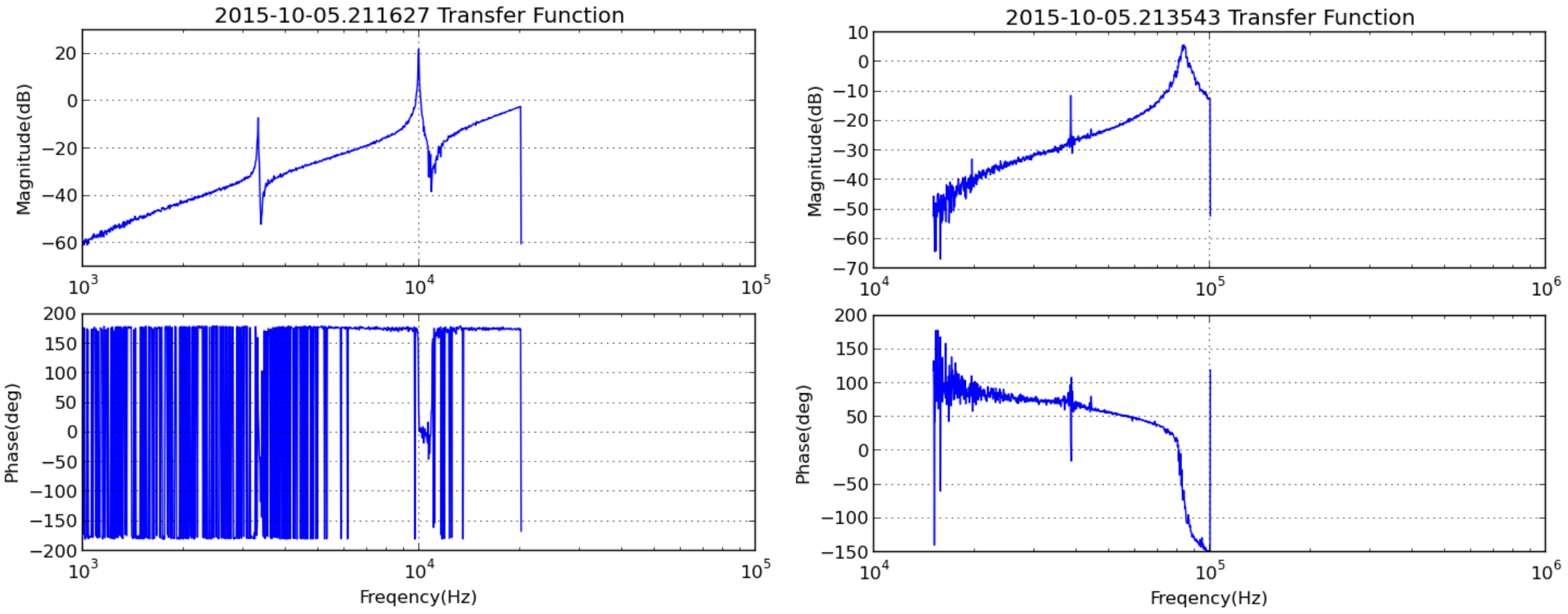

**(a)** **(b)**

**Figure 5-15:** Transfer functions used for calibration of cavity noise spectrum: **(a)** Transfer function of a drive on the PZT to the transmission photodiode. **(b)** Closed loop transfer function of the cavity locking loop taken for high-frequency calibration of the length spectrum (Fig. 5-16b). See text for details.

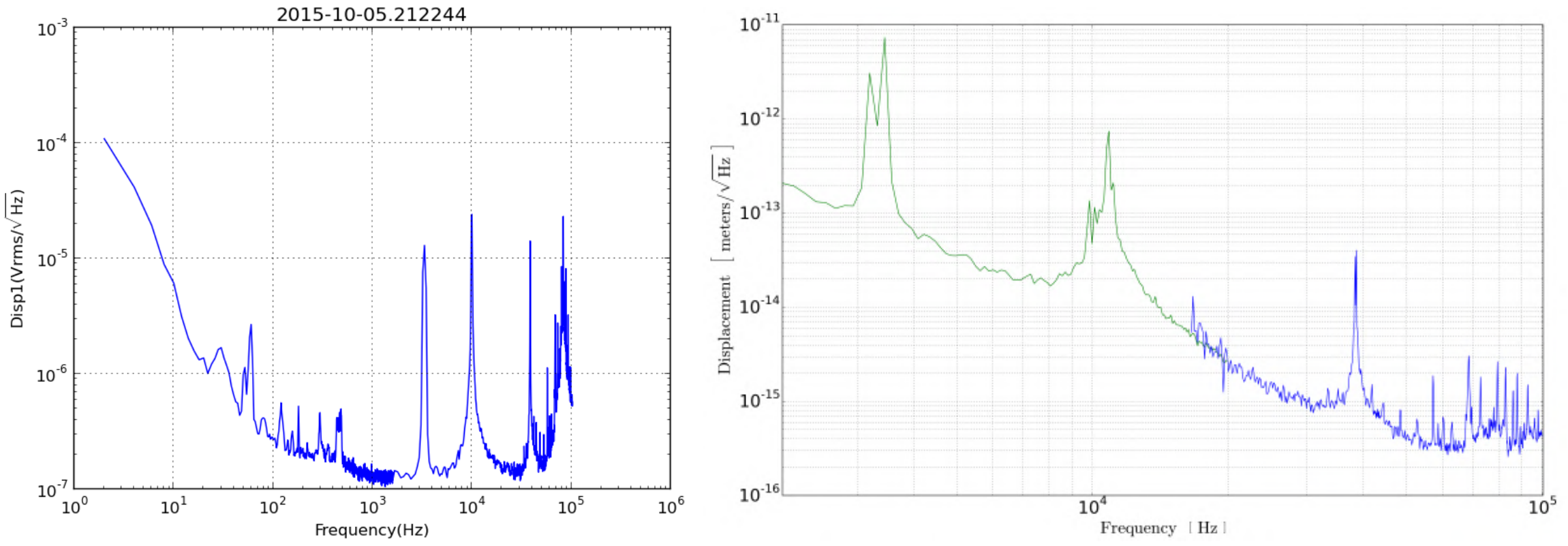

**(a)** **(b)**

**Figure 5-16:** (a) Raw spectrum of the fiber-mirror-cantilever-mirror cavity, as measured on the transmission PD. (b) Calibrated displacement spectrum in meters of the fiber-mirror-cantilever-mirror cavity, obtained by using a and Fig. 5-15. See text for details on the calibration procedure.

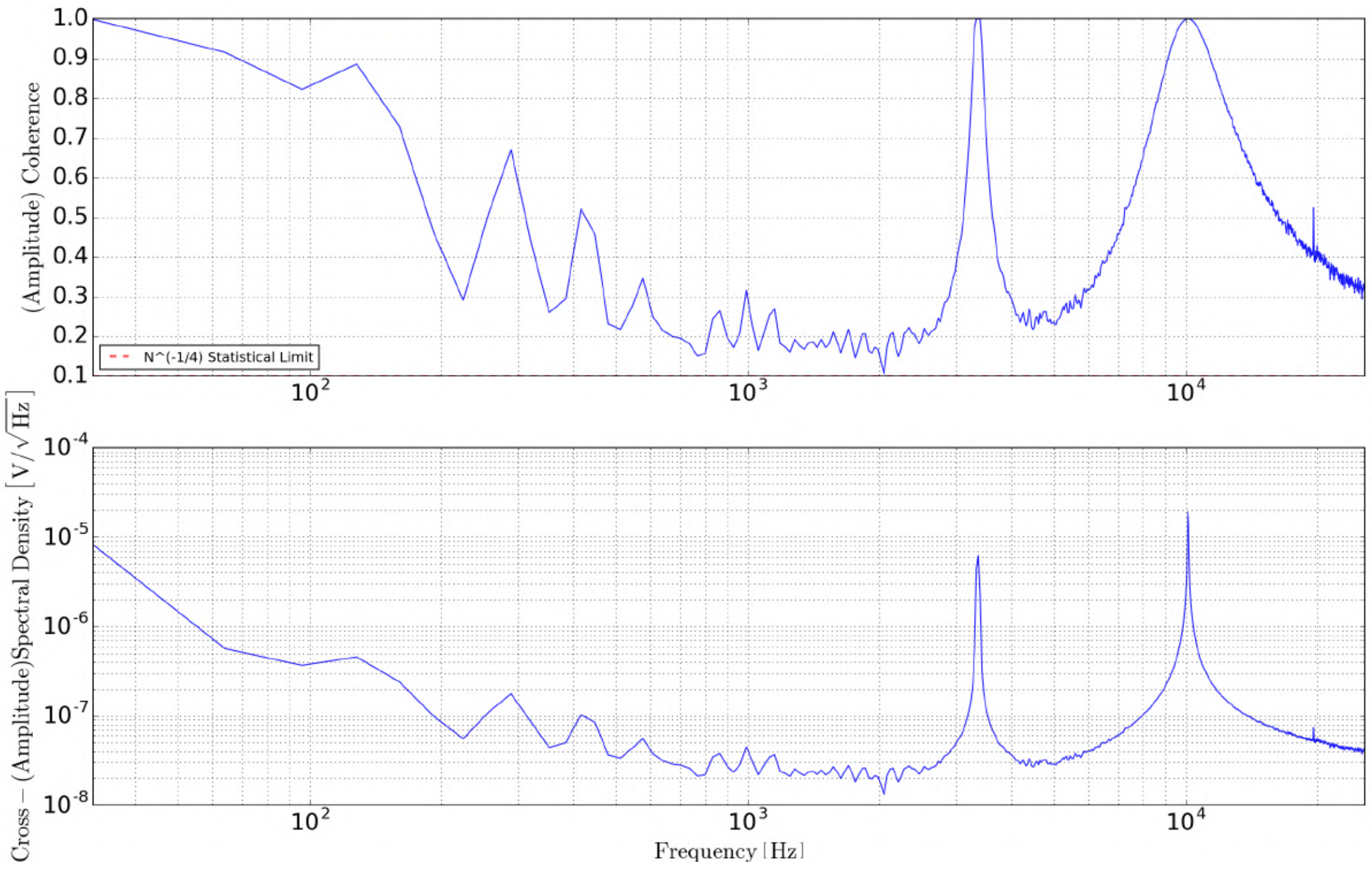

**Figure 5-17:** Coherence and cross spectral density of two photodetectors in transmission, used to see noise below shot noise and detector noise

Transmission PDs then integrates away the uncorrelated PD noises, revealing the underlying displacement spectrum and/or any other correlated noises lurking below (e.g. scattered light, if there is any).

Fig. 5-17 shows the coherence and cross-spectrum plots, showing that a sufficient number of averages have been measured to resolve all underlying correlated noises. The red dashed line at the bottom of the coherence plot is the mean value of the statistical sampling distribution for the coherence in the case when the two signals are uncorrelated Gaussian noise, and its value simply scales as $N^{-1/4}$ for N averages. For power coherence (i.e. the coherence between power spectral densities with units of, e.g., $V^2/Hz$) the red line scales as $N^-1/2$.

Fig. 5-18 shows the calibrated cantilever displacement cross-spectrum and the two auto-spectra of the transmission PDs (labeled TransT and TransR). The cross-spectrum noise is no more dominated by shot noise and detector noise.

### 5.7.2 Investigate contribution of scattered light

Next we use the same correlation measurement technique to determine the role of scattering in our system. There were two main motivations for performing this test. Firstly, in our transfer functions from the AM to transmission, we see scattered light take over the cavity TF at low frequencies. Secondly, we have had some difficulties fitting our measured spectrum to a reasonable thermal noise model which lead us to think we might be dominated by scattering in our spectrum (once we take out the dark noise by cross-correlation).

To investigate this, we took the cross-correlated spectra in transmission as shown in Figs. 5-17 and 5-18 while the cavity was unlocked and locked. The idea here is that we should see the noise from the scattered light both when the cavity is unlocked as well as locked. We don't know by how much would this scattered light change when we lock, but our zeroth order assumption is that that it stays the same between locked and unlocked. In order to measure the unlocked spectrum we turned the loops off and moved far away from resonance. This had to be far enough from the fringe so as to not have any fleeting resonances build up in the cavity. We ensured this by minimizing the FFT trace on the SR785 as we scanned the PZT.

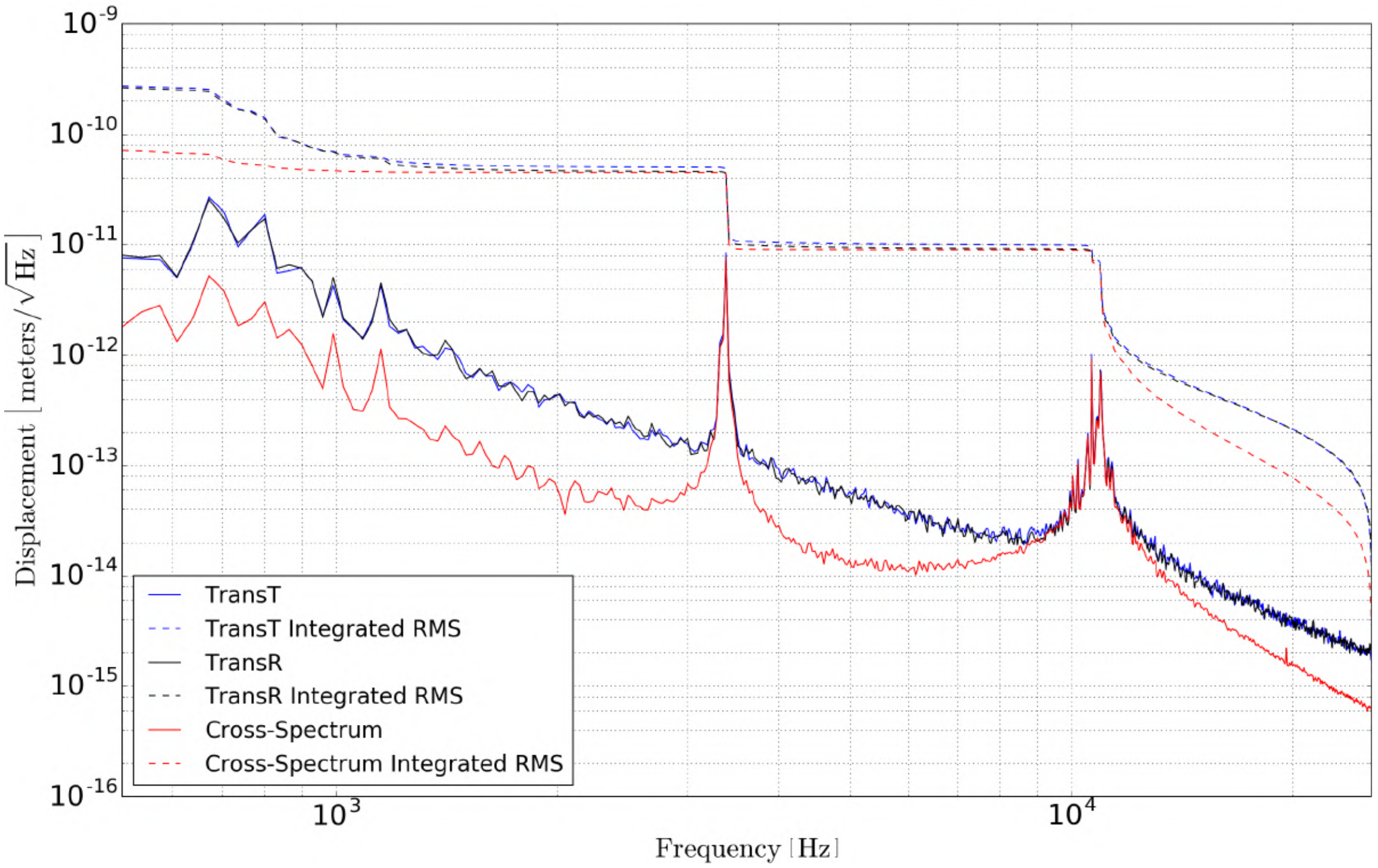

**Figure 5-18:** Cross spectra and auto spectra calibrated to cavity length

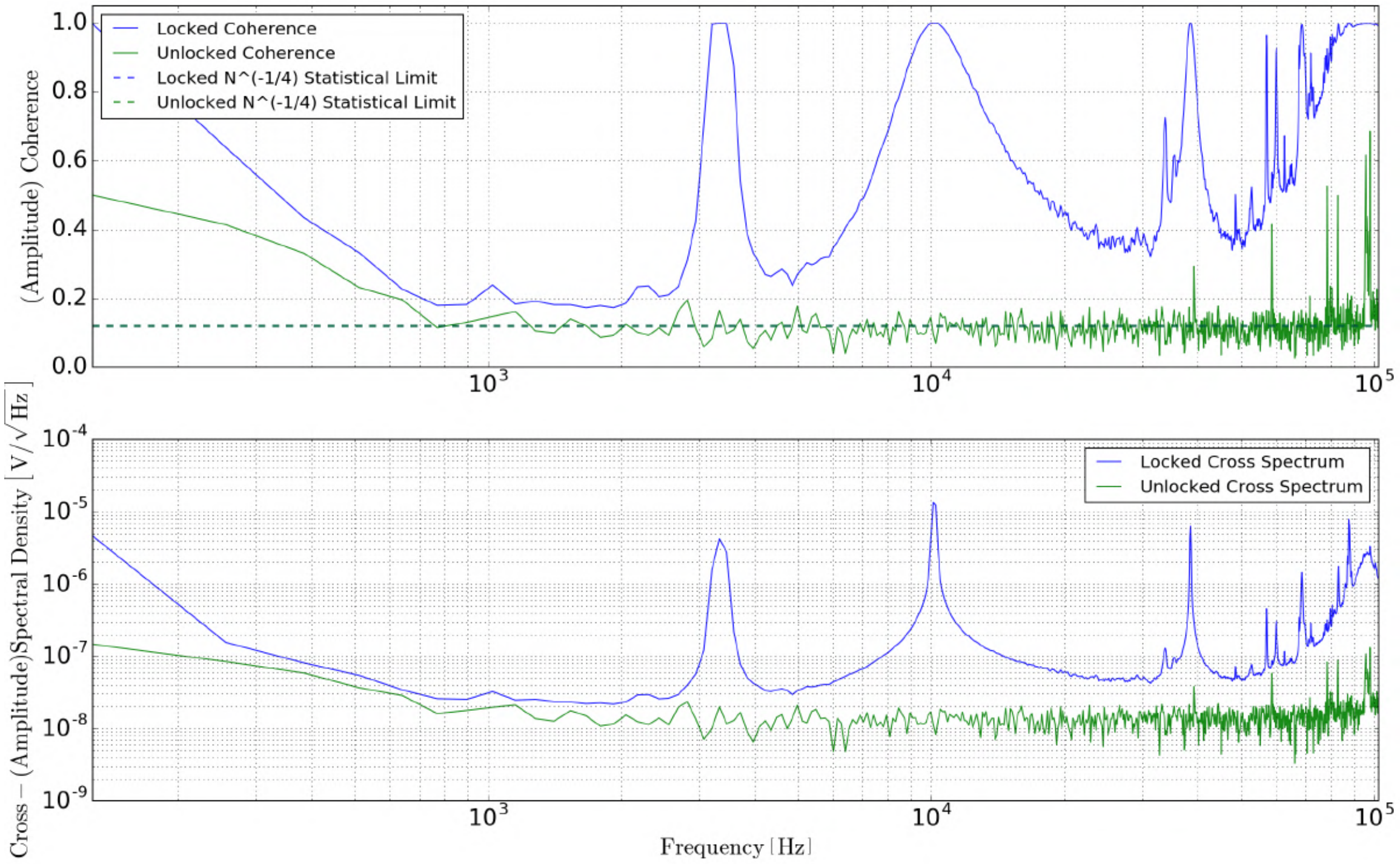

**Figure 5-19:** Cross-spectrum measurement to look for scattered light

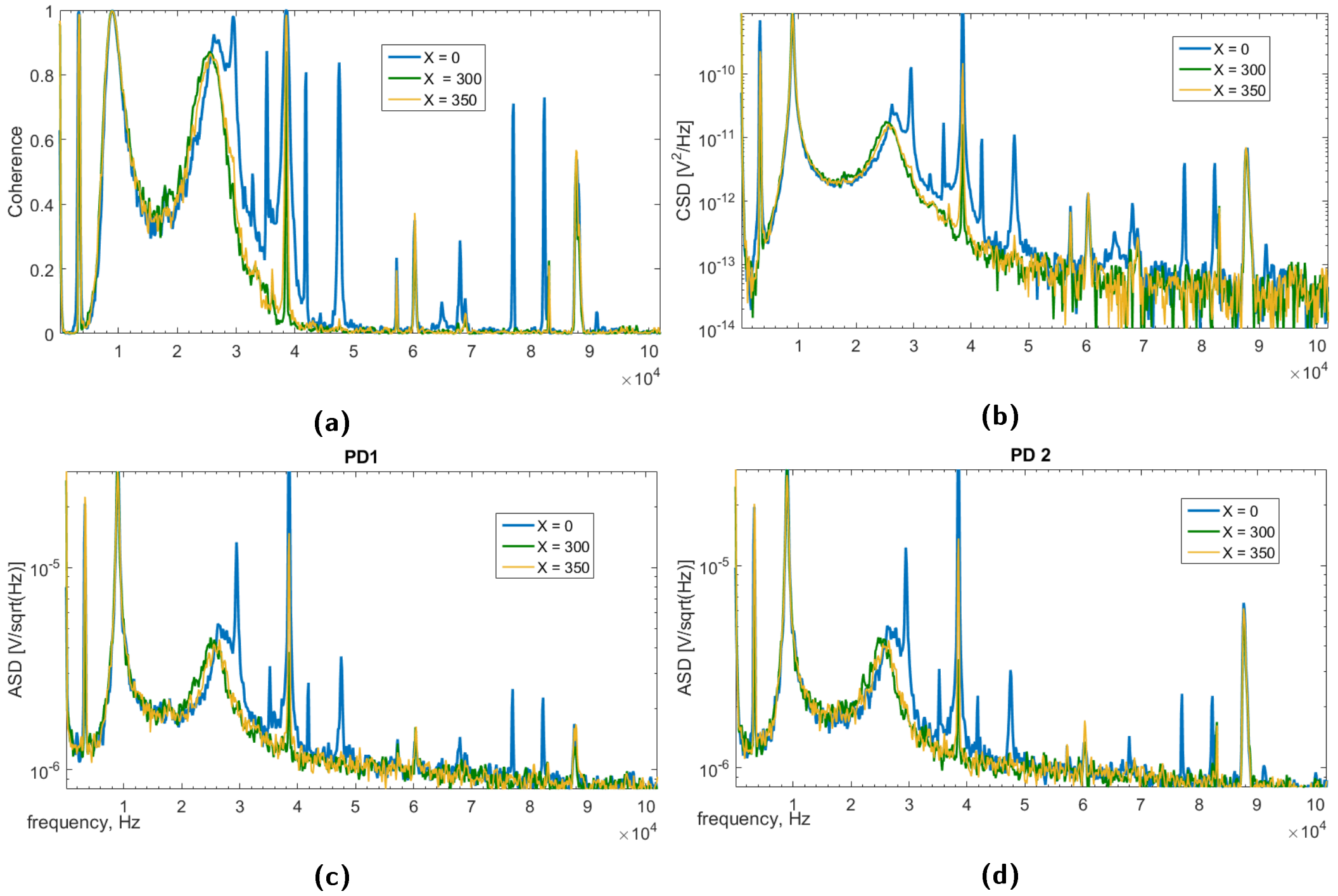

**Figure 5-20:** Cavity alignment using transmission spectra: a comparison of SR785 transmission spectrum measurement at X=0,300 and 350. Most of the change happened after 20 kHz, so it is plotted linear in frequency. (a) shows the coherence for the cross spectrum measurement. (b) shows the cross spectrum in power spectrum units. (c) shows the amplitude spectrum of first PD, and (d) shows the amplitude spectrum of second PD. Green and yellow (x=300,350) have lower noise than blue (x=0), some peaks became thinner and some even went under the noise floor.

The measurement is shown in Fig. 5-19. The coherence for both the spectra is above the statistical limit for 5000 averages. Since we used the same number of averages for both the measurements, this limit is same for them. In the cross spectral densities, the unlocked(green) trace is well below the locked(blue) trace. The green trace is an upper limit on the contribution of scattered light because it was the minimum the SR785 would give at these many averages. Some quick numbers tell us that scattered light amplitude should not contribute more than 10% to the total noise in frequencies above 30 kHz. In frequencies less than that, the upper limit on scattering contribution is 25%. This means that our limiting noise source is not scattered light.

### 5.7.3 Cavity alignment

After we re-attained a somewhat stable lock, we try to adjust the alignment of the cavity. To do this, we observe the transmission spectrum shown in Fig. 5-16a as we move the chip in a given direction, and minimize the peak heights in the spectrum. While doing this, we made sure to bring the OS the same every time we re-locked. We did this at about 25 kHz OS. We used the picomotor software to track the movements. As an example, we started with moving in positive x-direction. We found that some of the peaks started getting lower and lower. We saw the lowest at about x=300 to 350 picomotor steps, and then the peaks started coming back up after that.

Fig. 5-20 shows the individual spectra, cross-spectrum, and coherence of the two transmission PDs as a function of the x-position of the cantilever mirror. We see that the coupling to certain mechanical modes can be reduced by proper centering of the laser beam on the cantilever.

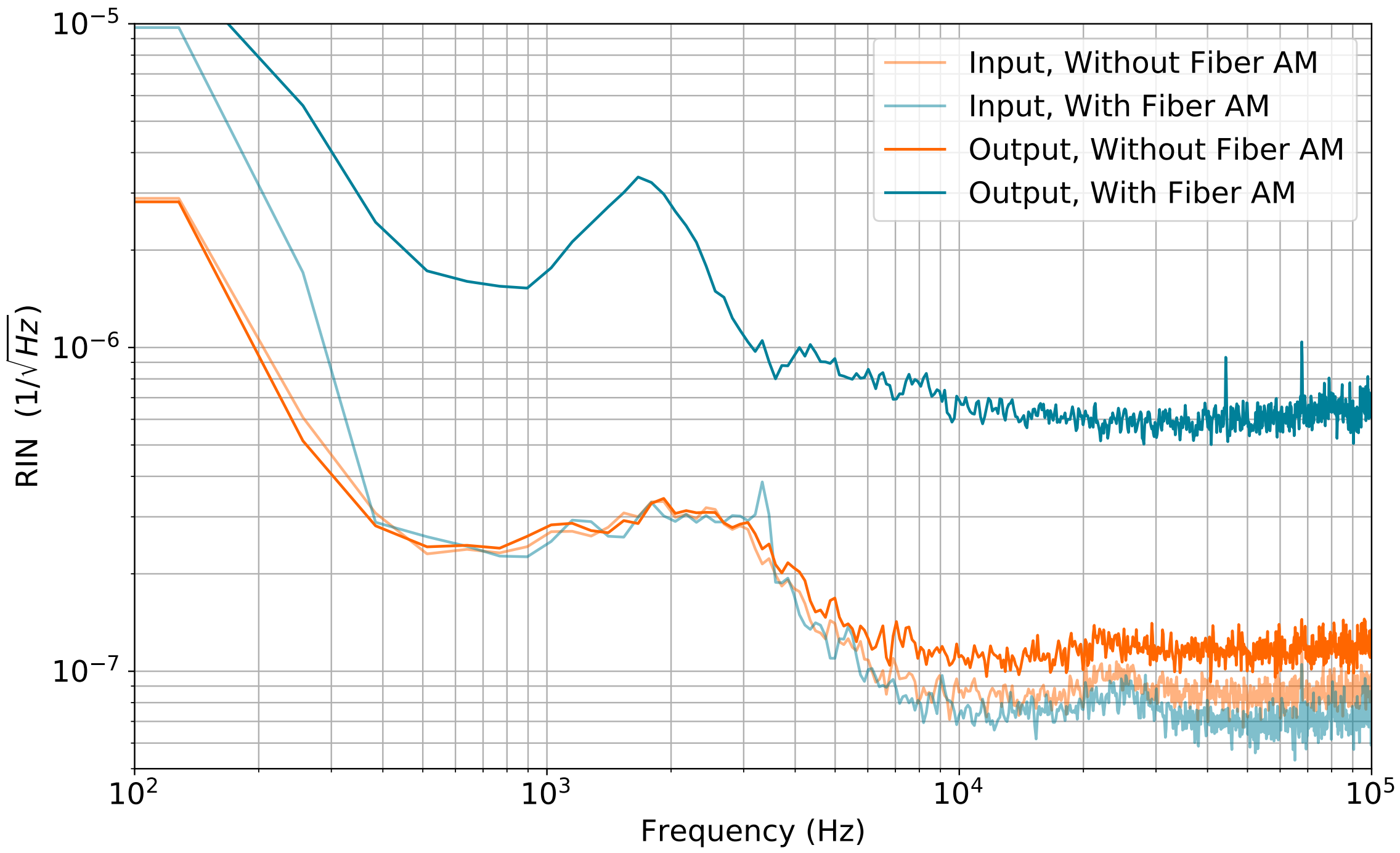

**Figure 5-21:** Intensity noise added by the in-fiber amplitude modulator

## 5.8 Future improvements

This experimental implementation was aimed at building a compact, fiber-coupled optomechanical squeezer. There were several shortcomings that can be improved in the future.

- **Fiber amplitude modulator** First among those was the intensity noise added by the in-fiber amplitude modulator. This is shown in Fig. 5-21. The fiber AM is based on a Mach-Zender interferometer, which is prone to converting all length and phase noise into amplitude modulation. It is also plausible this noise is coming from a fluctuating DC bias on the modulator. Both these problems can be addressed, by either controlling the environment in which the fiber AM is kept, or also wrapping a slow loop around the DC bias. One consideration is to move the ISS pickoff to after the fiber AM. In principle, this is okay since the ISS loop is mostly low-frequency feedback, while the cavity locking loop is a high-frequency feedback. But, the fiber AM is only capable of taking input power of 50 mW, which resulted in a limitation of power available downstream, for the ISS and local oscillator. In the future, a different manufacturer's fiber AM could be used that does not have this low power requirement.

- **Fiber mirror** The fiber mirrors themselves proved to have more scattering loss and mode-matching inefficiency than anticipated. This made harder to form stable cavities with these mirrors. In the future, new fiber mirrors can be tried to overcome this problem.

- **Imaging** Additionally, the side imaging is a crucial part of operation for a micro-cavity, so it must be improved to be easily readable. Doing this requires better illumination. While a copper holder is great for cooling, at least the room temperature iterations of this experiment can afford to have a transparent holder that allows for more light. We tried adding an LED inside the vacuum chamber for added illumination, but the combination of the current fiber mount and the gimball mount leave no solid angle for the light to enter, illuminate the cavity, and reach the camera. Achieving this requires a more open design.

- **PZT design** One possibility could be to switch to using a shear piezo instead of an annular piezo. This is a single solution to two problems. The fiber mirror would be glued on top of the shear pzt, so a one-eight part of the sphere is now open for illumination. Additionally, a shear pzt has a much higher unloaded mechanical resonance than an annular pzt, so could allow us to reach higher pzt resonances, to allow for using the pzt to lock the cavity or use the pzt to calibrate the full spectrum.

- **Omit the ferrule** Currently the fiber mirror is secured in a ferrule. This allows for somewhat easier handling of the fibers during ablation and coating. But threading the bare fiber into the ferrule requires stripping the protective glass coating. The lack of coating, along with the clamped nature of the millimeter-length bare-fiber piece makes the entire assembly extremely fragile, leading to a lot of breakage. Eliminating the ferrule would also eliminate the scattering problem accentuated by the reflective coating on the face of the ferrule, which gets deposited during the process of coating the fiber mirrors.
- **Monolithic design** Finally, a truly compact design can be achieved by building a monolithic system where each cantilever mirror is paired with a fiber mirror on a monolithic platform, making them much less susceptible to environmental noise.

Due to these reasons, we have switched to a traditional 1 cm cavity for the near future. The measurements shown in Chapters 6 and 7 have been taken at Louisiana State University in the lab of our collaborator, Prof. Thomas Corbitt, on such a 1 cm cavity [67], containing the same micro-cantilever devices, but a traditional macro-mirror as the input coupler.

# Chapter 6

# Quantum radiation pressure measurement

## Contents

This chapter is based on Ref. [36], and much of that manuscript is reproduced here verbatim.

## Abstract

Quantum mechanics places a fundamental limit on the precision of continuous measurements. The Heisenberg uncertainty principle dictates that as the precision of a measurement of an observable (e.g. position) increases, back action creates increased uncertainty in the conjugate variable (e.g. momentum). In interferometric gravitational-wave (GW) detectors, the laser power is increased as much as possible to reduce the position uncertainty created by shot noise, but necessarily at the expense of back action in the form of quantum radiation pressure noise (QRPN) [17]. Once at design sensitivity, Advanced LIGO [1], VIRGO [2], and KAGRA [3] will be limited by QRPN at frequencies between 10 Hz and 100 Hz. To improve the sensitivity of GW detectors, ideas have been proposed to mitigate the QRPN [68, 69, 19, 70, 71, 72], but until now there has been no platform to experimentally test these ideas. Here we present a broadband measurement of QRPN at room temperature at frequencies relevant to GW detectors. The measured noise spectrum shows effects from the QRPN between about 2 kHz to 100 kHz, and the measured magnitude of QRPN agrees with our model. We now have a testbed for studying techniques to mitigate quantum back action, such as variational readout and squeezed light injection [19], that could be used to improve the sensitivity of GW detectors.

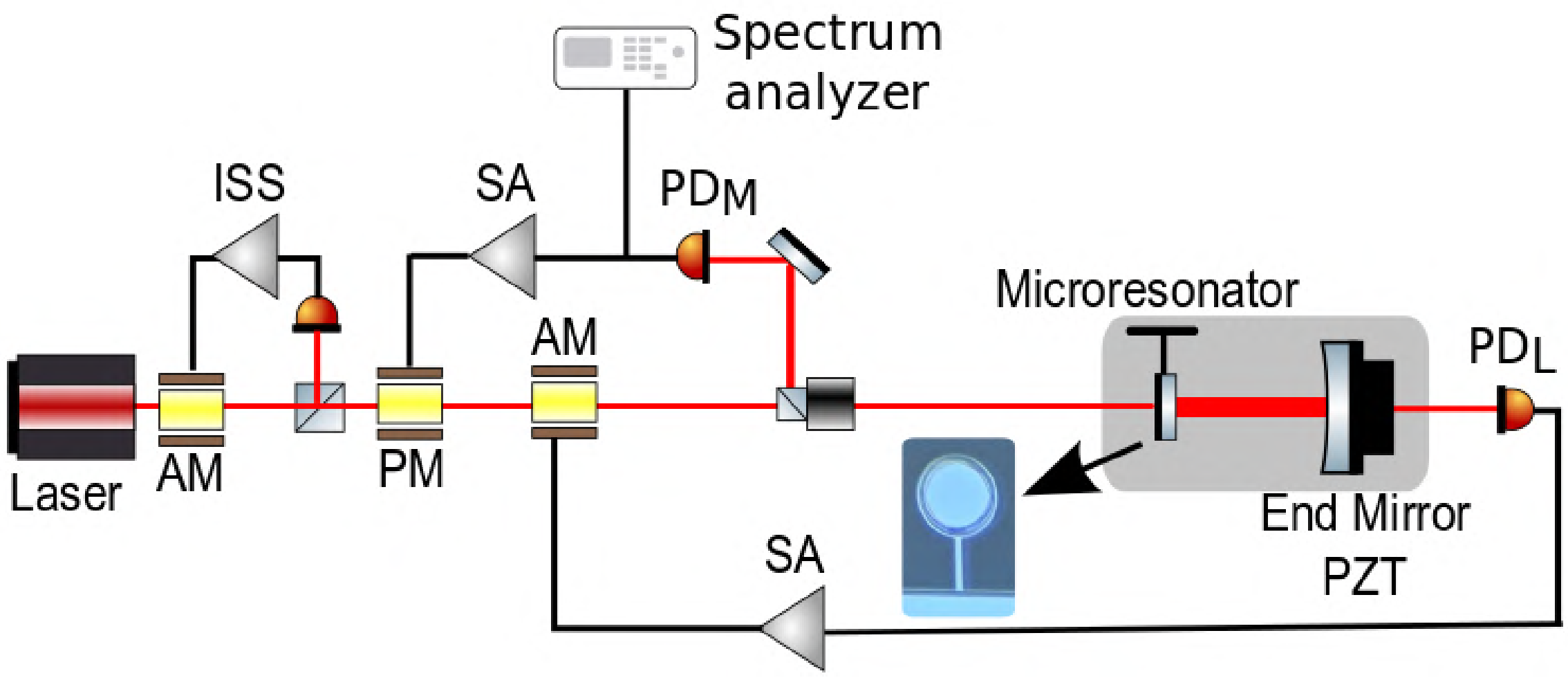

**Figure 6-1:** Experimental Setup. Light from a 1064 nm Nd:YAG laser is passed through an amplitude modulator (AM), phase modulator (PM), and second AM before being injected into the optomechanical cavity, which sits on a suspended optical breadboard to reduce seismic motion and is housed in a vacuum chamber at $10^{-7}$ Torr (shown in shaded gray). A micrograph of the single-crystal microresonator, comprising a 70-$\mu$m diameter GaAs/AlGaAs mirror pad supported by a GaAs cantilever, is included in the diagram. An intensity stabilization servo (ISS) is used to stabilize the laser power to shot noise by feeding back to the AM. The light transmitted through the optomechanical cavity is detected by photodetector $PD_L$. The signal from $PD_L$ is sent through a servo amplifier (SA) before being sent to the second AM to initiate the cavity lock sequence. The beam reflected by the cavity is detected by photodetector $PD_M$. The signal from $PD_M$ is used to lock the cavity by sending it through a separate SA feeding back to the PM. The $PD_L$ locking loop is turned off after the $PD_M$ locking loop is active. The signal from $PD_M$ is also sent to a spectrum analyzer for further analysis.

## 6.1 QRPN: Motivations and challenges

Gravitational wave detectors such as Advanced LIGO continuously monitor the position of test masses using electromagnetic radiation. The Heisenberg uncertainty principle limits the precision of such a continuous measurement due to the quantization of light. Uncertainty in the number of photons reflecting from a mirror exerts a fluctuating force due to radiation pressure on the mirror, causing mechanical motion[17, 73, 18]. This force leads to a noise source for GW measurements, called quantum radiation pressure noise (QRPN). GW interferometers typically use as much laser power as possible in order to minimize the shot noise and maximize the signal-to-noise ratio for GWs. Advanced LIGO and other second and third generation interferometers will be limited by QRPN at low frequency when running at their full laser power.

Given the imperative for more sensitive GW detectors, it is important to study the effects of QRPN in a system similar to Advanced LIGO, which will be limited by QRPN across a wide range of frequencies far from the mechanical resonance frequency of the test mass suspension. Studying quantum mechanical motion is challenging, however, due to the fact that classical noise sources such as environmental vibrations and thermally driven fluctuations [35] usually dominate over quantum effects. Previous observations of QRPN have observed such subtle quantum effects, even at room temperature, but these experiments have thus far been limited to high frequencies (MHz-GHz) and in a narrow band around a mechanical resonance [74, 75, 76, 77].

## 6.2 Experimental setup

In this work, we present a broadband and off-resonance measurement of QRPN in the audio frequency band. We have developed low-loss single-crystal microresonators with sufficiently minimized thermal noise such that the quantum effects can be observed at room temperature. The optomechanical system, shown in detail in Figure 6-1, is a Fabry-Pérot cavity with a mechanical oscillator as one of the cavity mirrors. The optomechanical cavity is just under 1 cm long and consists of a high-reflectivity single-crystal microresonator that serves as the input coupler and a macroscopic mirror with a 1 cm radius of curvature as the back reflector. The cavity is made slightly shorter than the 1 cm radius of curvature of the large mirror in order to achieve a small spot size on the microresonator while maintaining stable cavity modes. The microresonator consists of a roughly 70 $\mu$m diameter mirror pad suspended from a single-crystal GaAs cantilever with a thickness of 220 nm, width of 8 $\mu$m, and length of 55 $\mu$m. The mirror pad is made up of 23 pairs of quarter-wave optical thickness GaAs/$Al_{0.92}Ga_{0.08}As$ layers for a transmission of 250 ppm and exhibits both low optical losses and a high mechanical quality factor [78, 65, 61, 62, 55]. The microresonator has a mass of 50 ng, a natural mechanical frequency of 876 Hz, and a measured mechanical quality factor of 16,000 at room temperature (295 K). The cavity has a finesse of 13,000 and linewidth (HWHM) of 580 kHz.

A 1064 nm Nd:YAG laser beam is used to both stabilize the optomechanical cavity and to measure the mechanical motion of the microresonator. The cavity is detuned from resonance by 0.3 to 0.6 linewidths, and locked using a feedback loop that utilizes the restoring force produced by a strong optical spring [38], which shifts the mechanical resonance of the microresonator up to 145 kHz at high power. We choose to detune the cavity primarily because it is nearly impossible to avoid a significant optical spring effect owing to the weak restoring force provided by the cantilever supporting the microresonator. We would need to keep the cavity locked to resonance within $2 \times 10^{-5}$ linewidths, or about 10 Hz to avoid having an optical spring as stiff as the cantilever, and any deviations around this point would produce strong variations in the optical spring stiffness. Instead, by intentionally detuning the cavity to near 0.5 linewidths, we operate near the peak optical spring stiffness, where the cavity is relatively insensitive to variations in detuning, as described in Section 6.4. The error signal for the feedback loop is detected using photodetector $PD_L$ in transmission of the cavity and photodetector $PD_M$ in reflection. The error signal is fed back to an amplitude and phase modulator as shown in Figure 6-1. The final measurement configuration uses only the reflected light because the transmitted light has relatively large shot noise due to the small transmission (50 ppm) of the end mirror, which may pollute the measurement. Reflection locking with the phase modulator is less robust, and we are not able to directly acquire lock without first using the transmission locking and amplitude modulator. We measure the displacement noise spectrum by detecting the light that is reflected from the cavity. After the cavity is locked, the signal from $PD_M$ is sent to a spectrum analyzer for analysis. We measure an uncalibrated noise spectrum by first measuring the amplitude spectral density of $PD_M$'s output. We calibrate the spectrum by dividing it by the transfer function from the laser-cavity piezo to $PD_M$. This method treats the optical spring as a feedback loop as described in [38], and by factoring out its effect, we restore the observed displacement spectrum to what it would be in the absence of the optical spring and our electronic feedback. The laser piezo has been calibrated in frequency, which allows the resulting signal to be calibrated to displacement by using the cavity length.

## 6.3 Observation

### 6.3.1 Thermal Noise

In order to understand the resulting measurement of the microresonator motion, we must carefully account for various noise sources. Specifically, we consider QRPN, thermal noise, shot noise, dark noise of the photodiode readout, and classical intensity and frequency fluctuations of the laser. Thermal noise, governed by the fluctuation dissipation theorem, sets a limit on the precision of force and displacement measurements [79], and is also one of the main limitations in this experiment. We rely on direct measurements of thermal noise to quantify its effects. To measure thermal noise, we operate the cavity with about 10 mW of circulating power, a level at which the QRPN is small compared to the Brownian motion of the microresonator. One challenge in accurately accounting for the thermal noise is that as the circulating power in the cavity is increased, the beam position on the microresonator slightly shifts, and the coupling of the pitch and yaw modes of the microresonator changes. To account for this, we measure the thermal noise at different alignments with 10 mW of circulating power to match the desired alignment at higher power, and ultimately use these measurements to constrain a model. The measured thermal data is used at most frequencies, except those near the pitch, yaw, and side-to-side resonances, as described in Section 6.4. The observed thermal noise agrees with a structural damping model [35] from 200 Hz to 30 kHz. Modeled thermal noise is used near the resonances because it is difficult to reproduce the exact alignment at low and high powers. Structural damping models contain a frequency independent loss angle, $\phi$, and for a harmonic oscillator have a displacement amplitude spectral density of

$$\tilde{x}_{\rm th}(\omega) = \sqrt{\frac{4 k_B T \omega_m^2}{\omega m Q[(\omega_m^2 - \omega^2)^2 + \frac{\omega_m^4}{Q^2}]}} \tag{6.1}$$

where $k_B$ is the Boltzmann constant, $T$ is temperature, $m$ is mass, $Q = 1/\phi$ is the quality factor, $\omega = 2\pi \times f$, and $\omega_m$ is the angular frequency of the mechanical mode [35]. Above 30 kHz, we observe thermal noise that deviates from the structural damping model, which appears to be consistent with thermoelastic damping of the drumhead mode of the microresonator. In the noise budget, we use the measured thermal noise at these frequencies. The resulting thermal noise, which is used in our noise budget, is shown in Figure 6-2.

### 6.3.2 Quantum Noise

Quantum noise is the other dominant noise source in the experiment, and we use an input-output model for comparison to our measurements. The model calculates quantum noise using a set of equations that relate the output fields to the input fields [20, 46]. The model requires knowledge of the optical losses, detuning, and the power circulating in the cavity, in addition to the microresonator's mechanical susceptibility. The cavity losses, detuning and circulating power are constrained by measurements of the optical spring. The parameters for the microresonator are constrained by the thermal noise measurement. The details of these measurements are presented in Section 6.4, along with analysis of the effects of uncertainty in these parameters. The model then predicts the level of QRPN, as shown in Figure 6-2 for 220 mW of circulating power. To further verify the model of QRPN, we also measure the response of our system to intensity fluctuations of the input laser beam, scale that measurement with the level of shot noise for the input power, and calibrate the resulting projected noise level. This results in an independent measurement for the level of QRPN that agrees with the modeled result. The details of the measurement are presented in Section 6.4.

### 6.3.3 Scaling of QRPN with Power

In order to observe quantum back action, we measure the cavity displacement noise at four cavity circulating powers of 73 mW, 110 mW, 150 mW, and 220 mW. For each power level, the input power is scaled by the same factor. Due to the shifting alignment resulting from static radiation pressure, the optical losses in the cavity change as a function of power. This effect is quantified through measurements of the optical spring. The detuning varies from about 0.6 to 0.35 linewidths when going from low to high power, but this only effects the level of QRPN by $\pm 4\%$, as explained in Section 6.4. The measured noise at 220 mW, shown as the orange curve in Figure 6-2, shows that the measured noise agrees with the sum of all known noise sources. QRPN contributes significantly to the total noise at most frequencies between 2 kHz and 100 kHz. Below 10 kHz, thermal noise is the biggest contributor to the displacement noise, but the effect of QRPN is still visible in the displacement noise measurement down to 2 kHz, where it accounts for about 20% of the measured displacement noise. The measured classical radiation pressure noise from classical intensity fluctuations of the laser and laser frequency noise [80] are below the other noise sources across the measurement band, as shown in Fig. 6-4.

To demonstrate that the observed QRPN scales with the expected square root of power [17], we compare the noise at each power level. The data is shown in Figure 6-3, where the displacement noise spectrum has been integrated over a 1 kHz band between 21 kHz and 22 kHz. The observed data is consistent with the predicted scaling, and the QRPN is the largest noise source for circulating powers above 150 mW. For the measurement at 220 mW shown in Figure 6-3, QRPN represents 48% of the total noise, while the thermal noise accounts for 27%, with the remaining 25% comprised of the sum of the subdominant noise sources. We sum in quadrature the contribution of each of the noise sources to compute the total expected noise. We find that our four displacement noise measurements, shown as orange stars in Figure 6-3, agree with the total expected noise (black curve) with the statistical measurement error taken into account. The measurement error is calculated by repeating the measurement multiple times and is dominated by the fluctuations in the transfer function measurement that is used to calibrate the spectrum. The dashed black curve shown in Figure 6-3 is the predicted displacement noise without a contribution from QRPN. The measurements of the displacement noise rule out the model without QRPN.

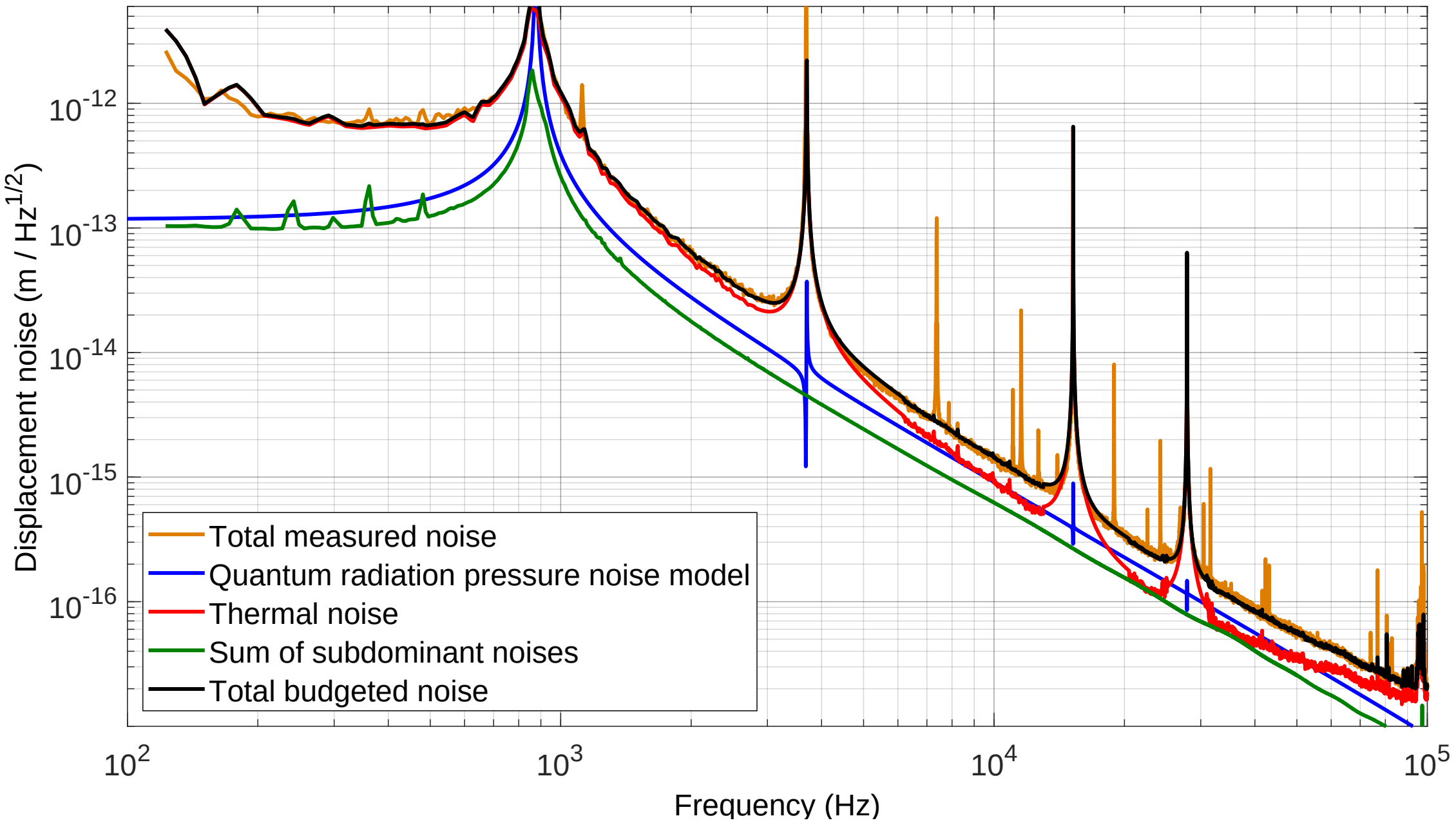

**Figure 6-2:** Measured and budgeted noise. Contributions of various noise sources to the displacement of the microresonator, shown as amplitude spectral densities, are shown. The orange curve is total noise measured with 220 mW of power circulating in the cavity. Also shown are spectra corresponding to QRPN (blue), thermal noise (red), and the quadrature sum of four subdominant noise sources (shot noise, dark noise, and laser intensity and frequency noises) (green). The black curve is the total noise that we can account for from these contributions. The resonances at 3.7 kHz, 15 kHz and 28 kHz are higher order mechanical modes of the microresonator. Each of these noise sources are discussed in detail in Section 6.4 and Fig. 6-4.

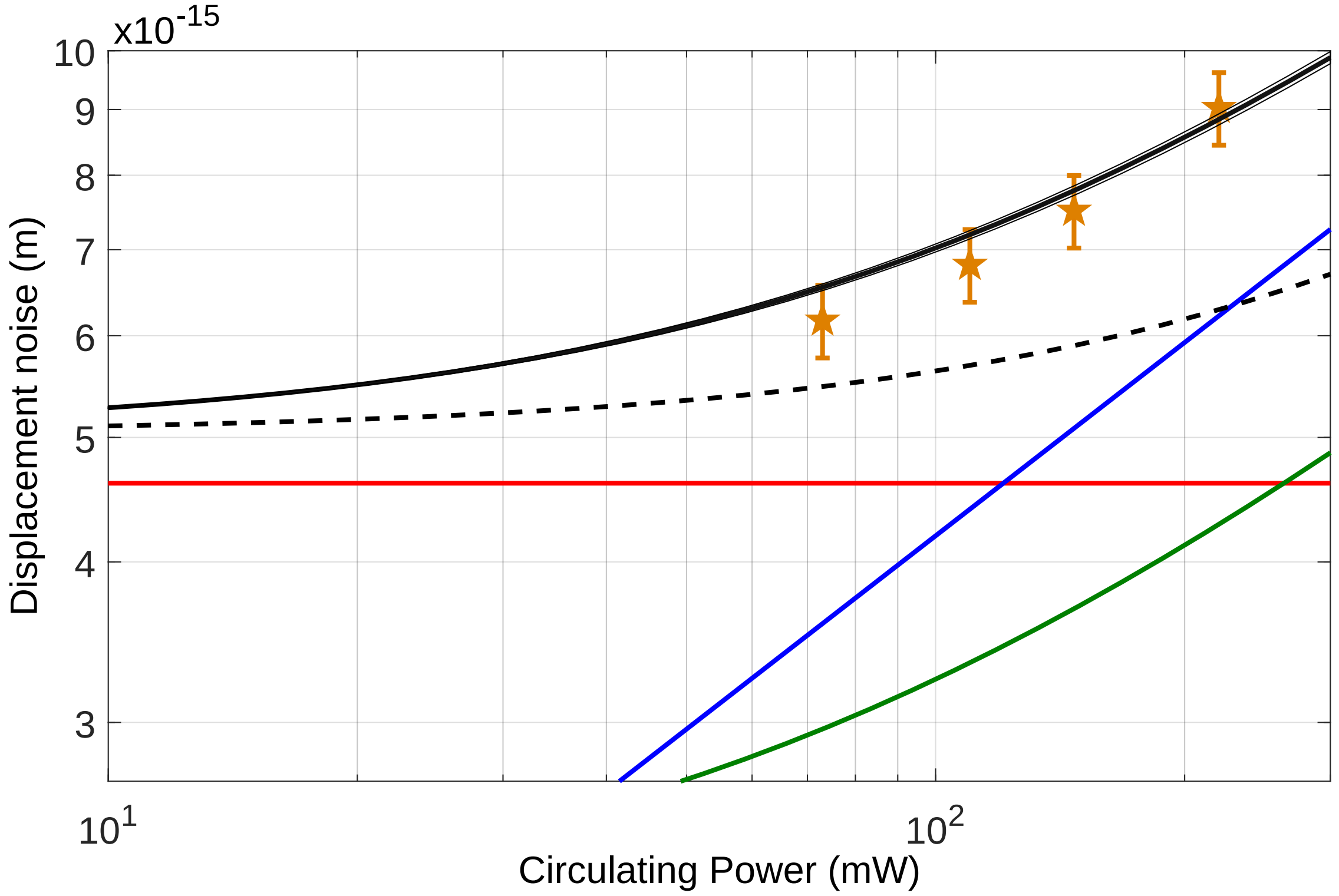

**Figure 6-3:** Power scaling. Four measurements at circulating powers of 73 mW, 110 mW, 150 mW, and 220 mW show how each of the noise sources scales with cavity circulating power. Each noise source shown in Figure 6-2 is integrated over a 1 kHz frequency band between 21 kHz and 22 kHz. The orange stars correspond to the displacement noise measured for the four power levels. The error bars on the measured data represent the measurement error based on the statistical uncertainty from multiple measurements. The blue line is the QRPN model, the red line is the measured thermal noise, and the green curve is the sum of the subdominant noise sources. The black curve is the expected total noise, which is calculated by adding the contributions of the QRPN, thermal noise, and subdominant noises, and is not a mathematical fit of the data points. The dashed black curve is the expected total noise without including the contribution from QRPN. The shaded gray region around the black curve represents the uncertainty in the total displacement noise due to the uncertainty in the level of QRPN from the model. More details on the uncertainty analysis are included in Section 6.4.

### 6.3.4 Additional Evidence of QRPN

In addition to showing that the noise scales correctly with optical power, a variety of other tests were performed to further verify that that we are observing QRPN. First, we put constraints on shot noise, dark noise, classical laser intensity and frequency noise, as seen in Fig. 6-4. Next, we fit the noise we attribute to QRPN to a power law in frequency, and require the resulting fit to match the measured data to within 10% in the 1 to 100 kHz frequency range. The resulting frequency scaling ($f^{-1.95\pm0.2}$) matches the expected frequency dependence ($f^{-2}$), and excludes the frequency dependence of thermal noise ($f^{-5/2}$). We also rule out that this could be an effect of excess bulk heating of the microresonator by verifying that the thermal noise at low frequencies remains the same within $\pm2\%$ measurement uncertainties (see Section 6.4). Using the optical spring measurement to constrain the cavity losses also allows us to rule out absorption photothermal effects [81] because any excess damping would be observed in mechanical response measurements. Finally, we perform a transfer function measurement by amplitude modulating the input light and measuring the response at $PD_M$. By comparing to a similar transfer function while the cavity is far from resonance, we show that the cavity coherently amplifies the amplitude modulations, indicating the presence of an optomechanical parametric process. Additionally, we can project the expected level of QRPN by multiplying this transfer function with the magnitude of the vacuum fluctuations that enter the cavity, and dividing by our calibration transfer function. We find that this projection matches the QRPN model.

## 6.4 Detailed methods

### 6.4.1 Noise budget

In addition to the QRPN and thermal noise shown in Figure 6-2, subdominant noise sources contribute to the measured displacement noise spectrum. A full noise budget for the 220 mW measurement is shown in Fig. 6-4. The largest of the subdominant noise sources is the shot noise and dark noise that is present on $PD_M$. Fig. 6-4 includes a measurement of the combined shot noise and dark noise. Factoring out the effect of the optical spring using the calibration discussed in the main text causes the white shot noise to have the frequency dependence shown in Fig. 6-4. Classical laser intensity noise and laser frequency noise lie below the other noise sources. Fig. 6-4 includes our measurement of the classical laser intensity noise, while the laser frequency noise level for the Nd:YAG ring laser (NPRO) is obtained from [80].

With all of the noise sources accounted for, we find that QRPN is the dominant noise source over a wide range of frequencies with 220 mW of light circulating in the cavity, as seen in Figs. 6-2 and 6-4. To quantify the effect of QRPN across our measurement band as a function of power, we provide a contour plot showing the ratio of QRPN to the total measured displacement noise in Fig. 6-5.

### 6.4.2 Thermal noise

As described in the main text, thermal noise sets a limit on the precision of mechanical experiments and can overwhelm attempts to measure quantum effects if it is too large. As one of the principal noise sources in this experiment, we must measure the thermal noise across our measurement band and account for it in our noise budget analysis. One difficulty in accounting for the thermal noise is that the cavity alignment shifts slightly as the circulating power is increased and the cantilever is deflected by radiation pressure. Even a small change in alignment can change the coupling of higher-order mechanical modes, specifically the yaw, pitch, and side-to-side modes, as shown in Fig. 6-6.

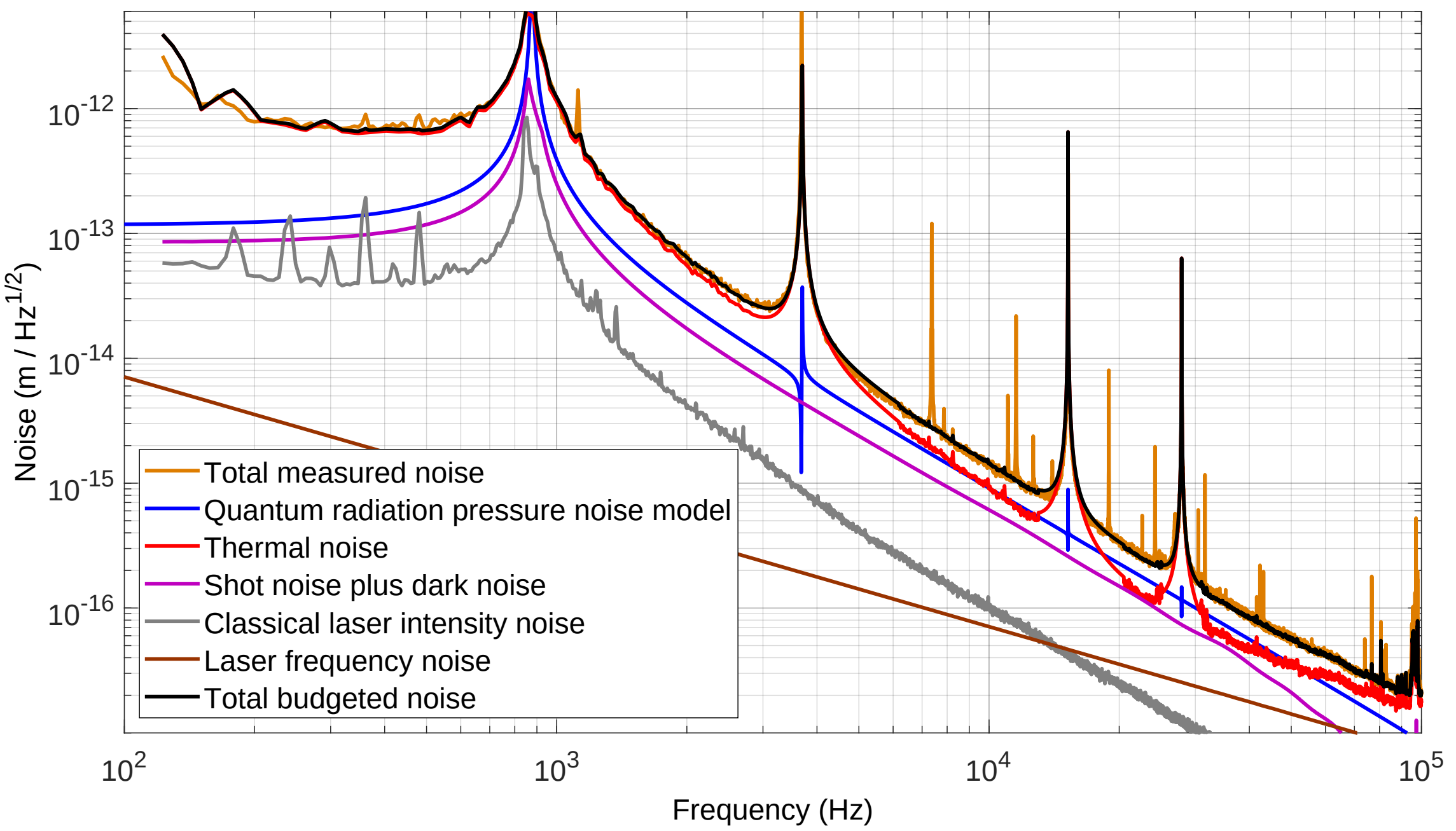

**Figure 6-4:** Full noise budget. For the 220 mW circulating power measurement, each noise source that contributed to the sum of subdominant noises in Figure 6-2 is shown. The narrow peaks in the displacement noise measurement are a result of parametric nonlinear coupling between various mechanical modes, and this coupling is negligible at low circulating powers.

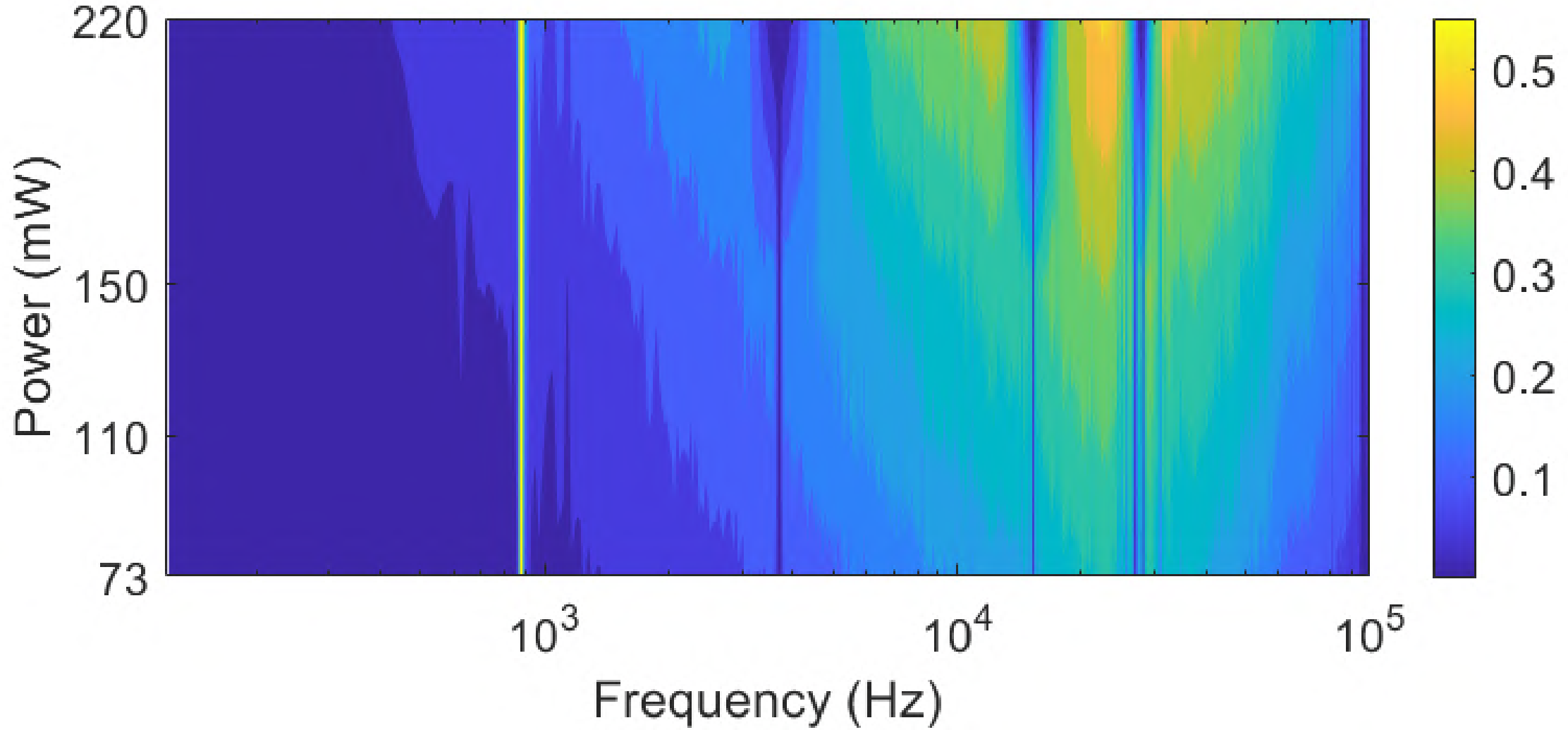

**Figure 6-5:** This contour plot shows what fraction of the total measured displacement noise (power spectral density) is contributed by QRPN (PSD), as a function of measurement frequency and circulating power. The color axis is the ratio of the power spectral densities of the QRPN model to the total measured noise. While in the rest of the chapter we present the data as amplitude spectral densities (in units of $\mathrm{m}/\sqrt{\mathrm{Hz}}$) in order to put it in perspective of GW measurements, we use PSDs to calculate percentage and ratios, and to make this plot because all the noises are added in quadrature to make up the total noise. We interpolate the data between the measurements at 73 mW, 110 mW, 150 mW, and 220 mW. The vertical stripe at 876 Hz is an artifact of the fundamental resonance not being perfectly resolved in the measurement. The blue vertical stripes at 3.7 kHz, 15 kHz, and 28 kHz are higher-order mechanical modes of the microresonator. The contours are at a spacing of 0.05 (5%).

| Parameter | Nominal value |
|---|---|
| Mechanical resonance frequency | $2\pi \times 876$ Hz |
| Mechanical quality factor | $1.6 \times 10^4$ |
| Mechanical damping rate | $2\pi \times 0.055$ Hz |
| Cavity decay rate (HWHM) | $\gamma = 2\pi \times 580$ kHz |
| Laser detuning from cavity resonance | $0.3 - 0.6\ \gamma$ |
| Optomechanical single-photon coupling strength | $2\pi \times 380$ kHz |
| Linearized light-enhanced optomechanical coupling | $2\pi \times 3.5$ MHz |
| Photon number circulating in cavity | $8 \times 10^7$ |
| Multiphoton cooperativity | $7.4 \times 10^8$ |
| Thermal phonon occupation | $7.4 \times 10^9$ |

**Table 6.1:** Standard optomechancial parameters: The table shows the measured parameters for our optomechanical system. The parameters in the first half of the table are used in our predictions of noise in the system. The second half of the table has the common optomechanical parameters for comparison with the current state-of-the-art optomechanical systems. As explained in the text, these parameters are used to characterize an on-resonance systems and are not used in our calculations.

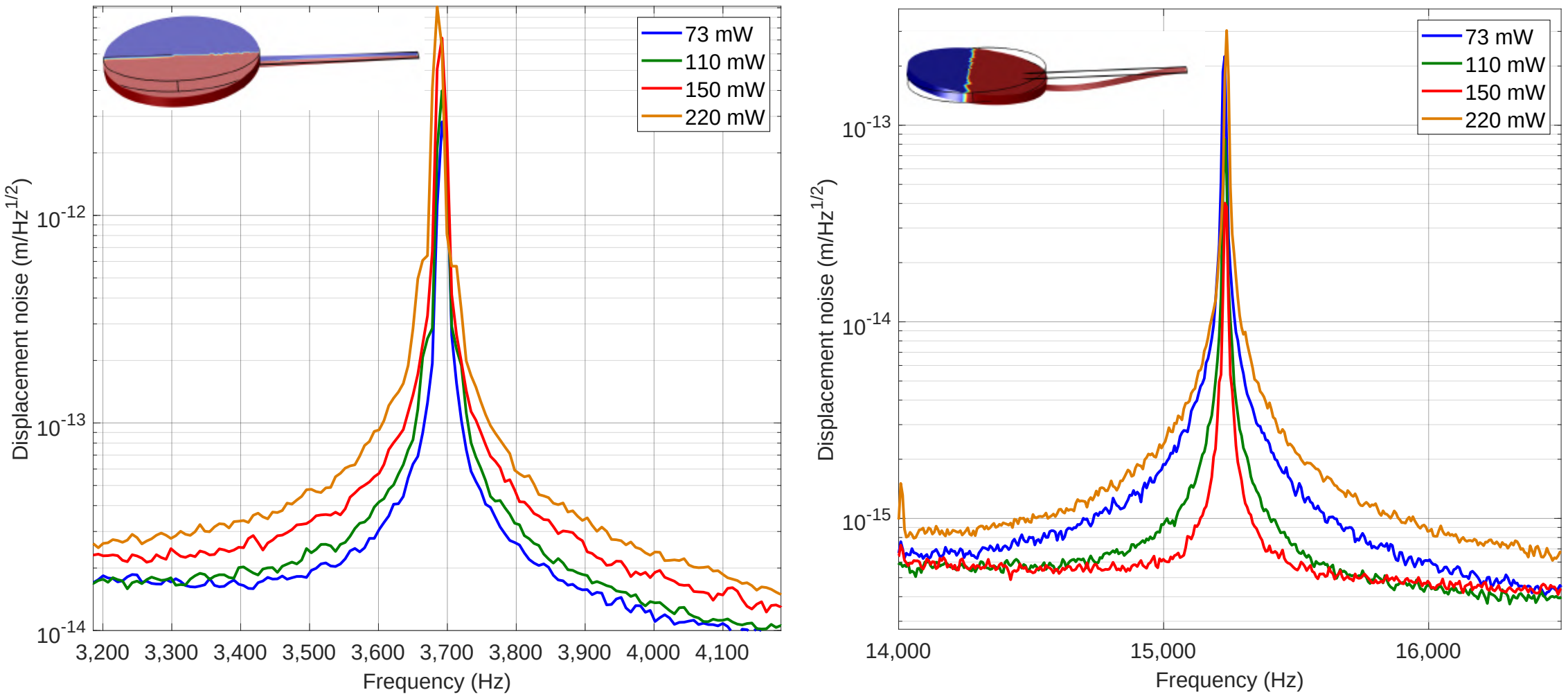

**Figure 6-6:** Dependence of thermal noise on circulating power (caused by change in beam position). The left plot shows the thermal noise around the yaw mechanical mode for each of the four circulating power levels. The right plot shows the thermal noise at frequencies centered around the pitch mechanical mode. The thermal noise around the pitch mode decreases going from 73 mW to 110 mW of circulating power, and then increases at 220 mW. This change is consistent with the cavity mode passing through the nodal point of this mode at an intermediate power level. Each plot includes an image from the finite element model depicting the motion associated with the mechanical mode. In both images, the blue portion represents a positive displacement from equilibrium (thin black outline), and the red area denotes a negative displacement. The nodal line for the mechanical modes is drawn in white.

We model the thermal noise using a finite element model of the microresonator that is based on dimensions obtained from a micrograph of the resonator and is further constrained by measurements of the frequencies and quality factors of the fundamental mode and the next three higher-order modes. We include the material properties of the GaAs and AlGaAs (such as density, YoungâĂŹs modulus, anisotropy etc.) in the model. The total thermal noise spectrum is then calculated using Equation 6.1 by summing

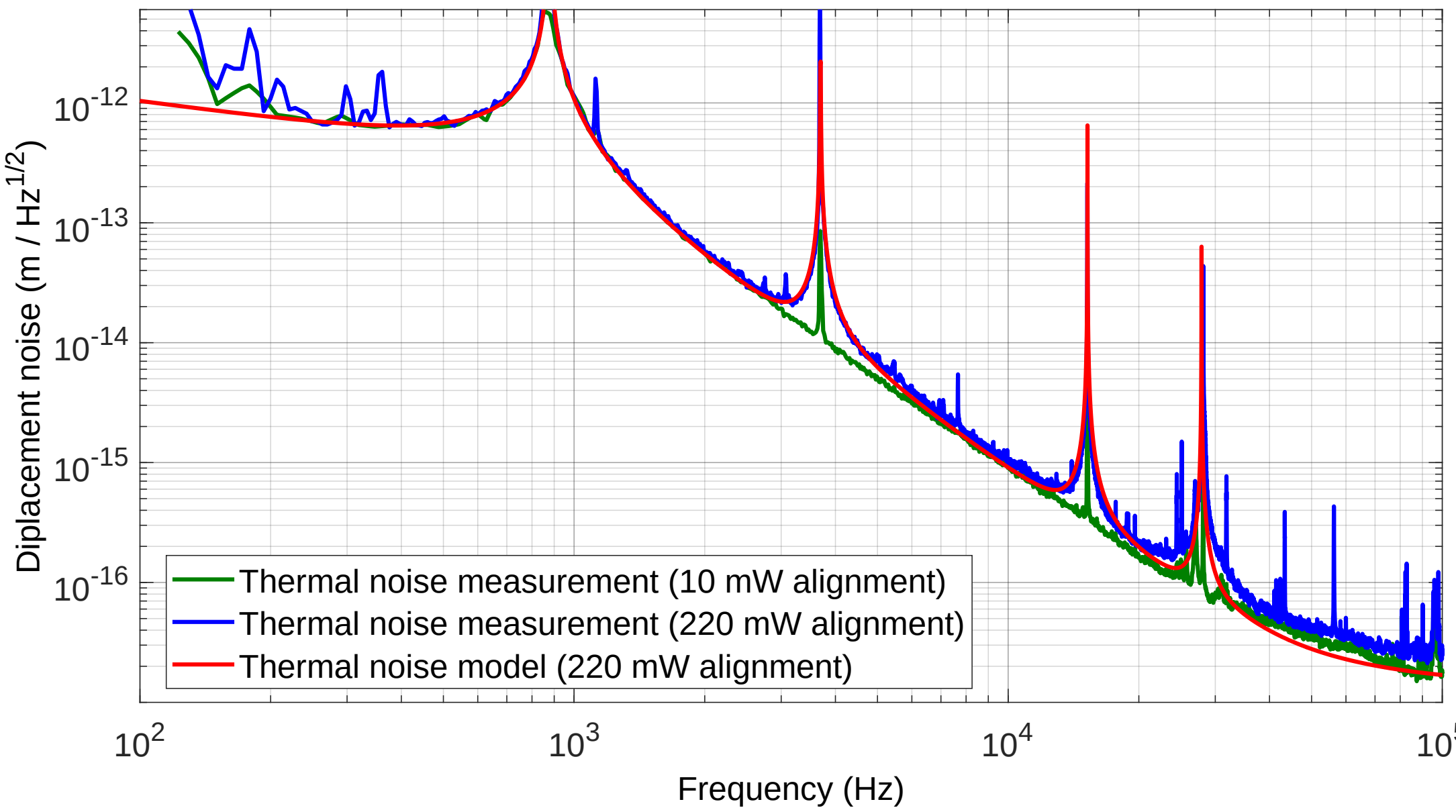

**Figure 6-7:** Comparison of thermal noise spectra at different alignments. The effect of the change in beam position is seen in the change of height of the peaks in the displacement spectrum at the frequencies of the higher-order mechanical modes. The blue curve is taken with 10 mW of circulating power with a cavity mode alignment similar to the QRPN measurement with 220 mW circulating power. The green curve is for an alignment to minimize the coupling of the pitch and yaw modes at 10 mW circulating power. The model curve is based on a model that sets the modal mass of the higher-order modes so that the higher-order mode peaks in the model matches those in the displacement measurement shown in Figure 6-2.

the contribution of each mechanical mode in quadrature as [35]

$$\tilde{x}^2_{th,total} = \sum_i \frac{4 k_B T \omega_i^2}{\omega m_i Q_i [(\omega_i^2 - \omega^2)^2 + \frac{\omega_i^4}{Q_i^2}]}, \tag{6.2}$$

where $\omega_i$, $Q_i$, and $m_i$ are the resonance frequency, quality factor and mass of each mode. We infer the modal mass for each mode by using the thermal noise measurement presented below and are able to reproduce the inferred modal masses by changing the beam position in the finite element model.

The thermal noise curve shown in Figure 6-2 in the main text is a combination of a measurement at frequencies away from the higher-order mechanical modes and a model at frequencies around the yaw, pitch and side-to-side modes. The modeled thermal noise is used around the mechanical modes because the thermal noise must be measured at a low circulating power of 10 mW, and it is difficult to reproduce precisely the same alignment at different power levels. The modal mass of the higher-order modes is set in the model by comparing the magnitude and width of the higher-order mode resonance peaks in the displacement noise measurement shown in orange in Figure 6-2 with those in the model, at frequencies dominated by thermal noise.

In order to demonstrate that the slight changes in alignment resulting from high circulating power do not introduce excess thermal or technical noise that could mask the effect of QRPN, we measure the displacement noise at 10 mW circulating power with the microresonator position shifted to be as close as possible to the alignment with 220 mW of circulating power, as determined by the observed peak height and width of the pitch and yaw modes. By comparing the results of that measurement with thermal noise measurements in the nominal alignment at lower power, shown in Fig. 6-7, we confirm that the thermal noise at frequencies away from the resonances is consistent with the model for structural damping ($f^{-5/2}$) and does not change in a way that is consistent with the observed QRPN ($f^{-2}$). Producing a thermal noise level at 20 kHz that would be as large as QRPN would require thermal noise with a frequency dependence inconsistent with our observed data in Fig. 6-7. Further, the degree of misalignment necessary would introduce cavity losses much greater than observed.

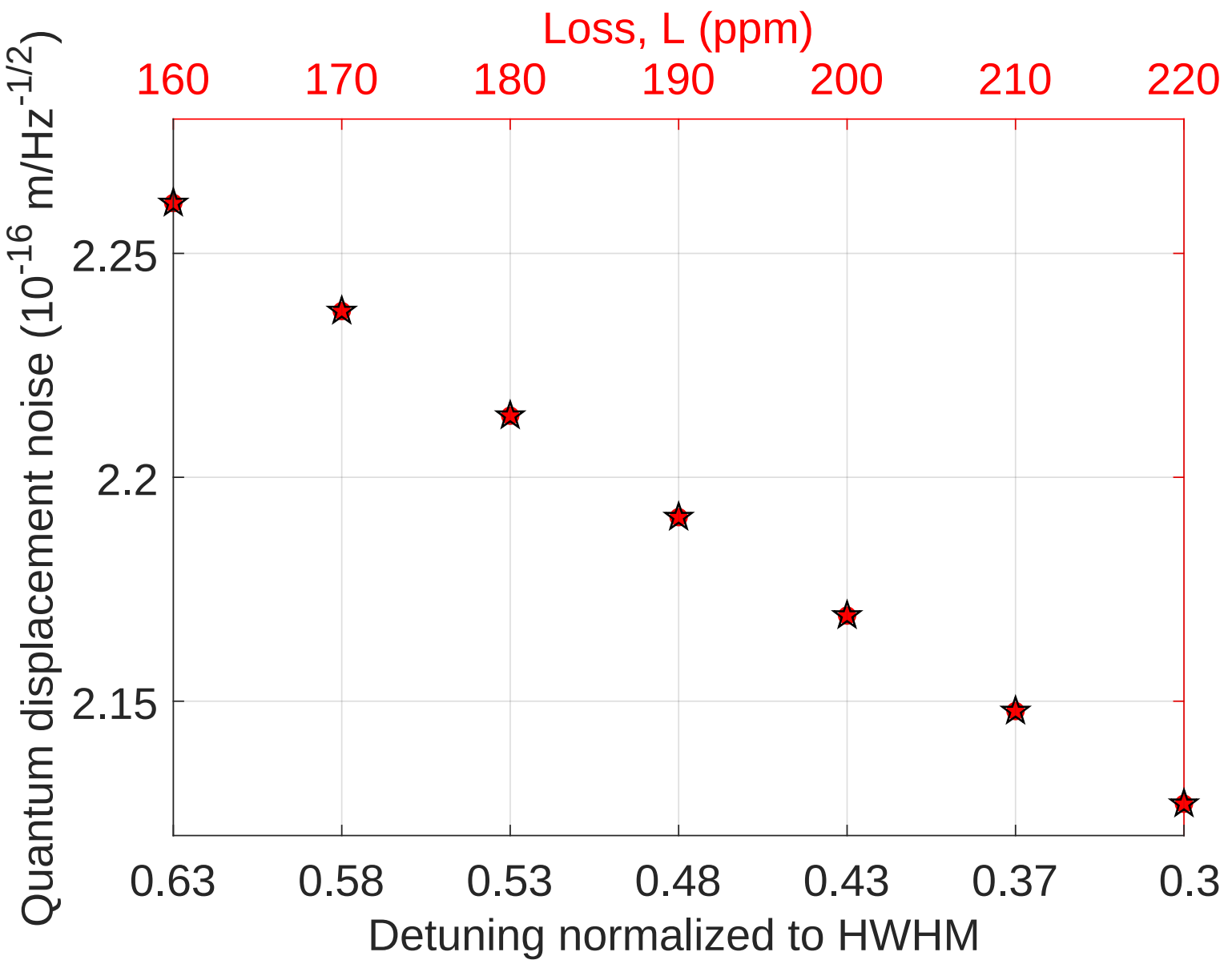

**Figure 6-8:** Effect of uncertainty in detuning and loss. Modeled QRPN at 20 kHz and 220 mW of circulating power as a function of intracavity loss and cavity detuning is shown. To be conservative, the range in values for the cavity loss and detuning in this figure are much larger than the constraints obtained by measurements of the optical spring.

### 6.4.3 Calibration and uncertainties

In order to properly model the expected level of QRPN, we must know key properties of the mechanical and optical system, specifically the mass of the microresonator, the power circulating in the cavity, the total optical losses in the cavity, the length of the cavity and its detuning from resonance. In this section, we describe how these quantities are measured, and the effects of uncertainty in those measurements.

The mass of the microresonator is measured to within $\pm 10\%$ using the measured thermal noise spectrum shown in Fig. 6-7 at frequencies near the fundamental resonance. The thermal noise, given in Equation 6.1, depends on the temperature, mechanical resonance frequency, quality factor, and mass. The temperature is well known, and there is insignificant heating of the cantilever with the small circulating power used in measuring the thermal noise. The mechanical resonance frequency and quality factor are measured to good precision using ringdown measurements with an optical lever setup. Therefore, we may use the thermal noise measurements to constrain the mass of the microresonator to 50 nanograms. The estimated $\pm 10$ % uncertainty is associated with systematic uncertainties arising in calibrating the measured thermal noise. The measured mass is consistent with a finite element model of the microresonator based on measured dimensions of the microresonator that are determined from a micrograph.

The cavity length is constrained by measurements of the geometric size of the cavity mode. Given the known radius of curvature of the back reflector mirror (1 cm $\pm$ 100 $\mu$m), the stability of the cavity puts an upper limit on the cavity length. We further constrain the cavity length by measuring the size of the cavity mode, which is imaged with a camera, and conclude that the overall cavity length must be within 100$\mu$ m of the radius of curvatuve of the reflector. Using this method, we conclude our cavity length is within 200 $\mu$m of 1 cm.

To constrain the optical parameters, we have found using measurements of the optical spring to be the most accurate technique. To measure the optical spring, we measure the frequency-dependent optical response of the system amplitude modulating the light injected into the cavity using the second amplitude modulator, as shown in Figure 6-1. We measure a swept sine transfer function between the injected modulation and the modulation detected in transmission of the cavity at $PD_L$ before the feedback is switched to the phase modulator. This provides a clear measurement of the optical spring frequency and damping, as described in [38].

To obtain the other parameters, we first hold the power incident on the cavity constant, while varying the cavity length by tuning the cavity piezo. By performing an optical spring measurement at each setting, we may find the cavity length and transmitted power level (as measured by $PD_L$) that has the highest frequency optical spring. This allows us to set the detuning to a known value of 0.6 with an accuracy of $\pm$7% for these settings, corresponding to a maximum in the optical spring constant given by [44]

$$K(\omega) = -\frac{32\pi P_{\text{circ}}}{c\lambda_0 T_{\text{total}}} \frac{\delta_\gamma}{(1+\delta_\gamma^2)} \times \frac{1+\delta_\gamma^2 - (\omega/\gamma)^2}{\left[1+\delta_\gamma^2 - (\omega/\gamma)^2\right]^2 + 4(\omega/\gamma)^2}, \tag{6.3}$$

where $P_{\text{circ}}$ is the circulating power in the cavity, $c$ is the speed of light in vacuum, $T_{\text{total}}$ is the fraction of light leaving the cavity in one round trip (including loss and mirror transmission), $\lambda_0$ is the laser wavelength, and $\delta_\gamma$ is the detuning of the cavity in units of the HWHM linewidth. This method is independent of knowledge of the optical power level and cavity losses.

Once the detuning is determined by the above method, we then constrain the circulating power and cavity losses by matching the optical spring frequency and quality factor using the measured transfer function of the optical response. This is possible because, while the real and imaginary parts of the optical spring constant vary identically with circulating power, they do not scale identically with the cavity linewidth (and hence losses) as described by [44]

$$\Gamma_{\text{os}} = \frac{2K_0/m}{\gamma\left[1+\delta_\gamma^2 - (\omega/\gamma)^2\right]}. \tag{6.4}$$

Using this method, we constrain the circulating power to be 73 mW with an accuracy of 10% and total cavity losses to 470 ppm $\pm$10 ppm. The cavity losses are composed transmission of the microresonator ($T_i \approx 250$ ppm), transmission of the end mirror ($T_o \approx 50$ ppm), with the remainder a combination of absorption, scattering and diffraction loss ($L = 170$ ppm). The observed total loss corresponds to a cavity linewidth of 560 kHz $\pm$10 kHz and is consistent with an independent measurement of the cavity linewidth. The independent measurement was performed by measuring the transfer function of amplitude fluctuations to $PD_L$ at frequencies between 500 kHz and 10 MHz with the cavity operated at a high detuning.

For the measurements at high circulating power levels, we scaled both the input power and transmitted power by identical factors. Nominally, one might expect this to keep the cavity detuning constant, but we found that the detuning varied by a small amount. This change in detuning is attributed to the changing alignment of the cavity beam on the microresonator as a result of the larger static radiation pressure as the power is increased. The slight shift in beam position on the microresonator can lead to changes in the cavity losses, likely due to diffraction loss around the edge. This change in losses will in turn be accompanied by a change in detuning in order to maintain the ratio if circulating power to input power. To take this effect into account, we measure the optical spring transfer function as described above. Since the input power and transmitted power are now already determined (because they were scaled from the 73 mW circulating power configuration), the optical detuning and optical losses are varied to match the observed optical spring transfer function to the model. From this analysis, we determine the cavity detuning to be 0.45 linewidths, 0.43 linewidths, and 0.35 linewidths to within $\pm$6 % and total losses ($T_{\text{total}} = T_i + T_o + L$) to be 490 ppm, 495 ppm, and 505 ppm to within $\pm$10 ppm for circulating power levels of 110 mW, 150 mW and 220 mW respectively. The total losses correspond to cavity linewidths of 585 kHz $\pm$10 kHz, 590 kHz $\pm$10 kHz, and 600 kHz $\pm$10 kHz for circulating power levels of 110 mW, 150 mW, and 220 mW respectively.

We further investigate the effect of uncertainty in the cavity detuning and cavity losses in our mea-surements and study its effect on the level of modeled QRPN. By holding the transmitted power and input power constant, the level of optical loss determines the required cavity detuning that will match the measurements. In Fig. 6-8 we show the modeled level of QRPN as we vary the loss and detuning for the 220 mW case. With our estimated uncertainties in detuning, we may conclude that the resulting uncertainty in the modeled QRPN is $\pm 4$ %, well within our measurement uncertainty.

Our calibration method of using the laser-cavity piezo to measure the response at $PD_M$ manifestly removes the effect of the optical spring and our electronic control loop. We calibrate by modulating the laser frequency and measuring its frequency dependent response to $PD_M$. This measurement relies on a change in the laser frequency acting equivalently to a change in cavity length, scaled by the factor $L/\omega_0$. Thus, this transfer function directly allows us to measure the effect of a given amount of displacement on our measurement PD, irrespective of the control system. This is equivalent to any experiment that uses feedback control to keep a system near its operational point.

### 6.4.4 Additional evidence of QRPN

To provide additional evidence that the displacement noise that we are measuring is a result of QRPN, we perform three more checks. First, we verify that this excess noise is not caused by optical heating of the microresonator. Secondly, we measure a transfer function from the amplitude modulations going into the cavity to our measurement photodetector $PD_M$ and show that the cavity acts as a parametric amplifier. Finally, we use the same transfer function to project how much displacement would be caused if the ingoing amplitude fluctuations would be the shot noise of the input light.

The effect of bulk heating of the cantilever, which could mimic QRPN, must be ruled out. Due to the structural damping observed in our device, the mirror motion is dominated by thermal noise below 10 kHz, while still being QRPN limited above. The low frequency thermal noise may be used as a thermometer to measure any heating as a result of higher circulating power. To explain the factor of two increase in noise observed at 20 kHz between low and high power as a result of heating, the temperature would have had to increase by a factor of 4. We can rule out this large increase in temperature by observing that the measured noise at frequencies dominated by thermal noise (between 1 kHz and 2 kHz for example) only increases by 2%, which is within measurement uncertainty.

To show that the cavity acts as an optical parametric amplifier, we measure the transfer function of amplitude modulations of the input light to the light detected at $PD_M$. This frequency dependent transfer function is written as

$$\mathrm{TF}_{\mathrm{AM}}(\omega) = \frac{\partial \mathrm{P}_{\mathrm{PD_M}}}{\partial \mathrm{P}_{\mathrm{in}}}. \tag{6.5}$$

To perform this measurement, we inject amplitude modulations on the input light using the second amplitude modulator in Figure 6-1 and measure the response at $PD_M$. We then perform this measurement again with the cavity unlocked (far from resonance), with the same amount of power incident on $PD_M$. The measurements show that the cavity acts to parametrically amplify the intensity fluctuations incident on the cavity by a constant factor of 4.2 at frequencies below the optical spring frequency. This parametric amplification is a result of the radiation pressure coupling.

Furthermore, to quantify the displacement noise resulting from this coupling, we multiply $TF_{AM}$ by the shot noise level of the effective input power to the cavity. We then apply our calibration transfer function ($TF_{cal}$) to the result to calibrate the displacement noise into units of length. This procedure is outlined in the equations below, where

$$\mathrm{TF}_{\mathrm{cal}}(\omega) = \frac{\partial \mathrm{P}_{\mathrm{PD_M}}}{\partial x}, \tag{6.6}$$

is the calibration transfer function, and

$$\tilde{P}_{\mathrm{eff}} = \sqrt{\frac{T_{\mathrm{total}}}{T_i} 2\hbar\omega_0 \mathrm{P}_{\mathrm{in}}} \tag{6.7}$$

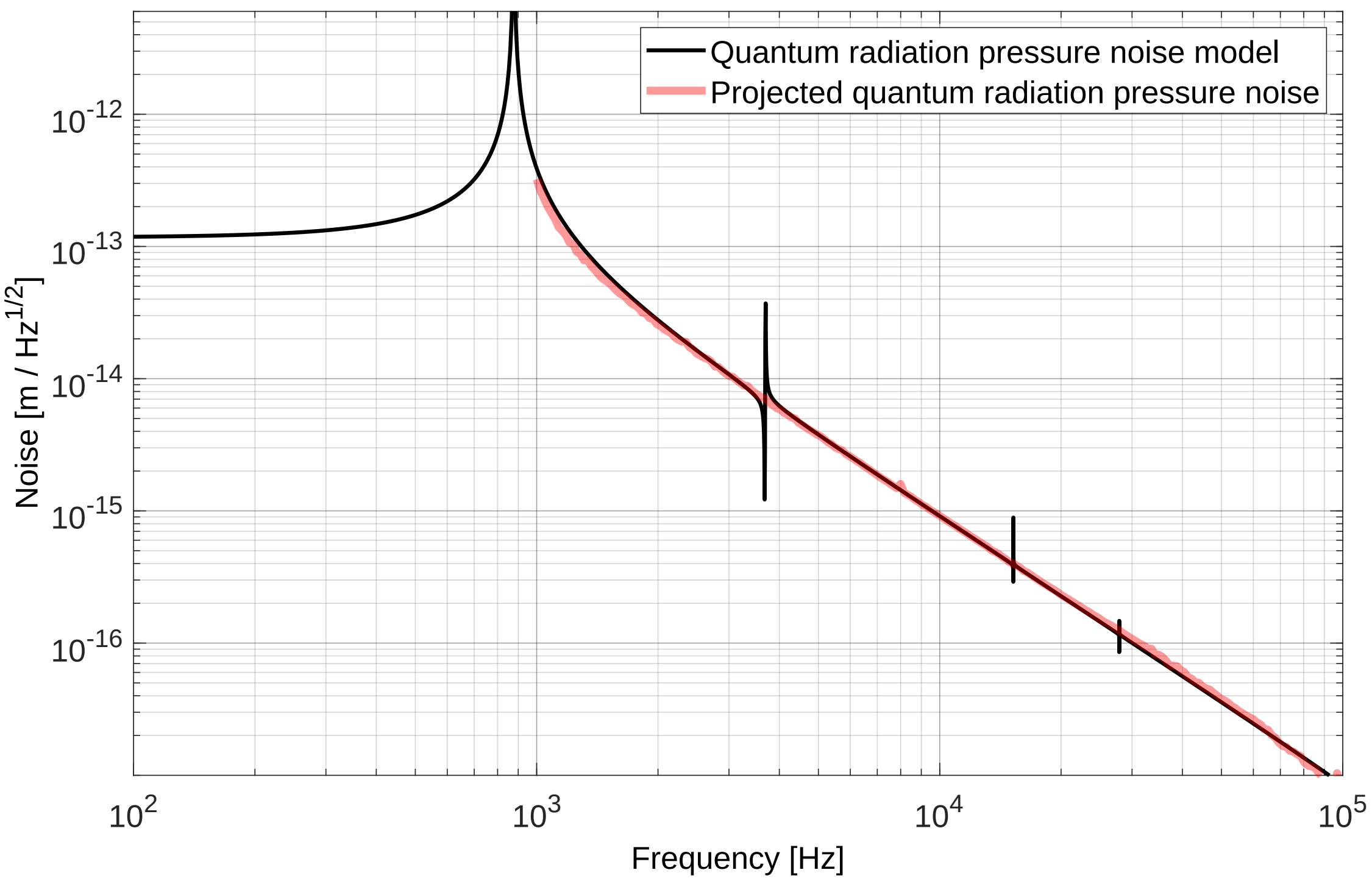

**Figure 6-9:** Projected QRPN. The measured QRPN level obtained by multiplying the transfer function measurement $\text{TF}_{\text{AM}}$ by the shot noise of the effective cavity input power and applying the calibration procedure, as described in the Section 6.4.4, is shown. The result of this calculation agrees with the modeled QRPN.

is the amount of effective vacuum fluctuations that enter the cavity. The projected displacement noise is calculated as

$$\tilde{x}_{\text{projected}}(\omega) = \frac{\tilde{P}_{\text{eff}}\text{TF}_{\text{AM}}}{\text{TF}_{\text{cal}}} \tag{6.8}$$

The power used to calculate $\tilde{P}_{\text{eff}}$ is scaled from the input power ($\text{P}_{\text{in}}$), because the input port only accounts for part of the vacuum fluctuations that enter the cavity; the rest enter from the other cavity losses. This calculation projects the coupling of shot noise to the measurement via radiation pressure on the cantilever. The result of this calculation, shown in Fig. 6-9, agrees with the modeled QRPN, and independently confirms the expected level of QRPN.

## 6.5 Comparison to standard cavity optomechanics

The device used in this experiment may not initially appear to be that useful for quantum cavity optomechanics because of its modest mechanical quality factor and low resonance frequency. In this section, we describe why standard metrics [21] are not effective in our case. We summarize common parameters in Section 6.4.1. Notably, our system does not satisfy the requirement of [74] for observing QRPN that $C/n_{th}\left(1+(2\omega_m/\kappa)^2\right)^{-1} > 1$, where $C$ is the multiphoton cooperativity, $n_{th}$ is the thermal phonon occupation, $\omega_m$ is the mechanical resonance frequency, and $\kappa$ is the FWHM cavity decay rate. This result should not be surprising given that we are in fact thermal noise dominated at the bare mechanical resonance frequency, and that these parameters are intended to be relevant in a viscously damped system. It also would not be logical to use the optical spring parameters in place of the bare mechanical parameters because the region in which we measure is below the optical spring resonance frequency. This requirement could be modified to account for the observed structural damping, and in the bad cavity limit as $C > n_{th} \times \frac{\omega_m}{\omega}$, which is satisfied in our system for sufficiently large $\omega$.

Here, we illustrate the key parameters of our system that make this measurement possible. The full calculation of quantum noise must include the effect of vacuum fluctuations that enter the system wherever there is a loss. We may simplify this calculation for illustrative purposes by assuming an ideal cavity for which the only loss is the transmission of the microresonator (i.e. $T_{\text{total}} = T_i$). To calculate the same level of QRPN as in our real system, we assume the circulating power and cavity decay rate are the same in both cases. For the $P_{\text{circ}} = 220$ mW circulating power level at a detuning of 0.35 linewidths, this is equivalent to $P_{\text{in}} = 31\mu$W incident on the cavity, obtained from the relationship

$$P_{\text{circ}} = \frac{4}{T_i} \times \frac{1}{1+\delta_\gamma^2} P_{\text{in}}, \tag{6.9}$$

where $T_i$ is the transmission of the microresonator. The amplitude spectral density for power fluctuations due to shot noise inside the cavity may then be calculated as

$$\tilde{P}_{\text{circ}} = \sqrt{2\hbar\omega_0 P_{\text{in}}} \times \frac{P_{\text{circ}}}{P_{\text{in}}}, \tag{6.10}$$

where $\omega_0$ is the laser frequency, and the first term is the power spectral density of power fluctuations for the incident light, and the second term accounts for the power amplification in the cavity. The power spectral density for the force on the microresonator is then $\tilde{F} = 2\tilde{P}_{\text{circ}}/c$, and the resulting QRPN for frequencies above the mechanical resonance is

$$\tilde{x}_{\text{QRPN}}(\omega) = 2\tilde{P}_{\text{circ}}/c \times \chi(\omega) \approx \frac{1}{mc\omega^2}\sqrt{\frac{32\hbar\omega_0 P_{\text{circ}}}{T_i\left(1+\delta_\gamma^2\right)}}, \tag{6.11}$$

where $\chi(\omega) \approx \frac{1}{m\omega^2}$ is the mechanical susceptibility, and $\omega$ is the measurement frequency. For our system, we obtain $\tilde{x}_{\text{QRPN}} \approx 8\times10^{16}\text{m}/\sqrt{\text{Hz}}\times\left(\frac{\omega/(2\pi)}{10\text{kHz}}\right)^{-2}$, which agrees to within 10% of our full model. Although this calculation was performed without consideration of the optical spring, its effects are removed in our calibration as previously discussed, and it does not modify the ratio of quantum to thermal noise at frequencies below the optical spring resonance.

In order to observe QRPN, we must make the thermal noise sufficiently small. For the structural damping model that we have found to be consistent with our observed data, given in Equation 6.1, the thermal noise displacement amplitude spectral density scales as $\omega^{-5/2}$, whereas the QRPN scales as $\omega^{-2}$. This difference indicates that the QRPN will dominate over thermal noise at high frequencies if these scaling laws hold. At frequencies much larger than the fundamental resonance, thermal noise from Equation 6.1 can be approximated to

$$\tilde{x}_{\text{th}}(\omega) = \sqrt{\frac{4k_B T\omega_m^2}{\omega^5 mQ}} \tag{6.12}$$

We may then write the ratio of QRPN to thermal noise as

$$\frac{\tilde{x}_{\text{QRPN}}}{\tilde{x}_{\text{th}}} = \sqrt{\frac{8\hbar\omega_0\omega P_{\text{circ}}}{k_B T m T_i\left(1+\delta_\gamma^2\right)}} \times \frac{1}{\omega_m c}. \tag{6.13}$$

Contrary to traditional cavity optomechanics, a low mechanical resonance frequency is advantageous. This calculation assumes a single mechanical degree of freedom, but as we can see from the measurements, coupling from other modes will also be important. In our system, the scaling law for thermal noise begins to break down at around 30 kHz due to the coupling of the drumhead mode of the microresonator. Despite that limitation, the scaling of thermal noise with frequency is a key factor that allows QRPN to be measured.

## 6.6 Conclusion

In conclusion, we present a measurement of quantum radiation pressure noise in a broad band of frequencies far from resonance of the mechanical oscillator. The observed noise spectrum shows the motion of the microresonator is affected by QRPN between about 2 kHz to 100 kHz. We show that the QRPN scales as the square root of the optical power, as expected for quantum noise, and also scales as $f^{-2}$ as expected in this frequency band. We find that the magnitude of the QRPN is consistent with the prediction from our model, as well as an independent projection of QRPN based on measurements. Since the first proposals of interferometric GW detectors, QRPN has been known to present a fundamental limit to the low frequency sensitivity of GW detectors. For the past two decades, the measurement of QRPN at frequencies relevant for GW detectors has eluded increasingly sensitive experiments. The ability to measure QRPN at frequencies in the GW band opens up the possibility of experimental tests of QRPN-reduction schemes [68, 69, 19, 70, 71, 72]. The capability reported here has already lead to measurements of ponderomotive squeezing (Chapter 7, Ref. [37]), cancellation of QRPN [82], and the suppression and amplification of QRPN with squeezed light [83]. From a fundamental standpoint, the measurement of QRPN amounts to observation of quantum vacuum fluctuations inducing motion of a macroscopic object.

# Chapter 7

# Room-temperature optomechanical squeezing measurement

## Contents

This chapter is based on Ref. [37], and much of that manuscript is reproduced here verbatim.

## Abstract

The radiation-pressure driven interaction of a coherent light field with a mechanical oscillator induces correlations between the amplitude and phase quadratures of the light. These correlations result in squeezed light – light with quantum noise lower than shot noise in some quadratures, and higher in others. Due to this lower quantum uncertainty, squeezed light can be used to improve the sensitivity of precision measurements. In particular, squeezed light sources based on nonlinear optical crystals are being used to improve the sensitivity of gravitational wave (GW) detectors. For optomechanical squeezers, thermally driven fluctuations of the mechanical oscillator's position makes it difficult to observe the quantum correlations at room temperature, and at low frequencies. Here we present a measurement of optomechanically (OM) squeezed light, performed at room-temperature, in a broad band near audio-frequency regions relevant to GW detectors. We observe sub-poissonian quantum noise in a frequency band of 30 kHz to 70 kHz with a maximum reduction of 0.7 $\pm$ 0.1 dB below shot noise at 45 kHz. We present two independent methods of measuring this squeezing, one of which does not rely on calibration of shot noise.

## 7.1 OM squeezing: Motivations and challenges

Measurements whose sensitivity is limited by quantum noise can be improved by modifying the distribution of quantum noise. For example, the shot noise limit of optical measurements using coherent states of light can be improved by using squeezed states [18, 17, 9]. Squeezing methods employ light with uncertainty below shot noise in the signal quadrature at the expense of increased noise in orthogonal quadratures. As a result of this redistribution of uncertainty, squeezed states can enhance the precision of measurements otherwise limited by quantum noise. The preeminent example is interferometric GW detectors where squeezed state injection lowers the noise floor below shot noise [19, 18, 17, 22, 23].

Squeezed states of light suitable for GW detectors have been successfully generated using nonlinear optical materials [84, 22, 23, 85, 86]. The OM interaction is similarly an effective nonlinearity [21] for the light field, which can squeeze its quantum fluctuations [87, 88, 19, 70, 20]. OM squeezing, also known as ponderomotive squeezing, has some advantages over squeezed state generation using nonlinear optical media. OM squeezing can be generated independent of the optical wavelength, with a tunable frequency dependence of the squeezing quadrature via the optical spring [20, 89], and in the long term OM squeezers have great potential to be miniaturized.

Previously, OM squeezing has been observed [30, 31, 32, 33] in systems operated close to the mechanical resonance (within an octave). While these experiments laid important foundations for OM squeezed light, some important challenges for practical OM squeezed light sources remained. For GW detection, for example, the squeezed light source needs to be broadband over three decades in the audio-frequency band, compact, and operating stably 24/7 at room temperature. Here, we present a measurement of squeezing produced by an OM system comprising a Fabry-Perot interferometer with a micro-scale mirror as a mechanical oscillator at room temperature, where for the first time OM squeezing has been observed in a room temperature system, at frequencies as low as tens of kilohertz and extending more than a decade away from the mechanical resonance. This observation of broadband OM squeezing at room temperature presents a new avenue for building quantum OM resources at room temperature that are independent of laser wavelength.

Overcoming thermal noise [35] has been a fundamental challenge to observing optomechanically generated squeezing beyond cryogenic temperatures. Reducing the quantum noise below shot noise in such a system is only possible if the motion of the oscillator has a significant contribution from quantum radiation pressure noise (QRPN), and is not overwhelmed by thermal fluctuations [89, 90]. Our mechanical oscillators are designed to have extremely low broadband thermal noise [90, 78, 65, 62, 55] and have been used to observe QRPN [36]. The thermal noise of these oscillators is sufficiently low to not overwhelm the effect of QRPN. Even so, thermal noise does limit the amount of measurable squeezing generated. In addition to the limitation set by thermal noise, our locking and detection scheme introduces losses that degrade some of the quantum correlations created by the OM coupling. Thermal noise, lossy detection, and cavity-feedback noise together limit the amount of squeezing and the frequency band in which it is observed.

A precise calibration of shot-noise has been the basis for all prior demonstrations of optomechanical squeezing. We demonstrate a technique based on photocurrent correlations that obviates the need for such a calibration. This technique may be useful on its own for future studies of squeezing in general.

## 7.2 Experimental setup

Our experimental setup consists of two main subsystems – the optomechanical cavity and the detection system, as shown in Fig. 7-1. The optomechanical system is a Fabry-Perot cavity pumped with a 1064 nm Nd:YAG NPRO laser. One of the two mirrors of this cavity is supported by a low-noise single-crystal micro-cantilever (similar to that employed in Chapter 6 or Ref. [36]), with a mass of 50 ng, a fundamental frequency of 876 Hz, and a mechanical quality factor of 16000. The other mirror is a 0.5 inch diameter mirror with radius of curvature 1 cm. The cavity is just under 1 cm long, has a finesse of around 11500, and a HWHM linewidth ($\gamma$) of 650 kHz.

We lock the cavity slightly blue detuned, at around $0.33\gamma$ away from resonance by using the strong optical spring (145 kHz) created by the detuned operation [20]. The optical spring by itself is unstable, so an electronic feedback in frequencies near the optical spring is used to stabilize the system using the transmitted light for the error signal. We use radiation pressure as the actuator for locking, as detailed in detail in Ref. [38], with one difference: in this experiment, we use a phase modulator as our actuator instead of an amplitude modulator. We can treat the instability of the optical spring in the same way, except for a slightly modified plant transfer function [66]. The open loop gain of the cavity locking loop is below one at all frequencies less than 140 kHz (Fig. 7-2). Since we must obtain a signal to stabilize the optical spring while leaving the squeezed light available to be independently measured, we split the light exiting the cavity at a beam-splitter ($BS_1$), using 15% of the total light to obtain the feedback error signal. This method introduces some common phase noise between the local oscillator and cavity field, which is included in our noise budget.

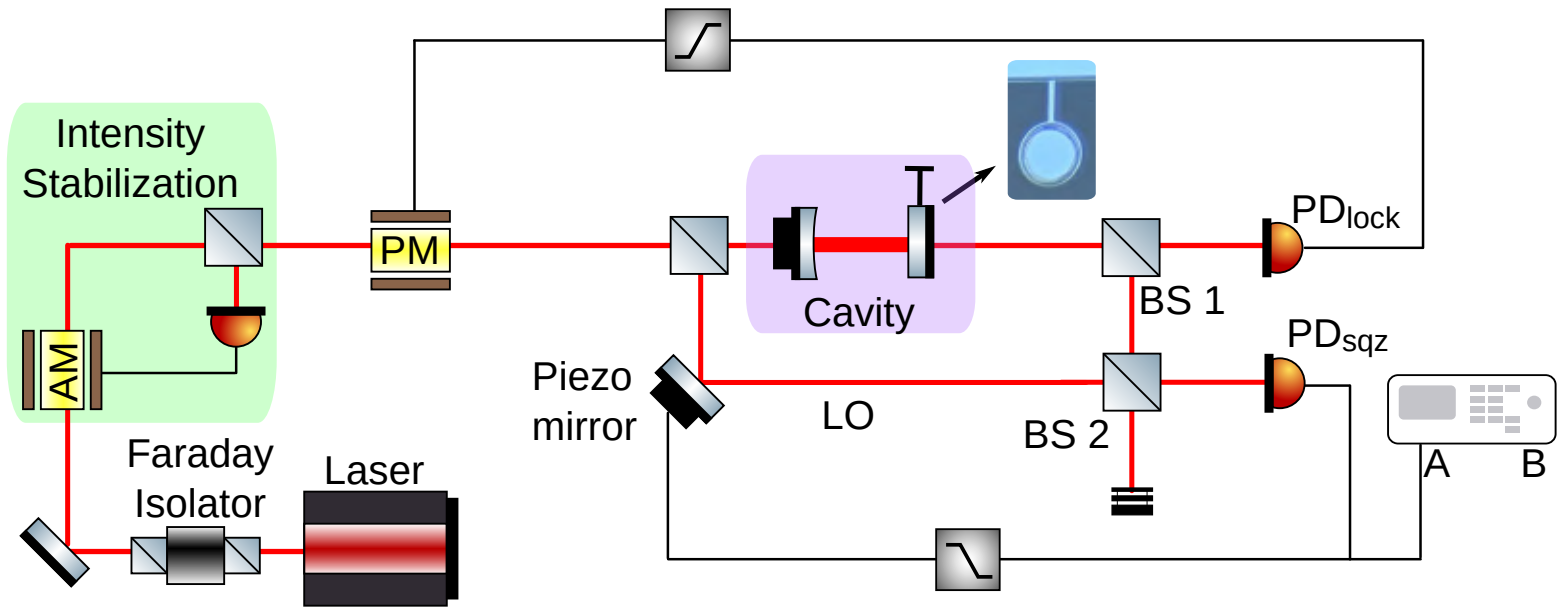

**Figure 7-1:** An overview of the main subsystems in the experiment. First, the classical intensity noise of an 1064 nm Nd:YAG laser is suppressed by an intensity stabilization servo using an amplitude modulator (AM) as the actuator. The light is then sent to an optomechanical cavity – the input mirror is a mechanically rigid macro-mirror, and the output mirror is a low-noise micro-scale mirror supported by a single-crystal micro-cantilever. The light inside this cavity gets squeezed due to the radiation pressure interaction between the circulating light and the movable micro-mirror. The cavity is locked by picking off 15% of the transmitted power through $BS_1$ on $PD_{lock}$, and feeding back that signal to a phase modulator (PM). The remaining 85% of the light is interfered with a local oscillator on $BS_2$ which reflects 96.5% and transmits 3.5% of the light. The phase between the LO and signal is locked by feeding back the DC part of the fringe detected on $PD_{sqz}$ to a piezo mirror in the LO path. The signal from $PD_{sqz}$ is also sent into the spectrum analyser for measurement.

## 7.2.1 Tunable homodyne detection

Traditionally, balanced homodyne detection is used to characterize squeezing, since it cancels classical intensity noise of the local oscillator and does not introduce loss. In our setup, however, we use a different method to measure the squeezing. This is because the classical intensity noise is sufficiently small to not require cancellation, and the level of squeezing we expect is low, making it insensitive to a small loss. The beam transmitted from the cavity (signal) is combined with a local oscillator (LO) beam on a 96.5%-3.5% beam splitter ($BS_2$), as shown in Fig. 7-1. We then measure the port that has 96.5% signal and 3.5% LO on a photodetector ($PD_{sqz}$). The output of $PD_{sqz}$ is low-pass filtered, amplified, and then fed back to a piezoelectric crystal driving the length of the LO path. This loop suppresses relative path length fluctuations between the signal and the LO, but only at frequencies well below the measurement band. The loop has a unity gain frequency of less than 1 kHz, and has an open loop gain of less than -40 dB at the measurement frequencies, as shown in Fig. 7-4. This eliminates the need to correct the squeezing spectrum for the response of the feedback loop. Additionally, there is no cross over between the homodyne loop and the cavity loop because their frequency regions of actuation are disjoint. Note that since $PD_{sqz}$ is an out-of-loop detector for both the cavity-locking as well as the homodyne-locking loop, a sub-shotnoise measurement on it is an indication of squeezing [91]. The lock maintains $PD_{sqz}$ at a constant DC voltage level, which we use to calibrate the shot noise level. The measurement quadrature is determined by the relative path length between the signal and LO. In the laboratory, the measurement quadrature can be tuned by changing either the lockpoint level, or the LO power, or both, see Fig. 7-3 and Eq. (7.1).

In our experiment, we opt to use a single-photodiode homodyne detection. Instead of combining the signal beam with the local oscillator (LO) on a 50-50 beam splitter, we combine it on a highly asymmetric beam splitter. We measure on the output port which transmits the larger signal fraction and reflects the smaller fraction of the LO. While this scheme introduces some loss of signal, it works with just a single photodetector and eliminates the need for performing perfect subtraction that is needed in a balanced homodyne. Since the amount of squeezing expected in this experiment is relatively low, the reduction in squeezing due to this beam splitter loss is small.

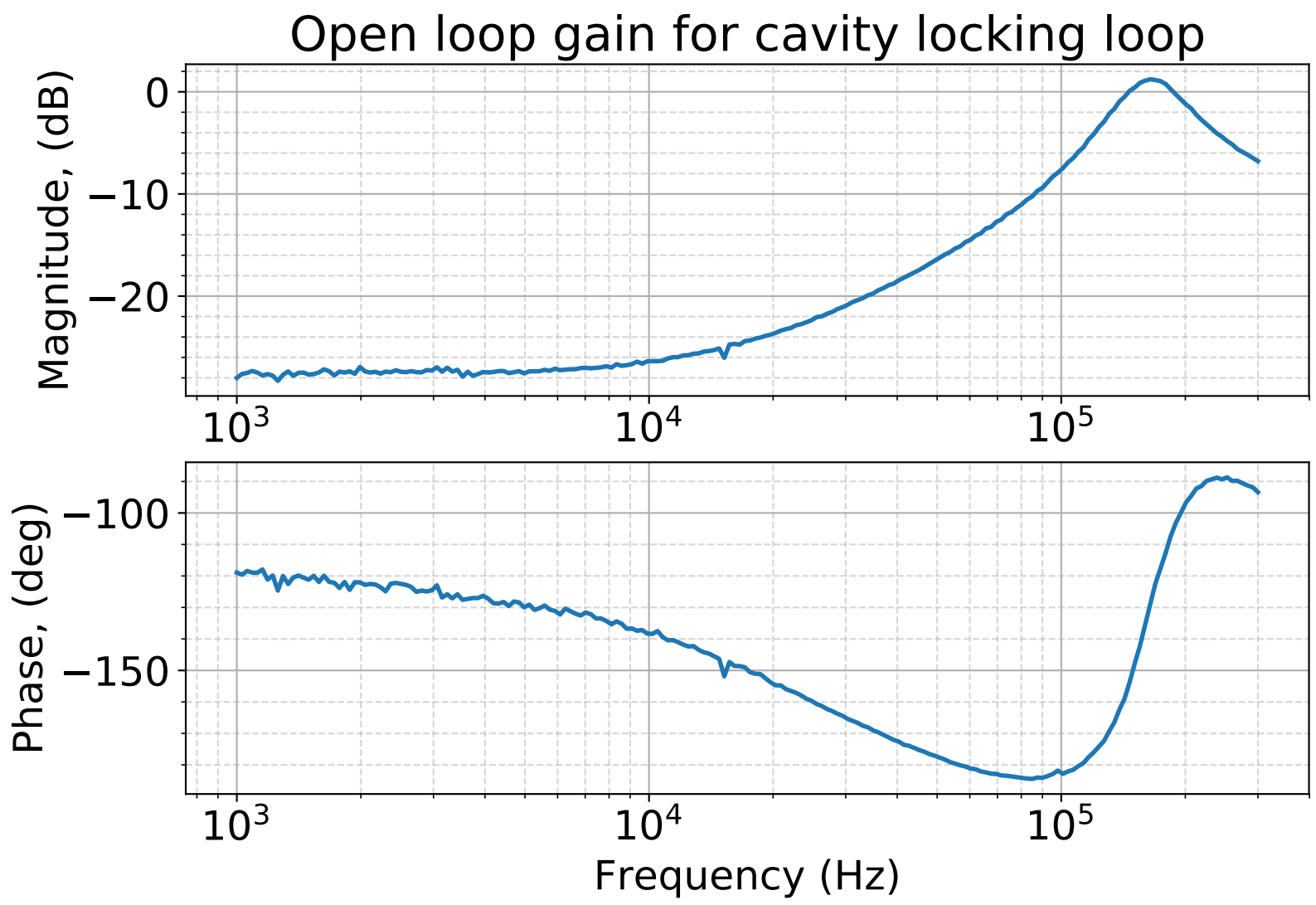

**Figure 7-2:** Open loop gain measurement of the cavity locking loop

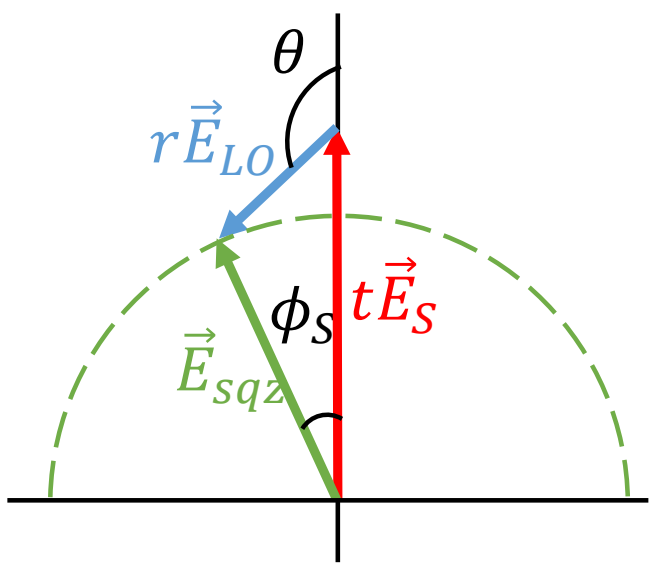

**Figure 7-3:** A phasor diagram showing how the tunable homodyne detector selects the measurement quadrature. The sum of the LO field (blue) and the signal field (red) selects the quadrature that is being measured (green). In the entire manuscript, we report this angle $\phi_S$ as the measurement quadrature. We determine the quadrature by knowing the power in all the three fields, and the visibility. The dashed green circle represents a contour of constant detection power. In order to keep the shot noise reference unchanged, we choose to always lock $PD_{sqz}$ with a constant total detected power, and vary the LO power to change the measurement quadrature. This has the effect of changing the angle $\theta$ of the LO.

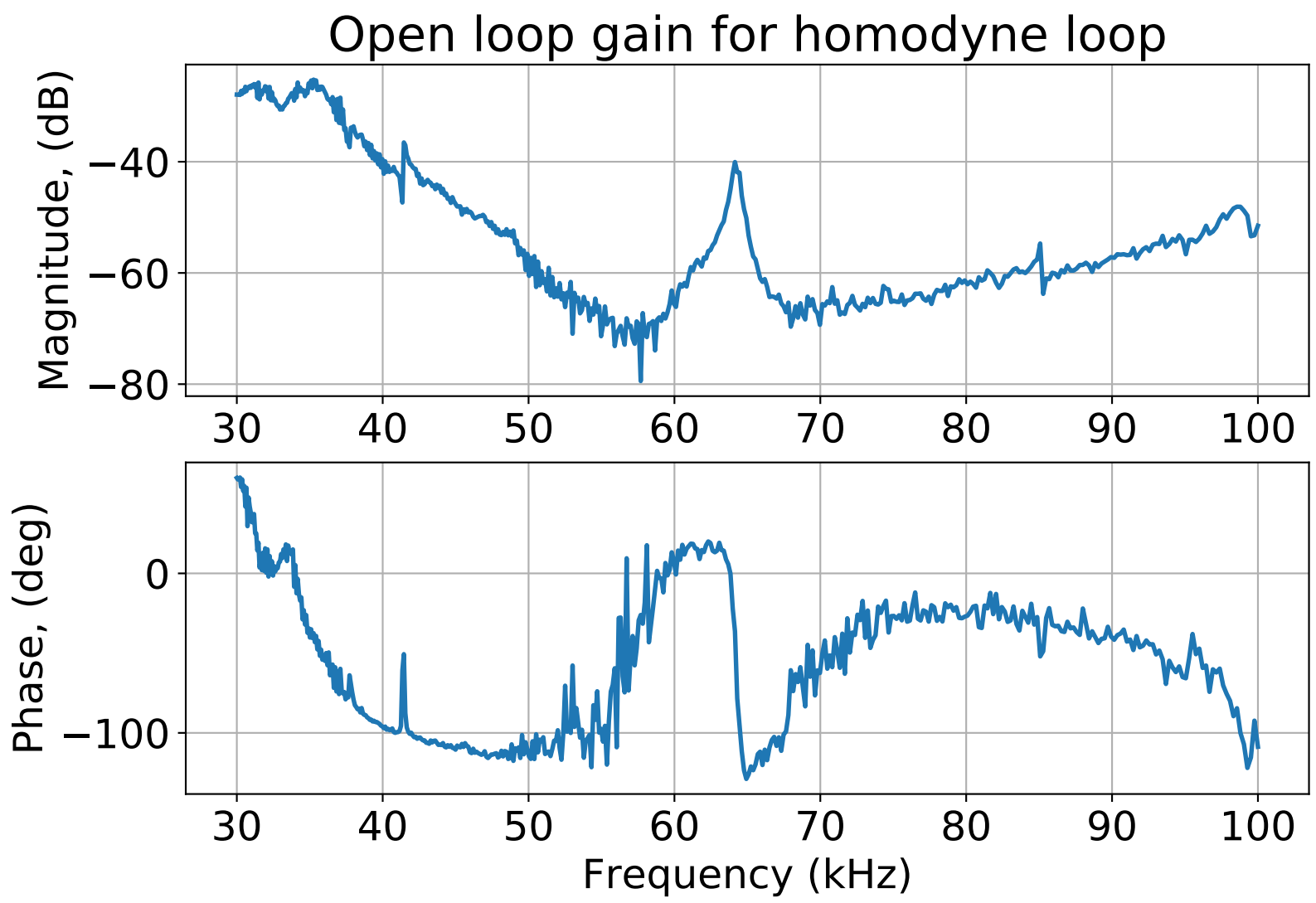

**Figure 7-4:** Measurement of the open loop transfer function of the homodyne locking loop around the squeezing frequency band. Since this loop is designed only to suppress large path length fluctuations between the local oscillator and the signal at low frequencies (<1 kHz), this loop has close to zero gain at our measurement frequencies.

We show below that the signal quadrature in which the measurement is performed is given by the angle made by the resultant of vector addition of the carrier of signal and LO with respect to the signal, as displayed in Fig. 7-3. Similarly, the LO quadrature measured is given by the angle this resultant makes with the LO.

$$\vec{E}_{\rm sqz} = t\vec{E}_{\rm S} + r\mathbb{R}(\theta)\,\vec{E}_{\rm LO} \tag{7.1a}$$

$$\tan\phi_{\rm S} = \frac{rE_{\rm LO}\sin\theta}{rE_{\rm LO}\cos\theta + tE_{\rm S}} \tag{7.1b}$$

$$\tan\phi_{\rm LO} = \frac{-tE_{\rm S}\sin\theta}{rE_{\rm LO} + tE_{\rm S}\cos\theta} \tag{7.1c}$$

$$\vec{E}_{\rm sqz} = |\vec{E}_{\rm sqz}|\,\vec{U}(\phi_{\rm S}) \tag{7.1d}$$

Here $\vec{E}$ represents the strength of the carrier of the signal(S), LO and the resultant (sqz). $t$ is amplitude transmitivity ($\sqrt{0.965}$) and $r$ is amplitude reflectivity ($\sqrt{0.035}$) of the beam splitter. We define a unit vector $\vec{U}(\phi_{\rm S})$ which represents a vector in the direction of the resultant, and determines the measured quadrature. Using [46], we can also calculate the loss effect of the beam splitter. We define $\vec{e}$ as the fluctuations on the field, normalized such that the shot noise is $|\vec{E}|^2$ [66]. We then propagate

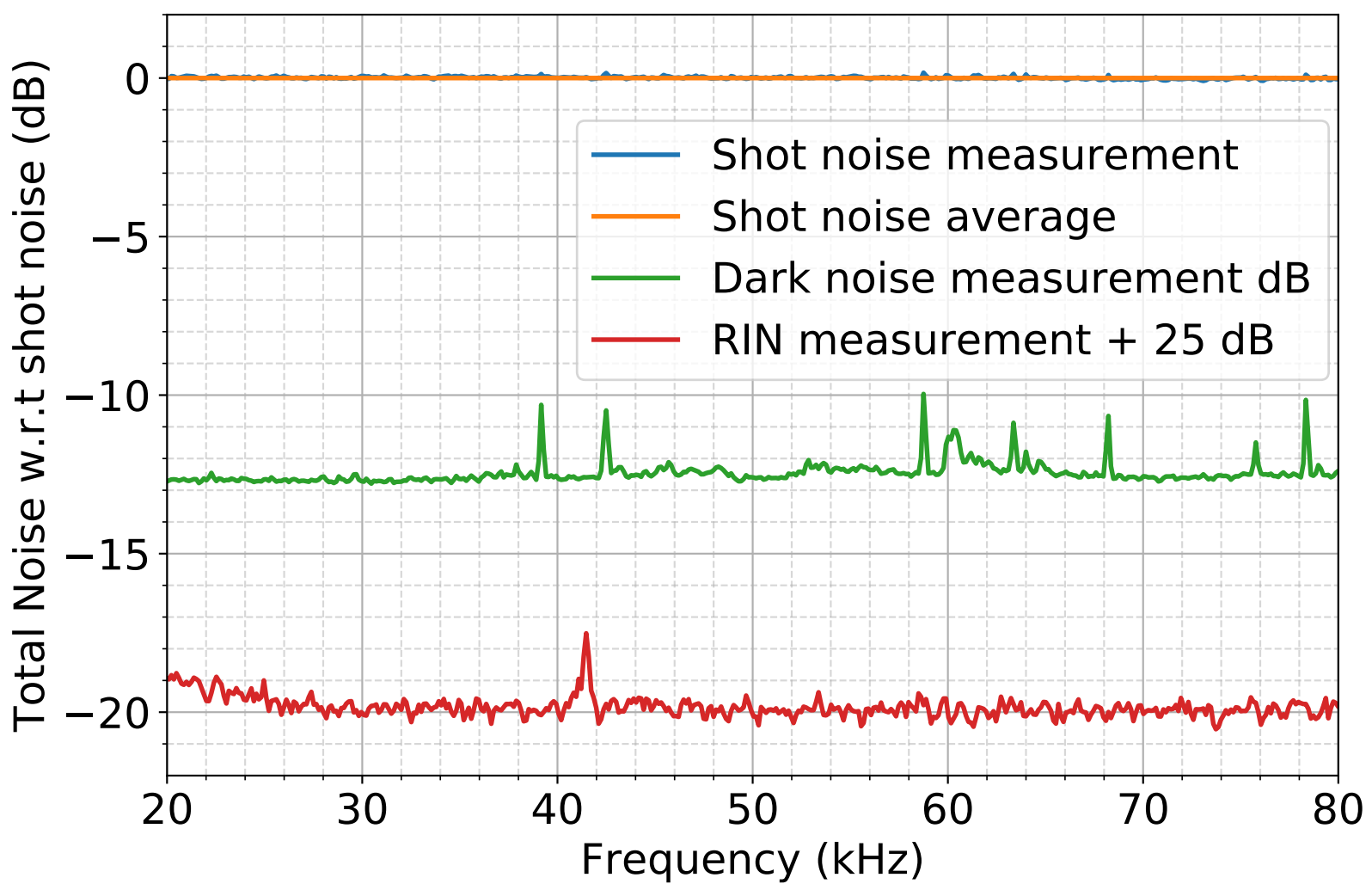

**Figure 7-5:** Classical laser intensity noise and dark noise, shown relative to shot noise. Since we always keep the total detected power on $PD_{sqz}$ constant (and just change the LO power to change the measurement quadrature), the relative dark noise and classical laser intensity noise can just be scaled to that power. In this figure, 25 dB is added to the measured RIN so that it is viewable on the same graph as the other noise measurements.

these fluctuations from the signal and the LO as they interfere on the beam splitter :

$$\vec{e}_{sqz} = t\vec{e}_S + re^{i\phi}\mathbb{R}(\theta)\,\vec{e}_{LO} \tag{7.2a}$$

$$\vec{E}_{sqz}.\vec{e}_{sqz} = |\vec{E}_{sqz}|\,\vec{U}(\phi_S)^\dagger.\vec{e}_{sqz} \tag{7.2b}$$

$$S_{sqz}(\phi_S) = \vec{U}(\phi_S)^\dagger.\vec{e}_{sqz}.\vec{e}_{sqz}^\dagger.\vec{U}(\phi_S) \tag{7.2c}$$

$$= \vec{U}(\phi_S)^\dagger.(t^2\vec{e}_S\vec{e}_S^\dagger + r^2\mathbb{R}(\theta)\,\vec{e}_{LO}\vec{e}_{LO}^\dagger\mathbb{R}(\theta)^\dagger).\vec{U}(\phi_S) \tag{7.2d}$$

$$= t^2 S_S(\phi_S) + r^2 \tag{7.2e}$$

where we have assumed that the LO is shot noise limited, and defined $S_S(\phi_S) = \vec{U}(\phi_S)^\dagger\cdot\vec{e}_S\vec{e}_S^\dagger\cdot\vec{U}(\phi_S)$ as the spectral density of the signal if measured perfectly with a balanced homodyne detector at the quadrature $\phi_S$. The above equations show that the total spectral density measured, $S_{sqz}(\phi_S)$ is a combination of $S_S(\phi_S)$ and 1, in the ratio of the beam splitter's reflectivity.

### 7.2.2 Shot noise calibration

In order to compare the measured noise to shot noise, we measure the shot noise level by turning off the homodyne lock, blocking the signal port, and tuning the LO power to get the same voltage on $PD_{sqz}$ as our lockpoint. This allows us to measure a spectrum of $PD_{sqz}$ that contains shot noise of the light, classical intensity noise, and the dark noise of $PD_{sqz}$. We then average this spectrum over our measurement band to obtain the reference level (0 dB). Classical relative intensity noise (RIN) is suppressed by an intensity stabilization servo (ISS) to about $8\times10^{-9}/\sqrt{\text{Hz}}$, and contributes less than $-40$ dB of the noise on $PD_{sqz}$ (Fig. 7-5). The RIN level is independently measured by performing a correlation measurement between $PD_{sqz}$ and another pick-off between the ISS and the PM. Dark noise accounts for about -12 dB of the shot noise level, and is not subtracted.

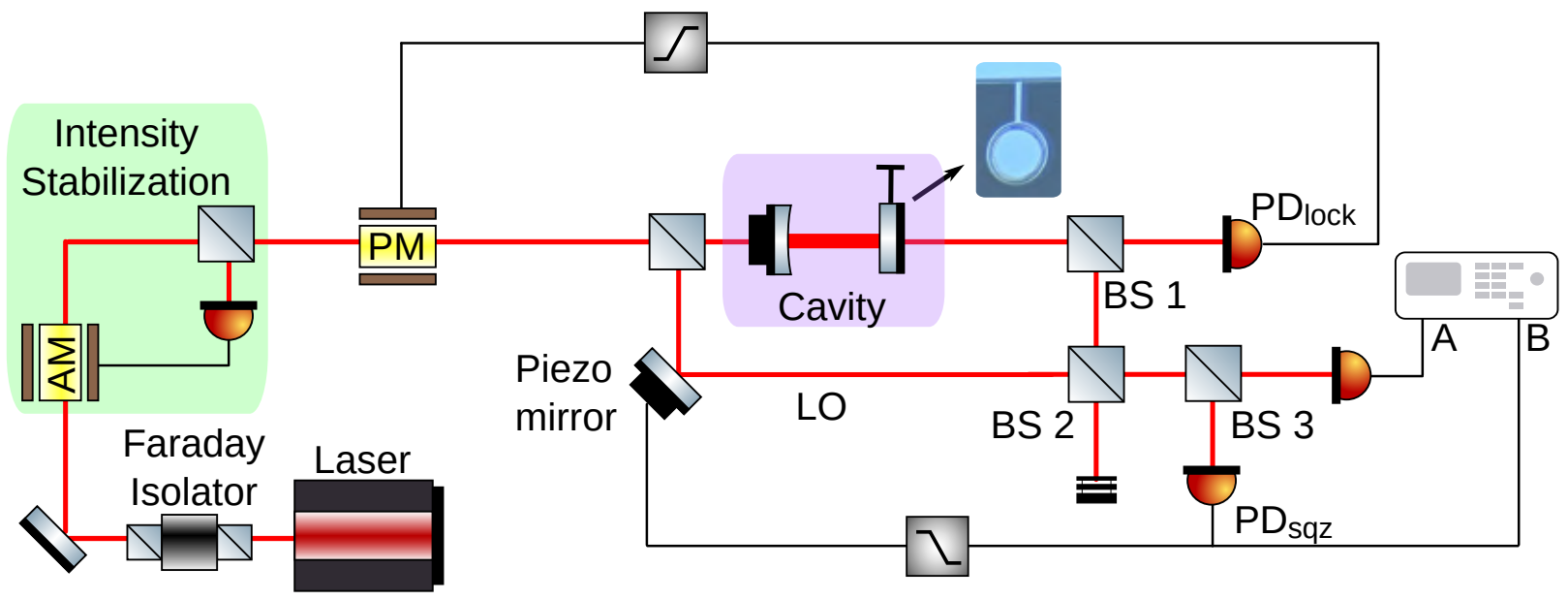

**Figure 7-6:** Setup for correlation measurement: We set the LO such that the field after $BS_2$ is amplitude squeezed and pass it through a 50-50 ($BS_3$). We then perform a cross-spectrum measurement of the two outputs and normalize it to the individual spectra. This quantity can only be negative if the input beam is squeezed in the amplitude quadrature (see Eq. (7.5)).

### 7.2.3 Correlations as a signature of squeezing

Consider splitting an intensity squeezed beam onto two photodetectors. For convenience, let's split it as 50%. The amplitude quadrature of the two fields hitting the photodetectors may then be written as

$$a_1 = \frac{e^{-r}x_1 + c - y_1}{\sqrt{2}} \tag{7.3}$$

$$b_1 = \frac{e^{-r}x_1 + c + y_1}{\sqrt{2}}, \tag{7.4}$$

where $x_1$ is the vacuum that has been squeezed by the factor $e^{-r}$, $c$ represents any classical noise that might be present, and $y_1$ is the vacuum that enters at the beamsplitter.

If we measure the the averaged cross power spectrum of the resulting photocurrents, but don't take the absolute value, we find

$$\langle S_{ab} \rangle = \frac{1}{2}\left(e^{-2r} + S_c - 1\right)\alpha\beta, \tag{7.5}$$

where we have normalized shot noise to 1, and assumed detector a has a relative gain of $\alpha$, and detector b has a relative gain $\beta$, and $S_c$ is the power spectrum of the classical noise scaled to shot noise. All the cross terms between $x_1$, $y_1$ and $c$ will average to 0, as they are uncorrelated. If the original field is squeezed, then that requires $e^{-2r} + S_c < 1$, which would then imply $\langle S_{ab} \rangle < 0$. Note that if this is not satisfied, such that we have classical noise that destroys the squeezing, then $e^{-2r} + S_c > 1$, which requires $\langle S_{ab} \rangle > 0$. Therefore, by looking at the sign of the average cross power spectrum, one can definitively prove whether squeezing is present or not.

To interpret this, when the beam is limited by classical noise, the power fluctuations hitting both PDs are identical and positively correlated. If the beam is exactly shot noise limited, the power fluctuations hitting the two PDs are uncorrelated. With a perfectly amplitude squeezed beam, the power fluctuations are exactly anti-correlated.

We may write the individual power spectra as

$$\frac{S_a}{\alpha^2} = \frac{S_b}{\beta^2} = \frac{e^{-2r} + S_c + 1}{2}. \tag{7.6}$$

Then define the normalized correlation as

$$C = \frac{\langle S_{ab} \rangle}{\sqrt{S_a S_b}} = \frac{e^{-2r} + S_c - 1}{e^{-2r} + S_c + 1}. \tag{7.7}$$

This is convenient because it supplies a unitless measure of the nature of the noise, and is independent of the relative gain of the photodetectors. This $C$ is similar to the square root of coherence, but retains phase information. In fact, the coherence may be written as $CC^*$.

We can see that if the field is entirely classical so that $S_c$ dominates, then $C = +1$. Likewise if the beam is exactly shot noise limited without squeezing, then $C = 0$. Finally, for an infinitely squeezed field with no classical noise, $C = -1$.

To simplify, let's call the total noise PSD of the original beam relative to shot noise $R = e^{-2r} + S_c$, in which case

$$C = \frac{\langle S_{ab} \rangle}{\sqrt{S_a S_b}} = \frac{R-1}{R+1}. \tag{7.8}$$

This leads to

$$R = \frac{1+C}{1-C}. \tag{7.9}$$

Thus, by measuring $C$, we have a method to measure the amount of noise relative to shot noise independent of our ability to calibrate shot noise.

This treatment is simplified by not propagating the DC carriers of the fields. The final physical result becomes invariant of the beam splitter convention if one keeps track of the DC carrier fields. The cross spectrum $\langle S_{ab} \rangle$ is negative for a squeezing beam $x$, irrespective of the beam splitter convention, as long as the carrier of the field $y$ is smaller than the carrier of the field $x$. This condition is trivially satisfied in our measurement, since $y$ is coming from vacuum fluctuations, with a zero DC field.

To include the effects of uncorrelated electronics noise on the photodetectors, we may rewrite the power spectrums for each detector as

$$S_a = \alpha^2 \frac{R+1}{2} + \alpha^2 S_{da} \tag{7.10}$$

$$S_b = \beta^2 \frac{R+1}{2} + \beta^2 S_{db}, \tag{7.11}$$

where $S_{da}$ and $S_{db}$ are the PSDs of each detector from electronics noise. The resulting normalized correlation is

$$\begin{aligned} C = \frac{\langle S_{ab} \rangle}{\sqrt{S_a S_b}} &= \frac{R-1}{\sqrt{(R+1+S_{da})(R+1+S_{db})}} \\ &= \frac{R-1}{R+1} \left[ \sqrt{1+\frac{S_{da}}{1+R}} \sqrt{1+\frac{S_{db}}{1+R}} \right]^{-1/2} \\ &= \eta \frac{R-1}{R+1}, \end{aligned} \tag{7.12}$$

where $\eta \leq 1$ is an effective efficiency of the measurement. In our experiment, since the dark noise is far below shot noise (Fig. 7-5), the efficiency $\eta$ is close to 1. If instead one had a lower efficiency, we can see from Eq. (7.12) that electronics noise will make observed correlations trend towards 0, and the inferred quantum noise level to shot noise.

## 7.3 Observation

### 7.3.1 Homodyne measurement

The result of the homodyne measurement of the signal is shown in Fig. 7-7. For a quadrature angle of 12.3° from the amplitude quadrature, we observe up to $0.7 \pm 0.1$ dB of squeezing (equivalent to a $15 \pm 2$ % reduction in the PSD), from 30 kHz to 60 kHz.

The distribution of squeezing is studied in detail by measurements of other quadratures of the homodyne signal. In order to do this without changing the locking loop or shot noise, we keep the homodyne locking offset the same, and vary the LO power. This allows us to change the measurement quadrature in a shot noise invariant way. In the top panel of Fig. 7-8, we show this measurement as a function of sideband frequency and quadrature.

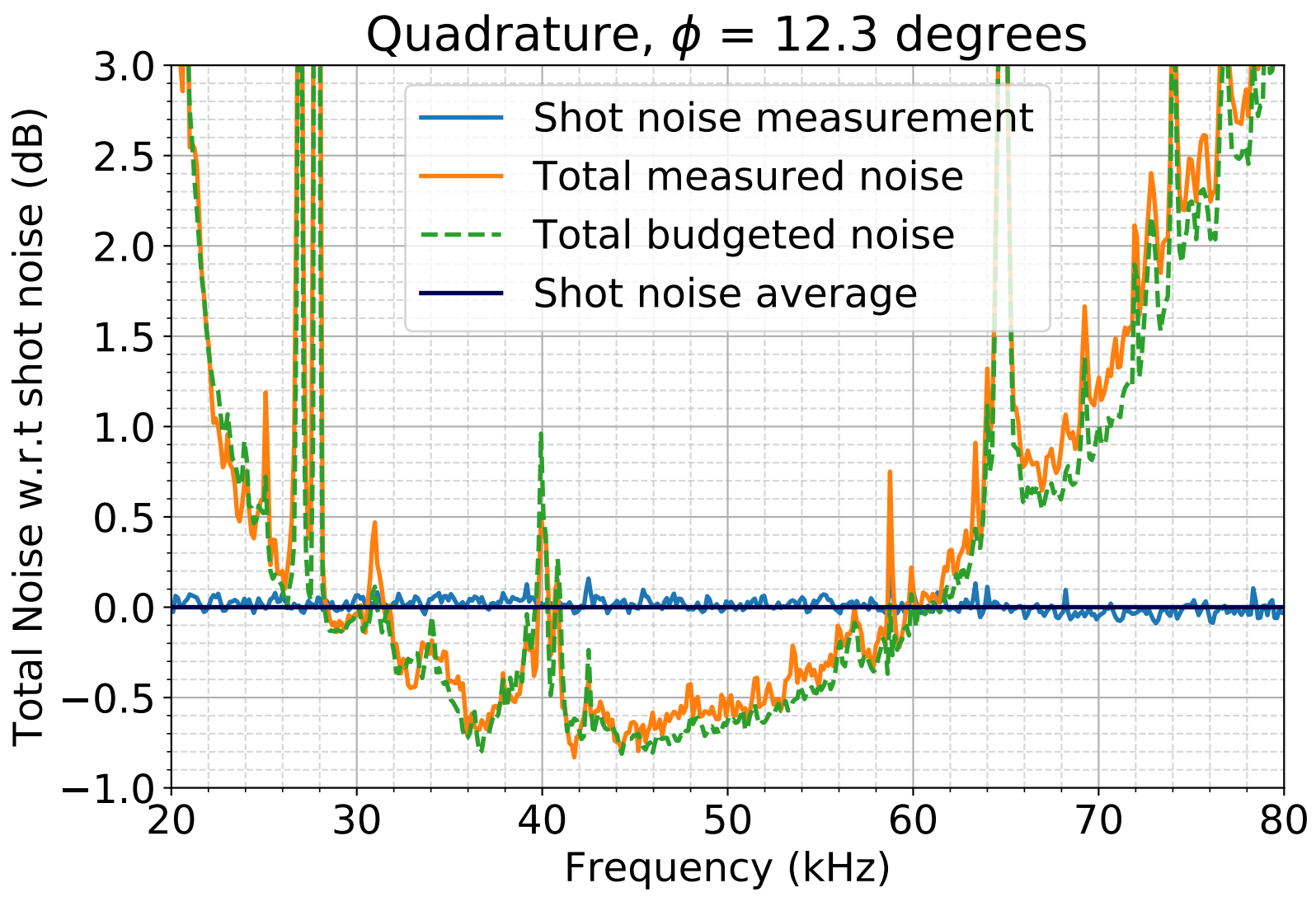

**Figure 7-7:** Measured spectrum and modeled noise budget at 12.3° quadrature. All quadrature angles are referenced so that 0° corresponds to the amplitude quadrature of the cavity transmission. This figure shows the measured spectrum relative to shot noise. We show the shot noise measurement in blue, which is used to obtain an average shot noise level. All the data in the work is scaled to this average shot noise level. The spectrum for total measured noise at 12.3° is shown in orange, showing squeezing from 30 kHz to 60 kHz, with maximum squeezing of $0.7 \pm 0.1$ dB (corresponding to a $15 \pm 2\%$ reduction in PSD) near 45 kHz. We also show the total budgeted noise in dashed green which is a quadrature-sum of quantum noise, thermal noise, classical laser noise, cavity-feedback noise, and differential phase noise.

To understand the observed squeezing, a detailed noise budget of the system is developed. The total budgeted noise in the squeezing quadrature is shown in Fig. 7-7, and in a quadrature dependent way in the bottom panel of Fig. 7-8. This noise budget includes a model [46] that predicts the contribution of quantum noise and previously measured thermal noise [36] for the measured cavity and homodyne parameters; measured cavity-feedback injected noise and differential phase noise between the LO and the cavity. Finally, the extra loss in the detection path is obtained by comparing measurement and noise budget at all frequencies and quadratures. Further details on the noise budget can be found in Section 7.4. As we see, the overall behavior of the system is similar in the measurement as well as noise budget, most importantly the squeezing quadrature.

### 7.3.2 Correlation measurement

For additional evidence of squeezing, we have also performed a correlation measurement on the squeezed light. Extending the approaches in Refs. [92, 93, 94, 95, 96], we demonstrate that these correlations are a way to characterize a squeezed light source without measuring shot noise. The light exiting the cavity, after combination with the LO, is split equally between two photodetectors, as shown in Fig. 7-6. As described in the SI, if the light is limited by classical noise, positive correlations should be observed in the two photocurrents. Shot noise limited light should produce zero correlations, and intensity squeezed light should produce negative correlations. We measure the cross power spectrum between the two photodetectors and confirm that negative correlations are observed, as shown in Fig. 7-9a. The cross-spectrum is negative from 33 kHz to 62 kHz, and positive elsewhere, which agrees with the measured spectrum in Fig. 7-7. For explicit comparison, we convert this correlation to the squeezing factor (using Eq. (7.9)). This squeezing factor is shown in Fig. 7-9b. This provides unconditional evidence that the light is squeezed at these frequencies and at this quadrature.

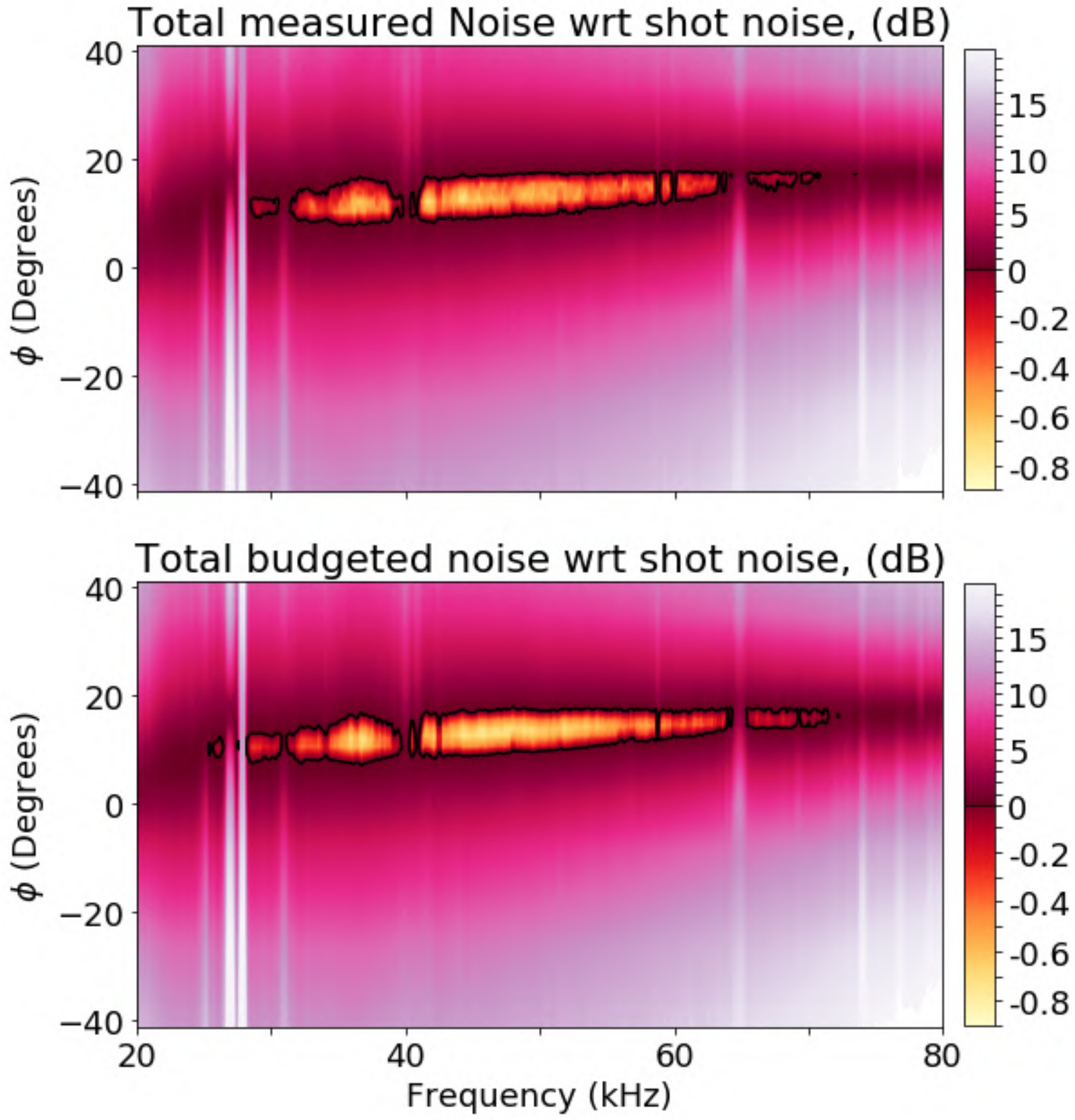

**Figure 7-8: Top:** Measured noises on $PD_{sqz}$ at fourteen different quadratures, distributed more densely near the squeezing quadrature, and sparsely elsewhere. The black contour line corresponds to shot noise. The regions inside it are squeezed (shown by the lower colorbar), and the region outside it are antisqueezed (shown by the upper colorbar). We observe squeezing from 10° to 17° and from 30 kHz to 70 kHz. We can also see one of the mechanical modes of the cantilever at 27 kHz.
**Bottom:** Budgeted noise relative to shot noise. The color scheme is same as the top panel. As is characteristic of OM squeezing below the optical spring frequency [89], the higher quadrature shot noise crossing for all frequencies occurs at the same quadrature. The upper part of the shot noise contour is nearly perfectly horizontal.

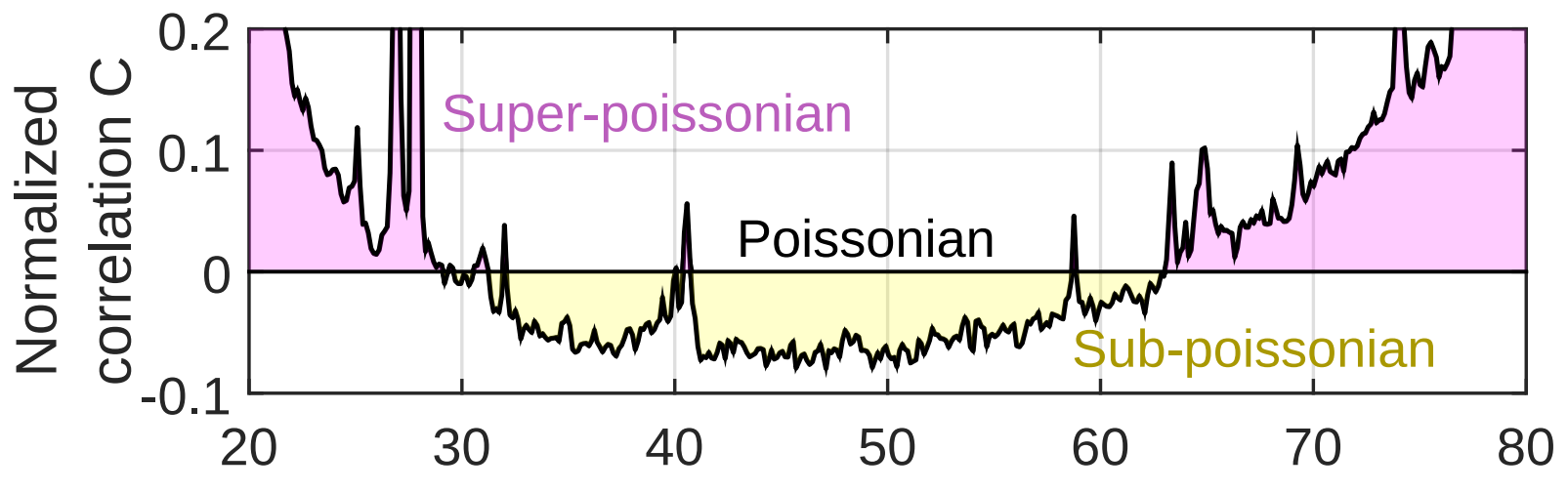

**(a)**

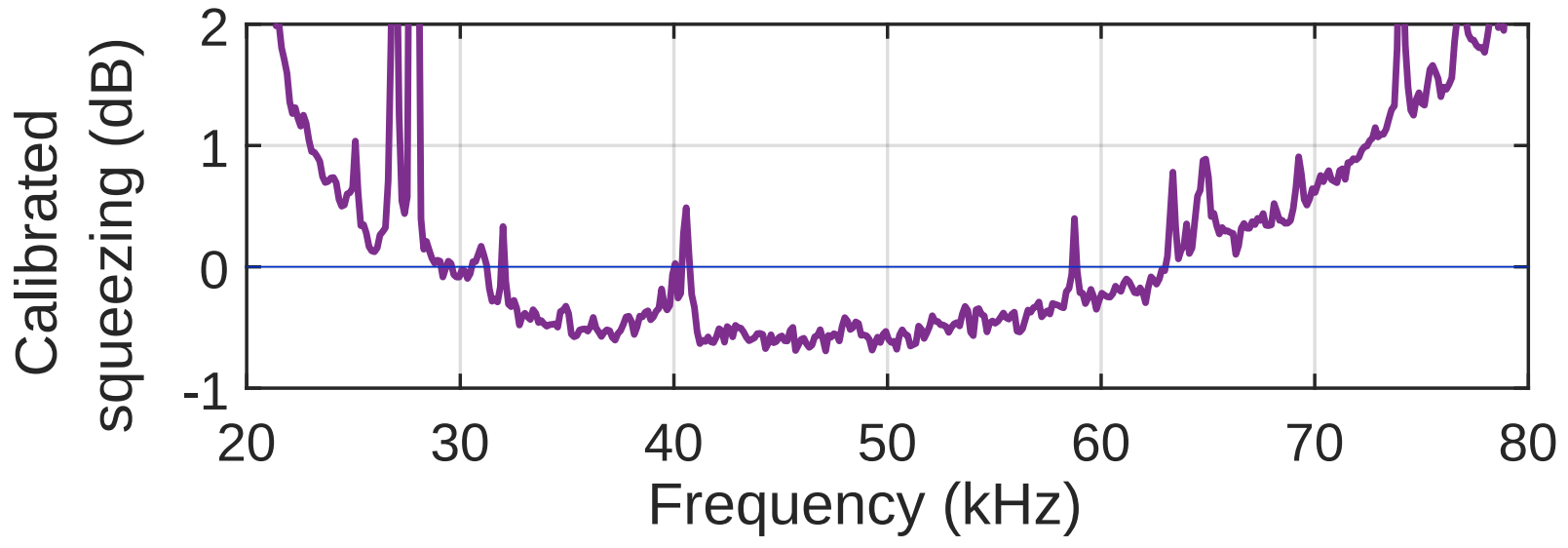

**(b)**

**Figure 7-9: (a)** Measurement of negative correlations. The existence of these negative correlations provides a verification of squeezing, and allows for a shot-noise independent way of verifying the existence of amplitude squeezing. **(b)** Squeezing spectrum obtained by using the negative correlations shown in (a) (see Eq. (7.9)). This spectrum is inherently calibrated to shot noise.

## 7.4 Noise budget

The measurement is also compared to a noise budget, shown in the bottom panel of Fig. 7-8. The total budgeted noise shown is the quadrature sum of individual contributions from measured thermal noise, quantum noise, classical laser noises, cavity-feedback noise, and differential phase noise between the signal and LO. It uses experimentally measured cavity parameters, thermal noise, $BS_1$ and $BS_2$ reflectivities and homodyne visibility, listed in Table 7.1. The quadratures for which squeezing is obtained depend on the various OM parameters of the cavity, such as the detuning, circulating power, losses, as well as the thermal noise. We measure the cavity detuning, intracavity power and losses by measuring transfer functions from amplitude modulations on input to transmitted light. The thermal noise is measured by a cross-spectrum measurement in the amplitude quadrature without the local oscillator [82]. We have also separately calibrated all the beam splitters, the mirror reflectivities, and the homodyne visibility. These measured quantities are then used to predict the squeezing using a numerical model based on Ref. [46]. In this model, we also include the effect of the unbalanced homodyne with an imperfect visibility and common-mode laser noises. The prediction is shown in Fig. 7-10. Once these technical noises have been suppressed, and the optical losses have been lowered, we would expect to see about 1.5 dB of squeezing from this system. This limit comes from a combination of escape efficiency and thermal noise [89].

We then characterize the impact of technical noises by measuring their contribution. First, we measure the contribution of noise injected by the cavity locking system. The dominant source of this noise is shot noise at $PD_{lock}$ due to 15% transmission of $BS_1$. In order to measure this feedback noise, we measure the coherence between $PD_{sqz}$ and the amplifier output that is fed to the phase modulator at input. This coherence when multiplied with the spectrum of $PD_{sqz}$ gives us the contribution of feedback noise. We do this at all measurement quadratures independently. We find that the impact of feedback noise is minimized at 17°, akin to other intracavity displacement noises like thermal noise.

In principle, this cavity-feedback noise could be subtracted from the final result, as it is a measured quantity, but we choose not to do so for the sake of simplicity. Instead, we chose to pick-off the LO beam just after the cavity-feedback phase modulator, so that there is common mode rejection of this locking loop phase noise at the homodyning stage at $BS_2$. The common mode rejection by the homodyne detection also allows us to cancel frequency noise originating from the NPRO laser, without requiring a frequency stabilization servo. Any scheme to measure squeezing not purely in the amplitude quadrature requires mixing the signal beam with an LO that is phase coherent with it, and so one always has the ability to reject common mode noise in this fashion. Also note that there is no risk of generating apparent squeezing after $BS_2$ by deriving the LO from the cavity locking field (e.g. from feedback-squashing of the in-loop field), because the LO and signal fields are both out-of-loop [91].

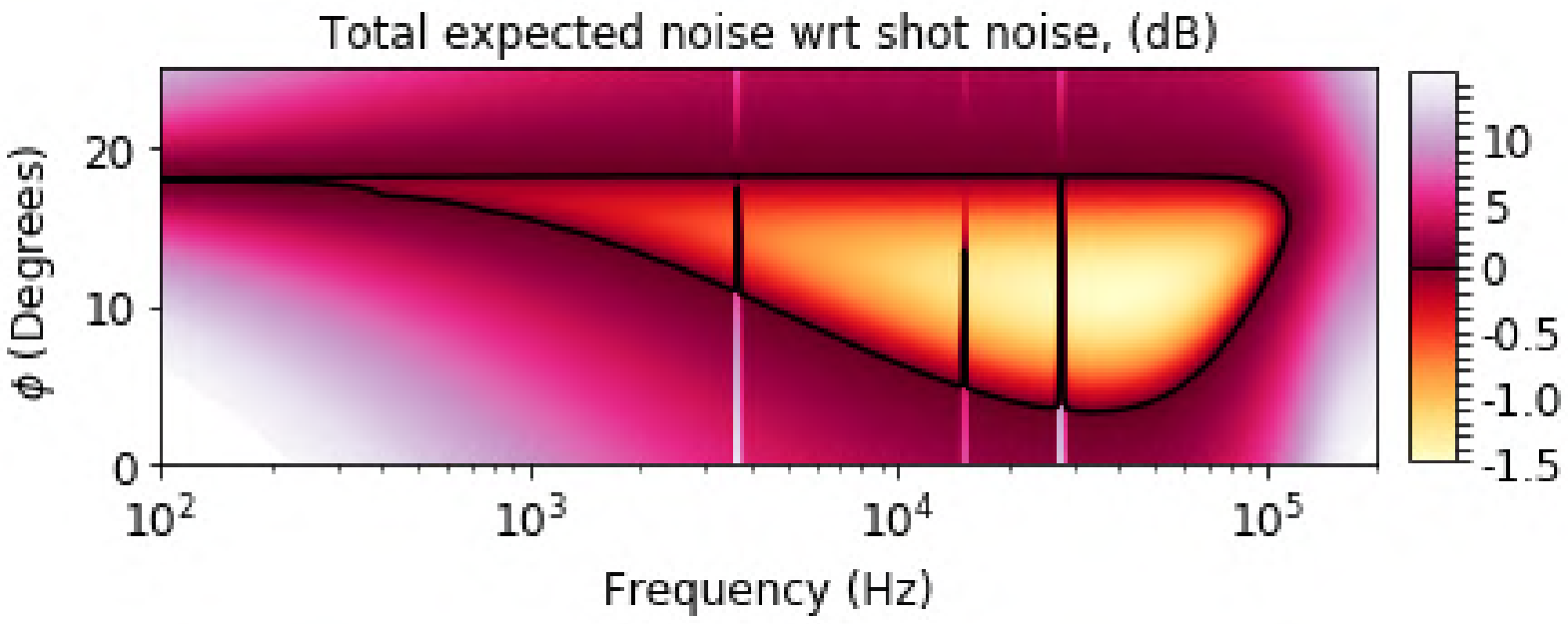

**Figure 7-10:** Expected squeezing with lower loss (100 ppm intracavity) and in the absence of technical noises. The differential phase noise masks the squeezing at low frequencies, whereas the noise injected by the cavity feedback electronics degrades the high frequency side of the correlations.

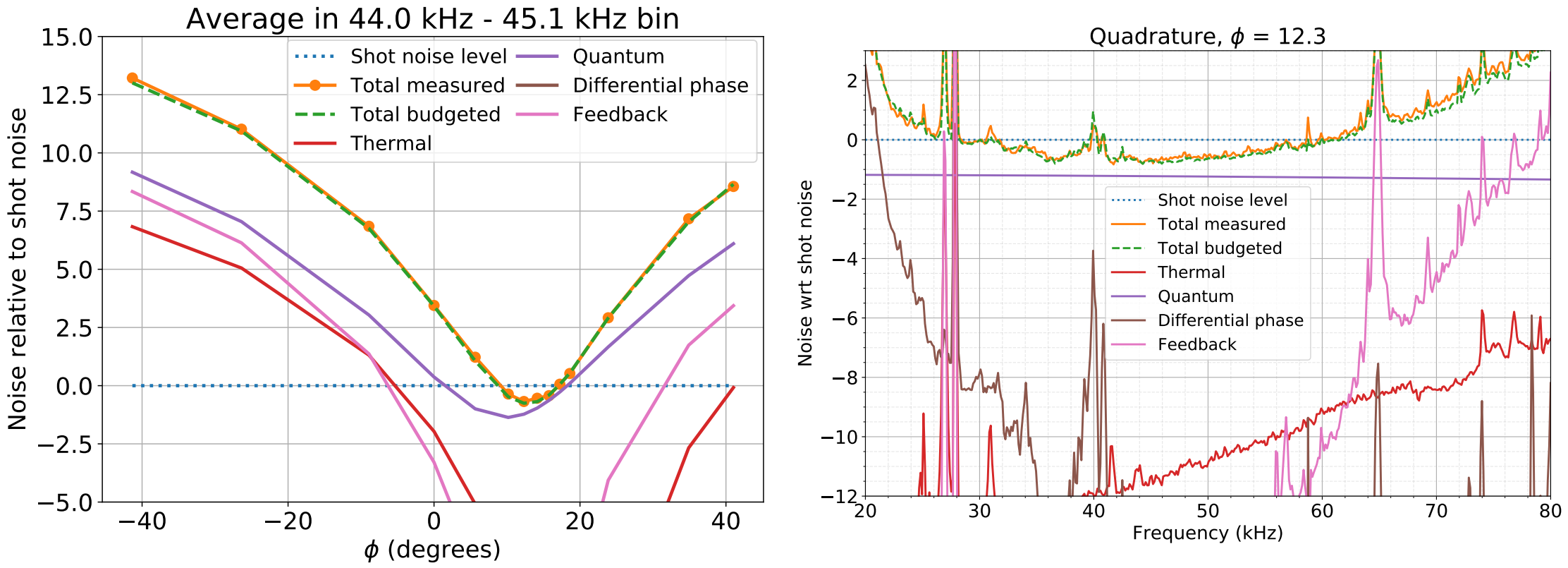

(a)

(b)

**Figure 7-11:** Noise budget: **(a)** Measurement and noise budget as a function of quadrature, averaged over a 1 kHz bin. **(b)** Measurement and full noise budget at the squeezing quadrature. The measured noise at 12.3° is shown in orange. Also shown are the contributions from quantum noise (with excess loss) in purple, thermal noise in red, differential phase noise in brown, and cavity-feedback noise in pink. The quadrature sum of all these contributions is shown in dashed green. All noises are relative to shot noise and are shown in dBs.

Additionally, displacement fluctuations that are relative between the LO and the signal path cause an effective phase noise in the measurement. We refer to this as the differential phase noise, and we measure it by analyzing the measured noise at 17°. At this quadrature, all displacement noises including the feedback noise are canceled, and the quantum noise contribution is at the shot noise level. So we attribute all noise above shot noise at 17° to this relative phase noise. We calculate the contribution of phase noise in all other quadratures by assuming that it is maximum at 90° quadrature and scaling it sinusoidally.

Finally, we are left with excess loss in the detection path. We fit this loss by adding a frequency and quadrature independent loss to the noise budget. We find an excess loss of $22 \pm 1\%$, which agrees with the measured loss of $21 \pm 8\%$. Note that optical loss is the only effect where a single scalar would be sufficient to explain the measurement over all quadratures and frequencies. All of the above contributions to the noise budget are shown in Fig. 7-11: as a function of measurement quadrature in Fig. 7-11a, and as a function of frequency in the squeezing quadrature in Fig. 7-11b. The total budgeted noise at each measured quadrature is compared to the measurement in Fig. 7-12

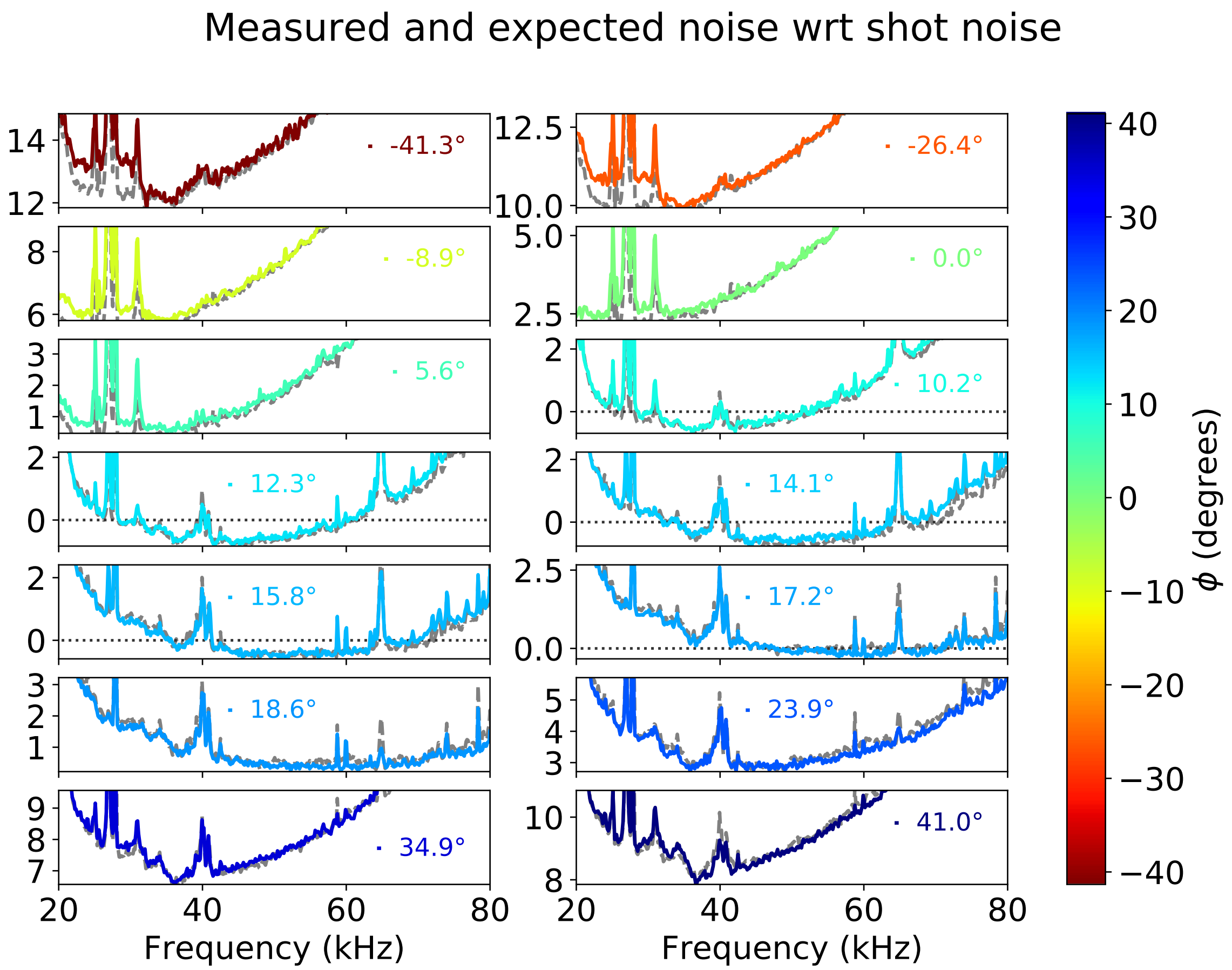

**Figure 7-12:** Measurement with total budgeted noise at various quadratures. Each subplot shows measurement at a particular quadrature. The color of the measurement refers to the respective quadrature, which can be inferred from the colorbar on the right. The grey spectrum on each subplot is the total budgeted noise in that quadrature.

## 7.5 Conclusion

To summarize, we report the first observation of room temperature, broadband optomechanical squeezing at frequencies as low as 30 kHz. This measurement not only paves the way for building miniature, wavelength-agnostic devices to improve the performance of quantum measurements like GW measurements, but also opens up possibilities of exploring broadband quantum correlations at room temperature.

| | |
|---|---|
| $BS_1$ reflectivity | 85 % |
| $BS_2$ reflectivity | 96.5 % |
| Input coupler transmission | 50 ppm |
| Cantilever mirror transmission | 250 ppm |
| Cavity losses | 250 $\pm$ 20 ppm |
| Cavity linewidth HWHM ($\gamma$) | 650 kHz |
| Cavity detuning | $0.33\gamma$ |
| Homodyne visibility | 0.93 |
| Intracavity power | 260 $\pm$ 30 mW |
| Signal power $\left(\left\|t\tilde{\mathrm{E}}_{\mathrm{S}}\right\|^2\right)$ | 58 $\pm$ 4 µW |
| LO power $\left(\left\|r\tilde{\mathrm{E}}_{\mathrm{LO}}\right\|^2\right)$ | 0 - 30 $\pm$ 3 µW |
| Detected power $\left(\left\|\mathrm{E}_{\mathrm{sqz}}\right\|^2\right)$ | 49 $\pm$ 3 µW |
| Detection inefficiency and extra losses | 21 $\pm$ 8 % |

**Table 7.1:** Experimental parameters determined from measurements in the lab

# Chapter 8

# Outlook

## Contents

## 8.1 Enhancing squeezing

To improve the squeezing performance, there are many things that are yet to be done:

- **Technical noise** While the current amount of squeezing in our setup (Fig. 7-7) is limited by extraneous technical noises, in the absence of those noises, the squeezing would be limited by intracavity losses. This expected squeezing in the absence of extraneous noise is shown in Fig. 7-10. After elimination of differential phase noise between the LO and cavity, the squeezing would extend to much lower frequencies. The differential phase noise injected into the system can be reduced by picking-off and recombining the LO inside the vacuum where the OM cavity is situated. The feedback noise can be subtracted by monitoring the actuator signal. A more optimized feedback scheme could also be implemented that injects lower noise, e.g. a feedback tuned for the measured quadrature's transfer function with respect to an appropriate input quadrature that would couple minimum noise to the squeezing signal [66].

- **Mechanical oscillators** We have around one hundred designs of mechanical oscillators on the latest generation chips (Table 4.4), only one of which has been explored experimentally. From the design studies, we suspect there are other devices on the chip that should have better performance – more squeezing and at lower frequencies (Fig. 7-10). We also have devices on the chip that will allow us to directly measure thermal noise due to thermoelastic damping (TED). TED increases thermal noise, but our current finite-element models are unable to predict TED correctly. So, experimental observation will help characterize TED and inform future designs.

- **Design optimization** In the future designs, we can also explore combined optimization of the optical and mechanical design, instead of keeping one fixed. Since we now have a better understanding of how optical and mechanical properties contribute to squeezing, and we have developed numerical toolboxes to handle each, one can combine the optimization to design a device that is optimized over the full parameter space instead of subspaces. For example, it might be possible that two devices of different mechanical design would have the same squeezing behavior for different mirror reflectivities. This kind of optimization has not been addressed in this work. We first find the best mirror reflectivities by using a fixed geometry, and then use those reflectivities to find best geometries. One can imagine an iterative version of this process leading to more gains. Another aspect of the devices that is insignificant for this first order optimization, but should be included in the future, is the change in thickness and hence mass of the oscillator due to change in mirror reflectivity (different reflectivities have different number of GaAs and AlGaAs layers). Finally, there might also be better configurations with respect to the beam spot size on the mirror. It could be that different geometries would be at their best for a different beam size (see Appendix A), which is currently kept fixed for all geometries in the optimization.

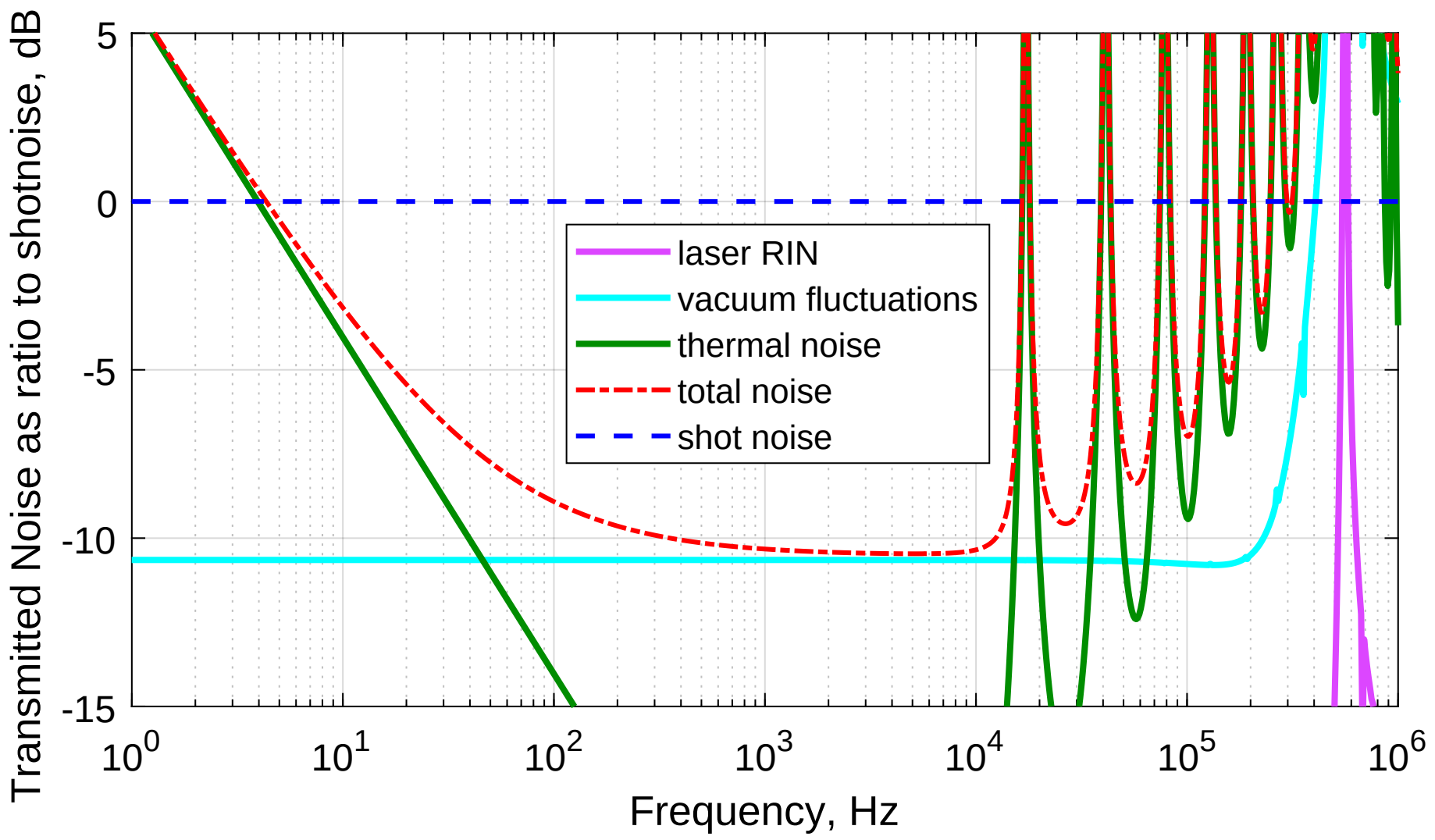

**Figure 8-1:** Example of squeezing prospect in the future: With technological advances in materials and resonator engineering, we can imagine lower mechanical loss and higher power handling. Assuming a ten times higher quality factor (200000) and ten times higher power handling (5 W), we see this system will be able to produce 10 dB squeezing. We have assumed a detuning of half-linewidth, an input mirror of transmission 10 ppm, end mirror of transmission 1200 ppm, and cavity losses of 30 ppm.

- **Materials and fabrication** Further advancements in material properties and fabrication techniques would also allow us to lower the thermal noise. Using techniques like stress dilution may also enable higher quality factors, leading to lower thermal noise [97, 98, 99, 100]. One such prospect for squeezing from such material advances is shown in Fig. 8-1, where we have assumed a 10 times higher quality factor and 10 times more power handling.

## 8.2 Application to precision measurement

- **Next generation GW detectors** As we get close to achieving the design sensitivity of Advanced LIGO, we must also conceptualize GW detectors that can achieve far beyond the sensitivity of LIGO. Three such design studies have been conducted so far, namely the Einstein Telescope [29], Voyager [27], and Cosmic Explorer [28]. These proposals study various aspects of optimizing for sensitivity by targeting reduction of the dominant noise sources in LIGO. Once such noise source is coating thermal noise (CTN), which is predicted to limit Advanced LIGO sensitivity near 100 Hz. In order to lower CTN, new coating as well as mirror materials must be researched. Changing the test-mass material could be accompanied with change of the laser wavelength, especially since some of the candidates like Silicon are not transparent at 1064 nm. There are also additional considerations of thermal distortion, detection efficiency, laser systems, scattering, and loss which play a role in the choice of laser wavelength [27]. An OM squeezer not only has the potential to be compact and robust, but also provides a wavelength-agnostic squeezed light source without requiring the search for a new non-linear optical material.

- **Bright squeezing** An OM squeezer innately generates a bright squeezed field. This means an OM squeezed field comes with its own internal phase reference, hence eliminating the need for an extra coherent phase locking field [101, 22]. For instance, if the state after $BS_2$ (Fig. 7-1) was sent to a precision measurement apparatus, it would include the carrier field which provides a self-referenced signal to allow locking to the correct squeezing quadrature. In contrast, vacuum squeezed states generated by a nonlinear crystal have to be accompanied by an extra, usually frequency-shifted optical beam that keeps track of the squeezing quadrature [101, 22].

This system, as a squeezed light source, or as a source of other quantum states like entangled states, makes a compact, robust quantum resource at room temperature. In addition to OM squeezing, this system also opens up a broad range of possibilities for quantum measurement on multimode optomechanical systems that exhibit quantum mechanical behavior in a thermal bath [102, 103, 104, 105], as well as quantum-mechanically calibrated temperature [41, 76] measurements. More generally, it brings us one step closer to experimentally investigating the interplay of quantum systems with classical decoherence mechanisms like thermal fluctuations and gravitational forces [106, 107, 108, 109, 110, 111, 112, 113].

# Appendix A

# Susceptibility

## Contents

In this appendix we will delve into some of the details of the mechanical behaviour of multimode oscillators. We will talk about their response to radiation pressure force and their contribution to thermal noise driven displacement noise. This is because most harmonic oscillators have more than one mechanical mode, and if the frequency band of interest is higher than the fundamental mode, the response of the oscillator is affected by the higher order modes. The analysis can be simplified by assigning an effective mass to each mechanical mode and then adding the response from each of the modes independently. This effective mass is called modal mass, the heavier a mode's modal mass, the lesser its contribution to the displacement for a given amount of force [114] [35]

## A.1 Generalised coordinates and forces

I will adopt the following notation : $\zeta_n(t)$ is the generalized normal mode of the motion, $Q_n(t)$ is the corresponding generalized force and $\vec{\psi}_n(\mathbf{r})$ is the displacement field associated with the $n^{th}$ normal mode where $\mathbf{r}$ is the position of a particular volume element and spans the body of the oscillator.

The displacement of a particular mass element located at $\mathbf{r}$ at a time $t$ is given by

$$\vec{R}(\mathbf{r}, t) = \sum_{n=0}^{N} \vec{\psi}_n(\mathbf{r})\zeta_n(t) \tag{A.1}$$

Assuming the wave functions $\psi's$ are mass normalized, i.e.

$$\int \rho(\mathbf{r})\ \vec{\psi}_m(\mathbf{r}) \cdot \vec{\psi}_n(\mathbf{r})\mathrm{d}^3\mathbf{r} = \delta_{nm}, \tag{A.2}$$

we can write the equations of motion for the generalized coordinates as

$$\ddot{\zeta}_n + \Omega_n^2\zeta_n = Q_n^e. \tag{A.3}$$

Here $\rho(\mathbf{r})$ is the mass density of the material and the generalized force $Q_n^e$ for the external force density

$\vec{F}^e(\mathbf{r},t)$ (in units of $N/m^3$) is given by

$$Q_n^e(t) = \int \vec{F}^e(\mathbf{r},t)\cdot\vec{\psi}_n(\mathbf{r})d^3\mathbf{r}. \tag{A.4}$$

The damping term is obtained by a generalized force of the dissipative force density for viscous damping,

$$\vec{F}^v(\mathbf{r},t) = -\rho(\mathbf{r})\sum \Omega_i\phi_i\vec{\psi}_i(\mathbf{r})\dot{\zeta}_i(t) \tag{A.5a}$$

$$Q_n^v(t) = \int \vec{F}^v(\mathbf{r},t)\cdot\frac{\partial\vec{R}}{\partial\zeta_n}\mathrm{d}^3\mathbf{r} \tag{A.5b}$$

$$= -\int \rho(\mathbf{r})\sum_{i=1}^{N}\Omega_i\phi_i\ \vec{\psi}_i(\mathbf{r})\dot{\zeta}_i(t)\cdot\vec{\psi}_n(\mathbf{r})\mathrm{d}^3\mathbf{r} \tag{A.5c}$$

$$= -\Omega_n\phi_n\ \dot{\zeta}_n(t). \tag{A.5d}$$

Similarly, the damping term for internal damping can be obtained by a generalized force of the dissipative force density for internal damping,

$$\vec{F}^s(\mathbf{r},t) = -i\rho(\mathbf{r})\sum \Omega_i^2\phi_i\ \vec{\psi}_i(\mathbf{r})\zeta_i(t) \tag{A.6a}$$

$$Q_n^s(t) = \int \vec{F}^s(\mathbf{r},t)\cdot\frac{\partial\vec{R}}{\partial\zeta_n}\mathrm{d}^3\mathbf{r} \tag{A.6b}$$

$$= -i\int \rho(\mathbf{r})\sum_{i=1}^{N}\Omega_i^2\phi_i\ \vec{\psi}_i(\mathbf{r})\zeta_i(t)\cdot\vec{\psi}_n(\mathbf{r})\mathrm{d}^3\mathbf{r} \tag{A.6c}$$

$$= -i\Omega_n^2\phi_n\ \zeta_n(t). \tag{A.6d}$$

In Eqs. (A.5d) and (A.6d), $\phi_n$ is the loss factor for the n-th mode, and the quality factor for each mode is $Q_n = 1/\phi_n$. Since the dominating damping mechanism in our system has been found to be structural (Figs. 1-7 and 6-7 and Section 6.3.1), let's add the damping term Eq. (A.6d) to Eq. (A.3) to get the full damped and forced equations of motion

$$\ddot{\zeta}_n + i\Omega_n^2\phi_n\zeta_n + \Omega_n^2\zeta_n = Q_n^e \tag{A.7}$$

We will work in the frequency domain so lets rewrite all of these equations in the Fourier domain,

$$\tilde{\mathbf{R}}(\mathbf{r},\Omega) = \sum_{n=0}^{\infty}\vec{\psi}_n(\mathbf{r})\tilde{\zeta}_n(\Omega), \tag{A.8a}$$

$$-\Omega^2\tilde{\zeta}_n + (1+i\phi_n)\Omega_n^2\tilde{\zeta}_n = \tilde{Q}_n^e, \tag{A.8b}$$

$$\tilde{Q}_n^e(\Omega) = \int \tilde{\mathbf{F}}^e(\mathbf{r},\Omega)\cdot\vec{\psi}_n(\mathbf{r})\mathrm{d}^3\mathbf{r}. \tag{A.8c}$$

Let's define a generalized susceptibility $\chi_n(\Omega)$ as the ratio of generalized normal mode to generalized external force,

$$\chi_n(\Omega) = \frac{\tilde{Q}_n^e(\Omega)}{\tilde{\zeta}_n(\Omega)}. \tag{A.9}$$

## A.2 Laser weighted displacement (LWD) measurement

Now that we have the formalism setup, we can start applying it to our particular case of measurement of cavity length change due to radiation pressure force and thermal noise. In a Fabry-Perot cavity, the field is going to have a spacial profile, let's call it $I(x,y)$ and one will measure the displacement as sensed by this beam. Since the beam is perpendicular to the oscillator, the length change sensed by the optical beam can be written as

$$\tilde{L}(\Omega) = \frac{1}{I_0}\int \tilde{Z}(\mathbf{r},\Omega)I(x,y)\delta(z-z_0)\mathrm{d}^3\mathbf{r} \tag{A.10}$$

where $z_0$ is the coordinate of the reflecting surface, $I_0 = \int I(x,y)dxdy$ is the total power incident, and $\tilde{Z}$ is the z-component of $\tilde{\mathbf{R}}$. Simplifying using Eq. (A.1),

$$\begin{aligned}\tilde{L}(\Omega) &= \frac{1}{I_0}\int \sum_n \psi_n^z(\mathbf{r})\tilde{\zeta}_n(\Omega)I(x,y)\delta(z-z_0)\mathrm{d}^3\mathbf{r} \\ &= \frac{1}{I_0}\sum_n \tilde{\zeta}_n(\Omega)\int \psi_n^z(\mathbf{r})I(x,y)\delta(z-z_0)\mathrm{d}^3\mathbf{r} \\ &= \sum_n \tilde{\zeta}_n(\Omega)N_n.\end{aligned} \tag{A.11}$$

Here we have defined the integral $N_n = \frac{1}{I_0}\int \psi_n^z(\mathbf{r})I(x,y)\delta(z-z_0)\mathrm{d}^3\mathbf{r}$ as the laser weighted displacement for the mode n.

## A.3 Radiation pressure

For radiation pressure, the profile of the applied force is the same as the beam profile, ie.

$$\tilde{\mathbf{F}}^{RP}(\mathbf{r},\Omega) = \frac{\tilde{F}_0(\Omega)}{I_0}I(x,y)\delta(z-z_0)\hat{z} \tag{A.12}$$

where $\tilde{F}_0(\Omega)$ is the total force applied, given by the optomechanical transfer functions.

$$\int \tilde{\mathbf{F}}^{RP}(\mathbf{r},\Omega)\mathrm{d}^3\mathbf{r} = \tilde{F}_0(\Omega)\hat{z} \tag{A.13}$$

Now from Eqs. (A.8c), (A.9) and (A.11) we can get the cavity length change caused by radiation pressure force as sensed by the optical beam:

$$\tilde{L}^{RP}(\Omega) = \sum_n N_n\tilde{Q}_n^{RP}(\Omega)\chi_n(\Omega) \tag{A.14a}$$

$$= \sum_n N_n\chi_n(\Omega)\int \tilde{\mathbf{F}}^{RP}(\mathbf{r},\Omega).\vec{\psi}_n(\mathbf{r})\mathrm{d}^3\mathbf{r} \tag{A.14b}$$

$$= \tilde{F}_0(\Omega)\sum_n N_n^2\chi_n(\Omega). \tag{A.14c}$$

## A.4 Thermal noise

From the fluctuation dissipation theorem we get the spectral density for fluctuations in the normal mode [35]

$$G_{\zeta_n}(\Omega) = \frac{-2kT}{\pi\Omega}Im[\chi_n(\Omega)]. \tag{A.15}$$

We can use this and Eq. (A.11) to find the effective fluctuations in cavity length. Since $\zeta's$ are normal modes, there is no cross terms between them and the power spectral density of length fluctuations is just the quadrature sum of the power spectral density of each mode scaled by the normalization factor for the measurement.

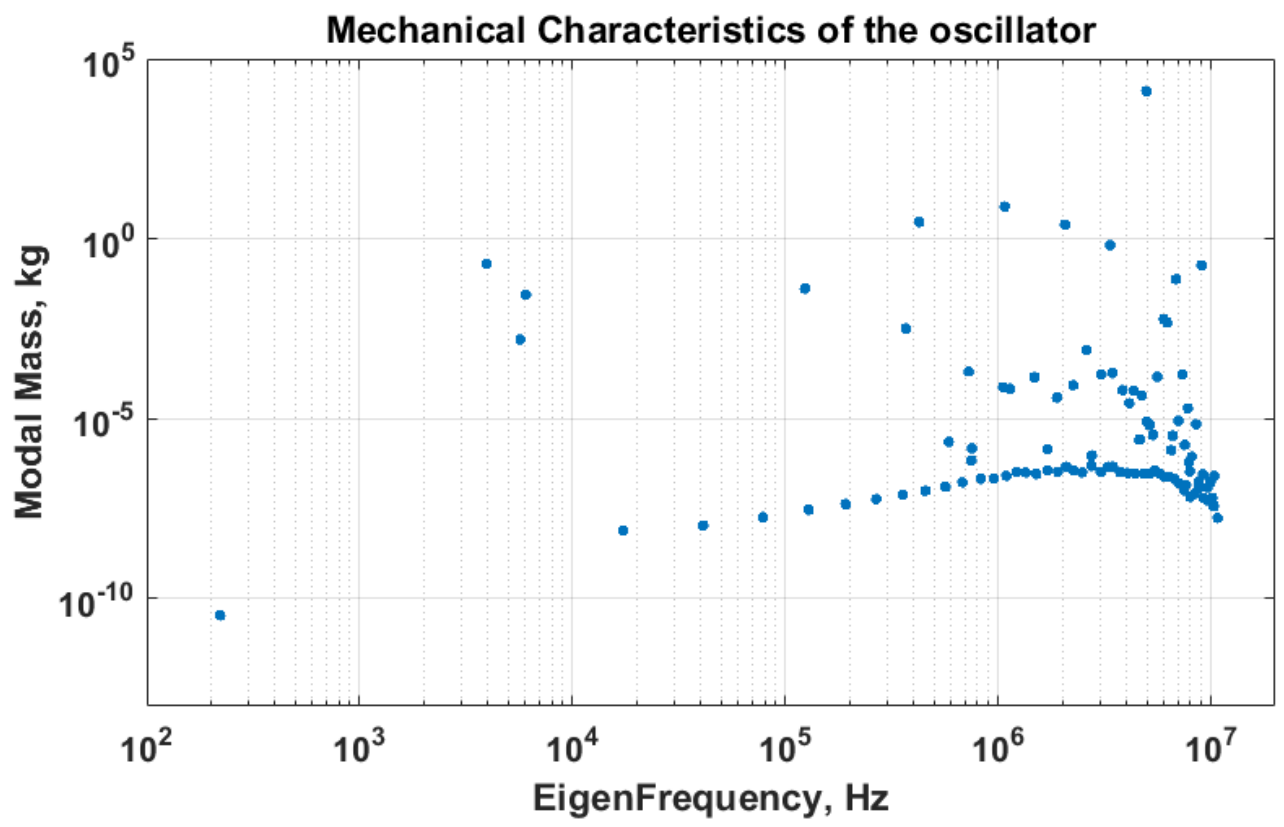

**Figure A-1:** Eigenmodes and their respective modal masses for a typical oscillator design.

$$
\begin{aligned}
G_L(\Omega) &= \sum_n G_{\zeta_n} N_n^2 && \text{(A.16a)}\\
&= -\sum_n \frac{2kT}{\pi\Omega} Im[\chi_n(\Omega)] N_n^2 && \text{(A.16b)}\\
&= \sum_n \frac{2kT}{\pi\Omega} N_n^2 \frac{\phi_n \Omega_n^2}{(\Omega_n^2-\Omega^2)^2+\Omega_n^4\phi_n^2} && \text{(A.16c)}
\end{aligned}
$$

## A.5 Modal Mas

We would now like to write the effective susceptibility for external force to displacement by using an effective mass for each mode. If the normalization in Eq. (A.2) is not given, we can renormalize the modes such that if we had wave functions that instead obeyed a general orthogonality condition,

$$\int \rho(\mathbf{r})\ \vec{\psi}_m(\mathbf{r}) \cdot \vec{\psi}_n(\mathbf{r}) \mathrm{d}^3\mathbf{r} = \delta_{mn} M_n, \tag{A.17}$$

we could define new wave functions $\vec{\psi\prime}_n = \frac{\vec{\psi}}{\sqrt{M_n}}$ that satisfy the nomalization in Eq. (A.2). Thus Eqs. (A.14c) and (A.16a) would get a factor of $\frac{1}{M_n}$ on the right hand side. Then we can write down the modal mass for each mechanical mode of the oscillator for both the radiation pressure and thermal cases, $m_n = \frac{M_n}{N_n^2}$. This modal mass enables us to write the following relations for the radiation pressure force and thermal noise spectral density:

$$
\begin{aligned}
L^{RP}(\Omega) &= F_0(\Omega)\chi(\Omega) && \text{(A.18a)}\\
\chi(\Omega) &= \sum_n \frac{1}{m_n(-\Omega^2+\Omega_n^2(1+i\phi_n))} && \text{(A.18b)}\\
G^{\mathrm{th}}(\Omega) &= \sum_n \frac{2kT}{\pi m_n \Omega} \frac{\phi_n \Omega_n^2}{(\Omega_n^2-\Omega^2)^2+\Omega_n^2\phi_n^2} && \text{(A.18c)}
\end{aligned}
$$

## A.6 Application to numerical simulations

A finite element analysis of each of the oscillator geometries gives us the frequencies $\Omega_n$ and displacement fields $\psi_n$ for each normal mode. We can then find their modal masses from the above treatment, by using the FEM solution to obtain

$$N_n = \frac{1}{I_0} \int \psi_n^z(\mathbf{r}) I(x, y) \delta(z - z_0) \mathrm{d}^3\mathbf{r} \tag{A.19a}$$

$$M_n = \int \rho(\mathbf{r}) \; \vec{\psi}_n(\mathbf{r}) \cdot \vec{\psi}_n(\mathbf{r}) \mathrm{d}^3\mathbf{r} \tag{A.19b}$$

$$m_n = \frac{M_n}{N_n^2}. \tag{A.19c}$$

For the particular oscillator used in this paper, the modal masses and eigen frequencies are shown in Fig A-1. We use a frequency independent loss $\phi_n = 1/20000$ with structural damping. Once we have $\Omega_n$, $\phi_n$, and $m_n$, we use those to calculate $\chi(\Omega)$ and $G^{\mathrm{th}}(\Omega)$ which are then in the numerical simulation in Chapters 3 and 4 to study the OM system.

# Appendix B

# Radiation Pressure Locking

## Contents

This chapter is based on Ref. [38], and much of that manuscript is reproduced here verbatim.

### Abstract

We describe and demonstrate a method to control a detuned movable-mirror Fabry-Pérot cavity using radiation pressure in the presence of a strong optical spring. At frequencies below the optical spring resonance, self-locking of the cavity is achieved intrinsically by the optomechanical (OM) interaction between the cavity field and the movable end mirror. The OM interaction results in a high rigidity and reduced susceptibility of the mirror to external forces. However, due to a finite delay time in the cavity, this enhanced rigidity is accompanied by an anti-damping force, which destabilizes the cavity. The cavity is stabilized by applying external feedback in a frequency band around the optical spring resonance. The error signal is sensed in the amplitude quadrature of the transmitted beam with a photodetector. An amplitude modulator in the input path to the cavity modulates the light intensity to provide the stabilizing radiation pressure force.

## B.1 Introduction

Cavity optomechanics, the interaction between electromagnetic radiation and mechanical motion, provides an ideal platform for measuring mechanical displacements and preparing and detecting mechanical resonators in the quantum regime [21]. In a simple cavity-coupled optomechanical system, the mechanical oscillator is driven by the radiation pressure force exerted by the probing laser field. The fluctuations in the radiation pressure force due to power fluctuations modulate the motion of the mechanical oscillator, effectively changing the length of the cavity and modifying the resonance condition of the cavity. This leads to changes in the optical power circulating inside the cavity, thus cyclically modulating the radiation pressure force exerted on the mechanical oscillator. This feedback results in the optical spring effect.

The optical spring effect was first discussed for Fabry-Perot cavities by Braginsky [115, 116]. Braginsky *et al.* [117], Buonanno and Chen [118], and Harms *et al.* [70] proposed using the optical bar and optical spring to enhance the sensitivity of gravitational wave detectors. Over the past two decades, many experiments have observed the optical spring in a variety of systems [119, 120, 121, 44, 122, 123, 124, 125, 126, 21, 55] and used it to optically cool mechanical resonators [127, 128, 129, 130, 131, 132, 133, 134, 135]. Furthermore, proposals to increase the sensitivity of Michelson-type gravitational wave detectors using the optical spring effect have included adding a signal-recycling cavity [136, 137, 39], using a detuned cavity to amplify the interferometric signal [138], adding a signal-extraction cavity or resonant sideband extraction [139, 140], and dynamically tuning the cavities to follow a gravitational wave chirp signal [141, 142]. Signal-recycling and signal-extraction cavities have been used in the GEO 600 [143] and Advanced LIGO [1] gravitational wave detectors, and are planned to be used in Advanced VIRGO [2], and KAGRA [3].

For a blue-detuned high-finesse optomechanical Fabry-Pérot cavity in which the cavity's resonance frequency is less than the laser frequency, the system's effective mechanical resonance frequency is shifted to a higher frequency than the mechanical oscillator's eigenfrequency via the addition of the optical spring constant. This leads to self-stabilization of the optomechanical system at frequencies below the optical spring frequency [20]. At the optical spring frequency, however, the lag in optical response due to the round trip optical delay leads to a dominating anti-damping force that renders the system unstable [40, 120, 121]. Such anti-damping forces normally require active feedback control to stabilize the optomechanical dynamics [40, 120].

Conventionally, detuned cavities are locked by using a simple "side of fringe" locking method. In this method the error signal is obtained from the slope of the cavity intensity profile on a transmission/reflection photodetector. This error signal is filtered and fed back to a piezoelectric actuator on the cavity mirror or to the frequency of the laser. The lock bandwidth is limited by the piezoelectric device's mechanical resonance frequency. The laser frequency modulation on the other hand has more bandwidth, but requires a large actuation range for short length cavities. As an alternative, in previous experiments, we have demonstrated the stabilization of the optomechanical cavity by utilizing the double optical spring effect [55].

In this work, we introduce a new feedback control method to lock a movable mirror Fabry-Pérot cavity using radiation pressure. We have implemented this scheme at two independent experiments at LSU and MIT. This scheme relies on the suppression of external disturbances by having a large in-loop optomechanical gain as a result of the large optical spring constant. This suppression, which is mediated via the radiation pressure force, lowers the fluctuations in cavity length and power. A schematic representation of the method is shown in Fig. B-1, where the error signal is derived from the transmitted power out of the cavity and is used to control the radiation pressure force acting on the cavity by modulating the intensity of the input laser field passing through an amplitude modulator (AM). An optimal error signal is extracted by passing the transmitted field through a filter. This filter comprises of a gain and a band-pass component. The gain and low-pass filter of the servo controller are to stabilize the anti damping on the optical spring. The high-pass filter is to avoid saturation of the AM actuator due to ambient/seismic fluctuations that are largest at low frequencies (below a few kHz for a typical lab environment). These seismically and acoustically driven fluctuations in cavity length are self-stabilized in the optomechanical dynamics due to the high OM gain at frequencies below optical spring.

## B.2 Theoretical framework

In understanding the noise stabilization of a strong optical spring system with feedback, it is informative to view the optical spring itself as a feedback mechanism. In this view, a closed-loop feedback system is formed between the mechanical oscillator and the optical cavity. The mechanical oscillator, with susceptibility $\chi_{\mathrm{m}}$, transduces a force into a displacement. The optical cavity, in turn, transduces the displacement back into a radiation pressure force, forming a closed loop. For simplicity, we consider the frequency dependent susceptibility of a single mechanical resonance at $\Omega_{\mathrm{m}}$, such that

$$\chi_{\mathrm{m}} = \frac{1}{m\left(\Omega_{\mathrm{m}}^2 - \Omega^2 + i\Omega\Gamma_{\mathrm{m}}\right)}, \tag{B.1}$$

where $\Omega$ is frequency, $m$ is the effective mass of the mode of oscillation, $\Gamma_{\mathrm{m}} = \Omega_{\mathrm{m}}/Q_{\mathrm{m}}$ with $\Gamma_{\mathrm{m}}$ and $Q_{\mathrm{m}}$ the mechanical damping and quality factor of the mechanical oscillator, respectively[1].

[1]A more realistic form including multiple resonances could be used instead, but the single resonance susceptibility works well for this analysis. This is because the higher-order modes have a larger effective mass than the fundamental mode due to their poor overlap with the cavity mode and hence don't contribute much to the broadband behavior.

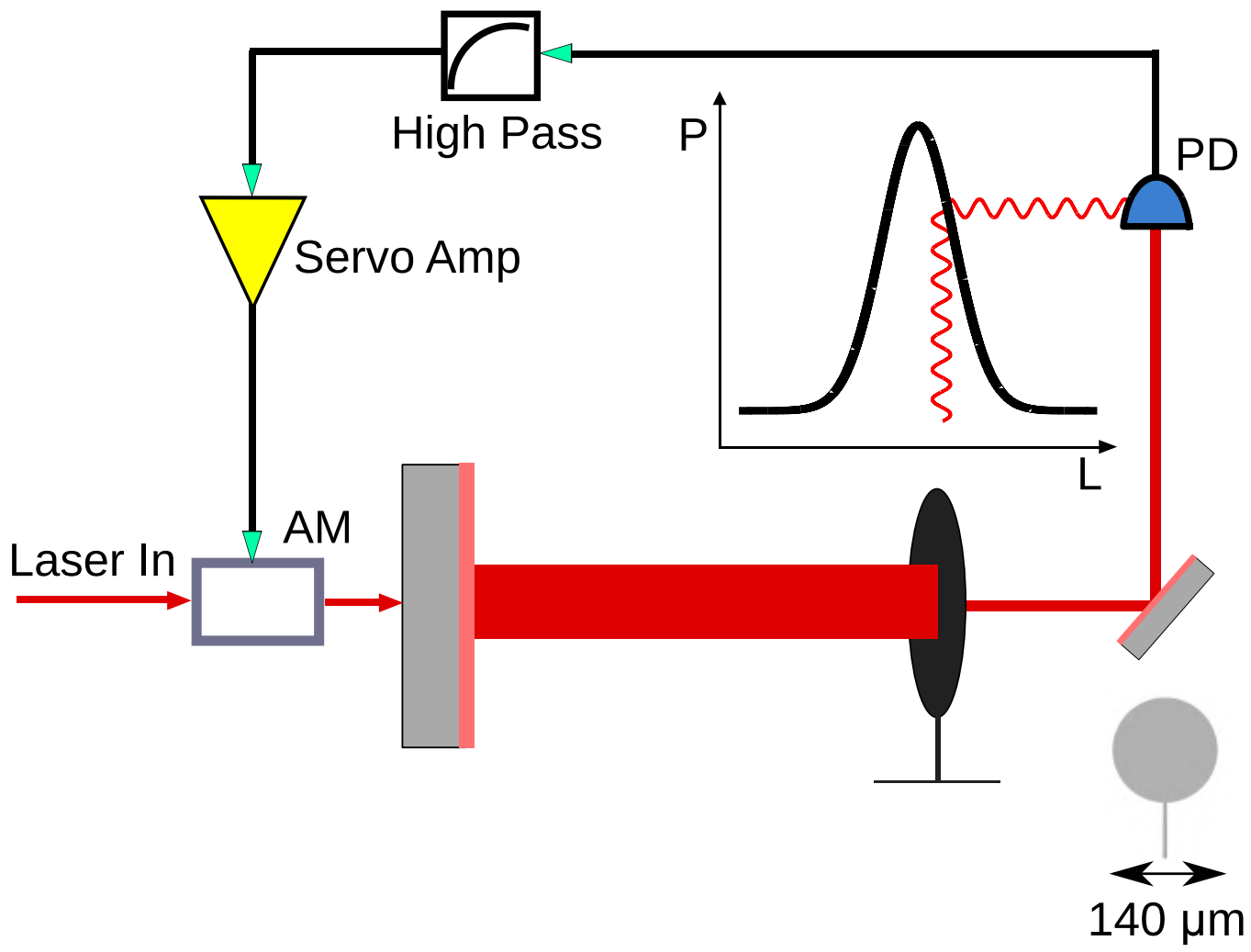

**Figure B-1:** Schematic of the experimental setup. The laser is passed through an amplitude modulator (AM), which modulates the intensity of the field before entering the detuned optomechanical cavity. The light is detected in transmission of the cavity and passed through a high pass filter and servo controller containing a gain and low pass filter component to obtain the error signal which is fed back to the AM. The inset shows the scale of the moveable mirror that forms one of the cavity mirrors.

The open-loop gain pertaining to the cavity's closed-loop response as shown in Fig. B-2 may be given in the limit $\Omega \ll \gamma$ as (see Ref. [66] and Chapter 2)

$$\begin{aligned} G_{\rm os} &= -\frac{32\pi\chi_{\rm m}P_{\rm cav}}{c\lambda_0 T}\frac{\delta_\gamma}{(1+\delta_\gamma^2)}\left(1-\frac{2i\Omega}{\gamma(1+\delta_\gamma^2)}\right) \\ &= -\chi_{\rm m}K_0\left(1-\frac{2i\Omega}{\gamma[1+\delta_\gamma^2]}\right) \\ &= -m\chi_{\rm m}\left(\Omega_{\rm os}^2 - i\Gamma_{\rm os}\Omega\right), \\ &= \frac{\Omega_{\rm os}^2}{(\Omega_{\rm m}^2-\Omega^2+i\Omega\Gamma_{\rm m})}\left(1-i\frac{\Gamma_{\rm os}\Omega}{\Omega_{\rm os}^2}\right) \end{aligned} \tag{B.2}$$

where $P_{\rm cav}$ is the intra-cavity power, $\lambda_o$ is the center wavelength of the laser, $c$ is the speed of light, T is the total fraction of light leaving the cavity via loss and mirror transmissions, $\gamma$ is the half width at half maximum (HWHM) for the cavity optical resonance in rad/s, $\delta_\gamma = \frac{\delta}{\gamma} = \frac{\omega_L-\omega_C}{\gamma}$ is the dimensionless detuning of the laser field from the cavity's resonance, and $K_0$ is the optical spring constant. The optical spring frequency is given by $m\Omega_{\rm os}^2 = K_0$, and its HWHM is $\Gamma_{\rm os} = 2\Omega_{\rm os}^2/\gamma/(1+\delta_\gamma^2)$. The real part of Eq. 2 corresponds to a position dependent restoring force and the imaginary part corresponds to a velocity dependent anti-damping force[2].

The effective susceptibility of the system to a force is then

$$\begin{aligned} \chi_{\rm os} &= \frac{x}{F_{\rm ext}} = \frac{\chi_{\rm m}}{1+G_{\rm os}} \\ &\approx \frac{1}{m}\frac{1}{\Omega_{\rm os}^2-\Omega^2-i\Gamma_{\rm os}\Omega}, \end{aligned} \tag{B.3}$$

where $x$ is the displacement of the resonator, $F_{\rm ext}$, is an external force, and in the last step we assume that the $\Omega_{\rm os} \gg \Omega_{\rm m}$ and $\Gamma_{\rm os} \gg \Gamma_{\rm m}$. At frequencies below the optical spring frequency the ambient motion

[2]This force will be anti-restoring and damping for a red detuned laser.

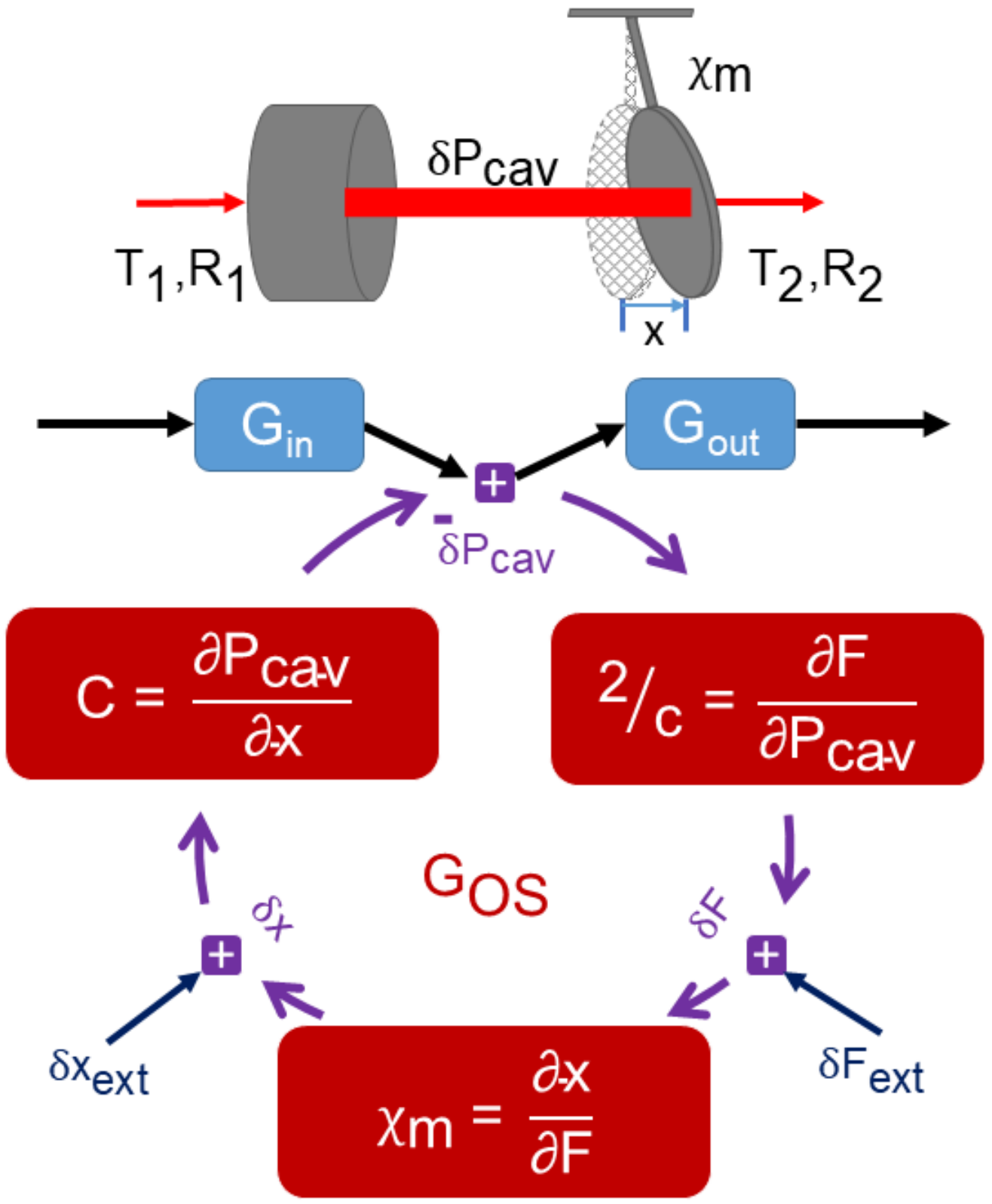

**Figure B-2:** Detailed loop diagram for the cavity's transfer function. The amplitude modulator adds to the intracavity power $P_{\text{cav}}$. The cavity power converts into radiation pressure force $F$, which then converts into cantilever displacement $x$ via its mechanical susceptibility $\chi_{\text{m}}$. The displacement causes a length change for the cavity, leading again to a change in the intracavity power via the cavity response $C$. This forms a closed loop system.

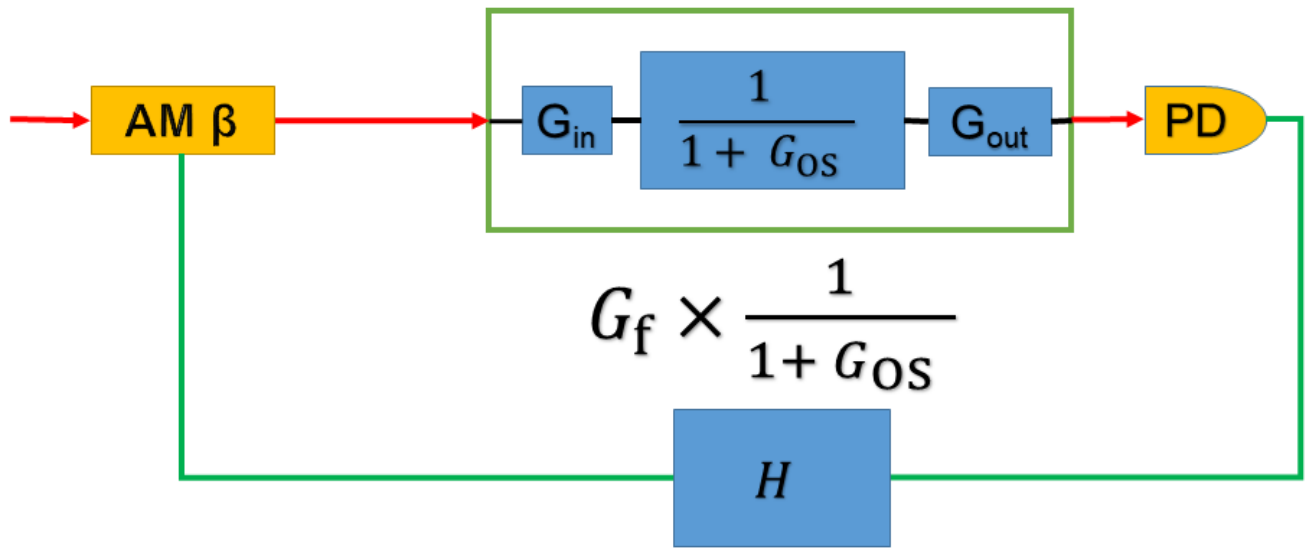

**Figure B-3:** Loop diagram for the feedback $G_{\text{f}}$. $H$ is the response of the high pass filter and servo controller, and $\beta$ is the response of the amplitude modulator. $\frac{1}{1+G_{\text{os}}}$ is the closed loop response of the cavity's optical spring system, as shown in Fig B-2. $G_{\text{in}} = \frac{4T_1}{T^2}\frac{1}{1+\delta_\gamma^2}$ is the transfer function of the power input to power inside the cavity with the effect of the detuning taken into account, and similarly $G_{\text{out}} = T_2$ is the transfer function from cavity power to transmitted power that is measured on the PD. Here $T_1$ is the transmission of the input mirror, $T_2$ is the transmission of the microresonator, and $T$ is the total loss of power from the cavity, in the form of transmission, scattering, absorption, etc. $G_{\text{f}} = G_{\text{out}} \times \text{PD} \times H \times \beta \times G_{\text{in}}$ is calculated by using the measurements of the individual transfer functions.

is therefore reduced by the factor

$$\left|\frac{\chi_{\rm m}}{\chi_{\rm os}}\right| \approx \left|\frac{\Omega_{\rm os}^2}{\Omega_{\rm m}^2 - \Omega^2 + i\Omega\Gamma_{\rm m}}\right|, \tag{B.4}$$

with the approximation assuming $\Omega_{\rm os} \gg \Omega$. This factor of suppression may be made very large if $\Omega_{\rm os} \gg \Omega_{\rm m}$.

In the limit of a large optical spring frequency, the optical spring provides sufficient stabilization to maintain cavity lock. Due to the negative damping (gain) of the optical spring feedback, however, the system is unstable on its own. This can be seen by writing the closed-loop gain in the s-domain by substituting $s = i\Omega$,. The closed-loop gain $G_{\rm cl}$ corresponding to this open-loop gain $G_{\rm os}$ is given by

$$\begin{aligned} G_{\rm cl} &= \frac{1}{1+G_{\rm os}} \\ &= \frac{\Omega_{\rm m}^2 + s^2 + s\Gamma_{\rm m}}{\Omega_{\rm m}^2 + s^2 + s\Gamma_{\rm m} + \Omega_{\rm os}^2 - s\Gamma_{\rm os}} \end{aligned} \tag{B.5}$$

From the above expression, one can see that this closed-loop gain has at least one right-half-plane pole[3] and will thus be unstable. This system must be stabilized by an external damping force. The feedback may be localized to frequencies near the optical spring resonance, and its only purpose is to stabilize the unity gain crossing.

The main purpose of the applied feedback $G_{\rm f}$ shown in Fig. B-3 is to change the shape of the phase response of the system so that the system is stable as well as has good stability margins.

Radiation pressure is a natural transducer to stabilize such a system because there is strong coupling of radiation pressure by assumption. In addition, amplitude modulators have higher response bandwidth than piezoelectric actuators and better range than laser frequency modulation. Furthermore, because these systems are typically operated detuned (within a few line-widths to achieve strong optical springs), the transmitted power through the cavity is a natural readout of the cavity motion.

## B.3 Experimental setup

The schematic shown in Fig. B-1 illustrates the experimental setup. The laser field from an NPRO Nd:YAG laser is passed through an amplitude modulator before passing through a half-wave plate and mode-matching lenses en route to the optomechanical cavity. The in-vacuum cavity is 1 cm long and consists of a 0.5-inch (12mm) diameter input mirror with a 1 cm radius of curvature and a microresonator as the second mirror. The input mirror is mounted on a piezoelectric actuator to allow for fine-tuning of the cavity length. The microresonator is fabricated from a stack of crystalline $Al_{0.92}Ga_{0.08}As/GaAs$ layers. It has a diameter of 140 µm and a mass of about 500 ng [78, 65, 61, 62, 55]. The microresonator has a natural mechanical frequency of $\Omega_{\rm m} = 2\pi \times 288$ Hz and a measured mechanical quality factor $Q_{\rm m} = 8000$, which gives $\Gamma_{\rm m} = 2\pi \times 36$ mHz.

The field transmitted through the cavity is detected by a photodetector. The photodetector signal is sent through a high-pass filter and servo controller before being used as the error signal to the amplitude modulator.

## B.4 Results and discussion

To help understand the feedback mechanisms and individual components of the feedback loops, Fig. B-2 and Fig. B-3 show the loop diagrams for the feedback loop $G_{\rm f}$ and the optical spring $G_{\rm os}$. Measurements of the open loop gain, plant transfer function, individual loop gains, and closed loop gain are described below.

In Fig. B-4, the blue curves show the plant transfer function, which is the system we would like to control. We see a peak corresponding to the optical spring at around 75 kHz in the magnitude.

[3]the terms in the denominator are not all of the same sign

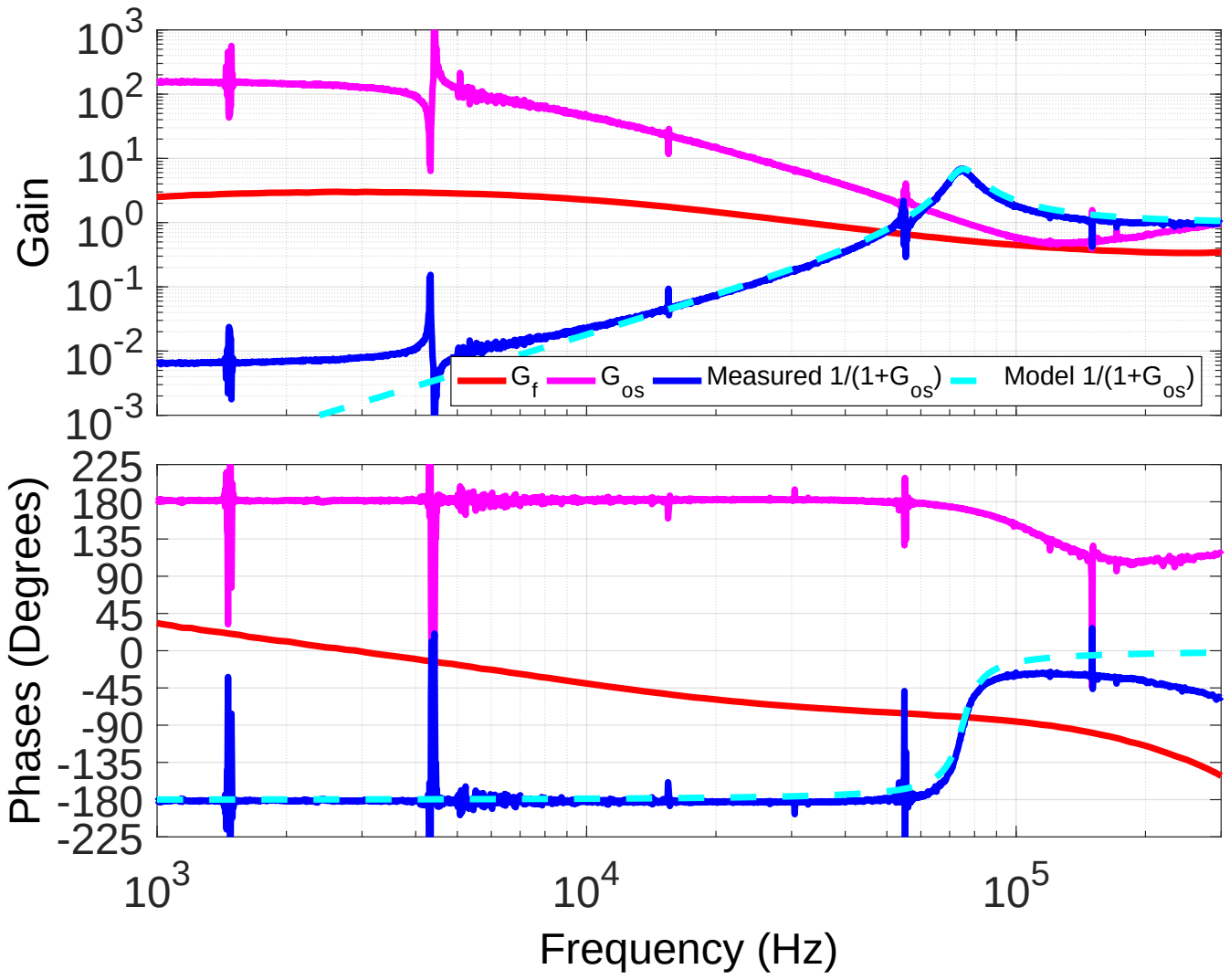

**Figure B-4:** Transfer function measurements of the plant, $G_{CL} = \frac{1}{1+G_{os}}$ shown in blue, the feedback, $G_f$ shown in red, and the open loop optical spring, $G_{os}$ shown in pink. A model for the plant, shown in dashed cyan, is calculated with Eq. B.2 using the measured values for $\Omega_{os}$, $\Omega_m$, and $\Gamma_m$ and setting $\Gamma_{os}$ so that the peak height and width match the measured data. $G_{cl}$ is obtained using the open loop gain, $\frac{G_f}{1+G_{os}}$, shown in Fig. B-5 and dividing out the measured $G_f$. The effect of the optical spring is visible with the peak at 75 kHz and a rise in phase of the plant transfer function. The measurement begins to flatten out below 5 kHz due to other circuitous signal couplings (eg scattered light). $G_{os}$ is then obtained from $G_{cl}$. The applied feedback loop is $G_f$. At 75 kHz, $G_f$ has a magnitude of 0.53 and a phase of $-80°$.

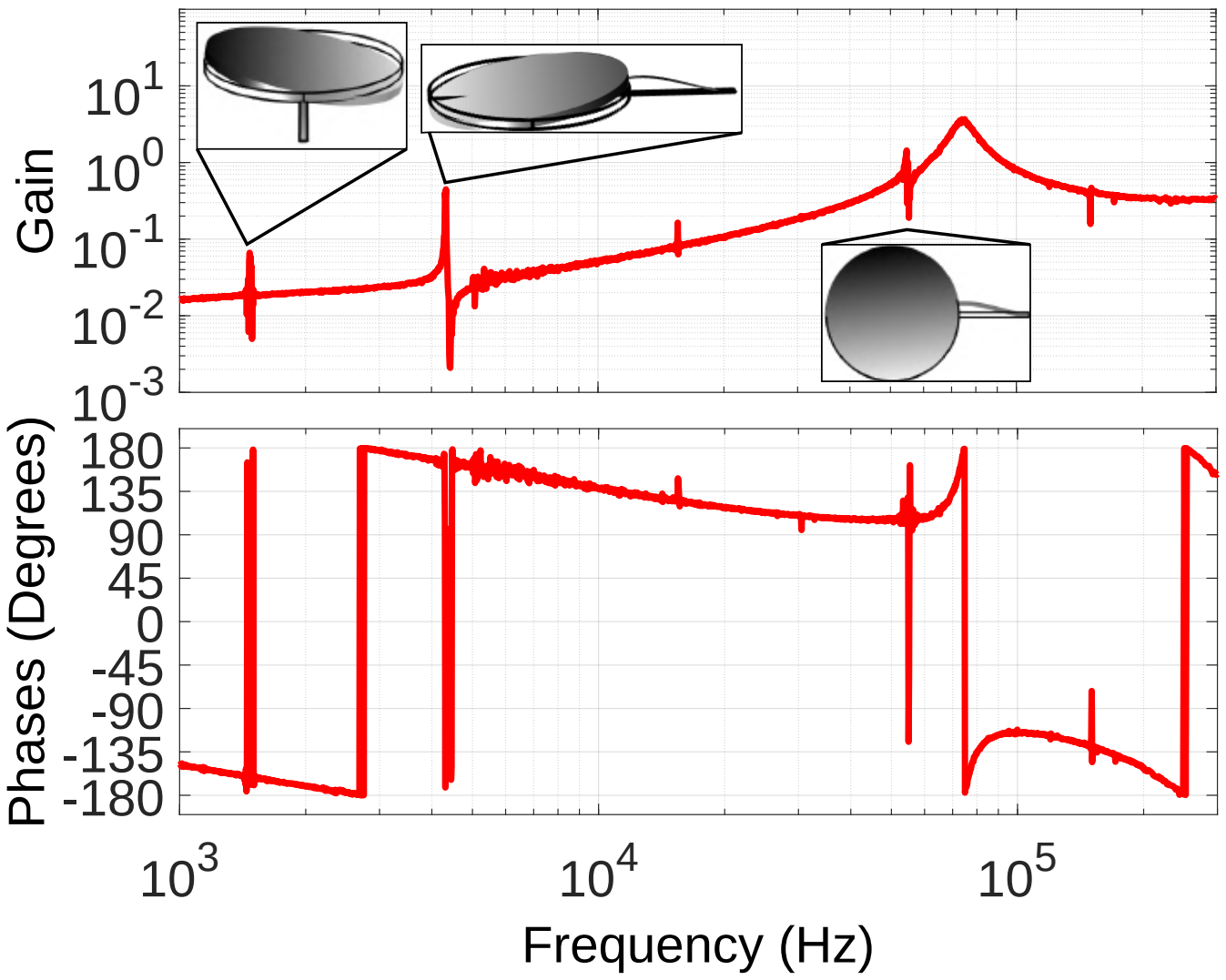

**Figure B-5:** Measurement of the open-loop gain or $\frac{G_f}{1+G_{os}}$ taken by injecting a signal before $H$, done with a circulating power of 0.2 W. Higher-order mechanical modes are visible at 1.4 kHz (yaw), 4.3 kHz (pitch), and 54 kHz (translation and yaw) are shown in the inset. Unity-gain crossings are at 61 kHz and 93 kHz with phases 109° and $-115°$. The gain is 0.34 at 250 kHz where the phase crosses $-180°$. Thus, the system is stable with phase margins of 71° and 65°, respectively, and a gain margin of 9.4 dB.

Since the system is unstable on its own, the plant transfer function is obtained using the open loop gain measurement with the feedback on. This open loop gain is shown in Fig. B-5, and we later divide it by the measured $G_f$, shown in red in Fig. B-4 to obtain the plant transfer function.

Fig. B-4 also shows the transfer function of the $G_{os}$ loop, which is obtained from the open loop transfer function shown in Fig. B-5. The large magnitude of $G_{os}$ at frequencies below the optical spring shows the large suppression that the system's internal response is providing.

The external electronic feedback loop, $G_f$, which is used to stabilize the system, is shown in Fig. B-4 in the red curves. The measurement of $G_f$ is obtained by measuring the response of individual elements in the loop, which includes the photodetector (PD), the high-pass filter and servo controller ($H$), and the amplitude modulator (AM), and multiplying them together. The high-pass filter has a corner frequency at 800 Hz and the servo controller has a P-I corner at 100 kHz with a low frequency gain limit of 20 dB. We chose these values to supply sufficient phase margin while also attenuating the feedback at low-frequencies to avoid saturating the AM actuator. The measurement of the elements of $G_f$ is done without using the cavity, so it gives the correct shape of $G_f$, but does not provide the absolute scaling of the loop because the effect of the cavity is not included. The calibrated $G_f$ is obtained by taking the effect of the cavity into account using the open-loop gain measurement above the optical spring peak where $\frac{1}{1+G_{os}} \approx 1$.

Fig. B-5 shows the measurement of the open loop gain taken by injecting a signal before $H$ (Fig. B-3) and measuring the response after PD. Since the measurement enters the $G_{os}$ loop, the open loop transfer function is given by $\frac{G_f}{1+G_{os}}$. The effect of the optical spring is also visible in Fig. B-5 with a resonance peak at 75 kHz and a falloff with $f^2$ below the optical spring. There are two unity-gain crossings at 61 kHz and 93 kHz with phases of 109° and $-115°$. The gain at 250 kHz where the phase crosses $-180°$ has a magnitude of 0.34. Thus, the system is stable with phase margins of 71° and 65°, respectively, and a gain margin of 9.4 dB. We note that while the $G_f$ shown in Fig. B-4 does produce a stable system, it is not a unique solution. While other solutions for $G_f$ may be more stable, the $G_f$ we use is simple and achieves our goal of stabilizing the system. We also note that the measurement of $G_{os}$ deviates from the expected $f^{-2}$ slope above $\sim$ 100 kHz. This is a result of imperfect measurements of the individual components of the loops, which leads to errors in the subtraction for the transfer functions of $G_{os}$ and $\frac{1}{1+G_{os}}$.

Another result of the dynamics of the optomechanical system is the reduced response to disturbances at frequencies below the optical spring frequency. Ambient motion causes the cavity length to change by $\Delta L \sim 10^{-7}$m, while the cavity linewdith is $\Delta\lambda \sim 10^{-11}$m. It is therefore necessary to suppress the ambient motion in order to operate the cavity. Fig. B-4 shows the optical spring resonance at 75 kHz. According to Eq. B.4, the ambient motion should be reduced at low frequencies by the factor $\left|\frac{\Omega_{os}^2}{\Omega_m^2-\Omega^2+i\Omega\Gamma_m}\right|$.

To verify this calculation, we modulate the laser frequency, which in effect, is the same as introducing a disturbance $\delta x_{ext}$ in Fig. B-2. Fig. B-6 shows a measurement of

$$-\frac{L_p}{f_0}\frac{G_{out} \times C \times PD}{1+G_f+G_{os}} \tag{B.6}$$

taken by modulating the laser frequency and measuring the output of the PD with $\frac{L_p}{f_0}$ the change in the laser frequency for a given change in length for the laser piezo. The amount that low-frequency vibrations are reduced by is calculated by taking the ratio of the value of the measurement above the optical spring frequency where the measurement is flat and the value of the measurement at low frequencies, yielding a suppression of at least 50,000.

The response of the system to an external force is

$$\frac{x}{F_{ext}} = \frac{\chi_m(1+G_f)}{1+G_{os}+G_f}. \tag{B.7}$$

According to Eq. B.7, ambient fluctuations are suppressed by the factor $1+G_{os}+G_f$. Since this factor is in common in Eq. B.6 and Eq. B.7, the laser frequency scan shown in Fig. B-6 is an accurate measure of the suppression of ambient fluctuations.

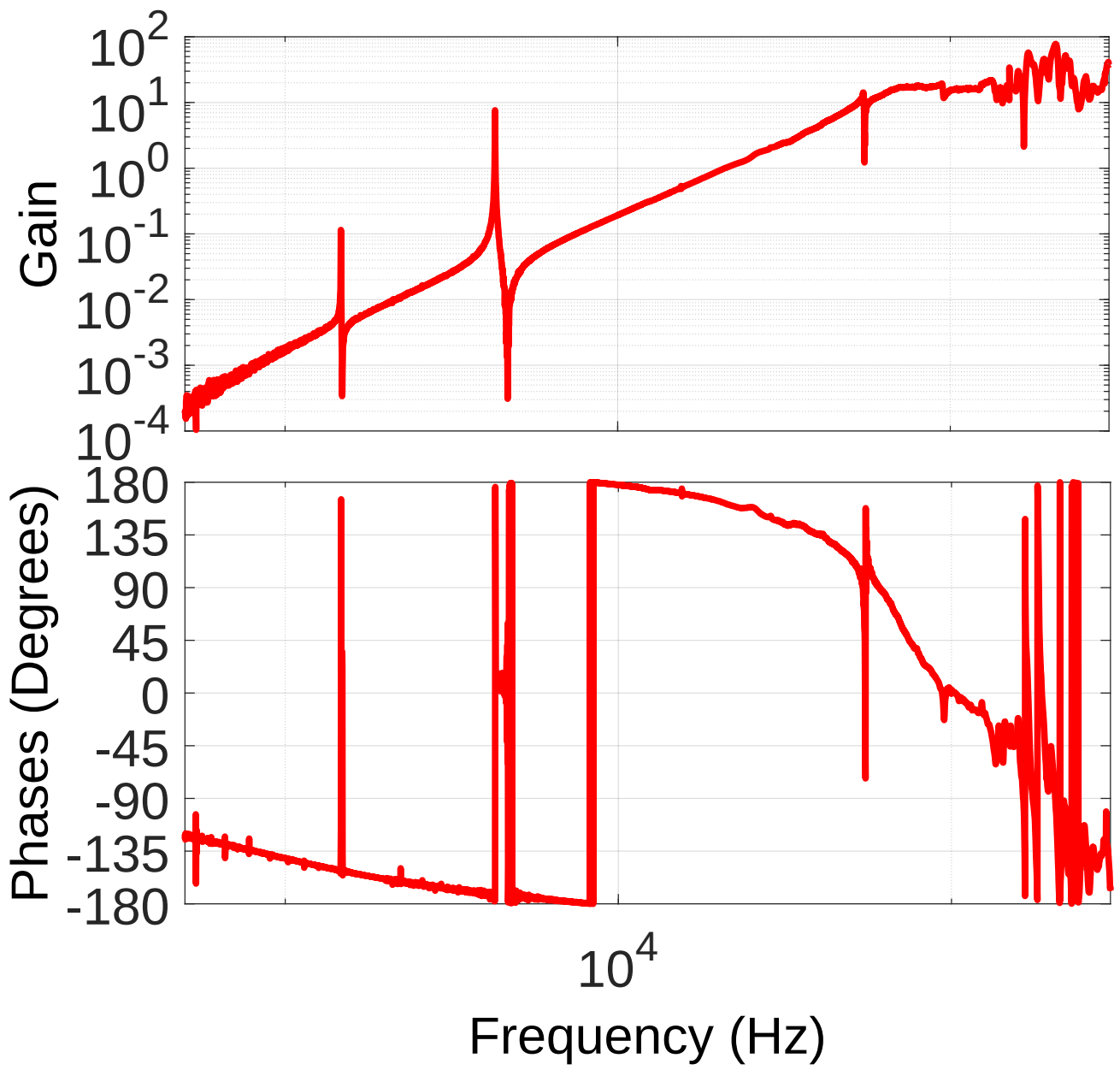

**Figure B-6:** Measurement of the closed-loop response performed by modulating the laser frequency. This plot shows the suppression of low-frequency frequency noise below the optical spring frequency at $\approx 75$ kHz. The amount of suppression is calculated by taking the ratio of the measurement above the optical spring and at a low frequency below the optical spring. Using the values at 100 kHz and 500 Hz, noise is suppressed by a factor of at least 50,000. Higher order mechanical modes are again visible at 1.4 kHz, 4.3 kHz, and 61 kHz.

## B.5 Conclusion

In conclusion, we have demonstrated a stable feedback control method to lock a moveable mirror Fabry-Pérot cavity using radiation pressure. In this scheme, the use of radiation pressure as an actuator provides a large locking bandwidth compared to a piezoelectric device used in the simple "side of fringe" locking. We have experimentally shown that the system is stable and reduces low-frequency disturbances by a factor of at least 50,000. The combination of the stable system and excellent low-frequency noise suppression allows the optomechanical cavity to be operated on time scales of hours to days without losing lock. With the low-frequency noise reduced, we aim to measure broadband quantum radiation pressure noise and ponderomotive squeezing at frequencies relevant to Advanced LIGO. In addition, since the quadrature of the field inside the cavity is actually rotated with respect to the input field, the feedback gain could be increased by modulating in a different quadrature. A modulation in an arbitrary quadrature can be achieved by stitching together two amplitude modulator crystals and adding a relative drive between them [144]. This configuration could be useful if the negative damping is too high to be compensated with a single amplitude modulator.

# Appendix C

# Analytical treatment for OM squeezing

Here I prove the results shown in Section 3.5. The Mathematica notebook used for this derivation is included below. The computation uses the formalism developed in [49], and converts it to the quadrature formalism. In this formalism, the state inside the cavity is a discrete quantized state, while all the input and output fields are under continuous quantization. We first compute all the transfer functions from each input field to each output field, and then use them to get the covariance matrices. Finally, there are a number off approximations made in this calculation, which make the task tenable:

- We assume that the frequency of measurement is much higher than the oscillator's resonance frequency, allowing us to write $F = -M\Omega^2 x$.
- Secondly, we assume that the frequency of measurement is much smaller than other relevant frequency scales in the problem, allowing us to set $\Omega \to 0$.
- Finally, all the $\lambda' s$ are assumed to be small, allowing us to use perturbation theory.

# Ponderomotive squeezing

### References

[1] 10.1103/PhysRevA.88.033805

part of PhD. thesis of Nancy Aggarwal, Department of Physics, Massachussets Institute of Technology, detailed explanations can be found in the thesis.

## Setup

## TFs

```
Qlist = {a, ad, ain3, ain3d, ain2, ain2d, aout2, aout2d, ain1,
   ain1d, aout1, aout1d, x, p, Frad, Fth}(*list of quantity*);
Qlen = Length[Qlist];
Do[Evaluate[Qlist[[i]]] = i, {i, 1, Qlen}];
```

```
TFlist = {
    {a, a} → - (γ - ⅈ Δ) (- ⅈ Ω)^-1 (*Δ=ω0-ωc*),
    {ad, ad} → - (γ + ⅈ Δ) (- ⅈ Ω)^-1,
    {a, x} → ⅈ g (- ⅈ Ω)^-1,
    {ad, x} → - ⅈ g (- ⅈ Ω)^-1,
    {a, ain3} → (2 γ3)^1/2 (- ⅈ Ω)^-1,
    {ad, ain3d} → (2 γ3)^1/2 (- ⅈ Ω)^-1,
    {a, ain2} → (2 γ2)^1/2 (- ⅈ Ω)^-1,
    {ad, ain2d} → (2 γ2)^1/2 (- ⅈ Ω)^-1,
    {a, ain1} → (2 γ1)^1/2 (- ⅈ Ω)^-1,
    {ad, ain1d} → (2 γ1)^1/2 (- ⅈ Ω)^-1,
    {aout2, ain2} → -1,
    {aout2d, ain2d} → -1,
    {aout2, a} → (2 γ2)^1/2,
    {aout2d, ad} → (2 γ2)^1/2,
    {aout1, ain1} → -1,
    {aout1d, ain1d} → -1,
    {aout1, a} → (2 γ1)^1/2,
    {aout1d, ad} → (2 γ1)^1/2,
    {x, Frad} → (M (-Ω^2))^-1,
    {x, Fth} → (M (-Ω^2))^-1,
    {Frad, a} → -ℏ g,
    {Frad, ad} → -ℏ g
  } /. Δ → δ γ;
DM = SparseArray[TFlist, {Qlen, Qlen}] (*Dynamical matrix*);
IMN = SparseArray[{Band[{1, 1}] → 1}, {Qlen, Qlen}];
TFM = Inverse[IMN - DM] (*Transfer function matrix*);
Ms2q = {{1, 1}, {-ⅈ, ⅈ}} (2)^-1/2 (*sideband to quadrature*);
```

```
σin1AtInput = {{1 + rin, 0}, {0, 1 + pn}}
 (*add rin and phase noise to the input noise - input laser basis*);
CavRotation = RotationMatrix[ArcTan[-δ]]; (*Undoing the rotation of
 arctan (-δ) because of the non radiation pressure case*)σin1AtCavity =
 ComplexExpand[CavRotation.σin1AtInput.ConjugateTranspose[CavRotation]] //
  Simplify(*add rin to the input noise - cavity basis*);
σin1AtCavity // MatrixForm
Simplify[σin1AtCavity /. {rin → λrin (1 + δ^2), pn → λpn (1 + δ^2)/δ^2}] // MatrixForm
```

$$\begin{pmatrix} \frac{1+\text{rin}+(1+\text{pn})\,\delta^2}{1+\delta^2} & \frac{(\text{pn}-\text{rin})\,\delta}{1+\delta^2} \\ \frac{(\text{pn}-\text{rin})\,\delta}{1+\delta^2} & \frac{1+\text{pn}+(1+\text{rin})\,\delta^2}{1+\delta^2} \end{pmatrix}$$

$$\begin{pmatrix} 1+\lambda\text{pn}+\lambda\text{rin} & \frac{\lambda\text{pn}}{\delta}-\delta\,\lambda\text{rin} \\ \frac{\lambda\text{pn}}{\delta}-\delta\,\lambda\text{rin} & 1+\frac{\lambda\text{pn}}{\delta^2}+\delta^2\,\lambda\text{rin} \end{pmatrix}$$

Total Input Laser Noise matrix, rewritten in the cavity basis.

$$\begin{pmatrix} \frac{1+\text{rin}+(1+\text{pn})\,\delta^2}{1+\delta^2} & \frac{(\text{pn}-\text{rin})\,\delta}{1+\delta^2} \\ \frac{(\text{pn}-\text{rin})\,\delta}{1+\delta^2} & \frac{1+\text{pn}+(1+\text{rin})\,\delta^2}{1+\delta^2} \end{pmatrix}$$

Total Input Laser noise matrix in the cavity basis,
expressed in terms of λrin and λpn (defined in next section) :

$$\begin{pmatrix} 1+\lambda\text{pn}+\lambda\text{rin} & \frac{\lambda\text{pn}}{\delta}-\delta\,\lambda\text{rin} \\ \frac{\lambda\text{pn}}{\delta}-\delta\,\lambda\text{rin} & 1+\frac{\lambda\text{pn}}{\delta^2}+\delta^2\,\lambda\text{rin} \end{pmatrix}$$

Total laser classical noise in the 11 term -- as we will see later, the coupling of laser classical noise to transmission only show up from this 11 term

```
(*LCN =Simplify[ σin1AtCavity[[1,1]]-1] *)
```

For this reason, let’s reassign the input covariance matrix some simpler terms:

```
(*σin1AtCavity = {{1+lcn11,lcn12},{lcn21,1+lcn22}};*)
```

# Cavity Parameters’ definitions

## Definitions of Ports

a and ad are annihilation and creation operators for fields inside the cavity (with discrete quantization). Hence they are dimensionless.

all other fields (inputs and outputs) are outside the cavity (with continuous quantization), hence they are of dimension $\frac{1}{\sqrt{\text{Hz}}}$

Port 1 is input laser and reflection, so $\gamma 1 = \frac{c\text{T1}}{4\text{ L}}$

Port 2 is transmission, $\gamma 2 = \frac{c\text{T2}}{4\text{ L}}$

Port 3 is loss, $\gamma 3 = \frac{c\text{ L2}}{4\text{ L}}$

And total cavity linewidth $\gamma = \gamma 1+\gamma 2+\gamma 3$

Escape efficiency, $\text{E1} = \frac{\text{T1}}{\text{T1}+\text{T2}+\text{L2}}$, and so on for E2 and E3.

## Converting g to nominal cavity parameters:

$$G_0 = \frac{\omega}{L}$$

$$g = G_0 \left| a \right|$$

Now we need to express $|a|$ in terms of normal quantities.

We know ac radiation pressure force is $\frac{2\,\delta P}{c}$.

$P = 1/2\,(E_0 + \delta E)^2 = 1/2\,E_0^2 + E_0\,\delta E$

In Ref[1], ac radiation pressure force is $\hbar\,G_0\,|a|\,(a + a^\dagger)$

Now we assume that $E_0 = N\,|a|$, and $2\,\delta E_0 = N\,(a + a^\dagger)$

Equating the two RP forces,

$$\frac{N^2\,|a|\,(a + a^\dagger)}{c} = \hbar\,G_0\,|a|\,(a + a^\dagger),$$

Giving,

$N^2 = \hbar\,G_0\,c$

Which then allows us to write

$$g^2 = G_0^2\,|a|^2 = \frac{1}{N^2}\,G_0^2\,E_0^2 = \frac{2\,P_0\,G_0}{\hbar\,c} = \frac{2\,P_0\,\omega}{\hbar\,c\,L} = \frac{4\,\pi\,P_0}{\hbar\,L\,\lambda} = \frac{16\,\pi\,P_0\,\gamma}{\hbar\,c\,\lambda\,T} = \frac{16\,\pi\,\gamma}{\hbar\,c\,\lambda\,T}\,\frac{P_{in}\,4\,T_1}{(1+\delta^2)\,T^2}$$

________________________________

So, let's summarize,

$$g^2 = \frac{4\,\pi\,P_0}{\hbar\,L\,\lambda},$$

where,

$P_0$ = intracavity power,

$L$ = cavity length

## Perturbation Parameters

### Thermal Noise

$$\lambda th = \frac{\gamma\,S_{F,th}}{g^2\,\hbar^2}$$

$$= \frac{\gamma\,\lambda\,L\,\hbar\,S_{F,th}}{4\,\pi\,P_0\,\hbar^2} = \frac{T\,(c\,\lambda)\,S_{F,th}}{P_0\,(16\,\pi\,\hbar)}$$

Here $S_{F,th}$ is the power spectral density of the force noise due to thermal fluctuations.

### Classical Laser Intensity Noise

$\lambda rin = \frac{rin}{(1 + \delta^2)}$, where rin is classical laser intensity noise relative to shotnoise in power spectral density relative to shotnoise. So,

$$\text{rin} = \frac{\text{RIN}^2\, P_{in}{}^2}{2\, \hbar\, \omega\, P_{in}} = \frac{\text{RIN}^2\, P_{in}}{2\, \hbar\, \omega} = \frac{\text{RIN}^2}{2\, \hbar\, \omega} \frac{P_0\, T^2\, (1+\delta^2)}{4\, T1}$$, resulting in

$$\lambda\text{rin} = \frac{\text{RIN}^2}{2\, \hbar\, \omega} \frac{P_0\, T^2}{4\, T1}$$

where RIN is now $\frac{\delta P}{P}$, the relative intensity noise $\left(\text{ASD, measured in } \frac{1}{\sqrt{\text{Hz}}}\right)$

Phase Noise

$$\lambda\text{pn} = \text{pn} \frac{\delta^2}{\delta^2 + 1},$$

where pn is classical laser phase noise in power spectral density, relative to shot noise.

# Port 2 (transmission)

## Individual TFs and covariances in transmission

Since we are working at low frequencies, we can make an approximation that is $\Omega \to 0$

### Transfer functions

#### Transfer function from ain1 to transmission (aout2)

```
TFMios21 = {{TFM[[aout2, ain1]], TFM[[aout2, ain1d]]},
    {TFM[[aout2d, ain1]], TFM[[aout2d, ain1d]]}} // Simplify;
TFMioq21 = Ms2q.TFMios21.Inverse[Ms2q] /. {√γ1 √γ2 → Sqrt[E1 E2] γ} /.
   {Ω → 0} // Simplify(*quadrature picture*)
```

$$\left\{\{0, 0\}, \left\{\frac{2\sqrt{E1\, E2}}{\delta}, 0\right\}\right\}$$

It is interesting to note that all elements of this transfer function are 0 except the 21 element. That means that

a) only the 11 component of σin1AtCavity will make it to the transmission covariance matrix.

b) It's contribution will be nonzero only to the 22 element of the transmission covariance matrix.

## Transfer function from ain2 to aout2

```
TFMios22 = {{TFM[[aout2, ain2]], TFM[[aout2, ain2d]]},
    {TFM[[aout2d, ain2]], TFM[[aout2d, ain2d]]}} // Simplify;
TFMioq22 = Ms2q.TFMios22.Inverse[Ms2q] /. {γ2 → E2 γ} /. {Ω → 0} //
  Simplify(*quadrature picture*)
```

$$\left\{\{-1, 0\}, \left\{\frac{2\,\text{E2}}{\delta}, -1\right\}\right\}$$

## Transfer function from ain3 to aout2

```
TFMios23 = {{TFM[[aout2, ain3]], TFM[[aout2, ain3d]]},
    {TFM[[aout2d, ain3]], TFM[[aout2d, ain3d]]}} // Simplify;
TFMioq23 = Ms2q.TFMios23.Inverse[Ms2q] /. {√γ3 √γ2 → Sqrt[E3 E2] γ} /.
   {Ω → 0} // FullSimplify(*quadrature picture*)
```

$$\left\{\{0, 0\}, \left\{\frac{2\sqrt{\text{E2 E3}}}{\delta}, 0\right\}\right\}$$

## Transfer function from Fth to aout2

```
TFvs2th = {{TFM[[aout2, Fth]]}, {TFM[[aout2d, Fth]]}}
 (*response to thermal noise*);
TFvq2th = Ms2q.TFvs2th /. {γ2 → E2 γ} /. {Ω → 0}
  (*qudrature picture*) // Simplify
```

$$\left\{\left\{\frac{\sqrt{\text{E2}\,\gamma}}{g\,\hbar}\right\}, \left\{-\frac{\sqrt{\text{E2}\,\gamma}}{g\,\delta\,\hbar}\right\}\right\}$$

# Covariances

```
σo21 = Simplify[
  TFMioq21.σin1AtCavity.ConjugateTranspose[TFMioq21] /. {rin → λrin (1 + δ^2),
     pn → λpn (1 + δ^2)/δ^2} // ComplexExpand, Assumptions → E1 > 0 && E2 > 0]
```

$$\left\{\{0, 0\}, \left\{0, \frac{4\,\text{E1 E2}\,(1 + \lambda\text{pn} + \lambda\text{rin})}{\delta^2}\right\}\right\}$$

```
σo22 = TFMioq22.ConjugateTranspose[TFMioq22] // ComplexExpand // Simplify
```

$$\left\{\left\{1, -\frac{2\,E2}{\delta}\right\}, \left\{-\frac{2\,E2}{\delta}, 1+\frac{4\,E2^2}{\delta^2}\right\}\right\}$$

```
σo23 = Simplify[TFMioq23.ConjugateTranspose[TFMioq23] // ComplexExpand,
  Assumptions → E2 > 0 && E3 > 0]
```

$$\left\{\{0, 0\}, \left\{0, \frac{4\,E2\,E3}{\delta^2}\right\}\right\}$$

```
σo2th = Simplify[Simplify[TFvq2th.ConjugateTranspose[TFvq2th]] (λth ℏ^2 g^2)/γ //
   ComplexExpand, γ > 0 && E2 > 0] (* add thermal noise*)
```

$$\left\{\left\{E2\,\lambda th, -\frac{E2\,\lambda th}{\delta}\right\}, \left\{-\frac{E2\,\lambda th}{\delta}, \frac{E2\,\lambda th}{\delta^2}\right\}\right\}$$

# Squeezing in Transmission

## Covariance Matrices

### All contributions

Add all the above matrices. (equivalent to adding independent noise sources in quadrature.) Also, eliminate E3 as a function of E1 and E2.

```
σo2Total[E2_, λth_, λrin_, λpn_, δ_] =
 Simplify[FullSimplify[(σo21 + σo22 + σo23 + σo2th)] /. {E3 → 1 - E1 - E2}]
```

$$\left\{\left\{1+E2\,\lambda th, -\frac{E2\,(2+\lambda th)}{\delta}\right\}, \left\{-\frac{E2\,(2+\lambda th)}{\delta}, 1+\frac{E2\,(4+4\,E1\,(\lambda pn+\lambda rin)+\lambda th)}{\delta^2}\right\}\right\}$$

### No Classical Noise, not full escape efficiency

```
σo2Quantum[E2_, δ_] =
 σo2Total[E2, 0, 0, 0, δ] (*no classical noise*) // FullSimplify
```

$$\left\{\left\{1, -\frac{2\,E2}{\delta}\right\}, \left\{-\frac{2\,E2}{\delta}, 1+\frac{4\,E2}{\delta^2}\right\}\right\}$$

```
Simplify[Det[σo2Quantum[E2, δ]]]
```

$$\frac{4\,E2 - 4\,E2^2 + \delta^2}{\delta^2}$$

This quantity goes to 1 for perfect escape efficiency. The determinant being greater than 1 is a non-minimum uncertainty state.

```
Plot[{Log10[Det[σo2Quantum[E2, δ]] /. δ → 0.2],
  Log10[Det[σo2Quantum[E2, δ]] /. δ → 1]}, {E2, 0, 1}]
```

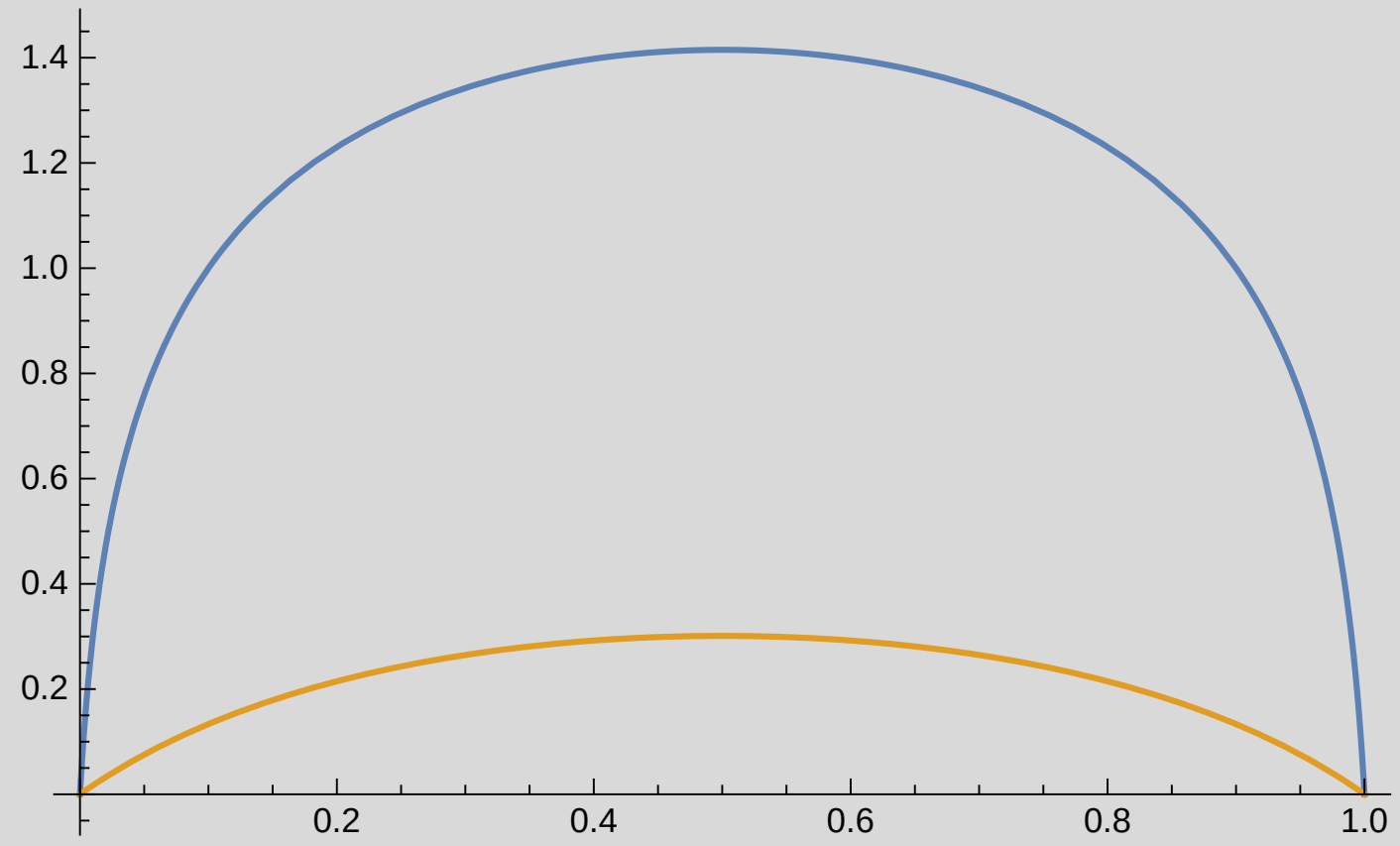

The above plot shows this determinant as a function of escape efficiency, at two detunings.

- The lower detuning (blue) has higher squeezing, so the effect of escape efficiency towards the total uncertainty is stronger.
- The higher detuning is shown in orange, and has lower squeezing. But it also stays to a total lower uncertainty than the other case.

In the next section it will be shown that in the perfect escape efficiency case, the squeezing is related to detuning as

$s^2 = \text{Tan}^2[\xi]$

$\delta = \text{Tan}[2\xi]$

$= \frac{2\,s}{1-s^2}$

Where the squeezing is $s^2$(equivalent to $e^{-2\,r}$), and anti-squeezing would be $\frac{1}{s^2}$

We can then make the same plot, where now the two traces are the amount of squeezing.

```
Plot[{Log10[Det[σo2Quantum[E2, 2 s/(1 - s^2)]] /. s → Sqrt[0.01]],
  Log10[Det[σo2Quantum[E2, 2 s/(1 - s^2)]] /. s → Sqrt[0.16]]}, {E2, 0, 1}]
```

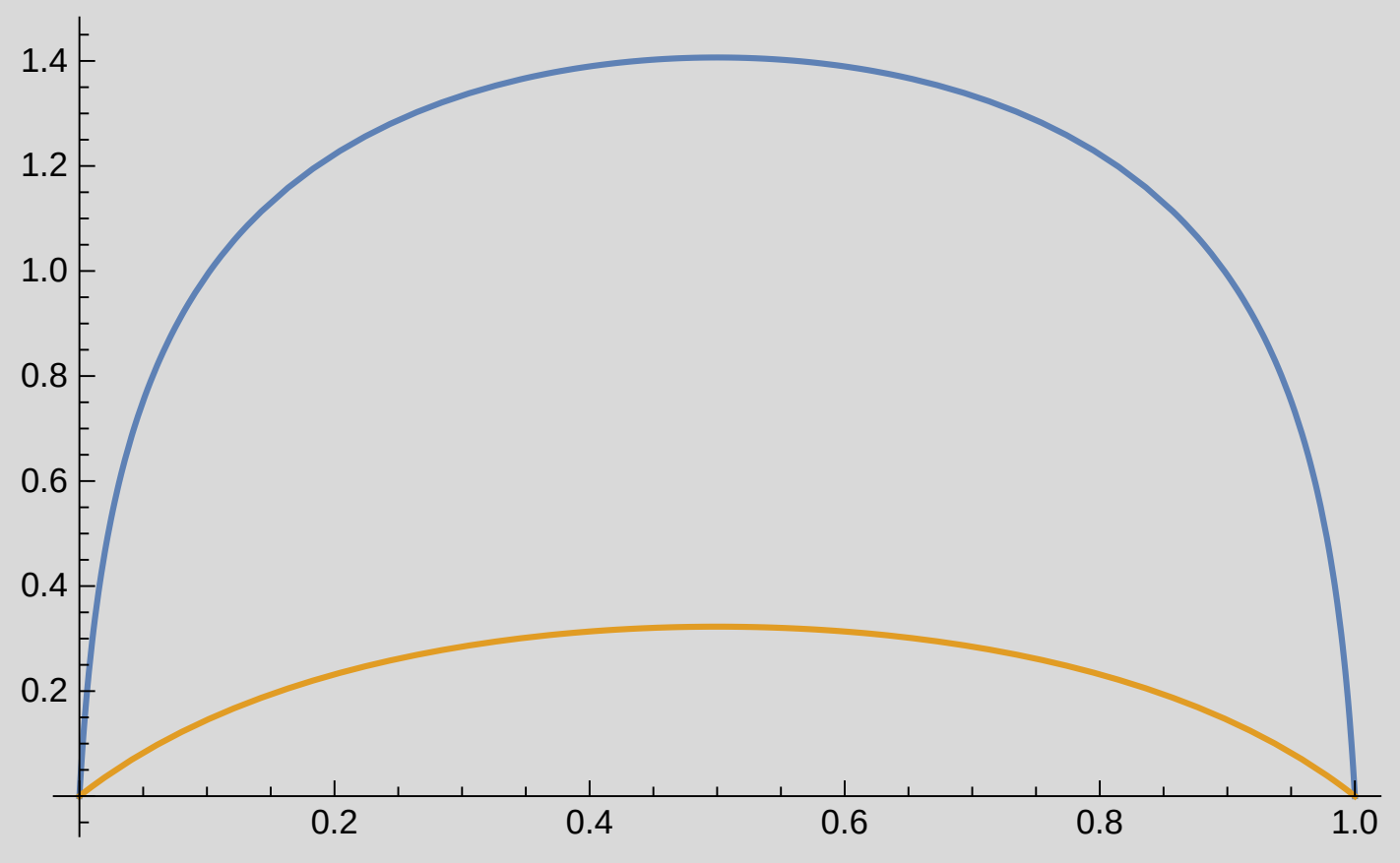

Similarly, we can show this effect of squeezing on the total uncertainty at a fixed escape efficiency

```
Plot[{Log10[Det[σo2Quantum[E2, 2 s/(1 - s^2)]] /. E2 → 0.1],
  Log10[Det[σo2Quantum[E2, 2 s/(1 - s^2)]] /. E2 → 0.5]}, {s, 0.01, 1}]
```

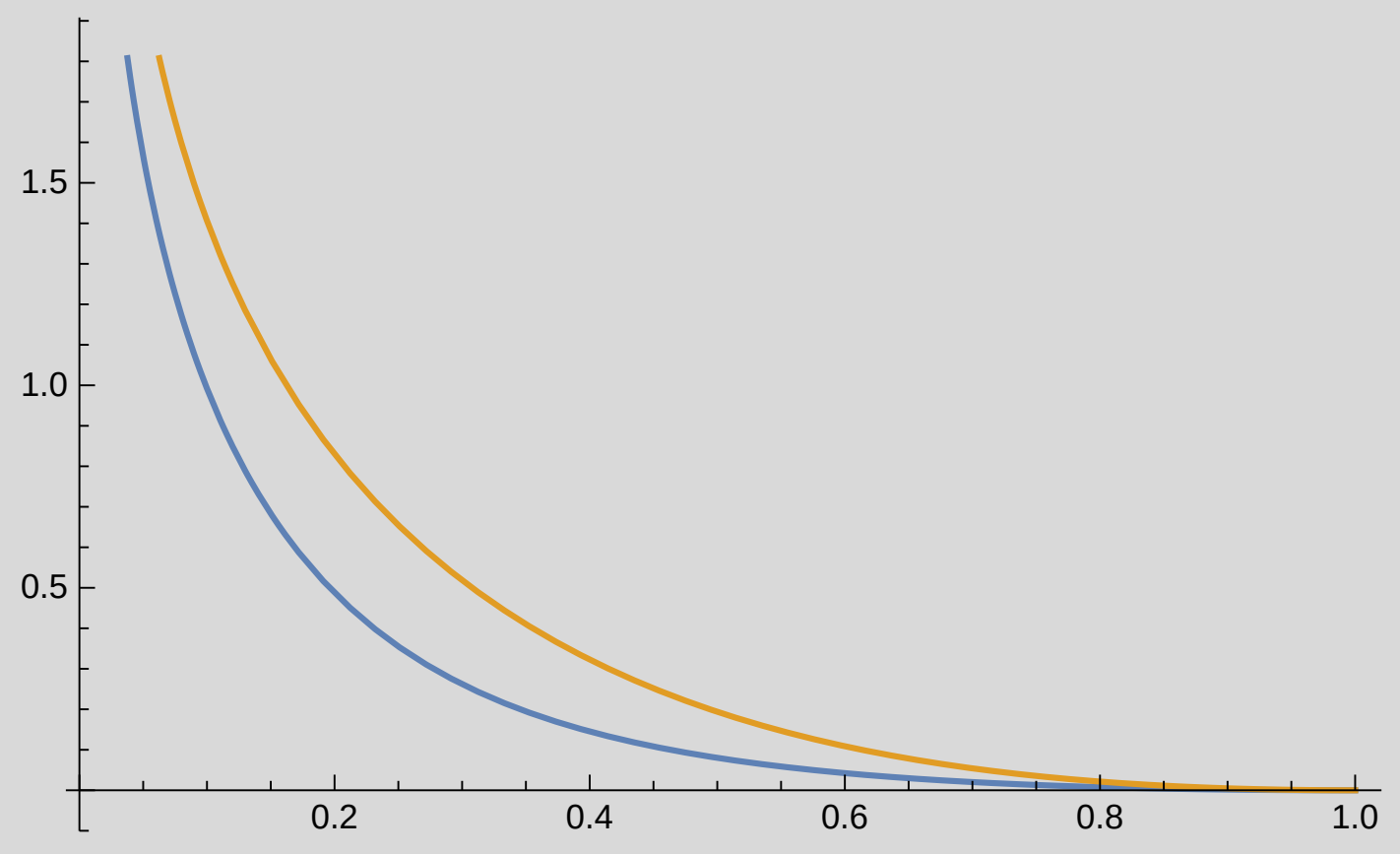

Here we see that we get a lot more total uncertainty at lower s (ie higher squeezing), and the total uncertainty converges to 1 for no squeezing (or s=1).

### Classical Laser Noise

```
σo2LCN[E2_, λcn_, δ_] = σo2Total[E2, 0, λrin, λcn - λrin, δ] (*RIN*) // Simplify
```

$$\left\{\left\{1, -\frac{2\,E2}{\delta}\right\}, \left\{-\frac{2\,E2}{\delta}, 1 + \frac{4\,E2\,(1 + E1\,\lambda cn)}{\delta^2}\right\}\right\}$$

Here we combine the classical intensity noise and phase noise on the laser, because only the 11 term of *σ*in1AtCavity shows up here.

```
σo2LCN[E2, λcn, δ] - σo2Quantum[E2, δ] // Simplify(*perturbation matrix*)
```

$$\left\{\{0, 0\}, \left\{0, \frac{4\,E1\,E2\,\lambda cn}{\delta^2}\right\}\right\}$$

### Thermal Noise

```
σo2Th[E2_, λth_, δ_] = σo2Total[E2, λth, 0, 0, δ] (*Thermal noise*) // Simplify
```

$$\left\{\left\{1 + E2\,\lambda th, -\frac{E2\,(2 + \lambda th)}{\delta}\right\}, \left\{-\frac{E2\,(2 + \lambda th)}{\delta}, 1 + \frac{E2\,(4 + \lambda th)}{\delta^2}\right\}\right\}$$

```
σo2Th[E2, λth, δ] - σo2Quantum[E2, δ] // Simplify (*perturbation matrix*)
```

$$\left\{\left\{E2\,\lambda th, -\frac{E2\,\lambda th}{\delta}\right\}, \left\{-\frac{E2\,\lambda th}{\delta}, \frac{E2\,\lambda th}{\delta^2}\right\}\right\}$$

## Eigenvalues and eigenvectors

We use the fact (which is also then self-derived here)  that the squeezing angle is related to detuning by
$\delta = \mathrm{Tan}[2\xi]$
So we substitute that in all the covariance matrices.

```
Simplify[σo2Quantum[E2, Tan[2 ξ]]] // MatrixForm
```

$$\begin{pmatrix} 1 & -2\,E2\,\mathrm{Cot}[2\,\xi] \\ -2\,E2\,\mathrm{Cot}[2\,\xi] & 1 + 4\,E2\,\mathrm{Cot}[2\,\xi]^2 \end{pmatrix}$$

```
EigenSysq2 = Simplify[FullSimplify[Eigensystem[σo2Quantum[E2, Tan[2 ξ]]]] /.
   {Csc[ξ]^2 → 1 + Cot[ξ]^2, Sec[ξ]^2 → 1 + Tan[ξ]^2}]
```

$$\left\{\left\{1 - E2 + E2\,\mathrm{Cot}[\xi]^2, 1 - E2 + E2\,\mathrm{Tan}[\xi]^2\right\}, \{\{-\mathrm{Tan}[\xi], 1\}, \{\mathrm{Cot}[\xi], 1\}\}\right\}$$

q is shorthand for quantum

```
Eq2 = EigenSysq2[[1]]
Vq2 = Simplify[{Cos[ξ], Sin[ξ]} FullSimplify[EigenSysq2[[2]]]]
Eq2AS = Eq2[[1]];
Eq2S = Eq2[[2]];
Vq2AS = Vq2[[1]];
Vq2S = Vq2[[2]];
```

```
{1 - E2 + E2 Cot[ξ]^2, 1 - E2 + E2 Tan[ξ]^2}
```

```
{{-Sin[ξ], Cos[ξ]}, {Cos[ξ], Sin[ξ]}}
```

E are eigenvalues (which are squeezing and antisqueezing).
V are eigenvectors (which give angle of squeezing and antisqueezing)

## Direct computation using Series[]

## Diagonalized Perturbation Theory

Here we approach the same problem using perturbation theory instead of Taylor series.
In addition to using the detuning as Tan[2ξ], we also look at these matrices in the basis rotated by ξ (because that should give us a diagonalized basis, and we have squeezing and antisqueezing there.)
We then apply perturbation theory in this new diagonalized basis - which is notated as p, standing for prime.

```
σp2Quantum[E2_, ξ_] =
 Simplify[FullSimplify[RotationMatrix[-ξ].σo2Quantum[E2, Tan[2 ξ]].
     RotationMatrix[ξ]] /. {Csc[ξ]^2 → 1 + Cot[ξ]^2, Sec[ξ]^2 → 1 + Tan[ξ]^2}]
```

```
{{1 - E2 + E2 Tan[ξ]^2, 0}, {0, 1 - E2 + E2 Cot[ξ]^2}}
```

(p stands for prime, ie in the rotated basis)

q is shorthand for quantum

```
Vqp2S = {1, 0};
Vqp2AS = {0, 1};
Eqp2S = Vqp2S.σp2Quantum[E2, ξ].Vqp2S
Eqp2AS = Vqp2AS.σp2Quantum[E2, ξ].Vqp2AS
```

```
1 - E2 + E2 Tan[ξ]^2
```

```
1 - E2 + E2 Cot[ξ]^2
```

- Thermal Noise

```
δσp2Th[E2_, λth_, ξ_] =
  Simplify[FullSimplify[RotationMatrix[-ξ].(σo2Total[E2, λth, 0, 0, Tan[2 ξ]] -
        σo2Quantum[E2, Tan[2 ξ]]).RotationMatrix[ξ]] /.
    {Csc[ξ]^2 → 1 + Cot[ξ]^2, Sec[ξ]^2 → 1 + Tan[ξ]^2}]
```

$$\left\{\left\{\frac{1}{4}\text{E2}\,\lambda\text{th}\,\text{Sec}[\xi]^2,\ -\frac{1}{4}\text{E2}\,\lambda\text{th}\,\text{Csc}[\xi]\,\text{Sec}[\xi]\right\},\ \left\{-\frac{1}{4}\text{E2}\,\lambda\text{th}\,\text{Csc}[\xi]\,\text{Sec}[\xi],\ \frac{1}{4}\text{E2}\,\lambda\text{th}\,\text{Csc}[\xi]^2\right\}\right\}$$

```
δEp2thS = Vqp2S.δσp2Th[E2, λth, ξ].Vqp2S
δEp2thAS = Vqp2AS.δσp2Th[E2, λth, ξ].Vqp2AS
```

$$\frac{1}{4}\text{E2}\,\lambda\text{th}\,\text{Sec}[\xi]^2$$

$$\frac{1}{4}\text{E2}\,\lambda\text{th}\,\text{Csc}[\xi]^2$$

$$\delta\text{Vp2thS} = \text{Simplify}\left[\frac{\text{Vqp2AS}.\delta\sigma\text{p2Th}[\text{E2}, \lambda\text{th}, \xi].\text{Vqp2S}}{\text{Eqp2S} - \text{Eqp2AS}}\ \text{Vqp2AS}\ /.\ \lambda\text{th} \to \frac{8\,\delta\theta^{T}{}_{\text{th}}}{\text{Tan}[2\,\xi]}\right]$$

$$\delta\text{Vp2thAS} = \text{Simplify}\left[\frac{\text{Vqp2S}.\delta\sigma\text{p2Th}[\text{E2}, \lambda\text{th}, \xi].\text{Vqp2AS}}{\text{Eqp2AS} - \text{Eqp2S}}\ \text{Vqp2S}\ /.\ \lambda\text{th} \to \frac{8\,\delta\theta^{T}{}_{\text{th}}}{\text{Tan}[2\,\xi]}\right]$$

$$\{0,\ \delta\theta^{T}{}_{\text{th}}\}$$

$$\{-\delta\theta^{T}{}_{\text{th}},\ 0\}$$

So the impact of thermal noise in transmission *is* an increase in the eigenvalues (ie higher noise in both squeezing and antisqueezing) given by δEp2thS and δEp2thAS, and *a* rotation of the squeezing ellipse, given by $\delta\theta^{T}{}_{\text{th}}$

$$\delta\theta^{T}{}_{\text{th}} = \frac{\lambda\text{th}\,\tan[2\,\xi]}{8}$$

$$\delta\text{Ep2thS} = \frac{1}{4}\text{E2}\,\lambda\text{th}\,\text{Sec}[\xi]^2$$

$$\delta\text{Ep2thAS} = \frac{1}{4}\text{E2}\,\lambda\text{th}\,\text{Csc}[\xi]^2$$

$$\lambda\text{th} = \frac{S_{F,\text{th}}\,\gamma}{\hbar^2\,g^2}$$

$$\delta = \text{Tan}[2\,\xi]$$

```
FullSimplify[δσp2Th[E2, λth, ξ] / (2 E2 Cot[2 ξ] δθ) /. λth → 8 δθ / Tan[2 ξ]] // MatrixForm
```

```
( Sec[ξ]^2           -Csc[ξ] Sec[ξ] )
( -Csc[ξ] Sec[ξ]     Csc[ξ]^2       )
```

- Classical Laser Noise

```
Simplify[FullSimplify[RotationMatrix[0].
    (σo2Total[E2, 0, λrin, λpn, Tan[2 ξ]] - σo2Quantum[E2, Tan[2 ξ]]).
    RotationMatrix[0]] /. {Csc[ξ]^2 → 1 + Cot[ξ]^2, Sec[ξ]^2 → 1 + Tan[ξ]^2}]
```

```
{{0, 0}, {0, 4 E1 E2 (λpn + λrin) Cot[2 ξ]^2}}
```

```
δσp2LCN[E1_, E2_, λcn_, ξ_] = FullSimplify[
  FullSimplify[RotationMatrix[-ξ].(σo2Total[E2, 0, λrin, λcn - λrin, Tan[2 ξ]] -
      σo2Quantum[E2, Tan[2 ξ]]).RotationMatrix[ξ]]]
```

```
{{E1 E2 λcn Cos[2 ξ]^2 Sec[ξ]^2, 2 E1 E2 λcn Cos[2 ξ] Cot[2 ξ]},
 {2 E1 E2 λcn Cos[2 ξ] Cot[2 ξ], E1 E2 λcn Cos[2 ξ]^2 Csc[ξ]^2}}
```

In transmission, only the 11 element of σin1AtCavity shows up. So we can simplify the intensity and phase noise into one quantity, called λcn

```
Simplify[δσp2LCN[E1, E2, λcn, ξ] / (4 λcn E1 E2 Cot[2 ξ]^2)]
```

```
{{Sin[ξ]^2, Cos[ξ] Sin[ξ]}, {Cos[ξ] Sin[ξ], Cos[ξ]^2}}
```

```
δEp2lcnS = Vqp2S.δσp2LCN[E1, E2, λcn, ξ].Vqp2S
δEp2lcnAS = Vqp2AS.δσp2LCN[E1, E2, λcn, ξ].Vqp2AS
```

```
E1 E2 λcn Cos[2 ξ]^2 Sec[ξ]^2
```

```
E1 E2 λcn Cos[2 ξ]^2 Csc[ξ]^2
```

```
δVp2LCNS = Simplify[
   (Vqp2AS.δσp2LCN[E1, E2, λcn, ξ].Vqp2S)/(Eqp2S - Eqp2AS) Vqp2AS /. λcn → (-(E1 Sin[4 ξ])/4)^-1 δθ^T_cn]
δVp2LCNAS = Simplify[(Vqp2S.δσp2LCN[E1, E2, λcn, ξ].Vqp2AS)/(Eqp2AS - Eqp2S) Vqp2S /.
    λcn → (-(E1 Sin[4 ξ])/4)^-1 δθ^T_cn]
```

$\{0, \delta\theta^{T}{}_{cn}\}$

$\{-\delta\theta^{T}{}_{cn}, 0\}$

So here we see that the effect of intensity noise is a small perturbation to the eigenvalues (given by δEp2lcnS and δEp2lcnAS for squeezing and antisqueezing respectively), in addition to rotation of the squeezing ellipse by

$$\delta\theta^{T}{}_{CN} = -(\lambda rin + \lambda pn)\ \frac{4}{E1\ Sin[4\ \xi]} = -\left(\frac{rin}{1+\delta^2} + \frac{pn\ \delta^2}{1+\delta^2}\right)\frac{4}{E1\ Sin[4\ \xi]}$$

$$= -\left(rin\ Cos^2[2\ \xi] + pn\ Sin^2[2\ \xi]\right)\ \frac{4}{E1\ Sin[4\ \xi]}$$

$$\delta\theta^{T}{}_{CN} = -\frac{1}{2\ E1}\ (rin\ Cot[2\ \xi] + pn\ Tan[2\ \xi])$$

$$\delta Ep2lcnS = E1\ E2\ \lambda cn\ Cos[2\ \xi]^2\ Sec[\xi]^2 = E1\ E2\ \left(rin\ Cos^2[2\ \xi] + pn\ Sin^2[2\ \xi]\right)\ Cos[2\ \xi]^2\ Sec[\xi]^2$$

$$\delta Ep2lcnAS = E1\ E2\ \lambda cn\ Cos[2\ \xi]^2\ Csc[\xi]^2 = E1\ E2\ \left(rin\ Cos^2[2\ \xi] + pn\ Sin^2[2\ \xi]\right)\ Cos[2\ \xi]^2\ Csc[\xi]^2$$

Note that the rotation angle is negative, so it rotates cloclwise unlike thermal noise

# Reflection (Port 1)

Refl port needs an additional rotation wrt the cavity field to bring the matrix in the reflection basis (transmission basis is conveniently the same as the cavity basis.)

Again, we make an approximation that is $\Omega \to 0$

# Individual Transfer Functions And Covariance Matrices

## Transfer functions

```
TFMios11 = {{TFM[[aout1, ain1]], TFM[[aout1, ain1d]]},
    {TFM[[aout1d, ain1]], TFM[[aout1d, ain1d]]}} // Simplify;
TFMioq11 = Ms2q.TFMios11.Inverse[Ms2q] /. {Ω → 0} /. {γ1 → E1 γ} //
  Simplify(*quadrature picture*)
```

$$\left\{\{-1, 0\}, \left\{\frac{2\, E1}{\delta}, -1\right\}\right\}$$

```
TFMios12 = {{TFM[[aout1, ain2]], TFM[[aout1, ain2d]]},
    {TFM[[aout1d, ain2]], TFM[[aout1d, ain2d]]}} // Simplify;
TFMioq12 = Ms2q.TFMios12.Inverse[Ms2q] /. {Ω → 0} /.
   {√γ1 √γ2 → Sqrt[E1 E2] γ} // Simplify(*quadrature picture*)
```

$$\left\{\{0, 0\}, \left\{\frac{2\sqrt{E1\, E2}}{\delta}, 0\right\}\right\}$$

```
TFMios13 = {{TFM[[aout1, ain3]], TFM[[aout1, ain3d]]},
    {TFM[[aout1d, ain3]], TFM[[aout1d, ain3d]]}} // Simplify;
TFMioq13 = Ms2q.TFMios13.Inverse[Ms2q] /. {√γ3 √γ1 → Sqrt[E1 E3] γ} /.
   {Ω → 0} // FullSimplify(*quadrature picture*)
```

$$\left\{\{0, 0\}, \left\{\frac{2\sqrt{E1\, E3}}{\delta}, 0\right\}\right\}$$

```
TFvs1th = {{TFM[[aout1, Fth]]}, {TFM[[aout1d, Fth]]}}
  (*response to thermal noise*);
TFvq1th = Ms2q.TFvs1th /. Ω → 0 /. {γ1 → E1 γ} // Simplify(*qudrature picture*)
```

$$\left\{\left\{\frac{\sqrt{E1\, \gamma}}{g\, \hbar}\right\}, \left\{-\frac{\sqrt{E1\, \gamma}}{g\, \delta\, \hbar}\right\}\right\}$$

## Covariances

```
σo11 = TFMioq11.σin1AtCavity.ConjugateTranspose[TFMioq11] /.
   {rin → λrin (1 + δ^2), pn → λpn (1 + δ^2)/δ^2} // ComplexExpand // Simplify
```

$$\left\{\left\{1+\lambda pn+\lambda rin,\ \frac{\lambda pn-\delta^2\,\lambda rin-2\,E1\,(1+\lambda pn+\lambda rin)}{\delta}\right\},\ \left\{\frac{\lambda pn-\delta^2\,\lambda rin-2\,E1\,(1+\lambda pn+\lambda rin)}{\delta},\ \frac{1}{\delta^2}\left(\delta^2+\lambda pn+\delta^4\,\lambda rin+4\,E1^2\,(1+\lambda pn+\lambda rin)-4\,E1\,(\lambda pn-\delta^2\,\lambda rin)\right)\right\}\right\}$$

```
σo12 = Simplify[TFMioq12.ConjugateTranspose[TFMioq12] // ComplexExpand,
  Assumptions → E1 > 0 && E2 > 0]
```

$$\left\{\{0,0\},\ \left\{0,\ \frac{4\,E1\,E2}{\delta^2}\right\}\right\}$$

```
σo13 = Simplify[TFMioq13.ConjugateTranspose[TFMioq13] // ComplexExpand,
  Assumptions → E1 > 0 && E3 > 0]
```

$$\left\{\{0,0\},\ \left\{0,\ \frac{4\,E1\,E3}{\delta^2}\right\}\right\}$$

```
σo1th = Simplify[ComplexExpand[TFvq1th.ConjugateTranspose[TFvq1th] λth ħ^2 g^2/γ],
  Assumptions → E1 > 0 && γ > 0]
```

$$\left\{\left\{E1\,\lambda th,\ -\frac{E1\,\lambda th}{\delta}\right\},\ \left\{-\frac{E1\,\lambda th}{\delta},\ \frac{E1\,\lambda th}{\delta^2}\right\}\right\}$$

# Squeezing in Reflection

## Covariance matrices

### Full

```
σo1Total[E1_, λth_, λrin_, λpn_, δ_] =
 FullSimplify[Simplify[(σo11 + σo12 + σo13 + σo1th)] /. {E3 → 1 - E1 - E2}]
```

$$\left\{\left\{1+\lambda pn+\lambda rin+E1\,\lambda th,\ \frac{\lambda pn-\delta^2\,\lambda rin-E1\,(2\,(1+\lambda pn+\lambda rin)+\lambda th)}{\delta}\right\},\right.$$
$$\left\{\frac{\lambda pn-\delta^2\,\lambda rin-E1\,(2\,(1+\lambda pn+\lambda rin)+\lambda th)}{\delta},\right.$$
$$\left.\left.\frac{1}{\delta^2}\left(\delta^2+\lambda pn+\delta^4\,\lambda rin+4\,E1^2\,(\lambda pn+\lambda rin)+E1\,(4-4\,\lambda pn+4\,\delta^2\,\lambda rin+\lambda th)\right)\right\}\right\}$$

### Quantum only

```
σo1Quantum[E1_, δ_] =
 σo1Total[E1, 0, 0, 0, δ] (*no classical noise*) // Simplify
```

$$\left\{\left\{1,-\frac{2\,E1}{\delta}\right\},\left\{-\frac{2\,E1}{\delta},1+\frac{4\,E1}{\delta^2}\right\}\right\}$$

```
Det[σo1Quantum[E1, δ]]
```

$$1+\frac{4\,E1}{\delta^2}-\frac{4\,E1^2}{\delta^2}$$

### RIN

```
FullSimplify[σo1Total[E1, 0, λrin, λcn - λrin, δ]]
```

$$\left\{\left\{1+\lambda cn,\ \frac{\lambda cn-2\,E1\,(1+\lambda cn)-(1+\delta^2)\,\lambda rin}{\delta}\right\},\left\{\frac{\lambda cn-2\,E1\,(1+\lambda cn)-(1+\delta^2)\,\lambda rin}{\delta},\right.\right.$$
$$\left.\left.\frac{1}{\delta^2}\left(\delta^2+\lambda cn+4\,E1^2\,\lambda cn+(-1+\delta^4)\,\lambda rin+4\,E1\,(1-\lambda cn+\lambda rin+\delta^2\,\lambda rin)\right)\right\}\right\}$$

As we can see, this matrix does not simplify as a function of λcn, it needs λrin, because λrin and λpn add differently towards the reflection covariance matrix.

```
σo1rin[E1_, λrin_, δ_] = σo1Total[E1, 0, λrin, 0, δ] (*RIN*) // Simplify
```

$$\left\{\left\{1+\lambda rin,\ -\delta\,\lambda rin-\frac{2\,E1\,(1+\lambda rin)}{\delta}\right\},\ \left\{-\delta\,\lambda rin-\frac{2\,E1\,(1+\lambda rin)}{\delta},\ 1+\frac{4\,E1^2\,\lambda rin}{\delta^2}+\delta^2\,\lambda rin+4\,E1\left(\frac{1}{\delta^2}+\lambda rin\right)\right\}\right\}$$

```
σo1rin[E1, λrin, δ] - σo1Quantum[E1, δ] // Simplify
```

$$\left\{\left\{\lambda rin,\ -\frac{(2\,E1+\delta^2)\,\lambda rin}{\delta}\right\},\ \left\{-\frac{(2\,E1+\delta^2)\,\lambda rin}{\delta},\ \frac{(2\,E1+\delta^2)^2\,\lambda rin}{\delta^2}\right\}\right\}$$

## Phase Noise

```
σo1pn[E1_, λpn_, δ_] = σo1Total[E1, 0, 0, λpn, δ] (*RIN*) // Simplify
```

$$\left\{\left\{1+\lambda pn,\ \frac{\lambda pn-2\,E1\,(1+\lambda pn)}{\delta}\right\},\ \left\{\frac{\lambda pn-2\,E1\,(1+\lambda pn)}{\delta},\ \frac{4\,E1+\delta^2+\lambda pn-4\,E1\,\lambda pn+4\,E1^2\,\lambda pn}{\delta^2}\right\}\right\}$$

```
σo1pn[E1, λpn, δ] - σo1Quantum[E1, δ] // Simplify
```

$$\left\{\left\{\lambda pn,\ \frac{\lambda pn-2\,E1\,\lambda pn}{\delta}\right\},\ \left\{\frac{\lambda pn-2\,E1\,\lambda pn}{\delta},\ \frac{(1-2\,E1)^2\,\lambda pn}{\delta^2}\right\}\right\}$$

## Thermal Noise

```
σo1Th[E1_, λth_, δ_] =
  σo1Total[E1, λth, 0, 0, δ] /. {rin → 0, pn → 0} (*Thermal noise*) // Simplify
```

$$\left\{\left\{1+E1\,\lambda th,\ -\frac{E1\,(2+\lambda th)}{\delta}\right\},\ \left\{-\frac{E1\,(2+\lambda th)}{\delta},\ 1+\frac{E1\,(4+\lambda th)}{\delta^2}\right\}\right\}$$

```
σo1Th[E1, λth, δ] - σo1Quantum[E1, δ] // Simplify
```

$$\left\{\left\{E1\,\lambda th,\ -\frac{E1\,\lambda th}{\delta}\right\},\ \left\{-\frac{E1\,\lambda th}{\delta},\ \frac{E1\,\lambda th}{\delta^2}\right\}\right\}$$

## EigenValues and Eigenvectors

```
Simplify[σo1Quantum[E1, Tan[2 ξ]]] // MatrixForm
```

$$\begin{pmatrix} 1 & -2\,\text{E1}\,\text{Cot}[2\,\xi] \\ -2\,\text{E1}\,\text{Cot}[2\,\xi] & 1+4\,\text{E1}\,\text{Cot}[2\,\xi]^2 \end{pmatrix}$$

```
EigenSysq1 = Simplify[FullSimplify[Eigensystem[σo1Quantum[E1, Tan[2 ξ]]]] /.
   {Csc[ξ]^2 → 1 + Cot[ξ]^2, Sec[ξ]^2 → 1 + Tan[ξ]^2}]
```

$\{\{1 - \text{E1} + \text{E1}\,\text{Cot}[\xi]^2,\ 1 - \text{E1} + \text{E1}\,\text{Tan}[\xi]^2\},\ \{\{-\text{Tan}[\xi],\ 1\},\ \{\text{Cot}[\xi],\ 1\}\}\}$

```
Eq1 = EigenSysq1[[1]]
Vq1  = Simplify[{Cos[ξ], Sin[ξ]} FullSimplify[EigenSysq1[[2]]]]
Eq1AS  =  Eq1[[1]];
Eq1S = Eq1[[2]];
Vq1AS  =  Vq1[[1]];
Vq1S = Vq1[[2]];
```

$\{1 - \text{E1} + \text{E1}\,\text{Cot}[\xi]^2,\ 1 - \text{E1} + \text{E1}\,\text{Tan}[\xi]^2\}$

$\{\{-\text{Sin}[\xi],\ \text{Cos}[\xi]\},\ \{\text{Cos}[\xi],\ \text{Sin}[\xi]\}\}$

## Direct computation using Series[]

## Diagonalized Perturbation Theory

```
σp1Quantum[E1_, ξ_] =
 Simplify[FullSimplify[RotationMatrix[-ξ].σo1Quantum[E1, Tan[2 ξ]].
     RotationMatrix[ξ]] /. {Csc[ξ]^2 → 1 + Cot[ξ]^2, Sec[ξ]^2 → 1 + Tan[ξ]^2}]
```

$\{\{1 - \text{E1} + \text{E1}\,\text{Tan}[\xi]^2,\ 0\},\ \{0,\ 1 - \text{E1} + \text{E1}\,\text{Cot}[\xi]^2\}\}$

(p stands for prime, ie in the rotated basis)

```
Vqp1S = {1, 0};
Vqp1AS = {0, 1};
Eqp1S  =  Vqp1S.σp1Quantum[E1, ξ].Vqp1S
Eqp1AS  =  Vqp1AS.σp1Quantum[E1, ξ].Vqp1AS
```

$1 - \text{E1} + \text{E1}\,\text{Tan}[\xi]^2$

$1 - \text{E1} + \text{E1}\,\text{Cot}[\xi]^2$

- Thermal Noise

```
δσp1Th[E1_, λth_, ξ_] =
  Simplify[FullSimplify[RotationMatrix[-ξ].(σo1Total[E1, λth, 0, 0, Tan[2 ξ]] -
        σo1Quantum[E1, Tan[2 ξ]]).RotationMatrix[ξ]] /.
     {Csc[ξ]^2 → 1 + Cot[ξ]^2, Sec[ξ]^2 → 1 + Tan[ξ]^2}]
```

$$\left\{\left\{\frac{1}{4}\,\text{E1}\,\lambda\text{th}\,\text{Sec}[\xi]^2,\ -\frac{1}{4}\,\text{E1}\,\lambda\text{th}\,\text{Csc}[\xi]\,\text{Sec}[\xi]\right\},\ \left\{-\frac{1}{4}\,\text{E1}\,\lambda\text{th}\,\text{Csc}[\xi]\,\text{Sec}[\xi],\ \frac{1}{4}\,\text{E1}\,\lambda\text{th}\,\text{Csc}[\xi]^2\right\}\right\}$$

```
δEp1thS = Vqp1S.δσp1Th[E1, λth, ξ].Vqp1S
δEp1thAS = Vqp1AS.δσp1Th[E1, λth, ξ].Vqp1AS
```

$$\frac{1}{4}\,\text{E1}\,\lambda\text{th}\,\text{Sec}[\xi]^2$$

$$\frac{1}{4}\,\text{E1}\,\lambda\text{th}\,\text{Csc}[\xi]^2$$

$$\delta\text{Vp1thS} = \text{Simplify}\left[\frac{\text{Vqp1AS}.\delta\sigma\text{p1Th}[\text{E1}, \lambda\text{th}, \xi].\text{Vqp1S}}{\text{Eqp1S} - \text{Eqp1AS}}\ \text{Vqp1AS}\ /.\ \lambda\text{th} \to \frac{8\,\delta\theta}{\text{Tan}[2\,\xi]}\right]$$

$$\delta\text{Vp1thAS} = \text{Simplify}\left[\frac{\text{Vqp1S}.\delta\sigma\text{p1Th}[\text{E1}, \lambda\text{th}, \xi].\text{Vqp1AS}}{\text{Eqp1AS} - \text{Eqp1S}}\ \text{Vqp1S}\ /.\ \lambda\text{th} \to \frac{8\,\delta\theta}{\text{Tan}[2\,\xi]}\right]$$

$\{0, \delta\theta\}$

$\{-\delta\theta, 0\}$

So here we see that the effect of thermal noise on the level of squeezing and antisqueezing, as well as the rotation caused by thermal noise, has the same form in reflection as the transmission.

$$\text{FullSimplify}\left[\frac{\delta\sigma\text{p1Th}[\text{E1}, \lambda\text{th}, \xi]}{2\,\text{E1}\,\text{Cot}[2\,\xi]\,\delta\theta}\ /.\ \lambda\text{th} \to \frac{8\,\delta\theta}{\text{Tan}[2\,\xi]}\right]\ //\ \text{MatrixForm}$$

$$\begin{pmatrix} \text{Sec}[\xi]^2 & -\text{Csc}[\xi]\,\text{Sec}[\xi] \\ -\text{Csc}[\xi]\,\text{Sec}[\xi] & \text{Csc}[\xi]^2 \end{pmatrix}$$

- Classical Laser Noise

```
Simplify[FullSimplify[RotationMatrix[θ].
   (σo1Total[E1, 0, λrin, λpn, Tan[2 ξ]] - σo1Quantum[E1, Tan[2 ξ]]).
   RotationMatrix[θ]]]
```

```
{{λpn + λrin, Cot[2 ξ] (λpn - 2 E1 λpn - 2 E1 λrin - λrin Tan[2 ξ]^2)},
 {Cot[2 ξ] (λpn - 2 E1 λpn - 2 E1 λrin - λrin Tan[2 ξ]^2),
  ((1 - 2 E1)^2 λpn + 4 E1^2 λrin) Cot[2 ξ]^2 + λrin (4 E1 + Tan[2 ξ]^2)}}
```

This shows we can't club rin and phase noise together for reflection.

- Classical Laser Intensity Noise

```
δσp1rin[E1_, λrin_, ξ_] = FullSimplify[
  FullSimplify[RotationMatrix[-ξ].(σo1Total[E1, 0, λrin, 0, Tan[2 ξ]] -
     σo1Quantum[E1, Tan[2 ξ]]).RotationMatrix[ξ]]]
```

```
{{λrin Cos[ξ]^2 (-2 + 2 E1 - E1 Sec[ξ]^2 + Sec[2 ξ])^2,
  1/4 λrin (E1 (-2 + 3 E1) + (-1 + 4 (-1 + E1) E1) Cos[4 ξ] + (-1 + E1)^2 Cos[8 ξ])
   Csc[2 ξ] Sec[2 ξ]^2},
 {1/4 λrin (E1 (-2 + 3 E1) + (-1 + 4 (-1 + E1) E1) Cos[4 ξ] + (-1 + E1)^2 Cos[8 ξ])
   Csc[2 ξ] Sec[2 ξ]^2,
  1/4 λrin (E1 + Cos[2 ξ] + (-1 + E1) Cos[4 ξ])^2 Csc[ξ]^2 Sec[2 ξ]^2}}
```

```
Simplify[δσp1rin[E1, λrin, ξ] / (λrin Csc[2 ξ] Sec[2 ξ]^2)] // MatrixForm
```

```
(                 1/2 (E1 - Cos[2 ξ] + (-1 + E1) Cos[4 ξ])^2 Tan[ξ]              1/4 (E1 (-2 + 3 E1
  1/4 (E1 (-2 + 3 E1) + (-1 + 4 (-1 + E1) E1) Cos[4 ξ] + (-1 + E1)^2 Cos[8 ξ])         1/2
```

```
δEp1rinS = Simplify[Vqp1S.δσp1rin[E1, λrin, ξ].Vqp1S]

δEp1rinAS = FullSimplify[Vqp1AS.δσp1rin[E1, λrin, ξ].Vqp1AS]
```

```
λrin Cos[ξ]^2 (-2 + 2 E1 - E1 Sec[ξ]^2 + Sec[2 ξ])^2
```

```
1/4 λrin (E1 + Cos[2 ξ] + (-1 + E1) Cos[4 ξ])^2 Csc[ξ]^2 Sec[2 ξ]^2
```

```
δVp1rinS  = Simplify[TrigExpand[(Vqp1AS.δσp1rin[E1, λrin, ξ].Vqp1S)/(Eqp1S - Eqp1AS) Vqp1AS],
    Assumptions → 0 < ξ < π/4 && 0 < E1 < 1] /. λrin → δθ^R_rin (8 E1)/(Tan[2 ξ]^3)
δVp1rinAS  = Simplify[TrigExpand[(Vqp1S.δσp1rin[E1, λrin, ξ].Vqp1AS)/(Eqp1AS - Eqp1S) Vqp1S],
    Assumptions → 0 < ξ < π/4 && 0 < E1 < 1] /. λrin → δθ^R_rin (8 E1)/(Tan[2 ξ]^3)
```

$$\left\{0, -\frac{1}{2}\left(E1\,(-2+3\,E1)+\left(-1-4\,E1+4\,E1^2\right)\,\mathrm{Cos}[4\,\xi]+(-1+E1)^2\,\mathrm{Cos}[8\,\xi]\right)\,\mathrm{Csc}[2\,\xi]^2\,\delta\theta^R{}_{rin}\right\}$$

$$\left\{\frac{1}{2}\left(E1\,(-2+3\,E1)+\left(-1-4\,E1+4\,E1^2\right)\,\mathrm{Cos}[4\,\xi]+(-1+E1)^2\,\mathrm{Cos}[8\,\xi]\right)\,\mathrm{Csc}[2\,\xi]^2\,\delta\theta^R{}_{rin}, 0\right\}$$

```
Simplify[δEp1rinS /. E1 → 1]
Simplify[δEp1rinS /. E1 → 1]
Simplify[δEp1rinS /. E1 → 1 /. λrin → rin/(1 + δ^2) /. δ → Tan[2 ξ]]
Simplify[δEp1rinS /. E1 → 1 /. λrin → rin/(1 + δ^2) /. δ → Tan[2 ξ]]
```

$16\,\lambda rin\,\mathrm{Csc}[4\,\xi]^2\,\mathrm{Sin}[\xi]^6$

$16\,\lambda rin\,\mathrm{Csc}[4\,\xi]^2\,\mathrm{Sin}[\xi]^6$

$rin\,\mathrm{Sin}[\xi]^2\,\mathrm{Tan}[\xi]^2$

$rin\,\mathrm{Sin}[\xi]^2\,\mathrm{Tan}[\xi]^2$

```
Simplify[δVp1rinS /. E1 → 1]
Simplify[δVp1rinS /. E1 → 1]
```

$\{0, -\delta\theta^R{}_{rin}\}$

$\{0, -\delta\theta^R{}_{rin}\}$

Maybe these can be further simplified if the rotation in reflection is incorporated. (it will probably be a function of E1 and $\delta$)

- Classical Laser phase noise

```
δσp1pn[E1_, λpn_, ξ_] = FullSimplify[
  FullSimplify[RotationMatrix[-ξ].(σo1Total[E1, 0, 0, λpn, Tan[2 ξ]] -
      σo1Quantum[E1, Tan[2 ξ]]).RotationMatrix[ξ]]]
```

$$\left\{\left\{\frac{1}{4}\lambda pn\left(1-2(-1+E1)\,Cos[2\xi]\right)^2 Sec[\xi]^2,\ \frac{1}{2}\lambda pn\left(1+2(-2+E1)E1+2(-1+E1)^2 Cos[4\xi]\right)Csc[2\xi]\right\},\ \left\{\frac{1}{2}\lambda pn\left(1+2(-2+E1)E1+2(-1+E1)^2 Cos[4\xi]\right)Csc[2\xi],\ \frac{1}{4}\lambda pn\left(1+2(-1+E1)\,Cos[2\xi]\right)^2 Csc[\xi]^2\right\}\right\}$$

```
Simplify[(2 δσp1pn[E1, λpn, ξ])/(λpn Csc[2 ξ])] // MatrixForm
```

$$\begin{pmatrix} \left(1-2(-1+E1)Cos[2\xi]\right)^2 Tan[\xi] & 1+2(-2+E1)E1+2(-1+E1)^2 Cos[4\xi] \\ 1+2(-2+E1)E1+2(-1+E1)^2 Cos[4\xi] & \left(1+2(-1+E1)Cos[2\xi]\right)^2 Cot[\xi] \end{pmatrix}$$

```
δEp1pnS = Simplify[Vqp1S.δσp1pn[E1, λpn, ξ].Vqp1S]

δEp1pnAS = FullSimplify[Vqp1AS.δσp1pn[E1, λpn, ξ].Vqp1AS]
```

$$\frac{1}{4}\lambda pn\left(1-2(-1+E1)Cos[2\xi]\right)^2 Sec[\xi]^2$$

$$\frac{1}{4}\lambda pn\left(1+2(-1+E1)Cos[2\xi]\right)^2 Csc[\xi]^2$$

```
δVp1pnS  = Simplify[TrigExpand[(Vqp1AS.δσp1pn[E1, λpn, ξ].Vqp1S)/(Eqp1S - Eqp1AS) Vqp1AS]] /.
  λpn → (Tan[2 ξ]/(8 E1))^-1 δθ^R_pn
δVp1pnAS  = Simplify[TrigExpand[(Vqp1S.δσp1pn[E1, λpn, ξ].Vqp1AS)/(Eqp1AS - Eqp1S) Vqp1S]] /.
  λpn → (Tan[2 ξ]/(8 E1))^-1 δθ^R_pn
```

$$\left\{0,\ -\left(1-4E1+2E1^2+2(-1+E1)^2 Cos[4\xi]\right)\delta\theta^R{}_{pn}\right\}$$

$$\left\{\left(1-4E1+2E1^2+2(-1+E1)^2 Cos[4\xi]\right)\delta\theta^R{}_{pn},\ 0\right\}$$

```
Simplify[δEp1pnS /. E1 → 1]
Simplify[δEp1pnS /. E1 → 1]
```

$$\text{Simplify}\left[\delta\text{Ep1pnS} \;/.\; \text{E1} \to 1 \;/.\; \lambda\text{pn} \to \text{pn}\,\frac{\delta^2}{1+\delta^2} \;/.\; \delta \to \text{Tan}[2\,\xi]\right]$$

$$\text{Simplify}\left[\delta\text{Ep1pnS} \;/.\; \text{E1} \to 1 \;/.\; \lambda\text{pn} \to \text{pn}\,\frac{\delta^2}{1+\delta^2} \;/.\; \delta \to \text{Tan}[2\,\xi]\right]$$

$$\frac{1}{4}\,\lambda\text{pn}\,\text{Sec}[\xi]^2$$

$$\frac{1}{4}\,\lambda\text{pn}\,\text{Sec}[\xi]^2$$

$$\text{pn}\,\text{Sin}[\xi]^2$$

$$\text{pn}\,\text{Sin}[\xi]^2$$

```
Simplify[δVp1pnS /. E1 → 1]
Simplify[δVp1pnS /. E1 → 1]
```

$$\{0,\ \delta\Theta^{R}{}_{\text{pn}}\}$$

$$\{0,\ \delta\Theta^{R}{}_{\text{pn}}\}$$

# Total Uncertainty

## Setup

We are interested in answering the following question:
If we take the thermal noise and laser noise out of the picture, but have an open cavity, do we still have a minimum uncertainty state if all the information is preserved.
In order to do this, we would like to get the determinant of a covariance matrix that is in the expanded Hilbert space, containing reflection, transmission, and loss.
For the sake of generality, loss and transmission can be interchangeable, so let’s assume we have no loss, and construct a 4 by 4 covariance matrix.
We already have the diagonal. Only need to calculate the off-diagonal.

```
σo121 =
 Simplify[ComplexExpand[ TFMioq11.σin1AtCavity.ConjugateTranspose[TFMioq21]] /.
   {rin → λrin (1 + δ^2), pn → λpn (1 + δ^2)/δ^2}, Assumptions → E1 > 0 && E2 > 0]
```

$$\left\{\left\{0, -\frac{2\sqrt{E1\,E2}\,(1+\lambda pn+\lambda rin)}{\delta}\right\}, \left\{0, \frac{2\sqrt{E1\,E2}\,\left(-\lambda pn+\delta^2\,\lambda rin+2\,E1\,(1+\lambda pn+\lambda rin)\right)}{\delta^2}\right\}\right\}$$

```
σo122 = Simplify[ComplexExpand[ TFMioq12.ConjugateTranspose[TFMioq22]],
  Assumptions → E1 > 0 && E2 > 0]
```

$$\left\{\{0, 0\}, \left\{-\frac{2\sqrt{E1\,E2}}{\delta}, \frac{4\,E2\sqrt{E1\,E2}}{\delta^2}\right\}\right\}$$

```
σo123 = Simplify[ComplexExpand[ TFMioq13.ConjugateTranspose[TFMioq23]],
   Assumptions → E1 > 0 && E2 > 0 && E3 > 0] /. E3 → 1 - E2 - E1
```

$$\left\{\{0, 0\}, \left\{0, \frac{4\,(1-E1-E2)\sqrt{E1\,E2}}{\delta^2}\right\}\right\}$$

```
σo12th = Simplify[ComplexExpand[ TFvq1th.ConjugateTranspose[TFvq2th]] λth ħ^2 g^2/γ,
  Assumptions → E1 > 0 && E2 > 0 && E3 > 0 && γ > 0]
```

$$\left\{\left\{\sqrt{E1\,E2}\,\lambda th, -\frac{\sqrt{E1\,E2}\,\lambda th}{\delta}\right\}, \left\{-\frac{\sqrt{E1\,E2}\,\lambda th}{\delta}, \frac{\sqrt{E1\,E2}\,\lambda th}{\delta^2}\right\}\right\}$$

# Quantum Noise only

```
σoTotalQuantum12 = Simplify[Simplify[σo121 + σo122 + σo123 + σo12th] /.
    {λth → 0, λrin → 0, λpn → 0}] /. E3 → 1 - E2 - E1
```

$$\left\{\left\{0, -\frac{2\sqrt{E1\,E2}}{\delta}\right\}, \left\{-\frac{2\sqrt{E1\,E2}}{\delta}, \frac{4\sqrt{E1\,E2}}{\delta^2}\right\}\right\}$$

```
σoTotalQuantum = ArrayFlatten[Simplify[ {{σo1Quantum[E1, δ], σoTotalQuantum12},
    {Transpose[σoTotalQuantum12], σo2Quantum[E2, δ]}}]]
```

$$\left\{\left\{1, -\frac{2\,\text{E1}}{\delta}, 0, -\frac{2\sqrt{\text{E1 E2}}}{\delta}\right\}, \left\{-\frac{2\,\text{E1}}{\delta}, 1+\frac{4\,\text{E1}}{\delta^2}, -\frac{2\sqrt{\text{E1 E2}}}{\delta}, \frac{4\sqrt{\text{E1 E2}}}{\delta^2}\right\}, \left\{0, -\frac{2\sqrt{\text{E1 E2}}}{\delta}, 1, -\frac{2\,\text{E2}}{\delta}\right\}, \left\{-\frac{2\sqrt{\text{E1 E2}}}{\delta}, \frac{4\sqrt{\text{E1 E2}}}{\delta^2}, -\frac{2\,\text{E2}}{\delta}, 1+\frac{4\,\text{E2}}{\delta^2}\right\}\right\}$$

```
σoTotalQuantum // MatrixForm
```

$$\begin{pmatrix} 1 & -\frac{2\,\text{E1}}{\delta} & 0 & -\frac{2\sqrt{\text{E1 E2}}}{\delta} \\ -\frac{2\,\text{E1}}{\delta} & 1+\frac{4\,\text{E1}}{\delta^2} & -\frac{2\sqrt{\text{E1 E2}}}{\delta} & \frac{4\sqrt{\text{E1 E2}}}{\delta^2} \\ 0 & -\frac{2\sqrt{\text{E1 E2}}}{\delta} & 1 & -\frac{2\,\text{E2}}{\delta} \\ -\frac{2\sqrt{\text{E1 E2}}}{\delta} & \frac{4\sqrt{\text{E1 E2}}}{\delta^2} & -\frac{2\,\text{E2}}{\delta} & 1+\frac{4\,\text{E2}}{\delta^2} \end{pmatrix}$$

```
DetTotalQuantum = Simplify[Det[σoTotalQuantum] /. E1 → 1 - E3 - E2]
```

$$\frac{4\,\text{E3} - 4\,\text{E3}^2 + \delta^2}{\delta^2}$$

E3 is any loss that is besides E1 and E2, and so is not included in the above Hilbert space. So that loss causes this reduced space to be more than minimum uncertainty state.

```
DetTotalQuantum /. E3 → 1
DetTotalQuantum /. E3 → 0
```

1

1

# Add classical Noise

Now we should check for the presence of classical noises, if the determinant of the combined state is a linear combination of (1, $\lambda$th, $\lambda$rin, $\lambda$pn)

```
σoTotal12 = Simplify[σo121 + σo122 + σo123 + σo12th,
   Assumptions → E1 > 0 && E2 > 0 && γ > 0] /. E3 → 1 - E2 - E1
```

$$\left\{\left\{\sqrt{E1\,E2}\;\lambda th,\;-\frac{\sqrt{E1\,E2}\;(2+2\,\lambda pn+2\,\lambda rin+\lambda th)}{\delta}\right\},\;\left\{-\frac{\sqrt{E1\,E2}\;(2+\lambda th)}{\delta},\;\frac{1}{\delta^2}\sqrt{E1\,E2}\;\left(4+(-2+4\,E1)\,\lambda pn+4\,E1\,\lambda rin+2\,\delta^2\,\lambda rin+\lambda th\right)\right\}\right\}$$

```
σoTotal =
  ArrayFlatten[FullSimplify[ {{σo1Total[E1, λth, λrin, λpn, δ], σoTotal12},
     {Transpose[σoTotal12], σo2Total[E2, λth, λrin, λpn, δ]}}]]
```

$$\Big\{\Big\{1+\lambda pn+\lambda rin+E1\,\lambda th,\;\frac{\lambda pn-\delta^2\,\lambda rin-E1\,(2\,(1+\lambda pn+\lambda rin)+\lambda th)}{\delta},\;\sqrt{E1\,E2}\;\lambda th,\;-\frac{\sqrt{E1\,E2}\;(2\,(1+\lambda pn+\lambda rin)+\lambda th)}{\delta}\Big\},$$
$$\Big\{\frac{\lambda pn-\delta^2\,\lambda rin-E1\,(2\,(1+\lambda pn+\lambda rin)+\lambda th)}{\delta},\;\frac{1}{\delta^2}\left(\delta^2+\lambda pn+\delta^4\,\lambda rin+4\,E1^2\,(\lambda pn+\lambda rin)+E1\,(4-4\,\lambda pn+4\,\delta^2\,\lambda rin+\lambda th)\right),$$
$$-\frac{\sqrt{E1\,E2}\;(2+\lambda th)}{\delta},\;\frac{1}{\delta^2}\sqrt{E1\,E2}\;\left(4+(-2+4\,E1)\,\lambda pn+4\,E1\,\lambda rin+2\,\delta^2\,\lambda rin+\lambda th\right)\Big\},$$
$$\Big\{\sqrt{E1\,E2}\;\lambda th,\;-\frac{\sqrt{E1\,E2}\;(2+\lambda th)}{\delta},\;1+E2\,\lambda th,\;-\frac{E2\,(2+\lambda th)}{\delta}\Big\},$$
$$\Big\{-\frac{\sqrt{E1\,E2}\;(2\,(1+\lambda pn+\lambda rin)+\lambda th)}{\delta},$$
$$\frac{1}{\delta^2}\sqrt{E1\,E2}\;\left(4+(-2+4\,E1)\,\lambda pn+4\,E1\,\lambda rin+2\,\delta^2\,\lambda rin+\lambda th\right),$$
$$-\frac{E2\,(2+\lambda th)}{\delta},\;1+\frac{E2\,(4+4\,E1\,(\lambda pn+\lambda rin)+\lambda th)}{\delta^2}\Big\}\Big\}$$

```
σoTotal // MatrixForm
```

$$\left(\begin{array}{ccc}
1+\lambda pn+\lambda rin+E1\,\lambda th & \frac{\lambda pn-\delta^2\,\lambda rin-E1\,(2\,(1+\lambda pn+\lambda rin)+\lambda th)}{\delta} & \sqrt{E1\,E2}\;\lambda t \\
\frac{\lambda pn-\delta^2\,\lambda rin-E1\,(2\,(1+\lambda pn+\lambda rin)+\lambda th)}{\delta} & \frac{\delta^2+\lambda pn+\delta^4\,\lambda rin+4\,E1^2\,(\lambda pn+\lambda rin)+E1\,(4-4\,\lambda pn+4\,\delta^2\,\lambda rin+\lambda th)}{\delta^2} & -\frac{\sqrt{E1\,E2}\;(2+\lambda t}{\delta} \\
\sqrt{E1\,E2}\;\lambda th & -\frac{\sqrt{E1\,E2}\;(2+\lambda th)}{\delta} & 1+E2\,\lambda th \\
-\frac{\sqrt{E1\,E2}\;(2\,(1+\lambda pn+\lambda rin)+\lambda th)}{\delta} & \frac{\sqrt{E1\,E2}\;(4+(-2+4\,E1)\,\lambda pn+4\,E1\,\lambda rin+2\,\delta^2\,\lambda rin+\lambda th)}{\delta^2} & -\frac{E2\,(2+\lambda th)}{\delta}
\end{array}\right.$$

```
DetTotal =
  FullSimplify[Det[σoTotal] /. {E1 → 1 - E3 - E2, λpn → pn δ^2/(1 + δ^2), λrin → rin/(1 + δ^2)}]
```

$$\frac{1}{\delta^2+\delta^4}\Big(-4\,\text{E3}^2\left(1+\text{rin}+(1+\text{pn})\,\delta^2\right)+4\,\text{E3}\left((1+\text{E2 pn})(1+\text{rin})+(1+\text{pn})(1+\text{E2 rin})\,\delta^2\right)-4\,\text{E3}^3\left(\text{rin}+\text{pn}\,\delta^2\right)\lambda\text{th}+4\,\text{E3}^2\left((2+\text{E2}(-1+\text{pn}))\,\text{rin}+(\text{pn}+\text{E2 pn}(-1+\text{rin})+\text{rin})\,\delta^2\right)\lambda\text{th}-\text{E3}\left(1+\delta^2\right)\left(1+\delta^2+\text{rin}\left(5-4\,\text{E2}\left(1+(-1+\text{E2})\,\text{pn}\right)+\delta^2\right)\right)\lambda\text{th}+(1+\text{rin})\left(1+\delta^2\right)\left(\lambda\text{th}+\text{E2 pn}\,\lambda\text{th}+\delta^2\left(1+\text{pn}+\lambda\text{th}+\text{E2 pn}\,\lambda\text{th}\right)\right)\Big)$$

```
DQ = Simplify[DetTotal /. {rin → 0, pn → 0, λth → 0}]
```

$$\frac{4\,\text{E3}-4\,\text{E3}^2+\delta^2}{\delta^2}$$

```
Simplify[Simplify[DetTotal /. {rin → 0, pn → 0}] - DQ] /. E3 → 0
Simplify[Simplify[Simplify[DetTotal /. {rin → 0, λth → 0}] - DQ] /. E3 → 0]
Simplify[Simplify[DetTotal /. {pn → 0, λth → 0}] - DQ] /. E3 → 0
```

$$\frac{\left(1+\delta^2\right)\lambda\text{th}}{\delta^2}$$

```
pn
```

```
rin
```

While in the total determinant, there are terms like λrin*λth, if we neglect those terms (they will be two small numbers multiplied), the contribution of each classical noise source gets added to the total uncertainty.

# Port 0 (intracavity)

## Transfer Functions and covariance matrices for each source

```
TFMios01 = {{TFM[[a, ain1]], TFM[[a, ain1d]]},
    {TFM[[ad, ain1]], TFM[[ad, ain1d]]}} // Simplify;
TFMioq01 = Ms2q.TFMios01.Inverse[Ms2q] /. {Ω → 0, γ1 → E1 γ} //
  Simplify(*quadrature picture*)
```

$$\left\{\{0, 0\}, \left\{\frac{\sqrt{2}\ E1}{\sqrt{E1\,\gamma}\ \delta}, 0\right\}\right\}$$

```
TFMios02 = {{TFM[[a, ain2]], TFM[[a, ain2d]]},
    {TFM[[ad, ain2]], TFM[[ad, ain2d]]}} // Simplify;
TFMioq02 = Simplify[Ms2q.TFMios02.Inverse[Ms2q] /. {Ω → 0, γ2 → E2 γ},
  Assumptions → 0 < E2 < 1 && γ > 0] (*quadrature picture*)
```

$$\left\{\{0, 0\}, \left\{\frac{\sqrt{2}\ E2}{\sqrt{E2\,\gamma}\ \delta}, 0\right\}\right\}$$

```
TFMios03 = {{TFM[[a, ain3]], TFM[[a, ain3d]]},
    {TFM[[ad, ain3]], TFM[[ad, ain3d]]}} // Simplify;
TFMioq03 = Simplify[Ms2q.TFMios03.Inverse[Ms2q] /. Ω → 0 /. γ3 → E3 γ,
  Assumptions → E3 > 0 && γ > 0] (*quadrature picture*)
```

$$\left\{\{0, 0\}, \left\{\frac{\sqrt{2}\ E3}{\sqrt{E3\,\gamma}\ \delta}, 0\right\}\right\}$$

```
TFvs0th = {{TFM[[a, Fth]]}, {TFM[[ad, Fth]]}} (*response to thermal noise*);
TFvq0th = Ms2q.TFvs0th /. Ω → 0 (*qudrature picture*)
```

$$\left\{\left\{\frac{-i\,\gamma + \gamma\,\delta}{2\sqrt{2}\ g\,\gamma\,\delta\,\hbar} + \frac{i\,g\,\gamma + g\,\gamma\,\delta}{2\sqrt{2}\ g^2\,\gamma\,\delta\,\hbar}\right\}, \left\{-\frac{i\left(-i\,\gamma + \gamma\,\delta\right)}{2\sqrt{2}\ g\,\gamma\,\delta\,\hbar} + \frac{i\left(i\,g\,\gamma + g\,\gamma\,\delta\right)}{2\sqrt{2}\ g^2\,\gamma\,\delta\,\hbar}\right\}\right\}$$

```
σ01 = TFMioq01.σin1AtCavity.Transpose[TFMioq01] /.
    {rin → λrin (1 + δ^2), pn → λpn (1 + δ^2)/δ^2} // Simplify
```

$$\left\{\{0, 0\}, \left\{0, \frac{2\,E1\,(1 + \lambda pn + \lambda rin)}{\gamma\,\delta^2}\right\}\right\}$$

```
σ02 = TFMioq02.Transpose[TFMioq02] // Simplify
```

$$\left\{\{0, 0\}, \left\{0, \frac{2\,E2}{\gamma\,\delta^2}\right\}\right\}$$

```
σ03 = TFMioq03.Transpose[TFMioq03] // Simplify
```

$$\left\{\{0, 0\}, \left\{0, \frac{2\,E3}{\gamma\,\delta^2}\right\}\right\}$$

```
σ0th = TFvq0th.Transpose[TFvq0th] (λth ħ^2 g^2)/γ // Simplify(* add thermal noise*)
```

$$\left\{\left\{\frac{\lambda th}{2\,\gamma}, -\frac{\lambda th}{2\,\gamma\,\delta}\right\}, \left\{-\frac{\lambda th}{2\,\gamma\,\delta}, \frac{\lambda th}{2\,\gamma\,\delta^2}\right\}\right\}$$

# Squeezing

## Covariance matrices

### Full Matrix

```
σtot0[γ_, λth_, λrin_, λpn_, δ_] =
  FullSimplify[(σ03 + σ02 + σ01 + σ0th) /. E3 → 1 - E1 - E2]
```

$$\left\{\left\{\frac{\lambda th}{2\,\gamma}, -\frac{\lambda th}{2\,\gamma\,\delta}\right\}, \left\{-\frac{\lambda th}{2\,\gamma\,\delta}, \frac{4 + 4\,E1\,(\lambda pn + \lambda rin) + \lambda th}{2\,\gamma\,\delta^2}\right\}\right\}$$

## No Classical Noise, not full escape efficiency

```
σtot0quantum[γ_, δ_] = σtot0[γ, λth, λrin, λpn, δ] /. {λth → 0, λrin → 0, λpn → 0}
  (*no classical noise*) // Simplify
```

$$\left\{\{0, 0\}, \left\{0, \frac{2}{\gamma\,\delta^2}\right\}\right\}$$

## Phase Noise

```
σtot0PN[γ_, λpn_, δ_] =
  σtot0[γ, λth, λrin, λpn, δ] /. {λth → 0, λrin → 0} (*RIN*) // Simplify
```

$$\left\{\{0, 0\}, \left\{0, \frac{2 + 2\,\mathrm{E1}\,\lambda pn}{\gamma\,\delta^2}\right\}\right\}$$

```
σtot0PN[γ, λpn, δ] - σtot0quantum[γ, δ] // Simplify
```

$$\left\{\{0, 0\}, \left\{0, \frac{2\,\mathrm{E1}\,\lambda pn}{\gamma\,\delta^2}\right\}\right\}$$

## RIN

```
σtot0RIN[γ_, λrin_, δ_] =
  σtot0[γ, λth, λrin, λpn, δ] /. {λth → 0, λpn → 0} (*RIN*) // Simplify
```

$$\left\{\{0, 0\}, \left\{0, \frac{2 + 2\,\mathrm{E1}\,\lambda rin}{\gamma\,\delta^2}\right\}\right\}$$

```
σtot0RIN[γ, λrin, δ] - σtot0quantum[γ, δ] // Simplify(*perturbation matrix*)
```

$$\left\{\{0, 0\}, \left\{0, \frac{2\,\mathrm{E1}\,\lambda rin}{\gamma\,\delta^2}\right\}\right\}$$

## Thermal Noise

```
σtot0Th[γ_, λth_, δ_] =
  σtot0[γ, λth, λrin, λpn, δ] /. {λrin → 0, λpn → 0}(*Thermal noise*) // Simplify
```

$$\left\{\left\{\frac{\lambda th}{2\,\gamma}, -\frac{\lambda th}{2\,\gamma\,\delta}\right\}, \left\{-\frac{\lambda th}{2\,\gamma\,\delta}, \frac{4 + \lambda th}{2\,\gamma\,\delta^2}\right\}\right\}$$

```
σtot0Th[γ, λSth, δ] - σtot0quantum[γ, δ] // Simplify (*perturbation matrix*)
```

$$\left\{\left\{\frac{\lambda Sth}{2\gamma}, -\frac{\lambda Sth}{2\gamma\delta}\right\}, \left\{-\frac{\lambda Sth}{2\gamma\delta}, \frac{\lambda Sth}{2\gamma\delta^2}\right\}\right\}$$

## Eigenvalues and eigenvectors

```
Simplify[σtot0quantum[γ, Tan[2 ξ]]] // MatrixForm
```

$$\begin{pmatrix} 0 & 0 \\ 0 & \frac{2\,\mathrm{Cot}[2\,\xi]^2}{\gamma} \end{pmatrix}$$

The intracavity state seems to violate the uncertainty principle here! Maybe it has something to do with the formalism assuming $\gamma \to \infty$?

```
EigenSysq0 = Simplify[FullSimplify[Eigensystem[σtot0quantum[γ, Tan[2 ξ]]]]]
```

$$\left\{\left\{\frac{2\,\mathrm{Cot}[2\,\xi]^2}{\gamma}, 0\right\}, \{\{0, 1\}, \{1, 0\}\}\right\}$$

```
Eq0 = EigenSysq0[[1]]
Vq0 = Simplify[ FullSimplify[EigenSysq0[[2]]]]
Eq0AS = Eq0[[1]];
Eq0S = Eq0[[2]];
Vq0AS = Vq0[[1]];
Vq0S = Vq0[[2]];
```

$$\left\{\frac{2\,\mathrm{Cot}[2\,\xi]^2}{\gamma}, 0\right\}$$

```
{{0, 1}, {1, 0}}
```

## Diagonalized Perturbation Theory

Here we approach the same problem using perturbation theory instead of Taylor series.
In addition to using the detuning as Tan[2$\xi$], we also look at these matrices in the basis rotated by $\xi$ (because that should give us a diagonalized basis, and we have squeezing and antisqueezing there.)
We then apply perturbation theory in this new diagonalized basis - which is notated as p, standing for prime.

```
σp0Quantum[γ_, ξ_] = Simplify[FullSimplify[
   RotationMatrix[-ξ].σtot0quantum[γ, Tan[2 ξ]].RotationMatrix[ξ]]]
```

$$\left\{\left\{\frac{2\,\text{Cot}[2\,\xi]^2\,\text{Sin}[\xi]^2}{\gamma},\ \frac{\text{Cos}[2\,\xi]\,\text{Cot}[2\,\xi]}{\gamma}\right\},\ \left\{\frac{\text{Cos}[2\,\xi]\,\text{Cot}[2\,\xi]}{\gamma},\ \frac{2\,\text{Cos}[\xi]^2\,\text{Cot}[2\,\xi]^2}{\gamma}\right\}\right\}$$

(p stands for prime, ie in the rotated basis)

q is shorthand for quantum

```
Vqp0S = {1, 0};
Vqp0AS = {0, 1};
Eqp0S  = Vqp0S.σp0Quantum[γ, ξ].Vqp0S
Eqp0AS  = Vqp0AS.σp0Quantum[γ, ξ].Vqp0AS
```

$$\frac{2\,\text{Cot}[2\,\xi]^2\,\text{Sin}[\xi]^2}{\gamma}$$

$$\frac{2\,\text{Cos}[\xi]^2\,\text{Cot}[2\,\xi]^2}{\gamma}$$

- Thermal Noise

```
δσp0Th[γ_, λth_, ξ_] = FullSimplify[RotationMatrix[-ξ].
   (σtot0[γ, λth, 0, 0, Tan[2 ξ]] - σtot0quantum[γ, Tan[2 ξ]]).RotationMatrix[ξ]]
```

$$\left\{\left\{\frac{\lambda\text{th}\,\text{Sec}[\xi]^2}{8\,\gamma},\ -\frac{\lambda\text{th}\,\text{Csc}[\xi]\,\text{Sec}[\xi]}{8\,\gamma}\right\},\ \left\{-\frac{\lambda\text{th}\,\text{Csc}[\xi]\,\text{Sec}[\xi]}{8\,\gamma},\ \frac{\lambda\text{th}\,\text{Csc}[\xi]^2}{8\,\gamma}\right\}\right\}$$

```
δEp0thS = Vqp0S.δσp0Th[γ, λth, ξ].Vqp0S
δEp0thAS = Vqp0AS.δσp0Th[γ, λth, ξ].Vqp0AS
```

$$\frac{\lambda\text{th}\,\text{Sec}[\xi]^2}{8\,\gamma}$$

$$\frac{\lambda\text{th}\,\text{Csc}[\xi]^2}{8\,\gamma}$$

```
δVp0thS =
  Simplify[(Vqp0AS.δσp0Th[γ, λth, ξ].Vqp0S)/(Eqp0S - Eqp0AS) Vqp0AS /. λth → (8 δθ^C_th)/(Tan[2 ξ] Sec[2 ξ]^2)]
δVp0thAS = Simplify[
  (Vqp0S.δσp0Th[γ, λth, ξ].Vqp0AS)/(Eqp0AS - Eqp0S) Vqp0S /. λth → 8 (δθ^C_th)/(Tan[2 ξ] Sec[2 ξ]^2)]
```

$\{0, \delta\theta^{C}{}_{th}\}$

$\{-\delta\theta^{C}{}_{th}, 0\}$

Rotation caused by thermal noise to the state inside the cavity is different from the one in transmission!

```
FullSimplify[(δσp0Th[γ, λth, ξ] γ 8)/λth] // MatrixForm
```

$$\begin{pmatrix} \mathrm{Sec}[\xi]^2 & -\mathrm{Csc}[\xi]\,\mathrm{Sec}[\xi] \\ -\mathrm{Csc}[\xi]\,\mathrm{Sec}[\xi] & \mathrm{Csc}[\xi]^2 \end{pmatrix}$$

- Classical Laser Noise

```
FullSimplify[RotationMatrix[0].
  (σtot0[γ, 0, λrin, λpn, Tan[2 ξ]] - σtot0quantum[γ, Tan[2 ξ]]).
  RotationMatrix[0]]
```

$$\left\{\{0, 0\}, \left\{0, \frac{2\,\mathrm{E1}\,(\lambda\mathrm{pn} + \lambda\mathrm{rin})\,\mathrm{Cot}[2\,\xi]^2}{\gamma}\right\}\right\}$$

```
δσp0LCN[γ_, E1_, λcn_, ξ_] = FullSimplify[
  FullSimplify[RotationMatrix[-ξ].(σtot0[γ, 0, λrin, λcn - λrin, Tan[2 ξ]] -
      σtot0quantum[γ, Tan[2 ξ]]).RotationMatrix[ξ]]]
```

$$\left\{\left\{\frac{2\,\mathrm{E1}\,\lambda\mathrm{cn}\,\mathrm{Cot}[2\,\xi]^2\,\mathrm{Sin}[\xi]^2}{\gamma}, \frac{\mathrm{E1}\,\lambda\mathrm{cn}\,\mathrm{Cos}[2\,\xi]\,\mathrm{Cot}[2\,\xi]}{\gamma}\right\}, \left\{\frac{\mathrm{E1}\,\lambda\mathrm{cn}\,\mathrm{Cos}[2\,\xi]\,\mathrm{Cot}[2\,\xi]}{\gamma}, \frac{2\,\mathrm{E1}\,\lambda\mathrm{cn}\,\mathrm{Cos}[\xi]^2\,\mathrm{Cot}[2\,\xi]^2}{\gamma}\right\}\right\}$$

Only the 11 element of $\sigma$in1AtCavity shows up. So we can simplify the intensity and phase noise into one quantity, called $\lambda$cn

```
Simplify[(δσp0LCN[γ, E1, λcn, ξ] γ)/(2 λcn E1 Cot[2 ξ]^2)]
```

$$\{\{\mathrm{Sin}[\xi]^2, \mathrm{Cos}[\xi]\,\mathrm{Sin}[\xi]\}, \{\mathrm{Cos}[\xi]\,\mathrm{Sin}[\xi], \mathrm{Cos}[\xi]^2\}\}$$

```
δEp0lcnS = Vqp0S.δσp0LCN[γ, E1, λcn, ξ].Vqp0S
δEp0lcnAS = Vqp0AS.δσp0LCN[γ, E1, λcn, ξ].Vqp0AS
```

$$\frac{2\,E1\,\lambda cn\,Cot[2\,\xi]^2\,Sin[\xi]^2}{\gamma}$$

$$\frac{2\,E1\,\lambda cn\,Cos[\xi]^2\,Cot[2\,\xi]^2}{\gamma}$$

```
δVp0LCNS = Simplify[
   (Vqp0AS.δσp0LCN[γ, E1, λcn, ξ].Vqp0S)/(Eqp0S - Eqp0AS) Vqp0AS /. λcn → (-(E1 Tan[2 ξ])/2)^-1 δθ^C_cn]
δVp0LCNAS = Simplify[(Vqp0S.δσp0LCN[γ, E1, λcn, ξ].Vqp0AS)/(Eqp0AS - Eqp0S) Vqp0S /.
    λcn → (-(E1 Tan[2 ξ])/2)^-1 δθ^C_cn]
```

$\{0, \delta\theta^{C}{}_{cn}\}$

$\{-\delta\theta^{C}{}_{cn}, 0\}$

# Covariance (squeezing) matrix visualization

## Function to Plot

```
plotcontour[mat_, sty_] := Module[{l1, l2, v1, v2},
   {{l1, l2}, {v1, v2}} = Eigensystem[mat];
   v1 = v1 / Sqrt[v1.v1];
   v2 = v2 / Sqrt[v2.v2];
   ParametricPlot[r (Sqrt[l1] v1 Cos[ϕ] + Sqrt[l2] v2 Sin[ϕ]), {ϕ, 0, 2 Pi},
    {r, 0, 1}, sty, Frame → True, Mesh → False, ImageSize → 100]
  ];
```

```
MVtest = {{√2, 1}, {1, √2}};
```

```
NMsqzvac := plotcontour[IdentityMatrix[2], {PlotStyle → Directive[Yellow],
    BoundaryStyle → Directive[Orange, Thickness[.015]]}];
```

```
NMsqztest = plotcontour[MVtest,
  PlotStyle -> Directive[Thickness[0.005], Red, Opacity[.50]]]
```

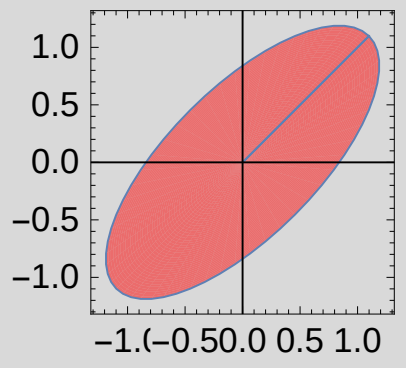

```
SetDirectory[NotebookDirectory[]]
SetDirectory["images"];
```

```
C:\Users\Nancy\Dropbox
  (MIT)\Aggarwal\micron\OptPaper\OptimizationPaperSharedFilesWithHaixing
```

## Plotting Transmission

Plot Quantum Only in Transmission

```
NMsqzq2[E2_, δ_] :=
  plotcontour[σo2Quantum[E2, δ], {PlotStyle → Directive[ Yellow, Opacity[0.1]],
    BoundaryStyle → Directive[Thickness[0.02], Yellow]}];
 (*,{PlotStyle→ Directive[ Lighter[Red],Opacity[0.3]],
   BoundaryStyle→Directive[Thickness[0.015], Lighter[Red]]}];*)
```

Plot Thermal Noise in Transmission

```
NMsqzth2[E2_, λth_, δ_] :=
  plotcontour[σo2Th[E2, λth, δ], {PlotStyle → Directive[Green, Opacity[0.2]],
    BoundaryStyle → Directive[Thickness[0.02], Green]}];
 (*,{PlotStyle→ Directive[ Lighter[Red]],
   BoundaryStyle→ Directive[Thickness[0.015],Lighter[Red]]}];*)
```

Plot RIN only in Transmission

```
NMsqzLCN2[Esc1_, E2_, λcn_, δ_] := plotcontour[
   σo2LCN[E2, λcn, δ] /. E1 → Esc1, {PlotStyle → Directive[Lighter[Cyan]],
    BoundaryStyle → Directive[Thickness[0.02], Lighter[Cyan]]}];
 (*,{PlotStyle→ Directive[Orange],BoundaryStyle→
    Directive[Thickness[0.015], Orange]}];*)
```

Plot Transmission with both RIN and Thermal noise

```
NMsqztot2[Esc1_, E2_, λth_, λcn_, δ_] :=
  plotcontour[σo2Total[E2, λth, λrin, λcn - λrin, δ] /. E1 → Esc1,
   {PlotStyle → Directive[Blend[{Blue, Green}, 0.5]],
    BoundaryStyle → Directive[Thickness[0.02], Blend[{Blue, Green}, 0.5]]}];
 (*,{PlotStyle→ Directive[ Orange],BoundaryStyle→
    Directive[Thickness[0.015],Orange]}];*)
```

```
NMsqzvac :=
  plotcontour[IdentityMatrix[2], {PlotStyle → Directive[Lighter[Yellow]],
    BoundaryStyle → Directive[Lighter[Orange, 0.7], Thickness[.02]]}];
```

Combine all

```
TransmissionSqz[E1_, E2_, λth_, λrin2_, δ_, ImgSize_] :=
  Show[NMsqzvac, NMsqztot2[E1, E2, λth, λrin2, δ], NMsqzLCN2[E1, E2, λrin2, δ],
   NMsqzth2[E2, λth, δ], NMsqzq2[E2, δ], PlotRange → All, ImageSize → ImgSize];
TransmissionSqz[.1, .75, 1, 1, .5, 100]
Export["Transmission_th_1_rin_1_d_0.5.pdf",
  TransmissionSqz[.1, .75, 1, 1, .5, 150], Background → None] // Timing
Export["Transmission_th_.5_rin_.5_d_0.5.pdf",
  TransmissionSqz[.1, .75, .5, .5, .5, 150], Background → None] // Timing
Export["Transmission_th_0.1_rin_0.1_d_0.5.pdf",
  TransmissionSqz[.1, .75, .1, .1, .5, 150], Background → None] // Timing
```

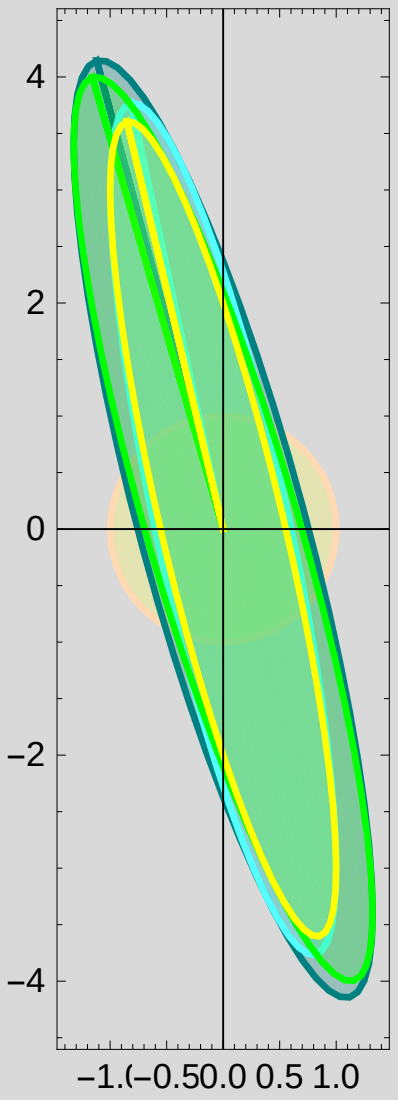

```
{0.21875, Transmission_th_1_rin_1_d_0.5.pdf}
```

```
{0.375, Transmission_th_.5_rin_.5_d_0.5.pdf}
```

```
{0.375, Transmission_th_0.1_rin_0.1_d_0.5.pdf}
```

```
TransmissionSqzTh[E2_, λth_, δ_, ImgSize_] := Show[NMsqzvac,
   NMsqzth2[E2, λth, δ], NMsqzq2[E2, δ], PlotRange → All, ImageSize → ImgSize];
TransmissionSqzTh[.75, .5, .5, 100]
Export["Transmission_th_.5_d_0.5.pdf",
  TransmissionSqzTh[.75, .5, .5, 100], Background → None] // Timing
Export["Transmission_th_1_d_0.5.pdf", TransmissionSqzTh[.75, 1, .5, 100],
  Background → None] // Timing
Export["Transmission_th_0.1_d_0.5.pdf",
  TransmissionSqzTh[.75, .1, .5, 100], Background → None] // Timing
```

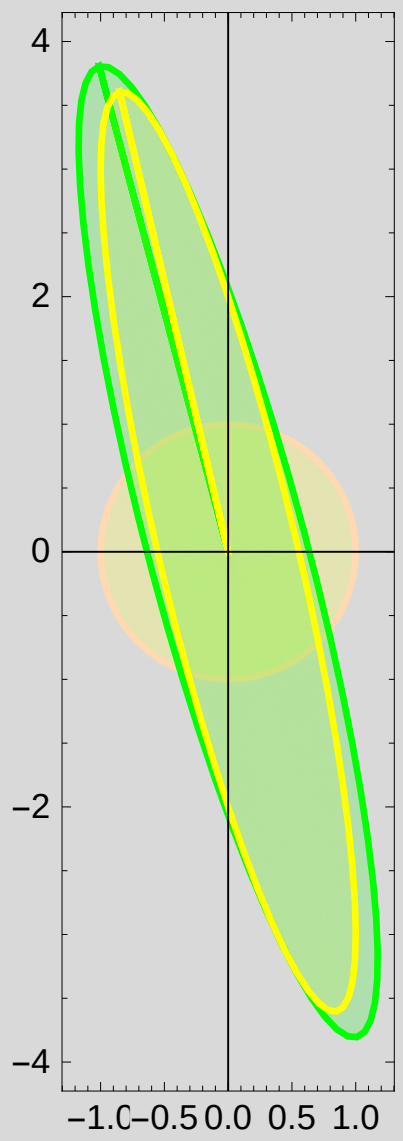

```
{0.234375, Transmission_th_.5_d_0.5.pdf}
```

```
{0.203125, Transmission_th_1_d_0.5.pdf}
```

```
{0.21875, Transmission_th_0.1_d_0.5.pdf}
```

```
TransmissionSqzRIN[E1_, E2_, λrin2_, δ_, ImgSize_] :=
  Show[NMsqzvac, NMsqzLCN2[E1, E2, λrin2, δ],
   NMsqzq2[E2, δ], PlotRange → All, ImageSize → ImgSize];
TransmissionSqzRIN[.1, .75, .5, .5, 100]
Export["Transmission_rin_0.5_d_0.5.pdf",
  TransmissionSqzRIN[.1, .75, .5, .5, 100], Background → None] // Timing
Export["Transmission_rin_1_d_0.5.pdf",
  TransmissionSqzRIN[.1, .75, 1, .5, 100], Background → None] // Timing
Export["Transmission_rin_0.1_d_0.5.pdf",
  TransmissionSqzRIN[.1, .75, .1, .5, 100], Background → None] // Timing
```

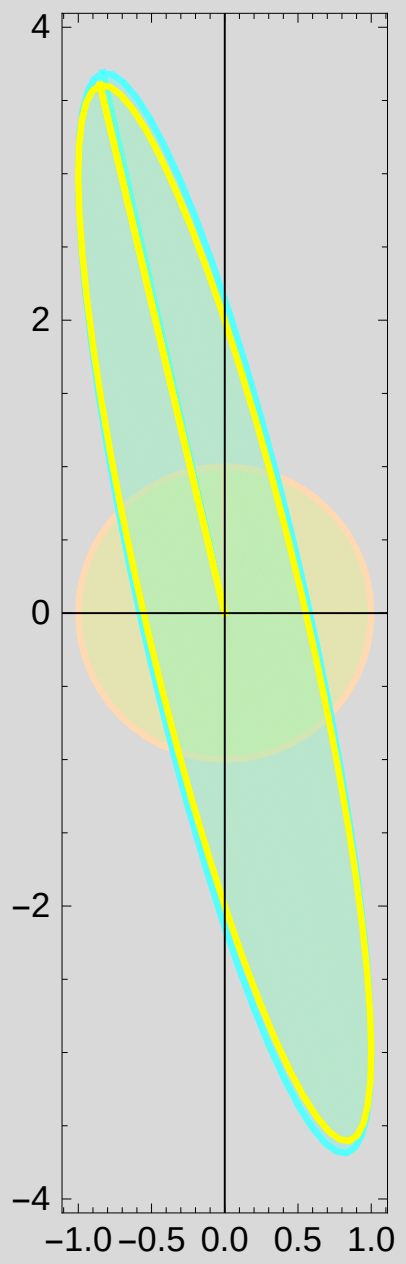

```
{0.171875, Transmission_rin_0.5_d_0.5.pdf}
```

```
{0.1875, Transmission_rin_1_d_0.5.pdf}
```

```
{0.140625, Transmission_rin_0.1_d_0.5.pdf}
```

## Plotting Reflection

Plot Quantum Only in Reflection

```
NMsqzq1[E1_, δ_] :=
  plotcontour[σo1Quantum[E1, δ], {PlotStyle → Directive[ Yellow, Opacity[0.1]],
    BoundaryStyle → Directive[Thickness[0.02], Yellow]}];
 (*,{PlotStyle→ Directive[ Lighter[Red],Opacity[0.3]],
   BoundaryStyle→Directive[Thickness[0.015], Lighter[Red]]}];*)
```

Plot Thermal Noise in Reflection

```
NMsqzth1[E1_, λth_, δ_] :=
  plotcontour[σo1Th[E1, λth, δ], {PlotStyle → Directive[Green, Opacity[0.2]],
    BoundaryStyle → Directive[Thickness[0.02], Green]}];
 (*,{PlotStyle→ Directive[ Lighter[Red]],
   BoundaryStyle→ Directive[Thickness[0.015],Lighter[Red]]}];*)
```

Plot RIN only in Reflection

```
NMsqzrin1[E1_, λrin_, δ_] :=
  plotcontour[σo1rin[E1, λrin, δ], {PlotStyle → Directive[Lighter[Cyan]],
    BoundaryStyle → Directive[Thickness[0.02], Lighter[Cyan]]}];
 (*,{PlotStyle→ Directive[Orange],BoundaryStyle→
    Directive[Thickness[0.015], Orange]}];*)
```

Plot Phase Noise only in Reflection

```
NMsqzpn1[E1_, λpn_, δ_] :=
  plotcontour[σo1pn[E1, λpn, δ], {PlotStyle → Directive[Lighter[Blue, 0.5]],
    BoundaryStyle → Directive[Thickness[0.02], Lighter[Blue, 0.5]]}];
 (*,{PlotStyle→ Directive[Orange],BoundaryStyle→
    Directive[Thickness[0.015], Orange]}];*)
```

Plot total noise in reflection

```
NMsqztot1[E1_, λth_, λrin_, λpn_, δ_] :=
  plotcontour[σo1Total[E1, λth, λrin, λpn, δ],
   {PlotStyle → Directive[Blend[{Blue, Green}, 0.5]],
    BoundaryStyle → Directive[Thickness[0.02], Blend[{Blue, Green}, 0.5]]}];
 (*,{PlotStyle→ Directive[ Orange],BoundaryStyle→
    Directive[Thickness[0.015],Orange]}];*)
```

```
NMsqzvac :=
  plotcontour[IdentityMatrix[2], {PlotStyle → Directive[Lighter[Yellow]],
    BoundaryStyle → Directive[Lighter[Orange, 0.7], Thickness[.02]]}];
```

Combine all

```
ReflectionSqz[E1_, λth_, λrin_, λpn_, δ_, ImgSize_] :=
  Show[NMsqzvac, NMsqztot1[E1, λth, λrin, λpn, δ],
   NMsqzrin1[E1, λrin, δ], NMsqzpn1[E1, λpn, δ], NMsqzth1[E1, λth, δ],
   NMsqzq1[E1, δ], PlotRange → All, ImageSize → ImgSize];
```

```
ReflectionSqz[.75, 1, 1, 1, .5, 100]
```

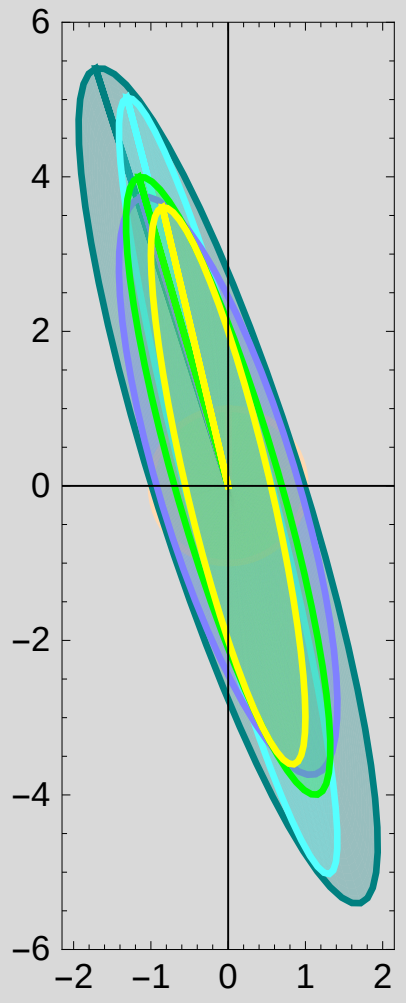

```
Export["Reflection_th_1_rin_1_pn_1_d_0.5.pdf",
  ReflectionSqz[.75, 1, 1, 1, .5, 150], Background → None] // Timing
Export["Reflection_th_.5_rin_.5_pn_.5_d_0.5.pdf",
  ReflectionSqz[.75, .5, .5, .5, .5, 150], Background → None] // Timing
Export["Reflection_th_0.1_rin_0.1_pn_0.1_d_0.5.pdf",
  ReflectionSqz[.75, .1, .1, .1, .5, 150], Background → None] // Timing
```

```
{0.328125, Reflection_th_1_rin_1_pn_1_d_0.5.pdf}
```

```
{0.40625, Reflection_th_.5_rin_.5_pn_.5_d_0.5.pdf}
```

```
{0.28125, Reflection_th_0.1_rin_0.1_pn_0.1_d_0.5.pdf}
```

```
ReflectionSqzTh[E1_, λth_, δ_, ImgSize_] := Show[NMsqzvac, NMsqzth1[E1, λth, δ],
   NMsqzq1[E1, δ], PlotRange → All, ImageSize → ImgSize];
ReflectionSqzTh[.75, .5, .5, 100]
Export["Reflection_th_.5_d_0.5.pdf",
  ReflectionSqzTh[.75, .5, .5, 100], Background → None] // Timing
Export["Reflection_th_1_d_0.5.pdf", ReflectionSqzTh[.75, 1, .5, 100],
  Background → None] // Timing
Export["Reflection_th_0.1_d_0.5.pdf",
  ReflectionSqzTh[.75, .1, .5, 100], Background → None] // Timing
```

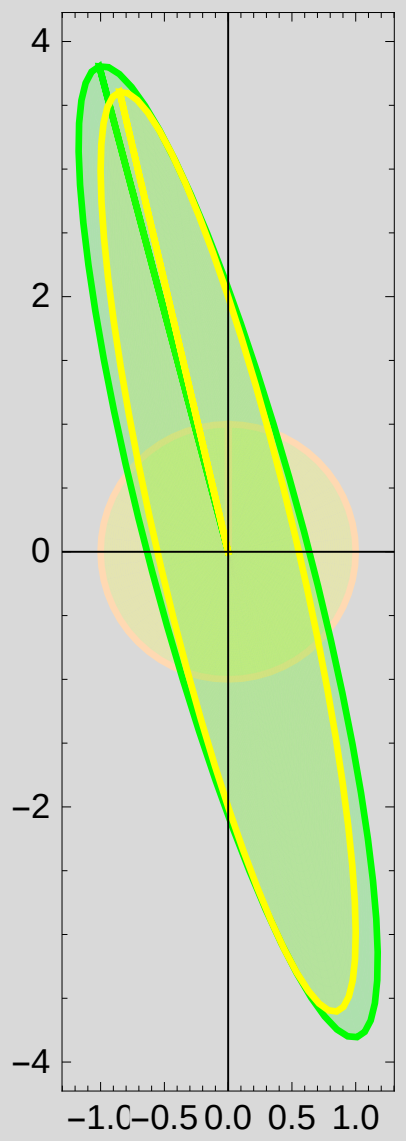

```
{0.15625, Reflection_th_.5_d_0.5.pdf}
```

```
{0.125, Reflection_th_1_d_0.5.pdf}
```

```
{0.21875, Reflection_th_0.1_d_0.5.pdf}
```

```
ReflectionSqzRIN[E1_, λrin_, δ_, ImgSize_] :=
  Show[NMsqzvac, NMsqzrin1[E1, λrin, δ],
   NMsqzq1[E1, δ], PlotRange → All, ImageSize → ImgSize];
ReflectionSqzRIN[.75, .5, .5, 100]
Export["Reflection_rin_0.5_d_0.5.pdf",
  ReflectionSqzRIN[.75, .5, .5, 100], Background → None] // Timing
Export["Reflection_rin_0.1_d_0.5.pdf",
  ReflectionSqzRIN[.75, .1, .5, 100], Background → None] // Timing
Export["Reflection_rin_1_d_0.5.pdf", ReflectionSqzRIN[.75, 1, .5, 100],
  Background → None] // Timing
```

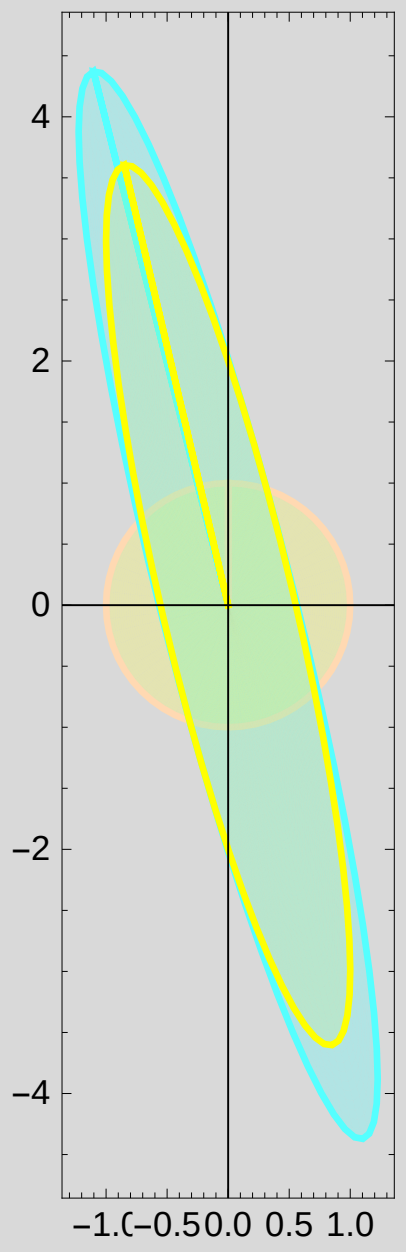

```
{0.1875, Reflection_rin_0.5_d_0.5.pdf}
```

```
{0.171875, Reflection_rin_0.1_d_0.5.pdf}
```

```
{0.203125, Reflection_rin_1_d_0.5.pdf}
```

```
ReflectionSqzPN[E1_, λpn_, δ_, ImgSize_] := Show[NMsqzvac, NMsqzpn1[E1, λpn, δ],
   NMsqzq1[E1, δ], PlotRange → All, ImageSize → ImgSize];
ReflectionSqzPN[.75, .5, .5, 100]
Export["Reflection_pn_0.5_d_0.5.pdf",
  ReflectionSqzPN[.75, .5, .5, 100], Background → None] // Timing
Export["Reflection_pn_0.1_d_0.5.pdf", ReflectionSqzPN[.75, .1, .5, 100],
  Background → None] // Timing
Export["Reflection_pn_1_d_0.5.pdf", ReflectionSqzPN[.75, 1, .5, 100],
  Background → None] // Timing
```

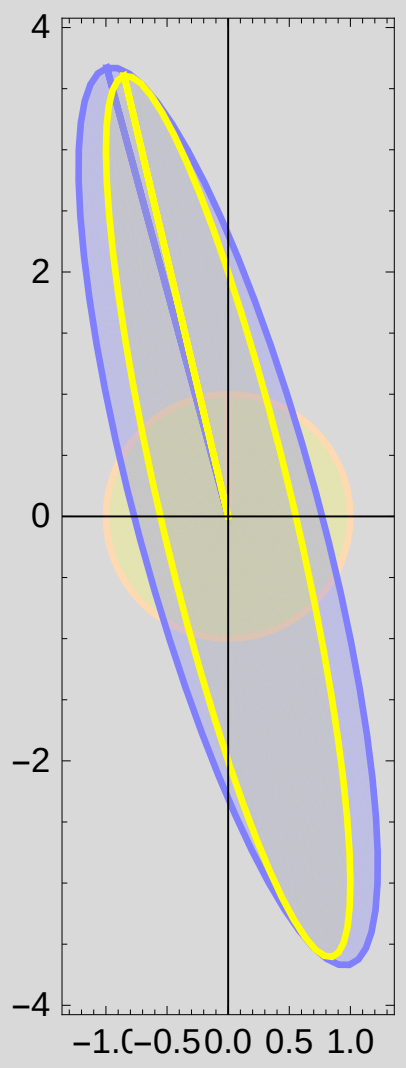

```
{0.125, Reflection_pn_0.5_d_0.5.pdf}
```

```
{0.203125, Reflection_pn_0.1_d_0.5.pdf}
```

```
{0.171875, Reflection_pn_1_d_0.5.pdf}
```